\documentclass[11pt,twoside]{book}
\usepackage{amstex,epsfig,latexsym,fancyheadings,amssymb,pifont,shadow}
\renewcommand{\chaptermark}[1]%
               {\markboth{#1}{#1}}
\renewcommand{\sectionmark}[1]%
               {\markright{\thesection\ #1}}
\newcommand{\beeq}{\begin{equation}}
\newcommand{\bea}{\begin{eqnarray}}
\newcommand{\ena}{\end{eqnarray}}
\newcommand{\no}{\noindent}

\newcommand{\T}{{\cal T}_k}
\newcommand{\op}{\oplus}
\newcommand{\nb}{\nonumber}
\newcommand{\Ra}{\Rightarrow}
\newcommand{\ra}{\rightarrow}
\newcommand{\path}{}
\newcommand{\bad}{\bar{\partial}}
\newcommand{\di}{\partial}

\begin{document}



\begin{titlepage}

\begin{picture}(250,300)
\put(-15,-340){\psfig{figure= \path 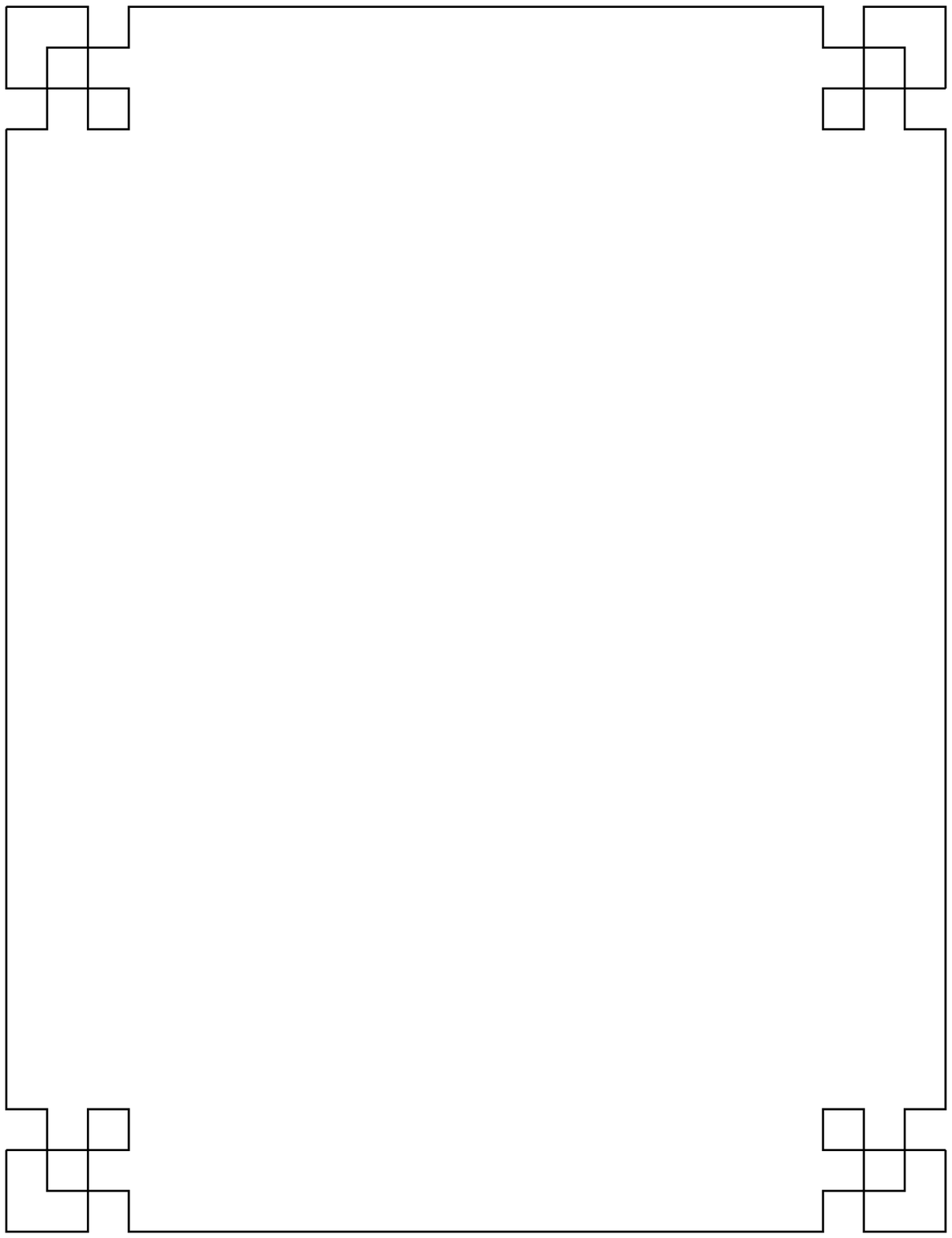,width=16cm,height=22.5cm}}
\end{picture}

\vspace{-9cm}

\begin{center}
{\LARGE \bf Luigi Pilo} \\[0.5cm]

\end{center}

\vskip 2truecm

\begin{center} 
\LARGE \bf
Chern-Simons Field Theory 

and

Invariants of 3-Manifolds
\end{center}

\vskip 3truecm
\begin{center}
{\large \rm  TESI DI PERFEZIONAMENTO } \\[3cm]
\end{center}

\vspace{4cm}

\begin{center}
{\Large \bf SCUOLA NORMALE SUPERIORE }

{\large    PISA - 1997 } \\[0.5cm]

\end{center}

\clearpage

\end{titlepage}


\mainmatter

\tableofcontents
\include{cit}
 \centerline{\Large \bf Introduction}
In the last decade the research field under the name of topological quantum field
theory (TQT) has undergone an impressive growth, since the early works by 
Schwarz 
\cite{swa} on the Abelian Chern-Simons theory and Witten \cite{wittf,wittd,witt1} 
a huge number of models has appeared in the literature (see \cite{blau,dij} for an 
extensive list of references). The adjective topological, deserved by all these 
models, stems from the goal they have been designed for: the representation 
and possibly the computation of topological and/or differential invariants of
a suitable space by using quantum field theory.  
For instance, the correlation functions of 
gauge invariant observables in the Chern-Simons theory studied in this thesis, 
represent topological
invariants of knots, and the modified partition function gives
a topological invariant for the three-manifold $\cal M$ in which the theory is
defined. The suggestion that quantum field theory might be useful to understand
recent results on four-dimensional manifolds obtained by Donaldson \cite{dona}
and on knot theory \cite{jon} obtained by Jones, was put 
forward by Atiyah and Schwarz \cite{atw,sch1} as a challenge for the 
theoretical physics community.
Witten in \cite{wittd,witt1} showed that those suggestions were viable and he 
produced two paradigmatic examples of topological quantum field theory.

We shall focus our attention on the topological quantum field theory introduced
by Witten in \cite{witt1}, namely the three-dimensional Chern-Simons field 
theory. 
Differently from Witten's original approach, we shall adopt a fully 
three-dimensional point of view; no result from two-dimensional 
conformal field 
theory  will be used. It turns out that the basic principles of quantum field 
theory, together with symmetry considerations are stringent enough to solve  
Chern-Simons theory (CS) in any closed connected and orientable three-manifold.
It is worth to stress that the solution presented here can be provided with a 
completely rigorous formulation. The mathematical foundation of our work 
relies 
on Atiyah's axiomatic approach to topological quantum field theory \cite{ata} and
in particular on the realization of these axioms studied  by Turaev \cite{tur}. 
Indeed, as a matter of fact, the link invariants and the three-manifold invariants
defined in terms of CS correlation functions coincide with the ``quantum 
invariants'' considered in \cite{tur}.

From the point of view of a theoretical physicist the CS model   
is just a ``strange''  gauge theory with a huge number of symmetries 
which constrain the dynamics. Actually, the constraints are so strong that there
is no room for standard ``physical'' excitations and, as a consequence, 
there is no dynamics at all. Moreover, the symmetries  plus some ``numerical'' 
input from perturbation theory allow to solve the theory in $\mathbb{R}^3$ and
in $S^3$. Remarkably, the same happens in any closed, connected and orientable
three-manifold $\cal M$.
The crucial fact is that one can represent the effect of the non-trivial 
topology of $\cal M$ on the expectation value $<W>_{\cal M}$ of an observable 
$W$ as a symmetry transformation on the corresponding expectation value in 
$S^3$. Producing the solution of the theory in $\cal M$ 
amounts to finding the operator $W({\cal L})$ whose insertion $<W({\cal L}) \, 
W>_{S^3}$ gives the correlation function
$<W>_{\cal M}$ in ${\cal M}$. It turns out that this leads to an implementation 
of Dehn's surgery at the level of CS gauge invariant correlation functions in $S^3$
which can be used to solve the theory in any closed, connected  and orientable
three-manifold. From topology it is known that any  closed, connected  and 
orientable three-manifold can be obtained from $S^3$ by using Dehn's surgery, 
which consists in a finite sequence of elementary surgery operations 
corresponding to cut a solid torus from $S^3$ and glue it back by using a 
suitable homeomorphism.

The goal of this thesis is two-fold: study the general properties of the 
surgery construction for a generic compact and semi-simple gauge group $G$ as 
a first instance, and, as a second instance, show how powerful the 
three-dimensional point of view is for practical computations. By Turaev's
or Witten's approaches only the case $G=SU(2)$ has been considered 
extensively; actually one of the original aspects of this work is a complete study
of the case $G=SU(3)$. Other original results concern the general construction
of the surgery operator under suitable conditions on the gauge group $G$, 
the derivation of some well known properties of rational conformal field theory in 
two dimensions by using three-dimensional CS theory, and finally a preliminary 
investigation on the relation between the fundamental group $\pi_1({\cal M})$ and 
the three-manifold invariant defined by using CS theory.

For the reader's convenience we shall outline the contents of this thesis.
The first two chapters contain preliminary material. In chapter 1, CS 
perturbation theory in $\mathbb{R}^3$ is reviewed; the observables of the theory, namely 
Wilson lines, are defined as quantum composite
operators and this leads to the concept of framed knots in $\mathbb{R}^3$.
The successive step is the study of CS theory in  $S^3$; it is shown that all the 
results obtained in $\mathbb{R}^3$ are still valid, provided one restricts
the CS coupling constant $k$ to integers values as a consequence of the gauge 
invariance constraint.
In chapter 2, the notion of tensor algebra is introduced and one of the most 
remarkable properties of CS observables is discussed: the satellite relations. 
In the last section of chapter 2, some useful calculation rules are summarized. 
With chapter 3 begins the 
original part of the work; the explicit solution of the theory in $S^3$ with the 
gauge group $G=SU(3)$ is given, many examples are also worked out in detail. 
Chapter 4 is devoted to the reduced tensor algebra, which encodes all the information
on the physical irreducible representations of the gauge group $G$. After
a general definition of the reduced tensor algebra, the case $G=SU(3)$ is studied
with full details. In addition, for completeness, also the construction of the 
reduced tensor algebra for $G=SU(2)$, already known in the literature, is 
presented. 
Chapter 5 contains a quick introduction to Dehn's surgery in preparation to
chapter 6, in which the surgery construction in the CS theory is introduced.
When the reduced tensor algebra is regular (this notion is defined in chapter
4) the general form for the surgery operator is produced and the surgery rules 
are given. As examples the manifolds $S^2 \times S^2$, $\Sigma_g \times S^1$ and
the Poincar\'e sphere with $G=SU(3)$ are considered. Some general properties of 
the surgery operator are also proved. Chapter 7 is devoted to the construction 
of a non-trivial three-manifold invariant from the CS partition function. After some
example, the relation between the fundamental group and the CS invariant is 
investigated. It is proved that, when $G=SU(2)$, the absolute value of the non-vanishing CS 
three-manifold invariant for the lens spaces $L_{p/r}$ depends only on the 
fundamental group of $L_{p/r}$; numerical evidence for the  case $G=SU(3)$ is 
also given.  In the last chapter, the relation between 
two-dimensional conformal field theory and three-dimensional CS theory is 
discussed . Appendix D contains a brief introduction to conformal field theory and a review 
of the original Witten's approach to CS theory.

\chapter{\bf The action principle and observables}
\section{\bf Perturbative quantization}

The field theory model which we are interested in is defined by the action \cite{witt1}
\beeq
S_{CS} \; = \; \frac {k}{4 \pi}  \int_{\mathbb{R}^3} \; {\rm Tr} \left ( A 
\wedge
d A \; + \; \frac {2}{3} i \,   A \wedge  A \wedge  A \right) \quad . 
\label{cs}
\end{equation}  
Eq.(\ref{cs}) defines a gauge theory for the compact and simple group  
$\, G \, $; the fundamental field is the connection 1-form
\beeq
A \; = \; A_{\mu} \, d x^{\mu} \, , \qquad 
A_{\mu} \; = \; A_\mu^a T_{a} \nb \quad .
\end{equation}
The generators of the gauge
group $\, G \, $ in some definite representation $\, \rho \, $ are associated 
with the Hermitian matrices $\, \{T^a \} \, $ 
\beeq
\left [\, T^a, \; T^b  \, \right  ] \; = \; i  f^{ab}_c \,  T^c \qquad 
a,b,c \; = \; 1, \cdots n  \quad n \; = \;  \mbox{dim(Lie G)} \quad .
\end{equation}
We use the following representation independent normalization   
\beeq
{\rm Tr} \left ( \, T^a \, T^b \, \right ) \; = \; \frac{1}{2} \, \delta^{ab}
\quad ;
\end{equation} 
the trace ${\rm Tr}_\rho$ in the representation $\rho$ is related to ${\rm Tr}$
by 
\beeq
{\rm Tr} \; = \; \frac{1}{4 \, {\cal X(\rho)}} {\rm Tr}_\rho \quad ,
\end{equation}
where ${\cal X}(\rho)$ is the Dynkin index of $\rho$.
In a local coordinate basis the action takes the form 
\beeq
S_{CS} \; = \; \frac{k}{4 \pi } \, \int_{\mathbb{R}^3} \, d^3x \> \epsilon^{\mu \nu 
\rho } 
\; {\rm Tr}  \left ( \, A_\mu \, \partial_\nu \, A_\rho \, + \, i \, \frac{2}
{3}  \, A_\mu \, A_\nu \, A_\rho \, \right ) \quad .  
\end{equation}
It is important to point out that, in order to define the classical theory, no 
metric structure in $\mathbb{R}^3$ is required.

\no
Let us consider a gauge transformation on $\, A \, $
\beeq 
A \;    \rightarrow  \; A^\prime \;  = \;  U^{-1}  A
U  \; - \; i  \, U^{-1} d  U  \quad ,
\label{gtra2}
\end{equation} 
where 
$\, U: \, \mathbb{R}^3  \rightarrow G \, $ is a $\, G \, $ valued function in $\, \mathbb{R}^3 \, $. 
Differently from  standard gauge theories like QED or QCD, the CS action is not gauge invariant, indeed
\beeq 
S_{CS}[A]  \; \rightarrow \; S_{CS}[A^\prime] \; = \; S_{CS}[A] \; + \; 
2 \pi k \, \Gamma_{\mathbb{R}^3} \left[U \right ] \quad ,
\end{equation}
where the Wess-Zumino functional
$\, \Gamma_M \, $ is defined in any orientable 
3-manifold $M$ as  
\bea 
&& \Gamma_M \left [U \right ] \; = \; \frac {k}{24 \pi^2}
\int_{M} \Omega \quad , \nb \\
&& \Omega \; = \; {\rm Tr} \left [ \left(U^{-1} d  U \right ) \wedge \left (U^{-1}  d  U \right ) \left ( U^{-1} d  U \right )
\right ] \ \quad .
\ena
In appendix A it is shown that $\Gamma_M \left [U \right ]$ is invariant under 
a small deformation of $U$ 
\beeq
\Gamma_M \left [U \, + \, \delta U \right ] \; = \;  \Gamma_M \left [U  \right ] \quad . \label{wzp}
\end{equation} 
Therefore, the CS action in $\mathbb{R}^3$ is invariant under infinitesimal 
gauge transformations.

The gauge invariance of the CS action is crucial to quantize the model in 
$\mathbb{R}^3$; actually one can use \cite{gm1,gau} the well experimented methods of 
BRS perturbative quantization \cite{brs}. As usual, the first step is the gauge fixing;  to 
carry out this task a metric on $\mathbb{R}^3$ is needed. A simple and 
practical choice is the 
following
\beeq 
g_{\mu \nu} \; = \; \delta _{\mu \nu} \qquad  \mu,\nu \; =\; 1, \, 2, \, 3
\quad .  
\end{equation}
The gauge fixing condition is the covariant (Lorenz) gauge
\beeq 
G(A) \; = \; \partial^{\mu} A_{\mu} \; = \; 0 \qquad . \label{lore}
\end{equation}
To avoid the over-counting of gauge equivalent configurations, the generating 
functional $Z$ is defined following the Faddev-Popov ansatz
\bea
&& Z[J, \eta, \bar{\eta}] \; = \; \int \,  D \left [A,\bar{c},c \right ] \; e^{i S_{CS}[A] + \, 2i  dx \, \int {\rm tr} \left(J^\mu A_\mu + \bar{c} \eta + \bar{\eta} c \right) } \nb \\
&& e^{\frac{ik}{4 \pi} \int  dx dy  \bar{c}^a(x) \, \frac {\delta
G^a}{\delta \omega^b} \, (x,y) \; c^b(y)} \; \delta \left [G^a \right ] \quad .
\label{inth}
\ena
We have introduced the sources for the gauge field $\, A \, $ and for the ghost fields
$\, \bar{c}, \; c \, $
\beeq
\bar{c} \; = \; \bar{c}^a T_a \, ,  \qquad  c \; = \; c^a T_a \, , \qquad  \eta \; = \;
\eta^a T_a \, ,  \qquad
\bar{\eta} \; = \; \bar{\eta}^a T_a \quad .
\end{equation}
By using 
\beeq 
\delta \left [G \right ] \; = \; \int  D \left [B \right ] \, \exp
\left (-i  \frac {k}{4  \pi} \: \int dx \, B^a(x) \, G^a(x)
\right ) \quad ,
\end{equation} 
and  eq.(\ref{inth}) one gets
\beeq 
Z \left [J, \eta, \bar{\eta} \right ] \, = \, \int  D \left [A,\bar{c},c
\right ] \,  \exp \left \{ i  S_{tot} \left [A,B,\bar{c},c \right ] \, + \, 2 i
\, \int  dx \;  {\rm tr}  \left(J^{\mu} A_{\mu} \, + \, \bar{c} \eta \, + \,
\bar{\eta} c \right) \right \} \, , 
\end{equation}
where
\bea
&& S_{tot}\; = \; S_{CS} \; - \; \frac{k}{4 \pi} \int dx \; B^a(x)  \partial^{\mu} A_{a \mu} \; + \; \frac{k}{4 \pi} \int dx  dy \;  \bar{c}_a(x)  \frac {\delta G^a}{\delta \omega^c}  (x,y)  c^c(y) \nb \\
&& \qquad = \; S_{CS} \; + \; S_{g.f.} \qquad . 
\ena
The operator $\,  \frac{\delta G^a}{\delta \omega^c}\, $ represents the variation of
$G^a$ under an infinitesimal gauge transformation
\beeq 
A_{\mu}^a \; \rightarrow \; A_{\mu}^a \; + \; \partial_{\mu}
\omega^a-A_{\mu}^b \omega^c {f^a}_{bc}\quad ,
\end{equation}
with
\beeq
U \left (x \right ) \; = \; \exp \left (i  \omega \right ),  \qquad   
 \omega \; = \; \omega^a T_a  \quad .
\end{equation} 
In particular, with the gauge choice (\ref{lore}) we have \cite{gm1}
\beeq 
S_{tot}\; = \; S_{CS} \; + \; \frac{k}{4  \pi} \int  dx \; A_{\mu}^a 
\partial^{\mu} B_a \; - \; \frac{k}{4 \pi} \int dx \; \partial^{\mu} 
\bar{c}_a(x) {\left (D_{\mu} \right )}^{ad} c_d \quad . \label{eff}
\end{equation}
The covariant derivative $\, {\left (D_{\mu} \right )}^{ad} \, $ in the adjoint 
representation is defined as 
\beeq
{\left (D_{\mu} \right)}^{ad} \; = \; \delta^{ad} \, \partial_{\mu}-f^{ad}_c 
\, A_{\mu}^c  \qquad . 
\end{equation}
The total action $\, S_{tot} \, $ in the Landau gauge comprising the classical 
action, the gauge-fixing term and ghost contribution  is given by equation 
(\ref{eff}).

The total action $\, S_{tot} \, $ is invariant under the following BRS 
transformations
\bea 
&& s\left (A_{\mu}^a \right ) \; = \; {\left (D_{\mu} c \right )}^a \nb \\
&& s \left (c^a \right) \; = \; - \frac {1}{2}  { \left [ c,c \right]}^a \nb \\
&& s \left (\bar{c}^a \right ) \; = \; B^a \nb \\
&& s \left (B^a \right) \; = \; 0 \qquad . \label{brst}
\ena

The CS theory is renormalizable by power counting; the beta function and the 
anomalous dimensions of the elementary fields
are vanishing \cite{del} to all orders in perturbation theory. The only free 
parameter of the model is the
renormalized coupling constant $\, k \, $ which is fixed  by the normalization
conditions. Let $\, \Gamma[A,B,\bar{c},c] \, $
be the effective action of the theory. Since there are no gauge anomalies in three dimensions, one
can always construct an effective action $\, \Gamma[A,B,\bar{c},c] \, $  which, 
in the limit in which the regularization cutoffs  are removed,
is BRS invariant.  BRS invariance will be used to fix the normalization of the fields; indeed, as a function of the renormalized fields $\, \Gamma[A,B,\bar{c},c] \, $ must satisfy the Slavnov-Taylor identity
\beeq
\left \{ \, s(A_\mu^a)  \> \frac{\delta}{\delta  A^a_\mu } \; + \; s(B)  \> 
\frac{\delta} { \delta 
B } \; + \;  s(c)  \> \frac{\delta}{ \delta  c } \; + \; s( {\overline c}) \> 
\frac{\delta}{ \delta 
{\overline c} }  \, \right \} \, \Gamma[A,B,\bar{c},c] \; = \;  0 
\quad . 
\label{renor}
\end{equation}
In eq.(\ref{renor}), the appropriate normalization for the composite operators 
is understood. The effective action admits an expansion in powers of the fields,
of course; let us consider for instance the terms $\, \Gamma^{(2)} [A_\mu ] \,$ 
and  $\, \Gamma^{(3)} [A_\mu ] \, $ of this expansion which are quadratic and cubic in the field $\, A_\mu \,$, 
\beeq
\Gamma^{(2)} [A_\mu ] \; = \; \int d^3x \, d^3y \, G^{\mu \nu}_{ab} (x,y) \, A^a_\mu (x,y) \,  A^b_\nu (y) \quad , 
\end{equation}
\beeq
\Gamma^{(3)} [A_\mu ] \; = \; \int d^3x \, d^3y \, d^3z \, H^{\mu \nu \rho }_{abc} (x,y,z) \,
A^a_\mu (x) \,  A^b_\nu (y) \,  A^c_\rho (z) \quad .  
\end{equation}
The BRS invariance (\ref{renor}) implies that the exact two-point function $\, 
G^{\mu \nu}_{ab} (x,y) \, $ is related to the three-point proper vertex 
$\, H^{\mu \nu \rho }_{abc} (x,y,z) \,  $  by 
\beeq
\partial_\mu^x \, H^{\mu \nu \rho }_{abc} (x,y,z) \; = \; \frac{2}{3} \, {f^d}_{ab} \> 
G^{\nu \rho  }_{dc} (x,z) \> \delta^3 (x-y) \quad . 
\label{eqdot}
\end{equation}
Our normalization of the fields is fixed by eq.(\ref{brst}); equivalently, 
our normalization of the elementary vector field $A_\mu$ is determined by
eq.(\ref{eqdot}). Let us now fix the normalization condition for $\, k \, $. 
As shown in \cite{del}, the exact two-point function $\, G^{\mu \nu}_{ab} (x,y)\, $ has the form 
\beeq
G^{\mu \nu}_{ab} (x,y) \; = \; Z \, \epsilon^{\mu \rho \nu} \, \partial^x_{\rho } \delta^3 (x-y) \,
\delta_{ab} \quad ,  \label{duep}
\end{equation}
where $\, Z \, $ is a real parameter. The parameter $\, Z\, $ which is computed in perturbation
theory is a function of the ``bare"  coupling constant $\, k_0 \, $.  The
functional dependence of $\, Z \, $ on $\, k_0 \, $ depends on the   
regularization prescriptions and/or by adding finite local counter-terms at each order
of the loop expansion. The value of the physical
coupling constant is also called the renormalized coupling constant and is
usually determined in terms of the effective action $\, \Gamma \, $ of the theory. We shall
follow the standard field theory procedure and the value of the renormalized coupling constant $\, k \, $ of the CS theory will be fixed by  
\beeq
k \; = \;  8 \pi \, Z \quad ,  \label{defco}
\end{equation}
where $\, Z \, $ enters the exact two-point function (\ref{duep}). With the field 
normalization (\ref{eqdot}), eq.(\ref{defco}) represents the normalization condition for $\, k \, $. 

Condition (\ref{defco}) gives a good definition of the renormalized coupling constant 
$\, k \, $ because eq.(\ref{defco}) it is in agreement with the classical expression (\ref{eff}) of the
action.  Consequently, when the correlation functions of the fields are
expressed in terms of $\, k \, $, the equations which follow from the action principle
are satisfied. Conversely, for the renormalized correlation functions the action
principle is valid only if the coupling constant $\, k \, $ satisfies condition
(\ref{defco}).  

All the results that we shall derive are consequences of  the action principle 
based on the functional (\ref{eff}); the expectation values of the gauge invariant
observables will be expressed in terms of the physical coupling constant 
$\, k\, $. 

A remarkable property of the quantum  CS theory is that, according to eq.(\ref{duep}), the two-point function does not receive radiative corrections; therefore, the
full (dressed) propagator coincides with the free one 
\beeq
\langle \, A^a_ \mu (x) \, A^b_\nu (y) \, \rangle \; = \; \frac{i}{ k} \, \delta^{ab} \, 
\epsilon_{\mu \nu \rho } \, 
\frac{(x-y)^\rho}{  |x-y|^3 }\quad .  \label{prop}
\end{equation}
The function defined by the propagator, integrated along two 
oriented, closed and non-intersecting paths $\, C_1 \, $ and $\, C_2 \, $ in
$\, \mathbb{R}^3 \, $, must
represent an invariant of ambient isotopy for a two-component link. The
invariant associated with the expression (\ref{prop}) is simply the linking 
number \cite{pol} of $\, C_1 \, $ and $\, C_2 \, $,  
\beeq
{\rm lk} (C_1, C_2) \; = \; \frac{1}{ 4 \pi } 
\oint_{C_1}  dx^\mu \, \oint_{C_2} dy^\nu \, \epsilon_{\mu \nu \rho }
 \, \frac{(x-y)^\rho}{ |x-y|^3 }  \quad . 
\end{equation}

The classical CS action was defined without referring to any metric structure in 
$\mathbb{R}^3 $. In addition, infinitesimal gauge invariance implies that $\, S_{CS} \, $ is also invariant under an infinitesimal diffeomorphism of $\mathbb{R}^3: \,  x^\mu \rightarrow x^\mu + \epsilon v^\mu \, $, generated by the vector 
field $\, v \, $. The ``general covariance'' property of $\, S_{CS} \, $ is
remarkable because  is achieved only with the differential structure of $\mathbb{R}^3$.
However, we have seen that in order to gauge fix the theory an ``external'' metric
is required. The question is: ``Does the external metric break general covariance ?''. In order to answer this question, let us compute the symmetric energy momentum tensor associated with  the total action (\ref{eff}). By definition we get 
\cite{gm1}
\bea
&& T_{\mu \nu} \; = \; - \frac{1}{2} \frac{1}{\sqrt{g}} \frac{\delta S_{tot}}{\delta g^{\mu \nu}} 
\;  = \; - \frac{k}{4 \pi} \left[A^a_{(\mu} \partial_{\nu)} B_a \; - \; 
g_{\mu \nu} \, A^a_\rho \partial^\rho A_a \right. \nb \\
&&\left. \qquad - \; \partial_{(\mu} \bar{c}^a \left
(D_{\nu)} c \right)_a \; + \; g_{\mu \nu}  \, \partial^\rho \bar{c}^a
\left(D^\rho c \right)_a \right] \; . \label{eten}
\ena
The operator $\, Q \, $, generating  the BRS transformation
in the Hilbert space of the theory, is defined  in such a way that its action on 
a field $\, \phi \, $ is the following 
\beeq 
\left [ Q, \phi \right ] \; = \; s \left ( \phi \right ) \quad . \label{qbrs}
\end{equation}
From eq.(\ref{eten}) and eq.(\ref{qbrs}) it follows that $\, T_{\mu \nu} \, $
is $\, Q $-exact; indeed
\beeq 
T_{\mu \nu } \; = \; \left [ Q, \Theta_{\mu \nu} \right ] \qquad ,
\end{equation}
where
\beeq
\Theta_{\mu \nu} \; = \;  \partial_{\{ \mu } \bar{c}_a A^a_{\nu \}} \; - \; 
g_{\mu \nu} \, \partial^\rho \bar{c}_a A^a_\rho \quad .
\end{equation}
The physical states of a gauge theory are those annihilated by the BRS charge, 
i.e.
\beeq
 Q \left | f \right \rangle \; = \; 0 \qquad . \label{com}
\end{equation}
As a consequence of eq.(\ref{com}), the expectation values of the energy momentum
tensor on physical states vanish
\beeq 
\left \langle f^{\prime} \left | T_{\mu \nu} \right | f \right \rangle \; = \; 0
\qquad  .
\end{equation}  
Actually, the BRS symmetry of the theory guarantees that the external metric required to 
gauge fix the theory is immaterial when   
physical states are considered. Thus, invariance under infinitesimal diffeomorphisms 
of $\, \mathbb{R}^3 \, $ is preserved also at the quantum 
level. 

\section{\bf Composite Wilson line operators}

Let $\, C \, $ be an oriented knot in $\mathbb{R}^3$ and $\rho $  an  
irreducible representation of the gauge group $G$. The Wilson line $W(C; \rho )$ 
is defined as
\beeq
W(C; \rho ) \; = \; {\rm Tr} \> {\rm P} \> {\rm exp} \left [ i \oint_C A^a_\mu (x) \, T^a_{(\rho )}
\, dx^\mu \right ] \quad , \label{wil}
\end{equation}
where the path-ordering is introduced according to the orientation of 
$\, C \, $ and $\{ \, T^a_{(\rho )} \, \}$ are the generators of $G$
in the representation $\rho $. The behaviour of $W(C; \rho )$ under a gauge 
transformation is the following
\beeq  
 P \exp {\oint_{\gamma} \tilde{A}}  \; \stackrel{Gauge}{\rightarrow} \; 
U \left (P_1 \right ) \left [ P \exp {\oint_{\gamma} \tilde{A}} \right ] U^{-1}
\left (P_2 \right )   \quad , \label{wilg} 
\end{equation} 
where  $\, P_1 \, $, $\, P_2 \,$ are the initial and final points of the curve 
$\, \gamma \, $. 
From eq.(\ref{wilg}), it follows that $\, W(C; \rho) \, $ is gauge invariant.

In the framework of quantum CS theory, the object defined by eq.(\ref{wil})  
gives rise to a composite operator, consequently, a suitable definition of the 
Wilson line must be provided. In other words one has to
specify the operative procedure used to compute the expectation values of this 
operator at any given order of the perturbative expansion of the theory. There 
is  only one known procedure which preserves general covariance:  
a framing for the closed curve $C$, which enters the definition of the Wilson line, 
must be introduced \cite{gm2,guad1}. Let us start from the ``classical'' 
expression for the Wilson line
\bea 
&& W_{classic}(C, \rho) \, = \, \left [1 \; + \; \int_0 ^1 A_{\mu}
\left (x(s) \right )  \frac {dx^{\mu}}{ds} ds \, - \, \int_0 ^1 ds \int_0 ^{s} ds 
\; A_{\mu} \left (x(s_1) \right)  A_{\nu} \left (x(s) \right ) \right. \nb \\
&&\qquad \qquad \qquad \left. \frac {dx^{\mu}}{ds_1} \frac {dx^{\nu}}{ds} \, + \, \cdots \right ] \nb \\
&& 0 \leq 1 \leq s \, , \label{wilcla}
\ena 
with knot $C$ parameterized by $\, x^\mu(s) \,$. The framing prescription for
the Wilson line is defined in terms of the framing contour $\, C_f\, $  given by
\beeq 
y^{\mu}(s) \; = \; x^{\mu}(s) \; + \; \epsilon \, n^{\mu}(s) \qquad  \epsilon >0
\quad  n^2=1 \quad  . 
\end{equation}
The vector field $\, n^\mu \,$ is chosen to be orthogonal to the tangent vector of $\, C \,$, i.e. $\, n^\mu \dot{x}^\nu \delta_{\mu \nu} = 0$.   
For each oriented knot $C$, we shall introduce a framing $C_f$. The framed Wilson line is defined in the following way. In eq.(\ref{wilcla}), each term
of the type
\beeq
\left[\dot{x}^{\mu_i}(s_i) A_{\mu_i}(x(s_i)) \right] \qquad ,
\end{equation}
is replaced by
\beeq
\left[\dot{x}^{\mu_i}(s_i) A_{\mu_i}(x(s_i)) \right]_f \; = \;
\left\{ [\dot{x}^{\mu_i}(s_i) \; + \; \epsilon_i \dot{n}^{\mu_i}(s_i) ]
 A_{\mu_i}\left(x(s_i) \; + \; \epsilon_i \, n(s_i) \right) \right\} \quad .
\end{equation}
The non-negative real numbers $\{ \epsilon_i \quad i=1,2, \, \cdots \}$ are chosen in such a way that $\epsilon_i \neq  \epsilon_j \quad \forall i,j$.
The composed Wilson line operators $W(C; \rho)$ associated with a link $C$ with colour $\rho$ is defined by replacing $C$ by its  
framed analog $C_{Framed}$, by taking the vacuum expectation value and, finally,
by letting $\epsilon_i$ go to zero
\beeq 
E \left (C; \rho \right ) \; = \; \lim_{\{\epsilon\} \rightarrow 0} \;  
\left \langle W \left (C_{Framed}; \rho \right ) \right \rangle \quad .
\end{equation}
The framing 
$\, C_f \, $ of $\, C \, $ is completely determined, up to smooth deformations
of $C$ and $C_f$ in the ambient space $\mathbb{R}^3$, by the linking
number $\, {\rm lk} (C, C_f) \, $  of $\, C \, $ and $\, C_f \, $.   When the linking number of
$\, C \,$ and $\, C_f \, $ is vanishing, $C_f$ is called a  preferred framing for $C$. 
Given an oriented framed knot $\, C \, $ in $\mathbb{R}^3$ and an irreducible 
representation $\, \rho \, $
of $G$, the associated Wilson line operator $\, W(C; \rho ) \,$ is
 well defined and is gauge invariant. We shall often call $\rho$ the colour of $C$.

Let us now consider a framed, oriented and coloured link $\, L \, $ in $\mathbb{R}^3
\, $ with $\, m\, $ components $\, \{ \, C_1 , C_2 , ... , C_m \, \}\, $.  Let the colour of each
component $\, C_i \, $ of $\, L \, $ be specified by an irreducible representation  $\, \rho_i\, $ of $G$. The Wilson line operator $W(L)$ associated with $L$ is
simply the product of  the Wilson operators defined for the single components 
\beeq
W(L)\; = \; W(C_1; \rho_1) \, W(C_2; \rho_2) \cdots W(C_m; \rho_m) \quad . 
\label{wilman}
\end{equation}
The set of expectation values 
\beeq
E(L) \; = \; \langle \, W(L) \, \rangle \Big |_{\mathbb{R}^3} \; = \; 
\frac{\langle \, 0 \, | \, W(L) \, |  
\, 0 \, \rangle \Big |_{\mathbb{R}^3}}{  
\langle \, 0 \, | \,  0 \, \rangle \Big |_{\mathbb{R}^3 }} \quad ,  \label{esp}
\end{equation}
which are defined for all the possible links $\{ \, L \,
\}$ which are framed, oriented and coloured,  is the set of the gauge invariant 
observables which we are interested in.  

The Wilson line operators (\ref{wilman}) are defined for links which are
contained inside some finite and closed domain of $\mathbb{R}^3$. For this kind of observables, the vacuum expectation values can be computed by means of the
standard BRS quantization procedure which is  based on the action (\ref{eff}). In $\mathbb{R}^3$ there is no constraint to be imposed on the value of the coupling constant $k$; thus, perturbation theory is well defined because the expansion parameter $\lambda \, $, given by  
\beeq
\lambda \; = \; (2\pi /k) \quad , 
\end{equation}
is a free parameter and the expectation values $E(L)$ are
well defined for arbitrary values of $\lambda$. This is why we chose $\mathbb{R}^3$ as starting manifold; in fact, all the perturbative aspects of the CS model refer to the theory defined in $\mathbb{R}^3$.

For any given link $L$ in $\mathbb{R}^3$, the expectation value (\ref{esp})
admits an  expansion in powers of $\lambda$; each term of this expansion 
can be computed by means of ordinary perturbation theory. In  order to verify
that the observable $E(L)$ is well defined, let us consider for instance the
first three terms of this expansion. 
Let us consider the unknot $U$ (simple circle) in $\mathbb{R}^3$ with framing $U_f$ 
and with colour given by the irreducible representation $\rho $ of $SU(3)$. The
first three terms of the $\lambda$-expansion of the vacuum expectation value of
the associated Wilson line operator  $W(U, U_f; \rho )$ have been computed \cite{gm2}
by means of ordinary Feynman  diagrams; the result is 
\bea
&&
\left. \langle \, W(U, U_f; \rho ) \, \rangle \right |_{\mathbb{R}^3 } \; =\; (\, {\rm dim} \> \rho \, ) \,
\left [ 1 - i\lambda \, Q(\rho )\; {\rm lk} (U,U_f) \right .\nb \\
&& \left. \qquad - \frac{1}{2} \lambda^2\, Q^2(\rho )\, 
\left (  {\rm lk} (U,U_f) \right )^2
  \; - \;  \frac{1}{4} \, \lambda^2\, Q(\rho )\,  \right ] + O(\lambda^3) 
\quad ,  \label{unper}
\ena
where $Q(\rho )$ is the value of the quadratic Casimir operator in 
the representation $\rho$, 
\beeq
T^a_{(\rho )} \, T^a_{(\rho )} \, = \, Q(\rho ) \cdot {\mathbb I} \quad .  
\end{equation}

\begin{figure}[h]
\vskip 0.9 truecm 
\centerline{\epsfig{file=\path 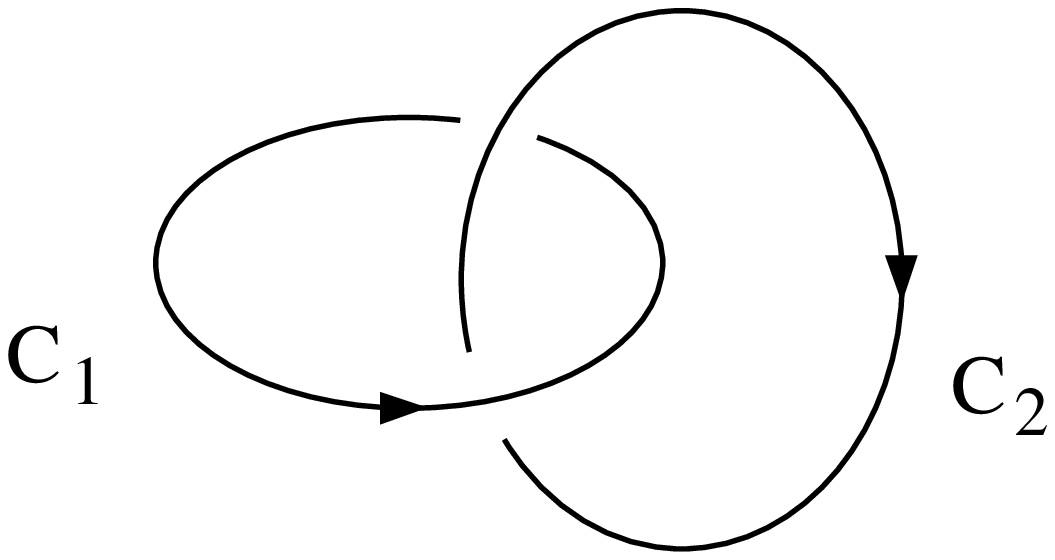,height=2cm,width=5cm}}
\vskip 0.9 truecm 
\centerline {{\bf Figure 1.1}}
\vskip 0.5 truecm 
\end{figure}

Let us now consider the Hopf link in $\mathbb{R}^3$  in which   
the two components $C_1$ and $C_2$ are oriented as it is shown in Fig.1.1.  
yLet $C_1$
and $C_2$   have framings $C_{1f}$ and  $C_{2f}$ respectively. When
the colours of $C_1$ and $C_2$ are given by the irreducible representations
$\rho$ and $\rho^\prime$ of $SU(3)$,  one finds \cite{gm2} 
\bea
&& \left. \langle \, W(C_1, C_{1f}; \rho ) \,  W(C_2, C_{2f}; \rho^\prime ) \, \rangle \right |_{\mathbb{R}^3 } 
\; =\; (\, {\rm dim} \> \rho \, ) \, (\, {\rm dim} \> \rho^\prime \, ) \nb \\
&& \left[ 1 - i\lambda \, Q(\rho ) \;  {\rm lk} (C_1,C_{1f})  
\; - \; i\lambda \, Q(\rho^\prime )\; {\rm lk} (C_2,C_{2f}) \right. \nb \\
&& - \; \frac{1}{2} \lambda^2\, Q^2(\rho )\; \left ( {\rm lk} (C_1,C_{1f})
\right )^2 \; - \; \frac{1}{2} \lambda^2\, Q^2(\rho^\prime )\; 
\left ( {\rm lk} (C_2,C_{2f}) \right )^2 \nb \\ 
&&\left. - \; \frac{1}{4} \, \lambda^2\, Q(\rho ) -  \frac{1}{4} \, \lambda^2\, Q(\rho^\prime ) \;  - \;  \frac{1}{4} \, \lambda^2\, Q(\rho )\, Q(\rho^\prime )
\,  \right ] + O(\lambda^3) \quad . \label{hope}
\ena
In conclusion, perturbation theory is well defined in $\mathbb{R}^3$ and the 
expectation values of the observables, which are defined by the framing 
procedure, represent ambient isotopy invariants of framed, oriented and 
coloured links in $\mathbb{R}^3$. 
\section{\bf Chern-Simons theory on the 3-sphere}
In order to study the theory in $S^3$, we need to discuss the non-invariance of 
$S_{CS}$ under ``large'', i.e. non-shrinkable to the identity, gauge transformations.  Let us introduce the concept of degree of map. Given a differentiable map $f :  N \rightarrow M$, with $N$ and $M$ closed, connected and orientable
manifolds of dimension $n$, the degree $Dgr_y[f]$ of $f$    
is defined as \cite{choq}
\beeq 
Dgr_y \left [f \right ] \; = \; \sum_{x_0 \epsilon f^{-1}(y)} \, sgn \,
det \left ( \frac {\partial f}{\partial x} \mid_{x_0} \right ) \quad  ,
\end{equation}
where $y \in M$ is a regular point.
The degree is an integer and satisfies the following properties
\begin{enumerate}
\item  $Dgr_y \left [ f \right ] $ does not depend on the regular point $y$
chosen.
\item If $f_1$ and  $f_2$ are homotopic maps then
\beeq 
Dgr \left [f_1 \right ] \; = \; Dgr \left [f_2 \right ] \quad .  
\end{equation}
\end{enumerate}
If $f_1$ and  $f_2$ are homotopic, one can continuously deform
$f_1$ to  $f_2$. Roughly speaking, $Dgr[f]$ measures the number of times that $N$
winds $M$. The degree admits the following integral representation 
\beeq 
\int_N \; f^{\ast} \;  \omega \; = \; Dgr \left [f \right ] \;
\int_M \; \omega \quad , \label{intf}
\end{equation}
where $\omega \;$ is a $n-$form on $M$, and  $f^{\ast}\omega $ is the $n-$form
$f-$related with $\omega$, i.e. $f^{\ast}\omega $ is the pull-back of $\omega$. 

Let us consider first the case when the gauge group $G$ is $SU(2)$ and the 
manifold $M$ where the theory is defined is $S^3$. The group
$SU(2)$ is homeomorphic to  $S^3$. Thus, a gauge transformation $U$ 
can be viewed as a map $U: \, S^3 \ra S^3$. By using the relation (\ref{intf}) in 
the expression for the Wess-Zumino functional, we get 
\bea 
&&\Gamma \left [U \right ] \; = \; \int_{S^3} \; \Omega \; = \; D \left
[U \right ] \; \int_G \; \omega \; = \; m \; \int_{S^3} 
\omega= m \, C  \nb \\
&&\Omega \; = \; \varphi^{\ast} \;  \omega \;  , \qquad  m \; = \; Dgr
\left [U \right ] \quad ,
\ena
where $C$ is a constant. One can verify by direct calculation (see appendix A) that 
\beeq
\Gamma[U] \; = \; m \quad . \label{su2}
\end{equation}
The integer $n$ labels the elements of $\pi_3(SU(2))$, indeed
$\pi_3(SU(2)) = \mathbb{Z}$. The multiplication law in $\pi_3(G)$ is
expressed in terms of $\Gamma[U]$ as (see appendix A)   
\beeq 
\Gamma \left [ U^{(1)} \; \cdot \; U^{(2)} \right ] \; = \; \Gamma
\left [U^{(1)} \right ] \; + \; \Gamma \left [U^{(2)} \right ] \quad . 
\label{group}
\end{equation}
If $U^{(1)} \;$ represents the generator of  $\mathbb{Z}$, one gets 
\beeq 
\Gamma \left [U^{(m)} \right ] \; = \; \Gamma \left [U^{(1)}  \cdots  U^{(1)}
\right ] \; = \; m \; \Gamma \left [U{(1)} \right ] \; \; \Rightarrow \; C 
\; = \; \Gamma \left [U^{(1)} \right ] \quad .  
\end{equation}
Let us now extend the previous result to a generic compact and simple group 
$G$. The crucial fact for our purpose is the following theorem due to  Bott \cite{bott}.

\bigskip

\shabox{\no {\bf Theorem 1.1}}
{\em Given a continuous map $f$ from $S^3$ into a simple and compact Lie group $G$, there exists a map $f^\prime$ from $S^3$ into $SU(2)\subset G$} homotopic to $f$. 

\bigskip

With the help of Theorem 1.1, eq.(\ref{su2}) can be extended to any 
compact a simple Lie group $G$. Thus, it follows
that the CS theory action in $S^3$  with a simple and compact
gauge group transforms under a gauge transformation according to 
\beeq  
S_{CS} \; \rightarrow \;  S_{CS} + 2  \pi \, k \, m \quad , 
\end{equation} 
where $n$ is an integer. As a result, the gauge invariance of $\exp(i \, S_{CS})$ in $S^3$ constrains $k$ to be an integer \cite{djt}. 

From a topological point of view, one can represent $S^3$ as $\mathbb{R}^3$ 
with the points at infinity identified. As in $\mathbb{R}^3$, any link $L$ in $S^3$ is contained 
in a three ball $B^3$. Thus links in $S^3$ and in $\mathbb{R}^3$ share the same topological properties. Let us now consider the CS quantum field theory in $S^3$; 
being $S^3$ and $\mathbb{R}^3$ locally the same, the theory in $S^3$ exhibits the
same  ultraviolet behavior as the corresponding theory in $\mathbb{R}^3$. 
All the general properties of the expectation values of the
observables in $\mathbb{R}^3$  are also valid in $S^3$.  A change in
the value of the coupling constant $k$ does not modify  the structure of the 
link polynomials $E(L)$,
it only modifies the numerical value of $\lambda$. Now, in order to preserve
gauge invariance, the coupling constant $k$ is not a free parameter as in $\mathbb{R}^3$, but it must be an integer. Given the expectation value $E(L)_{\mathbb{R}^3}$ of a Wilson line operator $W(L)$ in $\mathbb{R}^3$, the corresponding expectation value $E(L)_{S^3}$ in $S^3$ can be obtained simply by taking $k$ integer
\beeq
\left. \langle \, W(L) \, \rangle \right|_{S^3} \; = \;
 \left. \langle \, W(L) \, \rangle \right|_{\mathbb{R}^3} \qquad k \,  \text{ integer} \quad .
\end{equation}  
We shall fix the orientation of $S^3$ by adopting the usual right-handed rule to
compute linking numbers (see appendix B). Clearly, a modification of the orientation of $S^3$ is
equivalent to replace $k$ with $-\, k$ in the action (\ref{cs}). Moreover, 
the value $k=0$ must be excluded because, when $k=0$, the action (\ref{cs}) 
vanishes and thus a meaningful CS theory does not exist. Consequently, we may 
assume that $k$ is positive. Therefore, the relevant values of $k$ that we 
need to consider are 
\beeq
k \; = \; {\rm integer} \qquad , \qquad k \; = \; {\rm positive} \qquad . 
\end{equation}
Note that also in a generic three-manifold $M$, which is closed connected and orientable, $k$ is no more a free parameter; gauge invariance under large gauge
transformations in $M$  implies \cite{witt1} that $k$ must take certain integer
values. Consequently, ordinary perturbation theory in $M$ is expected to
be (in general) ill-defined.  Our way to solve the quantum CS theory 
in a generic manifold $M$ consists of three steps. Firstly, we will solve the 
theory in $\mathbb{R}^3$, where ordinary perturbation theory is reliable and 
defines the theory unambiguously.  Secondly, by taking into account the 
behavior of the action under large gauge transformations,  we will extend the 
results obtained in $\mathbb{R}^3$ to the case of  the three-sphere $S^3$.  Finally, 
we will use the symmetry properties of the topological theory to solve the 
model in a generic three-manifold $M$. In this context, the relevant symmetry 
we will need to consider is related to twist homeomorphisms of solid tori.

\chapter{\bf Properties of the observables}

\section{\bf Discrete Symmetries}

Let  $\rho$ be a representation of $G$; the generic element $U \in G$ in the
representation $\rho$ can be written as
\beeq 
U_{\rho} \; = \; \exp \left (i T_a[\rho] \theta^a \right ) \quad .
\end{equation}
The complex conjugate of $U_{\rho}^\ast$ gives a new representation 
$\rho^{\ast}$ of $G$, 
\beeq 
U_{\rho^\ast} \; = \; U_{\rho}^\ast \; = \; \exp \left (- i {T_a[\rho]}^\ast 
\theta^a \right ) \quad . \label{comp}
\end{equation}
From Eq.(\ref{comp}), it follows that the generators $T_a[\rho^\ast]$ in 
the representation $\rho^\ast$ are given by  $-{T_a[\rho]}^{\ast}$.
For a compact group any finite dimensional representation is equivalent to a unitary 
representation. Thus, without loss of generality, we shall consider only 
unitary representations of $G$, i.e. ${T_a[\rho]}^\dagger = T_a[\rho]$.  

Let $W(C; \rho)$ be a Wilson line associated with the knot $C$  and with colour 
$\rho$
\beeq 
W \left (C;\rho \right )\; = \; {\rm Tr} \, P \exp \left (i \int_C T_a[\rho]
A^a_{\mu} dx^{\mu} \right ) \quad .  
\end{equation}
As a first example of discrete symmetry, we shall study the behavior of 
$W \left (C;\rho \right )$ under the replacement of $\rho$  with  
$\rho^\ast$. By definition we have
\beeq 
W \left (C; \rho^{\ast} \right ) \; = \; {\rm Tr} \, P \exp \left (-i \int_C
T_a[\rho^\ast] A^a_{\mu} dx^{\mu} \right ) \; = \; {\rm Tr} \, P \exp \left (-i \int_C T[\rho]^t_a A^a_{\mu} dx^{\mu} \right ) \quad . \label{dis1}
\end{equation}
It is not difficult to show by direct inspection that  one can replace $T[\rho]^t_a$ by $T[\rho]_a$ in (\ref{dis1}) simply by reversing the orientation of $C$
and by introducing an overall minus sign in the exponent, i.e.
\beeq
W \left (C; \rho^{\ast} \right ) \; = \; {\rm Tr} \, P \exp \left (i \int_{C^{-1}}T[\rho]_a A^a_{\mu} dx^{\mu} \right ) \quad ,
\end{equation}
where $C^{-1}$ denotes  the knot obtained from $C$ by reversing its orientation.
As result we get \cite{guad1}
\beeq 
W \left (C; \rho^{\ast} \right ) \; = \; W \left (C^{-1}; \rho \right) \quad .
\label{conj}
\end{equation}
The extension of (\ref{conj}) to a generic link $L$ 
is straightforward
\beeq 
E \left (C_1^{-1}, \cdots ,C_m^{-1};\rho_1, \cdots ,\rho_m \right ) \; = \; E \left
(C_1,\cdots,C_m;\rho_1^{\ast},\cdots,\rho_m^{\ast} \right ) \quad . 
\label{inv}
\end{equation}
Clearly the distinction between a representation $\rho$ and its complex conjugate
is a matter of convention; as a consequence, given a link $L$ with components $\{C_1, \cdots, C_m \}$ with colours $\{\rho_1, \cdots, \rho_m \}$, the expectation 
value of the associated observable should be invariant under the replacement of
$\{\rho_1, \cdots, \rho_m \}$ with $\{\rho_1^\ast, \cdots, \rho_m^\ast \}$. More
explicitly
\beeq 
E \left (C_1, \cdots ,C_m;\rho_1, \cdots ,\rho_m \right ) \; = \; E \left
(C_1,\cdots,C_m;\rho_1^{\ast},\cdots,\rho_m^{\ast} \right ) \quad . 
\label{conj1}
\end{equation} 
From equations (\ref{conj1}) and (\ref{inv}) one gets
\beeq 
E \left (C_1,\cdots,C_m;\rho_1,\cdots,\rho_m \right )\; =\; E \left
(C_1^{-1},\cdots,C_m^{-1};\rho_1,\cdots,\rho_m \right ) \quad  . 
\label{orie}
\end{equation}
Equation (\ref{orie}) shows that for the expectation value of $W(L)$, only the 
relative orientation of the $L$ components is relevant. When a representation
$\rho$ is  real, i.e. $\rho \sim \rho^{\ast}$, one can find 
 a non-singular matrix $V$ such that
\beeq 
T[\rho]_a^{\ast} \; = \; - \; VT[\rho]_a V^{-1} \quad  . 
\label{equiv}
\end{equation}
By using $\exp \left( VBV^{-1} \right )=V \exp (B) V^{-1}$ and the cyclicity
of the trace one obtains \cite{guad1}
\beeq 
W \left (C;\rho^{\ast} \right) \; = \; W \left(C;\rho \right ) \; = \; W
\left(C^{-1};\rho \right ) \quad . 
\label{real}
\end{equation}
Equation (\ref{real}) implies that when $W(L)$ is associated with real 
representations, the regular isotopy invariant $E(L)$ does not depend
on the orientation of its components $\{C_1, \cdots, C_m \}$. 

Another useful  property can be deduced by taking the complex conjugation of the expectation value $E(L)$ of $W(L)$. By using the path integral representation 
of $E(L)$ it follows
\beeq E^{\ast}(L) \; = \; \int D \left [A,B,\bar{c},c \right ] \exp \left
(-iS_{tot} \right ) W^{\ast} \left(C_1,\cdots,C_m;\rho_1,\cdots,\rho_m
\right) \quad  . \label{c1}
\end{equation}
On the other hand, from equation (\ref{inv})
\bea &&W^{\ast} \left(C_1,\cdots,C_m;\rho_1,\cdots,\rho_m \right)
\nb \\ 
\qquad &&= \; W\left(C_1,\cdots,C_m;\rho_1^{\ast},\cdots,\rho_m^{\ast} \right)
\; = \; W \left(C_1^{-1},\cdots,C_m^{-1};\rho_1,\cdots,\rho_m \right )
 \quad . \label{c2}
\ena
The change in sign of $S_{eff.}$ is equivalent to reversing the orientation of 
$R^3$ ($S^3$), or to changing $k$ into $-k$.
Putting all the pieces together one gets \cite{guad1}
\beeq 
E^{\ast} \left(L,-k \right) \; = \; E \left (L^{-1},k \right ) \quad .
\end{equation}
\section{\bf Tensor algebra}
Let 
${\cal {R}}_G$ be the set of representations of a compact and 
simple  Lie group $G$. If $\rho_1,\rho_2 \in {\cal{R}}_G$, then  also $\rho_1 
\oplus \rho_2 \in {\cal{R}}_G$ and $\rho_1 \otimes \rho_2 \in {\cal{R}}_G$.
As a consequence ${\cal{R}}_G$ is actually a ring. ${\cal{R}}_G$ can be 
extended
to an associative and  commutative algebra ${\cal T}$ over the field 
$\mathbb{C}$ of the complex numbers \cite{guad2}. The product between two elements $\chi_1$ and $\chi_2$ of ${\cal T}$
will be denoted by $\chi_1 \chi_2$ and the identity element of ${\cal T}$ 
corresponding to the trivial representation of $G$ by $\chi[\boldmath{1}]$. 
The structure constants of $\cal T$ are determined by the
number of irreducible representations contained in the tensor
product of two given representations. To be more precise, for each irreducible
representation $\rho $ of the gauge group, we shall introduce an element 
$\chi [ \rho ] \in {\cal T}$; the set $\{ \, \chi [ \rho ] \, \}$ of all these
elements, which are defined for all the inequivalent irreducible
representations, is the set which contains the elements of the standard basis
of $\cal T$. The structure constants of $\cal T$ are given by 
\beeq
\chi [ \rho_1] \; \chi [\rho_2 ] \; = \; \sum_{\rho } \; 
F_{\rho_1 \, , \, \rho_2 \, , \, \rho}^{} \; \; \chi [\rho ] \quad , 
\end{equation}
where $F_{\rho_1 \, , \, \rho_2 \, , \, \rho}^{}$ is the multiplicity 
of the irreducible representation $\rho $ which is contained in the decomposition of the 
tensor product $\rho_1 \otimes \rho_2 $. If $\rho \notin \rho_1 \otimes \rho_2 $, then
the corresponding coefficient $F_{\rho_1 \, , \, \rho_2 \, , \, \rho}^{}$ is
vanishing.  In the standard basis of $\cal T$, the
structure constants take non-negative integer values. The symmetry properties 
of $\{ \, F_{\rho_1 \, , \, \rho_2 \, , \, \rho}^{}\, \}$ are fixed by the simple
Lie algebra structure of the gauge group; namely, one has 
\beeq
F_{\rho_1 \, , \, \rho_2 \, , \, \rho}^{} \; = \; 
F_{\rho_2 \, , \, \rho_1 \, , \, \rho}^{} \qquad , 
\label{lie1}
\end{equation}
\beeq
F_{\rho_1 \, , \, \rho_2 \, , \, \rho}^{} \; = \; 
F_{\rho_1^{*} \, , \, \rho_2^{*} \, , \, \rho^{*}}^{}  \qquad , 
\label{lie2}
\end{equation}
and 
\beeq
F_{\rho_1 \, , \, \rho_2 \, , \, \rho}^{} \; = \; 
F_{\rho_1^{*} \, , \, \rho \, , \, \rho_2}^{} \qquad , 
\label{lie3}
\end{equation}
where $\rho^{*}$ is the complex conjugate representation associated with the irreducible representation $\rho$.

Let the finite-dimensional representation $\rho$ be associated with the framed
and oriented knot $C$. The Wilson line operator $W(C;\rho )$ is also well
defined when $\rho $ is not irreducible. Indeed, in this case we can decompose $\rho $ into a direct sum of its irreducible components and, since $W(C ; \rho )$ is the 
trace of the quantum holonomy, $W(C ; \rho )$ can accordingly be written as a 
sum of the Wilson line operators defined for these irreducible components.
Consequently, for fixed knot $C$, $\, E(C)$ can be understood as a linear
function on the representation ring of the gauge group. Given a link $L$ with $m$ components, 
the expectation value  $E(L)$ in $\mathbb{R}^3(S^3)$ can be understood \cite{guad2} as a
multi-linear function on ${\cal T}^{\otimes n} $ 
\beeq 
E: \; \overbrace{{\cal{T}} \otimes \cdots \otimes {\cal{T}}}^{m} 
\rightarrow \mathbb{C} \quad . 
\end{equation}
Indeed, let $\chi \; = \;  \sum_{i} d(i) \; \chi_i$ be a generic element of
${\cal{T}}$, the linear extension of $E \left(C;\chi_i \right)$ into
${\cal{T}}$ is defined as 
\beeq E \left(C;\chi \right) \; = \; \sum_i d(i) \; E
\left(C;\chi_i \right) \quad , \end{equation} 
where $C$ is a generic knot. The case of $W(L)$ is similar: given a link $L$ with $m$ components $\left \{ C_1,\cdots ,C_m \right \}$, the expectation value 
$E(L)$, for fixed $L$ is a multi-linear functional on 
${\cal{T}}^{\otimes m} = \overbrace{{\cal{T}} \otimes \cdots
\otimes {\cal{T}}}^{m}$ defined as
\beeq E \left(L;,\chi^{(1)},\cdots ,\chi^{(m)} \right) \; = \; \prod_{a=1}^{m} 
\sum_{j_a} d_a \left(j_a \right) \; \chi_{j_a}
E\left(L;\chi_{j_1},\cdots ,\chi_{j_m}\right) \quad , 
\end{equation}
where 
\beeq \chi^{(a)} \; = \; \sum_{j_a} d_a(j_a) \, \chi_j \qquad  
a=1, \cdots ,m  
\end{equation}
is the colour of the $a$th component of $L$. 

\section{\bf{Satellite relations}}
Let us consider an oriented framed knot $C$ and the set $\{ \, W(C ; \rho ) \,
\}$ of the associated Wilson line operators which are defined for all the
possible inequivalent irreducible representations $\{ \, \rho \, \}$ of the
gauge group. The set $\{ \, W(C ; \rho ) \, \}$ is a complete set of gauge
invariant observables associated with the knot $C$ \cite{glib}. This means that, if 
${\cal O}(C)$ is a metric-independent gauge invariant observable of the CS theory 
which is defined in terms of the vector fields $A^a_\mu (x)$  and 
${\cal O}(C)$ is associated with the knot $C$, then ${\cal O}(C)$ can always be
written as 
\beeq
{\cal O}(C) \; = \; \sum_{\rho } \, \xi ( \rho ) \; W(C ; \rho ) \; = \; 
\sum_{\rho } \, \xi ( \rho ) \; W(C ; \chi [ \rho ] ) \quad , 
\label{eq47}
\end{equation}
where $\{ \, \xi ( \rho ) \, \}$ are numerical (complex)
coefficients. With a given choice of the framing of $C$, the coefficients 
$\{ \, \xi ( \rho ) \, \}$ are fixed and characterize the observable 
${\cal O}(C)$. Any function, which is defined on the equivalence classes of
conjugate elements of a compact and simple group, admits a linear decomposition in
terms of the characters of the group, of course \cite{seil}. Eq.(\ref{eq47}) is the 
analogue of
this decomposition for the gauge invariant observables which are associated
with the knot $C$. 

Eq.(\ref{eq47}) can be used to derive the generalized satellite relations which
\cite{glib} are satisfied by $E(L)$.  Let $V$ be a solid torus standardly embedded in
$S^3$; the oriented core of $V$ will be denoted by $K$ and its preferred framing
by $K_f$.  Consider now the oriented and framed component $C$ of a link $L$ in
$S^3$; let $C_f$ be the framing of $C$.  A tubular neighborhood $N$ of $C$ is a
solid torus embedded in $S^3$ whose core is $C$. The two solid tori $V$ and
$N$ are homeomorphic; we shall denote by $h^\diamond$ the homeomorphism 
$h^\diamond \, : \, V \, \rightarrow \, N$ which has the properties 
\beeq
h^\diamond (K) \; = \; C \quad , \quad {\rm and} \qquad h^\diamond (K_f) \; =
\; C_f \qquad \; .  
\end{equation}
Up to an ambient isotopy, the homeomorphism $h^\diamond$ is unique
and is determined by the orientation and by the framing of the link component
$C$.   

One can imagine that the framed component $C$ of $L$ is the image 
$h^\diamond (K)$ of the framed knot $K \subset V$ under the homeomorphism
$h^\diamond $. Starting from the link $L$, we shall now construct a new link
$L^\prime$;  $\, L^\prime$ is obtained by replacing the component $C$ of
$L$ by the image  $h^\diamond  (P)$ of a given (oriented and framed) link
$P$ in $V$. The link $L^\prime$ is called a generalized satellite of $L$;  
the link $L$ is a companion of $L^\prime$ and  
$P \subset V$ is called the pattern link. 

Suppose now that $L^\prime$ is a generalized satellite of $L$ defined in
terms of a given pattern link $P$. Furthermore, suppose that an irreducible
representation of the gauge group has been assigned to each component of
$P$.  We would like to know how the expectation value $E(L^\prime )$ is connected
with $E(L)$; the  precise relation between $E(L^\prime )$ and $E(L)$ is
called a generalized satellite relation. Remember that $L^\prime$ has been
obtained from $L$ by replacing the framed component $C$ with the image
$h^\diamond  (P)$ of the pattern link $P$. By definition, $P$ belongs to the
solid torus $V$ and $W(P)$, which denotes the  product of the Wilson line
operators associated with $P$, represents a gauge invariant observable defined
in $V$. Since the CS model is a topological field theory, the thickness  of $V$ 
is totally irrelevant. In the limit in which this thickness goes to zero, the
solid torus $V$ degenerates to a simple circle: the core $K$ of $V$. Thus,
$W(P)$ can be understood as a gauge invariant observable associated with the
knot $K$. Consequently, $W(P)$ admits an expansion of the type shown in 
eq.(\ref{eq47})
\beeq
W(\, P\, ) \; = \; \sum_{ \rho} \> \xi (\rho ) \; W(\, K \, ; \,  \chi [\rho ]\, )
\quad .       
\label{eq49}
\end{equation}
The complex coefficients $\{ \, \xi ( \rho ) \, \}$, appearing in
eq.(\ref{eq49}), depend on the pattern link $P$ and on the representations
assigned to its components. The composite Wilson line
operators $\{ \, W(K, \chi [\rho ]\, ) \, \}$ are defined for the framed knot $K$ (which
has preferred framing). Let us introduce the element $\> \chi $, 
\beeq
\chi \; = \; \sum_{ \rho } \> \xi (\rho ) \; \chi [ \rho ] \quad , 
\label{eq410}
\end{equation}
of the tensor algebra $\cal T$. Then, eq.(\ref{eq49}) can simply be
written as 
\beeq
W(\, P\, ) \; = \; W(\, K\, ; \, \chi \, ) \quad . 
\label{cart}
\end{equation}

\noindent At this stage, from the defining conditions (4.8) of $h^\diamond
$, it follows immediately \cite{guad1} that 
\beeq
E(L^\prime ) \; = \; E(L; \; {\rm with~the~component~} C {\rm ~of~} L {\rm
~associated ~with~} \chi \, ) \quad . 
\label{eq412}
\end{equation}
This equation is a consequence of two basic symmetry properties of the
CS theory. Firstly, gauge invariance implies that the non-Abelian ``electric"
flux associated with a meridinal disc of a solid torus must be conserved.
Secondly, general covariance implies that this flux can always be imagined to be concentrated on a single knot which coincides with the core of the solid torus. 

The satellite relation (\ref{eq412}) represents one of the main properties of the
expectation values $\{ \, E(L) \, \}$.  The $\xi$-coefficients,  appearing in
eq.(\ref{eq49}), can be determined by using several different methods; some of them have been presented in Ref.[3,4].  

Let us consider a particular example of
satellite relation which will be used frequently in the following sections.  
We shall consider a particular pattern link $B$ in the solid torus $V$. Since 
the complement of a tubular neighborhood $M$ of the circle $U$ in $S^3$ is a
solid torus standardly embedded in $S^3$, we can take $V = S^3 - \dot{M}$, where $\dot{M}$ is the interior of $M$. Consequently, we shall represent $B$ in the
complement of $U$ in $S^3$. The pattern link $B$, in which we are interested, is
the two component framed and oriented link shown in Fig.2.1; each  component  of
$B$ has preferred framing.

\begin{figure}[h]
\vskip 0.9 truecm 
\centerline{\epsfig{file=\path 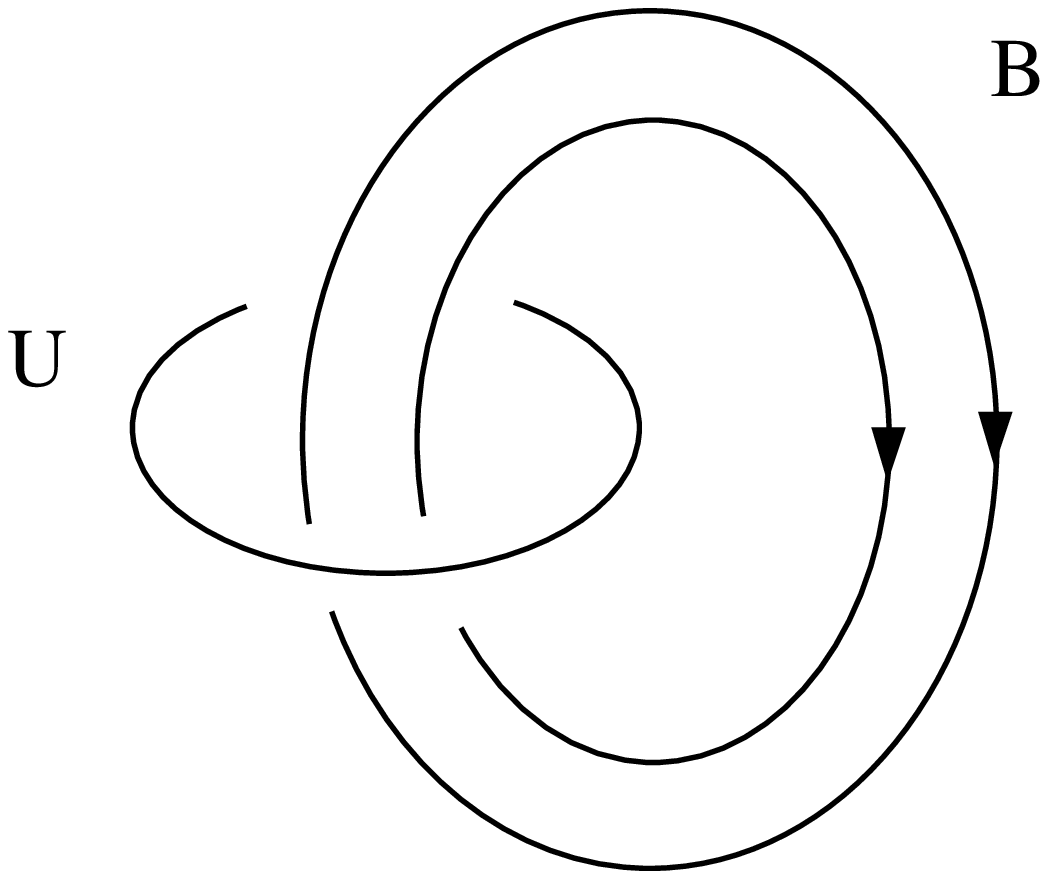,height=4cm,width=5cm}}
\vskip 0.5 truecm 
\centerline {{\bf Figure 2.1}}
\vskip 0.5 truecm 
\end{figure}

Let $\rho_1$ and $\rho_2$ be the irreducible representations which
are assigned to the two components $C_1$ and $C_2$ of $B$; we shall derive the explicit form the coefficient $\xi{\rho}$ appearing in the decomposition 
decomposition (\ref{eq49}). By definition $C_1$ e $C_2$ have preferred framing
i.e. 
\beeq \chi \left (C_1,C_{1f} \right ) \; = \; \chi \left (C_2,C_{2f} \right ) \; = \; \chi \left (C_1,C_2 \right ) \; = \; 0  \quad  . 
\label{fram1}
\end{equation}
In the limit in which the ``distance'' between $C_1$ and $C_2$ goes to zero,
because of (\ref{fram1}), the knot $C_2$ cannot be distinguished from the 
framing $C_{1f}$ of $C_1$. Thus in this limit we can replace the two Wilson
lines $W(C_1,\rho_1)$ and $W(C_2,\rho_2 )$ with a single Wilson line
$W(K,\rho^\prime)$, with $K$ is a knot ambient isotopic to $C_1$ (see figure 2.2).

\begin{figure}[h]
\begin{picture}(10,10)
\put(175,-120){$\rho_1$}
\put(195,-120){$ \rho_2$}
\put(278,-120){$\rho_1 \otimes \rho_2$}
\end{picture}
\vskip 0.5 truecm 
\centerline{\epsfig{file=\path 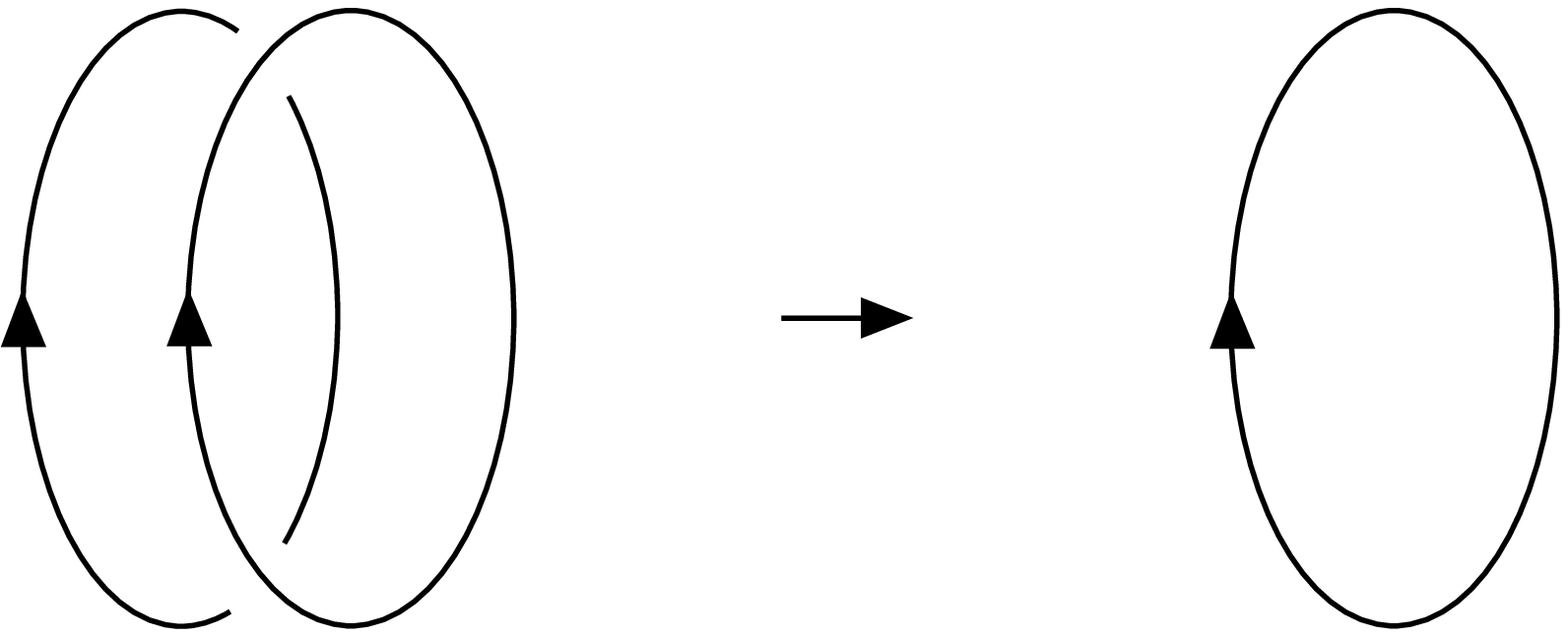,height=3cm,width=5cm}}
\vskip 0.9 truecm 
\centerline{{\bf Figure 2.2}}
\vskip 0.5 truecm 
\end{figure}

In particular
one can identify $K$ with the core of the solid torus $V$ in which the pattern 
link $B$ is contained. In order to determine $\rho^\prime \in {\cal T}$, we 
note that the representation of $G$ associated with $W(K)$ is contained in
$\rho_1 \otimes \rho_2$. Indeed  
\beeq tr({\rho_1}) \; tr({\rho_2}) \; = \; tr({\rho_1 \otimes \rho_2} )\quad  . 
\end{equation}
In conclusion the element $\rho^\prime$ has to be identified with $\chi[\rho_1]\chi[\rho_2]$, thus one gets
\bea
&&W( \, B\, ; \, \chi [\rho_1 ]\, , \, \chi [ \rho_2 ]\, ) \; = \; W(\, K \, ; \, \chi [\rho_1] \, \chi [\rho_2] \, ) \nb \\
&& \qquad \qquad \qquad \qquad \quad \; = \; \sum_{\rho} \, F_{\rho_1 \, , \, \rho_2 \, , \, \rho} \; \; W(\, K\, ; \,
\chi [\rho ] \, ) \quad .
\label{eq413}
\ena
where $K$ is the framed core of $V$ which has preferred framing and
is oriented as the components of $B$, see Fig.2.3.  In eq.(\ref{eq413}) one 
can easily recognize the structure constants of $\cal T$. 

\begin{figure}[h]
\vskip 0.9 truecm 
\centerline{\epsfig{file=\path 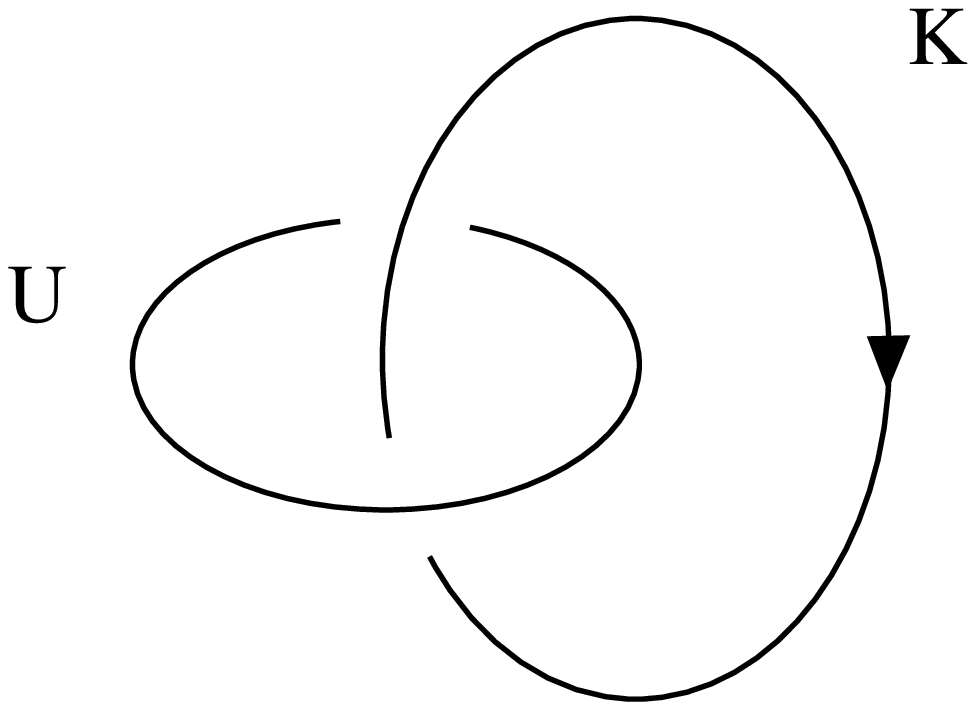,height=4cm,width=5cm}}
\vskip 0.5 truecm 
\centerline {{\bf Figure 2.3}}
\vskip 0.5 truecm 
\end{figure}

Thus, for the satellites constructed with the pattern link $B$, eq.(\ref{eq412}) takes the form 
\bea
&&E(\, L^\prime \, ) \; \; = \; \;  
E( \, h^\diamond (B)\, , C^\prime \, , \, ... \, ; \, \chi [\rho_1 ]\, , \, 
\chi [ \rho_2 ]\, , \, \chi [\rho^\prime ] \, , \, ... \, ) \nb \\  
&&\qquad \qquad = \;  \; E(\, C \, , C^\prime \, , \, ... \, ; \, \chi [\rho_1] \,  \chi [\rho_2] \, ,
\,  \chi [\rho^\prime ] \, , \,  ... \, ) \nb \\ 
&& \qquad \qquad = \;   \sum_{\rho} \, F_{\rho_1 \, , \, \rho_2 \, , \, \rho} \; \; E(\, C\, , \, 
C^\prime \, , \, ... \, ; \,  \chi [\rho ] \, , \, \chi [\rho^\prime ] \, ,
\, ...\, )   \quad . \label{satg}
\ena

\section{\bf Calculation rules}
Topological field theories are rigid in the sense that, due to the 
presence of a large symmetry, the expectation values of the observables are 
strongly constrained. These constraints can be put in the form of consistency 
relations that the observables must satisfy. In the case of Chern-Simons
theory, the consistency relations are stringent enough to determine uniquely 
the values of the observables. In other words, Chern-Simons theory is exactly
solvable. In this section we shall give a few rules \cite{guad1,glib} 
that will permit us to solve the theory.

\centerline{\bf (1) Projection decomposition} 
As we shown in appendix B, link diagrams can be understood as closures
of braids. 

\begin{figure}[t]
\begin{picture}(10,10)
\put(125,-165){$\rho_1$}
\put(145,-165){$ \rho_2$}
\put(315,-10){$\rho_1$}
\put(345,-10){$ \rho_2$}
\put(315,-165){$ \rho_1$}
\put(345,-165){$ \rho_2$}
\put(345,-90){$\rho(t) $}
\end{picture}
\vskip 0.5 truecm 
\centerline{\epsfig{file=\path 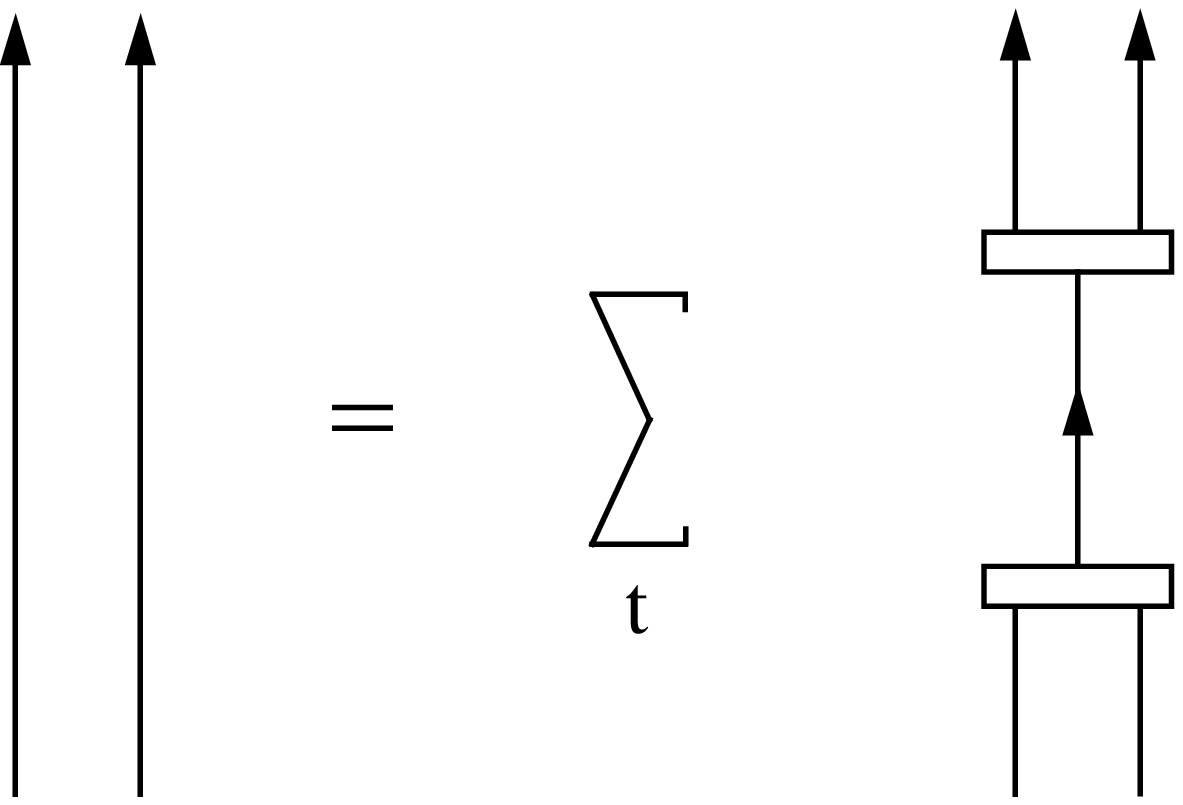,height=5cm,width=8cm}}
\vskip 0.5 truecm 
\centerline{{\bf Figure 2.4}}
\vskip 0.5 truecm 
\end{figure}
\begin{figure}[h]
\begin{picture}(10,10)
\put(150,-135){$\rho(t)$}
\put(150,-60){$ \rho(s)$}
\put(265,-100){$\delta_{s, \,t}$}
\put(350,-100){$ \rho(t)$}
\end{picture}
\vskip 0.5 truecm 
\centerline{\epsfig{file=\path 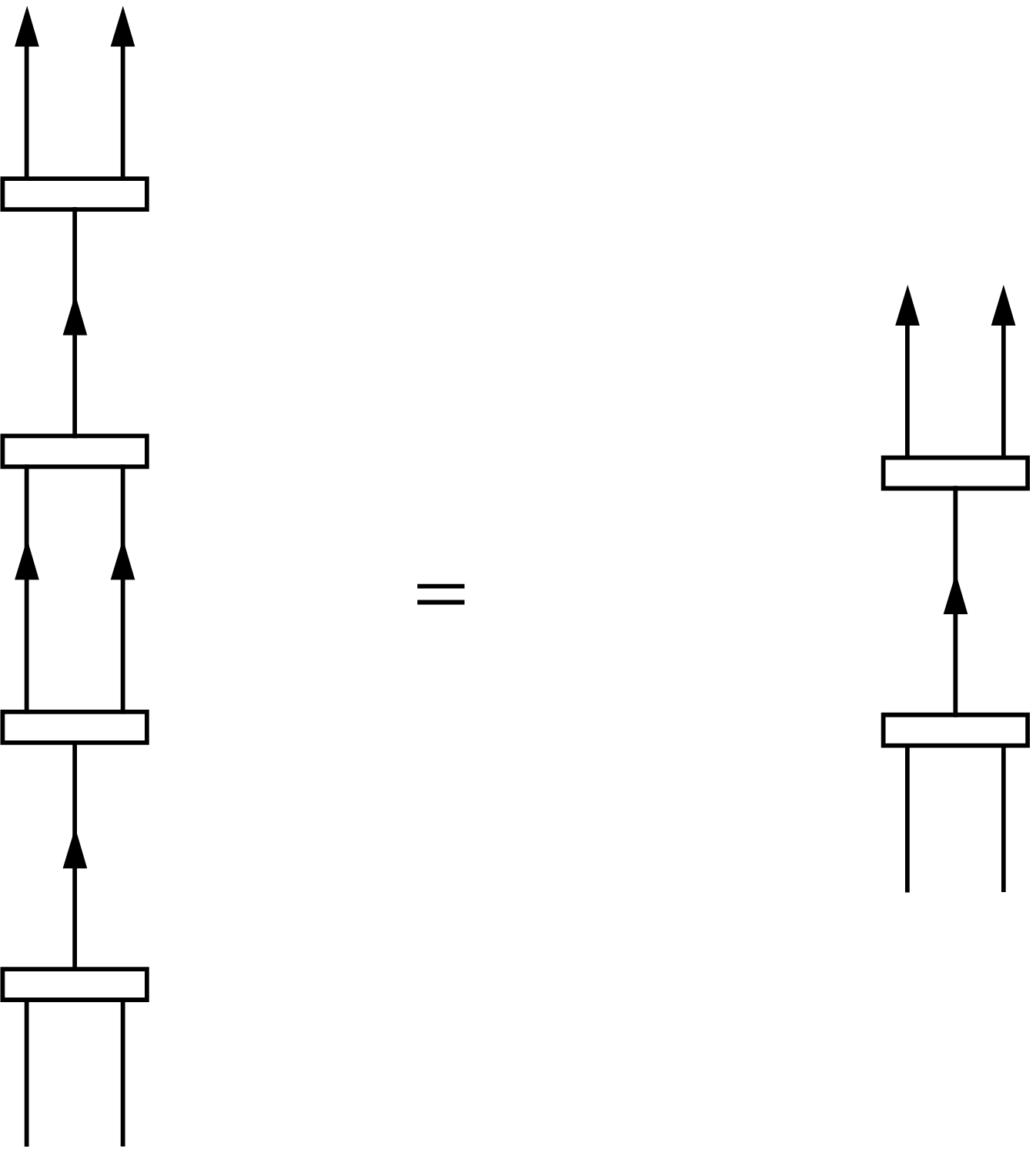,height=5.5cm,width=8cm}}
\vskip 0.5 truecm 
\centerline{{\bf Figure 2.5}}
\vskip 0.5 truecm 
\end{figure}

Let us consider the configuration corresponding to two parallel
strings associated with the irreducible representation $\rho_1$ and $\rho_2$
of $G$ satisfying
\beeq
\rho_1 \otimes \rho_2 \; = \; \sum_t \rho(t) \quad .
\label{pdec}
\end{equation}  
The projection decomposition rule involves a local modification of the link
diagram obtained by introducing a set of orthogonal projectors as shown 
in Fig.2.4. By means of the projection decomposition rule, the various 
irreducible representations entering (\ref{pdec}) are single out.
The orthogonality property of the projectors is expressed by the relation
shown in Fig.2.5.

\centerline{\bf (2) Crossing}
The Chern-Simons field theory can be quantized by using the canonical 
formalism rather than then the covariant BRS method. Although the method of 
canonical quantization is not viable in the calculation of expectation values 
of observables, it can be profitably used in the study of Chern-Simons 
monodromies. The configuration space $\mathbb{R}^3$ is sliced in space-like
surfaces, in any slice the equal-time canonical commutation relations are 
imposed. 

\begin{figure}[t]
\begin{picture}(10,10)
\put(57,-170){$\rho_1$}
\put(77,-170){$ \rho_2$}
\put(383,-8){$\rho_1$}
\put(403,-8){$ \rho_2$}
\put(406,-81){$ \rho(t)$}
\put(386,-170){$ \rho_1$}
\put(406,-170){$ \rho_2$}
\end{picture}
\vskip 0.5 truecm 
\centerline{\epsfig{file=\path 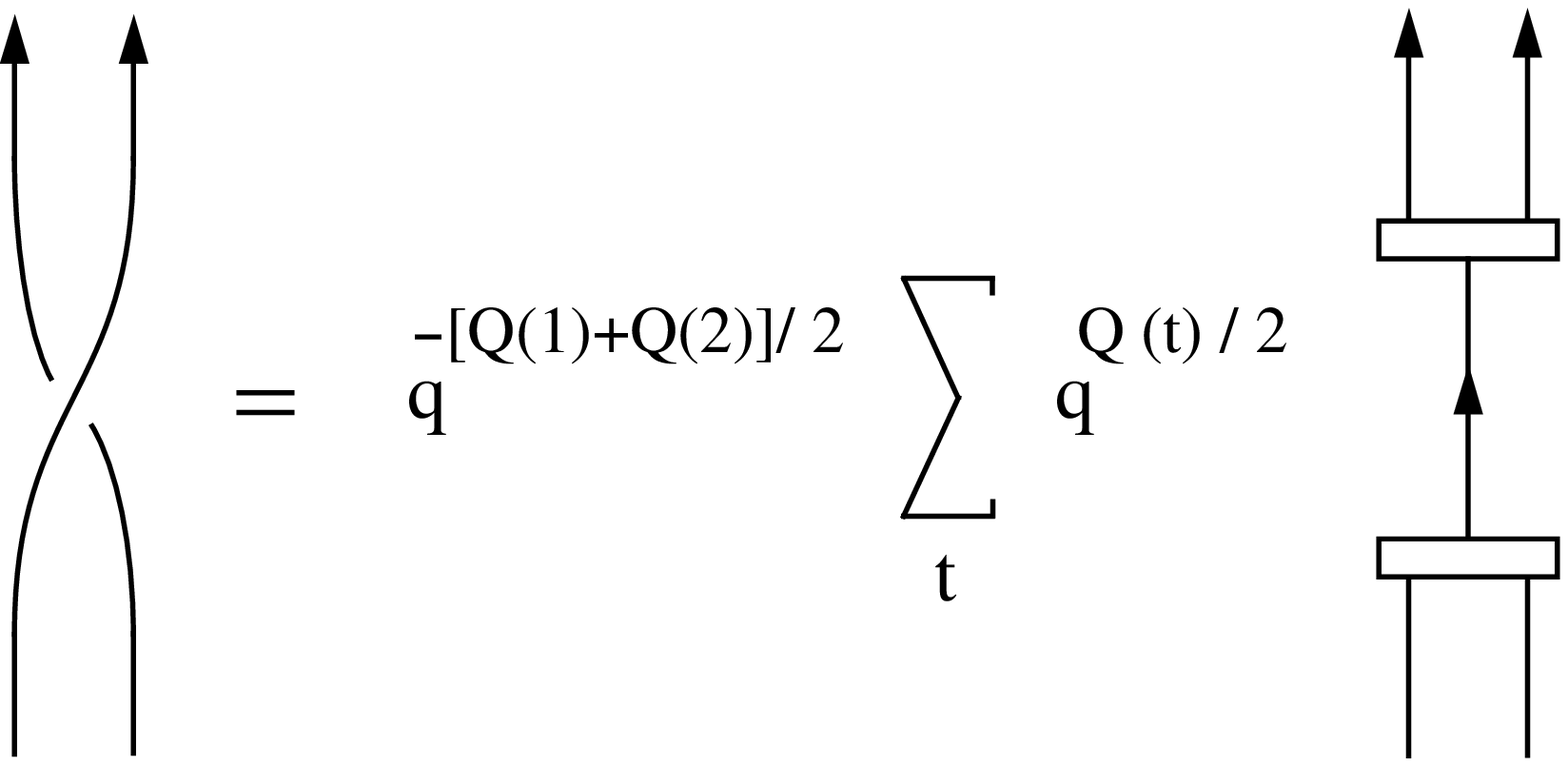,height=5cm,width=13cm}}
\vskip 0.5 truecm 
\centerline{{\bf Figure 2.6}}
\end{figure}
\begin{figure}[h]
\begin{picture}(10,10)
\put(57,-170){$\rho_1$}
\put(77,-170){$ \rho_2$}
\put(383,-8){$\rho_1$}
\put(403,-8){$ \rho_2$}
\put(406,-81){$ \rho(t)$}
\put(386,-170){$ \rho_1$}
\put(406,-170){$ \rho_2$}
\end{picture}
\vskip 0.5 truecm 
\centerline{\epsfig{file=\path 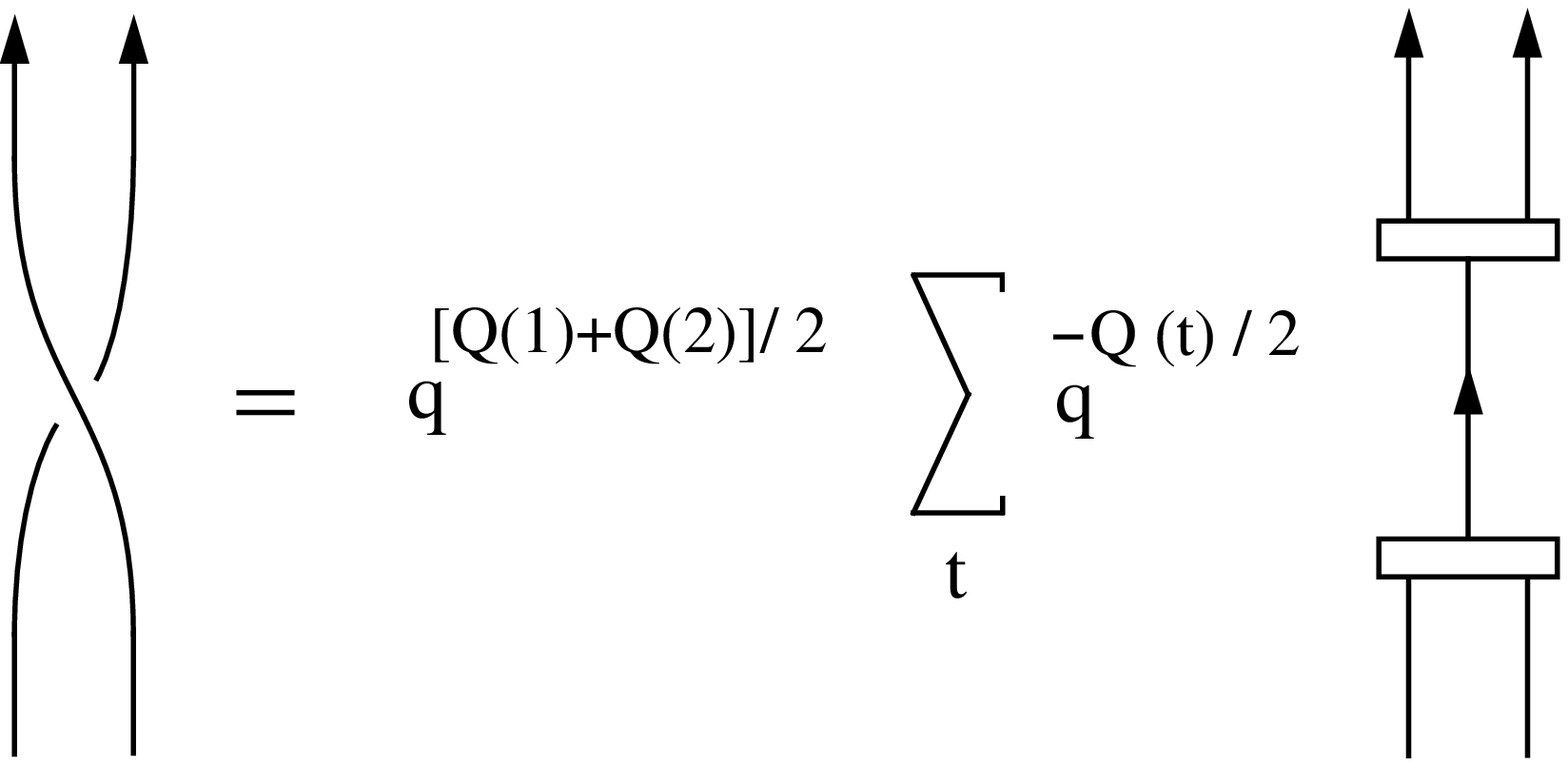,height=5cm,width=13cm}}
\vskip 0.9 truecm 
\centerline{{\bf Figure 2.7}}
\vskip 0.5 truecm 
\end{figure}

The state $\Psi$, corresponding to a configuration with two strings piercing
 the surface $t=0$ and associated with the irreducible representations 
$\rho_1, \, \rho_2$,  is interpreted as the introduction in the vacuum state 
of two static sources with quantum numbers $\rho_1$ and $\rho_2$. The problem 
of how the state $\Psi$ changes when the sources are exchanged can be analyzed
by using the canonical quantization in the Schroedinger representation
for the field. If the new state after the exchange is denoted by $\Psi^\prime$,
 one obtains
\beeq
\Psi^\prime \; = \; M \, \Psi  \quad ;
\end{equation}
for notation simplicity all indices have been suppressed. The matrix $M$ that encodes
the braiding properties of Chern-Simons theory is given by \cite{can}
\beeq
M \; = \; \Pi_{12} \, q^{T_{(\rho_1)}^a \otimes T_{(\rho_2)}^a} \quad ,
\label{brad}
\end{equation}
where $\Pi_{12}$ is the permutation operator of the two sources and 
\beeq
q \; = \; e^{-i 2 \pi/k} \quad 
\end{equation}
is the so-called deformation parameter. 
Eq.(\ref{brad}), togheter with the projection decomposition rule and the identity
\beeq
T_{(\rho_1)}^a \otimes T_{(\rho_2)}^a \; = \; \frac{1}{2} \left[Q(\rho_1 
\otimes \rho_2) \; - \; Q(\rho_1) \; - \; Q(\rho_2) \right] \, \mathbf{1} 
\otimes \mathbf{1} 
\end{equation}
lead to the crossing rule shown in Fig.2.6 and in Fig.2.7. 

\centerline{\bf (3) Twisting}
This rule refers to the behavior of a braid under a Markov move of type 2 
(see App.B) and is summarized in Fig.2.8

\begin{figure}[h]
\begin{picture}(10,10)
\put(122,-115){$\rho$}
\put(180,-73){$ = \; q^{+Q(\rho)}$}
\put(245,-115){$\rho$}
\put(122,-250){$\rho$}
\put(180,-210){$= \;  q^{-Q(\rho)}$}
\put(245,-250){$\rho$}
\end{picture}
\vskip 0.5 truecm 
\centerline{\epsfig{file=\path 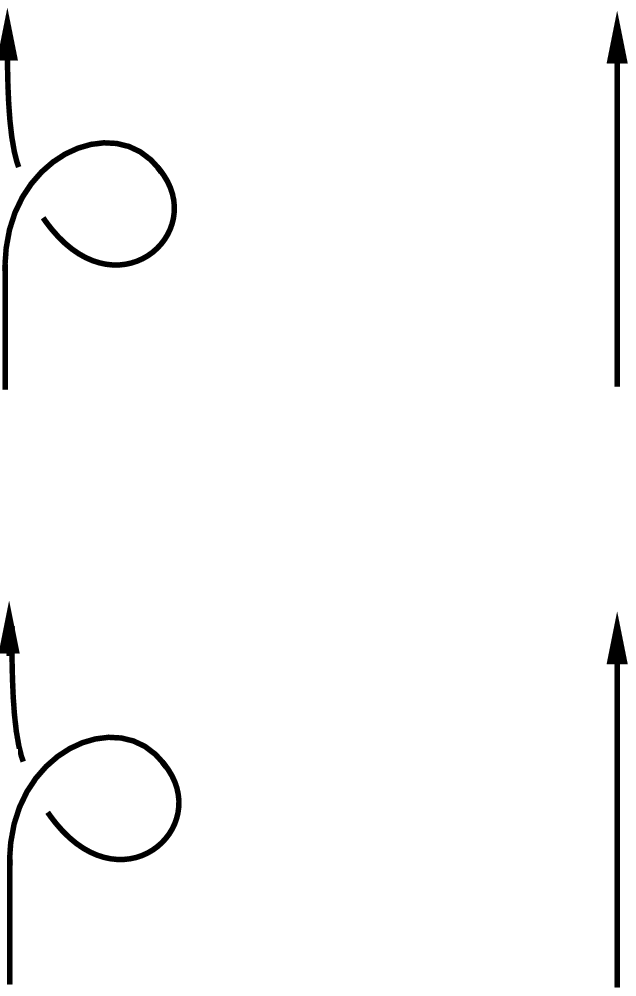,height=8cm,width=8cm}}
\vskip 0.9 truecm 
\centerline{{\bf Figure 2.8}}
\vskip 0.5 truecm 
\end{figure}

\no
The value $q^{\pm Q(\rho)}$ of the twist variable $\alpha(\rho)$ fits
the perturbative analysis given in \cite{gm2}. In addition, it can be 
shown \cite{glib} that is the only value with the right classical limit 
($\alpha \ra 1$ when $k \ra \infty$) and compatible with the satellite relations and with 
expression (\ref{brad}) for the braid matrix.

\begin{figure}[h]
\begin{picture}(10,10)
\put(125,-80){$\rho_1$}
\put(170,-80){$ \rho_2$}
\put(160,-115){$\rho$}
\put(125,-160){$\rho_1$}
\put(170,-160){$\rho_2$}
\put(330,-190){$\rho$}
\end{picture}
\vskip 0.5 truecm 
\centerline{\epsfig{file=\path 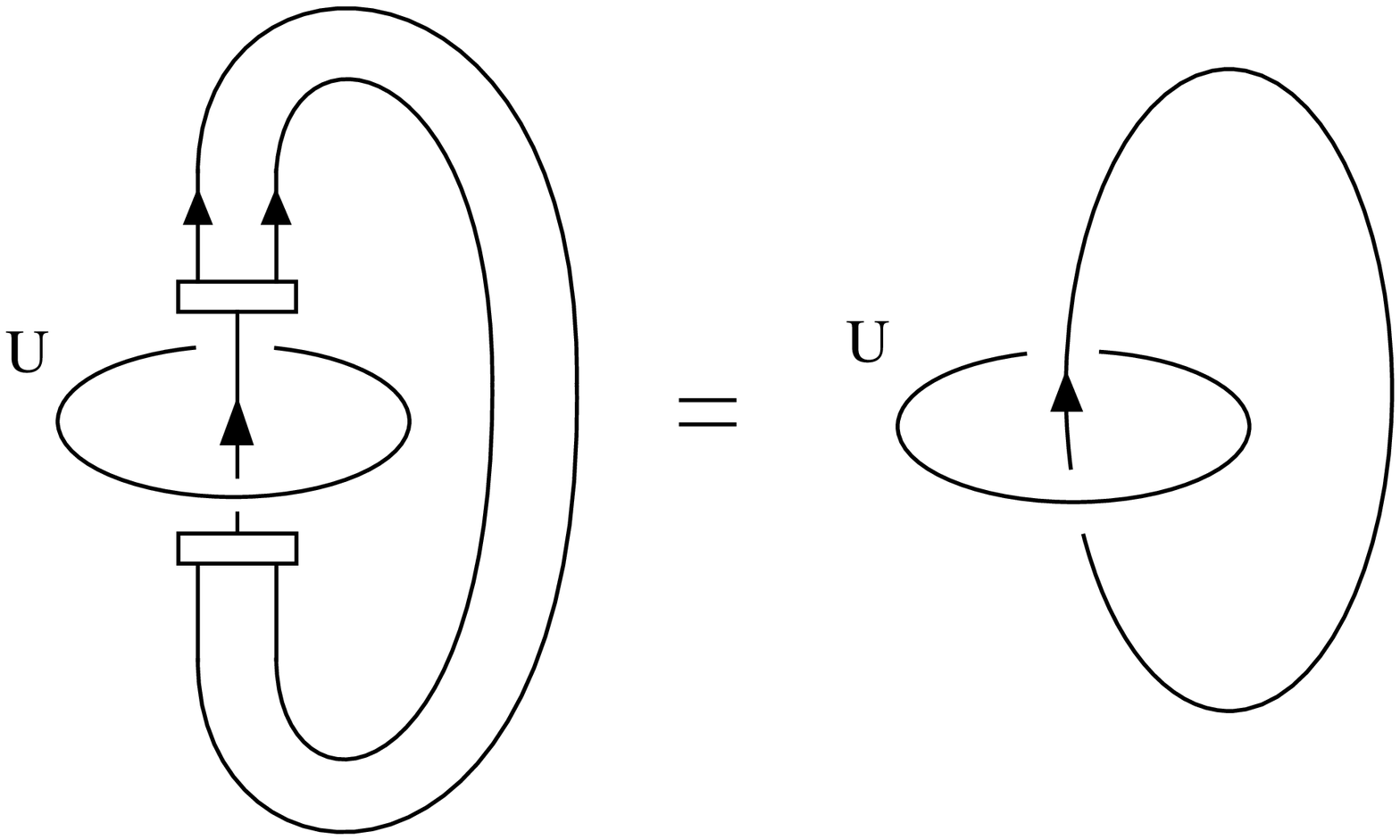,height=7cm,width=9cm}}
\vskip 0.2 truecm 
\centerline{{\bf Figure 2.9}}
\vskip 0.5 truecm 
\end{figure}
\begin{figure}[h]
\centerline{\epsfig{file=\path 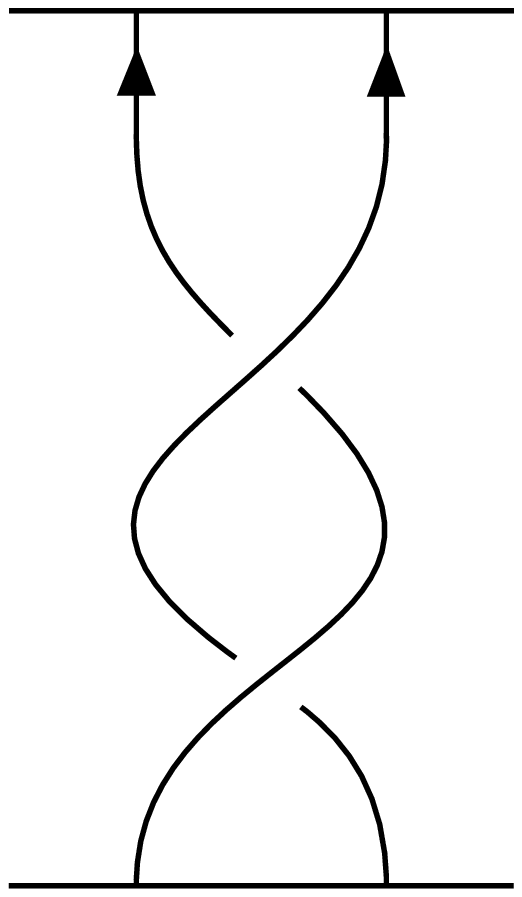,height=5cm,width=4cm}}
\vskip 0.5 truecm 
\centerline{{\bf Figure 2.10}}
\vskip 0.5 truecm 
\end{figure}

\centerline{\bf (4) Projection compatibility}

\no
This rule expressed by Fig.2.9 sets the natural compatibility between  the 
satellite relation (\ref{satg}) and the projectors introduced by rule 1.
Note that when rules 1 and 2 are used in a link diagram its character is altered 
due to the presence of projectors. Relations between link diagrams 
in which projectors are present cannot be used directly to obtain relations 
involving  expectation values of observables. The point is that projectors must be eliminated in favor of some suitable combination of strings. Rule (4) plays 
a key to eliminate projectors from a link diagram.

\centerline{\bf (5) Factorization}
If the link $L \subset \mathbb{R}^3(S^3)$ is the distant (disjoint)
union of the links $L_1$ and $L_2$, we shall write $L=L_1 \cup L_2$;  
general covariance implies
\cite{guad1} that  
\beeq
E(\, L_1 \cup L_2 \, ) \; = \; E(\, L_1\, ) \; E(\, L_2 \, ) \qquad . 
\label{dist}
\end{equation}

\no
Let us consider an example. The Hopf link shown in Fig.1.1 can be obtained 
form the closure of the braid shown in Fig.2.10. In oder to compute the expectation value of a Wilson line $W(C_1, C_1; \, \rho_1, \rho_2)$ associated with the Hopf link (by using the crossing rule) one
can replace the over-crossing configurations in  
the braid of Fig.2.10 by means of projectors. After that, the orthogonality property (see Fig.2.5) can be used  to eliminate one of two projectors. As final step, 
the projection compatibility is used in order to obtain a standard
link diagram. The result is
\beeq
H[\rho_1, \, \rho_2] \; = \; q^{- \left[ Q(\rho_1) \, + \, Q(\rho_2) \right]}   \sum_{\rho_t \in \rho_1 \otimes \rho_2}  q^{Q(\rho_t)} \; E_0[\rho_t] \quad .
\end{equation}
Thus, by using the calculation rules the expectation values of the observables  corresponding to the Hopf link can be expressed in terms of the observables associated with the unknot, 
i.e. a knot ambient isotopic to the standard circle in $\mathbb{R}^3$. 
   
\section{\bf Connected sum}
Let us consider the connected sum of links.  Suppose that
$L_1$ and $L_2$ are two disjoint links in $S^3$ and that the oriented components $C\in L_1$
and $C^\prime \in L_2$ are associated with the irreducible representation 
$\rho $ of the gauge group.  Starting from the distant union $L_1 \cup L_2$,
one can construct a new link which is called the connected sum of $L_1$ and
$L_2$. Inside some fixed three-ball in $S^3$,  the two components $C$ and
$C^\prime$ are cut and glued together as shown in Figure 2.4. The resulting new
link  is the connected  sum $L_1\# L_2[\rho ]$ of $L_1$ and $L_2$ which has been
obtained by acting on two link components with colour given by the irreducible
representation $\rho $ . 

\begin{figure}[h]
\begin{picture}(10,10)
\put(150,-100){$\rho$}
\put(167,-100){$\rho$}
\put(320,-120){$\rho$}
\put(320,-50){$\rho$}
\end{picture}
\vskip 0.9 truecm 
\centerline{\epsfig{file=\path 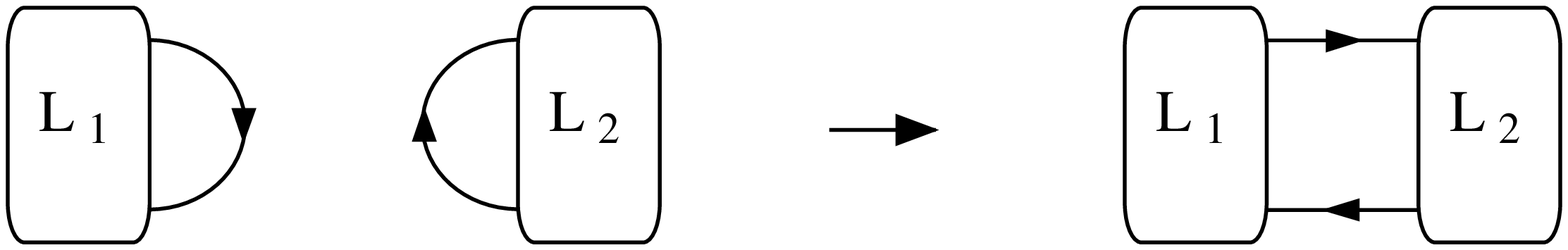,height=4cm,width=9cm}}
\vskip 0.5 truecm 
\centerline{{\bf Figure 2.11}}
\vskip 0.4 truecm 
\end{figure}

\noindent The expectation values for the connected sums of links satisfy the
relation \cite{guad1} 
\beeq
E( L_1\# L_2[\rho ]) \; = \; { E(L_1) \, E(L_2) \over E_0[\rho ]} \quad ,  
\label{eq:csr} 
\end{equation}
where $E_0[ \rho ]$ is the expectation value of the Wilson line
operator associated with the unknot (circle) in $S^3$ with preferred framing
and with colour given by the irreducible representation $\rho $ of $G$.

\chapter{\bf SU(2) and SU(3) Examples}

\section{\bf SU(2)}
For future convenience, in this section we shall collect the values of the unknot
and of the Hopf link when the gauge group is $G=SU(2)$. 
Detailed proofs of can be found in \cite{guad1,glib}. 

Any irreducible and finite dimensional representation of $SU(2)$ can be parametrized by a 
single label $J$ with $2J \in \mathbb{N}$. The value of the quadratic
Casimir operator $Q(J)$ in the representation $J$ is
\beeq
Q(J) \; = \; J(J \;+ \; 1) \quad ,
\end{equation}
the dimension $D(J)$ of the irreducible representation $J$ is
\beeq
D(J) \; = \; 2J \; + \; 1 \quad .
\end{equation}
The value of the unknot $E_0[J]$  in the representation $J$ is \cite{guad1}
\beeq
E_0[J] \; = \; \frac{q^{\left(2J+1 \right)/2} \; - \; q^{-\left(2J+1 \right)/2}}
{q^{1/2} \; - \; q^{-1/2}} \; = \; \frac{\sin \left[\left(2J \; + \; 1 \right) \pi/k \right]}{\sin \left(\pi/k \right)} \quad .
\label{unsu}
\end{equation} 
Finally, the expectation value of the Hopf link with colours $\chi[J_1]$ and $\chi[J_2]$ is given by \cite{guad1}
\beeq
H[J_1,J_2] \; = \; \frac{q^{\left(2J_1+1 \right)\left(2J_2 \; + \; 1 \right)/2}
\; - \; q^{-\left(2J_1+1 \right)\left(2J_2 \; + \; 1 \right)/2}}
{\sin \left(\pi/k \right)} \quad .
\label{hosu}
\end{equation}

\section{\bf SU(3)}

The satellite relation (\ref{satg}) will play a crucial role in our construction. 
In fact, we will use systematically the pattern link $B$, shown in Fig.2.1, to
construct satellites. We shall define a recursive procedure in order to
replace each link component, which is associated with  a higher dimensional
representations of $SU(3)$, with a suitable cabled component. In order to 
introduce the cabling procedure, however, we need to discuss some
basic properties of the representation ring of $SU(3)$. 

We shall use Dynkin labels to denote the irreducible representations of
$SU(3)$.  For each couple of non-negative integers $(m,n)$, the associated 
irreducible  representation corresponds to the highest weight $\mu$ given by  
\beeq
\mu \; = \; m \; \mu^{(1)} \, + \, n \; \mu^{(2)} \quad , 
\end{equation}
where $\mu^{(1)}$ and $\mu^{(2)}$ are the fundamental weights of $SU(3)$. 
Thus, $(0,0)$ is the trivial representation, $(1,0)$ denotes the
fundamental representation {\bf 3} and $(0,1)$ its complex conjugate {\bf 3*}. 
The representation $(m,n)$ has dimension  
\beeq
D(m,n) \; = \; {(m + 1) (n + 1) ( m + n + 2)\over 2} \quad , 
\label{eq52}
\end{equation}
and the value $Q(m,n)$ of the corresponding quadratic Casimir operator is 
given by 
\beeq
Q(m,n) \; = \; \frac{m^2 + n^2 + m n + 3 (m+n)}{ 3} \quad .  
\label{eq53}
\end{equation}

Given two representations $(m,n)$ and $(a,b)$, the decomposition of the
tensor product $(m,n) \otimes (a,b)$ into a sum of irreducible components can
be obtained, for instance, by means of the standard Young tableaux method. The
following relations will be useful for our discussion. 

\noindent For $m \geq 1$ and $n \geq 1$, one has 
\beeq
(m, n) \otimes (1, 0) \; = \; (m+1, n) \oplus (m-1, n+1) \oplus (m, n-1) 
\quad , 
\label{eq54}
\end{equation}
\beeq
(m, n) \otimes (0, 1) \; = \; (m, n+1) \oplus (m+1, n-1) \oplus (m-1, n) 
\quad .  
\end{equation}

\noindent For $m \geq 1$, one obtains 
\beeq
(m, 0) \otimes (1, 0) \; = \; (m+1, 0) \oplus (m-1, 1) 
\quad , \label{eq56}
\end{equation}
\beeq
(m, 0) \otimes (0, 1) \; = \; (m, 1) \oplus (m-1, 0) 
\quad . \label{eq57} 
\end{equation}

\noindent Finally, for $n \geq 1$, one gets 
\beeq
(0, n) \otimes (1, 0) \; = \; (1, n) \oplus (0, n-1) 
\quad , \label{eq58}
\end{equation}
\beeq
(0, n) \otimes (0, 1) \; = \; (0, n+1) \oplus (1, n-1)  
\quad .  
\label{eq59}
\end{equation}

Unlike the case of the group $SU(2)$,  an explicit 
formula which gives the decomposition of the tensor product  $(m,n)
\otimes (a,b)$, for arbitrary representations $(m,n)$ and $(a,b)$ of $SU(3)$, 
is not known. Thus, for $SU(3)$ one has to analyze, in general, each 
single tensor product separately. Even if the structure of the $SU(3)$ tensor
algebra $\cal T$ cannot be displayed in compact form,  we will show how
to derive the relevant properties of $\cal T$ and of its associated reduced
tensor algebra ${\cal T}_{(k)}$. In fact, in order to study the properties of 
the tensor algebra, we only need to consider the relations (\ref{eq54})-(\ref{eq59}). 

Let us denote by $R$ the representation ring of $SU(3)$ and by 
${\cal P}(y, {\overline y}\, )$ the ring of (finite) polynomials in the
two variables  $y$ and ${\overline y}\, $ with integer coefficients.  We
shall now show that $R$ admits a faithful representation in ${\cal P}(y,
{\overline y}\, )$.  This is a standard result of the theory of
simple Lie algebras and its validity  is based on the fact that the Lie algebra
associated with $SU(3)$ has rank two, of course. Here, we shall just recall  
the main arguments of the proof in order to illustrate how
the recursive use of eqs.(\ref{eq54})-(\ref{eq59}) determines the structure of $R$ and,
therefore, the structure of $\cal T$ \cite{gp1}. 

Each irreducible representation of $SU(3)$ is associated with an element of $R$;
the set of the elements which correspond to all the inequivalent irreducible
representations of $SU(3)$ is called the standard basis of $\cal T$. $R$ is a
commutative ring with identity;  consequently, for
each element $(m,n)$ of the standard basis, we only need to give the
corresponding  representative $[m,n]$ in ${\cal P}(y, {\overline y}\, )$. 
On the one hand, the ring $R$ is generated by the two elements associated
with the fundamental weights of $SU(3)$ plus the identity. On the other hand, 
${\cal P}(y, {\overline y}\, )$ is generated by the two elements $y$ and 
${\overline y}\,$ plus the identity. Thus, the
starting point is the obvious correspondence   
\beeq
(0,0) \; \longleftrightarrow \; [0,0] \; = \; 1 \qquad , 
\end{equation}
\beeq
(1,0) \; \longleftrightarrow  \; [1,0] \; = \; y \qquad , 
\end{equation}
\beeq
(0,1) \; \longleftrightarrow  \; [0,1] \; = \; {\overline y} \qquad .  
\end{equation}
Now, we need to find the representative $ [m,n] \in {\cal P}(y, {\overline y}\,
)$ of  a generic representation $(m,n)$. Let us consider firstly the set of
representations of the type $(m,0)$ with $m > 1$. From eq.{\ref{eq56}), one finds 
$$
[2,0] \; = \; [1,0] \cdot [1,0] - [0,1] \quad . 
\eqno(5.13)
$$
Therefore, 
$$
[2,0] \; = \; y^2 \, - \, {\overline y} \quad . 
\eqno(5.14)
$$
In general, eqs.(\ref{eq56}) and (\ref{eq57}) imply that 
\beeq
[m + 2, 0] \; = \; [m-1,0] + [m+1 , 0] \cdot [1,0] - [m,0] \cdot [0,1]
\qquad.  
\label{eq515}
\end{equation}
Equation (\ref{eq515}) must hold for any $m \geq 1$ and gives a recursive
relation for the elements $\{ \, [m,0]\, \}$. Since we already know
$[2,0]$, $[1,0]$, $[0,1]$ and $[0,0]$, this recursive relation 
determines $[m,0]$, for arbitrary $m$, uniquely. For example, one finds 
\beeq	
[3,0] \; = \;  y^3  - 2 y {\overline y} +1 \quad ,  
\end{equation}
\beeq
[4,0] \; = \; y^4 - 3 y^2 {\overline y} + {\overline y}^2 + 2 y \quad . 
\end{equation}

The same argument, based on eqs.(\ref{eq58}) and (\ref{eq59}), can be used to find the
polynomials $\{ \, [0,n] \, \}$ associated with the representations $\{ \,
(0,n) \, \}$, of course. Equivalently, since $(0,m)$ is the complex conjugate
of $(m,0)$, the polynomial $[0,m]$ can be obtained from $[m,0]$ simply by
exchanging $y$ with ${\overline y}\, $ and vice versa. 
Therefore, at this stage, all the polynomials of the type $\{ \, [m,0] \, \}$
and  $\{ \, [0,n] \, \}$ are uniquely determined for arbitrary $m$ and $n$. 

From eq.(\ref{eq57}), one obtains 
\beeq
[m,1] \; = \; [m,0] \cdot [0,1] - [m-1,0] \quad .
\end{equation}
This equation, which holds for any $m \geq1$, permits us to find all the
polynomials of the type $\{ \, [m,1] \, \}$. Similarly,  eq.(\ref{eq58}) defines the
recursive relation   
\beeq
[1,n] \; = \; [0,n] \cdot [1,0] - [0,n-1] \quad , 
\end{equation}
which uniquely fixes the polynomials $\{ \, [1,n] \, \}$ for arbitrary $n$. 
Finally, let us consider a generic representation $(m,n)$; the associated
polynomial $[m,n]\in {\cal P}(y, {\overline y}\, )$ can be determined, for 
instance, by using a recursive procedure in the values of the index $n$.
Indeed, eq.(\ref{eq54}) gives 
\beeq
[m, n+1] \; = \; [m,n] \cdot [0,1] - [m+1, n-1] - [m-1,n] \quad . 
\label{eq520}
\end{equation}
Assume, by induction, that the polynomials $\{ [m,n] \, \}$ are known for
arbitrary $m$ and for $n \leq n_0$. Then, eq.(\ref{eq520}) can be used to find
$[m,n_0 +1]$. On the other hand,  $[m,0]$ and $[m,1]$ are known; therefore,
eq.(\ref{eq520}) permits us to find $[m,n]$ for generic values of $m$ and $n$. 

To sum up, the representation ring $R$ of $SU(3)$ can conveniently be
described by ${\cal P}(y, {\overline y}\, )$; we have also proved that the
polynomial $[m,n]$, which is associated with the generic representation
$(m,n)$, is uniquely determined by the relations (\ref{eq54})-(\ref{eq59})
 \cite{gp1}. A 
few examples 
of representative polynomials are in order:  
\beeq
[0,2] \; = \; {\overline y}^2 -y \quad , \quad [1,1] \; = \; y {\overline y}
-1 \quad , 
\end{equation}
\beeq
[2,1] \; = \; y^2  {\overline y} - {\overline y}^2 - y \quad , 
\end{equation}
\beeq
[3,1] \; = \; y^3 {\overline y} - 2 y {\overline y}^2 - y^2 + 2 {\overline y}
\quad , 
\end{equation}
\beeq
[2,2] \; = \; y^2 {\overline y}^2 - y^3 -y {\overline y}^3 \quad . 
\end{equation}
In general, one finds 
\beeq
[m,n] \; = \; \sum_{i+j \leq m+n} \, a_{ij} \, y^i \, 
{\overline y}^j \qquad , 
\label{eq525}
\end{equation}
where $\{ \, a_{ij}\, \}$ are integer numbers and $a_{mn} =1$. The given
representation of $R$ in ${\cal P}(y, {\overline y}\, )$ is particularly
convenient for our purposes. Indeed, each polynomial (\ref{eq525}) provides the
explicit connection between the elements of the standard basis of $\cal T$
and the elements of the new basis defined in terms of the powers of $(1,0)$ and
$(0,1)$, which correspond to the monomials $\{ \, y^i  {\overline y}^j\, \}$. 
To be more precise, let  us denote by $\chi [m,n]$ the element of the standard basis
of $\cal T$ which is associated with the representation $(m,n)$ of $SU(3)$. 
Eq.(\ref{eq525})
implies that  
\beeq
\chi  [m,n] \; = \; \sum_{i+j \leq m+n} \; a_{ij} \;  
\left ( \, \chi [1,0] \, \right )^i \; \left ( \, \chi [0,1] \, \right )^j \quad ,  
\label{eq526}
\end{equation}
where the coefficients $\{ \, a_{ij} \, \}$ appearing in eq.(\ref{eq526}) coincide with the
coefficients $\{ \, a_{ij} \, \}$ entering eq.(\ref{eq525}). 

\subsection{\bf Computing the link polynomials} 

The quantum CS field theory is exactly solvable because, with a finite number 
of operations, on can find the exact expression of $E(L)$ for a generic link
$L$.  As we have already mentioned, $E(L)$ can easily be computed by using the
rules introduced in \cite{guad1,glib}
and are in agreement with the results obtained in perturbation theory, of course. 

In order to compute $E(L)$, one can use standard cabling; this method consists of two
steps. Firstly, we shall give the rules for the computation of a generic link
when each link component has colour specified by the fundamental representation of
$SU(3)$. Secondly,  we shall introduce a recursive procedure, which is based on the
use of the satellite relation (\ref{satg}), to compute the link invariants when the 
colours of the link components are arbitrary \cite{gp1}.  

\bigskip

\shabox{
\noindent {\bf Theorem 3.1}} {\em Let $\, L$ be a link diagram with components 
$\, \{ \, C_1, C_2, ..., C_n \, \}$ in  which each component is oriented and
has  a colour given by the fundamental representation $\,${\bf 3} of $\, SU(3)$.
The associated expectation value, }   
\beeq
E(L) \; = \; E(C_1,C_2,...,C_n ; \, {\bf 3}, {\bf 3} , ..., {\bf 3}\, ) \quad , 
\end{equation}
{\em is uniquely determined by }
\begin{enumerate}
\item  {\em regular isotopy invariance; } 
\item  {\em covariance under an elementary modification of the writhe; } 
\item  {\em skein relation; }  
\item  {\em value of the unknot with zero writhe. } 
\end{enumerate}
This theorem has been proved in \cite{glib}. Given link diagram $L$; let us
modify the writhe $w(C)$ of a single  component $C$ according to $w(C) \rightarrow
w^\prime (C) = w(C) \pm 1$.  Let us denote by $L^{(\pm)}$ the new link diagram
which has been obtained from $L$  according to the above procedure. Then, point (2)
means that \cite{guad1} 
\beeq
E(L^{(\pm)}) \; = \; q^{\pm 4/3} \> E(L) \quad . 
\label{eq63}
\end{equation}
The skein relation mentioned in point (3) is given by \cite{guad1} 
\beeq
q^{1/6} \, E(L_+) \, - \, q^{-1/6} \, E(L_-) \; = \; 
\left [ q^{1/2} - q^{-1/2}\, \right ]  \, E(L_0)  \qquad , 
\label{eq64}
\end{equation}
where the configurations $L_\pm$ and $L_0$ are shown in Fig.3.1. 

\begin{figure}[h]
\vskip 0.9 truecm 
\centerline{\epsfig{file=\path 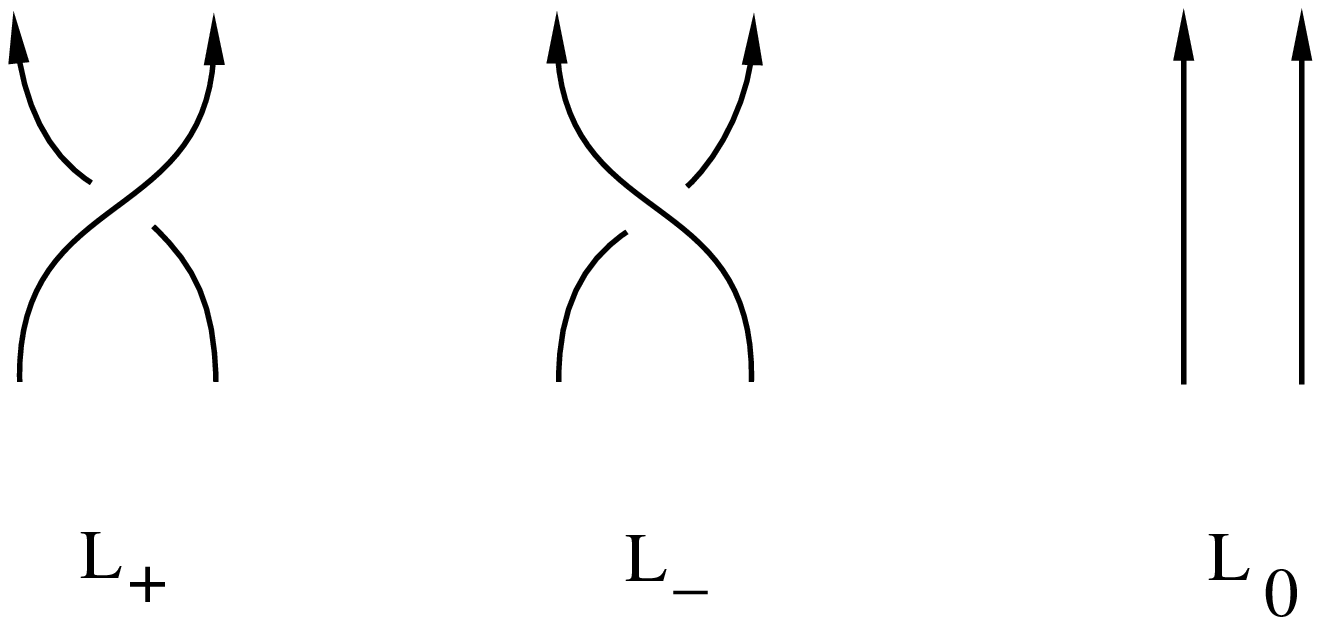,height=6cm,width=9cm}}
\vskip 0.5 truecm 
\centerline{{\bf Figure 3.1}}
\vskip 0.5 truecm 
\end{figure}

\noindent Finally,  the normalization of $E(L)$ is fixed by the value $E(U_0; {\bf
3})$ of the unknot $U_0$ with writhe equal to zero. This value cannot be chosen
arbitrarily but is uniquely fixed [2] by the field theory. For the gauge group
$SU(3)$, the value of the unknot is \cite{guad1}  
\beeq
E(U_0; {\bf 3}) \; = \; E_0[ {\bf 3}\, ] \; = \; q + 1 + q^{-1} \quad ; 
\label{eq65}
\end{equation}
the general expression for the unknot will be derived in the next section 
section 

\noindent One of the consequences of Theorem 3.1 is the following:   

\bigskip

\shabox{
\noindent {\bf Property 3.1}} {\em For any framed, oriented and coloured link 
$\, L$  with colours given by the fundamental representation $\,${\bf 3} of
$\, SU(3)$, the associated expectation value  $\, E(L)$ is a finite Laurent
polynomial in the variable} $\, x$ {\em given by} 
\beeq
x \; = \; q^{1/3} \quad . 
\label{eq66}
\end{equation}

\bigskip

\noindent {\bf Proof}  ~The value (\ref{eq65}) of the unknot with zero writhe
 is a finite Laurent polynomial in $x$ and, because of eq.(6.3), the value of the
unknot with arbitrary writhe also belongs to  $Z[ x^{\pm 1}]$.  By means of the
skein relation (\ref{eq64}) and of eqs.(\ref{eq63}) and (\ref{eq65}), one 
obtains that, for
the distant union of an arbitrary number of unknots, the associated invariant
belongs to  $Z[ x^{\pm 1}]$. This result is in perfect agreement with property
(\ref{dist}), of course.   Now, by using the skein relation recursively,  
$E(L)$ can be written as a finite linear combination of the invariants associated with the the
distant union of unknots in which each unknot may have a nontrivial writhe. Thus,
we only need to prove that the coefficients, entering this linear
combination, belong to $Z[ x^{\pm 1}]$.  Let us consider the standard ascending
method \cite{lic} to construct $E(L)$. This recursive method is based on the
observation that any given link diagram can be transformed into a distant union of
unknots provided some over-crossings are exchanged for under-crossings or vice versa. 
At each step of this recursive procedure, the skein relation (6.4) can be written in the
form      
\beeq
 E(L_+) \; = \;   q^{-1/3} \, E(L_-) \, + \,  
\left [ q^{1/3} - q^{-2/3}\, \right ]  \, E(L_0)  \qquad , 
\label{eq67} 
\end{equation}
or
\beeq
 E(L_-) \; = \; q^{1/3} \, E(L_+) \, -  \, 
\left [ q^{2/3} - q^{-1/3}\, \right ]  \, E(L_0)  \qquad .  
\label{eq68} 
\end{equation}
Since all the coefficients appearing in eqs.(\ref{eq67}) and (\ref{eq68}) 
belong to $Z[x^{\pm 1}]$, the inductive ascending procedure permits us to express $E(L)$ as
a linear combination of the values of the unknots in which all the coefficients
entering this linear combination belong to $Z[ x^{\pm 1}]$. Therefore, Property 3. 1 is
proved. {\hfill \ding{111}}
\vskip 0.5truecm

At this stage, we are able to compute $E(L)$ when the link components have  
colours which are given by the trivial representation, or the fundamental
representation {\bf 3} or its complex conjugate {\bf 3*}. Indeed, each
component associated with the trivial representation can be eliminated and
each oriented component, associated with {\bf 3*}, is equivalent to the same
component with opposite orientation associated with {\bf 3}. Consequently, by
means of Theorem 3.1, we can easily compute $E(L)$. 

Let us now consider the case in which the colours of the link components 
correspond to generic representations $\{ \, (m , n) \, \}$ of $SU(3)$. In order
to compute the expectation value of the associated Wilson line operator, one can use
standard cabling. This method is based on the satellite relations and on 
eq.(\ref{eq526}). Suppose that a given link component is characterized by an irreducible
representation $(m,n)$ of $SU(3)$ which is is different from $(0,0)$,  $(1,0)$ and
$(0,1)$. In this case, this component will be replaced by the image under
$h^\diamond$ of an appropriate linear combination of pattern links. 
We conclude this section by recalling that several properties of the link polynomials defined by 
Theorems 3.1 and 3.2 have been discussed in Ref.[4]. In particular, the computation of $E(L)$ can
be simplified by means of the rules introduced in [2]; for example, instead of using standard
cabling, $H[\, \rho_1 \, ; \, \rho_2 \, ]$ can be determined by means of eq.(8.13). It should
be noted that the covariance property of $E(L)$ under an elementary modification of the writhe 
can be expressed in the following general form. Let $L$ be an oriented link diagram in
which the component $C$ has writhe $w(C)$ and colour $\chi [m,n]$. Consider now the new link 
diagram $L^{(\pm )}$ which has been obtained from $L$ by means of an elementary modification 
of the writhe of $C$: 
\beeq
w(C) \; \rightarrow w^\prime (C) \; = \; w(C) \pm 1 \quad . 
\end{equation}
By using the twisting calculation rule of Chapter 2, one has 
\beeq
E(\, L^{(\pm )}\, ) \; = \; q^{\pm Q(m,n)} \; E( \, L \, ) \quad , 
\end{equation}
where $Q(m,n)$ is the value of the quadratic Casimir operator in the irreducible representation
$(m,n)$ of $SU(3)$. 

Pattern links are chosen to have all their components associated with $(1,0)$, or $(0,1)$, or
$(0,0)$. Such pattern links always exist and are determined precisely by 
eq.(\ref{eq526}). 
When all the link components which are associated with higher dimensional
representations of $SU(3)$ have been substituted, one gets a linear combination of
satellites in which all the link components have colours $(1,0)$, or $(0,1)$ or
$(0,0)$ and, by means of Theorem 3.1, one can finally compute the expectation value
of the observable.  We shall now give the details of this construction. 

Let us denote by $B(ij)$ the pattern link shown in Fig.3.2; $B(ij)$ is defined in
the solid torus which coincides with the complement of the circle $U$ in $S^3$. The
link  $B(ij)$ has $i+j$ oriented components and each component has preferred
framing. The satellite of a knot constructed with the pattern link $B(ij)$ is called
a cabled knot. Let us assume that $i$ components of $B(ij)$ have colour $\chi [1,0]$
and $j$ components have colour $\chi  [0,1]$; we shall denote by $W(B(ij))$ the
product of the associated Wilson line operators. According to eq.(\ref{eq49}),  $W(B(ij))$
admits an expansion of the type 
\beeq
W(B(ij)) \; = \; \sum_{\rho} \; \xi (\rho ) \; W(K; \chi [\rho ] )
\quad ,  
\label{eq6-9}
\end{equation}
where $K$ is the core of the solid torus (with preferred framing), shown in
Fig.2.3.  The coefficients $\{ \, \xi (\rho )\, \}$ entering eq.(\ref{eq6-9})  can be
determined by using eq.(\ref{eq413}) recursively. So, these coefficients are uniquely
determined by the structure constants of the tensor algebra $\cal T$. Consequently,
if the element $\chi [m,n]$ admits the presentation shown in eq.(\ref{eq526}), one obtains 
\beeq
W(K; \chi
[m,n] ) \; = \; \sum_{i+j \leq m+n} \, a_{ij} \; W(B(ij); \, \chi [1,0], ..., 
\chi [0,1] ,... ) \quad.   
\label{eq610}
\end{equation}

\begin{figure}[h]
\vskip 0.9 truecm 
\centerline{\epsfig{file=\path 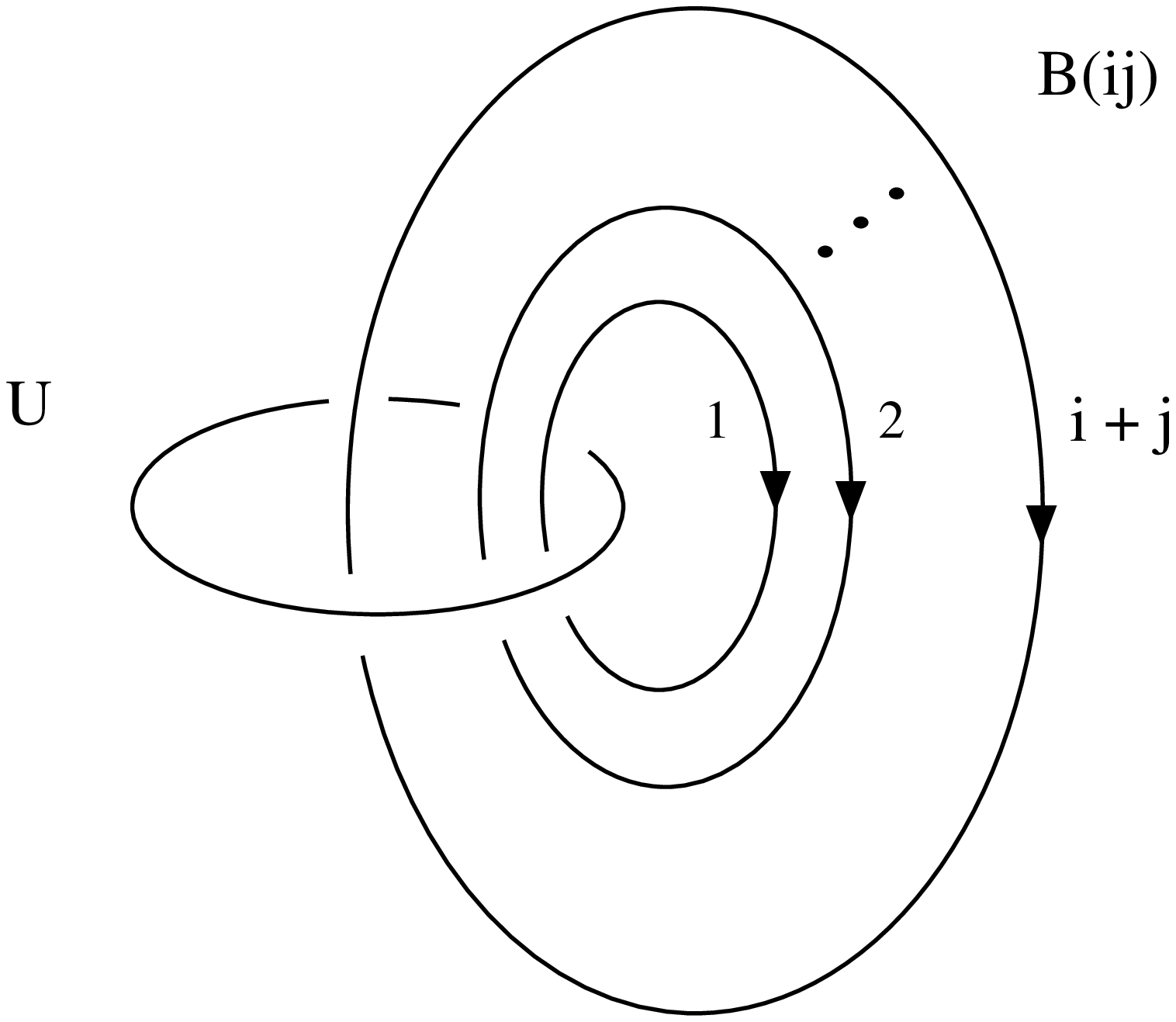,height=5cm,width=8cm}}
\vskip 0.5 truecm 
\centerline{{\bf Figure 3.2}}
\vskip 0.5 truecm 
\end{figure}

\noindent  Now, let $C$ be a generic component of a link with  colour $\chi [m,n]$.
As we have already mentioned, the oriented and framed component $C$ can be understood
to be the image $h^\diamond (K) $ of the oriented and framed knot $K$ (with colour
$\chi [m,n]$) under the homeomorphism $h^\diamond $ defined in section 2.3. 
Consequently, eq.(\ref{eq610}) implies that $C$ can be replaced by a linear combination of cabled
components   according to  
\beeq
C  {\rm ~~with~ colour~} ~[m,n] \; \longleftrightarrow \; \sum_{i+j \leq m+n} \,
a_{ij} \; h^\diamond ( \, B(ij)\, ) \quad .  
\label{eq611}
\end{equation}
Equation (\ref{eq611}) gives the desired relation which is
satisfied by the CS expectation values of the observables; each link
component, which is  associated with a higher dimensional representation of $SU(3)$,
is equivalent to a certain linear combination of cabled components which have  
colours given by the fundamental representation {\bf 3} or {\bf 3*}. 
The coefficients $\{ \, a_{ij} \, \}$ entering this linear combination are determined
by the structure constants of the tensor algebra of the gauge group and are
integer numbers.  

\bigskip 

\shabox{
\noindent {\bf Theorem 3.2}} {\em Let $\, L$ be an oriented link diagram  in  
which 
an irreducible representation  of $\, SU(3)$ is attached to each component. The
associated expectation value $\, E(L)$ can be computed by means of standard cabling.
$\, E(L)$ represents a regular isotopy invariant of oriented coloured link diagrams; 
moreover,  $\, E(L) \in Z[x^{\pm 1}]$}.     

\bigskip

\noindent {\bf Proof}  ~According to the standard cabling procedure,  each
link component  which is associated with a higher dimensional representation of
$SU(3)$  can be replaced by  the combination of cabled components shown in 
eq.(\ref{eq611}). Consequently, $E(L)$ can be written as a sum of satellites in which all
the components have colours given by the trivial representation, or the
representation {\bf 3} or its complex conjugate {\bf 3*}. Theorem 3.1 then implies
that $E(L)$ is a regular isotopy invariant of oriented and coloured link diagrams.
Finally, since the coefficients $\{ \, a_{ij} \, \}$ entering eq.(6.11) are integer
numbers, $E(L)$ is a linear combination with integer coefficients  of the invariants
defined in Theorem 3.1.  Therefore, according to Property  3.1, $E(L)$ belongs to 
$Z[x^{\pm 1}]$. {\hfill \ding{111}} 

\vskip 0.5truecm

Theorems 3.1 and 3.2 are the reconstruction theorems which  determine the
expectation values of the observables of the non-Abelian $SU(3)$  CS theory in
$\mathbb{R}^3$.  Let us consider a few examples of link polynomials. We shall denote by
$E_0[m,n]$ the value of the unknot with preferred framing and with colour $\chi
[m,n]$. One has 
\beeq
E_0 [2,0] \; = \; E_0 [0,2] \; = \; (1+q^{-2}) {(1-q^3)\over (1-q)} \quad , 
\end{equation}
\beeq
E_0 [1,1] \; = \; (1+q)^2 (1+q^{-2}) \quad , 
\end{equation}
\beeq
E_0 [4,1] \; = \; E_0 [1,4] \; = \;  {(q^{-5} -1)(1-q^7) \over (1-q)^2} \quad . 
\end{equation}

\noindent For the right-handed trefoil $3_1$ (with vertical framing convention), shown
in Fig.3.3, one finds 
\beeq
E(3_1 ; \chi [1,0])\; = \;  (q+1+q^{-1})\, (q^2 +1 -q^{-2}) 
\quad ,   
\end{equation}
\beeq
E(3_1 ; \chi [2,0] )\; = \;  { (1+q^{-2})\, (1-q^3) \over (1-q) } \, [ \, 
q^6 + q^3 + q^2 - q - q^{-2} - q^{-3} + q^{-5} \, ] 
\quad . 
\end{equation}

\begin{figure}[h]
\vskip 0.9 truecm 
\centerline{\epsfig{file=\path 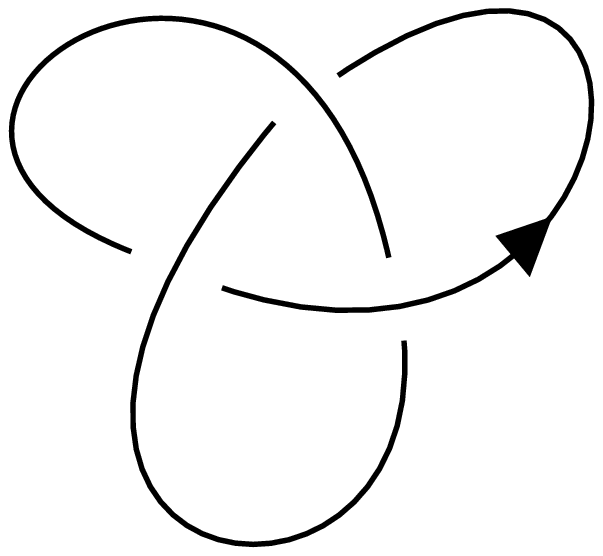,height=4cm,width=4cm}}
\vskip 0.5 truecm 
\centerline{{\bf Figure 3.3}}
\vskip 0.5 truecm 
\end{figure}

\noindent The figure-eight knot $4_1$ is shown in Fig.3.4; one gets 
\beeq
E(4_1 ; \chi [1,0] ) \; = \; q^4 + q^3 -1 + q^{-3} + q^{-4} \quad ,  
\end{equation}
\bea
E(4_1 ; \chi [2,0] ) && \; = \; q^{-10} \, (1+q^{18}) \, \frac{(1-q^3)}{ (1-q)}
\nb \\ 
&&\, - q^{-7} \, (1+ q^{11}) \, \frac{(1-q^4)}{ (1-q)} \, + q^{-2}\, (1+q^2) 
\, (1+q)^2
\quad . 
\ena

\begin{figure}[h]
\vskip 0.9 truecm 
\centerline{\epsfig{file=\path 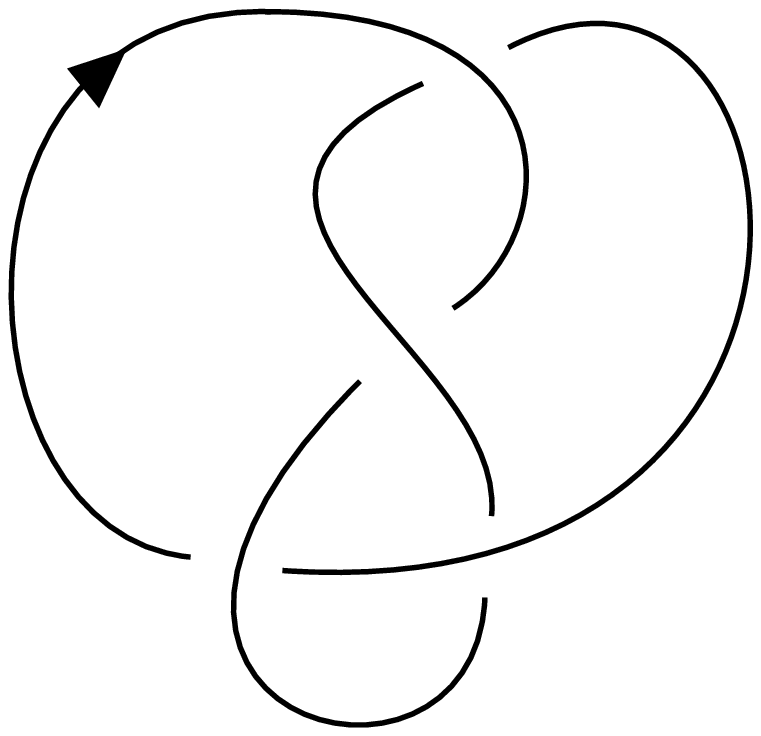,height=3.5cm,width=3.5cm}}
\vskip 0.5 truecm 
\centerline{{\bf Figure 3.4}}
\vskip 0.5 truecm 
\end{figure}

\noindent For the knot $5_1$, shown in Fig.3.5, one obtains 
\beeq
E(5_1 , \chi [1,0] ) \; = \; \frac{(1-q^3) \, (1+q^2 -q^{-4} )}{  (1-q) } 
\> q^{-1/3} \quad , 
\end{equation}
\bea
E(5_1 ; \chi [2,0] ) \; = \; && \frac{ (1-q^3) \, (1+q^2)}{  (1-q) }
\> q^{-31/3} \nb \\
&&+ \, \frac{(1-q^3)\, (1-q^5)}{ (1-q)^2} \, 
 \left [ \, {(1+q^3)\over (1+q)} - q^{-9} \, \right ] \>  q^{8/3} 
\quad . 
\ena

\begin{figure}[h]
\vskip 0.9 truecm 
\centerline{\epsfig{file=\path 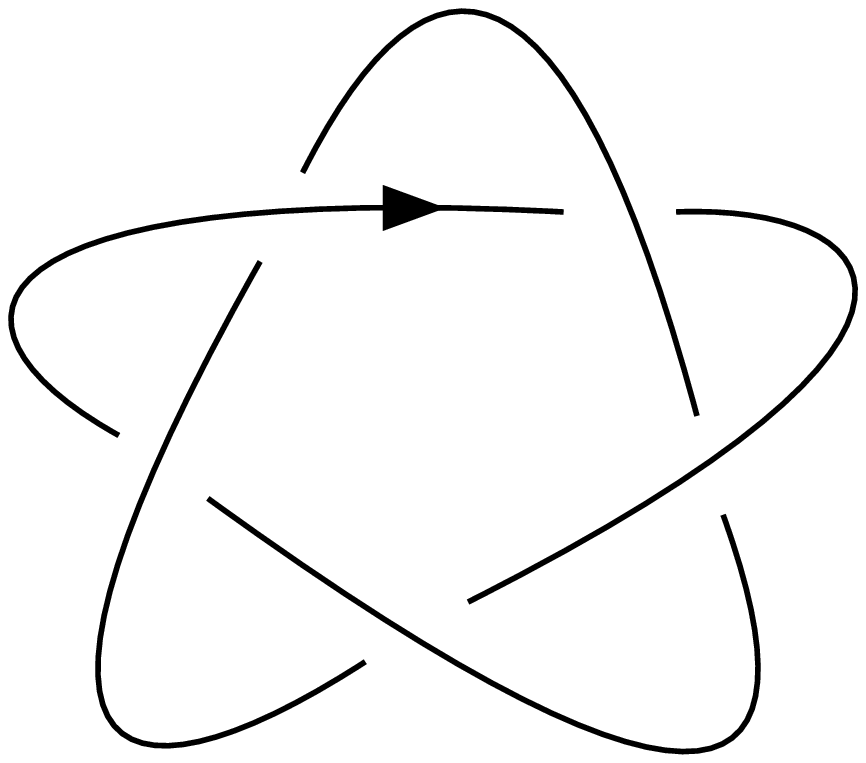,height=3.5cm,width=3.5cm}}
\vskip 0.5 truecm 
\centerline{{\bf Figure 3.5}}
\vskip 0.5 truecm 
\end{figure}

\noindent Finally, for the Whitehead link $5_2^1$, shown in Fig.3.6, one 
finds 
\bea
E(5^2_1; \chi [2,0] , \chi [1,0] ) \; &&= \; \frac{(1+q^2)\, (1-q^3)}{ (1-q)} \> q^{-1/3} \nb \\
&&\, + \, {(1-q^6)\over (1-q)} \left [ \, 
1-q^2 +2q^4-q^6 \, \right] \, (1+q^2) \> q^{-19/3} \quad . 
\ena

\begin{figure}[h]
\vskip 0.9 truecm 
\centerline{\epsfig{file=\path 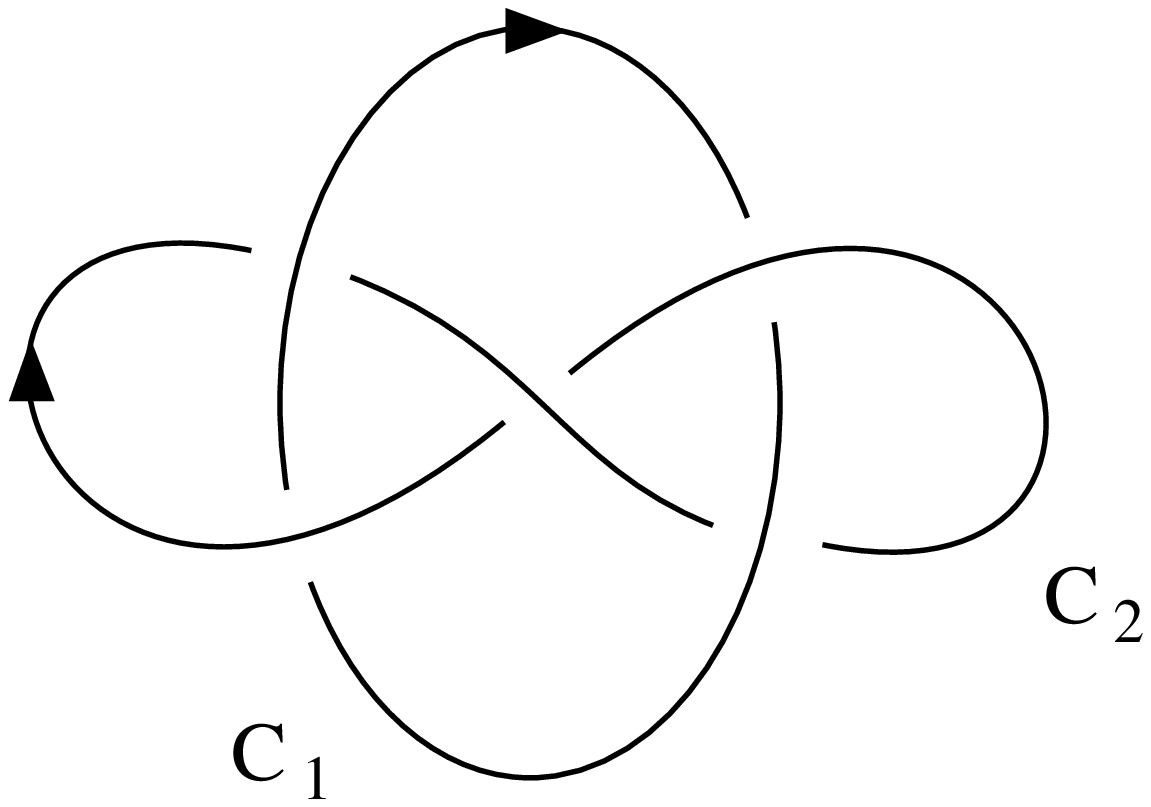,height=3.5cm,width=6cm}}
\vskip 0.5 truecm 
\centerline{{\bf Figure 3.6}}
\vskip 0.5 truecm 
\end{figure}

\subsection{\bf Values of the unknot} 

In this section we shall compute $E_0[m,n]$, which is the CS 
expectation value of the Wilson line operator associated with the unknot in
$\mathbb{R}^3$ (with preferred framing) with colour $\chi [m,n]$ \cite{gp1}. 
Our purpose is to
produce  the expression of $E_0[m,n]$ for arbitrary $m$ and $n\, $. In  
studying the properties of the link polynomials, the  values of the unknot play an
important role.  Indeed, in section 2.3 we have seen that, because of general covariance, 
any coloured link contained inside a solid torus determines a certain colour state
associated with the core of the solid torus.   Since the unknot is the companion of any
link, the values  of the unknot  represent the basic building blocks of the link
polynomials. In order to determine the structure of the reduced tensor algebra 
${\cal T}_{(k)}$, we shall use  the results of this section. 

In section 2.3 we have given the rules for the computation of a generic
coloured link. Thus, for fixed values of $m$ and $n$, the computation of
$E_0[m,n]$ is straightforward. The only nontrivial task is to find the
general dependence of $E_0[m,n]$ on the integers $m$ and $n$. In order 
to solve this problem, we shall use the symmetry properties of the CS theory and 
the recursive relations (\ref{eq54})-(\ref{eq59}).  

Let us denote by  $C$  the oriented unknot in $\mathbb{R}^3$ with preferred framing and
colour $\chi [m,n]$. Clearly, the orientation of $C$ can be modified by means of an
ambient isotopy transformation. Since a modification of the orientation
of $C$ is equivalent to replace $\chi [m,n]$ with $\chi [n,m]$,  one has 
\beeq
E_0[ m,n] \; = \; E_0[ n,m] \qquad . 
\label{eq71}
\end{equation}

Let us now consider the distant union of two oriented unknots $C_1$ and $C_2$ (both
with preferred framing) in $\mathbb{R}^3$. Suppose that $C_1$  has   colour $\chi
[\rho_1^{} ]$ and $C_2$  has colour $\chi [\rho_2^{}]$. Eq.(\ref{dist}) and the satellite
relation (\ref{satg}) imply that  
\beeq
E_0 [ \, \rho_1^{} \, ] \, E_0  [ \, \rho_2^{} \, ] \; = \; \sum_{\rho } \; 
F_{\rho_1 \, , \, \rho_2 \, , \, \rho}^{} \; \; E_0 [ \, \rho \, ] \quad , 
\label{eq72}
\end{equation}
where $\{ \, F_{\rho_1 \, , \, \rho_2 \, , \, \rho}^{} \, \}$ are the
structure constants of the tensor algebra $\cal T$ :  
\beeq
\chi [ \rho_1] \; \chi [\rho_2 ] \; = \; \sum_{\rho } \; 
F_{\rho_1 \, , \, \rho_2 \, , \, \rho}^{} \; \; \chi [\rho ] \quad .  
\end{equation}
Eq.(\ref{eq72}) shows that the set $\{ \, E_0 [m,n] \, \}$ of the possible values
of the unknot gives a representation in $Z[x^{\pm 1}]$ of the representation ring of
the group.  Consequently, one can use the relations (\ref{eq54})-(\ref{eq59}) to find 
the values of the unknot. In fact, the recursive argument that we presented
in section 3.2.1 determines the value of $E_0[m,n]$ uniquely.  
The result is summarized by
the following theorem \cite{gp1}. 

\bigskip 

\shabox{
\noindent {\bf Theorem 3.3}} {\em The expectation value $\, E_0 [m,n]$ of the unknot in
$\,$ $\mathbb{R}^3$ with preferred framing and colour $\, \chi [m,n]$ is given by}  
\beeq
E_0 [m,n] \; = \; q^{-(m+n)} \; \frac{(1-q^{m+1}) (1-q^{n+1}) (1-q^{m+n+2})}{
 (1+q) (1-q)^3} \qquad .   
\label{eq74}
\end{equation}

\bigskip

\noindent {\bf Proof} ~Clearly, eq.(\ref{eq74}) gives the correct value of the unknot 
for the representations $(0,0)$, $(1,0)$ and $(0,1)$. Moreover, one can easily
 verify that expression (\ref{eq74}) satisfies the recursive relations (\ref{eq54})-(\ref{eq59}). Consequently, 
eq.(\ref{eq74}) represents the unique solution of the recursive relations with the
correct initial data. Therefore, eq.(\ref{eq74}) gives the values of the unknot in the CS
theory. {\hfill \ding{111}}

\vskip 0.5truecm

In agreement with eq.(\ref{eq71}), the value of $E_0[m,n]$ shown in  equation 
(\ref{eq74}) is
symmetric in the indices $m$ and $n$.  Since the deformation parameter $q$
is given by $q= \exp(-i 2 \pi /k)$, the expression (\ref{eq74}) admits a Taylor expansion in
powers of $\lambda \; = \; (2\pi /k)$ around $\lambda =0$. Each term of this 
expansion
represents the value of the corresponding Feynman diagrams found in perturbation
theory. Let us now verify that the expression (\ref{eq54}) is really in 
agreement with the perturbative result (\ref{unper}). From eq.(\ref{eq74}) one
gets 
\bea
&&E_0[m,n] \; = \; \frac{(m+1)(n+1)(m+n+2)}{ 2} \, \left [ \, 1 
\, - \, \frac{\lambda^2}{ 12}( m^2 + n^2 + mn + 3m +3n) \, \right ] \nb \\ 
&& + O(\lambda^3  ) \quad . 
\ena
By taking into account eqs.(\ref{eq52}) and (\ref{eq53}), one finds that this expression
coincides precisely with the perturbative result (\ref{unper}). {\hfill \ding{111}}

Equation (\ref{eq74}) shows that $E_0[m,n]$ is actually a finite Laurent polynomial in the
deformation parameter $q$. Indeed, the factor $(1+q)(1-q)^3$ in the denominator of
the expression (\ref{eq74}) cancels out the roots of the numerator at $q=-1$ and $q=1$ . 
The dependence of  $E_0[m,n]$ on the CS
coupling constant $k$ can be explicitly displayed by writing the expression 
(\ref{eq74}) is
the  equivalent form  
\beeq
E_0 [m,n]\; = \; \frac{1}{ 2} \; \frac{\sin [ \pi (m+1)/k ] \; \sin [ \pi (n+1)/k ] \; 
\sin [ \pi (m+n+2)/k ]}{ \cos [ \pi /k ] \; \sin^3 [\pi /k] } \quad . 
\label{eq76}
\end{equation}

\subsection{\bf Values of the Hopf link} 

As we have already mentioned, we shall use Dehn's surgery on $S^3$ to solve the 
quantum CS theory in a generic three-manifold $M$.   To be more precise, 
we shall use the operator surgery method (see Chap.5-6) to compute the 
expectation values of the observables in  $M$  by means of the observables in $S^3$.  One of the basic ingredients in
the  construction of the surgery operators is the so-called Hopf matrix. 
Let us now  recall why the expectation value of the Hopf link plays a crucial role in
solving the topological field theory. 

Dehn's surgery method essentially consists \cite{rol} of removing and sewing 
solid tori in $S^3$; thus, we need to consider the properties of the CS expectation values which
are related to the different embeddings of solid tori in $S^3$.  Actually, because of
the so-called  Fundamental Theorem \cite{rol}, we really need to consider solid tori
standardly embedded in $S^3$.  Let us consider a solid torus $N$ standardly embedded
in $S^3$; clearly, its boundary  $\partial N \in S^3$ is a two-dimensional torus.  Now,
the crucial point to be noted is that $\partial N$ is  actually the boundary of two solid
tori which are both standardly embedded in $S^3$.  The first solid torus is $N$, of
course; the second solid torus is $S^3 - \dot{N} $ , where $\dot{N} $ is the interior of
$N$. 

In conclusion, a solid torus $N$ standardly embedded in $S^3$ really defines two
solid tori: $N$ itself and its complement in $S^3$.   Suppose now that a
certain coloured link $L_1$ is present in  $N$ and another coloured link $L_2$ is
contained in $S^3 - \dot{N} $;  we would like to study the properties of the associated
expectation value $\langle \,  W(L_1) \, W(L_2)\, \rangle $. The symmetry properties
of the CS expectation values, that we have introduced in chapter 2  are valid 
also in the three-sphere $S^3$.  In particular, the generalized satellite relations permit us to
find, for each link contained inside a solid torus,  the corresponding colour state
associated with the core of the solid torus.   Consequently, 
$\langle \,  W(L_1) \, W(L_2)\, \rangle $ can be expressed in terms of the values of
the Hopf link whose two oriented components represent the cores of the two solid tori
$N$ and  $S^3 - \dot{N} $.  

To sum up, the values of the Hopf link give us the pairing 
between the colour states which are associated with two complementary solid 
tori standardly embedded in $S^3$.  For this reason, the values of the Hopf link
are the fundamental ingredients in the construction of the surgery operators and,
together with the satellite relations,  characterize the topological
properties of the quantum CS field theory completely. Thus, our strategy is to produce
the values of the Hopf link in $S^3$. Since the CS expectation values in $S^3$
coincide with the expectation values in $\mathbb{R}^3$ with the constraint that $k$ is an
integer, we only need to determine the values of the Hopf link in $\mathbb{R}^3$.  

Let us consider the Hopf link in $\mathbb{R}^3$ whose two oriented components $\{ \, C_1
\, , \, C_2 \, \}$, shown in Fig.3.7, have preferred framings.  Let $C_1$ have
colour $\chi [m,n]$ and $C_2$ have colour $\chi [a,b]$. We are interested in the
associated expectation value 
\beeq
H[ \, ( m,n )\, ;\, ( a,b )\, ] \; = \; \langle \, W(\, C_1\, ;\, \chi [m,n] \, ) \>
W(\, C_2\, ;\,  \chi [a,b] \, ) \, \rangle \bigr |_{\mathbb{R}^3 } \quad .   
\label{eq81}
\end{equation}
By means of an ambient isotopy, the components $C_1$ and $C_2$ of the
Hopf link of Fig.3.7 can be exchanged; consequently, one has 
\beeq
H[ \, ( m,n )\, ;\, ( a,b )\, ] \; = \; H[ \, ( a,b ) \, ;\, ( m,n )\, ] \quad . 
\label{eq82}
\end{equation}
As we have mentioned in sect.4, $E(L)$ is invariant under a global ``charge
conjugation" transformation which consists of substituting each irreducible
representation $\rho $ for its complex conjugate $\rho^*$. Therefore, one 
finds 
\beeq
H[ \, ( m,n )\, ;\, ( a,b )\, ] \; = \; H[ \, ( n,m ) \, ;\, ( b,a )\, ] \quad . 
\label{eq83}
\end{equation}

\begin{figure}[h]
\vskip 0.9 truecm 
\centerline{\epsfig{file=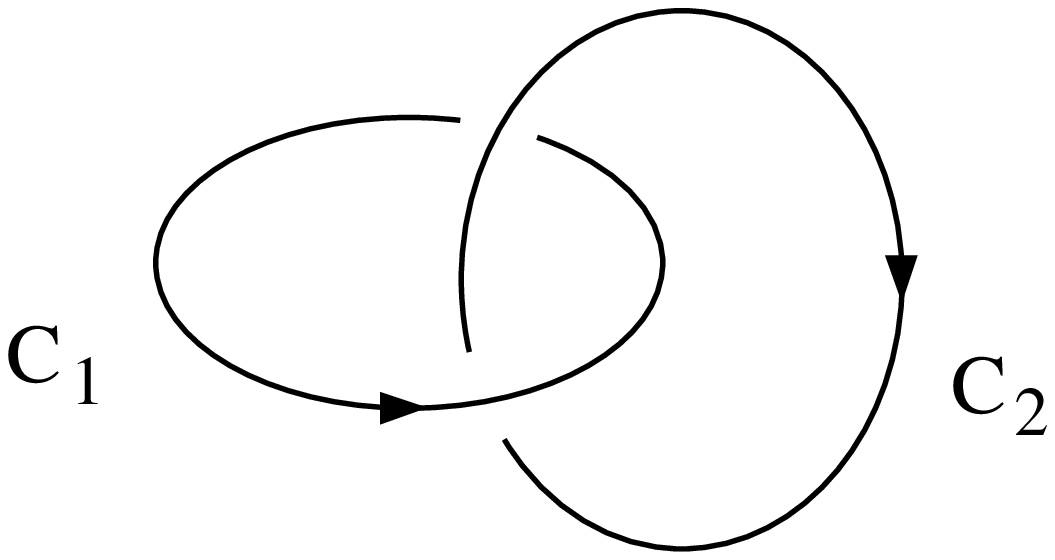,height=3.5cm,width=6cm}}
\vskip 0.5 truecm 
\centerline{{\bf Figure 3.7}}
\vskip 0.5 truecm 
\end{figure}

\noindent For fixed values of $(m,n)$ and $(a,b)$, $H[ \, ( m,n )\, ;\, ( a,b) \, ]$
can easily be computed by means of the rules specified by the reconstruction
theorems.  The general dependence of  $H[ \, (m,n )\, ;\, ( a,b )\, ]$  on the
irreducible representations $(m,n)$ and $(a,b)$ is summarized by the following
theorem. 

\bigskip

\shabox{
\noindent {\bf Theorem 3.4}} {\em The expectation value 
$\, H[ \, m,n \, ;\,  a,b \, ]$ of the Hopf link in
$\, \mathbb{R}^3$  is given by}  
\bea
&&H[ \, (m,n )\, ;\, ( a,b )\, ]\; = \;  q^{-[ (m+n) (a+b+3) + (m+3) b + (n+3) a]
/3}    \, \cdot \, {1 \over (1-q)^3 \, (1+q)} \, \cdot \, \nb \\ 
&&\left [ \, 1 + q^{(n+1)(a+b+2)+(m+1)(b+1)} + q^{(m+1)(a+b+2) + (n+1)(a+1)} 
 \right. \nb \\ 
&& \left. \quad - \, q^{(m+1)(b+1)} \, - \, q^{(n+1)(a+1)} \, - \, q^{(m+n+2)(a+b+2)} \, \right ] \quad . 
\label{eq84}
\ena
\vskip 0.5truecm

\noindent {\bf  Proof} ~In order to prove the validity of eq.(\ref{eq84}), we
shall follow
the same strategy as was adopted in the case of the unknot. Namely, we shall use
equations (\ref{eq54})-(\ref{eq59}) which, combined with the satellite relations and the
formula (\ref{eq:csr}) for the connected sums of links, permit us to find  
recursive relations for the values $H[ \, (m,n) \, ;\, ( a,b )\, ]$. Then, one can verify that
the expression (\ref{eq84}) satisfies these recursive defining relations and gives the
correct values of the initial data.  Thus, eq.(\ref{eq84}) represents the 
unique solution of the recursive relations which is consistent with the initial
values.  The algebraic part of the proof is more complicated than in the case of the
unknot, of course. 

Let us consider the link $L$ shown in Fig.3.8a. On the one hand, the link $L$ 
is a satellite of the Hopf link; therefore, by using the satellite formula 
(\ref{satg}), one
obtains  
\beeq
E(\, L \, ; \, \chi [\rho_1 ]\,  ,\,  \chi [ \rho_2 ]\,  , \, \chi [ \rho_3 ] \, ) \; = \; H [\,
\rho_2 \, ; \, \rho_1 \otimes \rho_3 \, ] \quad . 
\label{eq85}
\end{equation}

\begin{figure}
\begin{picture}(10,10)
\put(125,-7){$\rho_1$}
\put(180,-7){$\rho_3$}
\put(240,-7){$\rho_1$}
\put(345,-7){$\rho_3$}
\put(270,-70){$\rho_2$}
\put(310,-70){$\rho_2$}
\put(150,-70){$\rho_2$}
\end{picture}
\vskip 0.5 truecm 
\centerline{\epsfig{file=\path 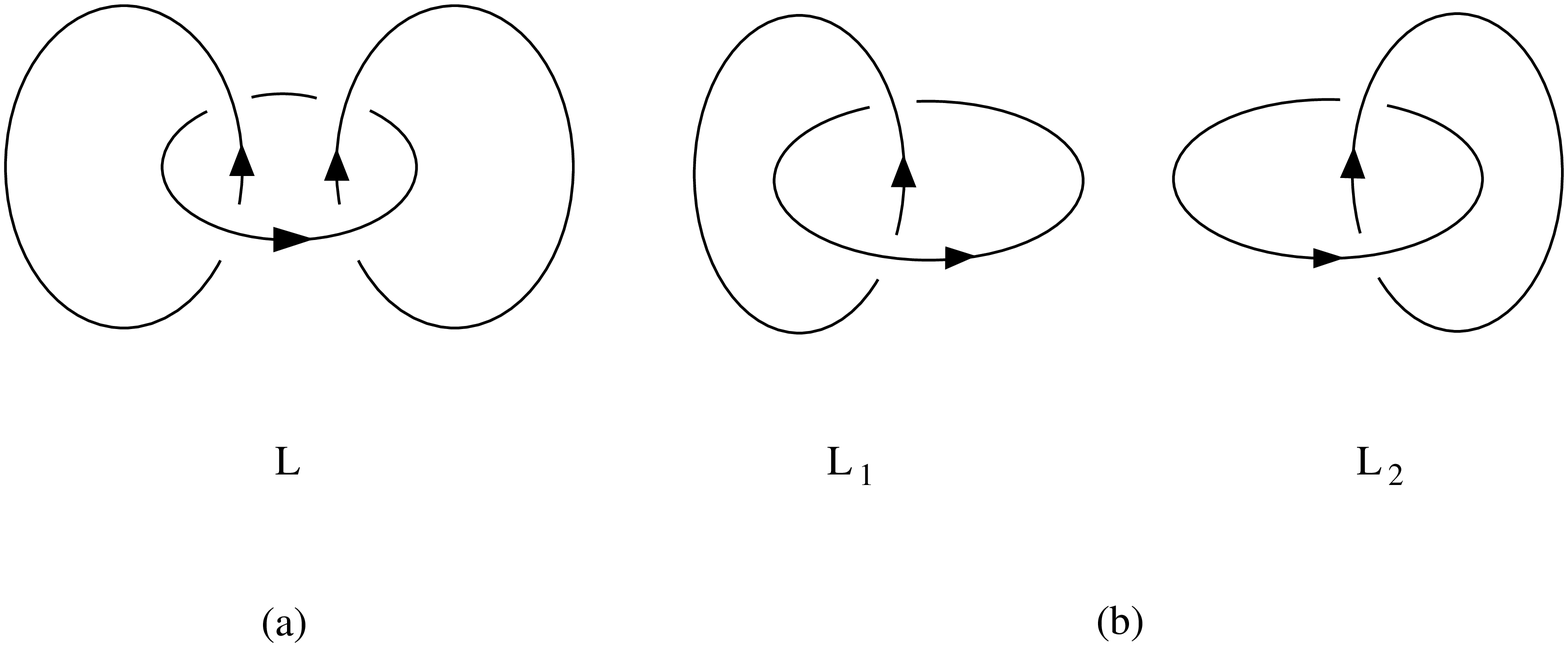,height=4cm,width=9cm}}
\vskip 0.5 truecm 
\centerline{{\bf Figure 3.8}}
\vskip 0.5 truecm 
\end{figure}

\noindent On the other hand, $L$ is the connected sum of the two Hopf links shown in Fig.3.8b. 
Thus, according to eq.(\ref{eq:csr}), one has (when $\rho_2$ is irreducible) 
\beeq
H [\, \rho_2 \, ; \, \rho_1 \otimes \rho_3 \, ]\; = \; \frac{ 
H [\, \rho_1 \, ; \, \rho_2 \, ]\;  H [\, \rho_2 \, ; \, \rho_3 \, ]}{ 
E_0 [ \, \rho_2 \, ] } \quad . 
\label{eq86}
\end{equation} 
Now, by setting $\rho_1=(m,n)$, $\rho_3=(1,0)$ and $\rho_2=(a,b)$, from 
eq.(\ref{eq86})
we  have
\bea
&&\frac{H [ \, (m,n)\, ;\, (a,b) \, ]\; \; H [\, (1,0)\, ;\, (a,b)\, ]}{E_0[ a,b]} \;   =
\; H [ \, (m+1,n)\, ;\, (a,b) \, ]\, + \nb \\
&&\quad \; \; + \; H  [\, (m-1,n+1)\, ; \,
(a,b)\, ] \, + \, H [\, (m,n-1)\, ;\, (a,b) \,  ] \quad . 
\label{eq87}
\ena
In the same way, if we set  $\rho_1=(m-1,n)$, $\rho_3=(0,1)$ and
$\rho_2=(a,b)$, we obtain
\bea
&&\frac{H [ \, (m-1,n)\, ;\, (a,b) \, ]\; \; H [\, (0,1)\, ;\, (a,b)\, ]}{ E_0[ a,b]} \;   =
\; H [ \, (m-1,n)\, ;\, (a,b) \, ] \, + \nb \\
&&\quad \; \; + \; H  [\, (m,n-1)\, ; \,
(a,b)\, ] \, + \, H [\, (m-2,n)\, ;\, (a,b) \,  ] \quad . 
\label{eq88}
\ena
Subtracting equations (\ref{eq87}) and (\ref{eq88}), we arrive at the promised
recursive relation 
\bea
&&H [\, (m+1,n)\, ;\, (a,b) \, ] \; = \; 
H [\, (m,n)\, ;\, (a,b) \, ] \; H [ \, (1,0)\, ; \, (a,b)\, ]\; E_0^{-1}[a,b] 
 \nb \\ 
&&- H [\, (m-1,n)\, ;\, (a,b) \, ] \; H [ \, (0,1)\, ; \, (a,b)\, ]\; E_0^{-1}[a,b] + 
H [\, (m-2,n)\, ;\, (a,b) \, ] \quad . 
\label{eq89}
\ena
Similarly, by using eq.(\ref{eq83}), we find 
\bea
&&H [\, (m,n+1)\, ;\, (a,b) \, ] \; = \; 
H [\, (m,n)\, ;\, (a,b) \, ] \; H [ \, (0,1)\, ; \, (a,b)\, ]\; E_0^{-1}[a,b] 
 \nb \\ 
&&- H [\, (m,n-1)\, ;\, (a,b) \, ] \; H [ \, (1,0)\, ; \, (a,b)\, ]\; E_0^{-1}[a,b] + 
H [\, (m,n-2)\, ;\, (a,b) \, ] \quad . 
\label{eq810}
\ena
By using equations (\ref{eq89}) and  (\ref{eq810}) recursively, one can 
determine the value  of $H[\, (m,n) \,
; \, (a,b)  \,]$ for $m,n \geq 2$  in terms of the initial values
\bea
&&H [\, (0,0) \, ; \, (a,b) \,] \nb \\
&&H [\, (1,0)  \, ; \, (a,b) \,] \nb \\
&&H [\,  (2,0) \, ; \, (a,b) \, ] \nb \\
&&H [\,  (1,1) \, ; \, (a,b) \, ] \nb \\
&&H [\,  (2,1) \, ; \, (a,b) \, ] \nb \\
&&H [\,  (2,2) \, ; \, (a,b) \, ]  \quad . 
\label{eq811}
\ena
Since any  link component with colour $\chi [0,0]$ can be eliminated, we have
\beeq
H [\, (0,0) \, ; \, (a,b) \, ]\; = \; E_0[a,b] \quad .
\label{eq812}
\end{equation}
The remaining values of the initial data can easily be calculated by using Theorems 3.1 and 3.2 
The result is \cite{gp1}
\bea
H [\, (1,0)  \, ; \, (a,b) \,]\; &&= \; q^{- \left(1+ {5b \over 3} + \frac{4a 
}{ 3}
\right)} {\left(1-q^{1+b} \right) \left(1-q^{1+a} \right) 
\left(1-q^{2+a+b} \right) \over (1-q)^3 (1+q)}  \nb \\
&&\; \left(1+q^{1+b}+q^{2+a+b} \right) \quad . 
\label{eq814}
\ena
\bea
H [\, (2,0) \, ; \, (a,b) \,] \; &&= \;  
{\left(1+q^{1+b}+q^{2+2b}+q^{2+a+b}+q^{2 \left(2+a+b \right)}+ q^{3+2b+a} \right) \over \left(1-q
\right)^3 (1+q)} \nb \\
&& \; q^{- \left(2+ \frac{7b }{ 3}+ \frac{5a }{ 3}
\right)}\left(1-q^{1+b} \right) \left(1-q^{1+a} \right) \left(1-q^{2+a+b}
\right) \quad . 
\label{eq815}
\ena
\bea
H [\, (1,1) \, ; \, (a,b) \,] \; &&= \; {\left(1-q^{1+b} \right)
\left(1-q^{1+a}
 \right) \left(1-q^{2+a+b} \right) \left(1+q^{1+b} \right) \left(1+q^{1+a}
\right) \over \left(1-q \right)^3 (1+q)} \nb \\
&&\; \left(1+q^{2+a+b} \right) q^{-2(1+a+b)}  \quad . 
\label{eq816}
\ena

\bea
&&H [\, (2,1) \, ; \, (a,b) \, ]\; = \; \frac{q^{-\left(3+ \frac{8 b }{ 3}+ 
\frac{7a} { 3} \right)} \left(1-q^{1+m} \right) \left(1-q^{1+n}
 \right) \left(1-q^{2+a+b} \right) }{ (1-q)^3 (1+q)}  \cdot \nb \\
&& (1+q^{1+b}+q^{2+2b}+q^{1+a}+2q^{2+b+a}+ \nb \\
&&+2q^{3+2b+a}+q^{4+3b+a}+q^{3+b+2a}+2q^{4+2b+2a}+q^{5+3b+2a}+ \nb \\
&&+q^{5+2b+3a}+ q^{6+3b+3a})  \quad . 
\label{eq817}
\ena
\bea
&&H [\, (2,2) \, ; \, (a,b) \,]\; =\; \frac{q^{-(4+3b+3a)}\left(1-q^{1+b} \right)
\left(1-q^{1+a}
 \right) \left(1-q^{2+a+b} \right)}{ (1-q)^3 (1+q)} \nb \\
&&(1+q^{1+b}+q^{2+2b}+q^{1+a}
+2q^{2+a+b}+2q^{3+2b+a}+q^{4+3b+a}+q^{2+2a}+ \nb \\
&&+3q^{4+2b+2a}
+2q^{5+3b+2a}+q^{6+4b+2a}+q^{4+b+3a}+2q^{5+2b+3a}+2q^{6+3b+3a}+ \nb \\
&&+q^{7+4b+3a}+q^{6+2b+4a}+q^{7+3b+4a}+q^{8+4b+4a} +2q^{3+b+2a})  \quad . 
\label{eq818}
\ena
Equations  (\ref{eq89}) and (\ref{eq810}) together with the initial 
data (\ref{eq811}) determine the values of
the Hopf link uniquely.  Now, one can verify \cite{gp1,io} that expression 
(\ref{eq84}) satisfies the recursive
relations  (\ref{eq89}) and (\ref{eq810}). Moreover, for $(m,n)$ equal to 
(0,0), (1,0), (2,0),  (1,1), (2,1), and (2,2), eq.(\ref{eq84}) reproduces the 
correct initial values. Consequently, eq.(\ref{eq84}) represents
the values of the Hopf link in the CS theory. {\hfill \ding{111}}     

\vskip 0.5truecm

The value of the Hopf link shown in eq.(\ref{eq84}) satisfies the symmetry 
properties (\ref{eq82}) and (\ref{eq83}).  
As usual, expression (8.4) admits a Taylor expansion in powers of $\lambda  =( 2\pi /k)$ around
$\lambda =0$. One can verifies that the first three terms of this expansion 
agree with the perturbative result (\ref{hope}).

\chapter{\bf Reduced tensor algebra}
\section{\bf Physically inequivalent representations}

The  Wilson line operators  $\{ \, W(\, L \, ) \, \}$ associated with all the possible links in $S^3$
form a complete set of gauge invariant observables in the CS theory. Each
link component is coloured with an irreducible representation of the gauge
group $G$. The labels, which are used to distinguish the inequivalent irreducible
representations of the gauge group, can be understood as gauge-invariant quantum
numbers assigned to the link components. In fact, as a linear
space, the tensor algebra $\cal T$ can be interpreted as a (gauge-invariant)
state space. This space is infinite dimensional since there are an
infinite number of inequivalent irreducible representations of $G$.  When $k$ is
generic, different elements of the standard basis of $\cal T$ represent
physically inequivalent (gauge-invariant) states associated with a knot (or a
solid torus). This means that any two different elements of $\cal T$ can be distinguished by the
values of the observables. 

When  $M= S^3$, we have seen in chapter 1 that gauge invariance 
implies that $k$ must be a non vanishing integer. Since $k$ multiplies the whole Lagrangian, a modification of the orientation of $S^3$ is
equivalent to change the sign of $k$; consequently, we only need to consider the case in which $k$
is a positive integer. For fixed integer $k$,  two different elements of $\cal
T$ do not necessarily correspond to different values of the observables.  In order
to determine the relevant quantum numbers associated with  a knot when $k$ is a
fixed integer, we shall introduce an equivalence relation  between the elements
of the tensor algebra.  Two elements  $\chi$ and $\chi^{\, \prime }$ of $\cal T$
are physically equivalent \cite{gp1} if the following equation  
\bea
&& \langle \,  W(\, C \, ; \, \chi \, )  \; W( \, C_1\, ; \, \chi_1 \, ) \,   
\cdots \, W (\, C_n \, ;\, \chi_n \, ) \, \rangle \Bigr |_{S^3} \; =  \nb \\ 
&& \qquad = \; \langle \,  W(\,  C  \, ;\,  \chi^{\, \prime } \, ) \; W( \, C_1 \, ; \, \chi_1 \, )
\,   \cdots \,  W (\, C_n \, ;\,  \chi_n \, ) \, \rangle \Bigr |_{S^3}  
\label{eq:fer}
\ena 

\noindent holds (with fixed $k$) for any link $L$ with components $\{ \, C, 
C_1, \cdots , C_n \,
\}$ and for arbitrary $\{ \, \chi_1 , \cdots , \chi_n \, \}$. This equivalence
relation between two elements of $\cal T$ will be denoted by $\chi \sim \chi^{\, \prime}$. 
Note that our definition of physical equivalence has a real physical meaning.
Indeed, all the properties of any gauge theory are determined by the set of the
gauge invariant observables exclusively. Thus if  $\chi \sim \chi^{\, \prime}$,
equation (\ref{eq:fer}) shows that there is no experiment which can  distinguish $\chi $
from $\chi^{\, \prime}$.  For fixed integer $k$, we shall decompose $\cal T$
into classes of physically equivalent elements; this classes will be denoted by $\psi$. The resulting
set of these classes has an algebra structure inherited from ${\cal T}$ and is called the reduced tensor algebra ${\cal T}_{(k)}$. 

In order to determine the reduced tensor algebra we need the following
property of CS observables \cite{gp1}.

\bigskip 

\shabox{\no {\bf Property 4.1}}{\em For any framed, oriented and coloured link 
$\, L$, in which one link component  has colour $\, \chi$, the associated
expectation value  $\, E(L)$ takes the form} 
\beeq
E(L)\;  \; = \; \; E_0[\chi] \; \; {\cal F} (L) \quad , 
\label{eq77}
\end{equation}
{\em where $\, {\cal F} (L)$ is a finite Laurent polynomial in the
variable $\, x = q^{\frac{1}{s}}$, with $s$ integer.}

\bigskip

\noindent {\bf Proof.}  ~Consider the connected 
sum $L_1\# L_2$ in which the two links $L_1$ and $L_2$ are copies of the link
$L$; in other words, $L_1$ and $L_2$ are separately ambient isotopic with
the framed link $L$. Let the connected sum $L_1\# L_2$ be obtained from $L_1$
and $L_2$ by acting on the link component associated with $\chi$. Then, from
eq.(\ref{eq:csr}) it follows that 
\beeq
E( L_1\# L_2 ) \; = \; {E(L) \> E(L) \over E_0[\chi]} \quad . 
\label{eq78}
\end{equation}
In general, one can show $E(L) \in Z[x^{\pm 1} ]$ \cite{glib} for any unitary 
group. For instance, in chapter 3, we have seen  that when $G=SU(3)$, $x=q^{1/3}$. One can factorize the maximal power of
$x^{-1}$ in $E(L)$ and write 
\beeq
E(L) \; = \;   x^{-a} \;  {\cal P} (x) \quad , 
\label{eq79}
\end{equation}
where $a$ is a non negative integer and ${\cal P} (x)$ is an ordinary finite
polynomial in $x$ with integer coefficients. Similarly, one has 
\beeq
E_0[\chi] \;  = \;  x^{-b}  \;  {\cal P}_0 \, (x)
\quad ,  
\label{eq710}
\end{equation}
where $b$ is a positive integer and $ {\cal P}_0 $ is a polynomial in $x$
with integer coefficients.  For instance, when $G=SU(3)$ and $\chi \equiv 
\chi[m,n]$, as shown in eq.(\ref{eq74}), one has $b=3(m+n)$. It should be
noted that $ {\cal P}_0 (x)$ is not vanishing for $x=0$. 
By using eqs.(\ref{eq79}) and (\ref{eq710}), eq.(\ref{eq78}) takes the form 
\beeq
E( L_1\# L_2 ) \; = \;  x^{- ( 2a - b)}  \; \;
{ {\cal P} (x)\; {\cal P} (x)\over {\cal P}_0 \, (x)} \quad . 
\label{eq711}
\end{equation}
Since $E( L_1\# L_2 ) \in Z[x^{\pm 1}]$, eq.(\ref{eq711}) implies that
all the roots of the polynomial ${\cal P}_0 (x)$ must also be roots of the
product ${\cal P}(x)\, {\cal P}(x)$.  A priori, there are now two 
possibilities: 
\begin{enumerate}
\item ${\cal P} (x)/ {\cal P}_0(x)$ is an ordinary polynomial in $x$ ; 
\item ${\cal P}(x) / {\cal P}_0(x)$ is not a polynomial in $x$ .  
\end{enumerate}
In case (1), ${\cal P}$ can be divided by ${\cal P}_0$ and one has 
\beeq
{\cal P} (x)\; = \;  {\cal P}_0 (x)\; {\cal G} (x)\quad , 
\end{equation}
where ${\cal G} (x)$ is a polynomial in $x$. Consequently, from eq.(\ref{eq79}) one obtains 
\beeq
E(L) \; = \; E_0 [\chi] \; x^{-a + b} \; {\cal G}(x) \quad .    
\end{equation}
Therefore, in case (1) one finds that eq.(\ref{eq77}) is satisfied with 
${\cal F} (L) = x^{-a + b} {\cal G}(x)$.  

In order to complete the
proof, we shall now show that possibility (2) is never realized; the point is 
that possibility (2)  is not consistent with the connected sum formula 
(\ref{eq:csr}). Indeed, if 
${\cal P}(x) / {\cal P}_0(x)$ is not a polynomial, then it has a pole 
$(x-x_0)^{-\beta }$  of a certain fixed order $\beta$. Since $ {\cal P}(x)$ 
multiplied by ${\cal P}(x) / {\cal P}_0(x)$ is a polynomial, this pole must be canceled by a root
of $ {\cal P}(x)$. Let us now consider the link $L_N$ which coincides with the connected
sum of $N$ copies  of the link $L$, as shown in Fig.4.1. Equation 
(\ref{eq:csr}) gives 
\beeq
E (\, L_N \, ) \; = \; E(L) \; \left ( \, \frac{E(L)}{  E_0[\chi]} \, \right )^{N -1}
\quad ,   
\label{eq714}
\end{equation}
which can be written as 
\beeq
E (\, L_N \, ) \; = \; x^{- [ Na - (N-1)b ]}  \; \;  {\cal P} (x) \; 
\left ( \, \frac{{\cal P} (x)}{ {\cal P}_0 \, (x)}\, \right )^{N-1} 
\quad . 
\end{equation}
The same argument that we have used before now implies that  ${\cal P} (x)$ 
must eliminate the pole $(x-x_0)^{- (N-1) \beta }$. 
Since $N$ can be chosen to be arbitrarily large, the  cancelation mechanism 
of the pole cannot take place because ${\cal P} (x)$ is a polynomial. 
Thus, possibility (2) is excluded and property 4.1 is proved. 
{\hfill \ding{111} 

\begin{figure}[h]
\vskip 0.9 truecm 
\centerline{\epsfig{file=\path 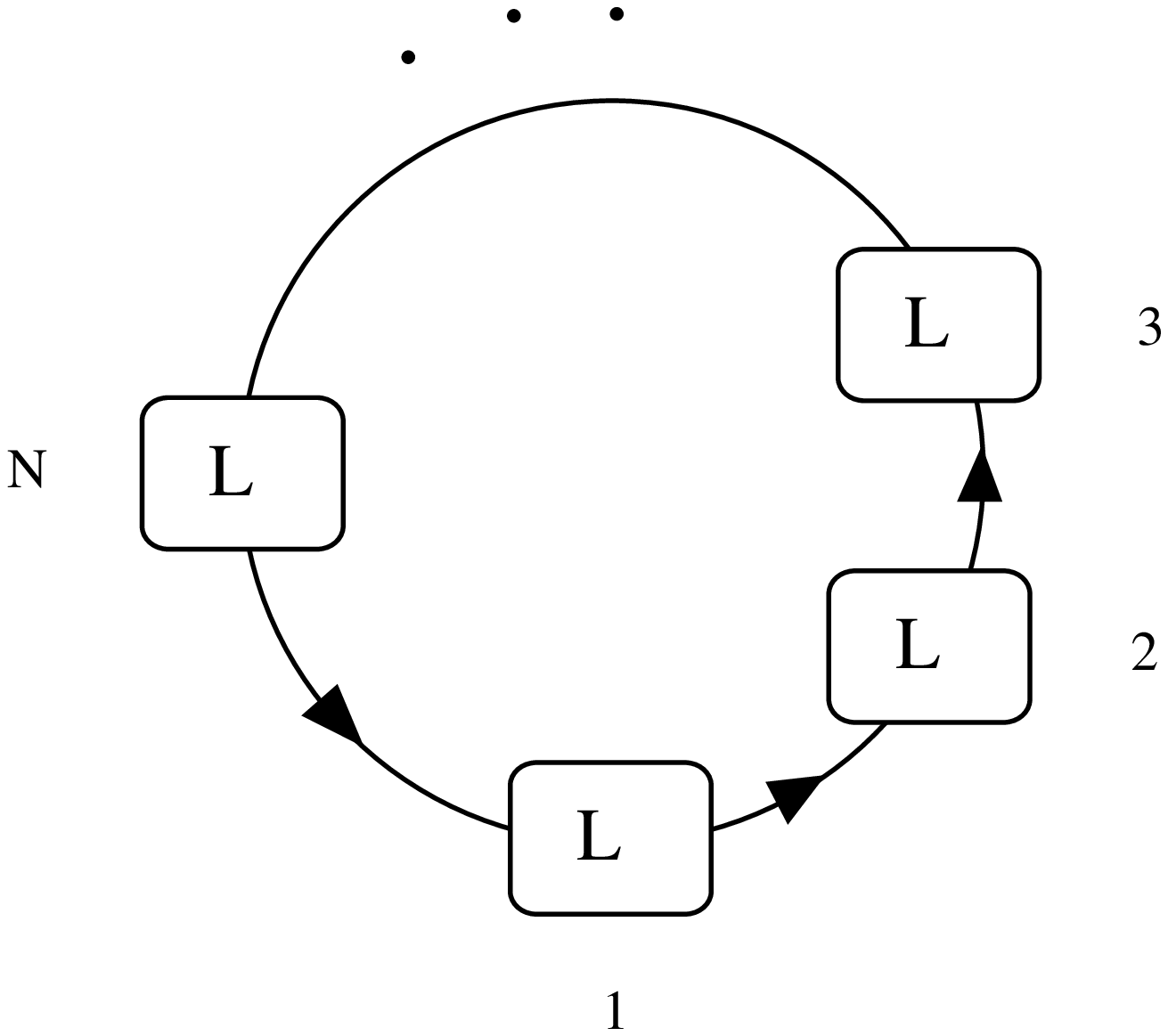,height=5cm,width=6cm}}
\vskip 0.5 truecm 
\centerline{{\bf Figure 4.1}}
\vskip 0.5 truecm 
\end{figure}

For fixed integer $k$, one can introduce the set  $I_{(k)}$ of elements of  
${\cal T}$ which are characterized by the following property;   $\zeta \in I_{(k)}$ if the
equation  
\beeq
\langle \,  W(\,  C \, ; \, \zeta \, ) \; W( \, C_1\, ; \, \chi_1 \, ) \,   \cdots \, 
W (\, C_m \, ;\, \chi_m \, ) \, \rangle \Bigr |_{S^3} \; = 0  \quad , 
\label{eq123}
\end{equation}
holds for any link $L$ in $S^3$ with components $\{ \, C, C_1, \cdots , C_m \,
\}$ and for arbitrary colour states $\{ \, \chi_1 , \cdots , \chi_m \, \} \subset {\cal T}$. In
other words,  $I_{(k)}$ is the set of elements of $\cal T$ which are physically equivalent to zero.
The following property is satisfied 

\bigskip 

\shabox{\no {\bf Property 4.2}} {\em The set $ I_{(k)}$ is an ideal of ${\cal T}$}

\bigskip

\no {\bf Proof}
Let $\chi_1 \in  I_{(k)}$ and $\chi \in {\cal T}$. In oder to prove property
4.2, we need to show that $\chi^\prime = \chi \chi_1 \in I_{(k)}$. Let us consider a link
$L$ whose component $C$  has colour $\chi^\prime$ and the link 
$L^\prime$ obtained from $L$ by replacing $C$ with $h^\diamond(B)$, where
$B$ is the pattern link shown in figure 2.1. By using the satellite formula
\ref{satg} one obtains
\beeq
E(L) \; = \; E(\cdots, \, h^\diamond(B), \cdots; \cdots, \, \chi, \chi_1, \cdots) \quad .
\label{temp}
\end{equation}
Property 4.2 follows by using property 4.1 in eq.(\ref{temp}). Indeed
\beeq
E(L) \; = \; E(\cdots, \, h^\diamond(B), \cdots; \cdots, \, \chi, \chi_1, \cdots) \; = \; E_0[\chi] {\cal P}(x) \; = \; 0 \quad .
\end{equation}   
{\hfill \ding{111}} 

By definition, the elements $\{ \, \Psi \, \}$ of the reduced tensor algebra ${\cal T}_{(k)}$ are the classes of physically
inequivalent elements; these coincide with the  elements of $\cal T$ modulo the ideal $I_{(k)}$. In other words, the reduced tensor algebra  ${\cal T}_{(k)}$
can be represented by as
\beeq 
{\cal T}_{(k)} \; = \; {\cal T}/ I_{(k)} \quad .
\end{equation}
Thus, the problem of finding ${\cal T}_{(k)}$ is  equivalent to the determination of  $I_{(k)}$.

The reduced tensor algebra represents the minimal set of the fundamental quantum
numbers which will be used, for fixed integer $k$, to colour the link 
components. Indeed, because of Property 4.2,
each element of $I_{(k)}$ is physically equivalent to the null element of $\T$.

As a 
linear space,  ${\cal T}_{(k)}$ admits a basis of physically inequivalent vectors $\{ \psi[\, i
\,], \;\; i \in {\cal D} \; \}$, where ${\cal D}$ is contained in the set of the
natural numbers. Since ${\cal T}_{(k)}$ is an algebra, one obtains 
\beeq
\psi[\, i \, ] \; \psi[\, j \,] \; = \; N_{ijm} \; \psi[\, m \,] \quad ; 
\end{equation}
the numbers $\{\,  N_{ijm} \,\}$ are called the structure constants of ${\cal
T}_{(k)}$. 
It is very useful to choose the elements $\{ \psi[i] \}$ in such a way that each
$\psi[i]$ represents the equivalence class of an irreducible representation of $G$ up to a
multiplicative factor. With this choice, the  connected sum formula 
(\ref{eq:csr}) is still valid with $\psi [i]$  playing the role of the 
irreducible representation $\rho $. 
Note that the unit in $\cal T$   corresponds to the one-dimensional trivial 
representation of $G$. The unit in ${\cal T}_{(k)}$ will be denoted by 
$\psi[1]$; $\psi[1]$ is the class defined by the 
trivial representation.
 
For fixed integer $k$, the decomposition (\ref{cart}) can
be rewritten as
\beeq
{\cal O}(V) \; = \; \sum_{i} \, \xi (i) \; W(C ; \psi[i] )  \quad . 
\label{eq:gid} 
\end{equation}
Since  ${\cal T}_{(k)} \; = \; {\cal T} \; / \; I_{(k)}$ and ${ \cal T}$ is a commutative algebra, the
structure constants $\{\,  N_{ijm} \,\}$ satisfy 
\beeq N_{ijm} \; = \; N_{jim} \quad . 
\end{equation}
The tensor algebra $\cal T$ admits a natural involution represented by complex conjugation,
$\ast: {\cal T} \ra {\cal T}$. Explicitly, $\chi[\rho] \stackrel{\ast}{\ra}
 \chi[\rho^{\ast}]$. The ideal $I_{(k)}$ is stable under the $\ast $ automorphism
of $\cal T$. Indeed, if $\eta \in I_{(k)}$ one has  
\beeq
\langle \,  W(\,  C \, ; \, \eta \, ) \; W( \, C_1\, ; \, \chi_1 \, ) \,   
\cdots \, W (\, C_n \, ;\, \chi_n \, ) \, \rangle \Bigr |_{S^3} \; = \; 0 \quad
,  
\end{equation}
for any choice of $\{  \chi_1, \chi_2, ..., \chi_n  \} \in {\cal T}$  and for 
any link  $L = \{C, C_1, ... ,C_n \}$. On the other hand, if $C$ has colour
$\eta^\ast $ one finds  
\bea
&&\langle \,  W(\, C \, ; \, \eta^\ast \, )  \; W( \, C_1\, ; \, \chi_1 \, ) \,   
\cdots \, W (\, C_n \, ;\, \chi_n \, ) \, \rangle \Bigr |_{S^3} \; = \nb
\\                      && \qquad = \; \langle \,  W(\,  C^{-1} \, ; \, \eta \,
) \; W( \, C_1\, ; \, \chi_1 \, ) \,    \cdots \, W (\, C_n \, ;\, \chi_n \, )
\, \rangle \Bigr |_{S^3} \; = \; 0 \quad  , 
\ena
where $C^{-1}$ denotes the knot $C$ with reversed orientation. Therefore, $\eta^\ast$ also belongs to
$I_{(k)}$. The action of $\ast$ on the elements   $\{ \, \psi[\, i \,] \, \}$   of ${\cal T}_{(k)}$ is
denoted by $\psi[i] \stackrel{\ast}{\ra} \psi[{i^ \ast}]$.

Since $F_{\rho_1 \rho_2 \rho} = F_{\rho_1^\ast \rho_2^\ast \rho^\ast}$, one can 
always choose the elements $\{ \psi [i] \}$ of the standard basis of ${\cal
T}_{(k)}$ in such way that  
\beeq
N_{ijm} \; = \; N_{i^\ast j^\ast m^\ast} \quad .  
\label{realn}
\end{equation}
Eq.(\ref{realn}) is a consequence of the fact that $\ast$ is an automorphism of
${\cal T}_{(k)}$.
The reduced tensor algebra 
${\cal T}_{(k)}$ is called regular if the following properties are fulfilled  \cite{gp2}: 
\vskip 0.5truecm
(i) ~$\, {\cal T}_{(k)}$ {\em is of finite order} $D$;

(ii) {\em the structure constants defined by the elements of the standard basis satisfy}
\beeq
N_{ijm} \; = \; N_{i^\ast m j} \quad ; \label{eq:fro} \end{equation}

(iii) {\em with respect to the the standard basis $\{\, \psi[i]   \; \; , \; \; i=1, 
\, 2 \, \cdots \, D \, \}$, one
has}     
\beeq
\lambda_+ \; = \;  \sum_{i=1}^D \, q^{Q(i)} \; E^2_0[i] \; \not= \; 0 
\label{eq:cas} \quad .  
\end{equation}
In equation  (\ref{eq:cas}), $Q(i)$ denotes the value of the quadratic
Casimir operator associated with an element of the class $\psi[i]$ and $E_0
[i]$ is the value of the unknot in $S^3$ with preferred framing and colour
$\psi [i]$.  

When the reduced tensor algebra is regular,  equation (\ref{eq:fro})
implies that 
\bea
&&N_{ij \boldmath{1}} \; =  \; N_{i^\ast \boldmath{1} j} \;  = \; \delta_{i^\ast \, j} \; = \; \delta_{i \, j^\ast}\quad ; \label{eq:1} \\  
&& N_{ij m^\ast} \; = \; N_{im j^\ast} \; = \; N_{jm i^\ast} \quad . 
\label{csim}
\ena
The meaning of equation (\ref{eq:1}) is clear: in the decomposition of the 
product $\psi[i] \, \psi[j]$ the unit element
$\psi[1]$ appears if and only if $\psi[i^\ast] = \psi[j]$. 

In the next sections we shall construct explicitly the reduced tensor algebra 
when $G=SU(2)$ and $G=SU(3)$, in both cases, ${\cal T}_{(k)}$ 
turns out to be of regular type. There are strong indications that this is
true also for any unitary group.  The reason  why we have introduced this
definition is that, when ${\cal T}_{(k)}$ is of regular type, an operator
realization of Dehn's surgery exists, see Chap.6  and \cite{gp2}. The operator 
surgery method will permit us to solve the CS theory in a generic three-manifold $M$. For our purposes, the existence of the surgery operator is 
crucial; so, we shall assume that ${\cal T}_{(k)}$ is regular. This seems to 
be a reasonable  assumption which can be checked directly for any concrete choice of $G$. 
  
It should be noted that the reduced tensor algebra ${\cal T}_{(k)}$ has a twofold origin. 
On the one hand, the algebraic structure of ${\cal T}_{(k)}$ is inherited
from the tensor algebra of the gauge group. On the other hand, the ideal $I_{(k)}$ which must be
factorized encodes several features of the CS field theory: 

(1) the ideal depends on the gauge group, of course;  

(2) the ideal depends on the  value of $k$; 

(3) the ideal strongly depends on the values  of the observables  of the CS theory. Indeed, the
physical equivalence relation (\ref{eq:fer}) involves all the observables of
the model. 

Let us consider the set ${\cal A} \subset {\cal R}_G$ characterized by the 
following property 
\beeq
\rho \in {\cal A} \Ra E_0[\rho] \; = \; 0 \quad .
\end{equation}
One can also consider ${\cal A}$ as contained in ${\cal T}$; indeed, by using
Property 4.1, it is  evident that ${\cal A} \subset I_{(k)}$. Because 
the knowledge of $I_{(k)}$ is equivalent to the knowledge of $\T$, our first
step in the construction of $\T$ will be the study of the set ${\cal A}$.

\section{{\bf Reduced tensor algebra for} $\mathbf{SU(3)}$}
\subsection{\bf Zeroes of the unknot} 

For fixed integer $k$, the value of the unknot $E_0[m,n]$ may vanish for certain values of $m$ and
$n$. In this section, we shall consider the set $ \cal A $ of elements $ \{ \, \chi [m,n] \, \}$ for 
which the associated value $E_0[m,n]$ is vanishing. As we have seen, $\cal A$     plays 
a crucial role in construction of   the reduced tensor algebra. 

Let us first consider the case in which $k \geq 3$. From eq.(\ref{eq76}) it 
follows that $E_0 [m,n]$ 
vanishes when  one of the following conditions is satisfied 
\begin{enumerate}
\item $\quad m\; = \; k\, a\, - \, 1 \, $ , 
\item $\quad n\; = \; k \, b \,  - \, 1  \, $ , 
\item ~$\quad m\, +\, n\; =\; k \, c \,  -\, 2  \, $ , 
\end{enumerate}
where $a$, $b$ and $c$ are integers. Let us represent the irreducible representations 
$\{(m,n)\}$ by the points of a two-dimensional square lattice. The elements of 
$\cal A$ determine a regular structure on this lattice. The zeroes described 
in point (1) lie on vertical lines which are equally spaced; those described 
in point (2) lie on (equally spaced) horizontal lines
and the zeroes in point (3) correspond to diagonal lines. The pattern of the 
zeroes is shown in
Fig.4.2. 

\begin{figure}[h]
\vskip 0.9 truecm 
\centerline{\epsfig{file=\path 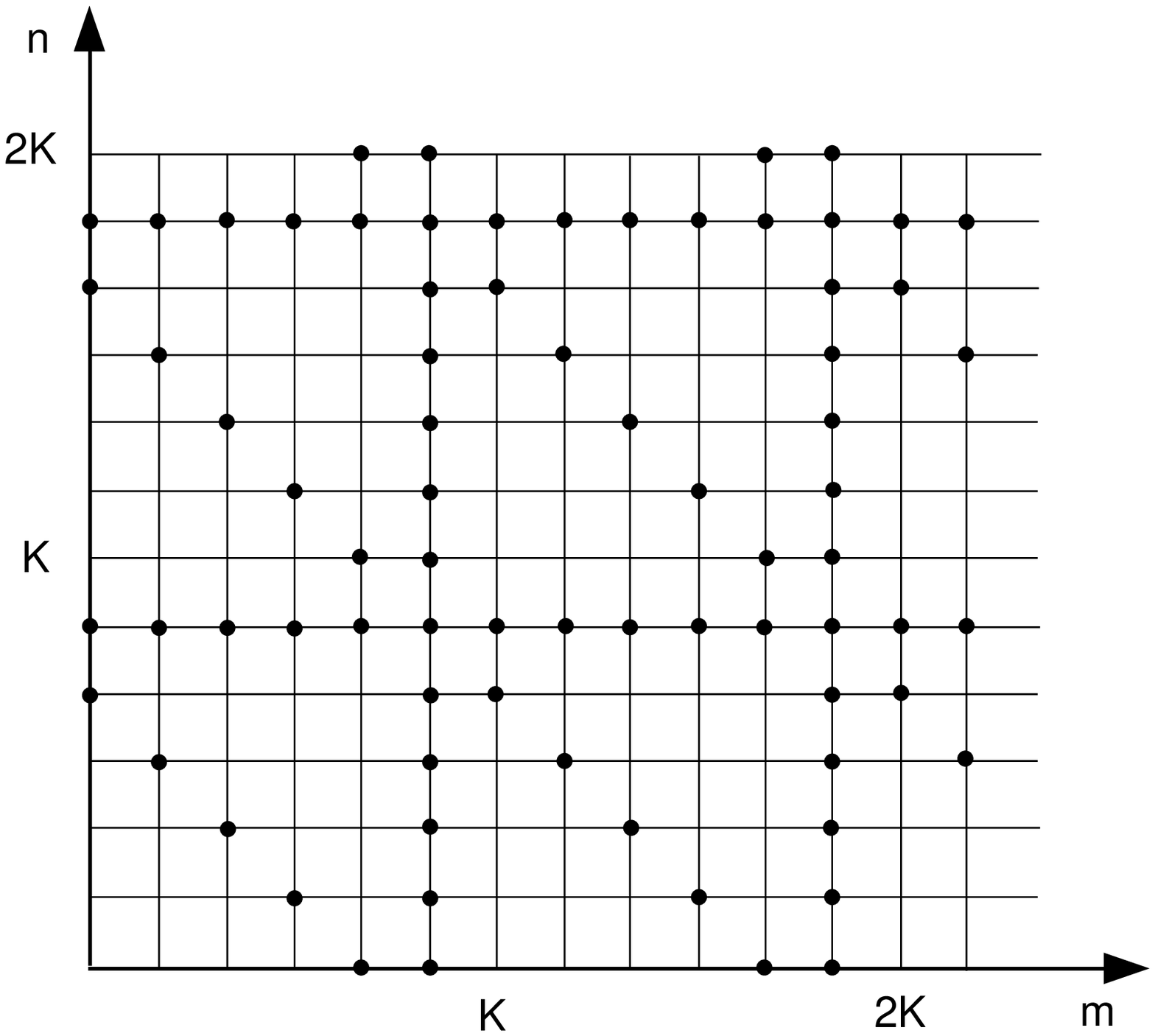,height=6.5cm,width=6.5cm}}
\vskip 0.5truecm 
\centerline{{\bf Figure 4.2}}
\vskip 0.5truecm 
\end{figure}

\noindent The region of the plane delimited by the two axes $m=0$, $n=0$ and the diagonal line $m+n
= k -  2$ does not contain zeroes of the unknot and will be denoted by 
$\Delta_k$. The region
$\Delta_k$ is called the fundamental domain; we will show that the classes 
represented by the
points in $\Delta_k$  form a complete basis of ${\cal T}_{(k)}$.

\noindent When $k=2$, eq.(\ref{eq76}) gives 
\beeq
\lim_{k \rightarrow 2} \, E_0 [m,n] \; = \; \left \{ \begin{array}{cc} 0 & 
\mbox{for } m \mbox{ and } n {\mbox \; odd} ; \\
-  (n+1)/2 & \mbox{for } n \mbox{ odd and } m \mbox{ even} ; \\ 
-  (m+1)/2 & \mbox{for } m \mbox{ odd and } n \mbox{ even} ;\\
-  (m+n+2)/2 & \mbox{for } m \mbox{ and } n \mbox{ even} . 
\end{array} \right.
\end{equation}
Finally, for $k=1$ one has 
\beeq
\lim_{k \rightarrow 1} \, E_0 [m,n] \; = \; \frac{1}{2} (m+1)(n+1)(m+n+2) \; = \; D(m,n) 
\quad . 
\label{eq102}
\end{equation}
It should be noted that the cases in which $k=2$ and $k=1$ present certain  
peculiarities with respect to the general situation that one has for $k\geq 3$. Consequently, in 
the construction of the reduced tensor algebra  we will need to distinguish  
the cases $k=1$, $k=2$ and $k \geq 3$.  

\subsection{\bf Null elements} 

In this section we shall prove that, when $k\geq 3$, all the elements of 
$\cal A$ are generated
\cite{io,gp1} by  two fundamental nontrivial ``null elements".  

\bigskip

\shabox{
\noindent  {\bf Property 4.3}} {\em For $\, k \geq 3$,  each element 
$\chi [m,n]$ of  $\, \cal A$ 
can be written  in the form 
\beeq
\chi [m,n] \; = \; \zeta_1 \> \chi_1 \, + \, \zeta_2 \> \chi_2   \quad ,
\label{eq111}
\end{equation}
where   
\beeq
\zeta_1 \;=\; \chi [ k-2 , 0] \quad , 
\label{eq112}
\end{equation}
\beeq
\zeta_2 \; =\; \chi [ k-1 , 0] \quad , 
\label{eq113}
\end{equation}
and the elements $\, \chi_1$ and $\, \chi_2$ belong to the representation 
ring of $\, SU(3)$. } 

\bigskip

\noindent {\bf Proof}  ~It is easy to verify that all the elements of $\cal R$ which can be written as in 
eq.(\ref{eq111}) form an ideal of $\cal R$ denoted by ${\cal B } (\zeta_1 , \zeta_2
)$. We need to prove that 
each element  of  $ \cal A $ belongs to  ${\cal B } (\zeta_1 , \zeta_2 )$. In 
order to do this, we
shall use the tensor product decompositions  
\beeq 
(m,n) \otimes  (1,0)\; =\; (m+1,n) \oplus (m-1,n+1) \oplus (m,n-1) \quad , 
\label{eq114}
\end{equation}
\beeq
(m,n) \otimes (0,1)\; =\; (m,n+1) \oplus (m+1,n-1) \oplus (m-1,n) \quad .   
\label{eq115} 
\end{equation}
Let us recall that the elements of $\cal A$ correspond to the points on diagonal, horizontal and vertical lines of the square lattice introduced in the previous section. First of all, we use eqs.(\ref{eq114}) and
(\ref{eq115}) to  prove that if three elements of  ${\cal B } (\zeta_1 , 
\zeta_2 ) $ are represented by 
three consecutive points in a diagonal line, then all the remaining points  
of the diagonal belong to ${\cal B } (\zeta_1 , \zeta_2 ) $.  Indeed, setting
$m \rightarrow m+1 \; \; , \; \;  n
\rightarrow n-1$ in eqs.(\ref{eq114})-(\ref{eq115}) and subtracting them, we 
have   
\bea
&&\chi [m,n] \; \chi [1,0] \;- \chi [m+1, n-1] \; \chi [0,1] \; = \nb \\
&&= \; \chi [m-1,n+1] \;- \chi [m+2,n-2] \quad . 
\label{eq116}
\ena
Thus, if  $\chi [m-1,n+1]$, $\chi [m,n]$ and $\chi [m+1,n-1]$ belong to 
${\cal B } (\zeta_1 , \zeta_2 )$, eq.(\ref{eq116}) implies that 
$\chi [m+2,n-2]$ also belongs to 
${\cal B } (\zeta_1 , \zeta_2 )$ (see Fig.4.3). 

\begin{figure}[h]
\vskip 0.9 truecm 
\centerline{\epsfig{file=\path 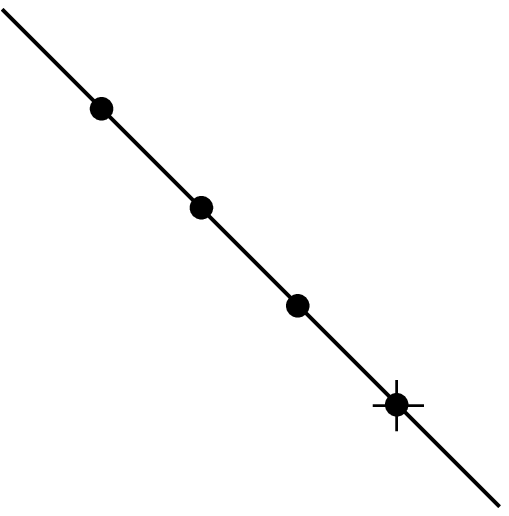,height=2cm,width=3cm}}
\vskip 0.5truecm 
\centerline{{\bf Figure 4.3}}
\vskip 0.5truecm 
\end{figure}

\noindent With the substitution $m \rightarrow m-1 $ and $n \rightarrow n-1 
\, $ in eq.(\ref{eq116}),  we
conclude that $\chi [m-2, n+2]$ also belongs to ${\cal B } (\zeta_1 , \zeta_2 )$, as shown in Fig.4.4.  

\begin{figure}[h]
\vskip 0.9 truecm 
\centerline{\epsfig{file=\path 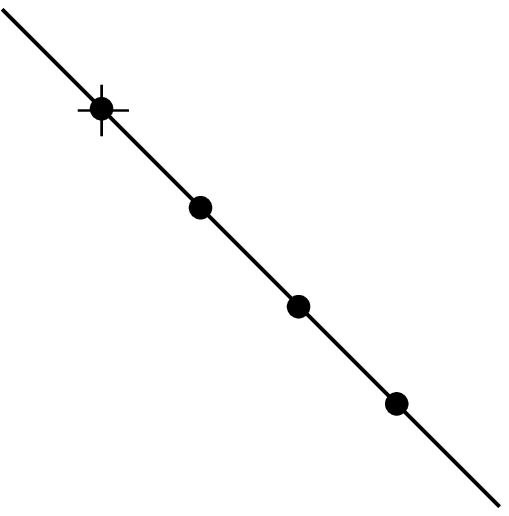,height=2cm,width=3cm}}
\vskip 0.5truecm 
\centerline{{\bf Figure 4.4}}
\vskip 0.5truecm 
\end{figure}

\noindent Clearly, all the remaining points on the diagonal can be obtained by
induction. The same argument can be used to prove that the presence of three 
consecutive elements of 
${\cal B } (\zeta_1 , \zeta_2 )$ in a vertical (or
horizontal line)  implies  that all the remaining points of the line are also elements of ${\cal B } (\zeta_1 , \zeta_2 )$. 
For a vertical line the relevant formula is:
\bea
&&\chi [m,n] \; \chi[ 1,0 ] \;- \chi [m,n+1] \; \chi [0,1] \;=  \nb \\
&&=\; \chi [m,n-1] \; - \chi [m,n+2]  \quad , 
\label{eq117}
\ena
whereas for an horizontal line the recursive formula is
\bea
&&\chi [m,n] \; \chi [1,0] \; - \chi [m-1,n] \;  \chi [0,1] \; = \nb \\
&&=\; \chi [m+1,n] \; - \chi [m-2,n]  \quad . 
\label{eq118} 
\ena
Now we are ready to express all the elements of $ \cal A $ in terms of $\zeta_1$ and $\zeta_2$. 
Let us consider the elements of the ``first" diagonal
$m+n = k-2$ (see figure 4.2); the fundamental null vector $\zeta_1 = \chi [k-2,0]$ corresponds to
the crossing point of the diagonal with the $m$-axis. From equation 
\beeq
\zeta_1 \; \chi [1,0] \; = \; \zeta_2  \; + \; \chi [ k-3 ,1] \quad , 
\end{equation}
it follows that $\chi [ k-3 ,1]$, which belongs to $\cal A$, also belongs 
to ${\cal B } (\zeta_1 , \zeta_2 )$.  The following equation   
\beeq
\chi [k-3,1] \; \chi [1,0 ]  - \zeta_1 \; \chi [0,1]  \; = \; \chi [k-4,2] \quad , 
\end{equation}
implies that $\chi [k-4,2] \in  {\cal B } (\zeta_1 , \zeta_2 )$. At this stage, we have shown that
three elements of $\cal A$, which are represented by three consecutive points on the first diagonal
line, belong to ${\cal B } (\zeta_1 , \zeta_2 )$. Consequently, all the
points of this diagonal line belong to ${\cal B } (\zeta_1 , \zeta_2 )$.  In other words, the
elements of $\cal A$ corresponding to the points of the first diagonal line (see Fig.4.2)) are
also elements of ${\cal B } (\zeta_1 , \zeta_2 )$. 

Let us consider now the points lying on the first vertical line $m=k-1$. The first base point is 
$\zeta_2 = \chi [k-1,0]$. The second point can
be obtained by using:  
\bea
&&\zeta_2 \; \chi [0,1] \; =\; \chi [k-1,1] \; + \;  \zeta_1  \quad , \nb \\
&& \quad \Rightarrow \chi [k-1,1] \in {\cal B } (\zeta_1 , \zeta_2 ) \quad . 
\ena
The third point can be expressed in terms of $\zeta_1$ and $\zeta_2$ according to  
\bea
&&\chi [k-1,1] \; \chi [0,1] \; = \; \chi [k-1,2]  + \chi [k,0] + \chi [k-2,1] \quad , \nb \\
&&\zeta_2 \; \chi [1,0] \; =\; \chi [k,0] + \chi [k-2,1] \quad , \nb \\
&& \quad \; \Rightarrow \chi [k-1,2] \; = \; \chi [k-1,1] \; \chi [0,1] - \zeta_2 \; \chi [1,0] \in 
{\cal B } (\zeta_1 , \zeta_2 ) \quad . 
\ena
Since $\chi [k-1,0]$, $\chi [k-1,1]$ and $\chi [k-1,2]$ belong to ${\cal B } (\zeta_1 ,
\zeta_2 ) $,  all the  points of the vertical line  $(k-1,n)$ belong
to ${\cal B } (\zeta_1 , \zeta_2 ) $. 

We now examine  the first horizontal line $n = k-1$; for the first point we 
have  
\bea
&&\chi [0,k-2] \; \chi [0,1] \; =\; \chi [0,k-1] + \chi [1 , k-3] \quad, \nb \\
&&\quad \; \Rightarrow \chi [0,k-1 ] \in {\cal B } (\zeta_1 , \zeta_2 ) 
\quad . 
\ena
The second point can be expressed as  
\bea
&& \chi [0,k-1] \; \chi [1,0] \; = \; \chi [1,k-1]  + \chi [0,k-2] \quad ,
\nb \\
&& \; \quad \Rightarrow \chi [1,k-1 ]\in 
{\cal B } (\zeta_1 , \zeta_2 )  \quad . 
\ena
For the third point we get 
\bea
&&\chi [1,k-1] \; \chi [1,0] \; =\; \chi [1,k-2] + \chi [0,k-2] + \chi [2,k-1] \quad , \nb \\
&& \chi [0,k-1] \; \chi [0,1] \; =\; \chi [1,k-2] + \chi [0,k-2] \quad , \nb \\
&& \; \quad \Rightarrow \chi [2,k-1 ] \; = \; \chi [1,k-1] \; \chi [1,0]  -  \chi [0,k-1] \; \chi
[0,1] 
 \in {\cal B } (\zeta_1 , \zeta_2 )   \quad . 
\ena
At this stage, by using eq.(\ref{eq118}) recursively, we find that all the points of the first
horizontal line belong to ${\cal B } (\zeta_1 , \zeta_2 )$. 

To sum up, in this first step we have shown that the elements of $\cal A$ which correspond to the
points of the first diagonal, vertical and horizontal lines belong to 
${\cal B } (\zeta_1 , \zeta_2 )$.

In the second step, we consider  all the remaining diagonal and horizontal lines.  Let us 
start with the base points of the second diagonal line. The element given by  $\chi [k-1,k-1]$
lies on the first horizontal line, therefore it belongs to ${\cal B } (\zeta_1 , \zeta_2 )$. 
The second base point corresponds to $\chi [k,k-2]$; indeed, 
\bea
&&\chi [k-1,k-2] \; \chi [1,0] \; =\; \chi [ k,k-2] + \chi [k-2,k-1] + \chi [k-1,k-3] \quad , \nb \\
&& \Rightarrow \chi [k,k-2] \; =\; \chi [k-1,k-2] \; \chi [1,0] - \chi [k-2,k-1] \nb \\
&&\;  \; -  \chi [k-1,k-3 \in  {\cal B } (\zeta_1 , \zeta_2 )\quad .
\ena
For the last element, we have 
\bea
&& \chi [k-2,k-1] \; \chi [0,1] \; =\; \chi [k-2,k] + \chi [k-1,k-2]  + \chi [k-3,k-1] \quad , \nb \\
&&\; \quad \Rightarrow \chi [ k-2,k]  \in {\cal B } (\zeta_1 , \zeta_2 ) \quad . 
\ena
By using eq.(\ref{eq116}) recursively, one  can complete the second diagonal line. The same argument 
that we have presented before can be used to analyze all diagonal and horizontal lines. Indeed, 
since all the points of the second diagonal belong to ${\cal B } (\zeta_1 , \zeta_2 )$, we can
construct the second horizontal line as we did for the first. Then,  we can construct the third
diagonal line and so on. It is clear that this recursive procedure shows that  all the
points on the diagonal and horizontal lines belong to ${\cal B } (\zeta_1 , \zeta_2 )$.

In the third and final step, we consider the vertical lines. 
Three base points for the second vertical line can be obtained by exploring the
crossing between the third diagonal and the first horizontal lines. The first element is
\bea
&&\chi [2k-1,k-1] \; \chi [1,0] \; =\; \chi [2k,k-1] + \chi [2k-2,k] + \chi [2k-1,k-2] \quad , \nb \\
&&\; \Rightarrow \chi [2k-1,k-2]  \in  {\cal B } (\zeta_1 , \zeta_2 )\quad . 
\ena
For the second element, we take 
\bea
&&\chi [2k-1,k-2 \; \chi [1,0] \; =\; \chi [2k,k-2]  + \chi [2k-2,k-1 +  \chi [2k-1,k-3] \quad , \nb \\
&& \; \Rightarrow \chi [2k-1,k-2 ] \in {\cal B } (\zeta_1 , \zeta_2 ) \quad . 
\ena
The third element, which corresponds to  $\chi [2k-1,k-1]$,  belongs to the first
horizontal line. Therefore, by using eq.(\ref{eq117}), we can complete the 
second vertical line. Repeating the same argument at every crossing point between the first
horizontal line and the diagonal lines, all the vertical lines can be
constructed. 

In conclusion, all the elements of $\cal A$ belong to ${\cal B } (\zeta_1 , \zeta_2 )$ or, in
other words, can be written in the form shown in eq.(\ref{eq111}). Consequently, Property 4.3 
is proved. {\hfill \ding{111}}

\subsection{\bf Reduced tensor algebra for $\mathbf{k \geq 3} $} 

In this section we will determine the structure of the reduced tensor algebra 
${\cal T}_{(k)}$ for
$k\geq 3$.  Let $L$  be a coloured and framed link in $S^3$ with components  
$\{ \, C ,
C^\prime , \cdots \, \}$ in which the component $C$  has colour $\eta  \in {\cal T}$.  If
(for fixed integer $k$) $\eta = \chi [m,n] \in {\cal A}$, Property 4.1 implies that  
$\langle \, W(L)
\, \rangle \bigr |_{S^3} = 0$  for any link $L$. As we have already mentioned, 
this means that $\eta =  \chi [m,n] \in {\cal A}$ is physically equivalent to 
the null element. 

\vskip 0.5truecm

\noindent {\bf Property 4.4} {\em If the component $\, C$ of a link $\, L$ has colour $\, \eta \; =
\; \chi_1 \; \chi_2$ with $\, \chi_1 \in {\cal A}$ and $\, \chi_2 \in {\cal T}$, one has} 
\beeq
\langle \, W(L) \, \rangle \bigr |_{S^3} \; = \;0
\label{eq121}
\end{equation}
\vskip 0.5truecm
\noindent {\bf Proof.}  ~By using the satellite formula
(\ref{satg}), the expectation value of $W(L)$ can be expressed as   
\bea
&&\langle \, W(L) \, \rangle \Bigr |_{S^3} \; \equiv 
\; \langle W(\, C, C^\prime , \cdots ; \chi_1 \> \chi_2,  \chi^{\, \prime } \cdots ) \rangle \Bigr
|_{S^3}\; = \nb \\
&& = \; \langle W(\, h^\diamond (B),  C^\prime , \cdots ; \chi_1 \, ,\,  \chi_2 ,  \, \chi^{\,
\prime } \cdots )\rangle \Bigr |_{S^3}   \; \equiv    \; \langle \, W(L^\prime ) \, \rangle \Bigr
|_{S^3}  \quad , 
\label{eq122}
\ena
where $h^\diamond $ is the homeomorphism defined in section 2.2 and $B$ is the 
two-component pattern link shown in Fig.2.3.  In eq.(\ref{eq122}),  $L^\prime $ denotes the satellite 
which has been obtained from $L$ by replacing the component $C$ with the image $h^\diamond (B)$ of
the pattern link. Note that $L^\prime $ has two components which have colours 
$\chi_1 $ and  $\chi_2 $. Therefore, if $\chi_1 \in {\cal A}$, by using 
Property 4.1 one
has  $\langle \, W(L^\prime ) \, \rangle \bigr |_{S^3} =0$ and, consequently, 
$\langle \, W(L) \, \rangle \Bigr |_{S^3} =0$. {\hfill \ding{111}} 

\vskip 0.5truecm

We shall now determine the ideal $I_{(k)}$.
Clearly, Property 4.1 implies that any element $\eta $ of the form $\eta \; = \; \chi_1 \; \chi_2$, 
with $\chi_1 \in {\cal A}$ and  $\chi_2 \in {\cal T}$, belongs to $I_{(k)}$. Since any element of
$\cal A$ is of the form (\ref{eq111}), it is natural to expect that each 
element $\zeta $  of
$I_{(k)}$ can be expressed in the form  
\beeq
\zeta \; = \; \zeta_1 \; \chi \; + \zeta_2 \; \chi^{\, \prime} \quad ,   
\label{eq124}                                   
\end{equation}
where $\chi $ and $\chi^{\, \prime }$ belong to $\cal T$. We will prove that this is indeed the
case; in other words, for fixed integer $k \geq 3$, $I_{(k)}$ is the ideal 
generated by the two null
vectors $\zeta_1$ and $\zeta_2$ shown in eqs.(\ref{eq112}) and (\ref{eq113}). 
The proof essentially consists of
two steps.  Firstly, assuming that each element of $I_{(k)}$  is of the form 
(\ref{eq124}), we shall
determine the corresponding set ${\cal T}_{(k)}$ of equivalence classes. 
Secondly, we will show that  this set is physically irreducible; i.e. $\Psi \sim 0$ implies $\Psi \equiv 0$. 

We shall now give three basic rules which connect, for fixed integer $k \geq 3$, physically
equivalent elements of $\cal T$.  Let us recall that each element $\chi [m,n]$ of the standard
basis of $\cal T$ is represented by a point in the square lattice shown in Fig.4.2. 
Furthermore, the elements of $\cal A$ are organized in diagonal, vertical and horizontal lines in
the same lattice. To each line is associated \cite{io,gp1} a correspondence 
rule. 

\bigskip

\shabox{
\noindent {\bf Rule 1}}  {\em Let us consider the element $\, \chi [m,n]$. 
Suppose that, for some
nonnegative integer $\, p$,  $\, \chi [m-p,n]$ belongs to $\, \cal A$ and is represented by a point
on a diagonal line; then   } 
\beeq
\chi [m ,n] \;  \sim \; - \; \chi [m- p,n-p] \quad . 
\label{eq125}
\end{equation}

\bigskip

\noindent {\bf Proof} ~When $p=1$, one has 
\beeq
0 \; \sim \;  \chi [m-1,n] \; \chi [1,0] \; = \; \chi [m,n] + \chi [m-2,n+1] + \chi [m-1, n-1] 
\quad .  
\label{eq126}
\end{equation}
Since $\chi [m-2, n+1] $ is represented by a point on the diagonal which is determined by $\chi
[m-1,n]$,  $\chi [m-2, n+1] $ belongs to $\cal A$ and then $\chi [m-2, n+1] \sim 0$. Therefore, eq.(\ref{eq126}) gives 
\beeq
\chi [m,n]\; \sim \; - \; \chi [m-1, n-1] \quad ,  
\end{equation}
which shows that eq.(\ref{eq125}) is satisfied for $p=1$. We now proceed by induction. Let us assume that 
eq.(\ref{eq125}) is valid for $p\leq {\overline p}$. We consider the following decompositions 
\beeq
\chi [m,n]\; \chi [1,0] \; = \; \chi [m+1,n] + \chi [m-1,n+1] + \chi [m, n-1] \quad , 
\label{eq128}
\end{equation}
\beeq
\chi [m-\overline p ,n-\overline p ]\; \chi [1,0] \; = \; \chi [m - \overline p +1,n - \overline
p] + \chi [m - \overline p -1,n - \overline p +1] + \chi [m - \overline p , n - \overline p -1]
\quad .   
\label{eq129} 
\end{equation}
By the induction hypothesis, one has 
\bea
&&\chi [m,n] \; \sim \; - \; \chi [m -\overline p , n - \overline p] \quad , 
\nb \\ 
&& \chi [m-1,n+1] \; \sim \; - \; \chi [m - 1 - \overline p , n + 1 - \overline p ] \quad , \nb \\
&& \chi [m,n-1] \; \sim \; - \; \chi [m - (\overline p -1), n - 1 - (\overline p -1)] \quad .  
\label{eq1210}
\ena
Therefore, by adding eqs.(\ref{eq128}) and (\ref{eq129}) and by using (\ref{eq1210}), one finds 
\beeq
\chi [m+1,n] \; \sim \; - \; \chi [m +1 - (\overline p +1) , n - (\overline p +1)] \quad .  
\label{eq1211}
\end{equation}
Eq.(\ref{eq1211}) shows that eq.(\ref{eq125}) holds also for $p = \overline p 
+1$;   this concludes the proof. {\hfill \ding{111}}  

\bigskip

\shabox{
\noindent {\bf Rule 2}}  {\em Let us consider the element $\, \chi [m,n]$. Suppose that, for some
nonnegative integer $\, p$,  $\, \chi [m-p,n]$ belongs to $\, \cal A$ and is represented by a point
on a vertical line; then   } 
\beeq
\chi [m ,n] \;  \sim \; - \; \chi [m- 2p,n+p] \quad .  
\label{eq1212}
\end{equation}

\bigskip 

\noindent {\bf Proof} ~When $p=1$, one has 
\beeq
0 \; \sim \;  \chi [m-1,n] \; \chi [1,0] \; = \; \chi [m,n] + \chi [m-2,n+1] + \chi [m-1, n-1] 
\quad .  
\label{eq1213}
\end{equation}
Since $\chi [m-1, n-1] $ is represented by a point on the vertical line which is determined by
$\chi [m-1,n]$,  $\chi [m-1, n-1] $ belongs to $\cal A$ and then $\chi [m-1, n-1] \sim 0$.
Therefore, eq.(\ref{eq1213}) gives 
\beeq
\chi [m,n]\; \sim \; - \; \chi [m-2, n+1] \quad ,  
\end{equation}
which shows that eq.(\ref{eq1212}) is satisfied for $p=1$. We now proceed by induction. Let us assume that 
eq.(\ref{eq1212}) is valid for $p\leq {\overline p}$. We consider the following decompositions 
\beeq
\chi [m,n]\; \chi [1,0] \; = \; \chi [m+1,n] + \chi [m-1,n+1] + \chi [m, n-1] \quad , 
\label{eq1215}
\end{equation}
\bea
&&\chi [m-2\overline p ,n+\overline p ]\; \chi [1,0] \; = \; \chi [m - 2\overline p +1,n +
 \overline p] + \nb \\ 
&&\; + \chi [m - 2\overline p -1,n + \overline p +1]  
 + \chi [m - 2\overline p , n +\overline p
-1] \quad .   
\label{eq1216} 
\ena
By the induction hypothesis, one has 
\bea
&&\chi [m,n] \; \sim \; - \; \chi [m -2\overline p , n + \overline p] \quad ,
 \nb \\ 
&&\chi [m-1,n+1] \; \sim \; - \; \chi [m - 1 - 2(\overline p -1), n + 1 + (\overline p -1)] \quad , \nb \\
&&  \chi [m,n-1] \; \sim \; - \; \chi [m - 2\overline p , n - 1 - \overline p ] \quad . 
\label{eq1217}
\ena
Therefore, by adding eqs.(\ref{eq1215}) and (\ref{eq1216}) and by using (\ref{eq1217}), one finds 
\beeq
\chi [m+1,n] \; \sim \; - \; \chi [m +1 - 2(\overline p +1) , n + (\overline p +1)] \quad .  
\label{eq1218}
\end{equation}
Eq.(\ref{eq1218}) shows that eq.(\ref{eq1212}) holds also for $p = \overline p +1$.    { \hfill \ding{111}} 

\bigskip

\shabox{
\noindent {\bf Rule 3}}  {\em Let us consider the element $\, \chi [m,n]$. Suppose that, for some
nonnegative integer $\, p$,  $\, \chi [m,n-p]$ belongs to $\, \cal A$ and is represented by a point
on a horizontal line; then   } 
\beeq
\chi [m ,n] \;  \sim \; - \; \chi [m + p, n -2p] \quad .  
\label{eq1219}
\end{equation}

\vskip 0.5truecm

\noindent {\bf Proof} ~The proof of eq.(\ref{eq1219}) is based on the same 
algebraic manipulations as those used
in  the proof of Rules 1 and 2. {\hfill \ding{111}}  

\vskip 0.5truecm

The equivalence relations described by Rules 1, 2 and 3 are shown in Fig.4.5. 
These rules must be
used to connect points of the physical region  ($m \geq0 $ and $n \geq 0$) of the square lattice. 
The points of the physical region of the lattice describe all the irreducible representations
$ \{ \, (m,n) \, \}$ of $SU(3)$ which label the elements $\{ \, \chi [m,n] \, \}$ of the standard
basis of $\cal T$. 

For fixed integer $k \geq 3$, let us consider a generic element $\chi [m,n]$. If $\chi [m,n] $
belongs to $\cal A$, then $\chi [m,n] \, \sim \, 0 $. If $\chi [m,n]\,  \not\in \, {\cal A}\, $, 
by using Rules 1, 2 and 3 recursively, it is easy to see that $\chi [m,n] $ is physically equivalent
(with a well determined sign) to an element $\chi [a,b]$ represented by a point in the fundamental
domain $\Delta_k$. The points of $\Delta_k$ have coordinates $(a,b)$ characterized by  
\beeq
\Delta_k \; \equiv \; \{ \; (a,b) \; \} \quad {\rm with} \quad \left \{\begin{array}{cc} 
 0 \; \leq \; a\; < \; k-2 \quad , & \; \\ 
 0 \; \leq \; b\; < \; - a + k-2  \quad . & \; \\
\end{array} \right.
\end{equation}
\vskip 0.8truecm 

\begin{figure}[h]
\vskip 0.9 truecm 
\centerline{\epsfig{file=\path 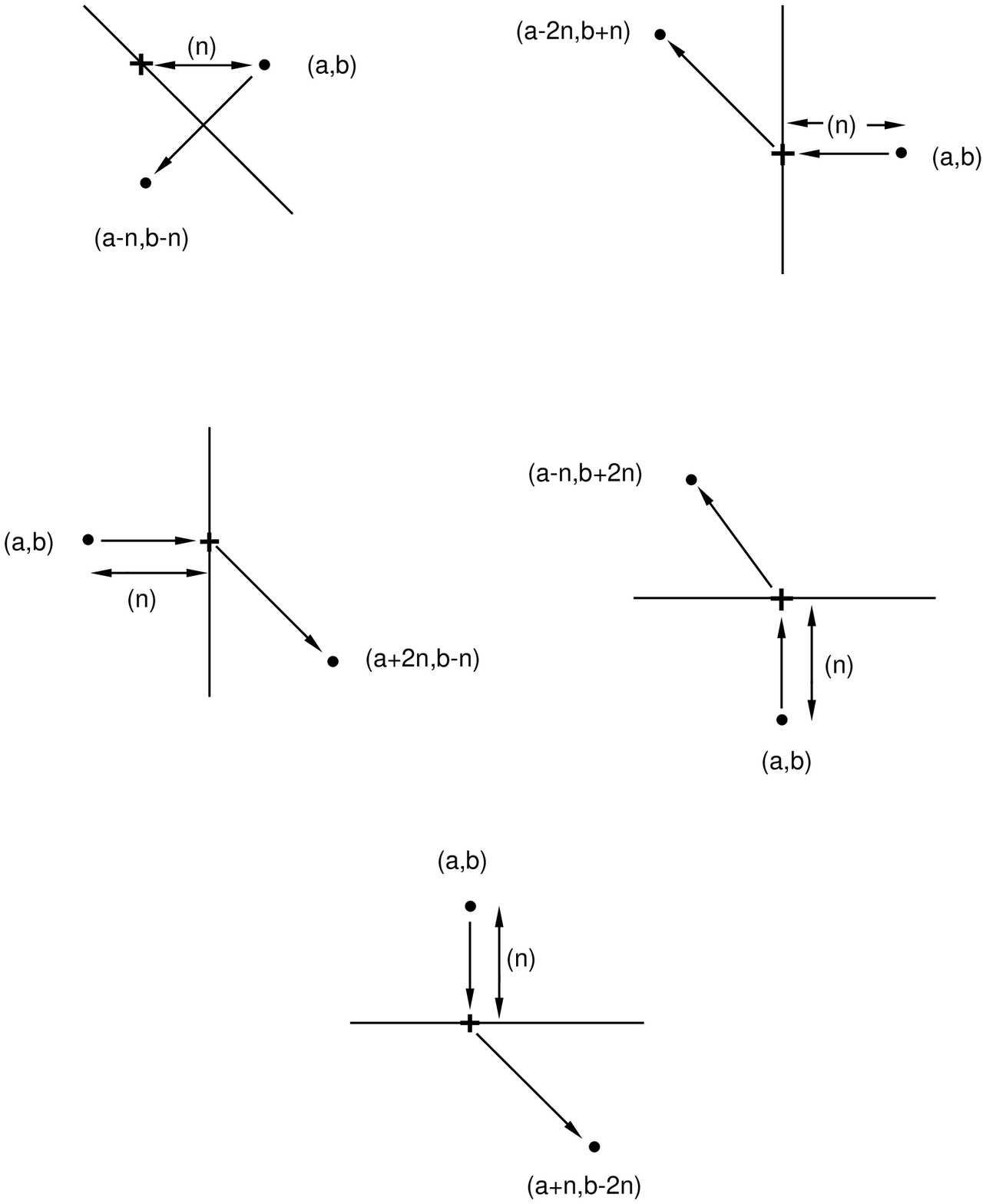,height=9.5cm,width=9.5cm}}
\vskip 0.5truecm 
\centerline{{\bf Figure 4.5}}
\vskip 0.5truecm 
\end{figure}

We shall denote by  $\Psi [a,b]$  the class associated with the irreducible representation
$(a,b)$ of $SU(3)$ with the point $(a,b) \in \Delta_k $. A generic element $\chi $ of $\cal T$
admits a linear decomposition in terms of the elements of the standard basis of $\cal T$ 
\beeq
\chi \; = \; \sum_{m,n} \,  \xi (m,n) \; \chi [m,n] \quad . 
\label{eq1221}
\end{equation}
Let $\Psi $ be the class of physical equivalent elements associated with $\chi  $. Since the
class corresponding to each $\chi [m,n]$ is the trivial class or a class $\Psi [a,b]$ with 
$(a,b) \in \Delta_k$, from eq.(\ref{eq1221}) it follows that   
\beeq
\Psi \; = \; \sum_{(a,b)\,  \in \, \Delta_k } \; \xi^{\, \prime } (a,b) \; 
\Psi [a,b] 
\quad ,  
\label{eq1222}
\end{equation}
where $\{ \, \xi^{\, \prime } (a,b) \, \}$ are linear combinations of the coefficients $\{ \, \xi (a,b) \, \}$.

To sum up \cite{gp1}, assuming that  for $k\geq 3$ each element of $I_{(k)}$ has the form
(\ref{eq124}), eq.(\ref{eq1222})
shows that the elements of $\cal T$ modulo the ideal $I_{(k)}$ admits a linear decomposition in
terms of the classes $\{ \Psi [a,b] \, \}$ with  $(a,b) \, \in \Delta_k $.
Therefore, $\{
\Psi [a,b] \, \}$ form a basis of ${\cal T}_{(k)}$ that we call the standard basis.  
It remains to be verified that ${\cal T}_{(k)}$, defined above, is physically irreducible. This
means that if  the element $\Psi_* \in {\cal T}_{(k)}$ 
is physically equivalent to zero   $ \Psi_* \, \sim \, 0$, then $\Psi_* \, = \,
0$. The proof is very simple.  Let us consider the Hopf link in which one component has colour
$ \Psi_* $ and the other component has colour given by a generic element $\Psi $ of  
${\cal T}_{(k)}$. If $\Psi_* \, \sim \, 0$, the associated expectation value vanishes  
\beeq
\langle \, W(C_1 ; \Psi_* ) \, W(C_2 ; \Psi ) \, \rangle \Bigr |_{S^3} \; =  \; 0 
\quad .  
\label{eq1223}
\end{equation}
Let us expand $\Psi_*$ as in (\ref{eq1222}),  
\beeq
\Psi_* \; = \; \sum_{(m,n) \, \in \, \Delta_k } \; \xi_* (m,n) \; \Psi [m,n] 
\quad . 
\end{equation}
Since eq.(\ref{eq1223})  holds for arbitrary $\Psi $, eq.(\ref{eq1223}) 
implies that,  for any $(a,b) \in
\Delta_k $,  one has 
\beeq
\sum_{(m,n) \, \in \, \Delta_k } \; \xi_* (m,n) \; H[(m,n) , (a,b)] \; = \; 0 
\quad . 
\label{eq1225}
\end{equation}
The complex numbers $\{ \, H[(m,n) , (a,b)] \, \}$, for $(m,n) \in \Delta_k 
$ and $(a,b) \in \Delta_k $, can be understood as the matrix elements of the 
the Hopf matrix $H$.  As shown in Appendix C, the Hopf matrix $H$ for $G=SU(3)$ is invertible; therefore, eq.(\ref{eq1225})
implies that 
\beeq
\xi_* (m,n) \; = \; 0 \quad , \quad {\rm for ~any} \quad (m,n) \in \Delta_k  \quad . 
\end{equation}
This shows that $\Psi_* \, = \, 0$. In conclusion, the results of this section are summarized by
the following theorem \cite{gp1}. 

\bigskip 

\shabox{
\noindent {\bf Theorem 4.1}} {\em For fixed integer $\, k \geq 3$, the reduced tensor algebra 
$\, {\cal T}_{(k)}$ coincides with the classes of elements of $\, \cal T$ 
modulo the ideal $\, I_{(k)}$ generated by the non-trivial null vectors 
$\, \zeta_1$ and $\, \zeta_2 $ defined in
}  sect.4.3.2.  

\vskip 0.5truecm 

Finally we note that, as a linear space, $\, {\cal T}_{(k)}$ has dimension which is equal to the
number of representative points on the square lattice which belong to the fundamental domain
$\Delta_k$. Consequently, for fixed $k\geq 3$, the reduced tensor algebra $\, {\cal T}_{(k)}$ is
of order $(k-1)(k-2)/2$.  When $k=3$, the order of $ {\cal T}_{(3)}$ is equal 
to unity;  in this case, the fundamental domain $\Delta_3$ contains a single 
point with coordinates (0,0). This means that, for $k=3$, there is only one class of physically equivalent states; this
class can be represented by $\chi [0,0]$. 

We note that according to the definition given in Sect.4.1, the reduced tensor
algebra  ${\cal T}_{(k)}$, with $k \geq 3$ is regular. Indeed, the only non trivial property to be verified is (\ref{eq:fro}), the proof can be found in 
App.C.2.
\subsection{\bf Reduced tensor algebra for $\mathbf{k=1}$}  

In this section we compute the reduced tensor algebra for $k=1 \, $.   
Since $E_0[m,n]$ does not vanish when $k=1\, $ (see eq.(\ref{eq102})), in the 
construction of the reduced
tensor algebra ${\cal T}_{(1)}$ one finds a situation which is quite different from the case $k
\geq 3$.  Clearly, when $k=1$ we cannot use the same argument that we presented in the previous
section; nevertheless,  we will show that the expectation value of a generic link $L$ 
has a beautiful periodicity in the colour state space $\cal T$.  

In order to find ${\cal T}_{(1)}$, we shall produce the explicit expression of 
$\langle \, W(L) \, \rangle \bigr |_{S^3}$ for a generic link $L$ when $k=1$.  As we shall show,
we only need to consider the double crossings between two lines of a link diagram. 
These crossings can be analyzed by using the graphical rules introduced in in 
chapter 2.
Let us consider a part of a link diagram (tangle) which describes a two-string configuration. 
The no-crossing tangle, representing two parallel lines, can be decomposed as shown in Fig.4.6. 

\begin{figure}[h]
\begin{picture}(10,10)
\put(210,-105){$\rho  \in \rho_1 \otimes \rho_2$}
\put(313,-90){$\rho$}
\put(150,-135){$\rho_1$}
\put(166,-135){$\rho_2$}
\put(298,-135){$\rho_1$}
\put(313,-135){$\rho_2$}
\end{picture}
\vskip 0.5 truecm 
\centerline{\epsfig{file=\path 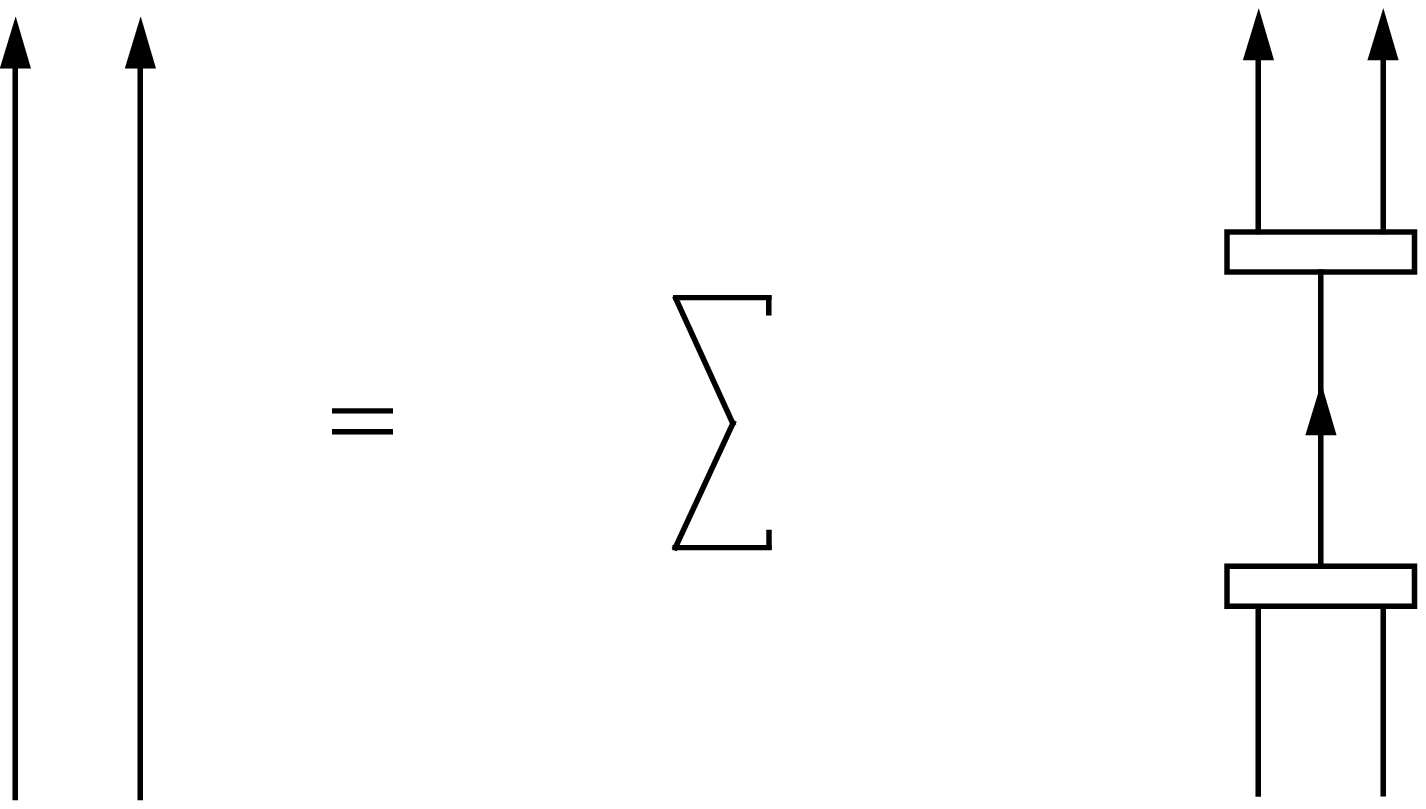,height=4cm,width=6cm}}
\vskip 0.5truecm 
\centerline{{\bf Figure 4.6}}
\end{figure}
\begin{figure}[h]
\begin{picture}(10,10)
\put(118,-160){$\rho_1$}
\put(147,-160){$\rho_2$}
\put(190,-100){$ \alpha(\rho_1)^{-1}  \; \alpha(\rho_2)^{-1}$}
\put(353,-100){$\rho$}
\put(270,-130){$\rho  \in \rho_1 \otimes \rho_2$}
\put(373,-163){$\rho_2$}
\put(353,-163){$\rho_1$}
\end{picture}
\vskip0.5truecm
\centerline{\epsfig{file=\path 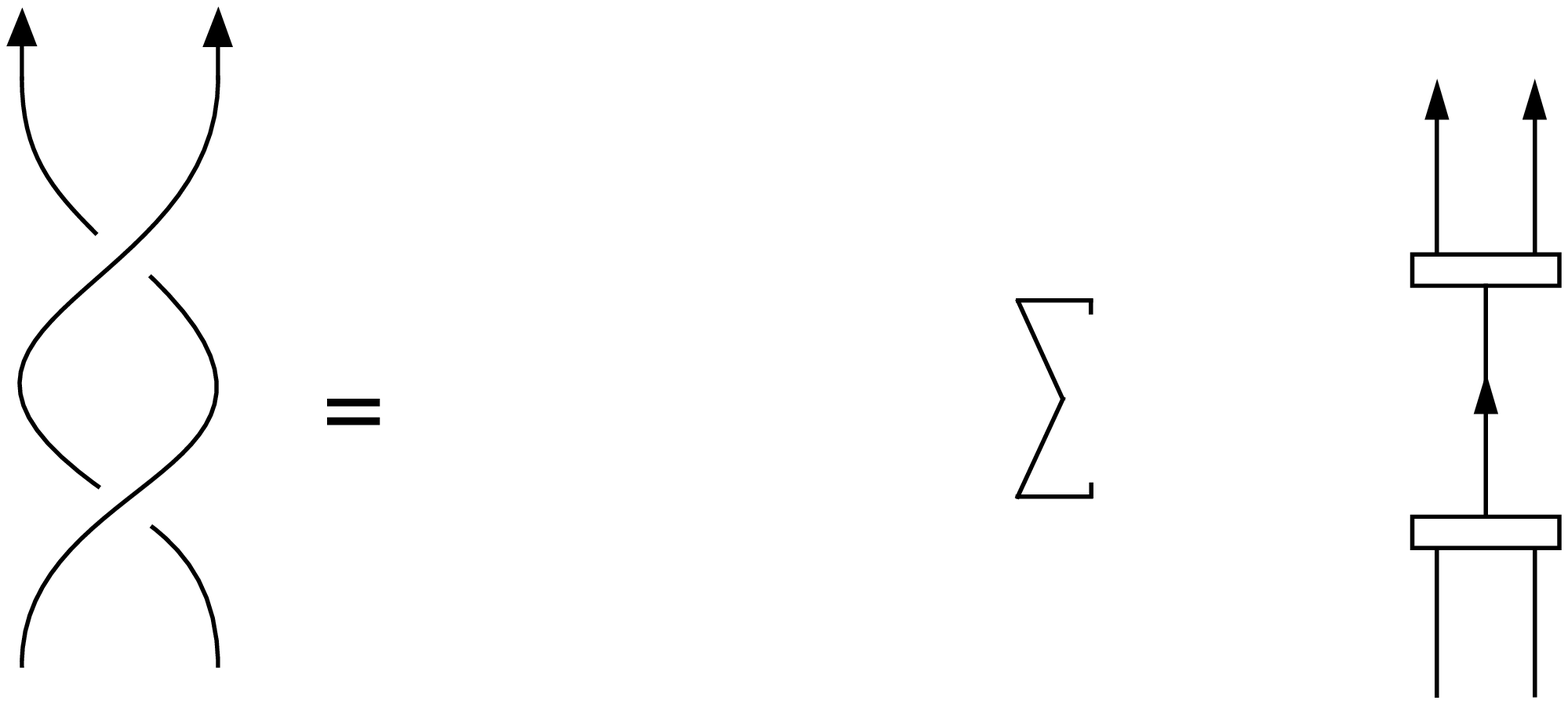,height=5cm,width=10cm}}
\vskip 0.5truecm 
\centerline{{\bf Figure 4.7}}
\vskip 0.5truecm 
\end{figure}

\noindent The ``projectors" appearing on the right-hand side of the relation 
shown in Fig.4.6
represent skein modules. One can always imagine that the part of the link, 
which is described by the tangle, is contained inside a solid torus.  Each projector simply selects one colour state of
the standard basis of $\cal T$  which flows along the core of this solid torus. 
The tangle corresponding to a double crossing can be decomposed \cite{guad1,glib} as shown in Fig.4.7, where the
twist variable $\alpha (a,b)$ is given by 
\beeq
\alpha(\, \rho \, ) \, = \, \alpha (\, a , b \, )\, = \, q^{Q( a , b)} \quad . 
\end{equation}

\begin{figure}[h]
\begin{picture}(10,10)
\put(240,-83){$Y_{ \rho_1  \rho_2}$}
\end{picture}
\vskip 0.5 truecm 
\centerline{\epsfig{file=\path 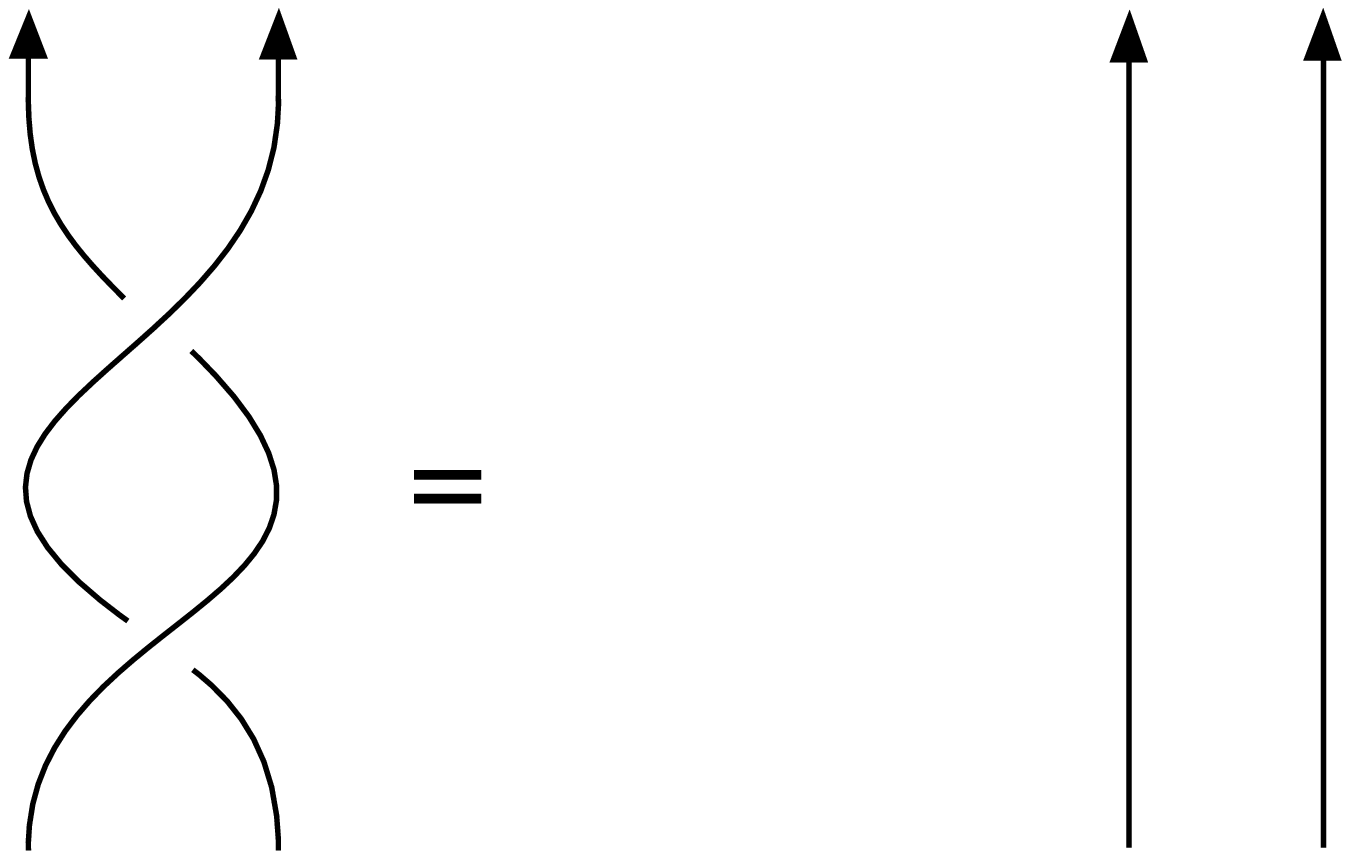,height=4cm,width=4cm}}
\vskip 0.5truecm 
\centerline{{\bf Figure 4.8}}
\end{figure}

When $k=1$, one has 
\beeq
\alpha (\, a ,  b \,)\, = \, e^{- {2 \pi i \over 3}(a^2+b^2+3a+3b+ab)} \quad .
\label{eq132}
\end{equation}
The algebraic structure of expression (\ref{eq132}) shows that $\alpha (\, a ,  b \,)$ depends on the 
values of $a$ and $b$ modulo 3. Let us introduce the triality $t_\rho$ of an irreducible
representation $\rho \simeq (a,b)$, defined as
\beeq
t_{\rho} \,= \, a-b \;\; \pmod{ 3} \quad .
\label{eq133}
\end{equation}
In our convention, the possible values taken by $t_\rho$ are $\{ \, -1, \, 0, \, 1\, \}$. 
The value of $\alpha$ variable  is
\bea
&&\alpha (\rho)=e^{- {2 \pi i \over 3}} \hskip 1truecm {\rm if} \quad t_{\rho} \neq 0 \; , \nb \\
&&\alpha (\rho)=1  \;\;\; \;\; \; \; \hskip 1truecm {\rm if} \quad t_{\rho}= 0 \; . 
\label{eq134}
\ena
Triality is conserved in the sense that all the irreducible representations $\{ \, \rho_i \, \}$
of $SU(3)$, which enter the decomposition of the tensor product $\rho_1 \otimes \rho_2$, have the
same triality:  
\bea
&& \rho_1 \otimes \rho_2 \, = \,  \bigoplus_i \rho_i \nb \\
&& t_{\rho_i} \; = \; t_{\rho_1} + \, t_{\rho_2} \;\; \; \pmod{ 3} \quad \quad \forall i 
\label{eq135}
\ena
Eqs.(\ref{eq134}) and (\ref{eq135}) imply  that, for $k=1$,  the decomposition illustrated in Fig.13.2 takes the
simple form shown in Fig.4.8. The complex coefficient $Y_{\rho_1 \, \rho_2}$ is given by
\beeq
Y_{\rho_1 \, \rho_2} \, = \, e^{\frac{2 \pi i}{  3} \; t_{\rho_1}\, t_{\rho_2} } \quad .
\end{equation}
Double under-crossing can be obtained simply by replacing
$Y_{\rho_1 \, \rho_2}$ with  its complex conjugate $Y^{\ast}_{\rho_1 \, \rho_2}$.

Let us now compute the  expectation value of a Wilson line operator associated
with a generic link $L$
with $n$ coloured components; the colour of the component $C_i$ is given by the irreducible
representation $\rho_i$. The relation shown in Fig.(4.8) permits us to
transform $L$ into a collection of $n$ disjoint knots. Consequently, one finds
\bea
&&\langle
W( \, C_1, \cdots ,C_n ; \, \rho_1,  \cdots , \rho_n \,) \rangle \Bigr |_{S^3} \; = \nb \\
&& \qquad =\,  \, F(\, t_{\rho_1}, \cdots , t_{\rho_n} \,) \;  \langle W(\, C_1; \, \rho_1 \, )
 \rangle \Bigr |_{S^3} \cdots  \langle W(\, C_n; \, \rho_n \, ) \rangle \Bigr |_{S^3} \quad , 
\ena
where 
\beeq
 F(\, t_{\rho_1}, \cdots , t_{\rho_n} \,) \, = \, \exp \left[ \, \frac{2 \pi i}{  3} \, \sum_ {i < j}
{\rm lk}(\, C_i, \, C_j \,) \; t_{\rho_i} \, t_{\rho_j} \right ]\quad . 
\end{equation}
At this stage,  by using the relation of Fig.4.7 and eq.(\ref{eq134}), we can 
transform  each knot
$C_i$  into the unknot. Therefore we have
\beeq
\langle W( \, C_1, \cdots ,C_n ; \, \rho_1,  \cdots , \rho_n \,) \rangle \Bigr |_{S^3}
\, = \, G(\, t_{\rho_1}, \cdots , t_{\rho_n} \,) \;  \left ( \prod_{i=1}^{n} \; E_0[ \, \rho_i \,
] \right )   \quad ,
\label{eq139}
\end{equation}
with 
\beeq
G \, = \, \exp \left \{ {- 2 \pi i \over 3} \left[ \sum_ {i <
j} 2 \, {\rm lk}(\, C_i, \, C_j \,) \, t_{\rho_i} \, t_{\rho_j}   \, + \, \sum_i {\rm lk} (\, C_i, \,C_{i_f} \,)
\, t_{\rho_i}^2 \right ] \right \} \quad .
\label{eq1310}
\end{equation}
At this point, from eq.(\ref{eq139}) and eq.(\ref{eq1310}) it is clear that 
the equivalence classes of physically
equivalent colour states are characterized only by triality. Therefore, the reduced tensor algebra 
${\cal T}_{(1)}$ is of order equal to three \cite{io,gp1}. The elements of the
 standard basis of 
${\cal T}_{(1)}$ are denoted by $\{ \, \Psi [0],\; \Psi [1], \; \Psi [-1] \, \}$; the 
structure constants are determined by triality conservation
\bea
&& \,\Psi [1], \;\;\; \Psi [0] \; \Psi [-1] \, = \, \Psi [-1] 
\nb \\
&&\Psi [1] \; \Psi [1] \, = \,  \Psi [-1], \;\;\; \Psi [1] \; \Psi [-1] \,= \, 
\Psi [0]
\nb \\
&&\Psi [-1] \; \Psi [-1] \,= \, \Psi [1]  \quad .
\ena
Since, for $k=1$, one has $E_0[a,b]  = D(a,b)$,
each element of $\cal T$ corresponds to an element of ${\cal T}_{(1)}$ according to 
\beeq
 \chi[a,b] \Ra \left \{\begin{array}{cc}
D(a,b) \; \Psi [0]  & \mbox{if } a-b=0 \quad (mod \, 3) \\
D(a,b) \; \Psi [1] & \mbox{if } a-b=1 \quad (mod \, 3) \\
D(a,b) \; \Psi [-1] & \mbox{if } a-b=-1 \quad (mod \, 3)  \quad .
\end{array} \right.
\end{equation}
As a check, let us verify that the Hopf matrix, defined in terms of basis elements of
${\cal T}_{(1)}$, is non singular. We consider the Hopf link, shown in Fig.8.1, in which each
component is characterized by an element of the standard basis  $\{ \, \Psi [0],\; \Psi [1],
\;  \Psi [-1]
\, \}$    of ${\cal T}_{(1)}$.  The expectation value of the associated Wilson line operator is 
\beeq
H[\, \Psi [i] , \Psi [\, j]\, ] \; = \; H_{ij} \quad , 
\end{equation}
where, according to eqs.(\ref{eq139}) and (\ref{eq1310}, the $3 \times 3$ 
matrix $H$ is given by 
\beeq
H \; = \;   \left ( \begin{array}{ccc}
1 & 1 &  1 \\
1 & e^{\frac{2 \pi i}{  3}} &  e^{\frac{i \pi}{   3}} \\ 
1 &  e^{\frac{-2 \pi i}{  3}} &  e^{\frac{2 \pi i}{ \ 3}}  
\end{array} \right ) 
\quad . 
\label{runo}
\end{equation}
Since the Hopf matrix $H$ is invertible, the set ${\cal T}_{(1)}$ is physically irreducible; this
concludes the construction of the reduced tensor algebra for $k=1$. 

\subsection{\bf Reduced tensor algebra for $\mathbf{k = 2}$}  

When $k=2$, the value of the unknot is
\beeq
E_{0} [m,n] \bigr |_{k=2} \; = \; \left \{\begin{array}{cc}
0 & \mbox{for } m \mbox{ and } n \mbox{ odd} ; \\ 
-  (n+1)/2 & \mbox{for } n \mbox{ odd and } n \mbox{ even} ; \\
-  (m+1)/2 & \mbox{for } m \mbox{ odd and } n \mbox{ even} ; \\ 
-  (m+n+2)/2 & \mbox{for } m \mbox{ and } n \mbox{ even} . \\ 
\end{array} \right.
\label{eq141}
\end{equation}
Property 2 and eq.(\ref{eq141}) imply that each irreducible representation 
$\chi[m,n]$, with $m$ and $n$
odd, is physically equivalent to the null element of ${\cal T}_{(2)}$. The set of representations 
$\{ \, \chi[m,n] \, \} $ with $m$ and $n$ odd is called ${\cal A}_2$. 
The value of the twist variable for $k=2$ is
\beeq
\alpha (a,b)\, = \,(-1)^{a+b+ab}\, e^{- \frac{i \pi}{ 3}(a-b)^2} \quad .
\label{eq142}
\end{equation}
Apart from the irreducible representations physically equivalent to the null element,
we find that the value of $\alpha (a,b)$ depends only on triality. More precisely, if the
irreducible representation $\rho$ does not belong to ${\cal A}_2$,  we have 
\bea
&&\alpha (\rho)=e^{ {2 \pi i \over 3}} \hskip 1truecm {\rm if} \quad t_{\rho} \neq 0 \; , \nb \\
&&\alpha (\rho)=1  \;\;\; \;\;  \hskip 1truecm {\rm if} \quad t_{\rho}= 0 \; . 
\label{eq143}
\ena
In analyzing the double crossings, we can ignore the representations contained in ${\cal A}_2$;
indeed, each projector on a representation of this kind gives a vanishing
contribution.  Therefore, eq.(\ref{eq143}) implies that the relation shown  in Fig.4.8 is valid also for
$k=2$; we only need to compute the new values of the coefficients $Y_{\rho_1 \, \rho_2}$. From
eq.(\ref{eq142}) and the relation shown in Fig.13.7, for $k=2$  one finds 
\beeq
Y_{\rho_1 \, \rho_2} \, = \, e^{\frac{-2 \pi i}{ 3} \; t_{\rho_1}\, t_{\rho_2} } \quad .
\end{equation}
The same argument that we used in the previous section now gives 
\beeq
\langle W( \, C_1, \cdots ,C_n ; \, \rho_1,  \cdots , \rho_n \,) \rangle \Bigr |_{S^3}
\, = \, G^\prime (\, t_{\rho_1}, \cdots , t_{\rho_n} \,) \;  \left ( \prod_{i=1}^{n} \; E_0[ \,
\rho_i \, ] \right )   \quad ,
\end{equation}
with 
\beeq
G^\prime \, = \, \exp \left \{ \frac{ 2 \pi i}{  3} \left[ \sum_ {i <
j} 2 \, {\rm lk}(\, C_i, \, C_j \,) \, t_{\rho_i} \, t_{\rho_j}   \, + \, \sum_i {\rm lk}
(\, C_i, \,C_{i_f} \,)
\, t_{\rho_i}^2 \right ] \right \} \quad .
\end{equation}
The equivalence classes of physically equivalent colour states are again characterized only by
triality. Therefore, the reduced tensor algebra  ${\cal T}_{(2)}$ is of order equal to three \cite{io,gp1}. The
elements of the standard basis of  ${\cal T}_{(2)}$ are denoted by $\{ \, \Psi [0],\; \Psi
[1], \; \Psi [-1] \, \}$ with  structure constants 
\bea
&&\Psi [0] \; \Psi [0]\, =\, \Psi [0], \;\;\; \Psi [0] \; \Psi [1] \,= \,
\Psi
[1], \;\;\; \Psi [0] \; \Psi [-1] \, = \, \Psi [-1] \nb \\
&& \Psi [1] \; \Psi [1] \, = \,  \Psi [-1], \;\;\; \Psi [1] \; \Psi [-1] \,= \, 
\Psi [0] \nb \\ 
&&\Psi [-1] \; \Psi [-1] \,= \, \Psi [1]  \quad .
\ena
Each element of $\cal T$ corresponds to an element of ${\cal T}_{(2)}$ according to 
\beeq
\chi [a,b] \Ra  \left \{ \begin{array}{cc}
E_0[a,b] \; \Psi [0] & \mbox{if } a-b=0 \quad (mod \, 3) \\
E_0[a,b] \; \Psi [1] & \mbox{if } a-b=1 \quad (mod \, 3) \\
E_0[a,b] \; \Psi [-1] & \mbox{if } a-b=-1 \quad (mod \,3)  \quad , 
\end{array} \right.
\end{equation}
where $E_0[a,b]$ is given in eq.(\ref{eq141}). 

When $k=2$ the Hopf matrix $H$, which is defined in terms of basis elements 
$\{ \, \Psi [0],\; 
\Psi [1], \; \Psi [-1] \, \}$  of ${\cal T}_{(2)}$, is given by 
\beeq
H \; = \; \left ( \begin{array}{ccc}  1 & 1 &  1 \\ 
1 & e^{\frac{-2 \pi i}{  3}} &  e^{\frac{2 \pi i}{  3}} \\ 1 &  e^{\frac{2 \pi i}{  3}} &  e^{\frac{-2 \pi i}{  3}} \end{array} 
\right ) 
\quad . 
\end{equation}
Clearly, the Hopf matrix $H$  is invertible and then  ${\cal T}_{(2)}$ is physically irreducible. 

We note that the reduced tensor algebras ${\cal T}_{(1)}$ and ${\cal T}_{(2)}$ are isomorphic and
coincide with the group algebra of the center $Z_3$ of $SU(3)$. 

\section{\bf Reduced tensor algebra for SU(2)}
\subsection{\bf Reduced tensor algebra for $\mathbf{k \geq 2}$}
The strategy to construct the reduced tensor algebra when $G=SU(2)$ is similar to the 
case $G=SU(3)$, however this case is simpler. In this section we shall 
consider the case $k \geq 2$.
From eq.(\ref{unsu}), it follows that for $k \geq 2$
\beeq
E_0[J] \; = \; 0 \qquad \mbox{ with } J \; = \; \frac{nk \; - \; 1}{2} \quad
n \in \mathbb{Z}^+, \; n \geq 1 \quad .
\label{zeros}
\end{equation}
Thus, $\chi[J] \in I_{(k)}$ when $J$ is of the form $J =  (nk-1)/2$, with
$n \in \mathbb{Z}^+, \; n \geq 1$. The lowest value $J_0$ of $J$ for which 
(\ref{zeros}) is satisfied is
\beeq
J_0 \; = \; \frac{k \; - \; 1}{2} \quad .
\end{equation}
By using the well known property of the tensor product decomposition into 
irreducible 
representations of $SU(2)$, it is quite easy to show that $I_{(k)}$ is
generated by $\chi[J_0]$ for $k \geq 2$. 
Indeed, the following property holds.

\bigskip

\shabox{
{\bf Property 4.4}} {\em Let $\chi[J]$ be such that $\chi[J-d] \sim 0$ with
$d=n/2 \qquad n \in \mathbb{N}$, i.e.
$\chi[J-d]$ is physically equivalent to the null element; then we have}
\beeq
\chi[J] \; \sim \; - \; \chi[J-2d] \quad .
\label{rule}
\end{equation}

\bigskip

{\bf Proof}

\no
We prove (\ref{rule}) by induction. Let us consider the following 
decomposition of representations
\beeq
\chi[J\; - \; 1] \; \chi[1] \; = \; \chi[J] \; + \; \chi[J \; - \; 1] \; + \; 
\chi[J \; - \; 2] \quad .
\label{s1}
\end{equation}
When $d=1$, by the hypothesis it follows that $\chi[J-1] \sim 0$, thus 
eq.(\ref{s1}) implies 
\beeq
\chi[J] \; \sim \; - \chi[J \; - \; 1] \qquad . \label{s2}
\end{equation}
Similarly, when $d=1/2$, by using the decomposition
\beeq
\chi[J\; - \; 1/2] \; \chi[1/2] \; = \; \chi[J] \; + \; 
\chi[J \; - \; 1] \quad ,
\label{s21}
\end{equation}
and the hypothesis $\chi[J -1/2] \sim 0$, it follows that
\beeq
\chi[J] \; \sim - \chi[J \; - \; 1] \qquad . \label{s3}
\end{equation}
Eqs.(\ref{s21}) and (\ref{s3}) show that property 4.4 holds when $d=1, \, 
1/2$. Now suppose that property 4.4 is true for a fixed $d$, we shall 
show that is also true for $d+1$ and $d+1/2$. Let us consider the following 
decompositions
\bea
&& \chi[J] \; \chi[1] \; = \; \chi[J \; + \; 1] \; = \; \chi[J] \; + \; \chi[J \; - \; 1] 
\label{s4} \quad ,\\
&&\chi[J \; - 2d] \; \chi[1] \; = \; \chi[J \; - \; 2d \; + \; 1] \; + \; \chi[J \; - \; 2d] \; + \; \chi[J \; - \; 2d \; - \; 1] \quad . \label{s5}
\ena
By the induction hypothesis
\beeq
\chi[J] \; \sim \; - \chi[J \; - \; 2d]\, , \qquad \chi[J \; - \; 1] \; \sim
\; - \chi[J \; - \; 2d \; + \; 1] \quad . \label{s6}
\end{equation}
Therefore, by adding eqs.(\ref{s4}) and (\ref{s5}) and by using eq.(\ref{s6}),
one gets
\beeq
\chi[J \; + \; 1] \; \sim \; - \chi[J \; - \; 2d \; - \; 1] \qquad .
\label{s7}
\end{equation}
Eq.(\ref{s7}) shows that property 4.4 holds also for $d+1$. Finally, let us consider 
the decompositions
\bea
&& \chi[J] \; \chi[1/2] \; = \; \chi[J \; + \; 1/2] \; + \; 
\chi[J \; - \; 1/2] \quad , \label{s8} \\
&&\chi[J \; - 2d] \; \chi[1/2] \; = \; \chi[J \; - \; 2d \; + \; 1/2] \; + \; 
\chi[J \; - \; 2d \; - \; 1/2] \quad . \label{s9}
\ena
By the hypothesis on has
\beeq
\chi[J] \; \sim \; - \chi[J \; - \; 2d] \, , \qquad \chi[J \; - \; 1/2]
 \; \sim \; \chi[J \; - \; 2d \; + \; 1/2] \quad .
\label{s10}
\end{equation}
By adding eqs.(\ref{s8}) and (\ref{s9}) and by using eq.(\ref{s10}),
it follows
\beeq
\chi[J \; + \; 1/2] \; \sim \; - \chi[J \; - \; 2d \; - \; 
1/2] \qquad .
\label{s13}
\end{equation}  
Eq.(\ref{s13}) shows that property 4.4 holds also for $d+1/2$. {\hfill \ding{111}}

\vskip 0.5truecm
\no
Property 4.4 determines the reduced tensor algebra completely. Let us 
introduce the ``fundamental'' domain $\Delta_k$ defined as
\beeq
\Delta_k \; = \; \left \{ J \mbox{ such that } 0 \leq J < J_0 \; = \;
\frac{k-1}{2} \right \} \qquad .
\end{equation}
Given $\chi[J] \in \T$, by using property 4.4 recursively, one can show that
either $\chi[J]$ is equivalent to an element of the fundamental domain or is 
physically  equivalent to the null element. Indeed, for any $J$, one can find
$n \in \mathbb{N}$ such that 
\beeq
J \; = \; n J_0 \; + \; \tilde{J} \quad , \label{s11}
\end{equation}
with $\tilde{J} \in \Delta_k$. In particular from eq.(\ref{s11}) and property
4.4 it follows that, when $\chi[J] \not\sim 0$, on has
\beeq
\chi[J] \; \sim \; (-1)^n \; \chi[\tilde{J}] \qquad .
\label{s12}
\end{equation}
Because any element $\chi \in \T$ can be decomposed as
\beeq
\chi \; = \; \sum_J \xi(J) \, \chi[J] \qquad ,
\end{equation}  
from eq.(\ref{s12}), it follows that the standard basis of $\T$ for
$k \geq 2$ is the following
\beeq
\left \{ \Psi[J] \qquad \mbox{with } \qquad \frac{k}{2} \; - \; 1 \geq J \geq 0 
\right \} \quad .
\label{sbas2}
\end{equation}    
In particular, any element $\Psi[J]$ of the standard basis admits as 
a representative an irreducible representation $\chi[J]$; the equivalence
relation is the following
\beeq
\Psi[J] \Leftrightarrow \chi[J] \, , \qquad  \mbox{with } \qquad \frac{k}{2}
 \; - \; 1 \geq J \geq 0 \quad .
\end{equation}
It follows from (\ref{sbas2}) that, for $k \geq 2$, $\T$ is of order $(k-1)$.
This results was originally obtained in \cite{guad1}.

\subsection{\bf Reduced tensor algebra for $\mathbf{k=1}$}

In this section we shall consider the special case $k=1$. From the general
expression (\ref{unsu}), it follows that $E_0[J]$ never vanishes when $k=1$.
Indeed
\beeq
\lim_{k \ra 1} \; E_0[J] \; = \; (-1)^{2J} \; (2J+1) \quad .
\end{equation}
When $k=1$ the deformation parameter is very simple $q(1)=q(1)^{-1}=1$.
Starting from this fact one can derive \cite{glib} a general expression for the expectation
value of a link $L$ with components $\{C_1, \cdots, C_m \}$ and colours 
$\{\chi[J_1], \cdots ,\chi[J_m] \}$; one has
\beeq
E(L) \; = \; \left(\prod_{i=1}^m E_0[J_i] \right) \, \exp \left[ \frac{i \pi}
{2} \left(\sum_a {\cal W}(C_a) \; + \; 2 \sum_{a<b} {\rm lk}(C_a,C_b) \right)
\right ] \quad .
\label{liuno}
\end{equation}
Eq.(\ref{liuno}) shows that there are only two physically independent classes ,
thus the reduced tensor algebra ${\cal T}_2$ is of order two. Let us introduce
the two elements $\Psi[0]$ and $\Psi[1/2]$ of the standard basis of ${\cal T}_2$. The structure constants are the following
\bea
&& \Psi[0] \; \Psi[0] \; = \; \Psi[0], \qquad \Psi[0] \; \Psi[1/2] \; = \; \Psi[1/2], \nb \\
&& \Psi[1/2] \; \Psi[1/2] \; = \; \Psi[0] \quad .
\label{redu} 
\ena
As a consequence of (\ref{redu}), ${\cal T}_2$ is isomorphic with the group algebra 
of the center of $SU(2)$, $Z_2$. Given $\chi[J] \in {\cal T}$,  it can 
be associated with an element of the standard basis of ${\cal T}_2$. Explicitly
\bea
&&\chi[J] \Ra (2J \; + \; 1) \Psi[0] \qquad  \; \; \mbox{ for J integer} \quad; \nb \\
&& \chi[J] \Ra (2J \; +\; 1) \Psi[1/2] \qquad  \mbox{for J not integer} \quad .
\ena

\chapter{\bf Dehn's surgery}
\section{\bf Introduction}
In this chapter we shall describe a powerful method which permits to 	
 construct every closed, connected and orientable 3-manifold starting from the 
reference 3-manifold $S^3$. We recall that a manifold is called closed
when 
it is compact and without boundary.  This technique, pioneered by
Dehn, is based on a finite number surgery operations. The single operations
consist in cutting from $S^3$ a solid torus $N=D^2 \times S^1$ and then in 
sewing back $N$ to $S^3 - N$ by identifying their boundaries with an 
appropriate homeomorphism $h$. Dehn's surgery will be the key ingredient in 
solving Chern-Simons field theory in any closed, connected and orientable 
3-manifold; it turns out that the surgery instructions in $S^3$ admit a 
realization as symmetry transformations on the observables of the Chern-Simons
theory in $S^3$.

In order to introduce Dehn's surgery, it is worth to recall some mathematical 
results concerning torus topology.
\section{\bf Knots in a torus}   

The two-dimensional torus $T^2$ is a manifold homeomorphic with $S^1 \times 
S^1$. A useful coordinates system  on $T^2$ is the following  
\beeq \left(e^{i \theta_1}, \; e^{i \theta_2} \right) \; \equiv \;
\left(\theta_1, \; \theta_2 \right) \qquad  0 \leq \theta_i,  \leq 2 \pi \;\;
i \, = \, 1, \, 2 \quad . 
\end{equation} 
The fundamental group of $T^2$ is $\pi_1(T^2)=\mathbb{Z} \op \mathbb{Z}$ \cite{rol}. Any element $g \in \pi_1(T^2)$ can be decomposed as
\beeq
g \; = \; {\gamma_L}^a \; \op \; {\gamma_M}^b \qquad a, \, b \in \mathbb{Z} 
\quad ,
\end{equation}
where $\gamma_{\scriptscriptstyle L}$ and $\gamma_{\scriptscriptstyle M}$ are 
the generators of  $\pi_1(T^2)$, 
respectively the standard longitude and the standard meridian of $T^2$; they 
corresponds to the following oriented curves in $T^2$
\bea 
&& \gamma_{\scriptscriptstyle L} \; = \; \left(e^{i \theta}, \, 1 \right) 
\qquad 0 \leq \theta 
\leq 2 \pi \nb \\
&& \gamma_{\scriptscriptstyle M} \; = \; \left(1, \, e^{i\theta} \right) \qquad  0 \leq \theta \leq
 2 \pi \quad . 
\ena
The standard longitude and the standard meridian are shown in Figure 5.1.

\begin{figure}[h]
\begin{picture}(10,10)
\put(150,-80){$\gamma_{\scriptscriptstyle L}$}
\put(265,-120){$  \gamma_{\scriptscriptstyle M}$}
\end{picture}
\vskip 0.9 truecm 
\centerline{\epsfig{file=\path 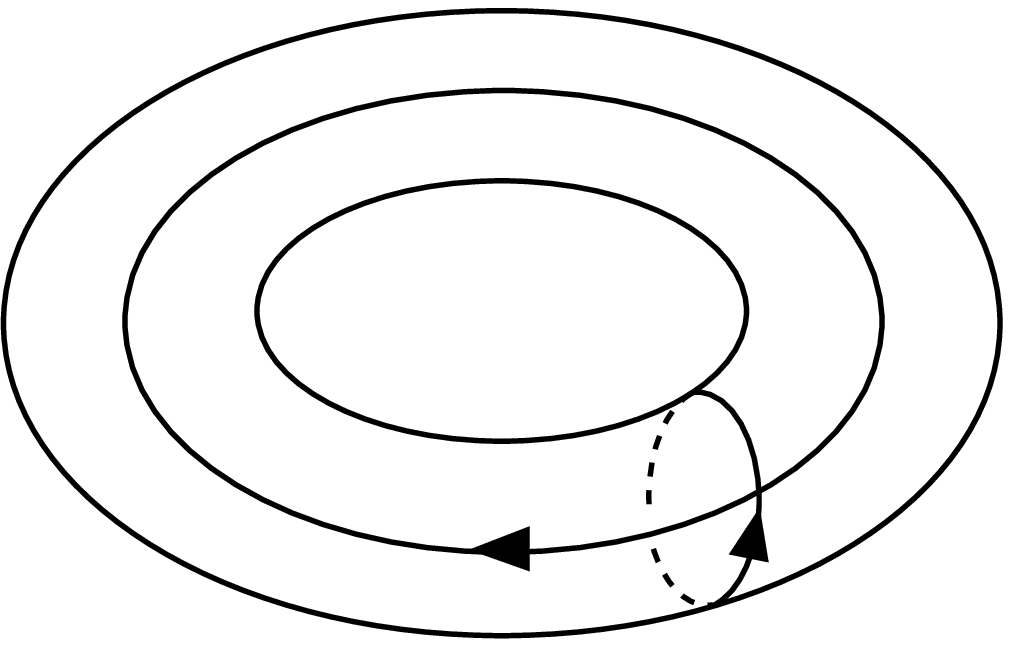,height=3cm,width=5cm}}
\vskip 0.9 truecm 
\centerline{{\bf Figure 5.1}}
\vskip 0.9 truecm 
\end{figure}

Roughly speaking, if one pictures $T^2$ as the surface of a doughnut, 
$\gamma_{\scriptscriptstyle L}$ winds the hole one time and 
$\gamma_{\scriptscriptstyle M}$ winds the interior one time.
Clearly, neither $\gamma_{\scriptscriptstyle L}$ nor 
$\gamma_{\scriptscriptstyle M}$ can be 
continuously shrunk to a point; moreover, $\gamma_{\scriptscriptstyle L}$ cannot be 
continuously deformed to $\gamma_{\scriptscriptstyle M}$. 

In general, given a continuous map  $f: \; S^1 \ra T^2$, its homotopy class $[f]$
is represented by two integers with sign \cite{rol}; that is
\beeq [f] \; = \;  \gamma_{\scriptscriptstyle L}^a
\; \op \;  \gamma_{\scriptscriptstyle M}^b \; \equiv ( a, \, b) \qquad . 
\end{equation} 
Is not difficult to show that if $K$ is a knot in $T^2$ and
\beeq 
[K] \; = \; ( a, \, b ) \qquad ,
\end{equation}
then $a, \, b$ must be coprime. It also can be shown that a knot  $K$ in 
$T^2$ is essential, i.e. homotopically non-trivial, if and only if $\left[K \right]
 \neq (0,0)$.

Let us introduce two important self-homeomorphisms of $T^2$ \cite{rol}. The longitudinal twist is defined as
\beeq 
h_L \left( e^{i \theta_1}, \, e^{i \theta_2} \right) \; = \; \left(e^{i \left(
\theta_1+ \theta_2 \right)}, \, e^{i \theta_2} \right) \qquad . 
\end{equation}
The meridinal twist is defined as
\beeq 
h_M \left(e^{i \theta_1},e^{i \theta_2} \right) \; = \; \left(e^{i
\theta_1}, \, e^{i \left(\theta_1+ \theta_2 \right)} \right) \qquad .
\end{equation} 
Clearly, the corresponding inverse maps will be given by 
\bea 
&&h^{-1}_L \left( e^{i \theta_1}, \, e^{i \theta_2} \right) \; =
\; \left( e^{i \left( \theta_1- \theta_2 \right)}, \, e^{i \theta_2} \right)
 \quad , \nb \\
&& h^{-1}_M \left( e^{i \theta_1}, \, e^{i \theta_2} \right) \; =
\; \left( e^{i \theta_1}, e^{i \left(\theta_2- \theta_1 \right)} \right) 
\qquad . 
\ena
The effects of the longitudinal and meridinal twists on the homotopy class of a 
map  $f: \, S^1 \ra T^2$ can be easily derived
\bea 
&&[f]_{h_L}\; = \; {h_L}^\ast \left( \gamma^a_L \op \gamma^b_M \right) \; =
\; \gamma_L^{(a+b)} \; \op \; \gamma_M^b \quad , \nb \\
&&[f]_{h_M} \; = \; \gamma_L^a \; \op \; \gamma_M^{(a+b)} \quad ,\nb \\
&&[f]_{h_L^{-1}} \; = \;  \gamma_L^{(a-b)} \; \op \; \gamma_M^b \quad ,\nb \\
&&[f]_{h_M^{-1}} \; = \; \gamma_L^a \; \op \; \gamma_M^{(b-a)} \qquad . 
\ena
The action of $h_L, \, h_M$ on $[f]$ is represented by $2 \times 2$ 
matrices
\bea 
&& ( a, \, b ) \stackrel{h_L}{\ra} ( a, \, b )_{h_L} \; = \; ( a, \, b)  \left(
\begin{array}{cc} 1 & 0 \\ 1 & 1 \end{array} \right) \; \; , \qquad  
( a, \, b)_{h_L^{-1}} \; = \; ( a, \, b)  \left( \begin{array}{cc} 1 & 0 \\ 
-1 &1 \end{array} \right) \quad , \nb \\ 
&&( a, \, b) \stackrel{h_M}{\ra}( a, \, b)_{h_M} \; = \; ( a, \, b)
\left( \begin{array}{cc} 1 & 1 \\ 0 & 1 \end{array} \right) \;\; , \qquad
( a, \, b)_{h_M^{-1}} \; = \; (a, \, b)
 \left(\begin{array}{cc} 1 & -1 \\ 0 & 1 \end{array} \right) \quad .
\ena
The standard longitude can be mapped into the standard meridian by
a sequence of twists
\bea 
&& ( 0, \, 1) \; = \; {{h_L}^\ast}^{-1} \, {h_{M}}^\ast \, 
( 1, \, 0) \quad , \nb \\ 
&& {h_L} \ra h_M \; = \; {h_{L}}^{-1} \, h_{M} \qquad  .
 \ena
The ambient isotopic class of a knot in $T^2$ is determined by its homotopy
class as stated by the following theorem: \cite{rol}

\bigskip 

\shabox{\no
{\bf Theorem 5.1}}
{\em Given two knots $J,$ $K \in T^2$, they are ambient isotopic if and only 
if $ [J ] \, = \,  \pm [K]$.}

\section{\bf The mapping class group of the torus}

Given a manifold $M$, the group of self-homeomorphisms, modulo an ambient isotopy, 
is called the mapping class group of $M$. Let us consider the case $M=T^2$.
Any homeomorphism $h: T^2 \ra T^2$ acts naturally on the elements of $\pi_1
\left(T^2 \right)$ represented by $( a, \, b)$ as 
\beeq
( a, \, b) \stackrel{h^\ast}{\ra} ( a, \, b)_h \; = \; h^\ast  \qquad .
\end{equation}
As we have seen, $h^{\ast}$ can be represented by $2 \times 2$ matrix
whose elements belong to $\mathbb{Z}$, i.e $h_{\ast} \in GL(2, \mathbb{Z})$.
On the other hand, because also ${h^{-1}}^\ast = {h^\ast}^{-1}$ must be in
$ GL(2, \mathbb{Z})$, it follows that $\det (h^\ast) = \pm 1$.  
Any element of $ GL(2, \mathbb{Z})$ can be written as a product of the following
matrices
\beeq h_M ^\ast \; = \;  \left( \begin{array}{cc} 1 & 1 \\ 0 & 1 \end{array}
\right) \; , \quad , h_L^ \ast \; = \;  \left( \begin{array}{cc} 1 & 0 \\ 1 & 0
\end{array} \right) \; ,  \quad  h_C ^\ast \; = \; \left( \begin{array}{cc} 
0 & 1 \\ 1 & 0 \end{array} \right) \quad . \label{eq:gen}
\end{equation}
The map $\ast$ gives a homomorphism between the group of
self-homeomorphisms Aut$(T^2)$ of $T^2$ and  $ GL(2, \mathbb{Z})$
\beeq \mbox{Aut}(T^2) \; \stackrel{\ast}{\ra} \; GL \left(2, \mathbb{Z}
\right)_{\ast}  \quad .
\end{equation}
The map $\ast$ is surjective and its kernel is non-trivial and consists
of the elements of $Aut(T^2)$ homotopically equivalent to the identity map;
these elements corresponds  under $\ast$ to the identity matrix in
 $ GL(2, \mathbb{Z})$ \cite{rol}. 
When all the elements of Aut$(T^2)$ 
homotopically
equivalent to the identity map are identified, the kernel of $\ast$ become
trivial, and the group of Aut$(T^2)$, modulo ambient isotopy, is isomorphic 
with $GL\left(2,{\mathbb{Z}}  \right)_{\ast}$. Conversely, the maps homotopically equivalent to the 
identity corresponds under $\ast$ to the identity matrix in
 $ GL(2, \mathbb{Z})$ \cite{rol}. 
Thus, the group of Aut$(T^2)$, modulo an ambient isotopy, is isomorphic with 
 $GL(2, \mathbb{Z})$.  In other words the mapping class
 group of $T^2$ is isomorphic with  $ GL(2, \mathbb{Z})$.

\section{\bf Solid torus}

Solid tori are the basic objects in Dehn's surgery.
A solid torus $V$ is a 3-manifold homeomorphic with $S^1 \times D^2$; its
boundary is a torus. The fundamental group of $V$ is $\mathbb{Z}$.
In $S^1 \times D^2$, the standard meridian of $T^2$ becomes homotopically 
trivial, only the standard longitude survives as a generator of $\pi_1(V)$. A 
point $p$ in
$S^1 \times D^2$ can be represented by the triple
\bea
&&p \equiv \left(e^{i \theta_1}, \, \rho \, e^{ i \theta_2} \right) \quad ;
\nb \\
&& 0 \leq \theta_1, \, \theta_2 \leq 2 \pi \qquad .
\label{coord}
\ena
where $\rho$ and $\theta_2$ are standard polar coordinates on the complex plane
and $D^2$ is embedded in it.
A meridian $\mu$ is  a simple closed curve which is essential in 
$\partial V$
but homotopically trivial  in $V$. A longitude  $\lambda$ is a simple 
closed curve in $V$
which intersect a meridian transversally.
The actual realization of topology of a solid torus $V$ is described by a
homeomorphism $h_{fr}: \, S^1 \times D^2 \ra V$ called framing of $V$
\cite{rol}. Given a framing $h_{fr}$, a longitude $\lambda$ is  naturally identified:
the simple closed curve in $\partial V=T^2$ 
\beeq 
\lambda \; = \; h_{fr} \left(  l  \right) \quad ,
\label{frami}
\end{equation}
where $l \subset S^1 \times D^2$ is the curve defined as
\beeq
 l \; = \; \left \{\left(e^{i \theta_1}, \,  1 \right);
\quad 0 \leq \theta_1 \leq 2 \pi \right \} \qquad .
\end{equation} 
Clearly, the definition of a longitude make use of the actual realization of
the topology of the solid torus as specified by the homeomorphism $h$. On the
contrary, the notion of meridian is intrinsic to $V$; indeed, given two meridians
of $V$ it can be shown that they are ambient isotopic.    

In general, given a knot $K$ in a manifold $M$, a tubular neighborhood of
$K$ is a homeomorphism  $h_{tub}$ defined by
\bea
&&h_{tub} :  S^1 \times D^2 \ra M \quad , \nb \\ 
&&h_{tub.}\left(S^1, \, 1 \right) \; = \; K \subset M \qquad . 
\label{core}
\ena
In other words, any knot in $M$ can be imagined as contained in a solid torus $V$
as shown in Fig. 5.3, the embedding of $V$ in $M$ depends on $K$. The
image of the $S^1$ ``component" under $h_{tub}$, see Eq. (\ref{core}), represents
the core of the solid torus $V$ embedded in $M$ and coincides with the knot
$K$ as shown in Fig. 5.2.

\begin{figure}[h]
\vskip 0.9 truecm 
\centerline{\epsfig{file=\path 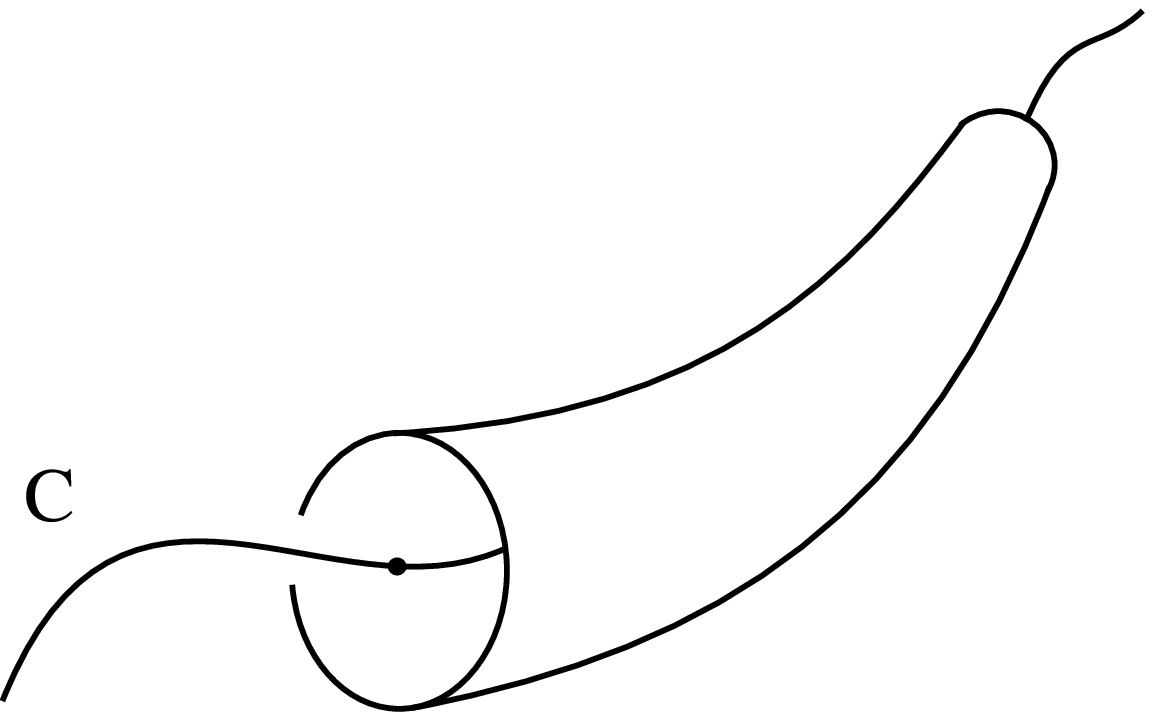,height=4cm,width=5cm}}
\vskip 0.9 truecm 
\centerline{{\bf Figure 5.2}}
\end{figure}
\vskip 0.9truecm

Let us consider a knot $C$ in $S^3$ and its tubular neighborhood $N \subset S^3$.
Given a framing  $h_{fr}$ of $N$, $h_{fr}$ defines a framing
$C_f$ for $C$ by the ``natural" longitude associated with $h_{fr}$, i.e.
\beeq
C_f \; = \; h_{fr} \left(S^1, \, 1 \right) \qquad .
\end{equation}
Conversely, given a framing $C_f$ of $C$, as introduced in Chap.1, a framing $h_{fr}$ of the 
tubular neighborhood $N$ of $C$ is defined. Indeed, once the orientation of
$C$ is chosen, the linking number between $C$ and $C_f$ fixes the framing $h_{fr}$ of $N$ up to 
an ambient isotopy. In particular, when the linking number
between $C$ and $C_f$ or between $C$ and $h_{fr} \left(S^1, \, 1 \right)$
is vanishing the framing is called preferred. By means of $h_{fr}$, one can introduce a
canonical set of generators for $\pi_1(\partial N)$ called the Rolfsen
basis. First of all, an orientation to $C$ is given.
The first generator will correspond to longitude $\lambda$ defined in
(\ref{frami}), where $h_{fr}$ is a preferred framing of $N$; the orientation of
$\lambda$ is induced by $C$. The second generator will correspond to the
meridian
\beeq
\mu = h_{fr} \left(1, \, \partial D^2 \right) \qquad .
\label{meri}
\end{equation}
 The orientation
of the meridian is chosen in a such way that
\beeq
{\rm lk}(\mu, \, \lambda) \; = \; 1 \qquad .
\end{equation} 
The Rolfsen basis for the tubular neighborhood of an unknot is shown in
Fig. 5.3.

\begin{figure}[h]
\begin{picture}(10,10)
\put(150,-80){$\lambda$}
\put(265,-120){$\mu$}
\end{picture}
\vskip 0.9 truecm 
\centerline{\epsfig{file=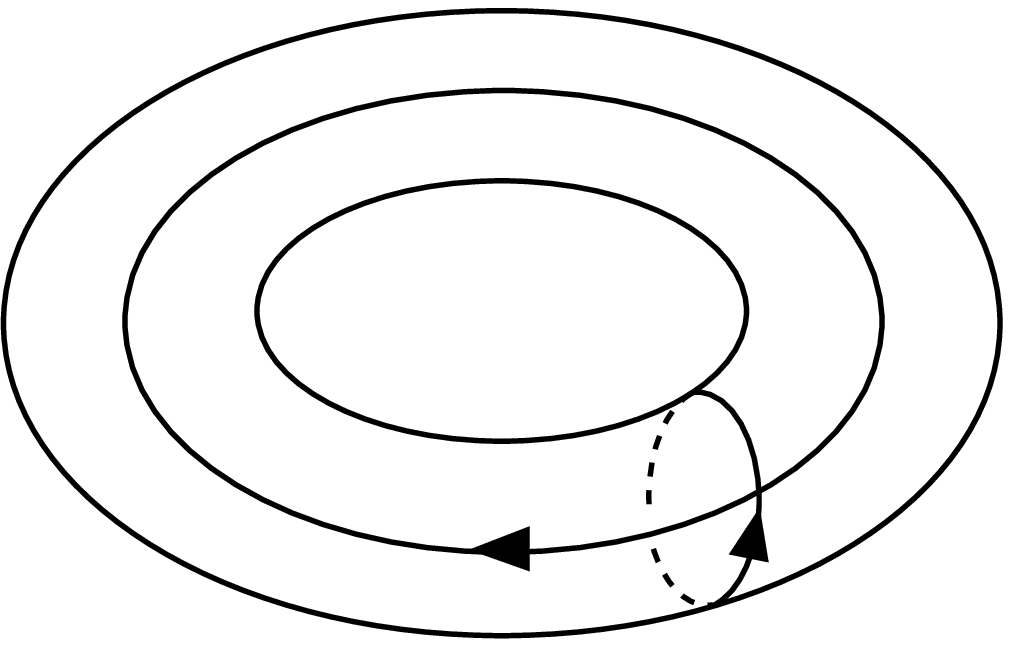,height=3cm,width=5cm}}
\vskip 0.9 truecm 
\centerline{{\bf Figure 5.3}}
\vskip 0.9 truecm 
\end{figure}

Given a solid torus $N$, a homeomorphism  $h: \partial N \ra \partial N$  
can be extended to a self homeomorphism 
$h^{\prime}:N \ra N$, if and only if the image of a meridian of $\partial N$ under $h$
is still a meridian of $\partial N$ \cite{rol}. A homeomorphism $H$ which satisfies this
property is represented  by an element $h^\ast \in SL(2, \, \mathbb{Z})$  of the
mapping class group of $\partial N$, and it is of the form
\beeq
h^\ast \; = \; \left(
\begin{array}{cc} 1 & n \\ 0 &1 \end{array} \right) \qquad n \in \mathbb{Z}
\quad .
\end{equation}
Thus, the self-homeomorphisms of $\partial V$ which can be extended as self-homeomorphisms
of $V$ are precisely  sequences of meridinal twists $h_M$ described in 
Sect. 5.2. 
The extensions $\tilde{\tau}_+$ and $\tilde{\tau}_-$ of respectively $h_M$ and 
of its inverse ${h_M}^{-1}$ on $S^1 \times D^2$ admit the following
presentation in terms of the coordinates (\ref{coord})
\beeq
\tau_\pm \; = \; \tilde{\tau}_\pm \left( e^{i \theta_1}, \, r e^{i \theta_2}
\right) \; = \; \left( e^{i \theta_1}, \, r e^{i (\theta_2 \pm \theta_1)}
\right) \qquad .
\end{equation}
Now, let us  consider the tubular neighborhood $N$ of a knot  $C$ and 
a preferred framing $h_{fr}$ of $N$, i.e. $h_{fr}: S^1 \times D^2 \ra N$. The right-handed twist
$\tau_+$ and the left-handed
twist $\tau_-$ on $N$ are obtained by the following composition of 
homeomorphisms
\beeq
\tau_\pm \; = \; h_{fr} \cdot \tilde{\tau}_\pm \cdot {h_{fr}}^{-1} \qquad .
\end{equation}
In the next section we shall see that $\tau_\pm$ are crucial in the Dehn's surgery;
in particular, a distinguished role is played by the action of $\tau_\pm$ on the solid torus
$S^3-\dot{N}$ obtained from  $S^3$ by removing the internal points of the tubular neighborhood 
of an unknot $U \subset S^3$. It should be noted that with respect to $N$ the role of
$\lambda$ and $\mu$ in $S^3-\dot{N}$ are exchanged. Actually, the curve
(\ref{frami})  is a meridian of $S^3-\dot{N}$ and the curve (\ref{meri}) is
a longitude of $S^3-\dot{N}$.
The effect of a right-handed (left-handed) twist on sample knots without framing in 
$S^3-\dot{N}$ depicted in Fig. 5.4 is shown in Fig. 5.5 and in Fig. 5.6. 
The effect of a right-handed twist on the same knots of those of Fig. 5.4,
now considered with preferred framing, is shown in Fig. 5.7.

\begin{figure}[h]
\vskip 0.5 truecm 
\centerline{\epsfig{file=\path 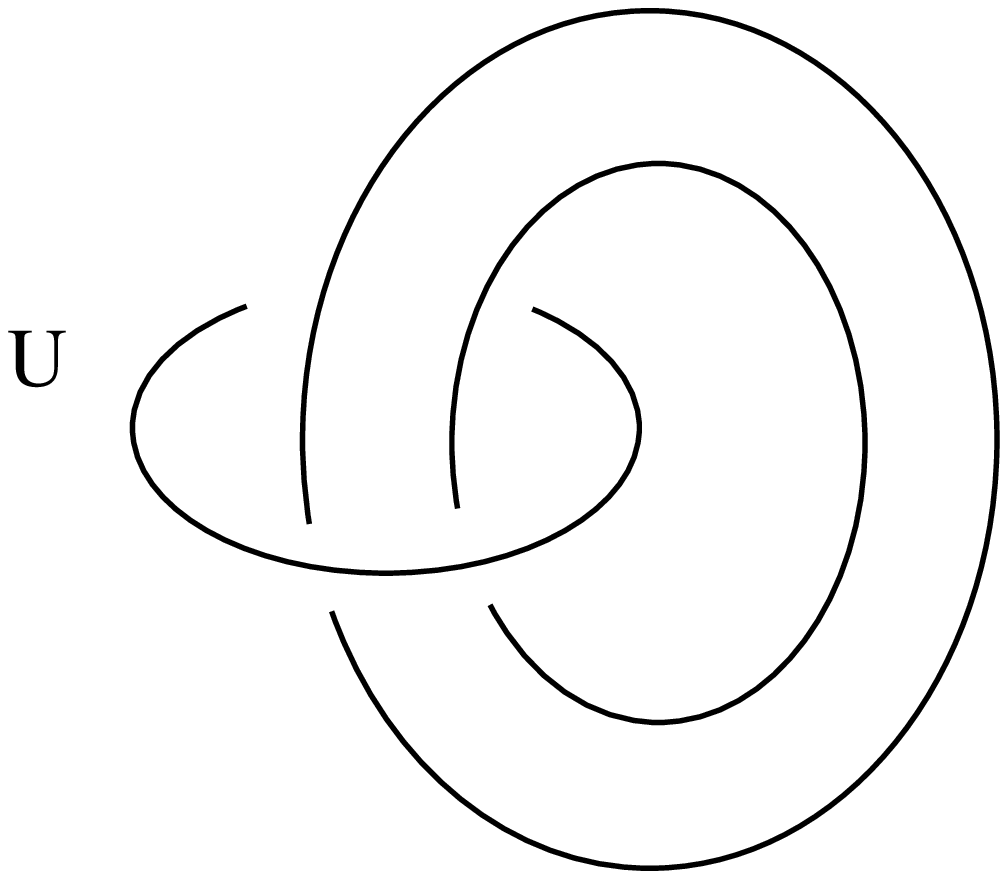,height=4cm,width=5cm}}
\vskip 0.5 truecm 
\centerline{{\bf Figure 5.4}}
\vskip 0.5 truecm 
\end{figure} 

\begin{figure}[h]
\vskip 0.5 truecm 
\centerline{\epsfig{file=\path 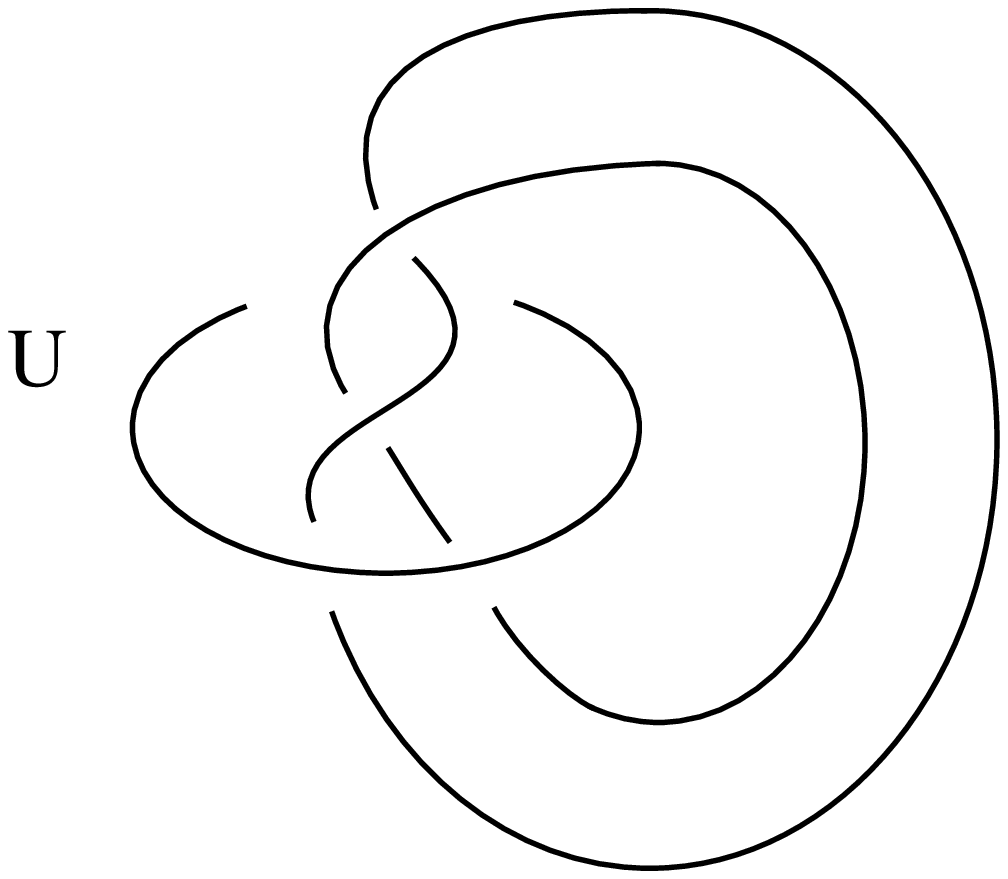,height=4cm,width=5cm}}
\vskip 0.5 truecm 
\centerline{{\bf Figure 5.5}}
\vskip 0.5 truecm 
\end{figure} 

\begin{figure}[h]
\vskip 0.5 truecm 
\centerline{\epsfig{file=\path 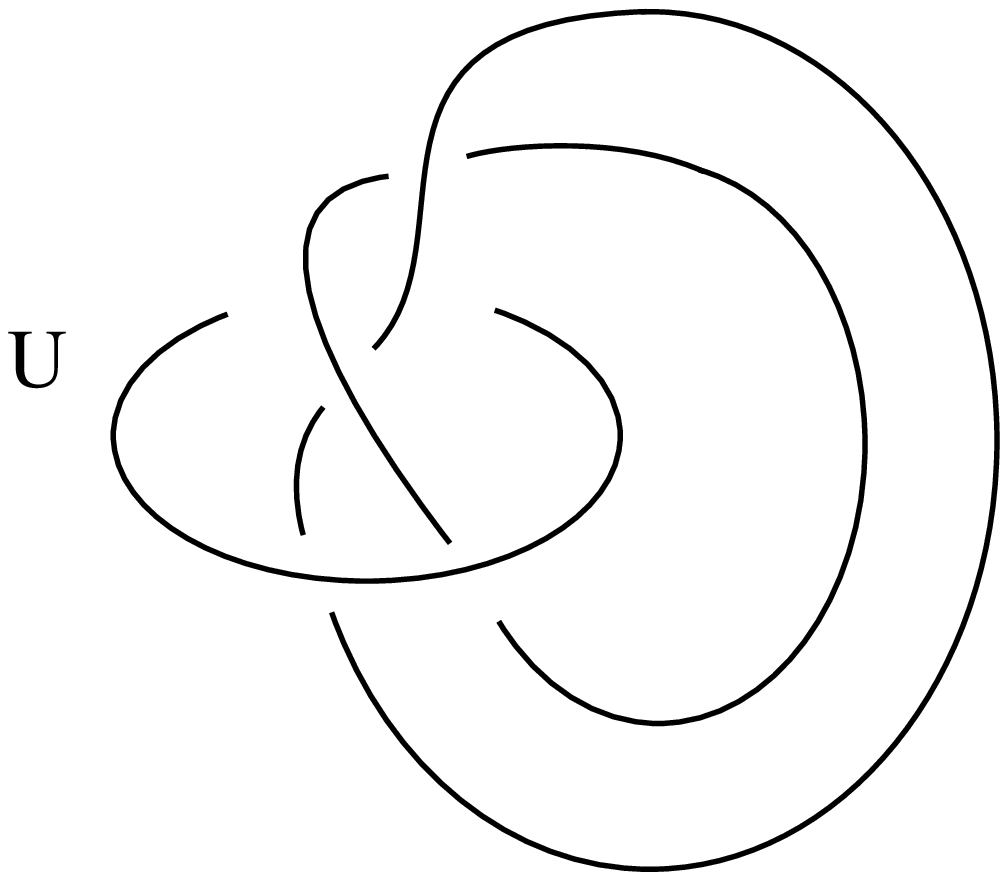,height=4cm,width=5cm}}
\vskip 0.5 truecm 
\centerline{{\bf Figure 5.6}}
\vskip 0.5 truecm 
\end{figure}

In general the effect of $\tau_\pm$ on the framing of a knot $K$ contained in $S^3 - \dot{N}$ 
is the following \cite{rol}
\beeq {\rm lk} \left(K^{\prime},K^{\prime}_f \right) \; = \; {\rm lk} \left(K,K_f \right)
\; \pm \;  \left[ \chi \left(K,U \right) \right]^2   \quad ,
\label{romo}
\end{equation}
where $K^{\prime}$ and $K^{\prime}_f$ are the images of $K$ and $K_f$ under $\tau_\pm$.
 
\begin{figure}[h]
\vskip 0.9 truecm 
\centerline{\epsfig{file=\path 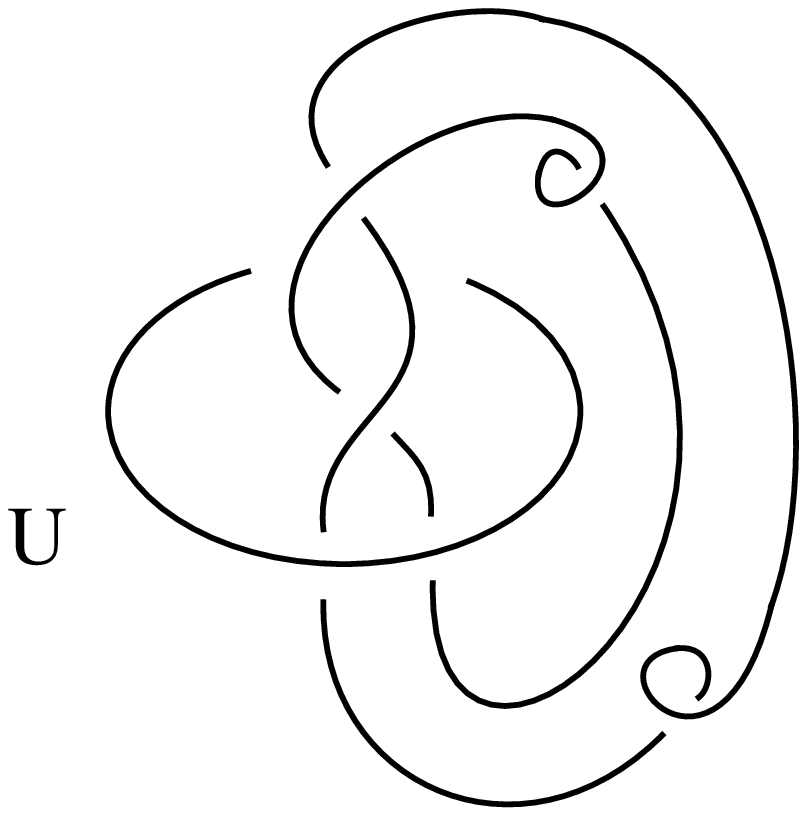,height=4cm,width=5cm}}
\vskip 0.9 truecm 
\centerline{{\bf Figure 5.7}}
\end{figure}

\section{\bf Dehn's surgery}
 
In this section we recall some basic definitions of surgery on three-manifolds. 
Let us consider surgery operations in $S^3$. 
A Dehn's surgery performed along a knot $\cal Z$ in $S^3$
consists of 

\begin{itemize}
\item  removing the interior $\dot{N}$ , of a tubular neighborhood
$N$ of the knot $\cal Z$, from $S^3$;  
\item  considering $S^3-\dot{N} $ and $N$ as distinct spaces whose
(distinct) boundaries $\partial (S^3-\dot{N} )$ and $\partial N$ are tori;  
\item gluing back $N$ and $S^3-\dot{N} $ by identifying the points
on their boundaries $\partial N$ and $\partial (S^3- \dot{N})$ according to a
given homeomorphism $h\, :\partial N \rightarrow \partial (S^3-\dot{N} )$ . 
\end{itemize}

\noindent The knot $\cal Z$ and the ``gluing" homeomorphism $h$ completely
specify the surgery operation and the resulting manifold is denoted by 
\beeq
{M } \; = \; (S^3-\dot{N} )\; \bigcup_h \; N \qquad . 
\label{eq2.1}
\end{equation}
Actually, the manifold (\ref{eq2.1}) depends \cite{rol}, up to homeomorphisms, 
only upon the homotopy class of $h(\mu )$ in $\partial (S^3-\dot{N} )$, where 
$\mu $ is a meridian of $N$. The surgery is then characterized by the knot
$\cal Z$ and by a closed curve $Y\in \partial N$ representing $h(\mu )$. 
The convention, introduced by Rolfsen \cite{rol}, which is used to codify the
surgery instruction is the following. The class $[\, Y\, ]\in \pi_1 (\partial N)$
is written as  
\beeq
[\, Y\, ] \; = \; a \cdot [\, \lambda \, ] \; + \; b \cdot [\, \mu \, ]
\quad , 
\label{eq2.2}
\end{equation}
where the generators $\lambda $ and $\mu $ are the longitude and the meridian
of a Rolfsen basis \cite{rol,guad3} in $\partial N$. Since $Y$ is a knot in $\partial N$, the integer 
coefficients $a$ and $b$ appearing in eq.(2.2) are relatively prime. The 
ratio 
\beeq
r \; = \; \frac{b}{ a }
\label{eq2.3}
\end{equation}
is called the surgery coefficient. In conclusion, the surgery
instruction is simply specified by the knot $\cal Z$ in $S^3$ and by the
rational surgery coefficient $r$.  

The knot $\cal Z$, along which surgery is performed, is not
oriented and $Y$ also is not oriented. Therefore, in a fixed
Rolfsen basis, the coefficients $a$ and $b$ appearing in
(\ref{eq2.2}) possess an overall ambiguity in their signs, depending on the 
choice
of the orientation of $Y$. Similarly, for a fixed orientation of $Y$, 
a different choice of the Rolfsen basis modifies both signs 
of the coefficients $a$ and $b$. 
This ambiguity does not affect the resulting manifold obtained by surgery
and, in fact, this ambiguity disappears in the surgery coefficient $r$. 

When $a=0$, necessarily $b=\pm 1$ and the surgery coefficient is indicated
by $r=\infty $. In this case, $Y$ is a meridian of $N$. This means that
the image, under the gluing homeomorphism $h$, 
of the meridian $\mu $ of $N$ is ambient isotopic with $\mu $ itself. 
Therefore, the resulting manifold is just $S^3$. In conclusion, the surgery
instruction specified by an arbitrary knot $\cal Z$ with surgery coefficient
$r=\infty $ corresponds to the identity.  

Clearly, the surgery operation of removing and sewing a solid torus can
be repeated several times. Therefore, a general  surgery instruction  
consists  of an unoriented link $\cal L$ in $S^3$ , called the  surgery
link, with given  surgery coefficients $\{ \, r_i\, \}$ assigned to its
components $\{ \,  {\cal L}_i \, \}$. 

For example, when $\cal L$ is the unknot with surgery coefficient $r=b/a$, 
the resulting space is homeomorphic with the lens space $L(b,a)$. In
particular, the unknot with surgery coefficient $r=0$ corresponds to 
$S^2\times S^1$. If $\cal L$ coincides with the Borromean Rings with all the
surgery coefficients equal to $+1$, the resulting manifold is the icosahedral
space or Poincar\'e manifold $\cal P$. The Borromean Rings with surgery
coefficients $r_i=0$, for $i=1,2,3$, represent $S^1\times S^1\times S^1$. 

\medskip 

Different surgery instructions do not necessarily correspond to different
manifolds. To be more precise, two manifolds associated with different
surgery instructions are homeomorphic if and only if the two surgery
instructions are related \cite{rol,kir} by a finite sequence of Rolfsen moves. 

A Rolfsen move of the first type is quite obvious: it states that one can
add or eliminate a component of the surgery link $\cal L$ with surgery
coefficient  $r=\infty $. A Rolfsen move of the second type 
describes the effects of 
an appropriate twist homeomorphism $\tau_{\pm }$ acting on a solid torus
which contains part of the surgery link.   Let $\cal
L$ be a  surgery link in which one of its components, say ${\cal L}_1$, is
the unknot with surgery  coefficient $r_1$. This means that all the remaining
components  $\{ \, {\cal L}_j\, \} $ (with $j\not= 1$) of $\cal L$ belong to
the complement solid torus  ${\cal V}_1$ 
of ${\cal L}_1$ in $S^3$. Under a twist homeomorphism $\tau_\pm $ of  
${\cal V}_1$, the component ${\cal L}_1$ is not modified, i.e. 
${\cal L}_1^\prime \, = \, {\cal L}_1$.    
The remaining components $\{ \, {\cal L}_j\, \} $ are transformed under the twist $\tau_\pm$, 
$\tau_\pm : {\cal L}_j \rightarrow {\cal L}_j^\prime $, 
according to the rules illustrated in the previous section (see also \cite{rol}).
Furthermore, the surgery coefficients also are modified. One
can show \cite{rol} that the new surgery coefficients are 
\beeq
r_1^{\, \prime} \; = \; \frac{1}{ (1/r_1) \pm 1} \quad , 
\label{eq2.4}
\end{equation}
and 
\beeq
r_j^{\, \prime} \; = \; r_j \pm \left [ {\rm lk} ({\cal L}_j, {\cal L}_1) \right
]^2  \quad {\rm for} \quad j\not=1 \quad ,   
\label{eq2.5}
\end{equation}
where ${\rm lk} ({\cal L}_j, {\cal L}_1)$ is the linking number of 
${\cal L}_j$ and ${\cal L}_1$. Thus, we have two surgery instructions
which are described, respectively, by 

\begin{description}
\item[(i)]  the surgery link $\cal L$ with surgery coefficients 
$\{ \, r_i \, \}$ ; 
\item[(ii)] the surgery link ${\cal L}^\prime $  with surgery coefficients
$\{ \, r_i^{\, \prime } \, \}$ .  
\end{description}
The surgery instructions (i) and (ii) are said to be related by a Rolfsen move
of the second kind and describe homeomorphic manifolds.   

For example,
the three surgery instructions shown in Fig.5.8 describe the manifold  $S^2\times S^1$. The first
and the second instructions are related by a Rolfsen move of the second
kind, whereas the second and the third instructions are related by a Rolfsen
move of the first kind.

\begin{figure}[h]
\vskip 0.9 truecm 
\centerline{\epsfig{file=\path 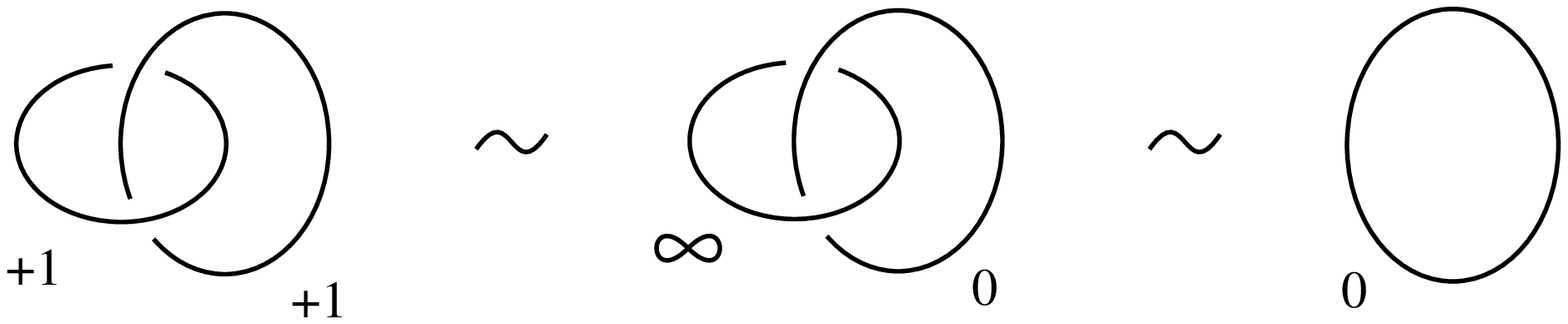,height=4cm,width=11cm}}
\vskip 0.9 truecm 
\centerline{{\bf Figure 5.8}}
\vskip 0.9 truecm 
\end{figure}

\noindent The surgery instructions, shown in Fig.2.2,  are
related by Rolfsen moves and correspond to the Poincar\'e manifold $\cal
P$. 

\begin{figure}[h]
\vskip 0.9 truecm 
\centerline{\epsfig{file=\path 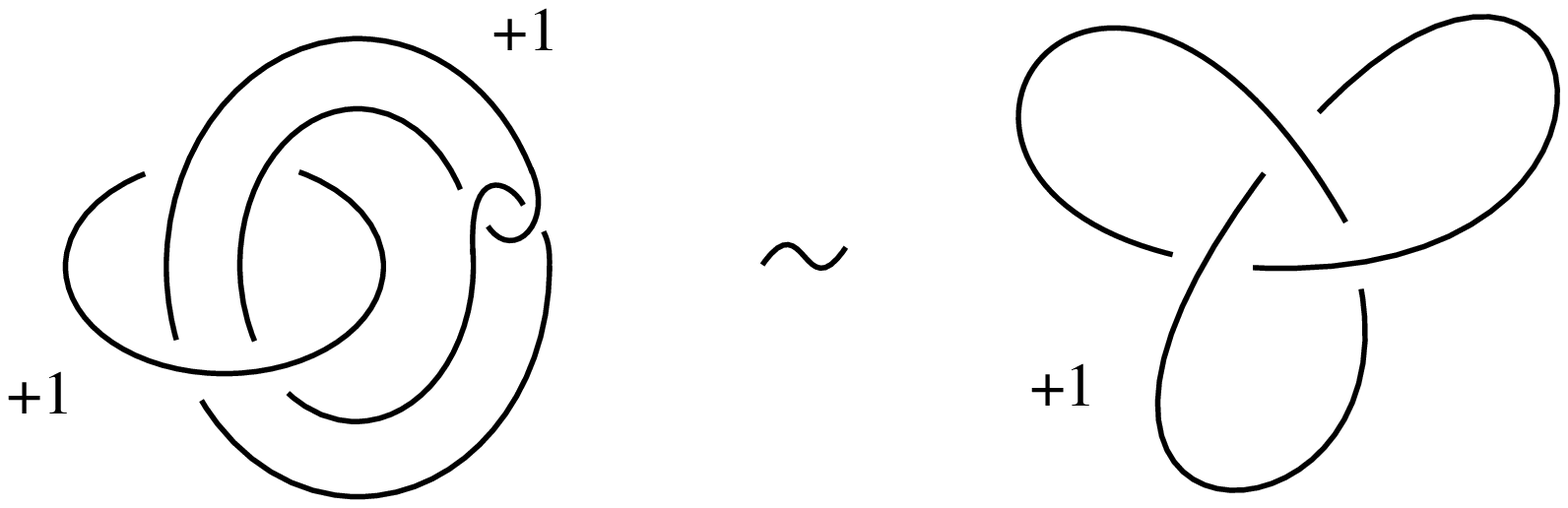,height=5cm,width=9cm}}
\vskip 0.9 truecm 
\centerline{{\bf Figure 5.9}}
\vskip 0.9 truecm 
\end{figure}

Two different surgery instructions, which are related by a finite sequence of 
Rolfsen moves, are called equivalent. The set of all possible surgery 
instructions can be decomposed into classes of equivalent instructions
and a function defined on these equivalence classes is called a
three-manifold invariant. 

It should be noted that finite sequences of Rolfsen moves
actually correspond to the set of orientation-preserving self-homeomorphisms
of a given three-manifold. If we wish to include self-homeomorphisms which
are not orientation-preserving, we simply need to add the inversion of $S^3$ (or 
mirror-reflection) as an admissible move.  

Let $M$ be the manifold described by the surgery link $\cal L $. The ambient isotopy class of a
given framed oriented link in $M$  will be represented by a framed oriented link $L$ in the
complement of $ \cal  L$ in $S^3$. 

Lickorish's (or Fundamental) Theorem \cite{lick} states that every closed, orientable and 
connected
three-manifold can be obtained by surgery in $S^3$; moreover one may always 
find such a surgery
presentation in which the surgery coefficients are all $\pm 1$ and the individual components of the
surgery link are unknotted.  Let us denote by $S_+ $ and $S_-$ the elementary surgery operations in
$S^3$ corresponding to the instructions described by the unknot $U$ in $S^3$ with surgery
coefficient  $r= + 1$ and $r= - 1$ respectively. The three-manifold obtained according to a single
surgery instruction $S_+$ or $S_-$ is homeomorphic with $S^3$.  
By  combining several surgeries $S_\pm$, any three-manifold $M$ can be obtained. 

Let us  introduce the so-called  ``honest"
surgeries  \cite{rol,kir,lick}.  It is not difficult to prove that one can 
always find a ``honest"   surgery presentation of a given three-manifold: this
 simply means that all the  surgery coefficients  $\{ r_i \}$ are integers. 
When the ratio $r$, appearing in eq.(\ref{eq2.3}), is an integer, one can take
$a=1$ and $b=r$ in eq.(\ref{eq2.2}). In this case, the curve $Y$ is a 
longitude of $N$ and
can be interpreted as a framing of the surgery knot  $\cal Z$. The linking
number ${\rm lk} ({\cal Z}, Y)$ of $\cal Z$ and $Y$ is given precisely by the
coefficient $b$, see eq.(\ref{eq2.2}), and then  
\beeq
{\rm lk} ({\cal Z}, Y) \; = \; r \quad . 
\end{equation}
Therefore, a surgery link $\cal L$ with integer surgery coefficients
$\{ r_i\}$ can be represented by a framed link $\cal L$ in which the linking
number of the component ${\cal L}_i$ and its framing ${\cal L}_{if}$ is equal
to $r_i$. 

 Let us now consider the set of all possible instructions
corresponding to ``honest" surgeries.  Two such different instructions
describe homeomorphic manifolds if and only if they are related [4] by a
finite sequence of Kirby moves.  A Kirby move is the analogue of a
Rolfsen move and can be defined as follows. 
Suppose that one component, say ${\cal L}_1$, of $\cal L$ is the unknot with
framing ${\cal L}_{1f}$ such that ${\rm lk} ({\cal L}_1,{\cal L}_{1f}) = \pm 1$. 
Then, this component can be eliminated provided that we perform a twist
$\tau_{\mp}$ on the complement solid torus ${\cal V}_1$ of ${\cal L}_1$ in
$S^3$. In general, let us consider a Kirby move when a
real link $L$ is also present in the manifold $M$. In this case, the
twist $\tau_{\mp}$ of solid torus ${\cal V}_1$ clearly acts on all the
remaining components of  $\cal L$ and, simultaneously, on the link $L$. 

By using  eqs.(\ref{eq2.4}) and (\ref{eq2.5}), entering the definition of the 
Rolfsen moves, and the transformation property \cite{rol} of framings under 
twist homeomorphisms, it is easy to prove that the invariance under Kirby 
moves is in fact equivalent to the invariance under Rolfsen moves.

\chapter{\bf Surgery in Chern-Simons field theory}
\section{\bf Elementary surgeries} 

In  order to solve the CS theory in a generic three-manifold $\cal M$, we need to compute the
expectation values of Wilson line operators in $\cal M$. As we have already mentioned, each link in
$\cal M$ can be represented by a link  in the complement of $\cal L$ in $S^3$, where $\cal L$ is a
surgery link corresponding to $\cal M$. According to Lickorish's Theorem, $\cal L$ can be taken to be a
collection of unknots with surgery coefficients $\pm 1$. Thus, we only need to consider the 
effect of a single elementary surgery operation $S_+$ or $S_-$.  

The surgery instruction associated with $S_\pm$ is represented by  the unknot $U$ in $S^3$ with
surgery coefficient $r=\pm 1$. From the invariance under Kirby moves, it follows that  $S_+ $ 
($S_-$) can be interpreted \cite{guad3} as the generator of a left-handed twist 
$\tau_-$ (right-handed twist $\tau_+$) in the complement solid torus of $U$ in $S^3$.  
Therefore, for any given framed link $L$ in the complement of $U$ in
$S^3$, the action of $S_\pm $ is given by   
\beeq
S_\pm \; : \; L \; \rightarrow \; L^{(\mp )} \quad , 
\end{equation}
where $L^{(\mp )} $ is the image of $L$ under a twist homeomorphism $\tau_\mp $
of the complement solid torus of $U$ in $S^3$. So, under surgery $S_\pm $, the
observables of the CS theory transform as 
\beeq
S_\pm \; : \; <W(L)>  |_{S^3} \; 
\rightarrow \; <W(L^{(\mp )} )>  |_{S^3} \quad .  
\end{equation}
At this point, our strategy \cite{guad3} is to find field theory operators 
$\widetilde W(U;\pm 1)$
which represent the surgery operations $S_\pm$ according to  
\beeq
\langle \, W(L^{(\mp )} ) \, \rangle   |_{S^3 }\; 
=  \langle \, W(L)\; \widetilde W(U;\pm 1)  \, \rangle   |_{S^3 } \; \Bigg/ \;  
\langle \, \widetilde W(U;\pm 1) \, \rangle   |_{S^3 } \quad .  
\label{3.3} 
\end{equation}
For the quantum CS field theory in $S^3$, with fixed integer $k$, the physically inequivalent gauge
invariant quantum numbers associated with a knot (or a solid torus) are given
(see Chap.4) by the elements of
the reduced tensor algebra ${\cal T}_{(k)}$. Let us denote by $\{ \, \psi_i \, \}$ the elements of
the standard basis of  ${\cal T}_{(k)}$, where the collective index $i$ runs from 1 to the dimension
of the reduced tensor algebra and $\psi_1$ denotes the identity element in ${\cal T}_{(k)}$. 

Let us recall (see Chap.2)   that any  gauge invariant observable 
${\cal O}(C)$  associated with the knot $C$ admits a decomposition 
\beeq
{\cal O}(C) \; = \; \sum_{\rho } \, \xi^\prime ( \rho ) \; W(C ; \chi [\rho ]\, ) 
\quad , 
\label{3.4}
\end{equation}
where $\{ \, \xi^\prime ( \rho ) \, \}$ are numerical (complex)
coefficients.  In equation (\ref{3.4}), the sum is over the inequivalent 
irreducible representations of the gauge group. The Wilson line operators, 
entering eq.(\ref{3.4}), are defined for the knot $C$ with  a
given choice of its framing and its orientation.   For fixed integer $k$,  different
elements $\chi [\rho ]$ and $\chi [\rho^\prime ]$ of the tensor algebra $\cal 
T$ do not necessarily
represent physically inequivalent colour states. In terms of the physically 
inequivalent colour
states, eq.(\ref{3.4}) becomes   
\beeq
{\cal O}(C) \; = \; \sum_{i} \, \xi ( i ) \; W(C ; \psi_i \, )  \quad , 
\label{3.5}
\end{equation}
where $\{ \, \psi_i \, \}$ are the elements of the standard basis of the reduced tensor algebra 
${\cal T}_{(k)}$ and the coefficients $\{ \, \xi ( i ) \, \}$ are linear combinations of 
$\{ \, \xi^\prime ( \rho ) \, \}$.  Therefore, the elementary surgery operators $\widetilde
W(U;\pm 1)$ can be written as 
\beeq	
\widetilde W(U;\pm 1)\; = \; \sum_{i} \, \phi_{\pm } (i) \;  
W(U; \psi_i \, ) \quad ,    
\label{3.6}
\end{equation}
where, in our convention, the unknot $U$ has preferred framing and a fixed 
orientation. 
Now, our purpose is to determine the coefficients $\{ \, \phi_{\pm } (i) \, \}$. 
Let us consider a generic framed $L$ which does
not intersect $U$. This means that $L$ belongs to the complement solid torus
$\cal V$ of $U$ in $S^3$. 
According to eq.(\ref{3.5}),  $W(L)$ can always be written as  
\beeq
W(L) \; = \; \sum_{j} \, \xi (\, j) \; W(C; \psi_j \, ) \quad ,  
\label{3.7}
\end{equation}
where $W(C;\psi_j \, )$ is the Wilson operator associated with the 
oriented core $C$ of $\cal V$ with preferred framing. 
Let $L^{(\pm)}$ ~($C^{(\pm )}$)~ be the image of $L$ ~($C$)~ under a twist 
homeomorphism $\tau_\pm $ of $\cal
V$. From eq.(\ref{3.7}) it follows that 
\beeq
W(L^{(\pm )}) \; = \; \sum_{j} \, \xi (\, j) \; W(C^{(\pm)}; \psi_j \, )  \quad . 
\end{equation}
The vacuum expectation value of both sides of this equation gives   
\beeq
\langle W(L^{(\pm )}) \rangle   |_{S^3} \; = \; \sum_{j} \, \xi (\, j) \;
q^{\pm Q(\, j)}\, E_0[\, j\, ]  \quad ,  
\label{3.9}
\end{equation}
where $Q(\, j)$ is the quadratic Casimir operator of an irreducible representation of $SU(3)$ which
belongs to the class $\psi_j$ and $ E_0[\, j\, ] $ is the value of the unknot 
in  $S^3$ with preferred
framing and with colour state $\psi_j$. 
Expression (\ref{3.9}) has been obtained by taking into account the transformation property of the expectation values under a modification of the
framing of the link components. 

If we denote by $\lambda_\pm $ the expectation value 
\beeq
\lambda_\pm \; = \; \langle \, \widetilde W(U;\pm 1) \, \rangle   |_{S^3 } \quad ,  
\end{equation}
equation (\ref{3.3}) takes the form 
\bea
&&\lambda_\pm \,  \sum_{j} \, \xi (\, j) \; q^{\mp Q(\, j)}\, E_0[\, j\, ]  \;
= \nb \\
&&= \>  \sum_{j}\; \sum_{i} \, \xi (\, j) \, \phi_{\pm
} (i) \;  \langle \, W(C; \psi_j ) \; W(U; \psi_i )\, \rangle   |_{S^3 } 
\quad , 
\label{3.11}
\ena
where $\langle \, W(C; \psi_j ) \; W(U; \psi_i )\, \rangle   |_{S^3 }=H_{ij}$ is the value of the Hopf  link in $S^3$ with components $U$ and $C$;  both components $U$ and $C$ of the
Hopf link have preferred framings and are coloured by $\psi_i$ and $\psi_j$
respectively. 

The coefficients $\{ \, \phi_\pm (i) \, \}$ must be chosen in such a
way that eq.(3.3) holds for any link $L$. This means that eq.(\ref{3.11}) must 
hold for arbitrary coefficients  $ \{ \, \xi (\, j) \, \}$. Therefore, for
any $j$,  $\{ \, \phi_\pm (i) \, \}$ must satisfy  
\beeq
\sum_{i} \, \phi_{\pm } (i) \;  H_{ij} \; = \; \lambda_\pm \> q^{\mp Q(\, j)}\>
E_0[\, j\, ] \quad .   
\label{3.12}
\end{equation}
The values $\{ \, H_{ij}\, \}$ of the Hopf link are interpreted as the matrix elements of a
symmetric matrix  $H$ called the Hopf matrix.  By definition of the reduced tensor algebra, 
this matrix is non-singular and the linear system (\ref{3.12}) determines the coefficients 
$\{ \, \phi_{\pm } (i) \, \}$
uniquely up to a multiplicative constant. Before working out the general case, we shall 
illustrate how to solve eq.(\ref{3.12}) in some simple cases. 

\section{\bf Elementary surgery operators for low values of k} 

As examples, in this section we shall solve (\ref{3.12}) for $G=SU(3)$ and $k=1,2$.
When $k=1$, the reduced tensor algebra  is of order three (see Sect.4.2.4); the 
elements of the
standard basis of ${\cal T}_{(1)}$  are denoted by $\{ \, \Psi [0] , \Psi , \Psi [-1] \,
\}$ where $\Psi [0]$ is the unit element. In this case, eq.(\ref{3.12}) 
takes the form 
\beeq
\left ( \begin{array}{ccc} 1 & 1 &  1 \\ 
1 & e^{\frac{2 \pi i}{ 3}} &  e^{\frac{-2 \pi i}{3}} \\
1 &  e^{\frac{-2 \pi i }{ 3}} &  e^{\frac{2 \pi i}{  3}}
\end{array}  \right ) \; 
\left ( \begin{array}{c} \phi_{\pm } (0) \\ \phi_{\pm } (1) \\ \phi_{\pm } (-1)\end{array} \right ) \; = \; 
\lambda_\pm \; 
\left ( \begin{array}{c} 1 \\ e^{ \pm \, 2\pi i /3} \\ e^{ \pm \, 2\pi i /3} 
\end{array} \right )
\quad . 
\label{4.1}
\end{equation}
The solution of this system is 
\bea
&&\phi_\pm (0) \; = \; \pm \frac{i}{\sqrt 3} \, \lambda_\pm \quad , \nb \\
&&\phi_\pm (-1) \; = \; \phi_\pm (1)\; = \; \pm \frac{i}{ \sqrt 3} \, \lambda_\pm \; 
e^{\mp i 2 \pi /3} \quad . 
\ena                                                                           The solution of eq.(\ref{4.1}) depends on the parameter $\lambda_{\pm }$ 
linearly; the (non-vanishing) value of $\lambda_{\pm }$ is free and the vacuum expectation values of the observables
will not depend on it. For later convenience, we fix $\lambda_{\pm }$ to be 
\beeq
\lambda_{\pm } \; = \; e^{\mp i \pi /2} \quad . 
\end{equation}
With this choice, the elementary surgery operators are 
\beeq
\widetilde W(U;\pm 1)\; = \; W(U ; \Psi_{\pm }) \quad , 
\end{equation}
where $U$ has preferred framing and 
\beeq
{\Psi }_{\pm} \; = \; \frac{1}{\sqrt 3} \left ( \, \Psi [0] + e^{\mp i 2 \pi /3} \Psi [1] + 
e^{\mp i 2 \pi /3} \Psi [-1] \, \right ) \quad . 
\end{equation}
When $k=2$, the reduced tensor algebra ${\cal T}_{(2)}$ is isomorphic (see Sect.4.2.3) 
with ${\cal T}_{(1)}$
and the Hopf matrix is the complex conjugate of the matrix shown in 
eq.(\ref{4.1}). Consequently, one has 
\bea
&&\lambda_{\pm } \; = \; e^{\pm i \pi /2} \quad , \nb \\
&&{\Psi }_{\pm} \; = \; \frac{1}{ \sqrt 3} \left ( \, \Psi [0] + e^{\pm i 2 \pi /3} \Psi [1] + 
e^{\pm i 2 \pi /3} \Psi [-1] \, \right ) \quad .
\ena
For $k=3$, the reduced tensor algebra is of order one; the basis element $\Psi [0,0]$ can be 
represented by the element $\chi [0,0]$ of $\cal T$ corresponding to the trivial representation of
$SU(3)$. When $k=3$, surgery is realized in a trivial way because the elementary surgery operators
coincide with the identity. 

When $k=4$, the reduced tensor algebra ${\cal T}_{(4)}$ is of order three with
basis elements 
\beeq
\psi_1 \; = \; \Psi [0,0] \quad , \quad \psi_2 \; = \; \Psi [1,0] \quad , \quad 
\psi_3 \; = \; \Psi [0,1] \quad ,  
\end{equation}
where $\{ \, \Psi [0,0], \Psi [1,0], \Psi [0,1]\, \}$ correspond (see Sect.4 ) to the three points of the fundamental domain $\Delta_4$.

\noindent Eq.(\ref{3.12}) takes the form 
\beeq
\left( \begin{array}{ccc} 1 & 1 &  1 \\ 
1 & e^{\frac{2 \pi i }{3}} &  e^{\frac{-2 \pi i}{  3}} \\ 
1 &  e^{\frac{-2 \pi i}{  3}} &  e^{\frac{2 \pi i}{ 3}}
\end{array} \right ) \; 
\left ( \begin{array}{c} \phi_{\pm } (1) \\ \phi_{\pm } (2) \\ \phi_{\pm } (3)
\end{array} \right ) \; = \; 
\lambda_\pm \; 
\left ( \begin{array}{c} 1 \\ e^{ \pm \, 2\pi i /3} \\ e^{ \pm \, 2\pi i /3} 
\end{array} \right )
\quad . 
\label{4.8}
\end{equation}
The solution of eq.(\ref{4.8}) is 
\bea
&&\lambda_{\pm } \; = \; e^{\mp i \pi /2} \quad , \nb \\  
&&{\Psi }_{\pm} \; = \; \frac{1}{ \sqrt 3} \left ( \, \psi_1 + e^{\mp i 2 \pi /3} \psi_2 + 
e^{\mp i 2 \pi /3} \psi_3 \, \right ) \quad .  
\ena
It is clear that, for every fixed value of $k$, the computation of the elementary surgery operators 
$\widetilde W(U;\pm 1)$ is  straightforward.  The main task of this chapter 
is to produce  the general expression of $\widetilde W(U;\pm 1)$ for any value
of $k$. This will be the subject of the next section. 

\section{\bf General expression for elementary surgery operators} 

In this section we shall determine the general solution of system (\ref{3.12}), which defines the elementary surgery operators, under the assumption that the reduced tensor algebra 
associated with the gauge group $G$ is 
regular (see Sect.4.1). 

In order to construct the elementary surgery operators, we need to prove the following Lemma. 

\bigskip 

\shabox{
\noindent {\bf Lemma 6.1}} ~{\em Let $\, {\cal F}(i,j)$ be a function on  
$\, {\cal T}_{(k)} \otimes {\cal T}_{(k)}$. Then, one has } 
\beeq
 \sum_i \sum_j \, N_{imj}\; \>  {\cal{F}}(i,j) \; = \; 
\sum_j  \sum_i \, N_{jmi} \; \> {\cal{F}}(i^{\ast}, j^{\ast}) \quad  . 
\label{5.12}
\end{equation}

\bigskip

\noindent {\bf Proof.} ~Since $\ast $ is an automorphism of ${\cal T}_{(k)}$ 
and the index $j$ is
summed over the whole algebra, the left-hand-side of eq.(\ref{5.12}) can be 
written as 
\beeq
\sum_i \sum_j \, N_{imj}\; \> {\cal{F}}(i,j) \; = \; 
\sum_i  \sum_j \,  N_{i m j^\ast }\; \>  
{\cal{F}} (i,j^{\ast})  \quad . 
\end{equation}
By using eq.(\ref{real}), one gets 
\beeq
\sum_i \sum_j \, N_{imj}\; \>  {\cal{F}}(i,j) \; = \; 
\sum_i  \sum_j \, N_{j m i^{\ast}} \; \> {\cal{F}}(i,j^{\ast}) \quad , 
\end{equation}
and, by using again the freedom to replace $i$ with $i^\ast $ in the sum, we 
have 
\beeq
\sum_i \sum_j \, N_{imj}\; \> {\cal{F}}(i,j) \; = \; 
\sum_j  \sum_i \, N_{jmi} \; \> {\cal{F}}(i^{\ast}, j^{\ast}) \quad  ,  
\end{equation}
which coincides with eq.(\ref{5.12}). {\hfill \ding{111}} 
\vskip 0.5truecm

Let us recall that the elementary surgery operators $\widetilde W(U;\pm 1)$ are written as 
\beeq
\widetilde W(U;\pm 1)\; = \; \sum_{i} \, \phi_{\pm } (i) \;  
W(U; \psi_i \, ) \quad ,    
\label{5.15}
\end{equation}
where a given orientation has been introduced for the unknot $U$ with
preferred framing. The
coefficients $ \{ \,  \phi_{\pm } (i) \, \}$ are fixed by the following 
theorem. 

\bigskip 

\shabox{
\noindent {\bf Theorem 6.1}} ~{\em When the reduced tensor algebra is regular, 
the elementary surgery operators are given by}  
$\,$eq.(\ref{5.15}) {\em with} 
\beeq
\phi_{\pm } (i) \; = \; a(k) \; q^{\pm Q(i)} \; E_0 [i] \quad , 
\label{5.16}
\end{equation}
\beeq
a(k) \; =  \; \frac{1}{|Z_{(+)}|} \qquad ; \quad Z_{(+)} \; = \; \sum_i  q^{Q(i)}\; 
\left(E_0[i] \right)^2 \quad .
\label{5.17}
\end{equation}
{\em Moreover, with the normalization  choice} $\,$(\ref{5.17}), {\em one has } 
\beeq
\lambda_\pm \; = \; \langle \widetilde W(U;\pm 1) \rangle   |_{S^3} \; = 
\; e^{\pm  i  \theta_k } 
\quad . 
\label{5.18}
\end{equation}

\bigskip 

\noindent {\bf Proof} ~We need to verify that the coefficients shown in 
eq.(\ref{5.16}) satisfy equation
(\ref{3.12}).  The Hopf matrix $H$ can be written (see Sect.2.3) as 
\beeq
H_{ij}\; =\; q^{-Q(i)-Q(\, j)} \sum_{m} \, N_{ijm } \; E_0 [ m ] \; q^{Q( m )} \quad . 
\end{equation}
Therefore, if one inserts the values (\ref{5.16}) for $\{ \, \phi_{+} (i)\, \}
$ in eq.(\ref{3.12}), one finds   
\beeq
\lambda_+ \, q^{-Q(\, j)} \; E_0 [\, j\, ]\; =\; a(k) \sum_i  \sum_{m} \, N_{ij m } \;
q^{-Q(\, j)} \, q^{Q( m )} \; E_0[ m ] \, E_0 [i ] \quad . 
\end{equation}
Since $E_0 [i] = E_0 [i^\ast ]$ and $Q(i) = Q(i^\ast )$, by using Lemma 6.1 we 
have 
\beeq
\lambda_+  \; E_0 [\, j\, ]\; =\; a(k) \sum_i  \sum_{m} \, N_{ m ji } \; 
q^{Q(m )} \; E_0[ m ] \, E_0 [i ] \quad . 
\label{5.21}
\end{equation}
On the other hand, as we have seen in Sect.2.3, the values of the unknot, 
for fixed integer $k$, give  a
representation of ${\cal T}_{(k)}$ in $\mathbb{R}$. Consequently, 
\beeq
\sum_i \, N_{ m ji } \;  E_0 [i ] \; = \; E_0 [ m ] \; E_0 [j] \quad . 
\label{5.22}
\end{equation}
Thus, equation (\ref{5.21}) takes the form 
\beeq
\lambda_+  \; =\; a(k) \; \sum_{ i } \, q^{Q( i )} \; E_0^2[ i ] \quad . 
\label{5.23}
\end{equation}
We are free to set $\lambda_+ = e^{i \theta_k}$ and $a(k)$ real. Actually the 
numerical values of $\phi_\pm(i)$ and $a(k)$ depend on the choice of the gauge group. 
Therefore, eqs.(\ref{5.17}) and (\ref{5.18}) are in agreement with eq.(\ref{5.23}). Clearly, the
coefficients $\{ \, \phi_- (i) \, \}$  and $\lambda_-$ can be obtained by taking the complex
conjugate of  $\{ \, \phi_+ (i) \, \}$ and $\lambda_+$.  
{\hfill \ding{111}}

\vskip 0.5truecm

Since the Hopf matrix is invertible by definition of the reduced tensor algebra, 
the coefficients 
$\{ \, \phi_\pm (i) \, \}$ are uniquely determined by eq.(\ref{3.12}) up to 
an overall numerical factor. 
In this sense, the solution (\ref{5.16}) is unique. Our normalization choice 
(\ref{5.17}) will simplify our
subsequent discussion on Kirby moves but  has no influence at all on the 
expectation values of the observables.     

As we have mentioned in Sect.5.3, surgery links $\cal L$ are not oriented. 
On the other hand, in the definition (\ref{5.15}) of the elementary surgery 
operators $\widetilde W(U;\pm 1)$ we have introduced
an orientation for the unknot $U$. Since $\widetilde W(U;\pm 1)$ represent the
 elementary surgery
operations $S_\pm$, the particular choice of this orientation must be 
irrelevant; let us now verify
that this is really the case. Let us recall that, in general, if the oriented 
not $C$ has colour $\psi_i$, a modification of the orientation of $C$ is 
equivalent to replace $\psi_i$ with
$\psi_{i^\ast }$ (see Sect.2.1).  Now, the coefficients $\{ \, \phi_\pm (i) \,
 \}$ given in eq.(\ref{5.16}) verify the
equality  $\phi_\pm (i) = \phi_\pm (i^\ast )$. Consequently, if we modify the 
orientation of the unknot $U$ in eq.(\ref{5.15}), the associated operators  
$\widetilde W(U;\pm 1)$ are not modified. Thus, 
the particular choice of the orientation of $U$ in eq.(\ref{5.15}) is totally 
irrelevant; as it should be. 

\section{\bf Surgery rules} 

In this section, we shall give the surgery rules in the quantum CS field
theories with gauge groups $G=SU(2)$ and $G=SU(3)$. We shall also discuss the 
invariance under Kirby moves of the results obtained according to these rules.

As we have shown in Chap.4, for fixed integer $k$ the physically inequivalent 
colour states associated with a knot are described by the elements of the 
reduced tensor algebra ${\cal T}_{(k)}$. 
In  Chap.4 we have also given, for each integer $k$, the correspondence rules 
between the elements of
$\cal T$ and the elements of ${\cal T}_{(k)}$. Therefore, for fixed integer 
$k$, we only need to consider the complete set of observables consisting of 
Wilson line operators associated with framed
oriented links whose components have colour states given by elements of  
${\cal T}_{(k)}$.     

Let us briefly summarize the results of Sects.6.2 and 6.4. The elementary 
surgery operators 
$\widetilde W(U;\pm 1) $ are given by   
\beeq
\widetilde W(U;\pm 1) \; = \; W(U; \Psi_\pm ) \quad , 
\label{6.1}
\end{equation}
where the unknot $U$ has preferred framing and  
\beeq
\Psi_\pm \; = \; a(k) \,  
\sum_{i}\,  q^{\pm Q(i)} \, E_0[i] \, \psi_i \quad .  
\label{6.2}
\end{equation}
In eq.(\ref{6.2}), the sum runs over all the elements of ${\cal T}_{(k)}$.
The calculation of the explicit values of $a(k)$ and of 
$\lambda_+= e^{i \phi}$ is presented in App.C. When
the gauge group $\, G \, $ is $\, SU(2)\, $, $\, a_k \, $ is given by 
\cite{glib}
\beeq
a(k) \, = \, \left \{ \begin{array}{cc} 1/\sqrt{2} & \quad k \, = \, 1 \quad \null \\
\sqrt{\frac{2}{k}} \sin{(\pi /k)} & \quad k \, \geq 2 \quad ;
\end{array} \right. \label{6.3}
\end{equation}
\beeq 
e^{i \theta_k} \, = \, \left\{ \begin{array}{cc} \exp \left (- i \pi /4 \right ) & \quad k = 1
\quad
\null
\\ \exp \left [i \pi 3(k-2)/(4k) \right ] & \quad k \geq 2 \quad ; \end{array} \right. \label{6.4}
\end{equation}
 whereas, when $\, G = SU(3)\, $, one has \cite{gp1,gp2}
\beeq
a(k) \, = \, \left \{ \begin{array}{cc} 1/\sqrt{3} & \quad k \, = \, 1,2 \\
16 \cos(\pi/k) \sin^3(\pi/k)/ (k \sqrt{3}) & \quad k\, \geq 3 \quad .
\end{array} \right.
\end{equation} 
\beeq 
e^{i \theta_k} \, = \, \left\{ \begin{array}{cc} \exp \left ( i \pi /2 \right ) & \quad k = 1
\quad
\null \\ \exp \left (- i \pi /2 \right ) & \quad k = 2 \quad \null \\
\exp \left (-i 6 \pi /k \right ) & \quad k \geq 3 \quad . \end{array} \right.
\label{6.4a}
\end{equation}

Eq.(\ref{3.12}) can be interpreted in the following way. 
The action of $\widetilde W(U;\pm 1)$ on the element  $\psi_j$ of the
reduced tensor algebra ${\cal T}_{(k)}$, associated with the core of the
complement solid  torus of $U$ in $S^3$, is 
\beeq
\widetilde W(U;\pm 1)\; : \; \psi_j\;  \rightarrow \; 
e^{\mp i \varphi } \, q^{\mp Q(\, j)} \; \psi_j \quad , 
\label{6.4n}
\end{equation}
Eq.(\ref{6.4n}) shows that, apart from the overall phase factor $e^{\pm i
\varphi }$, $\widetilde W(U;\pm 1)$ are the generators of twist homeomorphisms 
$\tau_\mp $ of the complement solid torus of $U$ in $S^3$. Since these phase 
factors are constants  (i.e., do not depend on the particular colour state 
under consideration), their
presence in eq.(\ref{6.4n}) is completely harmless. In fact, 
by introducing the correct vacuum normalization in the expectation values, as
shown in eq.(\ref{3.3}), these phase factors cancel out. 

By means of eqs.(\ref{6.1})-(\ref{6.4n}),  we can now compute explicitly the expectation 
values of Wilson line operators for the CS theory defined in any orientable, 
closed and connected three-manifold  $\cal M$ when the gauge group is $SU(2)$ or 
$SU(3)$. We shall firstly consider the surgery
rules in the case in which the surgery instructions have 
the form specified by the Fundamental Theorem. Then, we shall give the
surgery rules corresponding to a generic ``honest" surgery. 

According to Lickorish's Theorem, $\cal M$ has a presentation in
terms of Dehn's surgery in $S^3$. Moreover, it is always possible to
find a surgery description of $\cal  M$ corresponding to a surgery link $\cal
L$ in which all its components $\{ {\cal L}_a \}$ (with $1\leq a \leq p$) 
are simple circles and have surgery coefficients $\{ r_a \}$ equal to $+1$ or $-1$. 

Since we know how to represent in the field theory the elementary surgery
$S_\pm$,  associated with each single component of $\cal L$, we can construct
the operator $\widetilde W({\cal L}) $ representing the whole surgery. 
Indeed, for each component ${\cal L}_a$, we shall introduce a preferred
framing ${\cal L}_{af}$ and consider the operator $\widetilde W({\cal L}_a\,
; r_a )$, as defined in eq.(\ref{6.1}). The field theory operator   
$\widetilde W({\cal L}) $, associated with the 
whole surgery $\cal L$, is \cite{guad3} 
\beeq
\widetilde W({\cal L}) \; = \; \prod_{a=1}^p \; 
\widetilde W({\cal L}_a \, ;r_a ) \quad . 
\label{6.6}
\end{equation}
Let $L$ be a given framed link in $\cal M$. As we have already mentioned, 
the isotopy class of $L\subset \cal M$ can be described by a link 
(that we indicate by the same symbol $L$) in the complement of $\cal L$ in $S^3$. 

Following the approach of Sect.6.2, the expectation value 
$\langle \, W(L)\, \rangle  |_{\cal M}$ is defined to be \cite{guad3} 
\beeq
E(L)_{{\cal M}} \; = \; \langle \, W(L)\, \rangle  |_{\cal M}\; = \; \frac{
\langle \,  W(L) \; \widetilde W({\cal L}) 
\, \rangle  |_{S^3}}{  \langle \, \widetilde W({\cal L}) \, \rangle  |_{S^3}} 
\quad .  
\label{6.7} 
\end{equation}
By means of the surgery rules (\ref{6.6}) and (\ref{6.7}), one can
compute the expectation values $\{ \, \langle \, W(L)\, \rangle   |_{\cal M} \, \}$ in any
orientable, closed connected three-manifold $\cal M$. The origin of the
expectation value of the surgery operator in the denominator of (\ref{6.7}) is
two fold. Firstly, it provides a natural extension of the link invariant
$E(L)_{S^3}$ in $\cal M$. From (\ref{6.7}), it follows that when
the link $L$ can be embedded in a three ball $B^3 \subset {\cal M}$ then
\beeq
L \subset B^3 \subset {\cal M} \Ra E(L)_{{\cal M}} \; = \; E(L)_{S^3} \quad .
\label{natural}
\end{equation}
Indeed, the situation $L \subset B^3 \subset {\cal M}$ corresponds to the
following factorization of the surgery operator
\beeq
 \langle \,  W(L) \; \widetilde W({\cal L}) \, \rangle  |_{S^3} \; = \;
\langle \,  W(L) \, \rangle \; \langle \, \widetilde W({\cal L})  
\, \rangle  |_{S^3} \quad .
\end{equation}
Secondly, the presence of $\langle  \widetilde W({\cal L}) \rangle$ in the 
denominator, can be understood as the effect of the modification of the vacuum
to vacuum Feynman diagrams with respect of $S^3$.   

In order to discuss the invariance under
Kirby moves, it is convenient to give the surgery rules which
correspond to ``honest" surgeries in general. 

Any ``honest" surgery is described by a surgery link $\cal L$ whose
components $\{ {\cal L}_a \}$  
have integer surgery coefficients $\{ r_a \}$. In this case, 
each single component ${\cal L}_a$ is not necessarily ambient isotopic with
a simple circle, of course. For each component ${\cal L}_a$, 
we shall introduce a framing ${\cal L}_{af}$ such that the linking number of 
${\cal L}_a$ and ${\cal L}_{af}$ satisfies  
\beeq
 {\rm lk }({\cal L}_a, {\cal L}_{af}) \; = \; r_a \quad . 
\label{6.8}
\end{equation}
Then we shall consider the Wilson line operator 
\beeq
W({\cal L}_a\; ; \Psi_0 ) \quad , 
\label{6.9}
\end{equation}
which is associated with the framed component ${\cal L}_a$ with framing ${\cal
L}_{af}$ specified in eq.(\ref{6.8}).  The element  $\Psi_0$ (surgery 
colour state) is given by 
\beeq
\Psi_0 \; = \; a(k) \, 
\sum_{i} \, E_0[i] \, \psi_i \quad . 
\label{6.10}
\end{equation}
It should be noted that the framing choice (\ref{6.8}) is different from the 
preferred framing convention which is used in the definition (\ref{6.1}).  
Under a Kirby move, the integer surgery coefficient $r_a$,  
associated with a link component ${\cal L}_a$, transforms as the linking 
number 
${\rm lk }({\cal L}_a, {\cal L}_{af})$.  For this
reason, the framing choice (\ref{6.8}) has an intrinsic meaning. The 
information carried by the
integer surgery coefficient $r_a$ is now encoded in the framing of the 
component ${\cal L}_a$ and,
consequently, the colour state $\Psi_0$ of any surgery component is 
universal.  

\bigskip 

\shabox{
\noindent {\bf Theorem 6.2}} {\em The surgery operator, corresponding to the ``honest" surgery
described  by $\, {\cal L} = \{ {\cal L}_a \} $ with integer surgery 
coefficients $\, \{ r_a
\}$~ (with $\, 1\leq a \leq p$), is } 
\beeq
\widetilde W({\cal L}) \; = \; \prod_{a=1}^p \, W({\cal L}_a\; ; \Psi_0 ) 
\quad , 
\label{6.11}
\end{equation}
{\em where $\, \Psi_0$ is displayed in} $\,$eq.(\ref{6.10}). {\em 
The expectation value} $\,
\langle \, W(L)\, \rangle   |_{\cal M}$ {\em is given by } 
\beeq
\langle \, W(L)\, \rangle   |_{\cal M}\; = \; 
 {\langle \,  W(L) \; \widetilde W({\cal L}) \, \rangle   |_{S^3}  \over 
\langle \, \widetilde W({\cal L}) \, \rangle   |_{S^3}} 
\quad .  
\label{6.12} 
\end{equation}
{\em The results obtained according to} $\, $eq.(\ref{6.12}) {\em are 
invariant under Kirby
moves.} 

\bigskip

\noindent {\bf Proof.} ~By using the covariance properties  of the expectation
 values of the Wilson line operators under a change of framing (see Sect 2.), 
it is easy to verify that,  when  all the components $\{ {\cal L}_a \} $ are 
simple circles 
and the surgery coefficients $\{ r_a \}$ are equal to $+1$ or $-1$, eq.(\ref{6.12}) coincides  with eq.(\ref{6.7}).  In order to prove the consistency of 
eq.(\ref{6.12}), we need to demonstrate
that the results obtained according to eq.(\ref{6.12}) are invariant under 
Kirby moves. 

 Let $\cal L$ be the
instruction corresponding to a given ``honest" surgery. Suppose that one
component, say ${\cal L}_1$, of $\cal L$ is a simple circle with surgery
coefficient $r_1=\pm 1$. All the remaining components of $\cal L$ and the
given link $L$ belong to the complement solid torus ${\cal V}_1$ of ${\cal
L}_1$ in $S^3$. As we have stated above, $W({\cal L}_1\; ; \Psi_0 )$ is equivalent to 
$\widetilde W({\cal L}_1\; ; \pm 1)$.   Consider now the
numerator and, separately, the denominator of the expression (\ref{6.12}). From
eq.(\ref{6.4n}) it follows that ${\cal L}_1$ can be eliminated
provided that we perform a twist homeomorphism $\tau_\mp $ of ${\cal V}_1$
and, simultaneously, we multiply by the phase factor $e^{\mp i\varphi }$. 
This phase factor cancels out in the ratio (\ref{6.12}); therefore, only the 
effects of the twist
homeomorphism $\tau_\mp $ are relevant. Under this twist, $L$ is accordingly modified into its
homeomorphic image, as it should be. The same happens with the remaining components of
$\cal L$. It remains to be verified that the new surgery coefficients $\{
r^{\, \prime }_b \}$ (with $b\not= 1$) have the correct values displayed in
eq.(2.5); as a consequence of the transformation properties  of
framings under twist homeomorphisms (see Sect 5.4), this is indeed the case.  

In conclusion, the results obtained by means of the surgery rules (\ref{6.11})
 and (\ref{6.12}) 
are invariant under Kirby moves. On the other hand, with an appropriate
sequence of Kirby moves, any ``honest" surgery instruction can be transformed
into an equivalent surgery instruction of the type specified by the
Fundamental Theorem. In this case,  expressions (\ref{6.12})
and (\ref{6.7}) coincide and this concludes the proof. {\hfill \ding{111}}

\vskip 0.5truecm

\noindent With our definition of the elements $\Psi_\pm $ 
shown in eq.(6.2), $\Psi_-$ can be obtained from $\Psi_+$ by means of two
elementary changes of framing.  This is the main reason for our choice on their normalization.  

From the previous discussion on the
Kirby moves, it is clear that the presence of the phase factors 
$e^{\pm i\varphi }$ cannot be avoided. To be more
precise,  whatever the normalization of $\Psi_\pm $ 
is,  the numerator (and, separately, the denominator) of the
expression (\ref{6.12}) is  not invariant under Kirby moves. (The ratio 
(\ref{6.12}) is invariant under Kirby
moves, of course).  In other words,  a projective representation of the group of twist homeomorphisms 
of solid tori is realized on the state space of the CS theory. 

When $\langle \, \widetilde W({\cal L}) \, \rangle   |_{S^3}\not= 0$ the 
expression (6.12) is well defined. The expectation value 
$\langle \, \widetilde W({\cal L}) \, \rangle   |_{S^3}$ gives
information on the three-manifold $\cal M$, associated with the surgery link
$\cal L$ in $S^3$, and depends on the value of the coupling constant $k$.
For fixed $\cal M$, $\langle \, \widetilde W({\cal L}) \, \rangle   |_{S^3}$ may vanish
when $k$ takes values on a certain set of integers; in this
cases, eq.(6.12) is not well defined. Let us recall that, in
general, the internal consistency of the quantum CS theory defined in $\cal
M$ puts some restrictions \cite{guad3} on the possible values of $k$. 
It is natural to expect that $\langle \, \widetilde W({\cal L}) \, \rangle   |_{S^3}$ 
vanishes for precisely those values of $k$ for which the quantum CS theory
is not well defined in $\cal M$. 

\section{\bf General properties} 

In this section we shall focus our attention on some general aspects of the
construction of the elementary surgery operators presented in the previous sections. 
Moreover, we shall give another derivation of the solution (\ref{5.16}), 
(\ref{5.17}) of (\ref{3.12}). 

Even if the numerical results are different for different gauge groups, 
the underlying algebraic structure is universal.  It is remarkable that this 
universal structure is not a peculiarity of
topological quantum field theory but appears also in all the different 
approaches \cite{retu,tur,koh,lick1,mor} to the newly discovered three-manifold invariants. 
Thus, 
before considering explicit applications of the
surgery rules in the $SU(3)$ CS field theory, we would like to present here 
some general results concerning the construction of Dehn's surgery operators.   

Let us consider the CS field theory with compact simple Lie group $G$.  For 
generic values of the coupling constant $k$, the expectation values of the 
Wilson line operators are finite Laurent polynomials in a certain power of the
 deformation parameter $q = \exp (-i 2\pi /k)$.  
By construction, for any fixed link with $n$ components, these expectation 
values are multi-linear
functions on  ${\cal T}^{\otimes n}$, where $\cal T$ is the tensor algebra (or complexification of 
the representation ring) of $G$. As we have learned in Chapt.4, 
for fixed integer $k$, a complete set of gauge invariant observables is 
represented by 
the set of Wilson line operators associated with framed
oriented links whose components have colour states given by elements of ${\cal T}_{(k)}$.
In the considered examples where $G$ is a unitary group, ${\cal T}_{(k)}$ 
turns out to be of finite order.  It is natural to expect that ${\cal T}_{(k)}$ is  of finite order also for a generic (simple
compact Lie) group $G$. In what follows, we simply assume that this is indeed 
the case. 

For generic values of $k$, the generalized satellite relations \ref{satg}  are
 written in terms of the elements of $\cal T$. For fixed integer $k$, these 
relations can be expressed in terms of the
elements $\{ \, \psi_i \, \}$ of the standard basis of ${\cal T}_{(k)}$. 
For example, 
consider the knots $C_1$ and $C_2$ in the solid torus $V$ which coincides with
the complement of
the unknot $U$ in $S^3$, as shown in Fig.6.1. Let these knots have preferred 
framings and  colours 
$\psi_i$ and $\psi_j$ belonging to ${\cal T}_{(k)}$.  The product 
$W(C_1 ; \psi_i ) \, W(C_2 ; \psi_j )$ of the associated  Wilson line operators admits  the
decomposition   
\beeq
W(C_1 ; \psi_i ) \, W(C_2 ; \psi_j )\; = \; \sum_m \, N_{i j m } \; W(C_1 ; \psi_m ) 
\quad , 
\label{7.1}
\end{equation}
where $\{  \, N_{i j m } \, \}$ are the structure constants of 
${\cal T}_{(k)}$. 

At this point,  one should construct the surgery operators $\widetilde W(U;\pm 1) $ which are
determined by eqs.(\ref{3.6}) and (\ref{3.12}). Clearly, the Hopf matrix $H$ 
entering eq.(\ref{3.12}) depends on $G$; 
nevertheless, we shall show that the solution of eq.(\ref{3.12}), namely the 
form of $\widetilde W(U;\pm 1)$, does not depend on $G$. The proof is based 
exclusively on the topological properties of 
surgery; in particular, we will not use the structure of the Hopf matrix.

\begin{figure}[h]
\vskip 0.9 truecm 
\centerline{\epsfig{file=\path 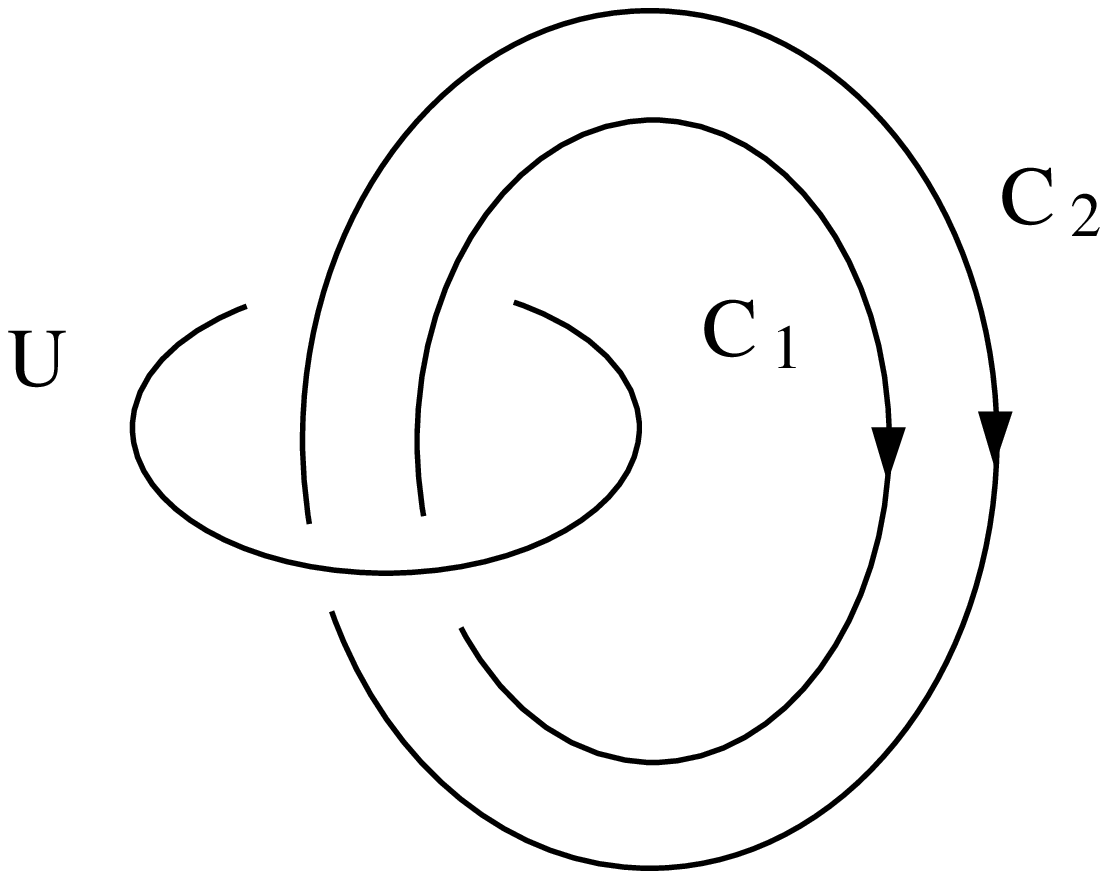,height=4cm,width=5cm}}
\vskip 0.9 truecm 
\centerline {{\bf Figure 6.1}}
\vskip 0.9 truecm 
\end{figure}

Let us consider the set of ``honest" surgeries.  Each surgery of this kind can be 
described by a framed unoriented link ${\cal L} = \{ {\cal L }_a \}$ in $S^3$; we use the standard
convention in which each surgery coefficient $r_a$ is given by the linking number of the component
${\cal L}_a$ and its framing ${\cal L}_{af}$, eq.(\ref{6.8}). As we have 
already mentioned, with the
choice (\ref{6.8}) of framing, the surgery operators are characterized by a 
universal colour state
$\Psi_0$, see eq.(\ref{6.9}).  The statement that the form of $\widetilde 
W(U;\pm 1) $ does not depend on
$G$ is equivalent to the statement that  $\Psi_0$
has the structure given in eq.(\ref{6.10}). 

\bigskip

\shabox{
\noindent {\bf Theorem 6.3}} {\em Let us assume that the reduced tensor algebra $\, {\cal T}_{(k)}$ in 
the Chern-Simons theory with  compact simple Lie gauge group $\, G$ is regular
(see Chapt.4).  The surgery
operator $\, \widetilde W( {\cal L})$, which is  associated with the surgery 
link $\, {\cal L} \, = \,
\{ \, {\cal L}_a \, \}$ with integer surgery coefficients  $\, \{ \, r_a \, \}$ and framings
specified by} $\,$eq.(\ref{6.8}), {\em is given by}  $\,$eq.(\ref{6.11})  
{\em with }   
\beeq
\Psi_0  \; = \; \, a(k) \; \sum_i \, E_0 [i] \; \psi_i  \quad , 
\label{7.2}
\end{equation}
{\em where  $\, a(k)$ is a non-vanishing normalization factor.}  

\bigskip

\noindent {\bf Proof} ~First of all we note that, in the computation of the 
expectation values (\ref{6.12}), the particular value of the normalization 
factor $a(k)$ is irrelevant. (The natural choice 
for the value of $a(k)$ will be described in a while.)  

Since the element $\Psi_0 \in {\cal T}_{(k)}$ does not
depend on the particular form of the surgery link ${\cal L} \in S^3$,  $\Psi_0
$ must be determined by general topological properties of surgery.  In order 
to describe these properties, we need to disentangle the action of a generic 
surgery operation, which is defined inside a solid torus $V$, from the particular embeddings of this solid torus in
$S^3$.   To be more precise, let us represent the solid torus $V$ by the 
complement of the unknot $U$
in $S^3$; we shall denote by $K$ the framed core of $V$ with preferred 
framing.  Suppose that $K$
represents the surgery instruction corresponding to the ``honest" Dehn's surgery
 with surgery
coefficient $r=0$.  Then, each framed component ${\cal L}_a$ of a surgery link
$\cal L$ in $S^3$ can be understood to be the image $h^\diamond (K)$ of $K$ 
under the homeomorphism $h^\diamond $  which has been defined in Sect.2.2.  

Since the element $\Psi_0 $ does not depend on $h^\diamond $, in order to 
find $\Psi_0 $ we
only need to consider the surgery operation described by $K$ in $V$.  By 
definition, this surgery
consists of  removing  a tubular neighbourhood $N$ of $K$ in $V$ and sewing it
back in such a
way that a meridian of $N$ is mapped into the curve $Y$ which is shown in 
Fig.6.2.  

\begin{figure}[h]
\vskip 0.9 truecm 
\centerline{\epsfig{file=\path 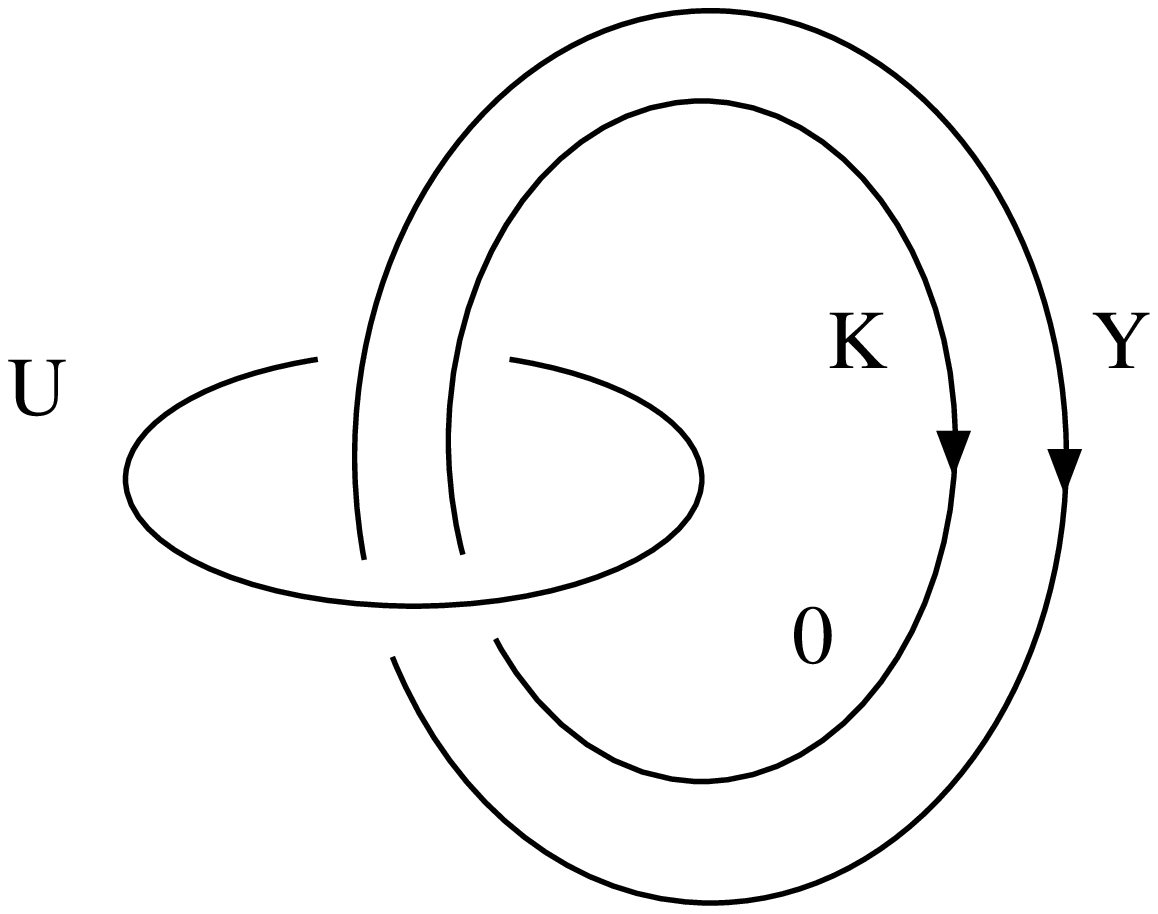,height=4cm,width=5cm}}
\vskip 0.9 truecm 
\centerline {{\bf Figure 6.2}}
\vskip 0.9 truecm 
\end{figure}

\noindent Since a meridian of $N$ bounds a disc (standardly embedded in a 
three-ball), the knot $Y$
shown in Fig.6.2  is really ambient isotopic with the unknot (simple circle) contained inside a
three-ball.   Clearly, if the knot $Y$ is framed, then $Y$ is ambient isotopic with the unknot with
preferred framing.  This is the desired property which characterizes completely the surgery operation described
by $K$ in $V$.  

Now we would like to represent this surgery by a Wilson line operator 
$W( K ; \Psi_0 )$ which is defined for $K$ with preferred framing.  As we have
 already mentioned, a given orientation for $K$ is also introduced, but the 
final results will not depend on
the choice of this orientation.  Suppose that a Wilson line operator $W(Y; 
\psi_j)$ is associated with
the knot $Y$ which is oriented and has preferred framing.  Since $Y$ is 
ambient isotopic with the unknot (contained inside a three-ball) with 
preferred framing, the Wilson line operator $W(Y; \psi_j)$
simply gives the contribution $E_0[\, j\, ]$.  Therefore, the element 
$\Psi_0 $ must be
determined in such a way that, inside the solid torus $V$, the insertion of  
$W(Y; \psi_j)$ (with
arbitrary $\psi_j$) is equivalent to the multiplication by $E_0[\, j\, ]$. We 
shall now verify that
the element $\Psi_0 $, given in eq.(\ref{7.2}),  has precisely this property. 

Let us consider the product $W(K; \Psi_0 ) \, W(Y; \psi_j)$ of the two Wilson 
line operators
associated with $K$ and $Y$ inside the solid torus $V$. By using  the 
decomposition (\ref{7.1}), one finds 
\beeq
W(K; \Psi_0 ) \, W(Y; \psi_j) \; = \; a(k) \, \sum_i \, E_0[i] \; \sum_m \, 
N_{i j m } \;W(K ; \psi_m ) \quad ,  
\end{equation}
and, by means of Lemma 6.1, one obtains 
\beeq
W(K; \Psi_0 ) \, W(Y; \psi_j) \; = \; a(k) \, \sum_i \, E_0[i^\ast ] \; \sum_m \, N_{m j i } \;
W(K ; \psi_{m^\ast} ) \quad .   
\label{7.4}
\end{equation}
Since the values of the unknot give a representation of ${\cal T}_{(k)}$, the 
relation (\ref{5.22}) is
valid; moreover,  $E_0[i^\ast ]= E_0 [i]$. Therefore,  eq.(\ref{7.4}) takes 
the form    
\bea
W(K; \Psi_0 ) \, W(Y; \psi_j) \; &&= \; E_0 [\, j\, ] \; a(k) \, 
\sum_m \, E_0[m] \; W(K ; \psi_{m} ) \nb \\
&&= \;  E_0 [\, j\, ] \; \, W(K; \Psi_0 ) \quad . 
\label{7.5}
\ena
This equation holds for any $\psi_j \in {\cal T}_{(k)}$. Therefore, 
eq.(\ref{7.5}) shows that the operator 
$W(K; \Psi_0 )$, with $\Psi_0$ given in eq.(\ref{7.2}), represents the surgery
operation which is described by the surgery knot $K$ with surgery coefficient 
$r=0$. 

In order to prove the invariance under Kirby moves, one has to consider the elementary surgery
operators $\widetilde W(U;\pm 1) $.  The operators $\widetilde W(U;\pm 1) $ can be obtained 
from $W(K; \Psi_0 )$ by using the satellite
relations. Let us consider a satellite of the unknot, with writhe $+ 1$ in $S^3$, which has been
obtained by means of the pattern link defined by the two knots $K$ and $Y$ in $V$.  This satellite is
shown in Fig.6.3.

\begin{figure}[h]
\vskip 0.9 truecm 
\centerline{\epsfig{file=\path 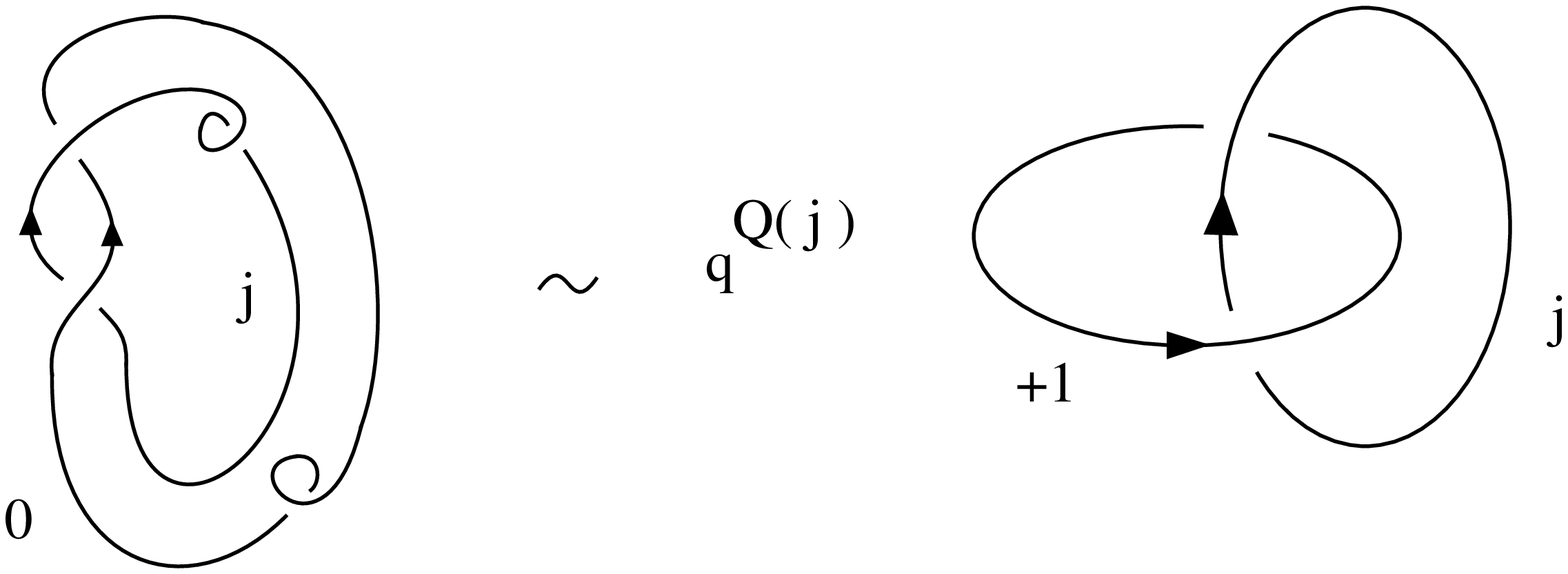,height=5cm,width=10cm}}
\vskip 0.9 truecm 
\centerline {{\bf Figure 6.3}}
\vskip 0.9 truecm 
\end{figure}

\noindent According to eq.(\ref{7.5}), the expectation value of the Wilson 
line operators associated with the
link of Fig.6.3 is equal to $E_0 [\, j \, ] \; \langle \, W(U; \Psi_+) \, 
\rangle   |_{S^3}$.
Therefore, by taking into account the behaviour of the expectation values under a change of
framing, one obtains  
\beeq
a(k) \, \sum_i \, q^{Q(i)} \; E_0[i] \; H_{ij} \; = \; q^{- Q(\, j)} \; E_0[\, j \, ] \; 
\left ( \, a(k) \, \sum_i q^{Q(i)} \; E^2_0[i] \, \right ) \quad . 
\label{7.6}
\end{equation}
This equation coincides with eq.(\ref{3.12}) with 
\beeq
\phi_+(i) \; = \; a(k) \; q^{Q(i)} \; E_0[i] \quad , 
\end{equation}
\beeq
\lambda_+ \; = \; a(k) \, \sum_i q^{Q(i)} \; E^2_0[i] \quad . 
\end{equation}
Clearly, $\phi_-$ and $\lambda_-$ can be obtained from $\phi_+$ and $\lambda_+$ by taking the
complex conjugates. By assumption, $\lambda_+ \not= 0$ and consequently 
$\lambda_- \not= 0$.  
Therefore, the invariance under Kirby moves can be proved by using the same argument as 
presented in
Sect.6.2.  The natural choice of the normalization factor $a(k)$ is to require
that $a(k) > 0$ and $|\lambda_\pm | =1$. This is the convention that we have 
adopted in the previous sections. {\hfill \ding{111}}

\vskip 0.5truecm

By definition of the reduced tensor algebra, ${\cal T}_{(k)}$ is physically 
irreducible ; this means
that if $\psi \in {\cal T}_{(k)}$ is physically equivalent to the null vector, then $\psi = 0$. 
It should be noted that, in the proof of Theorem 6.3, this property of 
${\cal T}_{(k)}$ has not been
used.  Theorem 6.3 provides an explicit solution to the problem of the 
surgery colour state $\Psi_0$;  the uniqueness of this solution follows from  
Theorem 6.4. 

\bigskip 

\shabox{
\noindent {\bf Theorem 6.4}} {\em Up to a normalization factor, the surgery 
colour state $\, \Psi_0$  is unique.} 

\bigskip

\noindent {\bf Proof} ~Suppose that we have found two surgery colour states 
$\Psi_0$ and $\Psi_0^\prime $ in ${\cal T}_{(k)}$. Let $\Psi_+$ and $\Psi_+^\prime $
be the corresponding colour states associated with the elementary surgery $S_+$. Let us fix the
normalization of  $\Psi_+$ and $\Psi_+^\prime $ (and, consequently, of $\Psi_0$ and
$\Psi_0^\prime$)  by imposing that eq.(\ref{3.12}) takes the form  
\beeq
\left ( \, H \cdot \Psi_+ \right )_i \; = \; q^{-Q(i)} \, E_0 [i] \; = \; 
\left ( \, H \cdot \Psi_+^\prime \right )_i \quad . 
\end{equation}
With this normalization, the colour state $\Psi = \Psi_+ - \Psi_+^\prime $ satisfies 
\beeq
H \cdot \Psi \; = \; 0 \quad . 
\label{7.10}
\end{equation}
Eq.(\ref{7.10}) implies that, for any link $L \subset S^3$ in which one of its
components $C$ has colour
$\Psi $, one has $\langle \, W(L) \, \rangle   |_{S^3} = 0$.  Indeed, by 
using the surgery
operators, one can find a surgery presentation \cite{rol} of $L\subset S^3$ in
 which $C$ is the unknot with
colour $\Psi $. Of course, the remaining components of $L$ and the surgery link belong to the
complement solid torus of $C$ in $S^3$. Therefore, by using the generalized satellite relations (see Sect.2.2),  $\langle \, W(L) \, \rangle   |_{S^3}$ 
can be expressed in terms of the values of the
Hopf link in which one of its components has colour $\Psi $. Consequently, 
eq.(\ref{7.10}) implies that 
$\langle \, W(L) \, \rangle   |_{S^3} = 0$. 

On the other hand, since ${\cal T}_{(k)}$ is physically 
irreducible,  one has   $\Psi =0$ and thus $\Psi_+ = \Psi_+^\prime $; this implies that 
$\Psi_0 =\Psi_0^\prime $. {\hfill \ding{111}}

\vskip 0.5truecm

The existence of an operator which represents surgery (Theorems 6.1-2) implies
that the Hopf matrix $H_{ij}$ is nonsingular. Indeed, suppose that 
eq.(\ref{7.10}) is satisfied with a
certain colour state $\Psi $. By using the method described in the proof of 
Theorem 6.3,
one can conclude that  either ${\cal T}_{(k)}$ is physically reducible or 
$\Psi =0 $. Since
the reduced tensor algebra ${\cal T}_{(k)}$ is (by definition) physically irreducible, one must
have $\Psi =0 $; this means that $H$ is nonsingular. 

\section{\bf Examples}

In this section we shall describe some concrete examples of computation of
observables in manifolds different from $S^3$. All the examples are worked out 
by using $G=SU(3)$ as the gauge group.

\subsection{{\bf The manifold} $\mathbf{S^2} \boldsymbol{\times} 
\mathbf{S^1}$} 

The manifold $S^2 \times S^1$ admits \cite{rol} a surgery presentation in which the 
surgery link is the unknot $U$ with surgery coefficient $r=0$. The manifold 
$S^2 \times S^1$ can also be represented by
the region of $\mathbb{R}^3$ delimited by two spherical surfaces which are 
centered at the origin and have
different radii; the points which have the same angular coordinates on the two
spheres are identified. 

The simplest knot $C$ which is homotopically nontrivial in $S^2\times S^1$ is 
shown in Fig.6.4. 
The link in $S^3$ shown in Fig.6.4a has two components; one component 
represents the surgery instruction corresponding to the Dehn's surgery on $S^3$ 
which gives $S^2 \times S^1$. The remaining
component, which has preferred framing, describes the knot $C$. Fig.6.4b gives
an equivalent description of $C$ in $S^2\times S^1$. 

\begin{figure}[h]
\vskip 0.9 truecm 
\centerline{\epsfig{file=\path 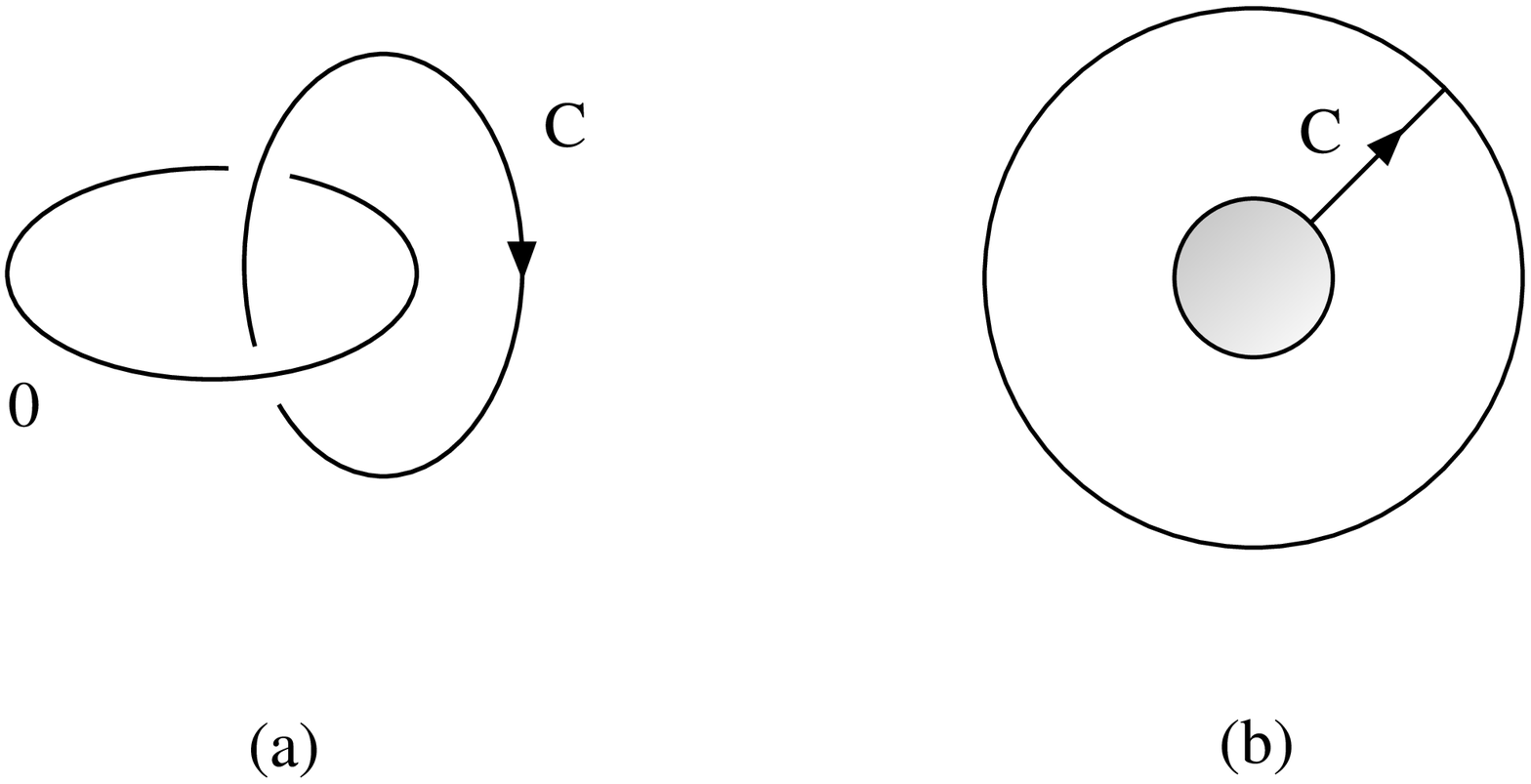,height=4.5cm,width=9cm}}
\vskip 0.9 truecm 
\centerline {{\bf Figure 6.4}}
\vskip 0.9 truecm 
\end{figure}

Let $C$ have colour $\psi_i$, the expectation value of the associated Wilson line operator in
$S^2\times S^1$ is given by Theorem 6.2, 
\beeq
\langle \, W(C ; \psi_i ) \, \rangle   |_{S^2\times S^1} \; = \; 
 \frac{\langle \,  W(C ; \psi_i ) \; \widetilde W(U ; \Psi_0 ) \, \rangle  
|_{S^3} }{ 
\langle \, \widetilde W(U ; \Psi_0) \, \rangle \ |_{S^3}} 
\quad .  
\end{equation}
On the one hand, one has 
\beeq
\langle \,  W(C ; \psi_i ) \; \widetilde W(U ; \Psi_0 ) \, \rangle |_{S^3} \; = \; 
 a(k) \; \sum_j \, E_0 [\, j \, ] \; H_{ji} \quad . 
\end{equation}
On the other hand, one gets 
\beeq
\langle \, \widetilde W(U ; \Psi_0) \, \rangle |_{S^3}\; = \; 
 a(k) \; \sum_j \, E^2_0 [\, j \, ]  \quad . 
\end{equation}
As shown in Appendix C, one finds 
\beeq
\sum_j \; E^2_0 [\, j \, ]\; = \; 
\frac{3k^2}{ 256 \, \sin^6
\left( {\pi /k} \right) \, \cos^2 \left({ \pi /k} \right)} \quad , 
\label{8.4}
\end{equation}
\beeq
\sum_j \; E_0 [\, j \, ] \; H_{ji} \; = \; \delta_{\mathbf{1} i} \;  \left [ \, 
\frac{3k^2}{ 256 \, \sin^6
\left( {\pi /k} \right) \, \cos^2 \left({ \pi /k} \right)} \, \right ]  \quad 
, 
\label{8.5}
\end{equation}
where $\psi_1$ denotes the unit element in ${\cal T}_{(k)}$; let us recall 
that the unit element of  ${\cal T}_{(k)}$  was indicated by  $\Psi[0] $  for 
$k=1$ and $k=2$, and by  $\Psi[0,0] $   for $k \geq 3$. From eqs.(\ref{8.4}) 
and (\ref{8.5}) it follows that 
\beeq
\langle \, W(C ; \psi_i ) \, \rangle  |_{S^2\times S^1} \; = \; \delta_{\mathbf{1} i} 
\quad . 
\label{8.6}
\end{equation}
In the next section we shall show that the result in (\ref{8.6}) is not 
peculiar to $SU(3)$ but it holds in general when the reduced tensor algebra
is regular.
Let us now consider the two components link shown in Fig.6.5; the two 
components $C_1$ and
$C_2$ shown in Fig.6.5a have preferred framings and colours $\psi_i$ and  
$\psi_j $ respectively. An
equivalent description of  $C_1$ and $C_2$ in $S^2 \times S^1$ is given in 
Fig.6.5b. 
By using the satellite formulae (\ref{satg}), from eq.(\ref{8.6}) one obtains
\beeq
\langle \, W(C_1, \, C_2 ; \psi_i, \, \psi_j ) \, \rangle  |_{S^2\times S^1} 
\; = \;
 \; N_{ij1} \; = \; \delta_{i j^*} \quad .
\label{8.7} 
\end{equation}
The three components of the link shown in Fig.6.6 have preferred framings and 
colours $\psi_i$, 
$\psi_j$ and $\psi_m$; by using the satellite relations one finds 
\beeq
\langle \, W(C_1, \, C_2, C_3 \, ; \psi_i, \, \psi_j, \psi_m  ) \, \rangle 
 |_{S^2\times S^1}
\; = \; \sum_n \, N_{ijn} \; \delta_{n m^*} \; = \; N_{ijm^*} \quad .
\label{8.8} 
\end{equation}

\begin{figure}[h]
\vskip 0.9 truecm 
\centerline{\epsfig{file=\path 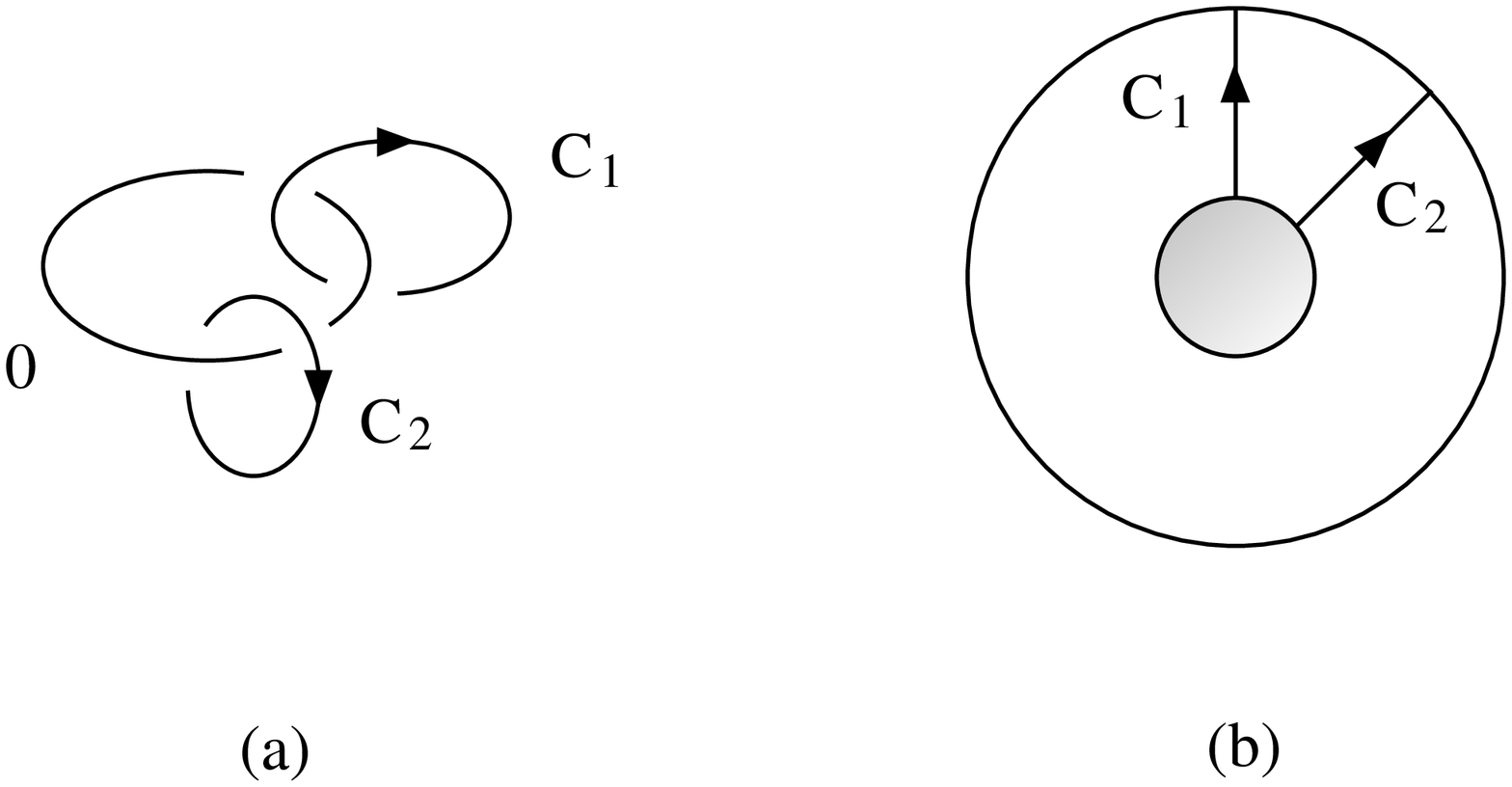,height=5cm,width=9cm}}
\vskip 0.9 truecm 
\centerline {{\bf Figure 6.5}}
\vskip 0.9 truecm 
\end{figure}

\begin{figure}[h]
\vskip 0.9 truecm 
\centerline{\epsfig{file=\path 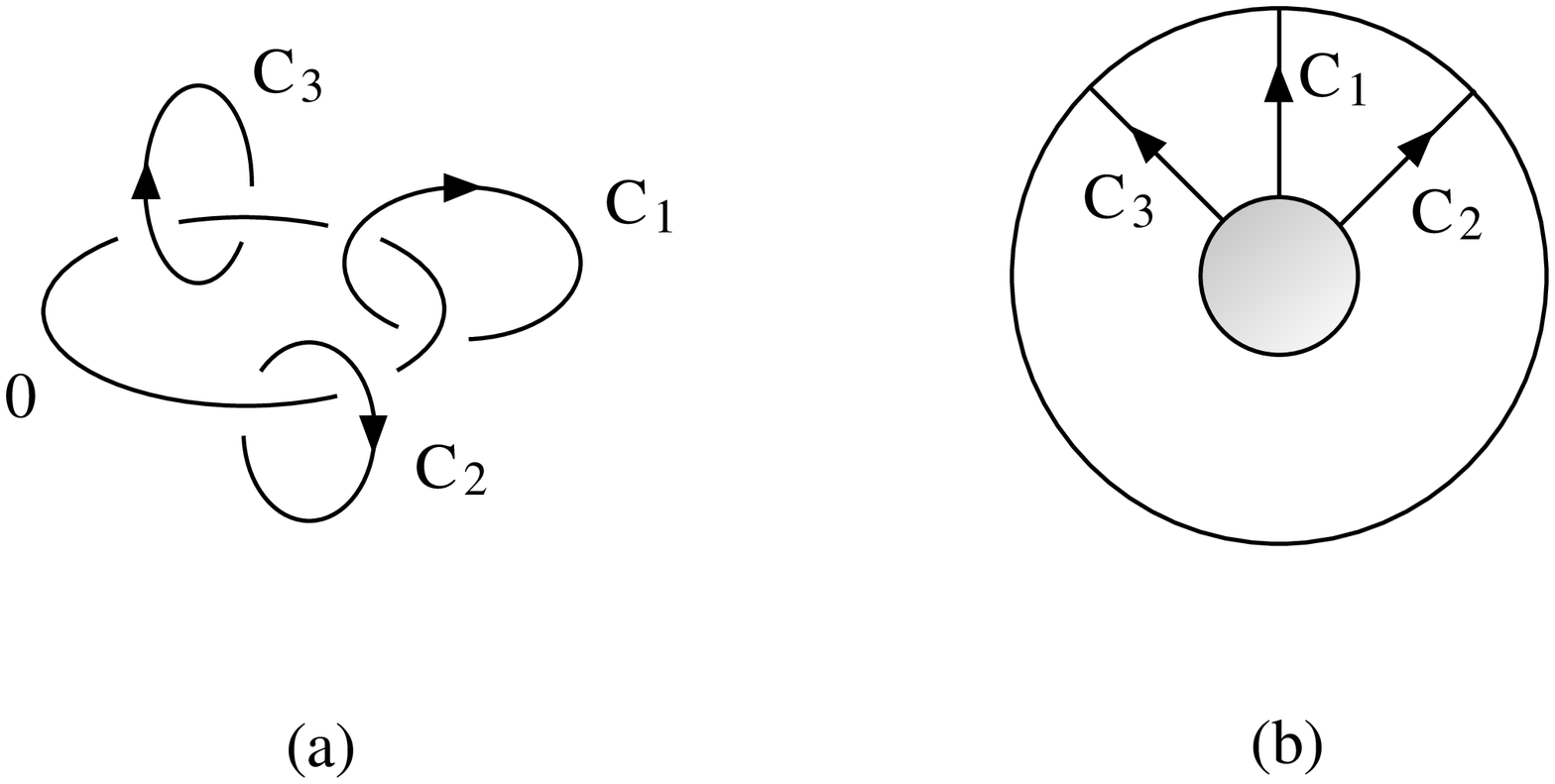,height=5cm,width=9cm}}
\vskip 0.9 truecm 
\centerline {{\bf Figure 6.6}}
\vskip 0.9 truecm 
\end{figure}

The expectation value of the three components link shown in Fig.6.6 provides a
 direct representation 
of the structure constants $\{ \,  N_{ijm^*} \, \}$ of the reduced tensor 
algebra ${\cal T}_{(k)}$ of $SU(3)$. 
Eq.(\ref{8.8}) shows that $ N_{ijm^*} $ is symmetric under a generic 
permutation of the indices; this is
in agreement with eq.(\ref{csim}). 

In order to verify the ambient isotopy invariance of the expectation values 
in  $S^2\times S^1$, let
us consider for example the link shown in Fig.6.7a; the components $C$ and $K$
have preferred framing
and colours  $\psi_i$ and $\psi_j$ respectively.  On the one hand,  according 
to the surgery rule (\ref{6.12}), one obtains   
\beeq
\langle \, W(C, ; \psi_i ) \, W(K; \psi_j ) \, \rangle  |_{S^2\times S^1} \; = \;
E_0 [\, j\, ] \; \delta_{i 1} \quad . 
\label{8.9}
\end{equation}
On the other hand,  by means of an isotopy in $S^2\times S^1$ (see Fig.6.7b)
one can move the knot $K$ inside a three-ball. Consequently,  $ W(K; \psi_j ) 
$ gives the
contribution  $E_0 [\, j\, ]$; thus  
\bea
\langle \, W(C, ; \psi_i ) \, W(K; \psi_j ) \, \rangle \ |_{S^2\times S^1} \; &&= \; 
E_0 [\, j\, ]\; \langle \, W(C, ; \psi_i ) \, \rangle   |_{S^2\times S^1}
\nb \\ 
&&=  \; E_0 [\, j\, ] \; \delta_{i 1} \quad , 
\ena
in agreement with eq.(\ref{8.9}). 

\begin{figure}[h]
\vskip 0.9 truecm 
\centerline{\epsfig{file=\path 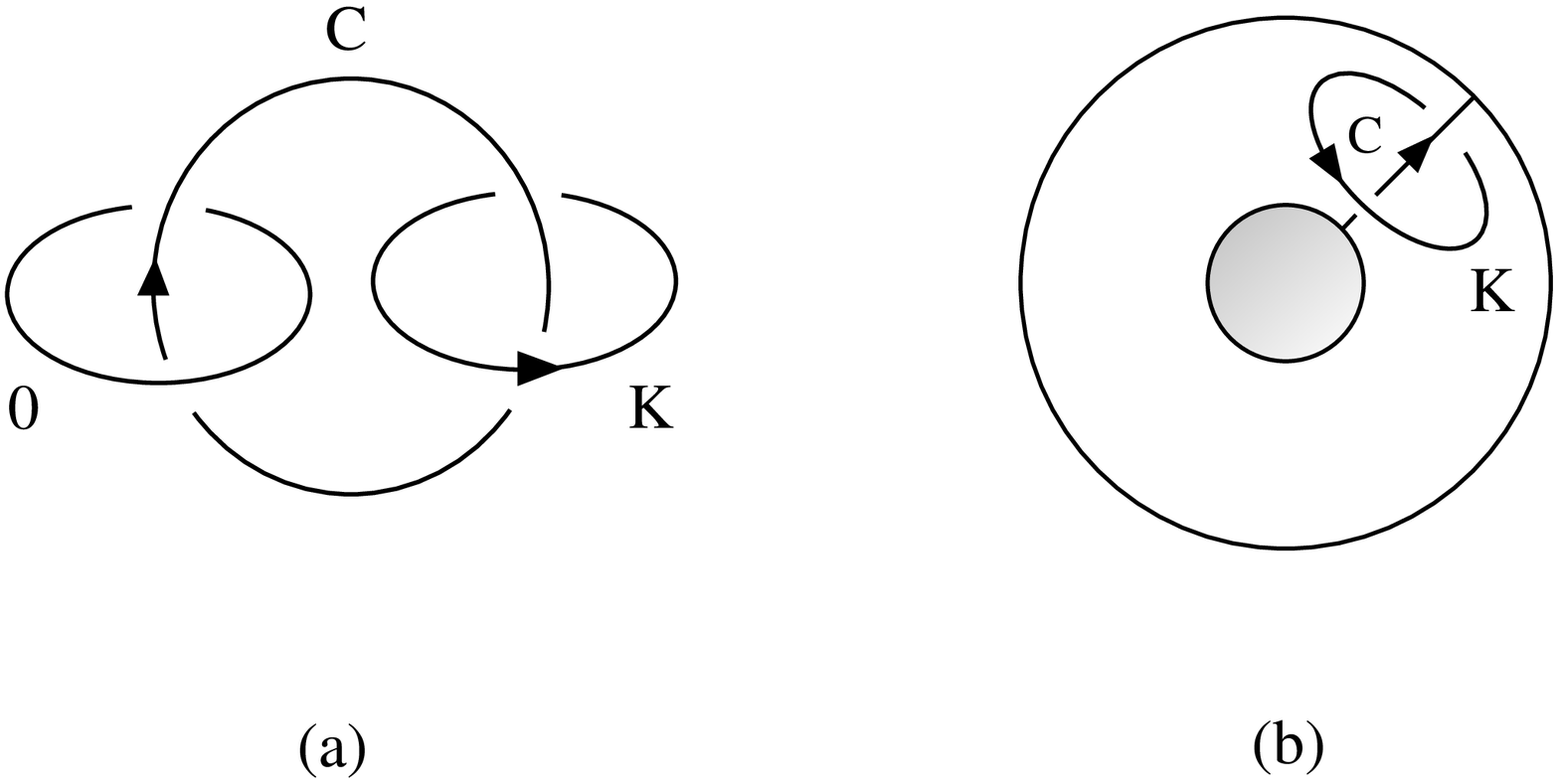,height=5cm,width=9cm}}
\vskip 0.9 truecm 
\centerline {{\bf Figure 6.7}}
\vskip 0.9 truecm 
\end{figure}

The two components $C_1$ and $C_2$ of the link shown in Fig.6.8 have preferred
framings and colours 
$\psi_i$ and $\psi_j$ respectively.  By using the surgery rules, one can 
compute the expectation value
of the associated Wilson line operators; one gets 
\beeq
\langle \, W(C_1, ; \psi_i ) \, W(C_2; \psi_j ) \, \rangle  |_{S^2\times S^1} \; = \; 
 q^{-2 Q(j)} \; \, \delta_{ij^*} \quad . 
\label{8.11}
\end{equation}
The result (\ref{8.11}) can also be derived by using of the invariance under 
Kirby moves. Indeed,
the  sequence of Kirby moves shown in Fig.6.9 and eq.(\ref{8.7}) imply 
\beeq
\langle \, W(C_1, ; \psi_i ) \, W(C_2; \psi_j ) \, \rangle  |_{S^2\times S^1} \; = \; 
q^{- Q(i) - Q(\, j)} \; \, \delta_{i j^*} \quad .  
\end{equation}

\begin{figure}[h]
\vskip 0.9 truecm 
\centerline{\epsfig{file=\path 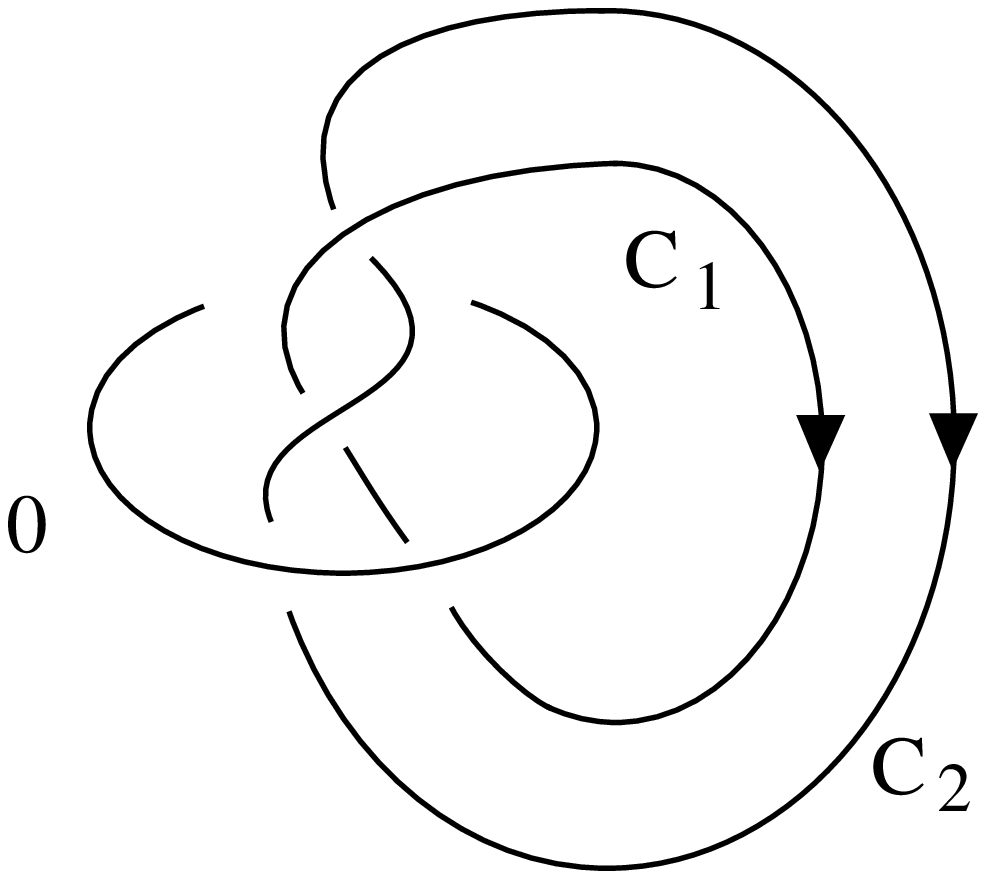,height=4cm,width=6cm}}
\vskip 0.9 truecm 
\centerline {{\bf Figure 6.8}}
\vskip 0.9 truecm 
\end{figure}

\begin{figure}[h]
\vskip 0.9 truecm 
\centerline{\epsfig{file=\path 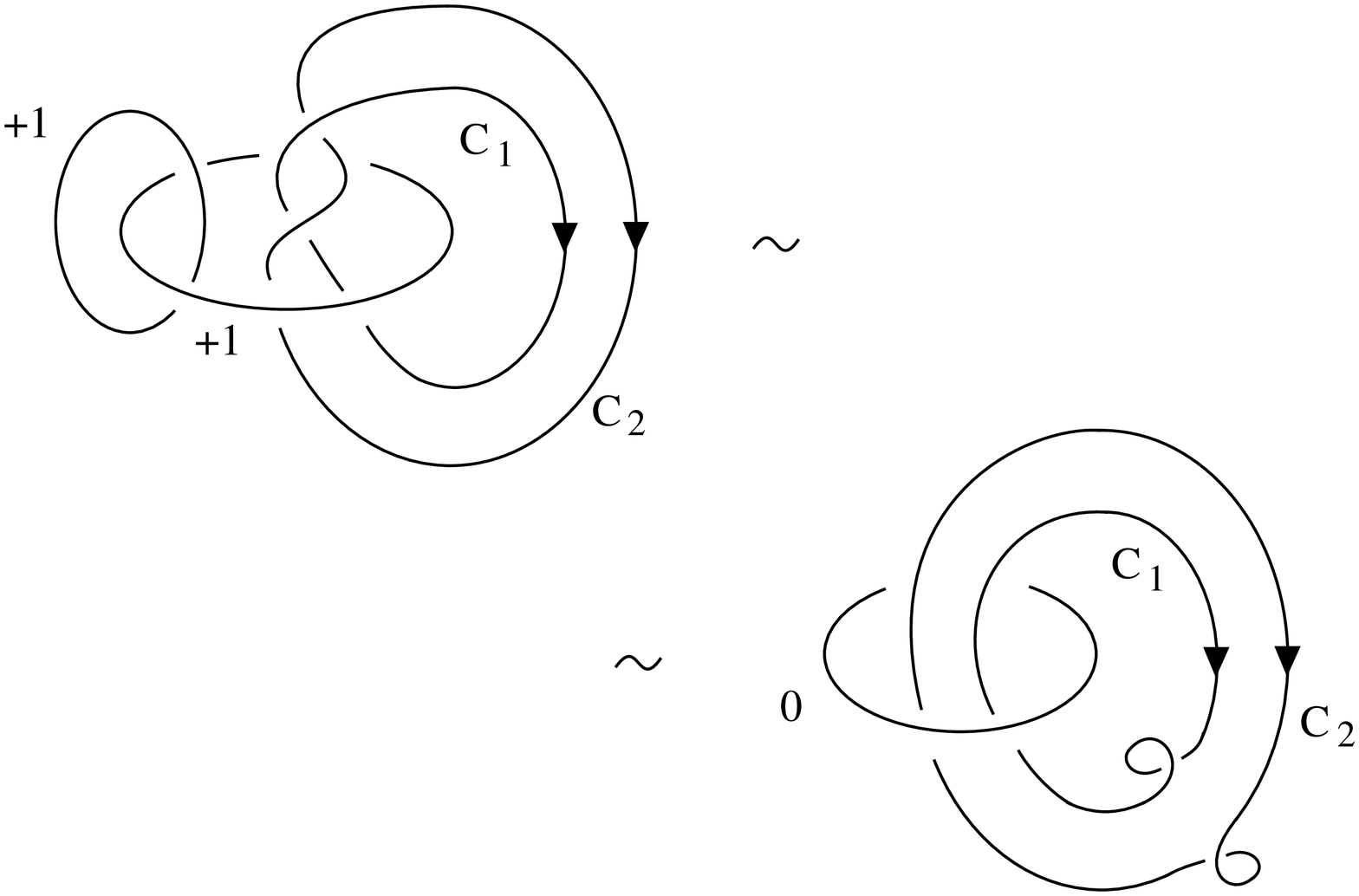,height=7cm,width=9cm}}
\vskip 0.9 truecm 
\centerline {{\bf Figure 6.9}}
\vskip 0.9 truecm 
\end{figure}

The vacuum expectation values of the Wilson line operators in $S^2\times S^1$ can be obtained by
means of a simple general rule. Let us consider the surgery presentation of 
$S^2\times S^1$ given by
the unknot $U$ in $S^3$ with surgery coefficient $r=0$. A generic link in $S^2\times S^1$ can be
represented by a link $L$ in the complement solid torus $N$ of $U$ in $S^3$. 
Within the solid
torus $N$, the associated Wilson line operator $W(L)$ admits the decomposition 
\beeq
W(L) \; = \; \sum_i \, \xi_L (i) \; W(C; \psi_i ) \quad ,  
\label{8.13}
\end{equation}
where $C$ is the oriented and framed core of $N$ shown in Fig.6.4a. Now, by 
using eq.(\ref{8.6}), one
finds 
\beeq
\langle \, W( L ) \,  \rangle  |_{S^2\times S^1} \; = \; \xi_L (1) \quad . 
\label{8.14}
\end{equation}
Since, for any link $L$, the $\xi$-coefficients can be determined uniquely (for example, by means of
the Hopf matrix),  eq.(\ref{8.14}) gives a compact description of 
$\langle \, W( L ) \,  \rangle   |_{S^2\times S^1}$. 

\subsection{\bf Lens spaces and Poincar\'e manifold} 

The  lens space $L(p,1)$ admits \cite{rol} a surgery presentation described by the unknot 
$U$ with surgery
coefficient $r=p$. In order to illustrate the use of the surgery rules, let us consider the case
in which $k=1$. The simplest nontrivial knot $C$ in $L(p,1)$ is shown in 
Fig.6.10; let $C$ have
preferred framing and colour $\psi_i$. Let us recall that ${\cal T}_{(1)}$ is 
of order three and the corresponding Hopf matrix is given in eq.(\ref{runo}). 
From the definition of surgery operator, one has 
\beeq
\langle \, W(C, ; \psi_i ) \,  \rangle  |_{L(p,1)} \; = \; 
\frac{ \sum_j \, q^{p \, Q(\, j)}\; \, E_0[\, j \, ] \; H_{ j i}}{
\sum_j \, q^{p \, Q(\, j)}\; \, E^2_0[\, j \, ] } \quad . 
\end{equation}
\beeq
\langle \, W(C, ; \psi_i ) \,  \rangle  |_{L(p,1)} \; = \; 
\left \{
\begin{array}{cc} 1 & \mbox{if } \psi_i = \Psi [0] \; ; \\ 
\left ( 1 - e^{- i 2 \pi \, p /3} \right ) \left ( 1 + 2 e^{- i 2 \pi \, p /3} \right )^{-1} & \mbox{if } 
\psi_i = \Psi [\pm 1]. \end{array} \right. 
\label{9.2}
\end{equation}
When $k=2$, $\langle \, W(C, ; \psi_i ) \,  \rangle  |_{L(p,1)}$ can be 
obtained by taking the
complex conjugate on the expression (\ref{9.2}). As we have already mentioned,
the case $k=3$ is trivial.
For $k=4$, one gets 
\beeq
\langle \, W(C ; \psi_i ) \,  \rangle  |_{L(p,1)} \; = \; 
\begin{cases}  1 & \mbox{if } \psi_i = \Psi [0] \; ; \\ 
\left ( 1 - e^{- i 2 \pi \, p /3} \right ) \left ( 1 + 2 e^{- i 2 \pi \, p /3}
 \right )^{-1} & \mbox{if } 
\psi_i = \Psi[1,0] \; ;  \\
\left ( 1 - e^{- i 2 \pi \, p /3} \right ) \left ( 1 + 2 e^{- i 2 \pi \, p /3}
\right )^{-1} & \mbox{if } 
\psi_i = \Psi [0,1] \;  . \end{cases} 
\end{equation}

\begin{figure}[h]
\vskip 0.9 truecm 
\centerline{\epsfig{file=\path 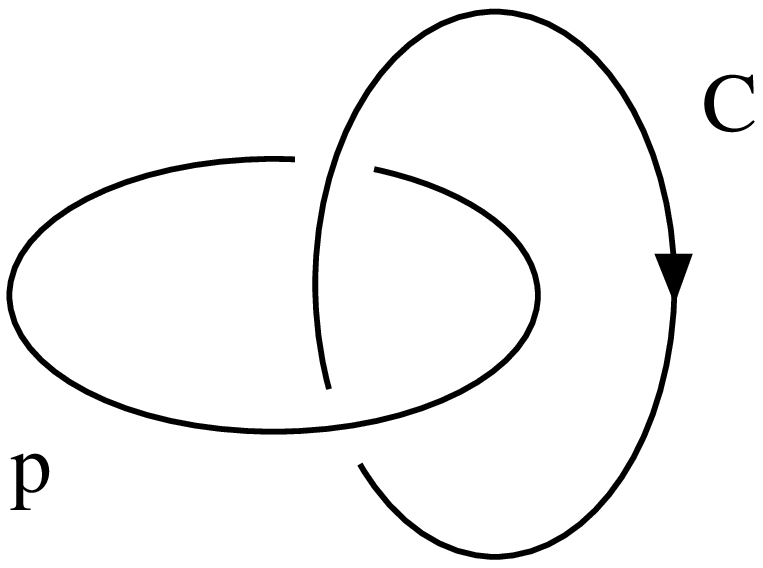,height=3cm,width=5cm}}
\vskip 0.9 truecm 
\centerline {{\bf Figure 6.10}}
\vskip 0.9 truecm 
\end{figure}

\noindent The general expression of $\langle \, W(L ) \,  \rangle  |_{L(p,1)}$
 can be obtained
by using the decomposition (\ref{8.13}) for $W(L)$, where $L$ belongs to the 
complement solid torus of $U$ in $S^3$. Similarly to the case of the manifold 
$S^2 \times S^1$, we only need to consider the knot
$C$ shown in Fig.6.10 with preferred framing and colour state $\psi_i$. 
Let us consider first $\langle \, W(L ) \, W(U, \Psi_0 ) \,  \rangle  |_{S^3}$, where the
unknot $U$ has framing specified by eq.(\ref{6.8}) with $r=p$. By means of 
two Kirby moves, the link shown
in Fig.6.10 can be transformed as shown in Fig.6.11. Therefore, by using 
eq.(\ref{8.8}) and by taking into
account the normalization factor, one finds 
\beeq
\langle \, W(C ; \psi_i) \, W(U, \Psi_0 ) \, \rangle |_{S^3}\; = \; 
a(k)\, \sum_{j m} \, N_{jmi^*} \; E_0[\, j\, ] \, E_0 [ m] \; q^{Q(\, j)} \,  
q^{(p-1) Q(m)} \, q^{-Q(i)} \quad . 
\end{equation}
Let us denote by $Z_{(p)}$ the expectation value 
\beeq
Z_{(p)} \; = \; \langle \, W(U, \Psi_0 \rangle ) \, |_{S^3}\;   
= \; a(k) \, \sum_i \, q^{p Q(i)}\; E_0^2 [i] \quad .  
\end{equation}
For $Z_{(p)} \not= 0$, the expectation value $\langle \, W(L ) \,  \rangle |_{L(p,1)}$ is
given by 
\beeq
\langle \, W(L ) \,  \rangle  |_{L(p,1)} \; = \;  Z^{-1}_{(p)} \; 
 \sum_{j m i} \, \xi_L (i) \; N_{jmi^*} \; E_0[\, j\, ] \, E_0 [ m] \; q^{Q(\, j)} \,  
q^{(p-1) Q(m)} \, q^{-Q(i)} \quad . 
\end{equation}

\begin{figure}[h]
\vskip 0.9 truecm 
\centerline{\epsfig{file=\path 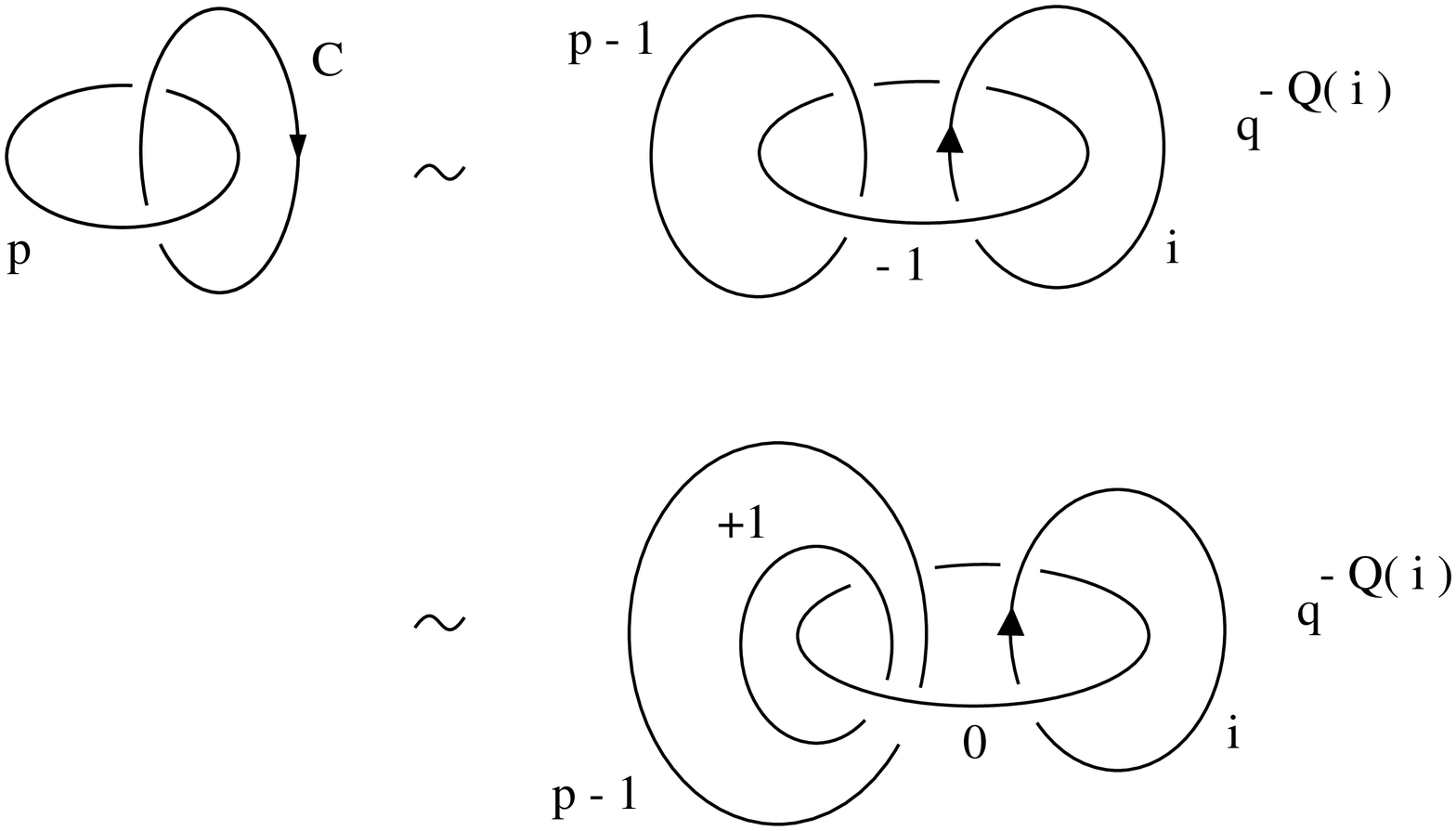,height=6cm,width=10cm}}
\vskip 0.9 truecm 
\centerline {{\bf Figure 6.11}}
\vskip 0.9 truecm 
\end{figure}

The last example of this section is the Poincar\'e manifold ${\cal P}$. This 
manifold is a homology
sphere but not a homotopy sphere. A surgery presentation of ${\cal P}$ is 
given by the right-handed trefoil knot $T$ in $S^3$ with surgery coefficient 
$r=1$ (and framing specified by
eq.(\ref{6.8})).  The knot $C \subset {\cal P}$  shown in Fig.6.11 has 
preferred framing. For
simplicity, we shall concentrate on  the case $k=1$.  By using the result 
(\ref{eq139}) of Chap.4, one obtains  
\beeq
\langle \, W(C ; \psi_i) \, W(T, \Psi_0 ) \, \rangle   |_{S^3}\; = \; 
\left \{ \begin{array}{cc}  - \, i & \text{for } \psi_i = \Psi[0] \; ; \\ 
(\sqrt 3 + i)/2 & \text{for }  \psi_i = \Psi [ \pm 1] \; . \end{array}
\right.
\end{equation}
Therefore, 
\beeq
\langle \, W(C ; \psi_i) \, \rangle   |_{\cal P}\; = \; 
\begin{cases}
1 & \text{for } \psi_i = \Psi [0] \; ; \\
( i \sqrt 3 - 1 )/2 & \text{for }  \psi_i = \Psi [ \pm 1] . \end{cases} 
\label{9.8}
\end{equation}
It is clear that the corresponding expressions for $k=2$ can be obtained from 
(\ref{9.8}) by taking the
complex conjugate and the results for $k=4$ coincide with (\ref{9.8}). For 
higher values of $k$, the
computation of $\langle \, W(L) \, \rangle   |_{\cal P}$ is 
straightforward. 

\begin{figure}[h]
\vskip 0.9 truecm 
\centerline{\epsfig{file=\path 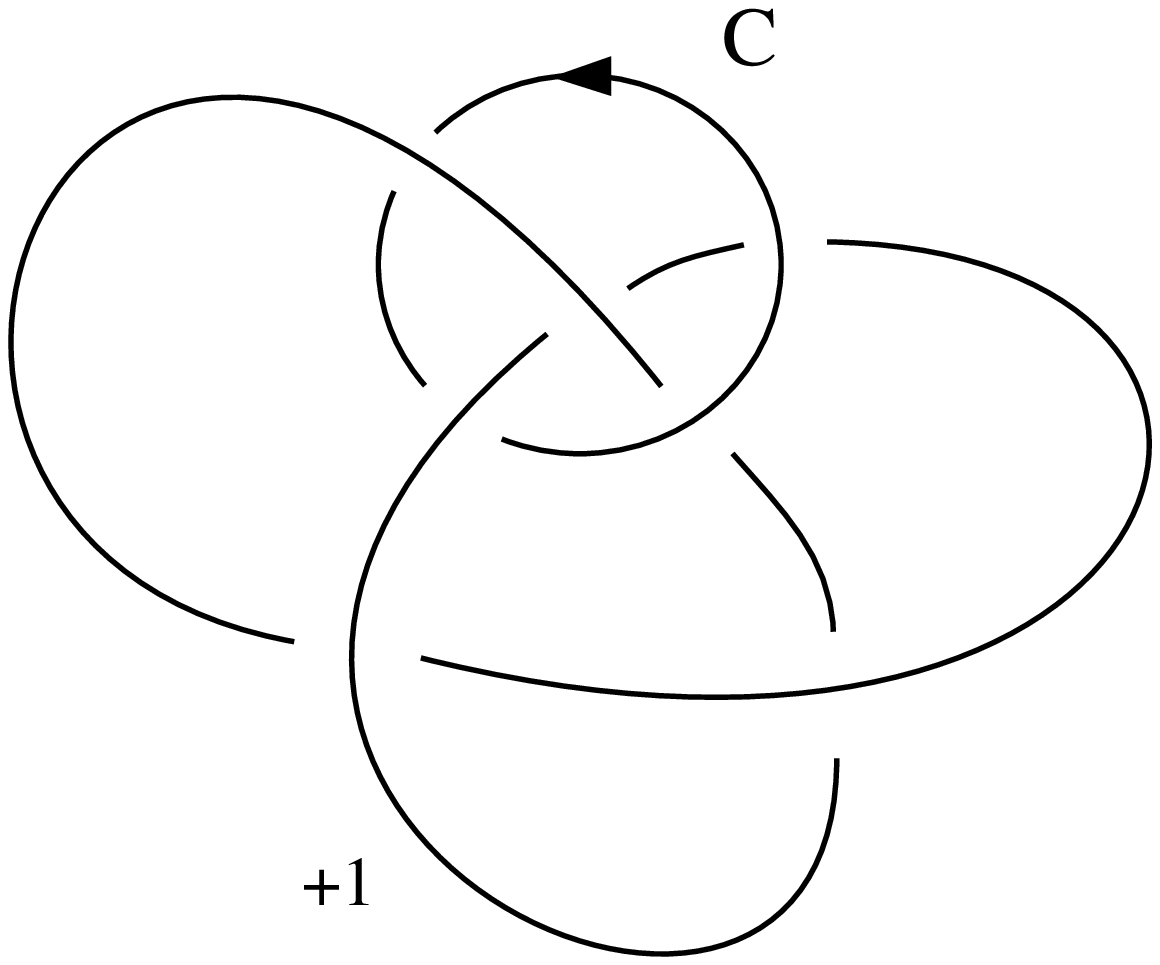,height=6cm,width=7cm}}
\vskip 0.9 truecm 
\centerline {{\bf Figure 6.12}}
\vskip 0.9 truecm 
\end{figure}

\subsection{{\bf Manifolds} $\mathbf{\boldsymbol{\Sigma}_g} {\boldsymbol 
\times} \mathbf{S^1}$} 

In this section we shall consider the set of manifolds $\Sigma_g \times S^1$, 
where 
$\Sigma_g $ is a Riemann surface of genus $g$.  Let us denote by ${\cal L} (g)
$ a surgery link
in $S^3$ corresponding to a surgery presentation of $\Sigma_g \times S^1$. 
We shall describe firstly how to construct ${\cal L} (g)$ for arbitrary $g$. 
Then, we shall
consider examples of links in  $\Sigma_g \times S^1$. 

The manifold $\Sigma_g $ can be obtained from
the two-sphere $S^2$ by adding $g$ handles, of course. Similarly, the 3-manifold 
$\Sigma_g \times S^1$ can be obtained from $S^2 \times S^1$ by ``adding $g$ handles" 
according to the prescription described in \cite{guad4}. 
Consider the link $L(g)$ in $S^3$ shown in Fig.6.13, where the component $U$ 
and the $g$ components
$\{ \, M_1 , ..., M_g \, \}$ have preferred framings. The surgery link ${\cal L} (g)$ is a satellite
of $L(g)$ which is obtained  by replacing each component $M_i$ (with $ 1 \leq i \leq g$ ) with
$h^\diamond (P)$. The homeomorphism  $h^\diamond$ has been defined in Sect2.2.
The pattern link
$P$, which is contained in the complement solid torus $N$ of the unknot $V$ in $S^3$, is shown in
Fig.6.14. The satellite  obtained according to this prescription is the  link ${\cal L} (g)$ in $S^3$
which has $2g+1$ components.  Each component of ${\cal L} (g)$ has  preferred framing and,
consequently, the associated surgery coefficient is $r=0$. 

\begin{figure}[h]
\vskip 0.9 truecm 
\centerline{\epsfig{file=\path 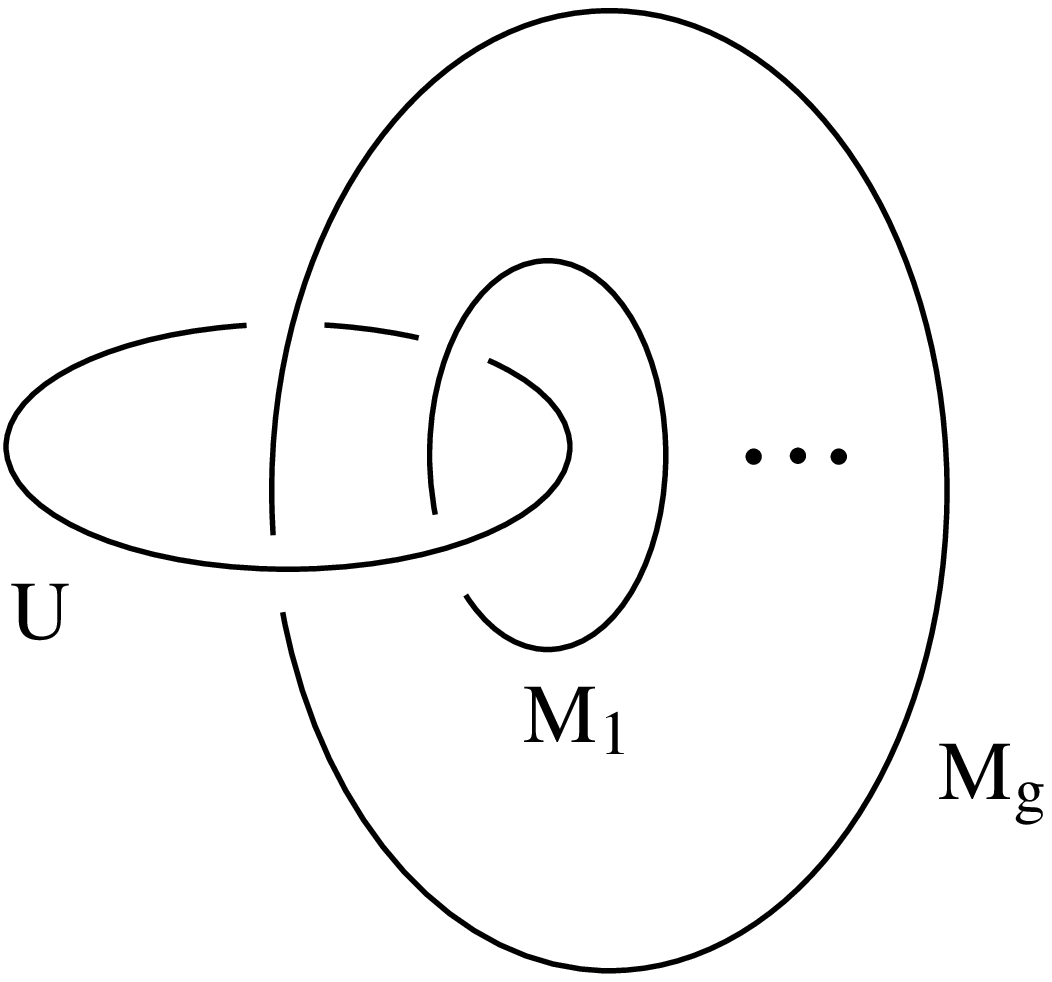,height=4cm,width=6cm}}
\vskip 0.9 truecm 
\centerline {{\bf Figure 6.13}}
\vskip 0.9 truecm 
\end{figure}

\begin{figure}[h]
\vskip 0.9 truecm 
\centerline{\epsfig{file=\path 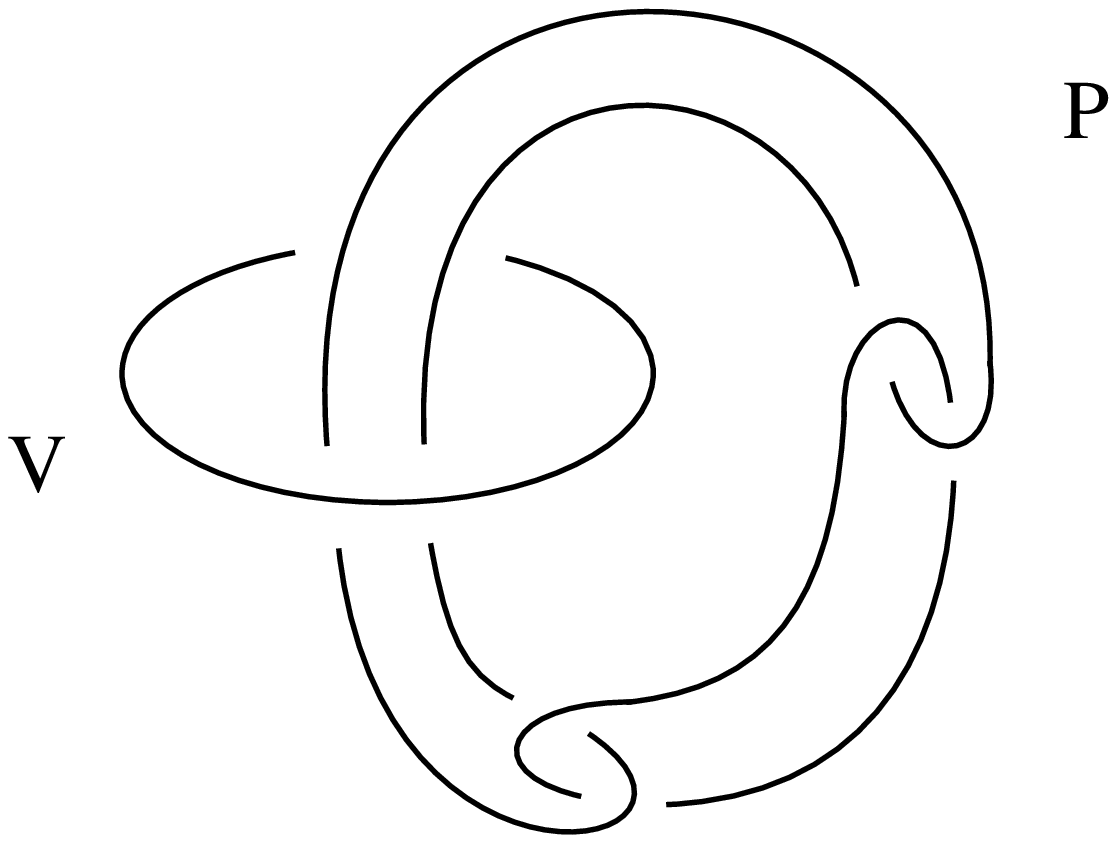,height=4cm,width=6cm}}
\vskip 0.9 truecm 
\centerline {{\bf Figure 6.14}}
\vskip 0.9 truecm 
\end{figure}

\noindent For example, the
surgery link ${\cal L}(1)$ is shown in Fig.6.15; ${\cal L}(1)$ is ambient 
isotopic with the Borromean rings and corresponds to the manifold 
$\Sigma_g \times S^1 \equiv T^2 \times S^1$. 

\begin{figure}[h]
\vskip 0.9 truecm 
\centerline{\epsfig{file=\path 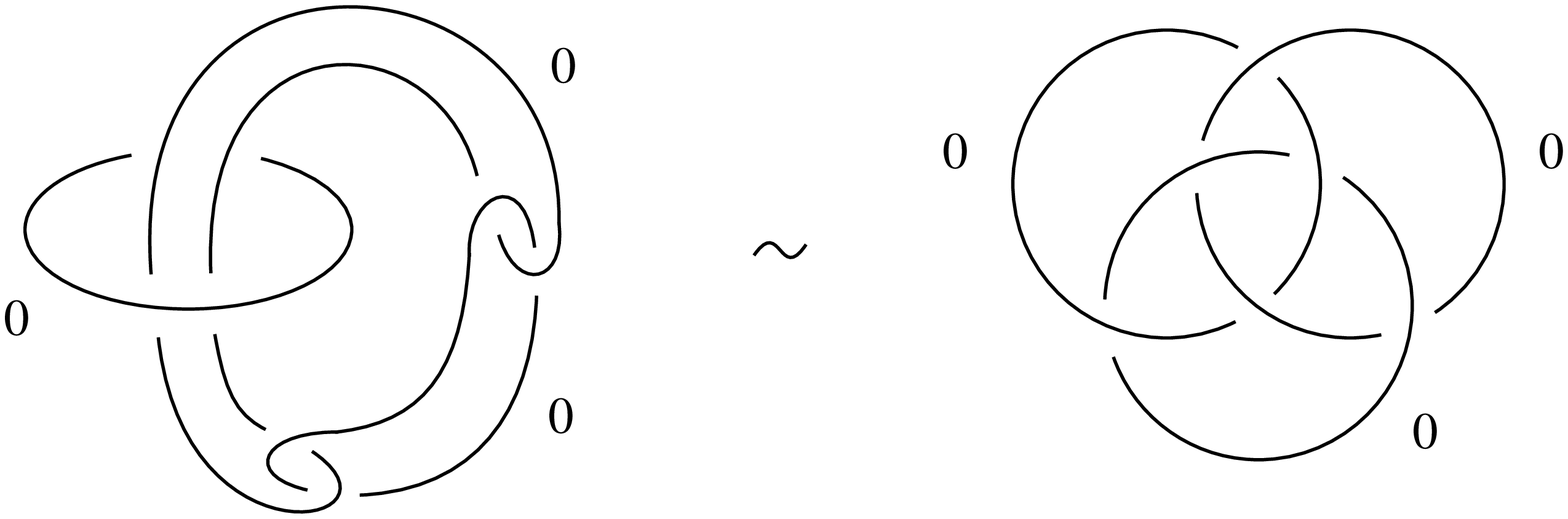,height=4cm,width=9cm}}
\vskip 0.9 truecm 
\centerline {{\bf Figure 6.15}}
\vskip 0.9 truecm 
\end{figure}

\noindent Suppose that both components of the pattern link $P$ have colour 
$\Psi_0$.   
We shall denote by $W(P; \Psi_0 , \Psi_0 )$ the product of the associated 
Wilson line
operators. Since $P$ is defined in $N$,  $W(P; \Psi_0 , \Psi_0 )$ is a
gauge invariant observable defined inside a solid torus. Consequently it 
admits a decomposition of the type \cite{gp1} see Sect.4. 
\beeq
W(P; \Psi_0 , \Psi_0 )\; = \; \sum_i \, \eta (i) \; W(K ; \psi_i ) \quad , 
\label{10.1}
\end{equation}
where $K$ is the core of $N$ with preferred framing. One
finds 
\beeq
\eta(i) \; = \; \sum_j \, E_0^{-1}[\, j \, ] \; H_{ji} \quad . 
\label{10.2}
\end{equation}
Details of the derivation of (\ref{10.2}) will be given in Chap.8.
Eq.(\ref{10.2}) is valid for any group $G$. Eq.(\ref{10.1}) gives the 
decomposition of the surgery operator, which
is associated  with ``one handle", in terms of a single coloured link component.
The above decomposition
will be useful in the computation of expectation values of Wilson line operators in $\Sigma_g \times S^1$.

Let consider for example the case $k=5$; the reduced tensor algebra ${\cal T}_{(5)}$ is of order six
and the elements of its standard basis are \cite{gp1}
\beeq
\{ \, \Psi [0,0], \, \Psi [1,0], \, \Psi [2,0], \, \Psi [0,1], \, \Psi [0,2],
\, \Psi [1,1] \, \} \quad .
\end{equation}

The nontrivial structure constants are given by 
\beeq
\Psi [1,0] \; \Psi [1,0] \; = \; \Psi [2,0] \; + \; \Psi [0,1] \qquad ,\qquad
\Psi [1,0] \; \Psi [0,1] \; = \; \Psi [0,0] \; + \; \Psi [1,1] \quad ,
\end{equation}
\beeq
\Psi [1,0] \; \Psi [2,0] \; = \; \Psi [1,1] \qquad ,\qquad 
\Psi [1,0] \; \Psi [0,2] \; = \; \Psi [0,1]  \; + \; \Psi [1,1] \quad ,
\end{equation}
\beeq
\Psi [1,0] \; \Psi [1,1] \; = \; \Psi [1,0]  \; + \; \Psi [0,2] \qquad , \qquad
\Psi [0,1] \; \Psi [0,1] \; = \; \Psi [0,2]  \; + \; \Psi [1,0] \quad  ,
\end{equation}
\beeq
\Psi [0,1] \; \Psi [2,0] \; = \; \Psi [1,0]  \; + \; \Psi [1,1] \qquad  , \qquad
\Psi [0,1] \; \Psi [0,2] \; =  \; \Psi [1,1] \quad  ,
\end{equation}
\beeq
\Psi [0,1] \; \Psi [1,1] \; = \; \Psi [0,1]  \; + \; \Psi [2,0] \qquad  , \qquad
\Psi [2,0] \; \Psi [2,0] \; = \; \Psi [0,2]   \quad  ,
\end{equation}
\beeq
\Psi [0,2] \; \Psi [0,2] \; = \; \Psi [2,0] \qquad  , \qquad
\Psi [0,2] \; \Psi [1,1] \; = \;  \Psi [1,0] \quad  ,
\end{equation}
\beeq
\Psi [1,1] \; \Psi [1,1] \; = \; \Psi [1,1]  \; + \; \Psi [0,0] \quad  .
\label{10.4}
\end{equation}
The non-vanishing $\eta$-coefficients for  $k=5$ are
\beeq
 \eta [0,0] \; = \; 6 \qquad , \qquad \eta [1,1] \; = \; 3 \quad .
\label{10.5}
\end{equation}
By using the decomposition (\ref{10.1}), one finds
\beeq
\langle \, \widetilde W({\cal L} (1) ) \, \rangle   |_{S^3 } \; = \; 6 \, \sqrt {3 (\sqrt 5 +1)
/2 } \quad .
\end{equation}
Let us now consider the knot $C \subset T^2  \times S^1$ shown in Fig.6.16;  
when $C$ has
preferred framing and colour $ \psi_i$, one gets
\begin{equation}
\langle \, \widetilde W( C; \, \psi_i ) \, \rangle   |_{T^2 \times S^1 } \; = \; 
\begin{cases}    1 & \text{for } \psi_i \; = \; \Psi [0,0]  \; ;
\\ 
 1/2 & \text{for }    \psi_i \; = \; \Psi [1,1]  \; . 
\end{cases}
\end{equation}
In a generic manifold of the type $\Sigma_g \times S^1$, we can interpret 
$S^1$ as a compactified ``time" interval, $\Sigma_g $ being the space-like surface. According to this 
interpretation, the
knot $C\subset T^2  \times S^1$ shown in Fig.9.16 describes the static 
situation in which a single
coloured puncture is present in $T^2$. 

Consider now the static case in which two coloured
punctures are present on $T^2$. Let $C_1$ and $C_2$ be the knots in $T^2  \times S^1$ which
describe these two punctures. From eq.(\ref{10.5}) it follows that 
\beeq
\langle \, \widetilde W( C_1; \, \psi_i ) \,  W( C_2; \, \psi_j ) \, \rangle   |_{T^2 \times S^1 }
\; = \; \delta_{i j^*} \; + \; \frac{1}{2} \, N_{ij \, [1,1]} \quad .
\label{10.8}
\end{equation}
It is clear that eq.(\ref{10.8}) can easily be generalized to the case in 
which several punctures are present
in $T^2$. Indeed, by means of the satellite formulae, we can replace the link 
associated with these 
punctures by a single knot. The same method can also be used to analyze the 
situation in which
the genus (number of handles) is greater than one. In fact, according to eq.(10.1), adding one
handle is equivalent to the introduction of  a puncture with colour $\psi_{(h)} \; = \; \sum_i \eta(i)
\psi_i$. For example, let us consider $\Sigma_2 \times S^1$ with one 
puncture having colour
$\psi_i$. By using  eqs.(\ref{10.5})  and (\ref{10.5}) one obtains
\beeq
\langle \, \widetilde W( C; \, \psi_i )  \, \rangle  |_{\Sigma_2 \times S^1 }
\; = \; \delta_{i\, [0,0]} \; + \; \delta_{i \, [1,1]} \quad .
\end{equation}

\begin{figure}[h]
\vskip 0.9 truecm 
\centerline{\epsfig{file=\path 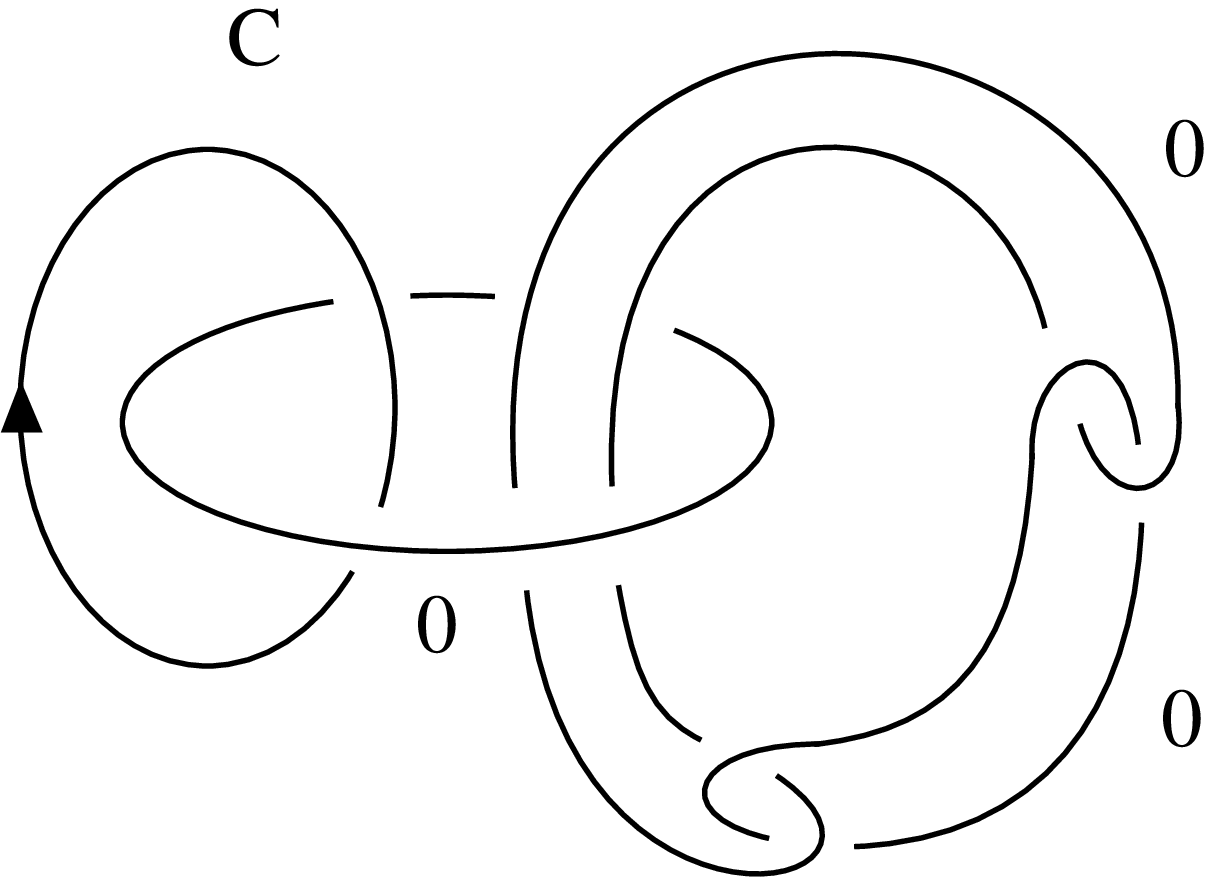,height=4cm,width=6cm}}
\vskip 0.9 truecm 
\centerline {{\bf Figure 6.16}}
\vskip 0.9 truecm 
\end{figure}

\section{\bf Properties of the Surgery}

After we have given some example of calculation of observables of $SU(3)$ 
CS theory in manifolds different form $S^3$, in this section we shall come
back to general properties of surgery. 

Let us consider the surgery instruction given by the unknot $U$ in $S^3$ 
with surgery coefficient $r = 0$. The corresponding surgery operation $S_0$ in
$S^3$ is the result of the following instructions \cite{rol}:
\begin{enumerate}
\item fix a tubular neighborhood $N$ of a unknot $U$ and remove the interior 
$\dot{N}$ of $N$ from $S^3$,  
\item consider $N$ and $S^3 - \dot{N}$ as distinct spaces, 
\item sew back $N$ with $S^3 - \dot{N}$ by identifying their boundaries according to
a homeomorphism $h \, : \, \partial N \ra \partial (S^3 - \dot{N} )$ which sends a meridian of $ N$ into a meridian of $ S^3 -\dot{N} $.
\end{enumerate}

\noindent The manifold corresponding to the surgery instruction given by the 
unknot $U$ in $S^3$  with surgery coefficient $r = 0$ 
is homeomorphic with $S^2 \times S^1$. 

We shall now prove \cite{gp3} two remarkable properties of the operator corresponding to the $S_0$ surgery. 
Let us consider the link in $S^3$ depicted in Fig.6.17 in which the component $U$
represents a surgery instruction with surgery coefficient $r=0$. This link 
gives a surgery description a knot $C$ in $S^2 \times S^1$ with nontrivial 
homotopy class. For the observable associated with $C$ in  $S^2 \times S^1$ 
the following property holds.  

\begin{figure}[h]
\vskip 0.9 truecm 
\centerline{\epsfig{file=\path 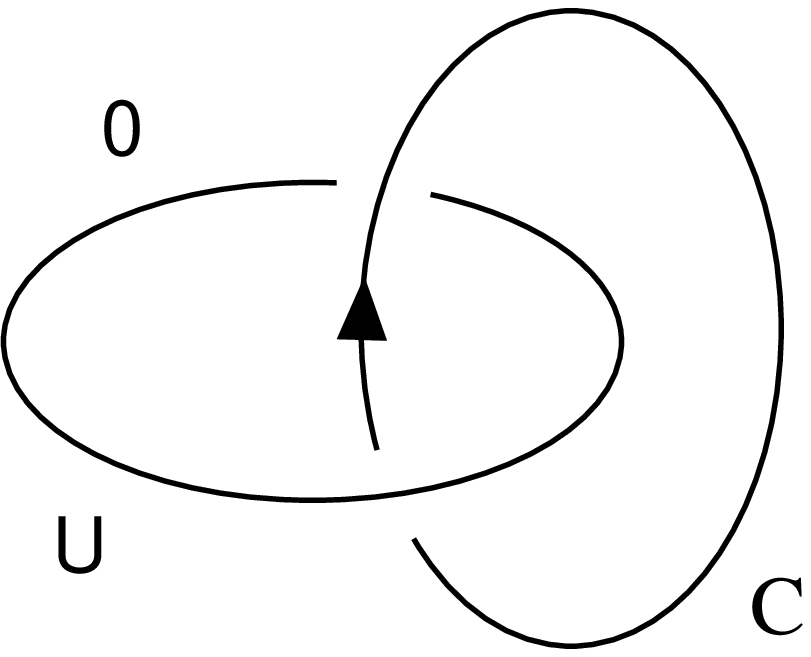,height=4cm,width=6cm}}
\vskip 0.9 truecm 
\centerline {{\bf Figure 6.17}}
\vskip 0.9 truecm 
\end{figure}

\bigskip

\shabox{
{\bf Theorem 6.5}}{\em ~The expectation value in $S^2 \times S^1$ of the Wilson line associated 
with the link $C$, shown in Fig.6.17, with colour $\psi[j]$ is} 
\beeq 
 \langle \, W(\, C; \; \psi[j] \, )  \, \rangle  \bigr |_{S^2 \times S^1}
\; = \; \frac{\langle \, W(\, U \, ; \, \Psi_0 \, ) \, W( \, C \, ;  \, \psi[j] \, )  \, \rangle 
\bigr |_{S^3}} {\langle \, W(\, U ; \; \Psi_0 \, )  \, \rangle  \bigr |_{S^3}} \; = \; \delta_{
1 \, j} \quad ,  
\end{equation}
{\em where ${\psi [1]} $ is the class defined by the one
dimensional trivial representation.}

\bigskip{ \bf ~Proof}
First of all, since 
\beeq
\langle \, W(\, U ; \; \Psi_0 \, )  \, \rangle  \bigr |_{S^3} \; = \; a(k)
\, \sum_m E^2_0[m] \; = \; a(k)^{-1} \; > \;  0 \quad ,  
\label{v42}
\end{equation}
by definition of the regular reduced tensor algebra, expression (\ref{v42}) 
is well defined.
Let us now consider the expectation value of the Wilson line associated with the
link shown in Fig.6.18. 

\begin{figure}[h]
\begin{picture}(10,10)
\put(225,-110){$\psi[i]$}
\put(280,-130){$\psi[j]$}
\end{picture}
\vskip 0.5 truecm 
\centerline{\epsfig{file=\path 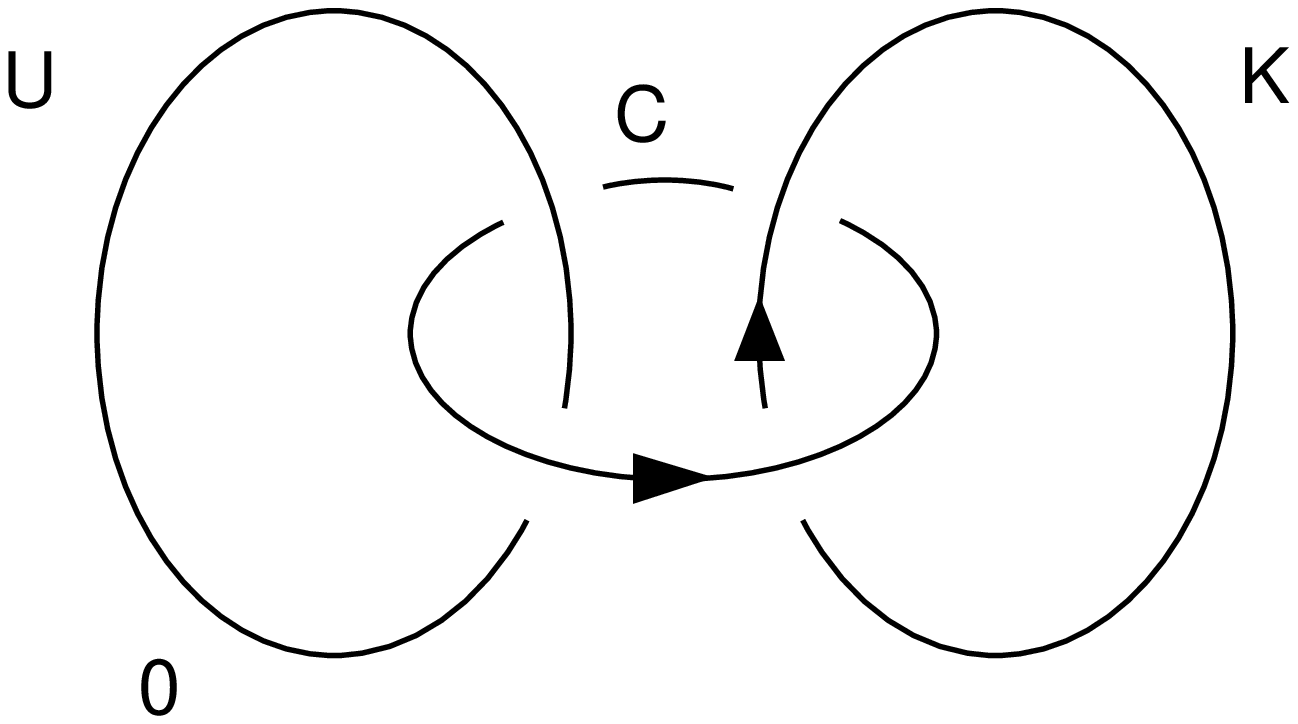,height=4cm,width=8cm}}
\vskip 0.5truecm 
\centerline{{\bf Figure 6.18}}
\end{figure}
 
On the one hand, by using the definition of the surgery operation $S_0$ one
finds 
\beeq 
\langle \, W(\, U,C,K; \, \Psi_0, \psi[j], \psi[i] \, )  \, \rangle  \bigr |_{S^3} \;
= \; E_0[i] \; \langle \, W(\, U, C ; \, \Psi_0 , \psi[j] \, )  \, \rangle  \bigr|_{S^3} \; .
\label{eq:vdue} 
\end{equation}
Indeed, since we are dealing with a topological field theory, by means of an isotopy we can move the
knot $K$  into the interior of the tubular neighborhood $N$ of $U$ which is involved in
the surgery operation $S_0$. By using the homeomorphism $h$ entering the definition of $S_0$, $K$
becomes homotopically trivial in $N$ and its  expectation value factorizes \cite{gp2} as we have seen in Sect.6.6. 
On the other hand, by using the
connected sum rule  one has 
\beeq 
\langle \, W(\, U,C,K; \, \Psi_0, \psi[j], \psi[i] \, )  \,
\rangle  \bigr |_{S^3} \; = \; \left( \, E_0[j] \, \right)^{-1} \; H[i,j] \; \langle \, W(\, U, C; \,
\Psi_0,\psi[j] \, )  \, \rangle   \bigr|_{S^3} \; .
\label{eq:vtre} 
\end{equation}
Since equations (\ref{eq:vdue}) and (\ref{eq:vtre}) are both satisfied, by setting 
\beeq
{\cal Z}(j) \; = \; \langle \, W(\, U \, ; \, \Psi_0 \, ) \, W( \, C \, ;  \, \psi[j] \, )  \, 
\rangle \bigr |_{S^3}
\quad ,
\end{equation} 
one finds  
\beeq
\left( \, E_0[i] \; E_0[j] \; - \; H[i,j] \, \right) \; {\cal Z}(j)  \; = \; 0
\qquad \;\; \forall \; \psi [i] \, , \, \psi[j] \, \in {\cal T}_{(k)} \; \; .
\label{eq:vcinq} 
\end{equation}
We now show that eq.(\ref{eq:vcinq}) implies that ${\cal Z}(j) = 0  \; \forall j \neq 1$. 

\no 
When $\psi[j] = \psi[1]$ the equality (\ref{eq:vcinq}) trivially holds. Indeed $E_0[1] \; = \;1$ and
$H[i,1] \; = \; E_0[i]$. Assume now that there is an element $\psi[m]$ of the standard basis, with
$\psi[m] \neq \psi[1]$, for which (\ref{eq:vcinq}) is satisfied with $j=m$ and ${\cal Z}(m) \neq 0$;
we need to  prove that, in  this case, one gets a contradiction. 
Indeed, from (\ref{eq:vcinq}) it follows that
\beeq
H[i, m] \; = \; E_0[i] \; E_0[ m] \qquad \; \; \; \forall \, \psi [i] \, \in \, 
{\cal T}_{(k)} \; .
\end{equation}
Let us consider a generic link $L_{\varphi}$ in $S^3 $ with one of its components
$A$ coloured with  $\varphi \in {\cal T}_{(k)}$. By using the surgery 
operators, one can find a surgery presentation \cite{rol} of $L_{\varphi}
\subset S^3$ in which $A$ is the unknot with colour  $\varphi$ and the remaining
components of $L_{\varphi}$  and the surgery link are contained in the
complement solid torus of $A$ in $S^3$. Let us denote by $B$ the core of the
complement solid torus of the unknot $A$ in $S^3$. By using the decomposition 
(\ref{eq:gid}), we obtain  \beeq  \langle \,  W(\, L_{\varphi} \,)  \, \rangle 
\bigr|_{S^3} \; = \; \sum_i \; \xi(i)  \;  \langle \,  W(\, A,B; \;
\varphi;\psi[i] \,)  \, \rangle  \bigr|_{S^3} \; . \end{equation}
The union of knots $A$ and $B$ is simply the  Hopf link in $S^3$. In general, $\varphi$
will be a linear combination of the standard basis of ${\cal T}_{(k)}$
\beeq
\varphi \; = \; \sum_n \; a_n \; \psi [n] \; . 
\end{equation}
Thus 
\beeq
 \langle \,  W(\, L_{\varphi} \,)  \, \rangle  \bigr|_{S^3} \; = \; \sum_{i,n} \; \xi(i)
\; a_n \; H[i,n] \; . \label{eq:33}
\end{equation}
If we choose $\varphi$ to be
\beeq
\varphi \; = \; \psi[m] \; - \; \psi[ 1] \; E_0[m] \; ;
\end{equation}
equation (\ref{eq:33}) becomes
\bea
 \langle \,  W(\, L_{\varphi}; \; \varphi \,)  \, \rangle  \bigr|_{S^3} \; && = \; \sum_i
\xi(i) \; \left( \, H[i,m] \; - \; E_0[m] \; E_0[i] \, \right) \nb \\
&& =\; \sum_i \xi(i) \; \left(\;  E_0[m] \; E_0[i] \; - \;  E_0[m] \;
E_0[i] \, \right) \nb \\
&&  = \; 0 \qquad .
\label{eq:tre} \ena
Equation (\ref{eq:tre}) means that $\varphi$ is physically equivalent to the null
element. Therefore,  $\varphi$ is the null element because ${\cal T}_{(k)}$ is 
(by definition) physically irreducible. This implies that $\psi[m] =
\psi[1]$. Thus, we have reached a contradiction. 

At this point, we have shown that 
\beeq
{\cal Z}(j) \; = \; f(k) \; \delta_{ 1\, j} \; .
\end{equation}
The normalization factor $f(k)$ can easily be obtained from the definition of
${\cal Z}(j)$. Indeed
\beeq
f(k) \; = \; {\cal Z}(\, j\, = \,  1 \, ) \; = \;  \langle \,   
W(\, U,K; \; \Psi_0 , \psi[ 1]\, ) \, \rangle    \bigr|_{S^3} \; = \;   \langle
\,  W(\, U; \; \Psi_0 \, )  \, \rangle   \bigr|_{S^3} \; = \; {a(k)}^{-1} \; \; . 
\end{equation}
Thus
\beeq 
\langle \,   W(\, C \, ; \, \psi[i] \, )  \, \rangle  \bigr |_{S^2 \times S^1}
\; = \; \frac{  \langle \,    \, W(\, U, \, C ; \Psi_0, \, \psi[i] \, )  \, \rangle  \bigr
|_{S^3}} {  \langle \,   \, W(\, U ; \Psi_0 \, )  \, \rangle  \bigr |_{S^3}} \; = \;
\delta_{  \mathbf{1} \, i} \; .  
\end{equation}
{\hfill \ding{111}}

\bigskip

\no
It should to be noted that Theorem 6.5 has a topological origin but is also 
strictly connected with the peculiar role played by the reduced tensor algebra
in the CS theory. Theorem 6.5 can be understood \cite{guad3} to be a 
consequence of the conservation of the non-Abelian electric flux originated by
 a static source in a
closed universe described by the 3-manifold $S^2 \times S^1$. 

We conclude this section by showing how the structure constants of 
${\cal T}_{(k)}$ can be expressed in terms of the CS gauge invariant 
observables. This will permit us to interpret the equations  
involving $\{ N_{ijm} \}$ in terms of relations between observables.

\bigskip 

\shabox{
\no
{\bf Theorem 6.6}}~{\em The structure constants  of ${\cal T}_{(k)}$ can be 
represented by the expectation value in $S^2 \times S^1$ of the Wilson line 
operator associated with the three-component link shown in Fig.6.19}
\beeq 
N_{i j m^\ast} \; = \;  \langle \, W(C_1;\psi[i]) \; W(C_2;\psi[j]) \; 
W(C_3;\psi[m]) \,
\rangle \bigr |_{S^2 \times S^1} \qquad .
\label{eq:pr2} 
\end{equation}

\begin{figure}[h]
\begin{picture}(10,10)
\put(305,-60){$\psi[i]$}
\put(320,-110){$\psi[j]$}
\put(155,-120){$\psi[m]$}
\end{picture}
\vskip 0.5 truecm 
\centerline{\epsfig{file=\path 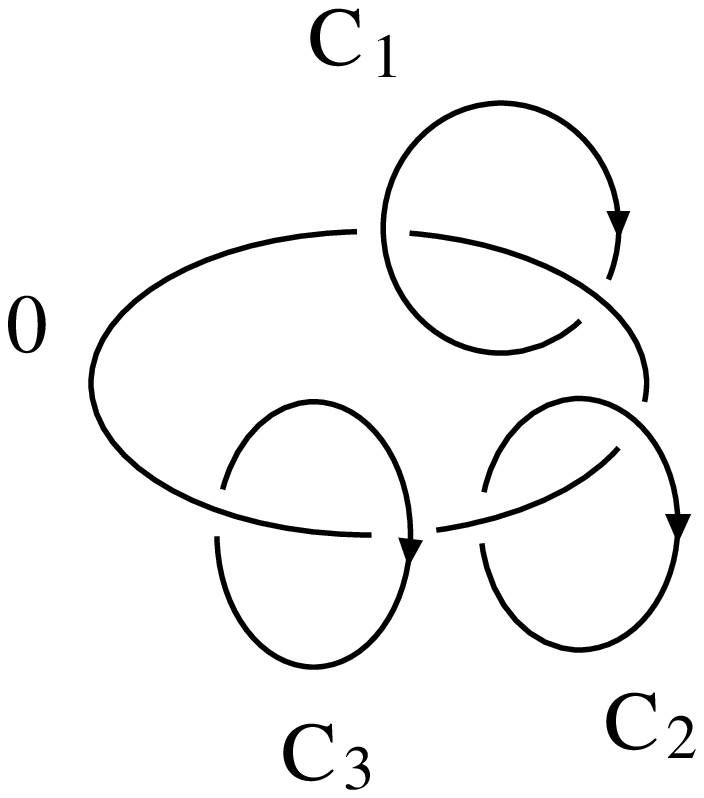,height=6cm,width=8cm}}
\vskip 0.5truecm 
\centerline{{\bf Figure 6.19}}
\end{figure}

{\bf Proof}

\no
By using the satellite formula (\ref{satg}), one obtains
\bea  
&&\langle \, W(C_1;\psi[i]) \: W(C_2;\psi[j]) \:W(C_3;\psi[m]) \,
\rangle \bigr|_{S^2 \times S^1}  \; = \nb \\ 
&& \qquad \qquad \qquad \qquad = \;  \langle \; W(C_1; \; \psi[i] \; \psi[j] \;
\psi[m])  \;  \rangle\bigr|_{S^2 \times S^1} \nb \\
&& \qquad \qquad \qquad \qquad = \;  \sum_t \, N_{ijt} \;  \langle
\; W(C_1;\psi[t] \; \psi[ m ])  \; \rangle\bigr|_{S^2 \times S^1} \nb \\
&& \qquad \qquad \qquad \qquad = \; \sum_t   \,  N_{ijt} \; N_{tm1} \; = \; \sum_t   \,  N_{ijt} \;
\delta_{t \, m^\ast } \nb \\ 
&& \qquad \qquad \qquad \qquad = \; N_{ij m^\ast} \; . 
\ena 
{\hfill \ding{111}}

\no
Equation (\ref{eq:pr2}) is in agreement with the symmetry properties of the 
structure constants $\{ \, N_{ijm} \, \}$ postulated in the  definition of a 
regular reduced tensor algebra. 

In Sect.6.4 we have introduced the quantities $Z_0$ and $Z_{(\pm)}$, which are related with the normalization of the elementary surgery operators 
\beeq
Z_0 \; = \; \sum_i \, E_0^2[i] \quad ,
\label{A11} 
\end{equation}
\beeq
Z_{(\pm 1)} \; =  \,  \sum_i  \, q^{\pm Q(i)} \, E_0^2 [i] \quad .
\label{A12}
\end{equation}
By definition, one  has $\, Z_{(-1)} = Z_{(+1)}^*$.
From the general properties of surgery it follows:

\bigskip

\shabox{{\bf Theorem 6.7}}{~\em The quantity $Z_0$ is related to the Hopf 
matrix by} 
\beeq
\left ( \, H^2 \, \right )_{i j} \; = \; Z_0 \; N_{ij \mathbf{1}} \quad . 
\end{equation}

\bigskip 

\no
{\bf Proof}~By definition we have
\beeq
\left(H^2 \right)_{ij} \; = \; \sum_m H_{im} \, H_{mj} \; = \; \sum_m E_0[m] \,
{E_0[m]}^{-1} \, H_{im} \, H_{mj} \quad .
\label{square}
\end{equation}
By using the connected sum formula (\ref{eq:csr}), Eq.(\ref{square}) can be 
written as 
\beeq
\left(H^2 \right)_{ij} \; = \; \sum_m E_0[m] \; \langle \, W(C_1, \, C_2, \, 
C_3; \, \psi[i], \, \psi[j], \, \psi[m] \, \rangle \bigr |_{S^3} \quad ,
\end{equation}
where $C_1, \, C_2$ and $C_3$ are the components of the link shown in 
Fig.16.20.

\begin{figure}[h]
\vskip 0.9 truecm 
\centerline{\epsfig{file=\path 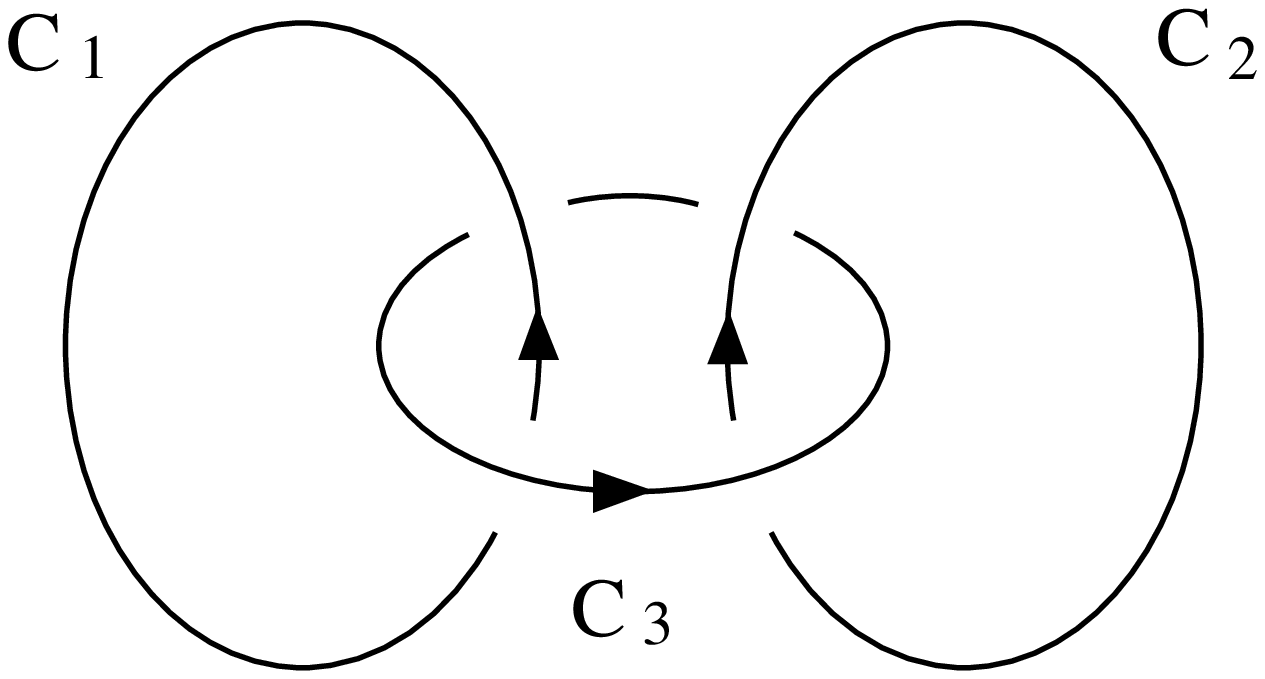,height=4cm,width=7cm}}
\vskip 0.9 truecm 
\centerline {{\bf Figure 6.20}}
\vskip 0.9 truecm 
\end{figure}

\no
The satellite relation (\ref{satg}) can be used to eliminate one of the components
of the link in Fig.6.20; we obtain
\beeq
\left(H^2 \right)_{ij} \; = \; \sum_{n, \, m} N_{ijn} \, E_0[m] \, H_{mn} 
\; = \; \sum_n {a(k)}^{-1} \, \langle \, W(U; \, \Psi_0) \, \rangle \bigr 
|_{S^3} \, N_{ijn} \, \langle \, W(U, \, C; \, \Psi_0, \, \psi[n]) \, \rangle 
\bigr |_{S^2 \times S^1} \quad .
\end{equation}
The knots $U$ and $C$ are the components of the link shown in Fig.6.17. From  
Theorem 6.5, it follows 
\beeq
\left(H^2 \right)_{ij} \; = \; {a(k)}^{-1} \, \langle \, W(U; \, \Psi_0) \, 
\rangle \bigr |_{S^3} \, N_{ij \mathbf 1} \; = \; Z_0 \, N_{ij \mathbf{1}}
\quad .
\end{equation}
{\hfill \ding{111}

By using Theorem 6.6, one can also show that $Z_0$ and $Z_{(\pm)}$ are not independent. Actually, the following relation exists.
   
\bigskip

\shabox{
\noindent {\bf Theorem 6.7}} {\em ~The complex numbers $\, Z_0$ , $Z_{(+1)}$ 
and $\, Z_{(-1)}$ satisfy}  
\beeq
Z_0 \; = \;  Z_{(+1)} \, Z_{(-1)}  \quad .
\end{equation}

\bigskip

\noindent {\bf Proof} ~On the one hand, from eq.(\ref{7.6}) one obtains 
\beeq
\sum_{ij} \, q^{Q(i)} \; E_0[i] \; E_0 [\, j\, ]\;  H_{ij} \; = \; 
\sum_{ij} \, q^{Q(i)} \; E_0[i] \, H_{ij} \, H_{j \mathbf{1}} \; = \; 
\sum_i \, q^{Q(i)} \; E_0[i] \, \left(H^2 \right)_{i \mathbf{1}} \; = \;
Z_{(+1)} \; Z_{(-1)} \quad . 
\label{A15}
\end{equation}
By using Theorem 6.6, from Eq.(\ref{A15}) one gets
\beeq
\sum_i \, q^{Q(i)} \, Z_0 \; E_0[i] \, N_{i \mathbf{1} \mathbf{1}} \; = \; 
Z_0 \; = \;  Z_{(+1)} \; Z_{(-1)} \quad .
\end{equation}
{\hfill \ding{111}}

When the gauge group is $SU(2)$ or $SU(3)$, from the explicit expressions of 
$H$, $Z_0$ and $Z_{(\pm)}$, the validity of the relations stated in 
Theorem 6.6 and in Theorem 6.7 can be verified by a direct calculation \cite{glib,gp2}.

\chapter{\bf A three-manifold invariant} 

\section{\bf Introduction}
Given a manifold $\cal M$, a topological invariant is a map $f: {\cal M} \ra 
\mathbb{C}$ which assigns to each manifold ${\cal M}$ a complex number 
$I({\cal M})$ in such way that, if ${\cal M}_1$ and ${\cal M}_2$ are homeomorphic,
then $I({\cal M}_1) =  I({\cal M}_2)$. Roughly speaking, two manifolds ${\cal M}_1$ 
and ${\cal M}_2$ are homeomorphic if ${\cal M}_1$ can be continuously deformed to
${\cal M}_2$. For example, the surface of an ellipsoid is homeomorphic with
a 2-sphere $S^2$; on the contrary, a torus cannot be continuously deformed to
$S^2$.

We shall construct a topological invariant of closed, connected and orientable
3-manifolds based on the CS partition function. In the framework of Dehn's surgery, any     
closed, connected and orientable 3-manifolds corresponds to a class of 
surgery instructions. By definition, given two sets of surgery instruction  
contained in the same class, they are related by a sequence of Kirby moves and
describe homeomorphic 3-manifolds. Thus, in the framework of Dehn's surgery, 
topological invariance is translated into invariance under Kirby moves.
   
\section{\bf Definition of the invariant}
As we have shown in the previous sections, the expectation value $\langle W(L) \rangle \bigr
|_{{\cal M}}$ represents a topological invariant of the link $L$ in the 
three-manifold $\cal M$. Our present goal is to construct a topological invariant of the 
manifold $\cal M$ itself. In the surgery presentation, each three-manifold $\cal M$ is 
characterized by a class
of equivalent surgery links  in $S^3$. Thus, it is natural to look for a
manifold invariant \cite{guad3} which is defined by the expectation value of the
Wilson line operators
associated with the surgery links $\{ \, {\cal L} \, \}$. Let us consider the 
partition function $Z({\cal M})$ for the CS theory in the 3-manifold $\cal M$ 
defined  as 
\beeq Z({\cal M}) \; = \; \frac{\langle \, 0|W({\cal{L}})|0 \, 
\rangle \bigr|_{S^3 }}{\langle \, 0|0 \, \rangle\bigr|_{|S^3}}
\; = \; \langle \, W({\cal{L}})
\rangle \bigr|_{|S^3} \; = \; \langle \, 0|0 \, \rangle \bigr|_{\cal{M}} \quad .
\label{partf}
\end{equation}
In the derivation of the surgery rules for 
the field theory, we have seen that $\langle \widetilde W({\cal L}) \rangle \bigr |_{S^3}$ 
carries informations on the manifold $\cal M$. Unfortunately, $\langle \widetilde W({\cal L})
 \rangle \bigr |_{S^3}$ is not invariant under Kirby moves and, consequently, the partition
function (\ref{partf}) cannot represent a three-manifold
invariant. In Sect.6.5 we have shown that, under a Kirby move, 
$\langle \widetilde W({\cal L}) \rangle \bigr
|_{S^3}$ gets multiplied by the phase factor $e^{\pm i\theta_k }$. Therefore, in 
order to define a topological invariant,  we simply need to introduce
\cite{retu,tur,koh,lick1,mor,km} a multiplicative term which cancels out this phase factor.   

Let $\cal M$ be a (closed, connected and orientable) three-manifold 
corresponding to the  ``honest" surgery
link $\cal L$ in $S^3$. Let us introduce an orientation for $\cal L$ and let us 
denote by $\sigma ({\cal L})$ the signature of the linking matrix $L_{ij}$ 
associated 
with $\cal L$, i.e
\beeq 
L_{ij}=  \begin{cases} {\rm lk}(C_i,C_j) & \text{for } \; i \neq j \\ 
{\rm lk}(C_i,C_{if}) & \text{for } \; i = j 
\end{cases} \quad .
\label{linkm}
\end{equation}

\bigskip

\shabox{
\noindent {\bf Theorem 7.1}} {\em The improved partition function} 
\beeq
I({\cal M}) \; = \;  \exp \left [\, i\, \theta_k \, \sigma ({\cal L})\, \right ] 
\; \, \langle \, \widetilde W({\cal L}) \, \rangle \bigr |_{S^3} \quad , 
\label{11.1}
\end{equation}
{\em where $\, \widetilde W({\cal L})$ has been defined in Sect.6.5, is
invariant under Kirby moves and represents a topological three-manifold 
invariant of} $\cal M$.

\bigskip

\noindent {\bf Proof.} ~Under a Kirby move, $\langle \widetilde W({\cal L}) \rangle \bigr |_{S^3} $
 transforms as 
\beeq
\langle \, \widetilde W({\cal L}) \, \rangle \bigr |_{S^3}  \; \rightarrow \; 
e^{\pm  i \, \theta} \; 
\langle \, \widetilde W({\cal L}) \, \rangle \bigr |_{S^3} \quad  ,
\label{11.2}
\end{equation}
whereas the signature $ \sigma ({\cal L})$ transforms as \cite{lick1,mor}
\beeq
\sigma ({\cal L}) \; \rightarrow \sigma ({\cal L}) \mp 1 \quad .
\label{11.3}
\end{equation}
Therefore $I ({\cal M})$ is invariant under Kirby moves. {\hfill \ding{111}} 

\bigskip

According to the surgery rules described in Sect.6.5, $I ({\cal M})$ can be 
interpreted \cite{guad3} as the value of 
the (improved) partition function of the Chern-Simons field theory in  
${\cal M}$. From the definition (\ref{11.1}), it follows that 
\beeq
I(S^3) \; = \; 1 \quad .
\label{norma}
\end{equation}
Eq.(\ref{norma}) shows $I({\cal M})$ is normalized in such way that $S^3$ plays
the role of a reference manifold.  
In the next section, as examples, we shall give the value of ${\cal I} ({\cal M})$ for 
some three-manifolds when the gauge group is $SU(3)$. 

\section{\bf Examples}

The manifold $S^2 \times S^1$ admits a surgery presentation described by the 
unknot with surgery
coefficient $r=0$. In this case,  $ \sigma ({\cal L}) \; = \; 0$ and then one 
gets 
\beeq
I(S^2 \times S^1) \; = \;  \begin{cases} \sqrt 3 & \text{for } k=1,2  \; , \\
\left ( \, k \, \sqrt 3 /16 \, \right ) \, \cos^{-1} \left ( \pi / k \right ) \, 
\sin^{-3} \left ( \pi
/ k \right ) & \text{for }k \geq 3 \; . 
\end{cases}
\end{equation}
Let us consider the lens spaces $L(p,1)$ with $p \geq 2$; a surgery link for 
these manifolds
is the  unknot with surgery coefficient $r=p$, the signature of the corresponding
 linking matrix is
equal to $+1$. By using the results of Sect.6.7.2, one has

\noindent $k=1$
\beeq
I (L(p,1)) \; = \;   \frac{ i}{ \sqrt 3 } \,  \left (1+2 \, e^{-i \pi \, 
p/3} \right ) \quad .
\end{equation}
\noindent $k=2$
\beeq
I (L(p,1)) \; = \;   \frac{- i}{  \sqrt 3 } \,  \left (1+2 \, e^{i \pi \, p/3} \right ) \quad .
\end{equation}
\noindent $k=3$
\beeq
I (L(p,1)) \; = \;  1  \quad .
\end{equation}
\noindent $k=4$
\beeq
I (L(p,1)) \; = \;   \frac{ i}{\sqrt 3 } \,  \left (1+2 \, e^{-i \pi \, p/3} \right ) \quad .
\end{equation}
$k=5$
\beeq
{\cal I} (L(p,1)) \; = \;  e^{- i  6 \pi /5 } \; \sqrt{\frac{ 2}{ 3  \left(  
\sqrt 5+1 \right) }} \
\left[1+2 \,  e^{ i \, 2 \pi \, p  /3 } \right  ] \left [ 1+ \frac{1}{2} (3+ 
\sqrt 5)  \; e^{- i  6 \pi \, p /5 } \right ] \quad .
\end{equation}

\noindent The Poincar\'e manifold admits a surgery presentation described by the 
right-handed trefoil with surgery coefficient $r=1$ as we have seen in 
Sect.7.6.3. In 1900, Poincar\'e conjectured that any 3-manifold having the same
homology groups of $S^3$ (homology spheres) is homeomorphic with $S^3$. Soon
after, the same Poincar\'e found a counter-example to his conjecture. Indeed
the Poincar\'e manifold $\cal P$ is a homology sphere, nevertheless it is not
homeomorphic with $S^3$, its fundamental group $\pi_1( {\cal P})$ being the icosahedral group.
      
The values of the associated invariant are

\no
$k=1,2,3,4$
\beeq
I ({\cal P}) \; = \; 1 \quad .
\end{equation}
$k=5$
\beeq
I ({\cal P}) \; = \;  \sqrt{ \frac{ 2}{  3 \left( \sqrt 5 +1 \right ) }}\, 
\left ( 2- i \sqrt 3 \right ) \left ( \frac{3 + \sqrt 5}{ 2 } \right ) \, \left
 ( 1- e^{-i \pi /5}
\right ) \quad .
\end{equation}
The manifold $T^2 \times S^1$ corresponds to the surgery link shown in Fig.6.15. 
From eq.(\ref{10.2}), it
follows that $\eta (1 ) = \sum_i 1 = {\rm dim}\> {\cal T}_{(k)}$. Consequently, 
one has 
\beeq
I (T^2 \times S^1) \; = \; {\cal I} (S^2 \times S^1)\; {\rm dim} \> {\cal T}_{(k)} \quad . 
\label{torus}
\end{equation}
Let us now consider the manifold $\Sigma_g \times S^1$; by using the handle
decomposition (\ref{10.1}), we find

\noindent $k=1,2,4$
\beeq
I (\Sigma_g \times S^1) \; = \; \sqrt 3 \, 3^g \quad .
\label{11.14}
\end{equation}
\noindent $k=3$
\beeq
I (\Sigma_g \times S^1) \; = \; 1 \quad .
\label{11.15}
\end{equation}
\noindent $k=5$
\beeq
I (\Sigma_g \times S^1) \; = \; \begin{cases}
3^g  \>  5^{g/2} \; F(g-1) \; \sqrt{ 3  \left(  \sqrt 5+1 \right) / 2 }
& \text{for } g \; \text{even} \; ; \\
3^g  \>  5^{(g-1)/2} \> \left[ F(g)+F(g-2) \right] \> \sqrt{ 3  \left(  \sqrt 5+1 \right) / 2 }
& \text{for } g \; \text{ odd} \end{cases} \quad .
\label{11.16}
\end{equation}
In eq.(\ref{11.16}), $F(g)$ denotes the $g$-th Fibonacci number; i.e. $F(g) $ is 
defined by 
$F(g) \; = \; F(g-1)  + F(g-2)$, with $F(1)=F(2)=1$. Details on the derivation 
of eq.(\ref{11.16}) can be found in  Appendix C.

\section{\bf Properties of the CS 3-manifold invariant}
\subsection{\bf Introduction}
The algebraic aspects of the  construction of the CS 3-manifold invariant, 
which is based on the structure of simple Lie groups, are well understood
\cite{tur,km,lick1,mor,kau,koh,guad3}. However, the topological meaning of 
this invariant is still
unclear. Let us recall that the invariant of the 3-manifold $\, I({\cal M}) \, $  defined  by means of the Chern-Simons quantum field theory \cite{wit,guad3} 
coincides with the Reshetikhin-Turaev invariant
\cite{retu,tur}.  In general, it is not known  how $\, I(M) \, $  is related, 
for instance, to the homotopy class of $\, {\cal M} \, $ or to the fundamental group 
of $\, {\cal M} \, $.  In this section we shall address this issue in the case 
${\cal M}= L_{p/r}$, where $L(p,r)$ are the lens spaces. In addition we shall formulate the 
following conjecture \cite{gp4}.

\bigskip

\shabox{\parbox{7cm}{{\bf Conjecture}: For non-vanishing  
$\, I({\cal M}) \, $, the absolute value $\, | \,  I({\cal M}) \, | \, $ only depends on the fundamental group $\pi_1 ({\cal M}) \, $.}} 

\bigskip

\noindent In the absence of a general proof, we shall verify the validity of 
the conjecture for
a particular class of manifolds: the lens spaces.  There are examples of lens 
spaces  $\, {\cal M}_1 \, $ and $\,{\cal M}_2 \, $ with the same fundamental group $\, \pi_1( {\cal M}_1)
\simeq \pi_1 ({\cal M}_2)\, $ which are not homeomorphic;  for all these manifolds, 
we shall prove
that  (for non-vanishing invariants) $\, | \,  I({\cal M}_1) \, | = | \,  I({\cal M}_2) \, |\, $  when
$\,   I({\cal M}) \,  $ is the invariant associated with the group $\, SU(2)\, $.  In the case in which the
gauge group is $\, SU(3) \, $, we will present numerical computations confirming the conjecture. This computation
is in agreement with the computer calculations for $\, SU(2) \, $ of Freed and Gompf
\cite{fregomp} and the expression of the $ \, SU(2) \, $ invariant obtained 
by Jeffrey \cite{jef}. 
Differently from
\cite{fregomp} and \cite{jef}, the present approach is based exclusively on 
the properties of 
3-dimensional Chern-Simons quantum field theory. 

\subsection{\bf Lens Spaces}
Lens spaces, which are characterized by two integers  $ \, p \, $ and $ \, r \, $, will be
denoted by  $\, \{ \,  L_{p/r} \, \} \, $. The fundamental group of $\, L_{p/r}\, $ is $\, Z_p\,
$. Two lens spaces $\, L_{p/r} \, $ and $\, L_{p^\prime/r^\prime}\, $ are homeomorphic if and
only if $\, |p| = |p^\prime| \, $ and $\, r =  \pm r^\prime \; (\mbox{mod } p)\, $ or $\, r
r^\prime = \pm 1 \; (\mbox{mod } p)\, $. Thus, we only need \cite{rol} to consider the case in
which $ \, p >1\, $  and $ \, 0 < r < p \, $; moreover, $ \, r \, $ and $ \, p \, $ are 
relatively prime.  The lens spaces $\, L_{p/r} \, $ and $\, L_{p^\prime/r^\prime}\, $ are
homotopic if and only if $|p|= |p^\prime|$ and $r r^\prime = \pm $ quadratic residue $ (\mbox{mod
} p)\, $.  Consequently, one can find examples of lens spaces which are homotopic  but are not
homeomorphic; for instance,  $\, L_{13/2}\, $ and $\, L_{13/5}\, $. One can also find examples of
lens spaces which are not homeomorphic and are not homotopic but have the same fundamental group;
for instance,  $\, L_{13/2}\, $ and $\, L_{13/3}\, $.

One possible surgery instruction corresponding to  the lens space $\, L_{p/r} \, $ is given by
the unknot \cite{rol} with rational surgery coefficient $\, (p/r) \, $. From this surgery
presentation one can derive \cite{rol} an ``honest" surgery presentation of $\, L_{p/r} \, $ by
using  a continued fraction decomposition of the ratio $\, (p/r) \,$ 
\beeq
\frac{p}{r} \, = \, z_d \, - \, \frac{1}{\; z_{d-1}\; -\;  \frac{1}{\; \ddots \; - \; 
\, \frac{1}{z_1}}} \qquad , 
\end{equation}   
where $\, \{z_1, \, z_2, \cdots, \, z_d \} \,$ are integers. The new surgery link
$\, {\cal L} \, $ corresponding to  a ``honest" surgery presentation of $\, L_{p/r} \, $ is a
chain with $ \, d \, $ linked components, as shown in Figure~6.21, and the integers $\, \{z_1, \,
z_2, \cdots, \, z_d \} \,$ are precisely the surgery coefficients. 

\begin{figure}[h]
\vskip 0.9 truecm 
\centerline{\epsfig{file=\path 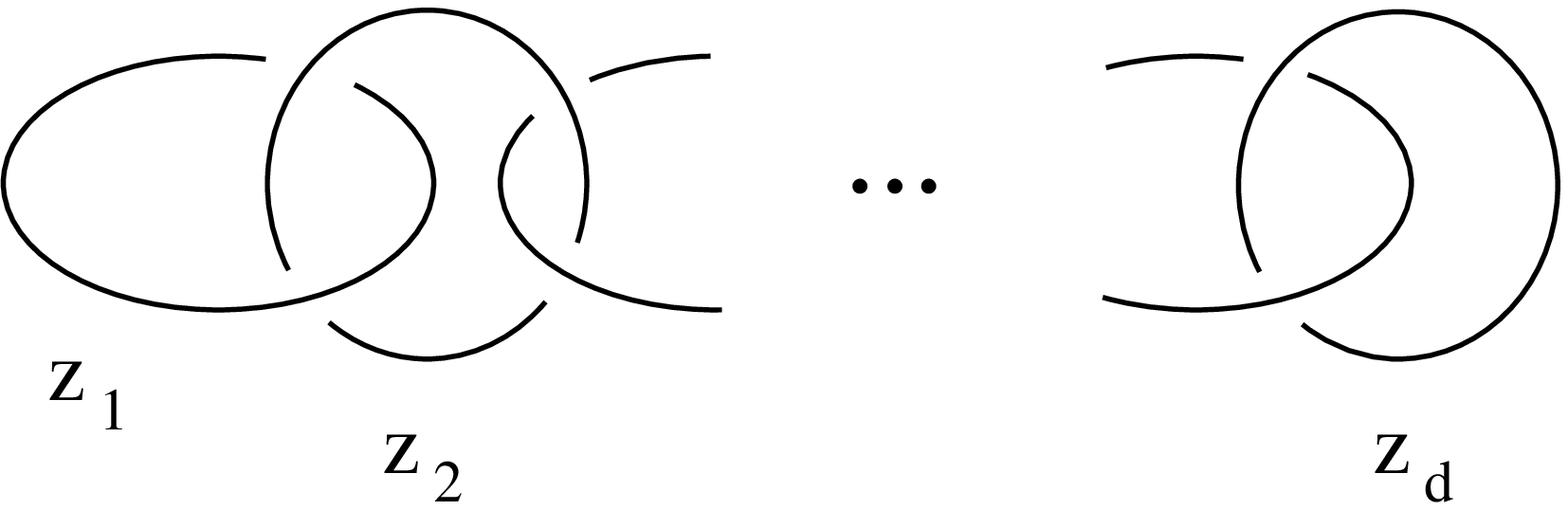,height=4cm,width=6cm}}
\vskip 0.5truecm 
\centerline {{\bf Figure 7.1}}      
\vskip 0.5truecm 
\end{figure}

\noindent According to the definition, the lens space invariant is given by 
\bea
&&I(L_{p/r}) \; = \; e^{i \theta_k \sigma ({\cal L}) } \, ( \, a_k \, )^d \, \sum_{j_1, \cdots ,
j_d \in \T} \;  \prod_{i=1}^d \left ( q^{z_i \, Q(j_i)} \right ) \, \times \nb  \\
&&
\qquad \qquad \qquad \times \; E_0[\, j_1] \cdots E_0[\, j_d]  \; \;  E\left (\, {\cal L}\, ; 
\psi [ \, j_1] \, , \cdots , \psi [\, j_d ] \right ) \qquad .
\label{iiq}
\ena

\noindent The link of Figure~6.21 can be understood as the connected sum of $\, ( d-1 ) \, $ Hopf
links $\, {\cal H} \, $, i.e. $ \, {\cal L} = {\cal H} \# {\cal H} \cdots \# {\cal H} \, $. 
Therefore, by using equation (\ref{eq:csr}}), expression (\ref{iiq}) can be written as 

\beeq
I(L_{p/r}) \, = \, e^{i \theta_k \sigma ({\cal L})} \, ( \, a_k \, )^d \, \sum_{j_1, \cdots
, j_d \in \T} \, q^{\left ( \, \sum_{i=1}^d z_i \, Q(j_i) \,  \right ) } \; H_{1 j_d}\,  H_{j_d
j_{d-1}}\,  \cdots \,  H_{j_2 j_1} \,  H_{j_1 1} \quad . \label{eq:def1}
\end{equation} 
Let us introduce the matrices
\beeq
(X)_{ij} \; = \; a(k) \, H_{ij} \, , \qquad (Y)_{ij} \; q^{Q(i)} \, \delta_{ij}
\quad .
\label{mod}
\end{equation}
As we shall see in the next Chapter, $X$ and $Y$ are associated with the modular group. In terms of the matrices (\ref{mod}), one finds \cite{gp4}
\beeq
I (L_{p/r}) \; = \; e^{i \theta_k \sigma ({\cal L})} \; ( a_k )^{-1}\;  \left[ \, F(p/r)
\right]_{11}
\qquad , \label{eq:lei}
\end{equation}
where $ \, \left[ \, F(p/r) \right]_{11} \, $ is the element corresponding to the first row and
the first column of the following matrix 
\beeq
F(p/r) \; = \; X Y^{z_d} XY^{z_{d-1}}X \cdots XY^{z_1}X \qquad . \label{matrix}
\end{equation}
 The invariant $\, I (L_{p/r})  \, $ given in equation (\ref{eq:lei})  is in agreement with
the expressions obtained in \cite{fregomp,jef} apart from an overall normalization factor. 

\subsection{\bf The SU(2) case}
In this section, we shall compute $ \, I(L_{p/r}) \, $  for the gauge group $ \, G=SU(2)
\, $. Then, we will show that in this case our conjecture is true; i.e. when $ \, I(L_{p/r})
\not= 0 \, $, the absolute value $ \, | \, I(L_{p/r}) \, | \, $ only depends on $ \, p \, $. 

For sake of notational simplicity, in this section, we shall use as label of the 
elements $\, \{ \, \psi [\, j\, ] \,\} \,$ of the standard basis of $\, {\cal T}_k\, $, for $\, k \geq 2\, $, the index $ \, j \, $ representing the 
dimension of the irreducible representation
described by  $ \, \psi [\, j\, ] \,  $ and  $  \,  1 \leq j \leq (k-1) \,$. 
Thus, by using the results of Chap.3 and Chap.4 and Chap.6,  the matrix elements of 
$\, X\, $ and $\, Y\, $ are given by  
\bea
&&\left(X \right)_{mn} \; = \; \frac{i}{\sqrt{2k}} \left[ \exp \left(-\, \frac{i \pi mn}{k} \right)
\, - \, \exp \left(\frac{i \pi mn}{k} \right) \right] \quad ; \nb \\
&&\left(Y \right)_{mn} \; = \; \xi \, \exp \left(- \, \frac{i \pi m^2}{2k} \right) \delta_{mn} \quad
; \ena
with 
\beeq
\xi \; = \; \exp(\, i \pi/2k\, ) \qquad . 
\end{equation}
When $ \, k=1 \, $,  one has 
\beeq
X \; = \;  \begin{pmatrix} 1 &1 \\ 1 & -1 \end{pmatrix} \quad , \quad Y \; = \; 
\begin{pmatrix} 1& 0 \\ 0 & i \end{pmatrix}
\quad .
\end{equation}
The algebra $\, {\cal T}_1\, $ is isomorphic with $\, {\cal T}_3\, $ and it 
is easy to verify
that 
\beeq  I_{k=1}(L_{p/r})  \; =\;  \left [ \, I_{k=3}(L_{p/r}) \, \right  ]^* \qquad . 
\end{equation}
Therefore, we only need to consider the case $\, k \geq 2\, $.  

In order to compute $ \, I(L_{p/r}) \, $ \cite{gp4}, we shall derive a recursive 
relation for the
matrix (\ref{matrix}); the argument that we shall use has been produced by 
Jeffrey 
\cite{jef} in a slightly different context. In fact, our final result for $ \, I(L_{p/r}) \, $ is
essentially in agreement with the formulae obtained by Jeffrey.  Since in our approach the
invariance under Kirby moves is satisfied, our derivation of $ \, I(L_{p/r}) \, $ proves that the
appropriate expressions given in  \cite{fregomp,jef} really correspond to the 
values of a
topological invariant of 3-manifolds.  

Let us introduce a few definitions; with the ordered set of integers $ \{z_1, \, z_2, \cdots,
\, z_d \} $ one can define the following partial continued fraction decompositions 
\beeq
\frac{\alpha_t}{\gamma_t} \; = \;  z_t \, - \, \frac{1}{\; z_{t-1}\; -\;  \frac{1}{\; \ddots \; -
\;  \, \frac{1}{z_1}}} \qquad , 
\end{equation}
where $ \, 1 \leq t \leq d \, $. The integers $ \, \alpha_t \, $ and $ \, \gamma_t \, $ satisfy
the recursive relations 
\bea
&&\alpha_{m+1} \; = \; z_{m+1} \, \alpha_m \, - \, \gamma_m, \quad , \quad \alpha_1 \, = \, z_1
\quad , \quad \alpha_0 \; = \; 1 \quad ;
\\ &&\gamma_{m+1} \; = \; \alpha_m  \qquad , \qquad \gamma_1 \, = \, 1  \qquad ,  
\ena
and, clearly, $\, \alpha_d / \gamma_d = p / r \, $. Finally, let $ \, F_t \, $ be the matrix 
\beeq
F_t \; = \; X Y^{z_t} XY^{z_{t-1}}X \cdots XY^{z_1}X \qquad ; \label{parma}
\end{equation}
by definition, one has $ \, F_d = F(p/r) \, $. 

\bigskip

\shabox{
{\bf Lemma 7.1}}~{\em The matrix element $\, (F_t)_{mn}\, $ is given by 
\bea
&&\left(F_t \right)_{mn} \; = \; B_t \, \sum_{s(m,k,|\alpha_t|) } \, \left[ \, e^{\frac{i
\pi\gamma_t }{2k \alpha_t} \left(s+\frac{n}{\gamma_t}
\right)^2} \, - \, e^{\frac{i \pi \gamma_t}{2k\alpha_t} \left(s- \frac{n}{\gamma_t}\right)^2} \,
\right]\quad ;
\label{eq:mat} \\ 
&&B_t \; = \; \frac{(-i)^{t+1}}{\sqrt{2k| \alpha_t | }} \, \xi^{z_1 + z_2 + \cdots + z_t} \, \exp
\left\{ - \frac{i \pi}{4} \left[\mbox{sign}(\alpha_0 \alpha_1) \, + \cdots +
\mbox{sign}(\alpha_{t-1} \alpha_t) \right] \right \} \nb \\
&&\qquad \null \qquad  \exp \left\{ \frac{i \pi n^2 }{2k} \left[\frac{1}{\alpha_0 \alpha_1} \, +
\cdots +
\frac{1}{\alpha_{t-2} \alpha_{t-1}} \right] \right\}\quad ; \label{bi}
\ena
where $\, s(m,k,|\alpha_t|)\, $ stands for the sum over a complete residue system modulo
$\, (2k|\alpha_t| ) \, $ with the additional constraint $\, s \equiv m \; (mod \, 2k)\, $}.

\bigskip

\no
{\bf Proof} ~The proof is based on induction. First of all we need to verify the validity of
equations (\ref{eq:mat}) and (\ref{bi}) when $\, t=1\, $. In this case, from the definition
(\ref{parma}) one gets
\beeq
\left( F_1 \right)_{mn} \; = \; - \frac{1}{2k}\,  \frac{1}{2} \, \xi^{z_1} \, \sum_{s =
1}^{2k}e^{-i  \pi s^2   z_1/(2 k)}\,  \left[\, e^{-i \pi s(m+n)/k} \, - \, e^{-i \pi s(m-n)/k} \,
+  \, \mbox{c. c.} \, \right]  \quad .
\label{eq:sum1} 
\end{equation}
Since the sum (\ref{eq:sum1}) covers twice a complete residue system modulo $\, k \, $, i.e.  $ \,
1 \leq s \le  2k\, $,  a multiplicative factor $\, 1/2\, $ has been introduced in
(\ref{eq:sum1}). The change of variables  $\, s \rightarrow - s \, $ shows that the last two
terms in (\ref{eq:sum1}) are equal to the first two terms. Therefore, equation (\ref{eq:sum1})
can be written as 
\beeq
\left( F_1 \right)_{mn} \; = \; - \frac{1}{2k} \,  \xi^{z_1} \, \sum_{s = 1}^{2k}\,  e^{-i 
\pi s^2   z_1/(2 k)}\,  \left [ \, e^{-i \pi s(m+n)/k} \, - \, e^{-i \pi s(m-n)/k} \,  \right ] 
\label{eq:sum2}\qquad .
\end{equation}   
At this point, one can use the reciprocity formula (\ref{B7}) reported in the 
appendix and one
gets 
\bea
&&\left( F_1 \right)_{mn} \; = \; \frac{-1}{\sqrt{2k| z_1 | }} \, \xi^{z_1} \, \exp
\left\{ - \frac{i \pi}{4} \, sign (\alpha_0 \alpha_1) \right \} \, \times \nb \\  
&&\qquad \qquad \null \qquad \times \, \sum_{v=0}^{|z_1|-1} \, \left [ \, e^{\frac{i \pi}{2k z_1}
\left ( 2 k v + m +n \right)^2} \, - \, e^{\frac{i \pi}{2k z_1} \left ( 2k v + m -n \right)^2} \,
\right] \quad . \label{inter}
\ena
By introducing the new variable $ \, s = 2k v + m \, $, one finds that in equation (\ref{inter})
the variable $\, s \, $ covers a complete residue system modulo $\, ( 2k|z_1| ) \, $ with the
constraint that $\, s \equiv m \; (mod \, 2k)\, $. Therefore, equation (\ref{inter}) can be
written in the form 
\beeq
\left( F_1 \right)_{mn} \; = \; B_1 \, \sum_{s(m,k,|z_1|)}\, 
\left [\, e^{\frac{i \pi}{2k z_1} \left(s +n \right)^2} \, - \, e^{\frac{i \pi}{2k z_1}  \left(s
-n \right)^2} \, \right] \qquad .
\end{equation}
This confirms the validity of equation (\ref{eq:mat}) when $ \, t = 1 \, $. In order to complete
the proof, suppose now that (\ref{eq:mat}) is true for a given $\, t\, $; we shall show that 
(\ref{eq:mat}) is true also in the case $\, t+1\, $. Indeed, one has    
\beeq
\left(F_{t+1} \right)_{mn} \; = \; \sum_{v = 1}^{k} \, \left( \, X \,  Y^{z_{t+1}} \,
\right)_{mv} \, \left( \, F_t \, \right)_{vn} \qquad .
\end{equation}
From equation (\ref{eq:mat}) one gets  
\bea
&&\left(F_{t+1} \right)_{mn} \; = \; - B_t \, \frac{i \xi^{z_{t+1}}}{\sqrt{2k}}\,  \frac{1}{2}\, 
\sum_{v =1}^{2k}\, \sum_{s(v,k,|\alpha_t|)} \, e^{- \frac{i\pi}{2k} v^2 z_{t+1}} \nb \\
&&\qquad \qquad \left[ \, e^{\frac{i \pi \gamma_t}{2 k \alpha_t} \left( s- \frac{n}{\gamma_t}
\right)^2 } \, e^{- i\pi m v/k} \, - \, e^{\frac{i \pi \gamma_t}{2 k \alpha_t} \left( s+ \frac{n}
{\gamma_t}\right)^2 }\,  e^{- i\pi mv/k} \right. \nb \\ 
&&\qquad \qquad \; \; \left. - \, e^{\frac{i \pi \gamma_t}{2 k \alpha_t} \left( s-
\frac{n}{\gamma_t} \right)^2 }\,  e^{ i \pi mv/k} \, + \, e^{\frac{i \pi \gamma_t}{2 k \alpha_t}
\left(s +\frac{n}{\gamma_t}
\right)^2 } \, e^{ i \pi mv/k} \right ] \quad .
\ena
Again, the last two terms can be omitted provided one introduces a multiplicative factor $ \, 2
\, $. Moreover, because of the constraint $\, v = s \, (mod \, 2k)$, one can set $\, v =
s\, $, thus
\bea
&&\left(F_{t+1} \right)_{mn} \; = \; - B_t \,  \frac{i \, \xi^{z_{t+1}}}{\sqrt{2k}} \, e^{\frac{i
\pi n^2}{2k \alpha_t \gamma_t}}\, \sum_{s = 0}^{2k|\alpha_t|-1} \nb \\
&&\qquad \left \{ \, e ^{-\frac{i \pi}{2k \alpha_t} \left[ \alpha_{t+1} s^2 + 2 \left(\gamma_{t+1}
m + n \right)s \right]} \, - \, e ^{-\frac{i \pi}{2k \alpha_t} \left[ \alpha_{t+1} s^2 + 2
\left(\gamma_{t+1} m - n
\right) s \right]} \, \right \} \quad .
\ena
By using the reciprocity formula, one obtains the final expression for $\, (F_{t+1})_{mn}\, $
\bea
&&(F_{t+1})_{mn}\; = \; - i \, B_t \,  \xi^{z_{t+1}}\,  \sqrt{\frac{|\alpha_t|}{|\alpha_{t+1}|} }
\, e^{\left(\frac{i \pi n^2}{2k \alpha_t \alpha_{t+1}} \right)} \, e^{ \frac{- i \pi}{4}{ 
sign}(\alpha_t \alpha_{t+1})} \nb \\ 
&& \qquad \; \sum_{v=1}^{|\alpha_{t+1}|} \, \left \{ \, e^{\frac{i \pi \alpha_t}{2k\alpha_{t+1}}
\left ( \, 2kv \, + \, m \, + \, \frac{n}{\alpha_t} \, \right)^2 } \, - \,   
e^{ \frac{i \pi \alpha_t}{2k\alpha_{t+1}} \left(\, 2kv \, + \, m \, - \, \frac{n}{\alpha_t}\, 
\right)^2 } \, \right \} \quad . \label{final}
\ena
In terms of the variable $ \, s = 2kv + m \, $, equation (\ref{final}) can be rewritten in the
form (\ref{eq:mat}) and this concludes the proof. {\hfill \ding{111}}

\bigskip 

From the definition (\ref{eq:lei}) and Lemma 7.1 it follows 

\bigskip

\shabox{{\bf Theorem 7.2}}~{\em Let $SU(2)$  be the gauge group, the 3-manifolds invariant $\, I_k (L_{p/r})\, $ for
$\, k \geq 2\, $ is given by}
\bea
&&I_k(L_{p/r}) \; = \;   \sum_{s \, (mod \;  p)} \left \{\, \exp{ \left[ \frac{i
\pi (r+1)^2}{2p k r} \right]}  \,  \exp \left [ \frac{i 2 \pi}{p}
\left[r ks^2 \, + \,  (r+1)s \right]  \right] \right.   \nb \\
&& \quad - \left. \, \exp{\left[ \frac{i \pi (r-1)^2}{2p k r} \right]}
 \, \exp \left[{ \frac{i 2 \pi}{p}
\left[r ks^2 \,  + \,  (r -1)s \right]} \right] \right \} \, \frac{e^{i
\theta_k \sigma ({\cal L}) }\> B_d}{ a_k}
\,  \quad . \label{invlens}
\ena

\bigskip 

\no
{\bf Proof}~According to equation (\ref{eq:lei}), the expression for the matrix element $ \, \left[ \,
F(p/r) \right]_{11} \, $ has been written by means of a sum over a complete residue system
modulo $\, p \, $. {\hfill \ding{111}} 

\bigskip 

\no
As shown in equation (\ref{invlens}),  the expression for $\, I_k ( L_{p/r} ) \, $ is
rather involved; nevertheless, $\, | \,  I_k ( L_{p/r} ) \, |^2 \, $ can be computed
explicitly. Let us introduce the modulo-$ p\, $ Kroneker delta symbol defined by  
\beeq
\delta_p \left ( x \right) \; = \; \left \{ \begin{array}{cc} 0 \qquad & x \, \not \equiv \,  0 \; \;
( \,  mod \, p \, ) 
\\ 1 \qquad & x \equiv 0 \; \; (\,  mod \, p \, )  
\end{array} \right. \qquad ;
\end{equation}
where $\, p \, $ and $\, x\, $ are integers. One can easily verify that, for integer $ \, n \,
$,  
\beeq 
\left \{ \begin{array}{cc} \delta_p(xn) \; = \; \delta_p(x) \qquad  &{\rm if } \quad  (n,p) \, = \, 1
\quad ; \\
\delta_{pn}(xn) \; = \; \delta_p(x)  \qquad . & \qquad \end{array}\right. 
 \label{proper} 
\end{equation}
Finally, we shall denote by $\, \phi (n) \, $ the Euler function \cite{cin} which is equal to the
number of residue classes modulo $\, n \, $ which are coprime with 
$ \, n \, $. 

\bigskip 

\shabox{
{\bf Theorem 7.3}}
{\em The square of the absolute value of $\, I_k  ( L_{p/r}  ) \, $ is given in the following
list~; 

\no for $\, p =2\, $
\beeq
\left | \, I_k  ( L_{2/1} ) \, \right |^2 \; =  \; \left[ \, 1+(-1)^k \, \right] \, 
\frac{\sin^2 \left[\pi/(2k) \right]}{\sin^2 \left[\pi/k \right]} \qquad ; \label{facile} 
\end{equation}
for $\, p  > 2\, $ one has : 

\no when $\, p \, $ and $ \, k \, $ are coprime integers, i.e. $\, (k,p) \, = \, 1\, $,  
\bea 
&& \left | \, I_k ( L_{p/r} ) \, \right |^2 \; =  \; \frac{1}{2}\, \left[ \, 1-(-1)^p \, \right]
\, \frac{ \sin^2 \left[\, \pi \left( \, k^{\phi(p)} -1 \, \right)/(kp) \, \right] }{\sin^2
(\pi/k)}  \, + \nb \\ 
&& \qquad \frac{1}{2}\, \left[\, 1+(-1)^p \, \right] \, \left[ \, 1+(-1)^{p/2} \, \right] \, 
\frac{  \sin^2 \left[ \, \pi \left( \, k^{\phi(p/2)}-1\, \right)/(kp) \right] }{\sin^2(\pi/k)}
\qquad ; 
\ena
when the greatest common divisor of $ \, p \, $ and $\, k \, $ is greater than unity, i.e. $\,
(k,p) \, = \, g >1\, $ and $\, p /g \, $ is odd
\beeq
\left | \, I_k ( L_{p/r} )\, \right |^2  \; =  \; \frac{g}{4\, \sin^2( \pi/k) } \, 
\left[\, \delta_g(\, r-1\, ) \, + \, \delta_g(\, r+1\, ) \, \right] \qquad ; \label{44}
\end{equation}
when $\, (k,p) \, = \, g >1 \, $ and $\, p /g \, $ is even
\bea
&& \left | \, I_k ( L_{p/r} ) \, \right |^2 \; =  \; \frac{g}{4\, \sin^2( \pi/k) } \, 
\left \{ \, \delta_g(r+1) \, \left [ \, 1 + (-1)^{kp/2g^2} \, (-1)^{(r+1)/g} \, \right ] \, +
\right. \nb \\  
&& \left. \qquad \delta_g(r-1) \, \left [ \, 1 + (-1)^{kp/2g^2} \, (-1)^{(r-1)/g} \, \right ] \,
\right \} \qquad . \label{45}
\ena }

\bigskip

\no
{\bf Proof}~From Theorem~7.2 it follows that the square of the absolute value of the lens space invariant is 
\beeq
\left |\, I_k (L_{p/r} )\,  \right |^2 \; =  \; a(k)^{-2} \, \left( 2kp \right)^{-1} \,
{\cal S}(k,p,r) \qquad , 
\end{equation}
with
\bea
&&{\cal S}(k,p,r) \; = \;  \sum_{s,t \, \left(mod \, p \right)} \left \{ \, \exp \left\{\frac{i 2
\pi}{p} \left[ kr \left(s^2-t^2 \right) \, + \, \left(r+1 \right) \left(s-t \right) \right] \right \}
\right. \nb \\ 
&&\qquad - \, \exp\left(\frac{i 2 \pi}{kp} \right) \, \exp \left\{\frac{i 2 \pi}{p}
\left[ kr \left(s^2-t^2 \right) \, + \, r \left( s-t \right) \, +\, s\,  +\, t\, \right]
\right \} \nb \\
&& \qquad - \exp\left(- \frac{i 2 \pi}{kp} \right) \, \exp \left\{\frac{i 2 \pi}{p}
\left[ kr \left(s^2-t^2 \right) \, + \, r \left( s-t \right) \, - \, s \, - \, t \,  \right]
\right
\} \nb \\
&& \qquad \qquad + \left. \, \exp \left \{ \frac{i 2 \pi}{p} 
\left[ kr \left(s^2-t^2 \right) \, + \, \left( r-1 \right) \left( s-t \right) \right] \right \} 
\right \} \qquad . \label{123}
\ena
The indices  $\, s\, $ and $ \, t \, $ run over a complete residue system modulo $\, p\, $. When $\,
p \, = \, 2\, $,  each sum contains only two terms and the evaluation of (\ref{123}) is
straightforward; the corresponding result is shown in equation (\ref{facile}). Let us now consider
the case in which $ \, p > 2 \, $. By means of the change of variables $\, s \rightarrow s+t\, $, 
the sum in $\, t\, $ becomes a geometric sum and one obtains 
\bea 
&&{\cal S}(k,p,r) \; = \; p \,  \sum_{s \, \left(mod \, p \right)} \left \{ \exp \left\{\frac{i 
2 \pi}{p} \left[ kr s^2  \, + \, \left(r+1 \right)s \right ] \right \}  \, \delta_p \left(2krs
\right) \right. \nb \\
&&\qquad - \, \exp \left(\frac{i 2 \pi}{kp} \right) \, \exp \left\{\frac{i 2 \pi}{p}
\left[ krs^2 \, + \, \left(r+1 \right)s  \right] \right \} \, \delta_p (2krs \, + \, 2)
\nb
\\
&&\qquad - \, \exp\left(\frac{- i 2 \pi}{kp} \right) \, \exp \left\{\frac{i 2 \pi}{p}
\left[ krs^2 \, + \, \left(r-1 \right)s  \right] \right \}  \, \delta_p (2krs \, - \, 2) \nb
\\
&&\qquad + \, \left.   \exp \left\{\frac{i 2 \pi}{p}
\left[ kr s^2  \, + \, \left(r-1 \right)s \right ] \right \} \, \delta_p \left(2krs \right)
\right \} \qquad . \label{eq:sum10}
\ena
By using properties (\ref{proper}), one can determine the values of $\, s\, $
which give contribution to (\ref{eq:sum10}). Let us start with $\, (k,p) \, =  \, 1\, $.  Clearly,
in this case   one has  
\beeq
\delta_p(2rks) \, \neq \, 0 \Rightarrow \left \{ \begin{array}{cc} s \, = \, p & \qquad p \; \mbox{
odd} \\ s \, = \, p, \, p/2 & \qquad p \; \mbox{ even} \end{array}  \right. \quad . 
\end{equation}
When $\, (k,p) \, =  \, 1\, $ and $\, p\, $ is odd, one gets
\beeq
\delta_p (2krs \; \mp \; 2) \, = \, \delta_p (krs \; \mp \; 1) \qquad .
\end{equation}
The delta function gives a non-vanishing contribution if and only if the following congruence is satisfied
\beeq
rks \, = \, \pm 1 \; \; (mod \, p) \quad . \label{eq:con}
\end{equation}
The unique solution \cite{cin} to (\ref{eq:con}) is given
 by 
\beeq
s \; = \; \pm {(rk)}^{\phi(p)-1} \qquad . 
\end{equation}
When $\, (k,p) \, =  \, 1\, $ and $ \, p \, $ is even, one finds two solutions 
\beeq 
s_1 \; = \; \pm {(rk)}^{\phi(p/2)-1} \quad , \quad  s_2 \; = \; \pm  {(rk)}^{\phi(p/2)-1} \, +
\, p/2   \qquad . 
\end{equation}
Let us now examine the case $\, (p,k)  =   g > 1\, $. We introduce the integer $\, \beta \, $ defined
by $\, p  = g \beta \, $. For $\, \beta \, $ odd, one has
\beeq
\delta_p(2krs) \;  = \; \delta_\beta (s) \qquad .
\end{equation}
Within the  residues of a complete system modulo $\, p\, $, the values of $\, s\, $ giving
non-vanishing contribution are of the form $\, s = \alpha \beta \, $ with $ \, 1 \leq \alpha
\leq g\, $. When $\, \beta \, $ is even, one gets 
\beeq
\delta_p(2krs) \;  = \; \delta_\beta (2s) \; = \; \delta_{\beta /2} (s) \qquad .
\end{equation}
The solutions of the associated congruence are 
\beeq
s \; = \; \alpha \, \frac{\beta}{2} \qquad  1 \leq \alpha \leq 2g \qquad .   
\end{equation}   
When $\, (k,p) \, = \, g >1\, $ and $\, p\, $ is odd, $\, \delta_p \left[2r(ks \pm 1) \right]\, $
does not contribute because $\, rks  =  \pm 1 \; (mod \, p)\, $ has no solutions. On the other hand,
if $\, p\, $ is even we have
\beeq
\delta_p \left [ \, \left( 2rks \; \pm \; 2 \right )\,  \right] \; = \; \delta_{p/2}(rks \pm 1)
\qquad . \label{134}
\end{equation}
The delta function (\ref{134}) is non-vanishing when $\, (p/2,k) \, = \, 1\, $ and, in this case, 
the two solutions are $\, s_1  =   \pm {(rk)}^{\phi(p/2)-1}\, $ and $ \, s_2 = s_1 +
p/2 \, $.  This exhausts the analysis of the modulo $\, p\, $ Kroneker deltas when $\, p >2\, $. 

\no
At this stage, Theorem 7.3 simply follows from the substitution of the values of $\, s \,$ for which
the various Kroneker deltas modulo $ \, p \, $ are non vanishing. In the case $ \, (k,p) =1 \, $ and
$\, p \, $ odd, the algebraic manipulations are straightforward. When $ \, (k,p) =1 \, $ and
$\, p \, $ even, the evaluation of (\ref{eq:sum10}) needs some care. In this case, one has to deal
with factors of the form 
\beeq
\exp \left[\frac{i \pi}{b} \left (a^{\phi(b)} -1 \right) \right ] \qquad ; \label{eli}
\end{equation}
with $\, b > 2 \, $ even and $\, (a, b)   = 1 \, $. In Appendix C.6, it is 
shown that terms of the
type  (\ref{eli}) are trivial because actually
\beeq
a^{\phi(b)} \; \equiv \; 1 \quad (\, mod \; 2b) \qquad .
\label{numb}
\end{equation}
Finally, the derivation of equations (\ref{44}) and (\ref{45}) is straightforward.   {\hfill \ding{111}}

\bigskip

\no
Let us now consider the dependence of $\, | \, I (L_{p/r})\, |^2 \, $ on $ \, r \, $. As shown in
equations (\ref{44}) and (\ref{45}), $\, | \, I (L_{p/r})\, |^2 \, $ depends on $\, r \, $.
However, this dependence is rather peculiar: when $ \, I (L_{p/r}) \not= 0 \,  $, $\, | \, I
(L_{p/r})\, |^2 \, $ does not depend on $\, r \, $. Indeed, when expression
(\ref{44}) is different from zero, its values are given by 
\beeq 
0 \; \not= \; \mbox{(\ref{44})} \; = \; \left \{ \begin{array} {cc} \sin^{-2}(\pi /k) & \quad \mbox{
for}
\; g =2
\; ;
\\  (g/4) \, \sin^{-2}(\pi/k)  & \quad \mbox{ for} \; g >2 \; . 
\end{array} \right.
\end{equation}
Similarly, when expression (\ref{45}) is different from zero, its value is given by
\beeq
0 \; \not= \; \mbox{(\ref{45})} \; = \; \frac{g}{2 \, \sin^2(\pi/k)} \qquad .
\end{equation} 
To sum up, when $ \, I (L_{p/r}) \not= 0 \,  $, $\, | \, I (L_{p/r})\, |^2 \, $ only depends on 
$\, p\, $ and, therefore, it is a function of the fundamental group $\, \pi_1(L_{p/r})\, =
\, Z_p \, $. Thus, Theorem~7.3 proves the validity of our conjecture for the 
lens spaces when the gauge group is $\, SU(2)\, $.

\subsection{\bf The SU(3) case}
In this section we shall present numerical computations confirming the validity of our conjecture for
lens spaces when the gauge group is $ \, SU(3) \, $.  As in the $\, SU(2)\, $ case, the $\, SU(3)\, $
Chern-Simons field theory can be  solved explicitly in any closed, connected and orientable
three-manifold \cite{gp1,gp2}.  The general surgery rules for $ \, SU(3) \, $ and for any integer
 $ \, k \, $ have been derived in \cite{gp2}. In particular, it turns out that 
\beeq  
I_{k=1}(L_{p/r})  \; =\;  \left [ \, I_{k=2}(L_{p/r}) \, \right  ]^* \; = \; I_{k=4}(L_{p/r}) 
\qquad . 
\end{equation}
Therefore, we only need to consider the case $\, k \geq 3\, $.  For $ \, k \geq 3 \,
$, the matrices  which give a projective representation  of the modular group have the following form
\bea
&&X_{(m,n) \, (a,b)} \; = \;  \frac{i}{k \sqrt{3}} q^{-2} q^{-[ (m+n) (a+b+3) + (m+3) b + (n+3)
a] /3}  \nb \\
&&\qquad  \left[ \, 1 + q^{(n+1)(a+b+2)+(m+1)(b+1)} + q^{(m+1)(a+b+2) + (n+1)(a+1)}  \right. \nb \\
&&\qquad \left. - \, q^{(m+1)(b+1)} \, - \, q^{(n+1)(a+1)} \, - \, q^{(m+n+2)(a+b+2)} \right] \quad ;
\ena
\beeq
Y_{(a,b) \, (m,n)} \; = \; q^{[m^2 + n^2 + m n + 3 (m+n)]/3} \delta_{am} \, \delta_{bn}
\qquad ;
\end{equation}
\beeq 
C_{(a,b) \, (m,n)} \; = \; \delta_{an} \delta_{bm} \qquad ;
\end{equation}
where each irreducible representation of $\, SU(3)\, $ has been denoted by  a
couple of nonnegative integers $ \, (m,n)\, $  (Dynkin labels).  

\no
By using equation (\ref{eq:def1}), we have computed $\, I_k(L_{p/r})\, $ numerically for some
examples of lens spaces. In particular, we have worked out the value of the invariant for the lens
spaces $\, L_{p/r}\, $, with $\, p \leq 20\, $ and $\, 3 \leq k \leq 50 \, $.  In all these cases,
the results are in agreement with our conjecture. 

Our calculations have been performed on a Pentium based PC running Linux. 
For instance, the results of the computations for the cases $\, L_{8/1}\, ,
\, L_{8/3}\, , \, L_{15/1}\, , \, L_{15/2}\, , \, L_{15/4} \, $ with $\, 3 \leq k \leq 50 \, $ are
shown in Tables 1, 2, 3, 4, 5.   The spaces $\, L_{8/1}\, $ and $\, L_{8/3}\, $ are not homotopically
equivalent; as shown in Tables~1 and 2, the phase of the invariant distinguishes these
two manifolds.  The case in which $ \, p=15 \, $ is more interesting because there are two different
spaces belonging to the same homotopy class;  $\, L_{15/1}\, $ and $\, L_{15/4}\, $ are
homotopically equivalent and $\, L_{15/2}\, $ represents the other homotopy class. The phase
of the invariant distinguishes the manifolds of the same homotopy class. 

\subsection{\bf Discussion}

In this Section, we have presented some arguments and numerical results 
supporting the conjecture that, for non-vanishing  $\, I({\cal M}) \, $, the absolute 
value $\, | \,  I({\cal M}) \, | \, $ only depends on
the fundamental group $\pi_1 ({\cal M}) \, $. Since the Turaev-Viro invariant 
\cite{tuvi} coincides
\cite{tur} with  $\, | \,  I({\cal M}) \, |^2 \, $,  our conjecture gives some
hints 
on the topological interpretation of the Turaev-Viro invariant.  For the gauge
group $ \, SU(2) \, $,  $\, | \,  I({\cal M}) \, |^2 \, $ can be understood as
the 
improved partition function of the Euclidean version of (2+1) gravity with positive cosmological constant \cite{arwil,tom}. In this case, our
conjecture  suggests that, for instance,  the semi-classical limit is uniquely 
determined by the fundamental group of the universe.  

Finally, one may ask for which values of $ \, k \, $ the equality  $\, I_k(M) =0\, $ is  satisfied
and what is the meaning of this fact. The complete solution to this problem is not known. From the
field theory point of view,  gauge invariance of the factor $\, \exp \left ( i S_{_{CS}} \right ) \,
$ (where $\, S_{_{SC}} \, $ is the  Chern-Simons action) in the functional measure gives nontrivial
constraints on the admissible values of $ \, k \, $ in a given manifold $ \,{\cal M} \, $. In certain
cases one finds that, in correspondence with the ``forbidden" values of  $\, k \, $, the
invariant $\, I_k({\cal M})\, $ vanishes. So, it is natural to expect that  $\, I_k({\cal M}) =0\, $ is related to
a breaking of gauge invariance for large gauge transformations. From the mathematical point of view,
$\, I_k({\cal M}) =0\, $ signals the absence of the natural extension of $\, E( {\cal L})  \, $ to an
invariant $\, E_{_{{\cal M}}}( {\cal L})  \, $ of links in the manifold $ \, {\cal M} \, $. More precisely, when 
$\, I_k({\cal M}) \not= 0\, $ for fixed integer $ \, k \, $, one can define an invariant  $\,
E_{_{{\cal M}}}( {\cal L})  \, $ of oriented, framed and coloured links $ \, \{ \, {\cal L} \subset {\cal M} \,
\}\, $ with the following property: if the link $ \, {\cal L} \, $ belongs to a three-ball embedded
in $ \, {\cal M} \, $, then one has   $\, E_{_{{\cal M}}}( {\cal L})  = E( {\cal L}) \, $. The values of
the invariant $\, E_{_{{\cal M}}}( {\cal L})  \, $ correspond to the vacuum expectation values of the 
Wilson line operators associated with links in the manifold $ \, {\cal M} \, $. When $\, I_k({\cal M}) =0\, $,
the invariant $\, E_{_{{\cal M}}}( {\cal L})  \, $ cannot be constructed; consequently, for these 
particular values of $ \, k \, $,   the quantum Chern-Simons field theory is  not well
defined in $ \, {\cal M} \, $.

\chapter{\bf Three-dimensional field theory vs two-dimensional conformal 
field theory}
\section{\bf Introduction}
In his remarkable paper \cite{wit} on quantum field theory and the Jones polynomials, Witten argued
how to solve the three-dimensional Chern-Simons (CS) theory by using certain 
results of two-dimensional conformal field theory (CFT). The basic ingredient 
entering his construction is the $S$ matrix representing the modular transformation $\tau \rightarrow - 1/ \tau$  in the 
WZNW model \cite{wzm,kiz}. 

Alternatively, one can solve, following the way discussed in the previous 
chapters, the CS model by using the 
properties of the quantum field theory which are determined by the three-dimensional action principle. In this second approach,
the existing connections between the three-dimensional topological field 
theory and certain properties
of two-dimensional CFT are found a posteriori.   In particular, all the 
relations satisfied by the $S$
matrix admit an interpretation in terms of three-dimensional topology. In this
chapter we shall describe the topological origin of these relations. 

Let us briefly recall some  results obtained in CFT, a more detailed discussion can be found in 
App.D and in the references contained therein. For simplicity,
we shall concentrate on the
two-dimensional conformal WZNW model. The fusion rules, which are defined in
terms of the OPE of primary fields, are summarized by \cite{apz}   
\beeq 
\phi_i \, \times \, \phi_j \; = \; \sum_m {C_{ij}}^m \; \phi_m \quad .
\end{equation}
Following the notations of Ref. \cite{ver}, we shall denote by $S$ and $T$ 
the generators of the
modular group acting on the linear space of the characters of the chiral 
symmetry algebra.
Verlinde conjectured \cite{ver} that the modular transformation represented by
$S$ diagonalizes the
the fusion rules,
\beeq
{C_{ij}}^m \; = \; \sum_n \frac{S_i^n \, S_j^n \, S^{\dagger m}_n}{S_0^n}
\label{fusio}
\quad . 
\end{equation}
The analytic conformal blocks can be considered \cite{frd} as holomorphic sections of a vector bundle 
$V_{g,n}$ over the moduli space of the n-punctured Riemann surface of genus $g$. As shown  by
Verlinde in the case $n=0$, the dimension of  $V_{g,0}$ can be expressed as
\beeq
{\rm dim }\, V_{g,0} \; = \; \sum_n \left( S_{n0} \right )^{2(1-g)} \quad .
\label{bund}
\end{equation}
This equation can be generalized to the case of punctured Riemann surface; in particular, one has
\cite{mos} \beeq
{\rm dim } \, V_{g,1}(i) \; = \; \sum_n \left( S_{n0} \right )^{(1-2g)} \, S_{in} \quad .
\label{dimc}
\end{equation}

We shall use three-dimensional quantum field theory arguments exclusively; all the
results we shall obtain are consequences of the properties of the vacuum expectation values of the CS
observables.  Firstly, we shall explain why the fusion algebra of two-dimensional CFT is isomorphic
with the algebra defined by the gauge invariant observables of the CS theory.   Secondly, we
shall derive equations (\ref{fusio})-(\ref{dimc}). 

\section{\bf Fusion algebra and reduced tensor algebra}

The analytic correlation functions of primary fields in the conformal WZNW model have monodromy
properties described by the  Knizhnik-Zamolodchikov equation \cite{kiz}.  The 
associated braid group representations have been studied in great details; as 
shown by Kohno \cite{koh}, these
monodromy representations are equivalent to the $R$-matrix representations 
defined in terms of the so-called quantum deformations of the simple Lie 
algebras. On the other hand, the skein relations
satisfied by the expectation values of the CS theory can be associated with braid group
representations which, again, are equivalent to the $R$-matrix representations \cite{glib}. Thus, the
braiding structures which are found in conformal field theory, in the quantum group approach and in
the CS theory coincide. In fact, as shown by Drinfeld, the braid representations determined by the
quasi-tensor category associated with the quasi-triangular quasi-Hopf algebras are universal
\cite{drin}. 

\begin{figure}[h]
\begin{picture}(10,10)
\put(120,-40){$\psi[i]$}
\put(183,-35){$\psi[j]$}
\put(337,-30){$\psi[i] \, \psi[j]$}
\put(130,-135){$\phi_i$}
\put(160,-130){$\phi_j$}
\put(310,-135){$\phi_i \times \phi_j$}
\end{picture}
\vskip 0.5truecm 
\centerline{\epsfig{file= \path 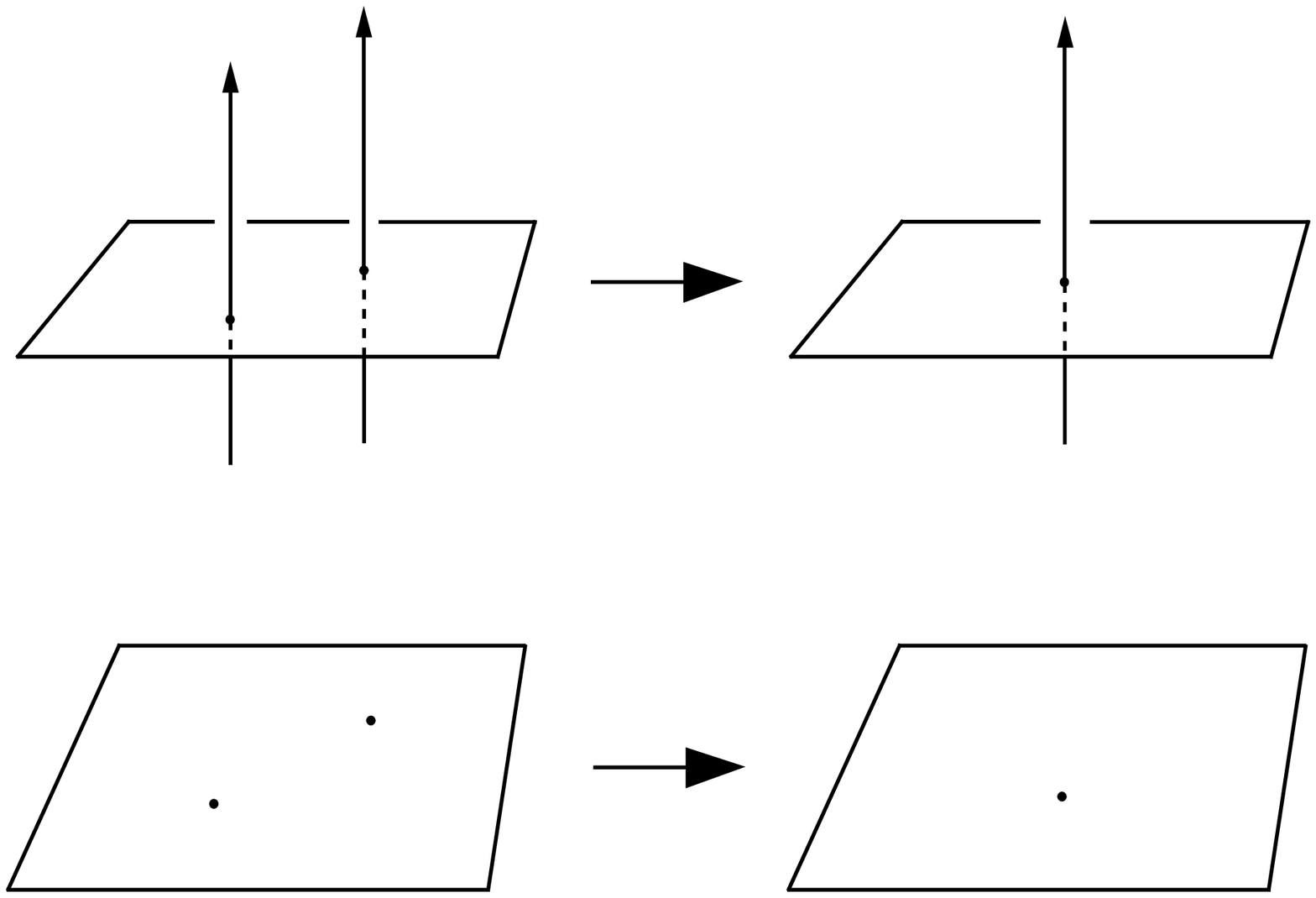,height=5cm,width=10cm}}
\vskip 0.5truecm 
\centerline{\bf Figure 8.1}      
\vskip 0.5truecm 
\end{figure}

As shown in Fig.8.1, the fusion rules of the CFT correspond to the
algebraic structure of the satellite relations in the three-dimensional CS
theory. 

\noindent In our context, Drinfeld's universality theorem implies that the 
fusion
algebra of the  $SU(N)_\ell $ ~WZNW model of level $\ell $ is isomorphic with the
reduced tensor algebra ${\cal T}_{(k)}$ of the $SU(N)$ CS theory with $k = \ell
+ N$. The proof is based on the following considerations. 
The Knizhnik-Zamolodchikov  equation \cite{kiz} in the WZNW model determines a
flat connection of the type considered by Kohno; moreover, the Sugawara form of
the energy-momentum  tensor implies that the resulting deformation parameter
$q_W$ is given by  \beeq 
q_W\; = \; \exp \left [ -i 2\pi /(\ell + N) \right ] \quad . 
\label{eq:def}
\end{equation}
The fusion algebra is generated by a primary field $\phi_N$ associated with the fundamental
representation of $SU(N)$; therefore, the monodromy properties of a generic correlator of primary
fields are determined by the correlators of $\phi_N$. With the value (\ref{eq:def}) of the
deformation parameter, the fusion rules close on a finite set of conformal blocks which are labeled
by certain irreducible representations of $SU(N)$.

In the $SU(N)$ CS theory, the deformation parameter $q$ is 
\beeq 
q \, = \, \exp ( -i 2\pi /k) \quad . 
\end{equation}
The components of the links are labeled by the irreducible
representations of $SU(N)$. In Sect. we have seen that the value of any 
observable associated with generic representations of
$SU(N)$ is uniquely determined  by the set of observables 
associated with the fundamental
representation of $SU(N)$. The three-dimensional counterpart of the fusion 
property of the
primary fields is represented by the structure of the satellite relations. By comparing the
deformation parameters of the conformal theory and of the CS theory, one finds that they coincide
when $k = \ell + N$. With this fixed integer value of $k$, see Chap.4, the
algebraic structures associated with the satellite relations of the CS theory 
close on a finite set
of physically inequivalent representations. These representations
identify the equivalence classes belonging to the reduced tensor algebra ${\cal T}_{(k)}$. The
structure constants of ${\cal T}_{(k)}$ characterize the satellite relations 
and then they correspond
to the fusion rules of the conformal theory.  Therefore, the fusion algebra 
of the conformal theory
with level $\ell $ must be isomorphic with the reduced tensor algebra ${\cal T}_{(k=\ell +N)}$. 

In the case in which the gauge group is $SU(2)$, the equivalence between the 
structure
constants of the reduced tensor algebra ${\cal T}_{(k=\ell +2)}$ and the fusion rules of the
$SU(2)_\ell $ ~WZNW model  has been established explicitly \cite{guad2}. When 
$G=SU(3)$, one can verify
that,  for any fixed level $\ell $,  the fusion algebra of the WZNW model is isomorphic with
the $SU(3)$ reduced tensor algebra ${\cal T}_{(k=\ell +3)}$ \cite{gp1}. 

It should be noted that the relation between the two-dimensional WZNW model 
and the
three-dimensional CS theory only concerns the algebraic structure of the 
monodromies or,
equivalently, of the braid group representations. In fact, the physical 
contents of the two theories
are completely different. For example, the Hamiltonian of the  WZNW model is 
not vanishing, whereas
the Hamiltonian of the CS theory is vanishing (after gauge fixing, the Hamiltonian of the CS theory
is a  BRS-variation). Differently from the WZNW model, in the CS theory there are no propagating 
physical degrees of freedom.  

To sum up, the isomorphism between the fusion algebras of the conformal models and the reduced
tensor algebras of the CS theory makes  clear the role played by the representation ring of the groups
in the derivation of the fusion rules. Actually, this isomorphism can also be used to compute the
fusion rules. 

\section{\bf Modular group}
 
In this section, we shall show that a projective representation of the  
modular group is defined  
on ${\cal T}_{(k)}$; we shall also derive the Verlinde formula (\ref{fusio}). 
The  two $D\times D$-matrices
representing the generators of the modular group will be denoted by  $\{ X \, , \, Y \,
\}$, where $X $ corresponds to the $S$ matrix of the  conformal models and $Y $ is the analogue of
the  $T$ matrix.  The $X $ matrix is introduced according to  
\beeq X_{ij} \; = \; \frac{H[i,j]}{\langle \; W( \, U;\Psi_0 \, )
\;\rangle \bigr|_{S^3}} \; = \; a^(k) \; H[ i,j ] \qquad . 
\label{eq:mo1}
\end{equation}  As we have already mentioned, an elementary $+1$ modification 
of the framing of a link
component with colour $\psi [i]$ is represented in ${\cal T}_{(k)}$ by 
\beeq
\psi[i] \; \ra \; q^{Q(i)} \; \psi[i] \quad .
\end{equation} Therefore, we shall define the $Y$ matrix as
\beeq Y_{ij}  \; = \;  q^{Q(i)}  \; \delta_{i\, j}\quad . \label{eq:mo2}
\end{equation} 
Finally, we shall denote by $C$ the ``charge conjugation" matrix 
\beeq C_{ij} \; = \; \delta_{i\, j^\ast} \quad .
\end{equation}

\bigskip

\shabox{{\bf Theorem 8.1}} ~{\em The matrices
$\{ \, X \, , \, Y \, , \, C \, \} $ satisfy the relations}
\beeq
 X^2  \; = \; C \qquad;
\label{eq:f1} 
\end{equation}
\beeq
\left( \, X \> Y \, \right )^3 \; = \; e^{- \, i \, \theta} \> C \quad .
\label{eq:f2} 
\end{equation} 

\bigskip

\no {\bf Proof} Equation (\ref{eq:f1}) follows from Theorem 6.6. Indeed, by 
setting $m=1$  in equation
(\ref{eq:pr2}),   one finds 
\beeq N_{ij1} \; = \; a(k) \; \langle \, W( \, U \, ; \, \Psi_0 \, )\, 
 W( \, C_1 \,   ; \, \psi[i] \, ) \, W( \, C_2 \, ; \, \psi[j] \, )\,  \rangle \bigr|_{S^3 }
\quad , 
 \label{eq:f3} 
\end{equation} where the link components $\{ \, U, \,  C_1,\,  C_2 \, \}$ are 
shown in Figure 8.2. 

\begin{figure}[h]
\begin{picture}(10,10)
\put(295,-145){$\psi[j]$}
\put(175,-145){$\psi[i]$}
\end{picture}
\vskip 0.5 truecm 
\centerline{\epsfig{file=\path 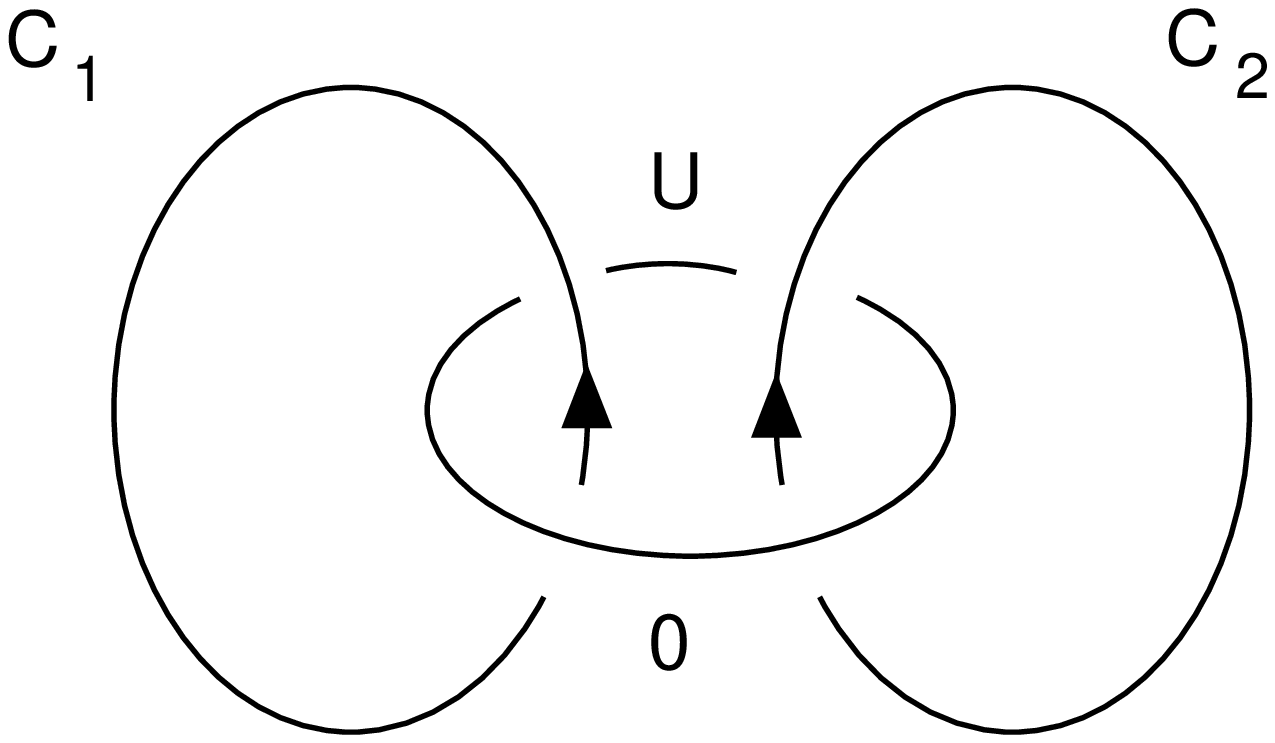,height=4cm,width=8cm}}
\vskip 0.9truecm 
\centerline{{\bf Figure 8.2}}
\end{figure}

By using the
connected sum rule, Eq. (\ref{eq:f3}) can be rewritten as
\bea N_{ij1} \; = \; \delta_{i \, j^\ast } \; = \; C_{ij} \; &&= \; {a(k)}^2
 \, \sum_m H[i, \, m]
\; H[m,
\, j] \nb \\ && = \;  \left( \, X^2 \, \right )_{ij}  \quad ,  
\ena 
which shows that eq. (\ref{eq:f1}) is satisfied. 

In order to prove equation (\ref{eq:f2}), let us decompose $(\, X  \,  Y \, )^3$ as
\beeq
\left( \, X   \, Y \, \right)^3_{ij} \; = \; \sum_m 
\left( \, X   \, Y \, X \, \right)_{im} \;  \left( \, Y  \,
 X \, Y \, \right)_{mj}
\quad .  
\label{eq:90} 
\end{equation} 
By using the connected sum rule, one can represent  $\, a(k)  \left( \, X   \, Y 
\, X \, \right)_{im}$ by means of the CS observable associated with the link 
in $S^3$  shown in
Figure 8.3. 

\begin{figure}[h]
\begin{picture}(10,10)
\put(310,-45){$\psi[m]$}
\put(135,-45){$\psi[i]$}
\end{picture}
\vskip 0.5 truecm 
\centerline{\epsfig{file=\path 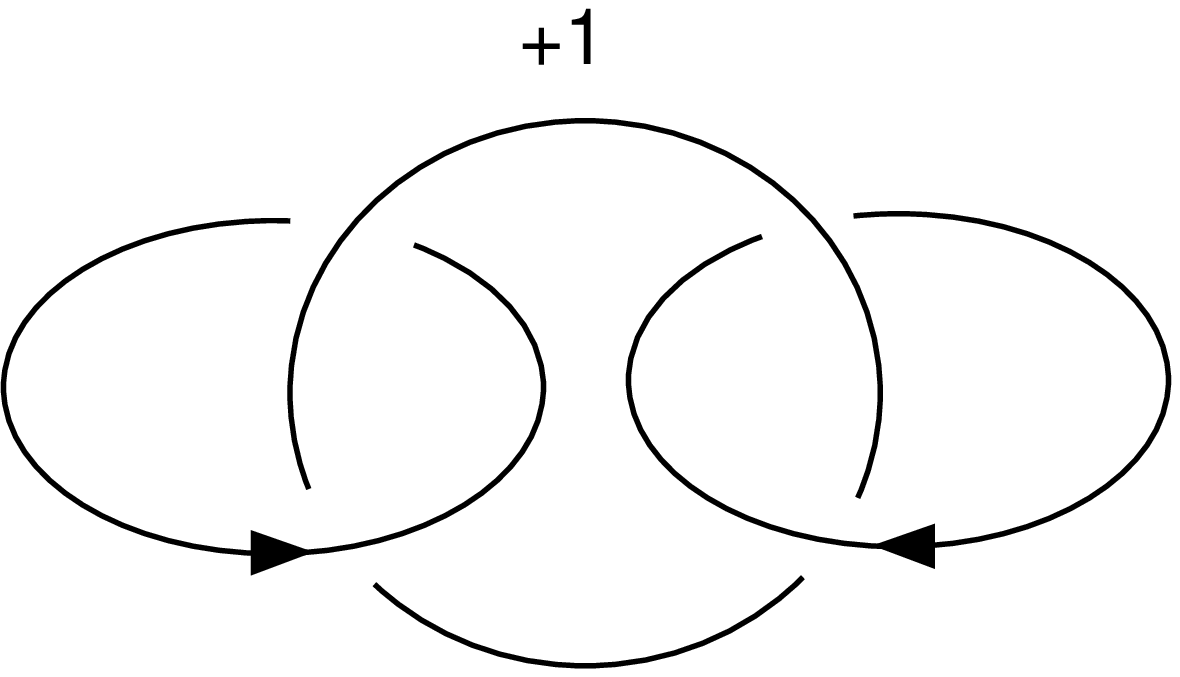,height=4cm,width=8cm}}
\vskip 0.5truecm 
\centerline{{\bf Figure 8.3}}
\end{figure}

\noindent From the definitions (\ref{eq:mo1}) and (\ref{eq:mo2}) it follows 
that  
\beeq
\left( \, Y  \, X \, Y \, \right )_{mj} \; = \; a(k) \; q^{\left[ Q(m) + Q(j)
\right]} \; H[m,\, j] 
\quad . 
\label{eq:91} 
\end{equation}

\noindent Consequently, the whole matrix $a(k) \left ( X  \, Y \, \right )_{ij}^3$ can be
represented by the expectation value of the observable depicted in Figure 8.4.

\begin{figure}[h]
\begin{picture}(10,10)
\put(300,-135){$\psi[j]$}
\put(135,-115){$\psi[i]$}
\end{picture}
\vskip 0.5 truecm 
\centerline{\epsfig{file=\path 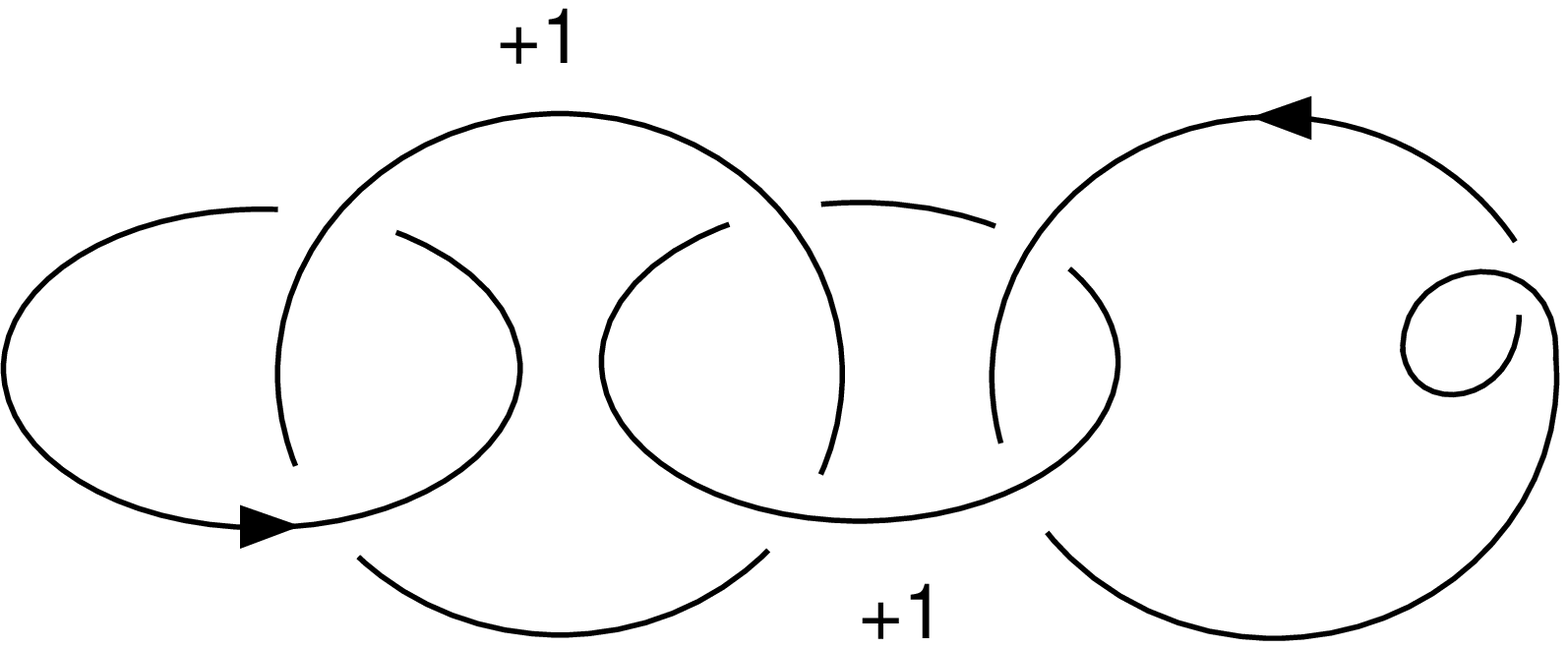,height=4cm,width=8cm}}
\vskip 0.5truecm 
\centerline{{\bf Figure 8.4}}
\end{figure}

\noindent From the property of the surgery discussed in Sect.6-2 and Sect.6.5, it follows that the effect produced by the surgery operator associated with the 
unknot can be represented as in Fig.8.5. Thus, by using the relation shown in 
Fig.8.5, one finds that the link of Figure 8.4 is
equivalent to the link shown in Figure 8.6. 

\begin{figure}[h]
\begin{picture}(10,10)
\put(310,-20){$\psi[j]$}
\put(170,-30){$\psi[i]$}
\put(230,-63){$e^{-i \theta_k}$}
\end{picture}
\vskip 0.5 truecm 
\centerline{\epsfig{file= \path 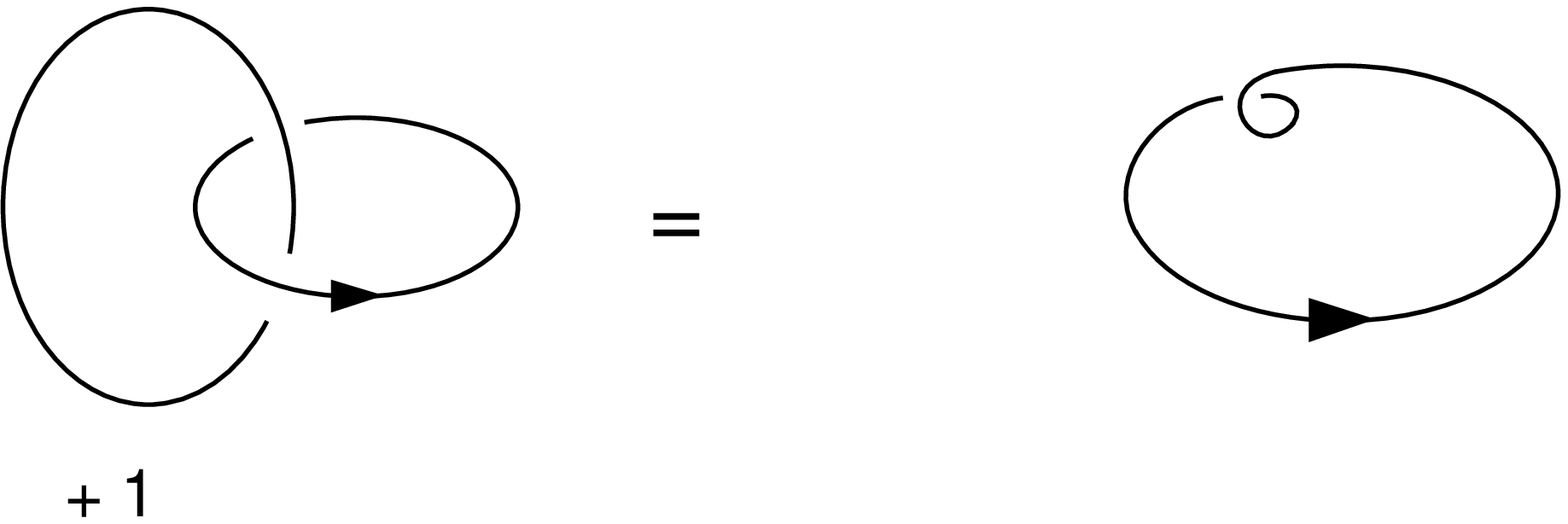,height=4cm,width=8cm}}
\vskip 0.2 truecm 
\centerline {{\bf Figure 8.5}}      
\end{figure}
\begin{figure}[h]
\begin{picture}(10,10)
\put(320,-115){$\psi[j]$}
\put(135,-115){$\psi[i]$}
\put(100,-80){$e^{-i \theta_k}$}
\end{picture}
\vskip 0.5 truecm 
\centerline{\epsfig{file= \path 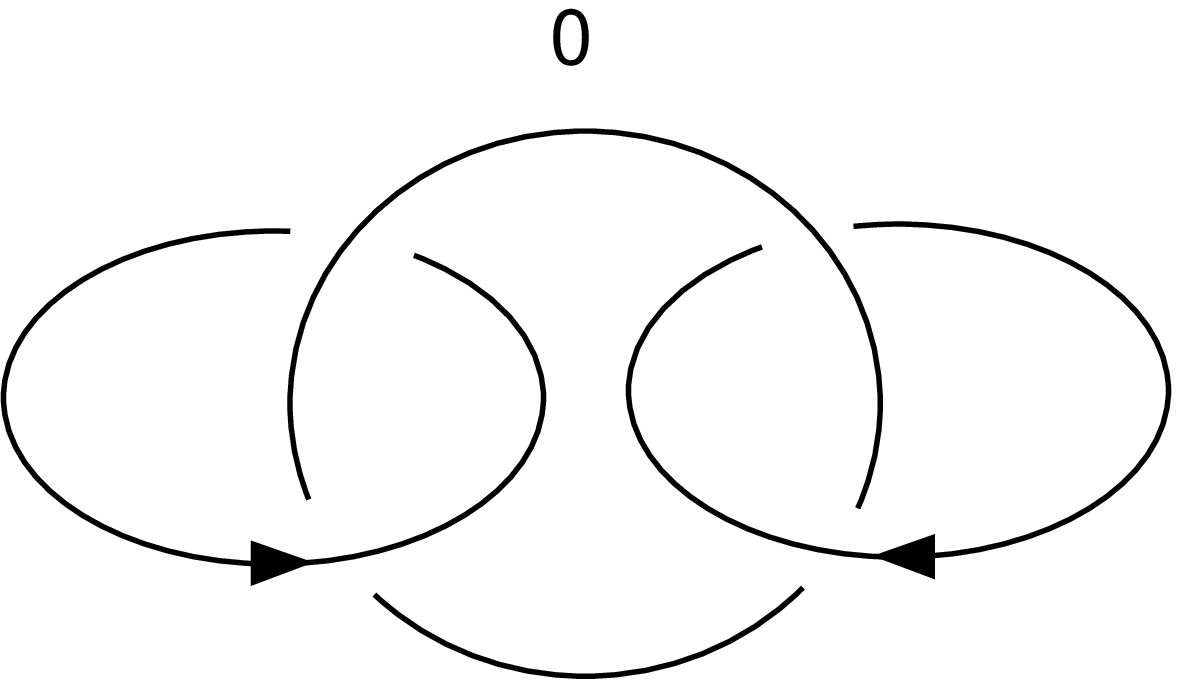,height=4cm,width=7cm}}
\vskip 0.5 truecm 
\centerline {{\bf Figure 8.6}}      
\vskip 0.5truecm 
\end{figure}

At this stage, from the satellite relation
(\ref{satg}) and Theorem 6.5, it follows that 
\beeq 
\left( \, X \> Y \, \right )_{ij}^3 \; = \; e^{- \, i \, \theta_k} \> \delta_{i \, j^\ast }
\; = \; e^{- \, i \, \theta} \> C_{ij} \quad ;  
\end{equation} 
this concludes the proof of equation (\ref{eq:f2}). {\hfill \ding{111}}
\bigskip

By means of the matrices  $\{ \, X \, , \, Y \, , \, C \, \} $, which give 
the projective
representation (\ref{eq:f1}) and (\ref{eq:f2}) of the modular group, one can 
easily construct a
representation of the modular group in agreement with the standard conventions
of CFT.    

Let us now derive the Verlinde formula (\ref{fusio}). According to 
Theorem 6.6, the structure constants $\{ \,
N_{ijm} \, \}$ of the reduced tensor algebra ${\cal T}_{(k)}$ are given by the expectation value  in
$S^2 \times S^1$ of the CS observable associated with the link  components  
$ \{ \, C_1 , \, C_2 , \,
C_3 \, \}$ shown in Figure 6.19.   The whole link  of Figure 6.19 is defined 
in $S^3$ and has four
components because one component represents a surgery instruction; this link  
can be understood as a
(double) connected sum of three Hopf links. Thus, by using the connected sum 
rule (\ref{eq:csr}), one finds  
\beeq  N_{ijm^\ast } \; = \; {a(k)}^2 \, \sum_n \left(\, E_0[n] \,\right)^{-1} \; H[i,n] \; H[j,n]
\; H[m,n] \quad . \
\label{eq:mo7}
\end{equation}
\noindent Let us recall that each link component with colour $\psi [\mathbf{}1] $ can be eliminated;
consequently, one has $H[1,n] \, = \, E_0 [n]$. By taking into account the definition (\ref{eq:mo1})
of the
$X $ matrix, equation (\ref{eq:mo7}) can be written as 
\beeq  N_{ijm^\ast } \; = \;   \sum_n \frac{X_{in} \, X_{jn} \, X_{mn}}{X_{1n}}
\quad .  \label{eq:999}
\end{equation} Equation (\ref{eq:999}) coincides with the Verlinde formula 
(\ref{fusio}). 
\section{\bf Punctured Riemann surfaces}

In order to derive equations (\ref{bund}) and (\ref{dimc}), one needs to find 
the three-dimensional counterpart of the
dimensions of the bundles $V_{g,n}$. The braiding properties of the 
correlators of primary fields on
a Riemann surface  $\Sigma_g$ of genus $g$  coincide with the braiding 
properties of  Wilson line
operators associated with links which are defined in the three-manifold 
$\Sigma_g \times S^1$ and
whose components run along the ``time" direction determined by the $S^1$ 
component of the manifold.  
By using Dehn's surgery on $S^3$, one can easily produce a surgery presentation 
of these links in 
$\Sigma_g \times S^1$. The analogue of ${\rm dim } \, V_{g,0}$ is represented by the CS ``partition
function" in $\Sigma_g \times S^1$; whereas,  ${\rm dim } \, V_{g,n}$ corresponds to the (properly
normalized) expectation value in $\Sigma_g \times S^1$ of the Wilson operator associated with links
describing $n$ static punctures on $\Sigma_g$. Let us now recall how the manifold $\Sigma_g
\times S^1$ can be represented \cite{guad4} by means of Dehn's surgery on $S^3$.
Since we wish to
consider arbitrary values of the genus $g$, it is convenient to start with the manifold 
$S^2 \times S^1 \equiv \Sigma_0 \times S^1$ and give a constructive method to
``add handles" to it. 

In two dimensions,  the Riemann surface $\Sigma_g$ can be constructed by 
adding $g$ handles to the
sphere $S^2$. As shown in Figure 8.7, adding one handle to a Riemann surface 
is equivalent to
introduce two surgery punctures on it. Equivalently, by fusing these two punctures, adding one handle
to $\Sigma_g$ is equivalent to introduce a single surgery puncture on $\Sigma_g$.   

\begin{figure}[h]
\vskip 0.9 truecm 
\centerline{\epsfig{file= \path 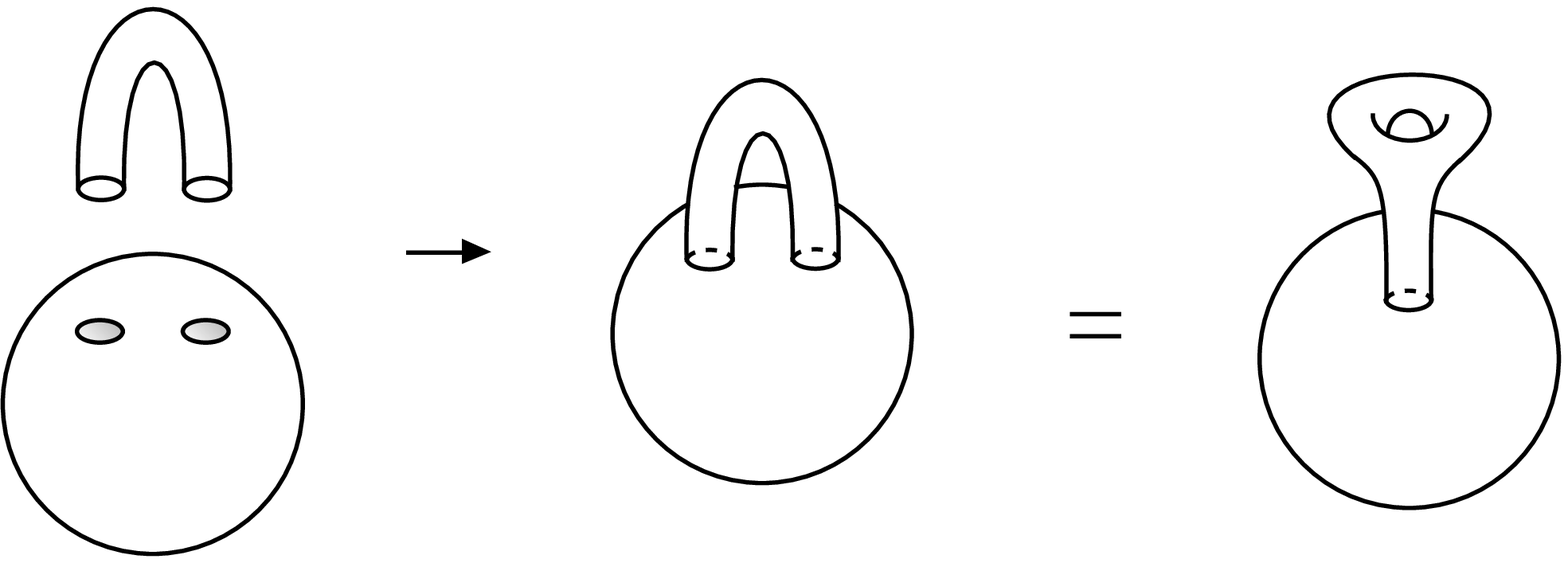,height=4cm,width=9cm}}
\vskip 0.5truecm 
\centerline {{\bf Figure 8.7}}      
\vskip 0.5truecm 
\end{figure}

\no   Let us now consider the construction of $\Sigma_g \times S^1$ by adding 
handles in the
three-dimensional context.  The starting manifold is $S^2 \times S^1$  which 
corresponds to the
surgery link given by the unknot $U$ in $S^3$ with surgery coefficient $r=0$. 
The operation of
adding one handle can be  represented \cite{guad4} 
(in the three-dimensional context) by the
introduction of two new surgery components with vanishing surgery coefficients; these two components
must be added to the unknot $U$ as shown in Figure 8.8.  The surgery link of 
Figure 8.8  is defined in $S^3$ and corresponds to  manifold $\Sigma_1 \times 
S^1$.   

\begin{figure}[h]
\vskip 0.9 truecm 
\centerline{\epsfig{file= 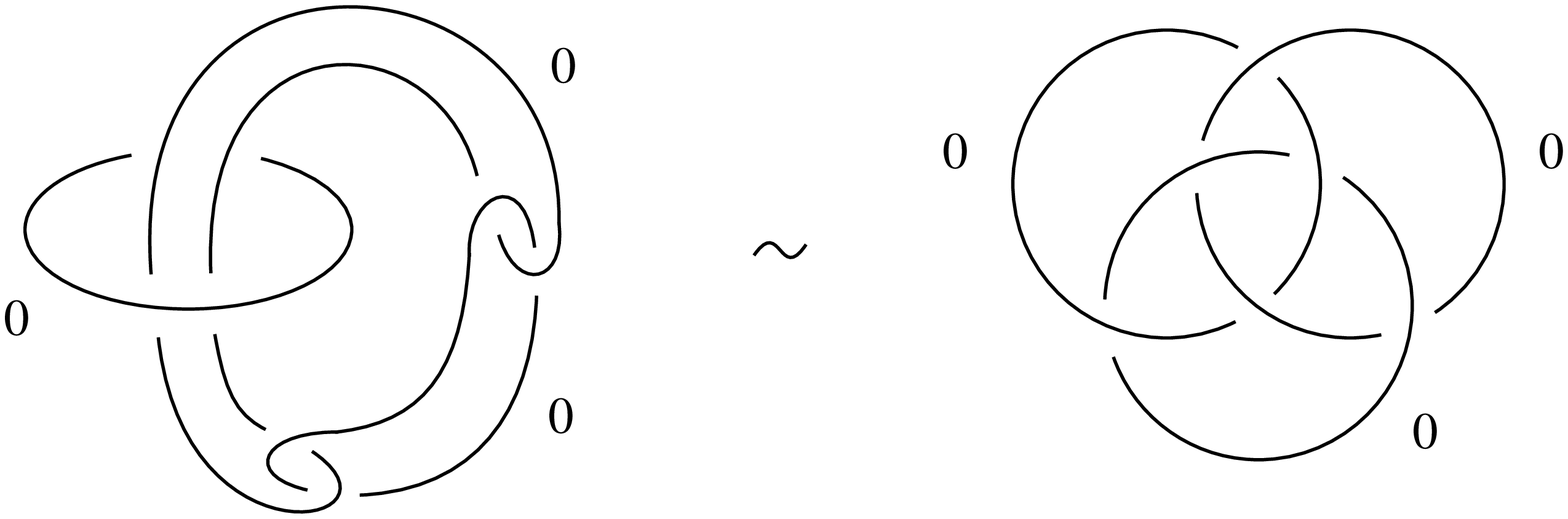,height=5cm,width=8cm}}
\vskip 0.5 truecm 
\centerline {{\bf Figure 8.8}}      
\vskip 0.5truecm 
\end{figure}

\noindent For each handle, one has to introduce a copy of the two surgery components (with vanishing
surgery coefficients) mentioned before.  One can imagine that these two surgery components belong 
to a solid torus standardly embedded in $S^3$; inside this solid torus, these two components define
a link which will be called $P$. The two components of $P$ have colour given by $\Psi_0$.   In
agreement with equation (\ref{eq:gid}), the Wilson line operator associated 
with  the coloured link
$P$, which belongs to the complement solid torus $N$ of the unknot $U$ in $S^3$, can be decomposed
as 
\beeq  W(\, P\, ;\, \Psi_0 \, , \, \Psi_0 \, ) \; = \; W(\, K\, ;\, \psi_h \, ) \quad , 
\label{eq:xz1}
\end{equation} where $K$ is the core of $N$, shown in Figure 8.9, and 
\beeq 
\psi_h \; = \; \sum_i \, \eta (i) \, \psi [i] \quad . 
\end{equation}

\begin{figure}[h]
\begin{picture}(10,10)
\put(310,-150){$\psi_h$}
\end{picture}
\vskip 0.5truecm 
\centerline{\epsfig{file= \path 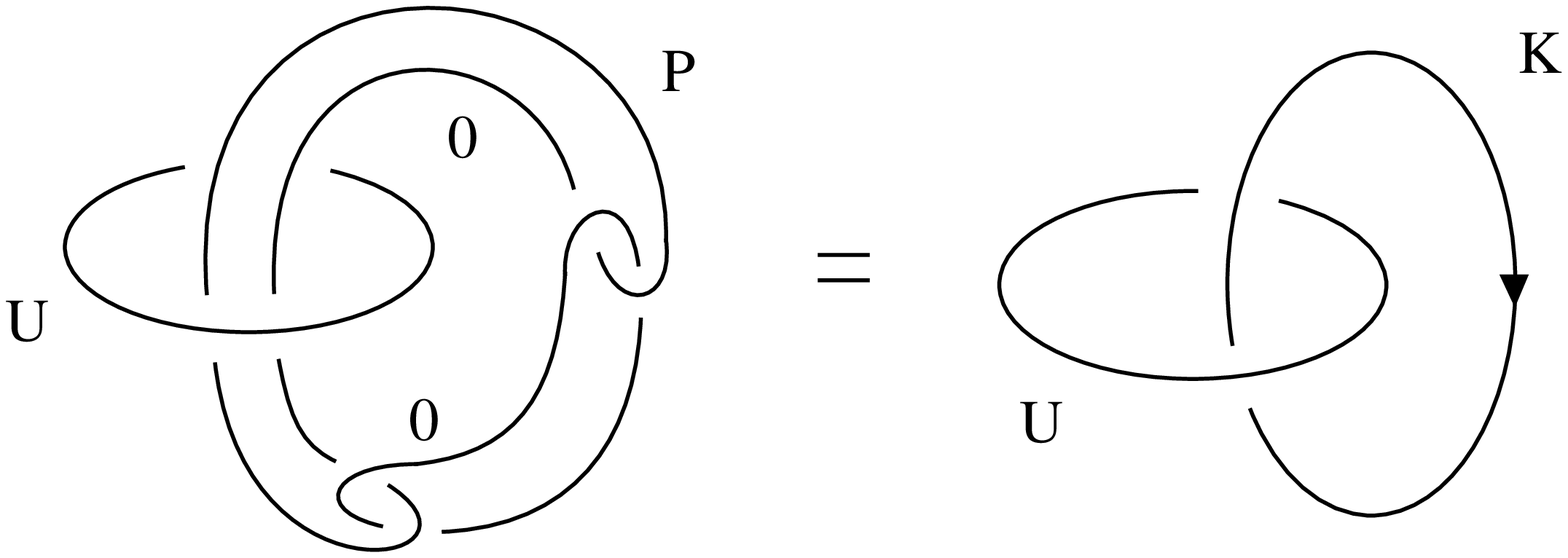,height=5cm,width=8cm}}
\vskip 0.5truecm 
\centerline{\bf Figure 8.9}      
\vskip 0.5truecm 
\end{figure}

\no The coefficients $\{ \, \eta (i) \, \}$ give the decomposition of the 
surgery operator, which
represent one handle, in terms of a single coloured link component. Consequently, adding $g$ handles
to $S^2 \times S^1$ is equivalent to adding $g$ new components with colour $\psi_h$ to the unknot
with surgery coefficient $r=0$.  The introduction of the surgery operator 
$W(\, P\, ;\, \Psi_0 \, , \, \Psi_0 \, )$ by means the decomposition (\ref{eq:xz1}) is the analogue
of the introduction (in two dimensions) of a single surgery puncture  on $S^2$. In order to compute
the expectation values of observables in 
$\Sigma_g \times S^1$, the knowledge of $\{ \, \eta (i) \, \}$ is crucial.   Quite remarkably, the
topological properties of the one handle surgery operator are encoded entirely in the structure
constants of ${\cal T}_{(k)}$.  

\bigskip

\shabox{{\bf Theorem 8.2}}~
{\em The coefficients $\{ \, \eta (i) \, \}$ are given by} 
\beeq 
 \eta (i) \; = \; \sum_j \, N_{j j^\ast i} \quad . 
\label{eq:xz2}
\end{equation}

\bigskip

\no {\bf Proof}~Suppose that a Wilson line operator is associated with the component $U$, shown in Figure 8.9, with
colour $\psi [j]$. The expectation value in $S^3$ of the observable associated with $P$ and $U$ can
be expressed as shown in Figure 8.10. 

\begin{figure}[h]
\begin{picture}(10,10)
\put(90,-57){$\sum_i \eta(i) \; H[i, \, j]$}
\put(240,-75){$\psi[j]$}
\put(130,-152){$\frac{1}{a(k)} \; \sum_m E_0[m]$}
\put(240,-165){$\psi[j]$}
\put(350,-120){$\psi[m]$}
\put(180,-240){$\sum_m$}
\put(250,-263){$\psi[j]$}
\put(370,-240){$\psi[m]$}
\put(340,-280){$\psi[m]$}
\end{picture}
\vskip 0.5truecm 
\centerline{\epsfig{file= \path 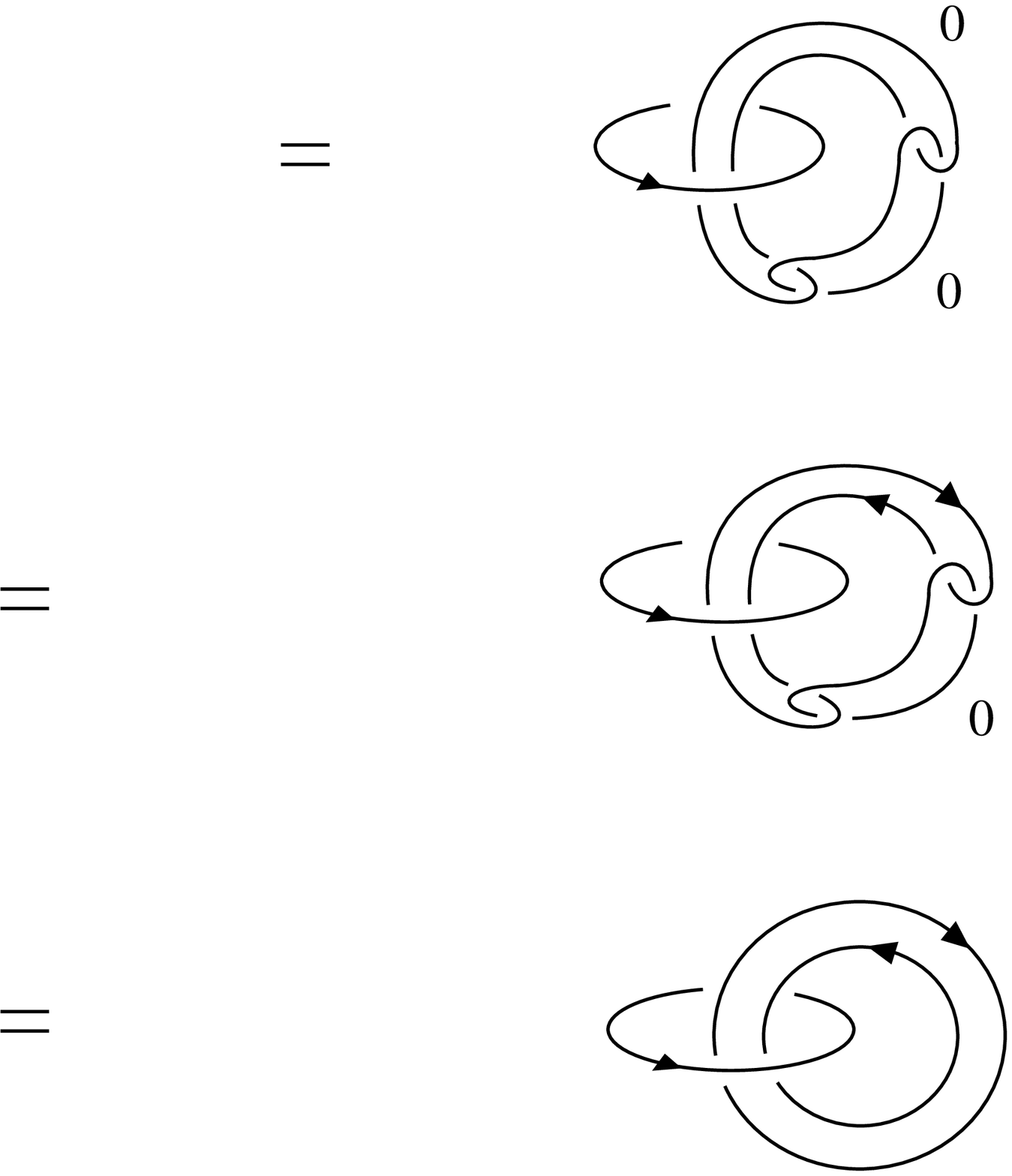,height=9cm,width=10cm}}
\vskip 0.5truecm 
\centerline{\bf Figure 8.10}      
\vskip 0.5truecm 
\end{figure}
\begin{figure}[h]
\begin{picture}(10,10)
\put(280,-80){$\frac{a(k)}{E_0[i]}$}
\put(120,-20){$\psi[i]$}
\put(180,-20){$\psi[i]$}
\put(350,-10){$\psi[i]$}
\put(350,-130){$\psi[i]$}
\end{picture}
\vskip 0.5truecm 
\centerline{\epsfig{file= \path 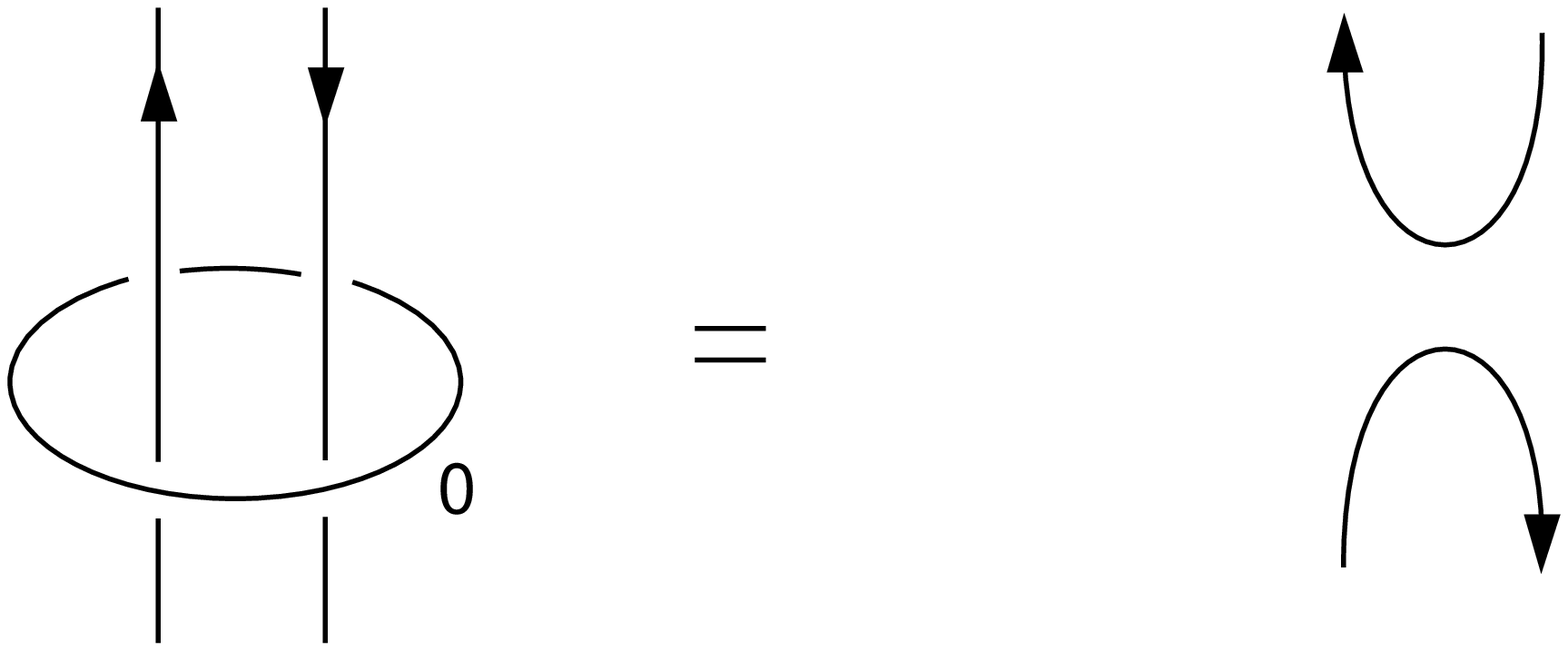,height=4cm,width=8cm}}
\vskip 0.5truecm 
\centerline{\bf Figure 8.11}      
\vskip 0.5truecm 
\end{figure}

\noindent In the last step of  Fig.~8.10, we have used the relation between 
tangles shown in Figure~8.11;
this relation can easily be understood by using the following argument.  

\no Let us first consider the tangle on the left-hand-side of Figure 8.11. 
From Theorem 6.5, we know
that the operator associated with the unknot with colour $\Psi_0$ selects, 
among all the possible 
physical states running along the core of the solid torus $N$ determined by the complement of the
unknot in $S^3$, the element $\psi[\mathbf{1}]$.  The tangle appearing on the 
right-hand-side of Figure 8.11 
(understood as subset of the same solid torus $N$) contains, as propagating state along the core of
the solid torus, only $\psi[\mathbf{1}]$. Therefore, the two tangles must be proportional. The normalization
constant can be determined by taking the closure of both tangles; the result is shown in Figure 8.11. 

From Figure 8.10 and the satellite relation (\ref{satg}), it follows that 
\beeq
\sum_i \, \eta(i) \; H[i,j] \; = \; \sum_{m,n} \, N_{m m^\ast n} H[n,j] 
\quad .\label{eq:eta}
\end{equation} Let us recall that, when the reduced tensor algebra is regular, the Hopf matrix $H$ is
invertible and, in particular, from eq.(\ref{eq:f1}) one obtains 
\beeq  H^{-1}[i,j] \; = \; {a(k)}^2 \, H[i,j^\ast] \quad .
\end{equation} Thus, the solution of the linear system (\ref{eq:eta}) is given by eq.~(\ref{eq:xz2}). {\hfill \ding{111}}
\bigskip
 
By using the Verlinde formula (\ref{eq:999}) and $X^2 = C$, the $\eta$-coefficients can also be
expressed as
\beeq
\eta(i) \; = \; \sum_n \frac{X_{ni^\ast}}{X_{1 n}} \; = \; \sum_n \frac{H[n, \, i^\ast]}{E_0[n]} 
\quad .
\end{equation}

In conclusion, the surgery link in $S^3$ corresponding to the manifold 
$\Sigma_g \times S^1$ is
shown in Figure 8.12. This link has $(g+1)$ components; one component is the unknot with surgery
coefficient $r=0$ and the remaining $g$ components, which have colour $\psi_h$,  represent the $g$
handles which must be added to $S^2 \times S^1$. 

\begin{figure}[h]
\begin{picture}(10,10)
\put(315,-35){$\psi_h$}
\put(280,-45){$\psi_h$}
\end{picture}
\vskip 0.5truecm 
\centerline{\epsfig{file= \path 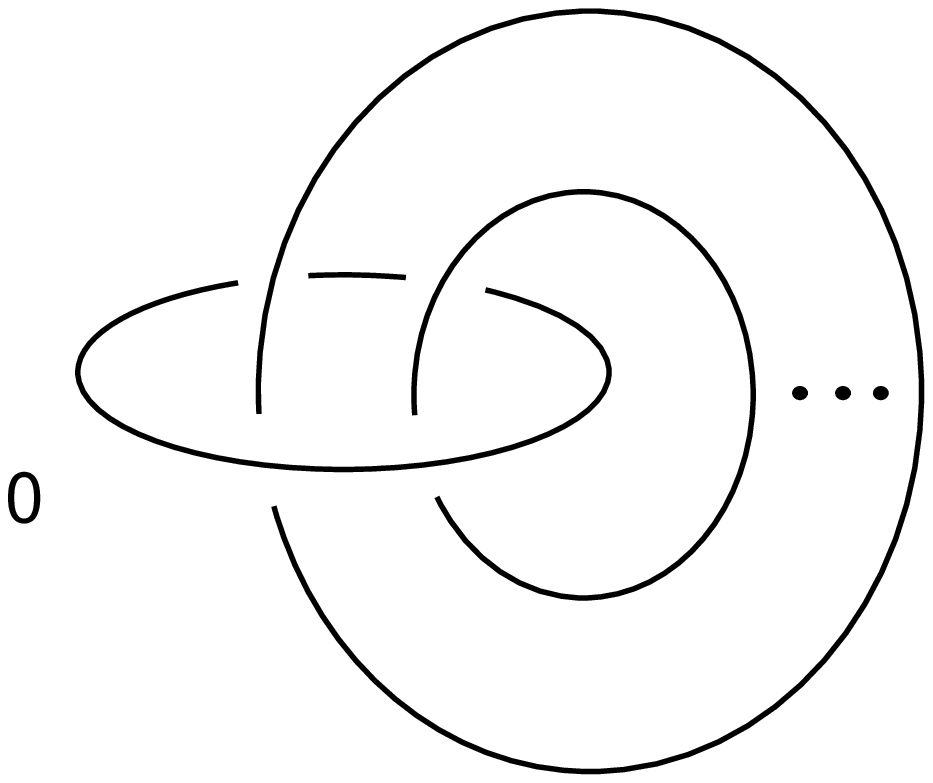,height=4cm,width=6cm}}
\vskip 0.5truecm 
\centerline{\bf Figure 8.12}      
\vskip 0.5truecm 
\end{figure}
 
\section{\bf Dimensions of the physical state spaces}

Let us consider the CS theory defined in $\Sigma_g \times S^1$; the $S^1$ 
component of the
manifold can be interpreted as a compactified time-interval.   Since the CS Hamiltonian is vanishing
on the {\it physical} state space ${\cal H}(g)$ of the theory, the partition function of the CS
theory defined in  $\Sigma_g \times S^1$ is proportional to the dimension  of ${\cal H}(g)$. The
dimension of ${\cal H}(g)$ is equal \cite{guad4} to  the number of 
independent generalized characters
of the corresponding CFT on $\Sigma_g$, i.e.  ${\rm dim }\, {\cal H}(g) = {\rm dim } \, V_{g,0}$. 
In order to find the proportionality constant between  ${\rm dim }\, {\cal H} 
(g)$ and 
${\cal I}(\Sigma_g \times S^1)$, we note that, as a consequence of 
Theorem 6.5, one has
\beeq  {\rm dim }\, {\cal H}(g = 0)\;  =\; 1 \quad .
\end{equation} Therefore, there exists the following relation between 
 ${\rm dim }\, {\cal H}(g) $ and the 3-manifold invariant $I(\Sigma_g \times
S^1)$ introduced in Chap.7
\bea {\rm dim }\, {\cal H}(g) &&  = \frac {{\cal I}(\Sigma_g \times S^1)}{{\cal I}(S^2 \times S^1)}
 \nb \\ && = \;  \frac {\langle \, W(U; \,
\Psi_0) \, W(C_1\, ;  \, \psi_h , ) \, W(C_2\, ;  \, \psi_h , ) 
\cdots  W(C_g \, ;  \, \psi_h \, ) 
\, \rangle \bigr |_{S^3}}{\langle \, W(U \, ; \, \Psi_0)
\, \rangle \bigr |_{S^3}}  \quad .  \label{eq:cise}
\ena  
The numerator appearing in eq.(\ref{eq:cise}) is the expectation value of the 
surgery operator
describing $\Sigma_g \times S^1$ and the denominator is the expectation value 
of the surgery
operator  associated with $S^2 \times S^1$. 
We have denoted by $\{ \, C_1\, , \, C_2 \, , ..., \, C_g
\, \}$ the surgery components, representing the $g$ handles, shown in 
Figure 8.12.
 
In order to obtain a general formula for ${\rm dim } {\cal H}(g)$, we shall now derive a  rule 
which allows us to eliminate the surgery components $\{ \, C_1 \, , \,  C_2 \, , ..., \, C_g \,
\}$.  Let us consider the Hopf link in $S^3$ with components $U$ and $C$. Let the $U$ component be
coloured with a generic element $\zeta \in {\cal T}_{(k)}$ and the $C$ component with $\psi_h$. By
using the connected sum rule (\ref{eq:csr}), we have 
\bea
\langle \, W(\, U,C; \, \zeta, \psi_h \, ) \, \rangle \bigr |_{S^3}\; && = \; \sum_{i,j} H[i^\ast,j]
\;  E_0^{-1}[j] \;  \langle \, W(\, U,C; \, \zeta, \psi[i] \, )
 \,  \rangle \bigr  |_{S^3} \nb \\ && = \; \sum_{i,j} \langle \, W(\, H \# H; \, \zeta, \psi[i],
\psi[j]  \, )
\, \rangle \bigr |_{S^3} \;  E_0[i]  \;   E_0^{-1}[j]  \; \nb \\ && = \; a(k) \; \sum_{j} \langle \,
W(\, H \# H; \, \zeta, \Psi_0,
\psi[j]  \, ) \, \rangle \bigr |_{S^3} \;  E_0^{-1}[j]  \nb \\ && = \; \langle \, W(\, H \# H; \,
\zeta, \Psi_0, \varphi  \, )  \, \rangle  \bigr |_{S^3} \quad ;
\label{eq:cnov} \ena where $\varphi \; = \; a(k) \; \sum_m  E_0^{-1}[m]  \; \psi[m]$. The link
$H \# H$ is the connected sum of two Hopf links shown in Figure 8.13.

\begin{figure}[h]
\begin{picture}(10,10)
\put(180,-145){$\zeta$}
\put(285,-145){$\varphi$}
\end{picture}
\vskip 0.5 truecm 
\centerline{\epsfig{file= \path 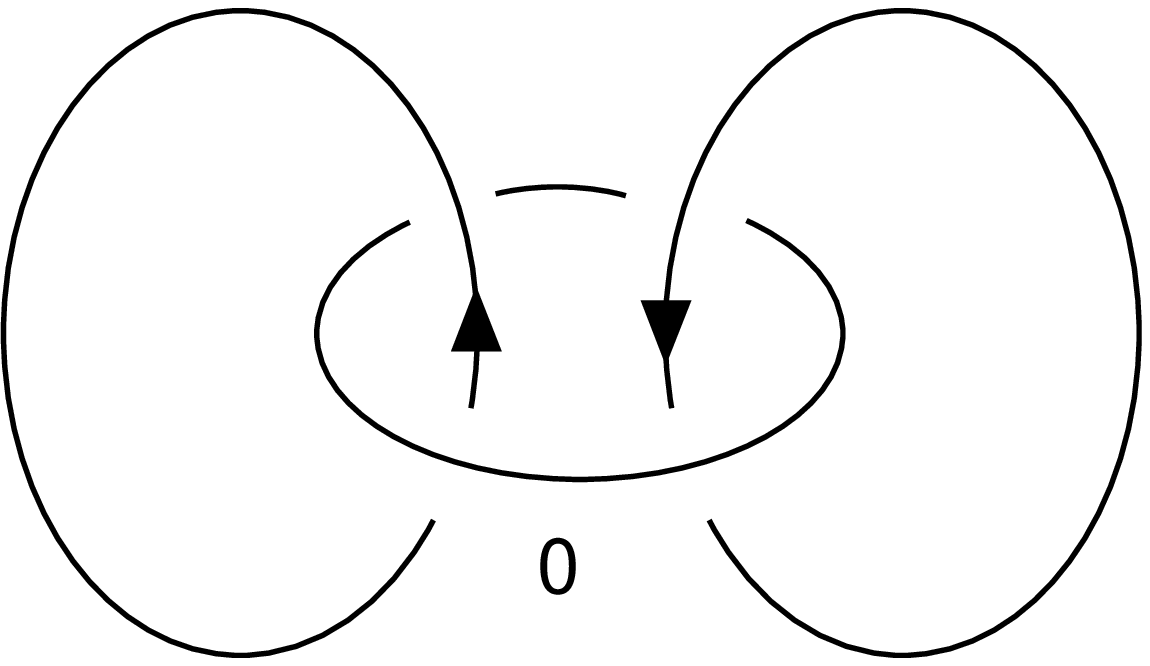,height=4cm,width=6cm}}
\vskip 0.8truecm 
\centerline {{\bf Figure 8.13}}      
\vskip 0.5truecm 
\end{figure}

\no Equation (\ref{eq:cnov}) implies \cite{gp2}) the relation between tangles 
shown in Figure 8.14. 

\begin{figure}[h]
\begin{picture}(10,10)
\put(177,-20){$\psi[i]$}
\put(290,-20){$\psi[j]$}
\put(190,-95){$\psi_h$}
\put(320,-120){$\varphi$}
\end{picture}
\vskip 0.5 truecm 
\centerline{\epsfig{file= \path 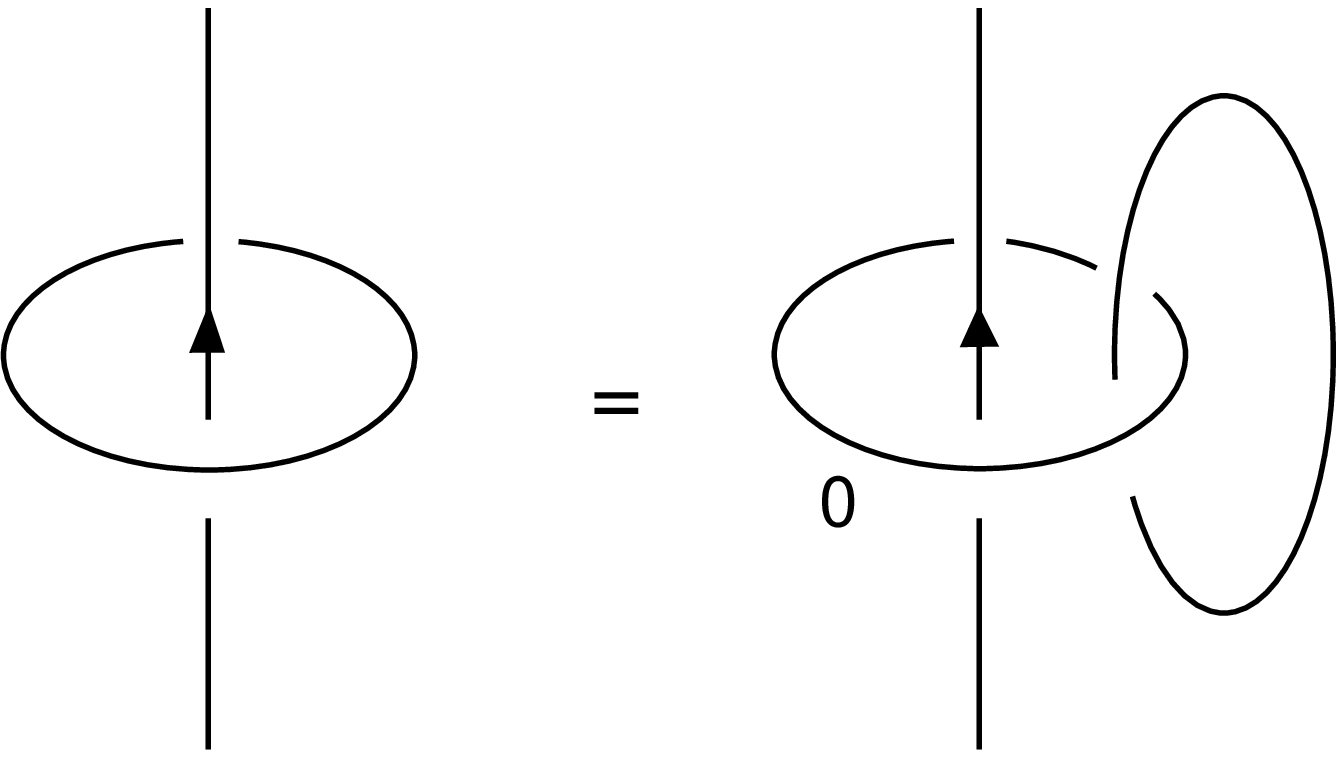,height=4cm,width=7cm}}
\vskip 0.5truecm 
\centerline {{\bf Figure 8.14}}      
\vskip 0.5truecm 
\end{figure}
\begin{figure}[h]
\begin{picture}(10,10)
\put(172,-20){$\psi[i]$}
\put(355,-20){$\psi[j]$}
\put(205,-120){$\psi_h$}
\end{picture}
\vskip 0.5 truecm 
\centerline{\epsfig{file= \path 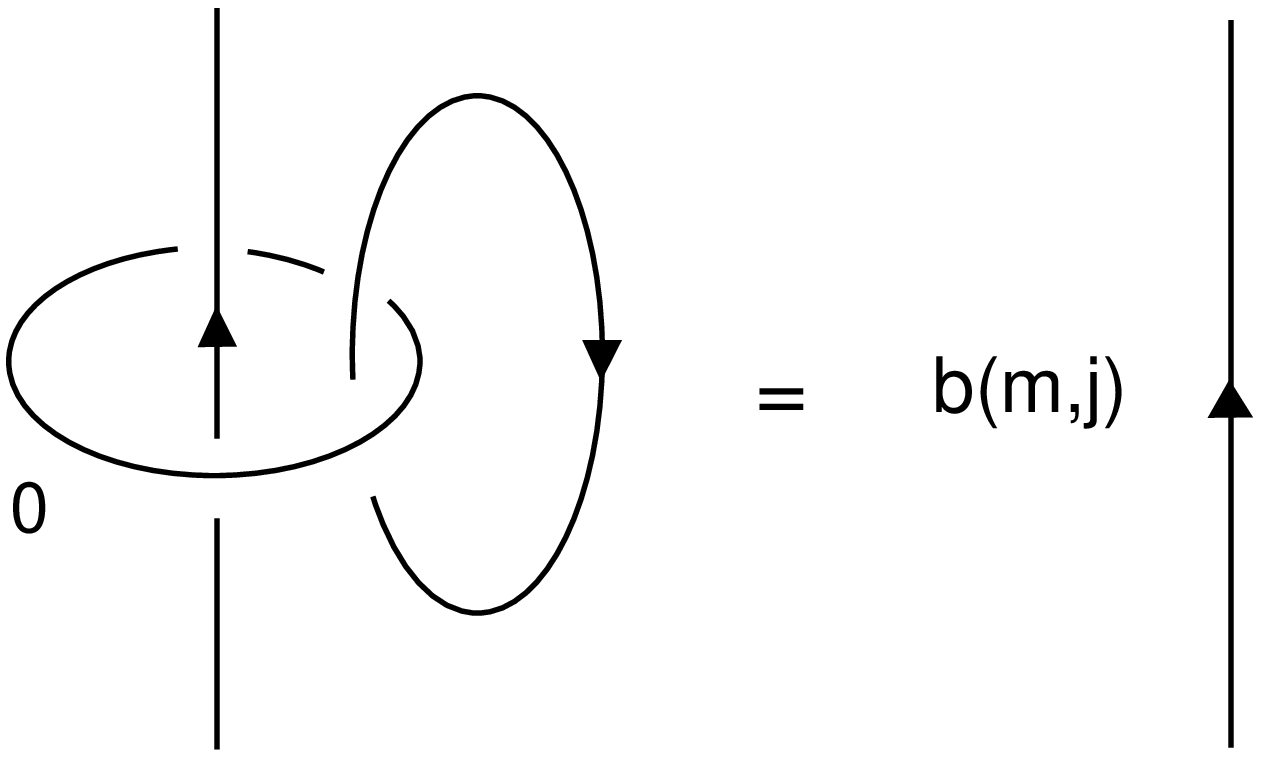,height=4cm,width=8cm}}
\vskip 0.5truecm 
\centerline {{\bf Figure 8.15}}      
\vskip 0.5truecm 
\end{figure}

\no  On the other hand, the tangle identity of Figure 8.15 also holds 
\cite{gp2}

\no  The coefficient $b(m,j)$ can be easily determined by closing the tangles
shown in Figure 8.15 and by using Theorem 6.5. Thus one gets
\beeq 
b(m,j) \; = \;  a(k) \; E_0^{-1}[j] \;  \delta_{jm^\ast} \; \quad .  
\end{equation}  
We can now use the decomposition of Figure 8.15 to recast the tangle shown in 
Figure 8.14 in  the form shown in Figure 8.16. 

\begin{figure}[h]
\begin{picture}(10,10)
\put(178,-20){$\psi[i]$}
\put(355,-20){$\psi[j]$}
\put(190,-100){$\psi_h$}
\end{picture}
\vskip 0.5 truecm 
\centerline{\epsfig{file= \path 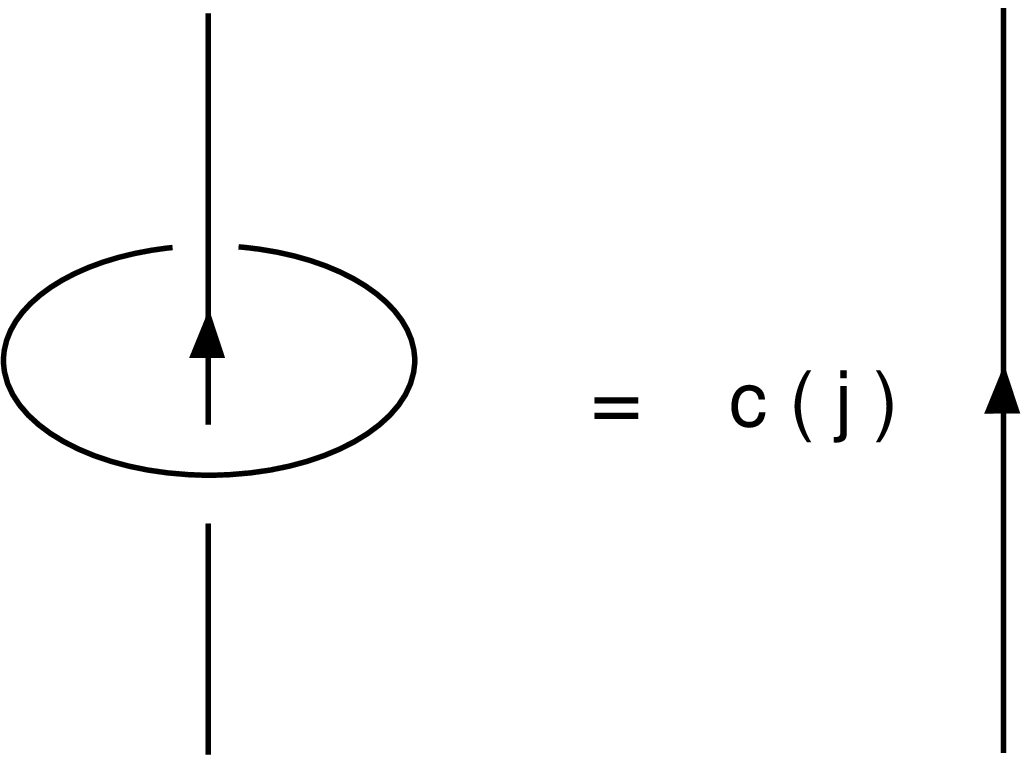,height=4cm,width=8cm}}
\vskip 0.5truecm 
\centerline {{\bf Figure 8.16}}      
\vskip 0.5truecm 
\end{figure}
  
\no The coefficient $c(j)$ which appears in Figure 8.16 is given by
\beeq 
c(j) \; = \; \frac{a^2(k)}{E_0^2[j]} \qquad . 
\label{eq:coe}
\end{equation} 
To sum up, the relation shown in Figure8.16 means that each link component 
$C_i$ with colour
$\psi_h$ (which is associated with one handle) can be eliminated provided one multiplies the resulting
new link with an appropriate factor; if the component $C_i$ is linked with a knot with colour $\psi
[j] $, this factor $c(j)$ is given in equation (\ref{eq:coe}). 

At this point, one can derive the general expression for ${\rm dim }\, {\cal H}(g)$; in fact, each
component of the surgery link with colour $\psi_h$ (representing one handle) can be eliminated and
substituted with the multiplicative factor $c(j)$.  Consequently, from equation  (\ref{eq:cise}) one
finds 
\bea &&{\rm dim }\, {\cal H}(g) \; = \; \nb \\  &&\quad = \; a^{-2}(k)\, \sum_j \, E_0[j] \, 
\langle \, W(U; \, \psi [j]) \, W(C_1\, ;  \, \psi_h , ) \, W(C_2\, ;  \, \psi_h , ) 
\cdots  W(C_g \, ;  \, \psi_h \, ) \, \rangle \bigr |_{S^3}    \nb \\  &&\quad = \; a^{-2}(k) \,
\sum_j \, E_0 [j] \, \left ( \frac {a^2(k)}{E^2_0[j]} \right)^g \, 
\langle \, W(U; \, \psi [j])  \, \rangle \bigr |_{S^3} \nb \\  &&\quad = \; \sum_j \, \left ( \frac
{a^2(k)}{E^2_0[j]} \right)^{(g-1)} \nb \\  &&\quad = \;  \sum_j \, X_{{\bf 1}j}^{2(1-g)} \qquad .
\label{eq:bun} 
\ena Equation (\ref{eq:bun}) coincides with equation (\ref{bund}).

Finally, let us derive equation (\ref{dimc}). In this case, we must compute 
\cite{guad4} the dimension of the
physical state space of the CS theory defined in the manifold $\Sigma_g \times S^1$ in the presence
of a static puncture on $\Sigma_g$. Let us now consider the knot $C$ in 
$\Sigma_g \times S^1$ shown
in Figure 8.17. 

\begin{figure}[h]
\begin{picture}(10,10)
\put(180,-10){$\psi[i]$}
\put(320,-25){$\psi_h$}
\put(300,-60){$\psi_h$}
\end{picture}
\vskip 0.5 truecm 
\centerline{\epsfig{file= \path 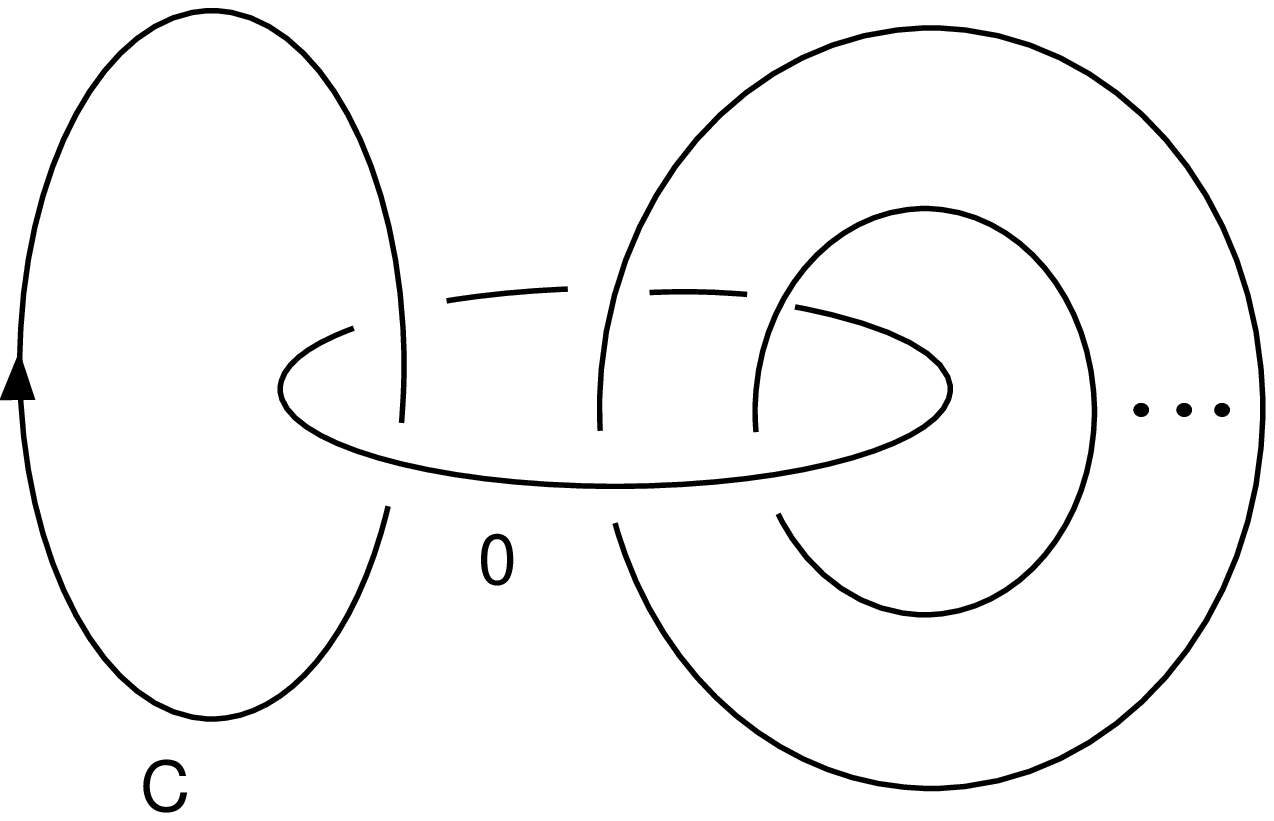,height=7cm,width=7cm}}
\vskip 0.5truecm 
\centerline {{\bf Figure 8.17}}      
\vskip 0.5truecm 
\end{figure}
    
\no From a 2-dimensional point of view, the knot $C$ corresponds to a static puncture on $\Sigma_g$.
The dimension of the physical state space ${\rm dim }\, {\cal H}(g,\psi[i])$ is given by the analogue
of equation (\ref{eq:cise}), namely  
\bea  
&&\hskip-40pt{\rm dim }\, {\cal H}(g,\psi[i]) \; = \nb \\   &&\hskip-40pt \quad = \;   \frac
{\langle \, W(U; \,
\Psi_0) \, W(C \, ; \, \psi [i] \, ) \,  W(C_1\, ;  \, \psi_h , ) \, W(C_2\, ;  \, \psi_h , ) 
\cdots  W(C_g \, ;  \, \psi_h \, ) 
\, \rangle \bigr |_{S^3}}{\langle \, W(U \, ; \, \Psi_0)
\, \rangle \bigr |_{S^3}} \quad .  \label{eq:cis}
\ena 
By using the same method illustrated before, one finds 
\beeq {\rm dim }\, {\cal H}(g, \psi[i]) \; = \; \sum_j X_{ 1j}^{(1-2g)} \, X_{ji} \qquad , 
\label{eq:bun1} 
\end{equation} which corresponds to equation (4). This concludes our topological derivation of
the Verlinde formulae (2)-(4). 

The expression appearing in (\ref{eq:bun1}) can be generalized to the case of 
n punctures with
colours $\left \{ \psi [i_1], \; \psi [i_2], \; \cdots , \; \psi [i_n] \right \}$. Indeed, by using
the satellite relation of eq.(\ref{satg}), the set of $n$ punctures can be 
replaced with one
puncture with colour $\psi$ given by
\beeq
\psi \; = \; \psi[i_1] \; \psi[i_2] \; \cdots \; \psi[i_n] \quad ;
\end{equation} the element $\psi \in  {\cal T}_{(k)}$ admits the following 
decomposition
\beeq
\psi \; = \; \sum_{m} c_m \left(i_1, \, \cdots , \, i_n \right) \; \psi[m] \quad , 
\end{equation} 
where the coefficients $\{ c_m \}$ are determined by the ${\cal T}_{(k)}$ 
structure constants.
Therefore, the dimension of the CS physical state space associated with the 
manifold $\Sigma_g \times S^1$ in the presence of $n$ static punctures on $\Sigma_g$ is
\bea 
{\rm dim }\, {\cal H}(g, \psi[i_1], \, \cdots, \, \psi[i_n]) \; && = \; \sum_{m} c_m \left(i_1,
\,
\cdots , \, i_n \right) \, {\rm dim }\, {\cal H}(g, \psi[m]) \\ \nb && = \; \sum_{m,j} c_m
\left(i_1, \,
\cdots , \, i_n \right) \, X_{1j}^{(1-2g)} \, X_{jm}   \quad .
\ena
\section{\bf Discussion}

In this Chapter we have shown that the fusion rules of conformal field theory 
and the associated
Verlinde formulae admit an interpretation in terms of three-dimensional topology. In our approach,
the existing connections between certain aspects of two-dimensional CFT and the topology of
three-manifolds have not been assumed; these connections have been derived from the properties of the
quantum CS field theory. 

The starting point of our analysis has been the
three-dimensional CS theory and the existence  of a set of link invariants defined on framed,
oriented and coloured links in $S^3$. As a consequence of general covariance of the CS theory  
(combined with the  framing dependence of the composite Wilson line operators), these link
invariants satisfy a  certain number of relations.  The connected sum rule (9) and the satellite
relation (\ref{satg}) summarize the relevant properties of the link 
polynomials. 

Because of the satellite relations, an algebra structure is defined in the colour state space
associated with the link components.  This algebra, called the tensor algebra, is defined by the
structure constants of the representation ring of the gauge group.  For integer values of the
coupling constant $k$, not all the different colour states are distinguished by the values of the
observables. Two different elements of the tensor algebra  are called physically equivalent when the
values of their associated observables are equal.  For fixed integer $k$,  the tensor algebra can be
decomposed into classes of physically equivalent elements.  In general, the number of linearly
independent classes is finite; consequently, the algebra defined by these classes (called the 
reduced tensor algebra ${\cal T}_{(k)}\, $) is of finite order. Since the braid group
representations obtained in CFT and in the CS theory are equivalent, the fusion algebra of certain two-dimensional conformal models  is isomorphic with the reduced tensor algebra of the
corresponding three-dimensional CS theories.  

The fusion rules, which correspond to the structure constants of ${\cal T}_{(k)} $,  can be
represented by means of the expectation values of Wilson line operators in the manifold $S^2 \times
S^1$. We have shown that, when the reduced tensor algebra is regular, a projective representation of
the modular group is defined on ${\cal T}_{(k)}$;  the Verlinde formula (2) is then a  consequence
of the connected sum rule which is satisfied by the expectation values of the CS observables.  

The operator surgery method has been used to implement Dehn's surgery in the field theory context. 
The topological properties of surgery can be expressed in terms of relations between the expectation
values of the CS observables.  We have  seen that a simple surgery presentation of the manifolds $\{
\, \Sigma_g \times S^1\, \}$ can be given; in fact, we have derived a general rule for adding
handles (in the three-dimensional context) to the manifold $S^2 \times S^1$.  We have shown that the
surgery operator, associated with one handle, is determined by the structure constants of ${\cal
T}_{(k)}$; this result  has been used to derive  equations (\ref{bund}) and 
(\ref{dimc}). Indeed, by using the
operator surgery method, we have computed the dimensions of the physical state spaces of the CS
theory defined in  $\Sigma_g \times S^1$ and we have proved that the resulting expressions coincide
with equations (\ref{bund}) and (\ref{dimc}).

\chapter{\bf Conclusion}
In this thesis we have studied the Chern-Simons (CS) field theory with a 
non-Abelian 
gauge group $G$. When the theory is defined in $S^3$ gauge invariance constrains
the renormalized coupling constant $k$ to be an integer. As a consequence, 
there are elements of the tensor algebra $\cal T$ of $G$ which cannot be 
distinguished by using the CS observables. We have given the general rules to 
find the reduced tensor algebra 
${\cal T}_{(k)}$ of $\cal T$, whose elements give the physical inequivalent
quantum numbers that characterize uniquely the CS observables. Some general 
properties of the reduced tensor algebra have been given, and the important 
notion of regular reduced tensor algebra has been defined.   

When $G=SU(3)$ we have produced an explicit
expression for the value of the unknot and for the Hopf link, for arbitrary 
irreducible representations of $SU(3)$. We also have worked out the reduced tensor
algebra by giving a set of ``equivalence rules'' among the elements of the 
$SU(3)$ tensor algebra. By using this result, we have solved explicitly 
the CS theory in any closed, connected and orientable 
three-manifold when the gauge group is $SU(3)$; in addition, many examples 
have been examined in detail.

When the reduced tensor algebra is regular, we have found the general 
expression for the surgery operator and we have given 
the surgery rules to compute a generic observable in any closed, connected and 
orientable three-manifold; the invariance of the observables under Kirby moves 
has been proved explicitly.  

Starting from the partition function $Z({\cal M})$ of the Chern-Simons field 
theory, i.e. the sum of the vacuum to vacuum Feynman's diagrams, one can show
that $I({\cal M}) = \exp(i \varphi) \, Z({\cal M})$, with a 
suitable choice of the phase
$\varphi$, is a topological invariant of ${\cal M}$.  We have computed the CS 
topological invariant $I({\cal M})$ for various three-manifolds. Unfortunately, 
up to now, the topological interpretation of $I({\cal M})$ is not known. Toward 
this goal, we have investigated the relation between $I({\cal M})$ and the 
fundamental group $\pi_1({\cal M})$ of $\cal M$. In the case $G=SU(2)$, we have 
proved that when $|I(L_{p/r})|\neq 0 $ 
depends only on $\pi_1(L_{p/r})$; also strong numerical evidence 
suggesting the same result holds when $G=SU(3)$ has been produced. The 
generalization  of the above result to any three-manifold is an open problem 
equivalent  to proving or disproving the following conjecture: for non-vanishing  
$\, I({\cal M}) \, $, the absolute value $\, | \,  I({\cal M}) \, | \, $ only 
depends on the fundamental group $\pi_1 ({\cal M}) \, $.

The last chapter is devoted to the relation between two-dimensional conformal 
field
theory and three-dimensional Chern-Simons field theory. The point of view 
usually 
adopted in the existing literature is reversed. We do not follow the approach 
pioneered by Witten in which two-dimensional conformal field theory is the key 
tool in order to compute Chern-Simons observables, on the contrary, 
the attention is focused on three-dimensional field theory. We have shown that 
the celebrated results 
concerning the fusion rules for primary fields obtained by H.~Verlinde can be 
proved by using the general properties of the surgery operator.       

In Appendix D we have included a quick review of some basic ideas of 
two-dimensional conformal field theory, with these tools in hand, a sketch of 
the original Witten's solution
of Chern-Simons theory has been given. As we have stressed many times, the two 
approaches are quite different, both from a theoretical point of view and from 
the point of view of practical computations. As shown in Appendix D, Witten's 
approach has a drawback: the relation between the renormalized CS coupling 
constant $k$ and the
level $\kappa$ of two-dimensional current algebra is ambiguous, indeed, the 
original Witten's guess is inconsistent. The easiest way to find the correct 
relation, 
perhaps the only one, is perturbation theory. On the contrary, the 
three-dimensional field theoretic approach presented in this thesis is free 
from ambiguities.
Nevertheless, Witten's idea is fascinating, it gives a new insight on the 
relation 
between conformal invariance in two dimensions and general covariance in three 
dimensions, providing a beautiful connection between the WZW model and CS 
theory.
As a matter of fact, the method that we have used in this thesis is really 
powerful and provides an excellent way to compute the CS observables; 
in particular, the surgery construction is transparent and many properties can 
be derived.
     
Finally let us end with an open problem. Given a 3-manifold there are 
values of $k$ for which $I_k =0 $; we argued that this problem is related to a 
breaking of the gauge invariance, however a rule to find the ``forbidden''
values of $k$ is still missing.

\begin{appendix}
\chapter{\bf The Wess-Zumino functional}
\section{\bf Behaviour under smooth deformation}
In this appendix we shall discuss some useful properties of the Wess-Zumino
functional $\Gamma$ defined as 
\bea 
&& \Gamma_M \left [U \right ] \; = \; \frac {1}{24 \pi^2}
\int_{M} \Omega \quad , \nb \\
&& \Omega \; = \; {\rm Tr} \left [ \left(U^{-1} d  U \right ) \wedge \left (U^{-1}  d  U \right ) \left ( U^{-1} d  U \right )
\right ] \ \quad ,
\ena
where $M$ is an orientable 3-manifold and $U$ is a smooth
map from $M$ to a compact and simple Lie group $G$. In addition, we have 
used a matrix representation of $G$.   
We shall first prove the $\Gamma$ is invariant under a small deformation of
the map $U$, i.e.
\beeq
\Gamma_M \left [U \, + \, \delta U \right ] \; = \;  \Gamma_M \left [U  \right ] \quad . 
\end{equation} 
By using the cyclicity of the trace one gets
\beeq
\delta \Gamma_M \left [U \right ] \; = \; \frac{1}{24 \pi^2}
\int_{M} 3 \, {\rm Tr} \left[ 
\left(U^{-1}d \delta U \; - \; U^{-1} \delta U U^{-1} d U \right) \wedge
\left(U^{-1}d U \right) \wedge \left(U^{-1}d U \right) \right] \; .
\label{uno}
\end{equation} 
Let us note that 
\bea
&& d \left[U^{-1} \delta U  \left(U^{-1}d U \right) \wedge \left(U^{-1}d U
 \right)  \right] \; = \nb \\ 
&& \left[\left(U^{-1}d \delta U \; - \; U^{-1} (dU) U^{-1} \delta U \right) \wedge
\left(U^{-1}d U \right) \wedge \left(U^{-1}d U \right) \right]  \qquad , 
\label{due}
\ena
and 
\bea
&&{\rm Tr} \left[ U^{-1} (dU) U^{-1} \delta U  \wedge
\left(U^{-1}d U \right) \wedge \left(U^{-1}d U \right) \right] \; = \nb \\
&& {\rm Tr} \left[ U^{-1} \delta U U^{-1} d U  \wedge
\left(U^{-1}d U \right) \wedge \left(U^{-1}d U \right) \right] \qquad .
\label{tre}
\ena
By using equations (\ref{due}) and (\ref{tre}), the variation (\ref{uno}) of 
$\Gamma$ can be written as
\beeq
\delta \Gamma_M \left [U \right ] \; = \; \frac{1}{24 \pi^2}
\int_{M} 3 \, d \, {\rm Tr}   \left[U^{-1} \delta U  \left(U^{-1}d U \right) \wedge \left(U^{-1}d U
 \right)  \right] \quad .
\label{qua}
\end{equation}
Stokes's theorems implies that 
\beeq
\delta \Gamma_M \left [U \right ] \; = \; \frac{1}{24 \pi^2}
\int_{\partial M} 3 \, {\rm Tr}\left[U^{-1} \delta U  \left(U^{-1}d U \right)
\wedge \left(U^{-1}d U \right)  \right]  \quad .
\label{cin}
\end{equation}
In our case $\partial M = \emptyset$, thus equation (\ref{cin}) leads to  
$\delta \Gamma[U]=0$. 

\section{\bf Composition of maps}
We shall now proof the validity of equation (\ref{group}) in Chapter 1.
After a straightforward calculation one obtains
\beeq
\Gamma[UV] \; = \; \Gamma[U] \; + \; \Gamma[V] \; + \; \Delta[U, \, V] \quad ,
\label{cin1}
\end{equation}
where $\Delta$ is given by
\beeq
\Delta[U, \, V] \; = \; \frac{3}{24 \pi^2} \int_M {\rm Tr} \left[V^{-1} \Phi(U)
\wedge \Phi(U) V \wedge \Phi(V) \; + \; V  \Phi(V)
\wedge \Phi(V) V^{-1} \wedge \Phi(V) \right ] \, .
\end{equation}
For sake of simplicity we have introduced the Lie algebra valued 1-form 
$\Phi(U)$ defined as
\beeq
\Phi(U) \; = \; U^{-1} d U \quad .
\end{equation}
It should be noted that $\Phi(U)$ is the pull-back of the Maurer-Cartan form
on $G$ under the map $U$. One can easily verify that the 3-form appearing in
$\Delta$ is exact; indeed one gets
\beeq
 \left[V^{-1} \Phi(U)
\wedge \Phi(U) V \wedge \Phi(V) \, + \, V  \Phi(V)
\wedge \Phi(V) V^{-1} \wedge \Phi(V) \right ] \, =  \, 
d \left[U^{-1}dU \wedge dV V^{-1} \right]  . \label{sei}
\end{equation}
By taking into account equation (\ref{sei}) and by using Stokes's theorem, 
equation 
(\ref{cin1}) can be written as 
\beeq
\Gamma[UV] \; = \; \Gamma[U] \;  + \; \Gamma[V] \; + \; \frac{3}{24 \pi^2} 
\int_{\partial M}
{\rm Tr} \left[U^{-1}dU \wedge dV^{-1} V \right] \quad . \label{wpol}
\end{equation}
In our case $\partial M \, = \, \emptyset$, thus equation (\ref{wpol}) leads to
equation (\ref{group}).  

\section{\bf Normalization: $\mathbf{G=SU(2)}$}
In order to determine the normalization for the Wess-Zumino functional when
$M=S^3$, we only need to consider the case $G=SU(2)$. One can represent $S^3$
as $R^3$ in which the points on the surface $|\vec{x}|^2 = a$ are identified
in the limit $a \ra \infty$. The map $U$ must satisfy the following 
condition
\beeq
\lim_{|\vec{x}| \ra \infty} \, U(\vec{x}) \; = \; U_{\infty} \quad . \label{con}
\end{equation}
The map $U$ can be chosen in the following way
\beeq
U(\vec{x})  \; = \; \exp \left[ i \, \vec{\sigma} \cdot \vec{n} \, f(|\vec{x}|)
\right] \quad ,
\end{equation}
where $\vec{n} =\vec{x}/|\vec{x}|$, $\vec{\sigma}= \{\sigma_1, \sigma_2, \sigma_3 \}$ are the Pauli matrices, and $f$ is a smooth function. By using
the identity $\vec{\sigma} (\vec{\sigma} \cdot \vec{n}) \, = \,\vec{n} \, 
\mathbf{1} \, + \, i \, \vec{n} \wedge \vec{\sigma}$, one obtains  
\beeq
U^{-1} d  U \, = \, \left\{i \frac{\cos(f) \sin(f)}{|\vec{x}|} 
\left[\sigma_\mu  \, - \, n_\mu \, (\vec{\sigma} \cdot \vec{n}) \right]
\, - \,   (\vec{\sigma} \cdot  \vec{n}) f^\prime n_\nu
\, - \, i \frac{\sin^2(f)}{|\vec{x}|} ( \vec{n} \wedge \vec{\sigma})_\mu
\right\} dx^\mu \, ,
\end{equation}
where $f^\prime$ is the first derivative of $f$ with the respect to its 
argument. After a rather tedious calculation we arrive at the following 
expression for the Wess-Zumino functional
\beeq
\Gamma[U] \; = \; \frac{1}{2 \pi^2} \int \frac{f^\prime \sin^2(f)}{|\vec{x}|^2} \, d^3x
\quad . \label{inte}
\end{equation}
By introducing the radial coordinate $r= |\vec{x}|$, the integral in (\ref{inte}) is easily computed with the help of  spherical coordinates. Indeed
\beeq
\Gamma[U] \; = \; \frac{2}{\pi} \int_0^\infty \left[ \frac{d}{dr} f(r) \right] \sin^2(f)
 \, dr \; = \; \frac{2}{\pi} \int_{f(0)}^{f(\infty)} \sin^2 y \, dy \quad .
\end{equation}
The final result is the following
\beeq
\Gamma[U] \; = \; \pi^{-1} \left[f(\infty) \; - \; f(0) \; - \; \frac{1}{2} \sin\left(
2 f(\infty) \right)  \; + \; \frac{1}{2} \sin\left(2 f(0) \right) \right ]
\quad .
\end{equation}
Condition (\ref{con}) implies the constraint
\beeq
\lim_{r \ra \infty} f(r) \; = \; m \pi \qquad m \in \mathbb{Z} \quad ,
\end{equation}
in addition we also need that $f(0)=0 $. These conditions lead to
\beeq
\Gamma[U] \; = \; m \quad .
\end{equation}

\chapter{\bf Elements of  knot theory}
\section{\bf Ambient and regular isotopy} 
In this appendix some basic notions of knot theory are introduced, a more 
complete treatment can be found in \cite{kau,rol}.

A knot $C$ in a manifold $M$ is a subset of $M$ homeomorphic with $S^1$, i.e. 
there exists a continuous map  $f$ with a continuous inverse, such that $f: \, 
C \ra S^1$. A link $L$ in $M$ with $m$ components is a collection of $m$ non 
intersecting knots in $M$. 
In this appendix we shall deal with knots in $R^3$ or $S^3$. For notational
simplicity we shall refer to knots in $R^3$, it is understood that everything 
applies also to the $S^3$case. 

\begin{figure}[h]
\vskip 0.9 truecm 
\centerline{\epsfig{file=\path 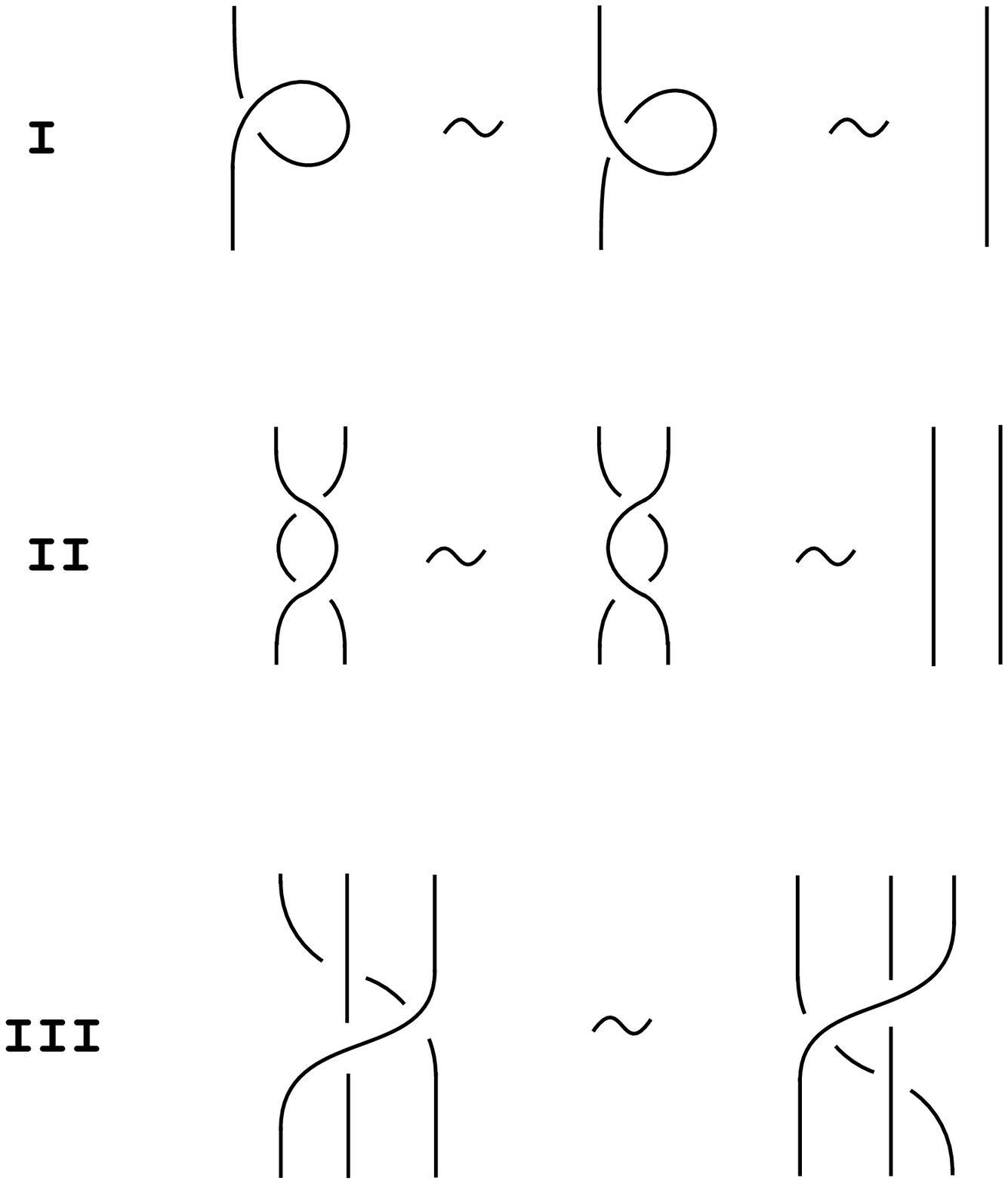,height=7cm,width=7cm}}
\vskip 0.9 truecm 
\centerline{{\bf Figure B1}}
\vskip 0.9 truecm 
\end{figure}

Given two knots $C_1$ and $C_2$, they are ambient isotopic if there
exists a homeomorphism $h$ such that $h(C_1) \, = \, C_2$ and in addition
$h$ is the end map of the following homotopy
\bea
&&h_t: \; R^3 \ra R^3 \qquad t \in [0, \, 1] \quad , \nb \\
&&h_0 \; = \; {\rm identity} \quad , \nb \\
&&h_1 \; = \; h \quad .
\ena
The property of $C_1$ and $C_2$ to be ambient isotopic is denoted by
$C_1 \equiv C_2$. We shall call $C$ unknot when it is ambient isotopic to a 
knot having the same homotopy type as a point, i.e. $C$  can be shrunk 
continuously to a point. 

A useful graphical representation of links in $\mathbb{R}^3$ in terms of diagrams is 
obtained by projecting the links on a plane. Conventionally, admissible link
diagrams contain simple crossings only. By means of link diagrams the ambient
isotopic equivalence relation can be reformulated in terms of the Reidemeister
moves shown in figure B1.  

Given two link diagrams $Dl_1$ and $Dl_2$, they represents ambient isotopic 
links $L_1$ and $L_2$ if and only if one can obtain $Dl_1$ from $Dl_2$ by 
using a finite sequence of Reidemeister moves. For our purposes, we need to 
consider oriented links. In a link
diagram the orientation is introduced by considering oriented strings. The 
over-crossing $L_+$ and under-crossing $L_-$ configurations occurring in a allowed link
diagram are shown in figure B2.     

\begin{figure}[h]
\vskip 0.9 truecm 
\centerline{\epsfig{file=\path 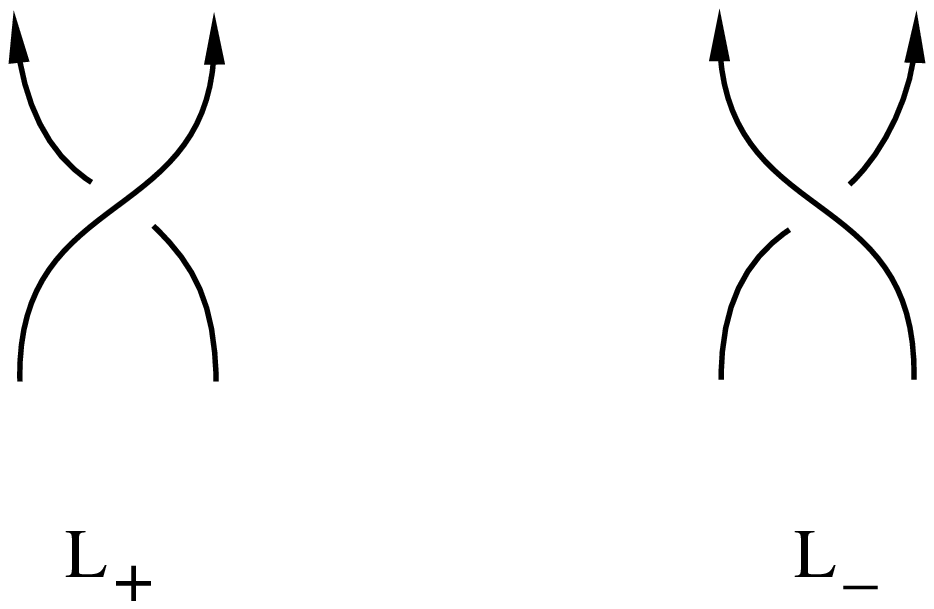,height=4cm,width=4cm}}
\vskip 0.5 truecm 
\centerline{{\bf Figure B2}}
\vskip 0.9 truecm 
\end{figure}

It is worth to point out that Reidemeister move of type I is peculiar. Indeed, 
differently from types II and III, only a single string is involved. 
Starting from this consideration, one introduces the notion of regular isotopic
links which will play a crucial role in the Chern-Simons theory.
Let us consider two link diagrams $Dl_1$ and $Dl_2$. They are
regular isotopic if $Dl_1$ can be obtained from $Dl_2$ by using a finite 
sequence of Reidemeister moves of type II and III only.

From the definition, the notion of ambient isotopy admits a clear topological
interpretation and it introduces an equivalence relation on the set of links
in $\mathbb{R}^3$. On the contrary, the notion of regular isotopy is inherited
from link diagrams; the question is: ``does there exist a geometric 
interpretation of regular isotopy invariance beyond the link diagram level ? ''

\no
The answer to this question is positive: regular isotopic link diagrams
can be interpreted as representing ambient isotopic bands. Indeed, one can 
imagine to replace the strings composing a link by oriented bands  
(see figure B3).
The topology of the link does not change, however, for each component $C_i$
one needs to specify an additional bit of information: the twist $T(C_i)$. The 
twist variable $T(C_i)$ represents how many times the band corresponding to
$C_i$ is twisted. Clearly, the twist is an ambient isotopy invariant for bands. 

\begin{figure}[h]
\vskip 0.9 truecm 
\centerline{\epsfig{file=\path 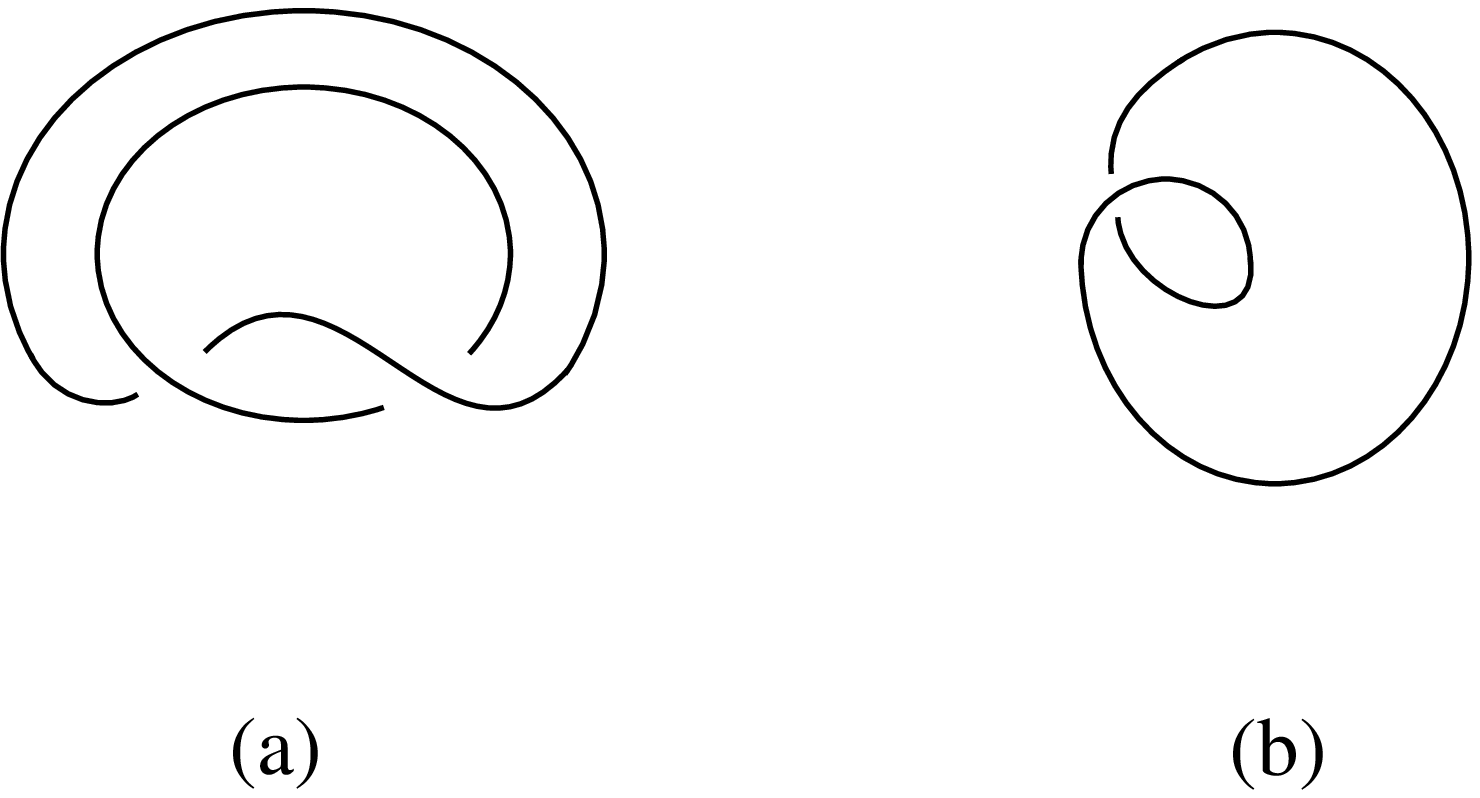,height=5cm,width=6cm}}
\vskip 0.5 truecm 
\centerline{{\bf Figure B3}}
\vskip 0.9 truecm 
\end{figure}
  
In order to complete the connection between ambient isotopic links made of 
bands and regular link diagrams, we have to specify the twist $T(C)$ of knot 
$C$ in terms of some regular ambient isotopy invariant of $C$.  
Let us introduce the writhe of the link diagram $Dl$ as
\beeq 
{\rm WR} \left (Dl \right ) \; = \; \sum_{x \epsilon L} \, \epsilon \left (x
\right ) \quad , \label{wri}
\end{equation}
where $x$ represents a generic over-crossing or under-crossing in $Dl$, and
\beeq 
\epsilon \left (L_+ \right ) \; = \; 1 \;, \qquad \epsilon \left (L_-
\right ) \; = \; -1 \quad  . 
\end{equation}
It is straightforward to verify from the definition (\ref{wri}) that the writhe
of a link diagram is a regular isotopy invariant. Indeed, the writhe can 
be written as 
\beeq 
{\rm WR} \left (Dl \right ) \; = \; n_+ \; -  \; n_-  \quad , 
\end{equation}
where $n_+$ and $n_-$ are the number of over-crossings and under-crossings.
The relation between the writhe of a link diagram $Dl$ associated with the knot $C$
is prescribed to be
\beeq 
{\rm WR}(Dl) \; = \; T(C) \quad . \label{verti}
\end{equation}
In chapter 1 we have introduced the framing prescription in order to properly
define composite Wilson operators. The knot $C$ and its framing $C_f$ can be
considered as the boundary of a band, and  one can naturally interpret 
framed links as links made of bands. With the vertical framing convention (\ref{verti}), oriented framed links and links made of bands are in one to one 
correspondence. A representative element of the class of ambient 
isotopic knots with a framed and oriented knot $C$  is 
uniquely associated with an element of the class of regular isotopic diagram 
$Dl$ of $C$ in the following way
\beeq  
lk(C, \, C_f) \; = \; {\rm WR}(Dl) \quad .
\end{equation}
For sake of simplicity, we shall denote the writhe of an element of the class 
of regular isotopic link diagrams associated with a framed and oriented knot
$C$ simply by ${\rm WR}(C)$. In general the linking number between two 
framed and oriented knots $C_1$ and $C_2$ can be expressed as
\beeq 
lk \left (C_1, \, C_2 \right ) \; = \; \frac {1}{2} \left [ {\rm WR} \left (L
\right ) \; - \; {\rm WR} \left (C_1 \right ) \; - \; {\rm WR} \left (C_2 
\right ) \right ] \quad ,
\end{equation}
where $L$ is the  link obtained by the union of  $C_1$ and  $C_2$.
The linking number is an ambient isotopy invariant for oriented knots.

\section{\bf Link invariants}
A link invariant is a function 
\beeq
L \stackrel{f}{\ra} f(L)
\end{equation}
which assigns to each link $L$ an object $f(L)$ in such way that, if $L_1$ and
$L_2$ are ambient isotopic, then $f(L_1) \, = \, f(L_2)$. By using a similar
definition one can also consider invariants for regular isotopic links. In the
 previous section, a 
numerical invariant for ambient isotopic links, the linking number, and a 
numerical invariant for regular isotopic links, the writhe, have been defined.

The Jones polynomials  $V \left (L; \, q \right )$ \cite{jon} in the variable
$q^{\pm \frac{1}{2}}$ with integer coefficients  satisfy the following 
properties
\begin{description}
\item{i)} ambient isotopy invariance;

\item{ii)} $V \left (C_u; \, q \right ) \; =1 \; $, where $C_u$ is the unknot; 

\item{iii)}
$ q \, V \left (L_+ \right ) \; - \; q^{-1} \, V \left (L_- \right ) \; = \;
\left (q^{\frac{1}{2}} \; - \; q^{- \frac{1}{2}} \right ) \, V \left (L_0
\right ) $.
\end{description}
Property iii) is called skein relation; it relates the invariants of skein
related links. The links  $L_+, \; L_-$ and $L_0$ are skein related if they 
can be represented by link diagrams which are identical except for a small
part contained inside a fixed open disc as shown in figure B4.   

\begin{figure}[h]
\vskip 0.9 truecm 
\centerline{\epsfig{file=\path 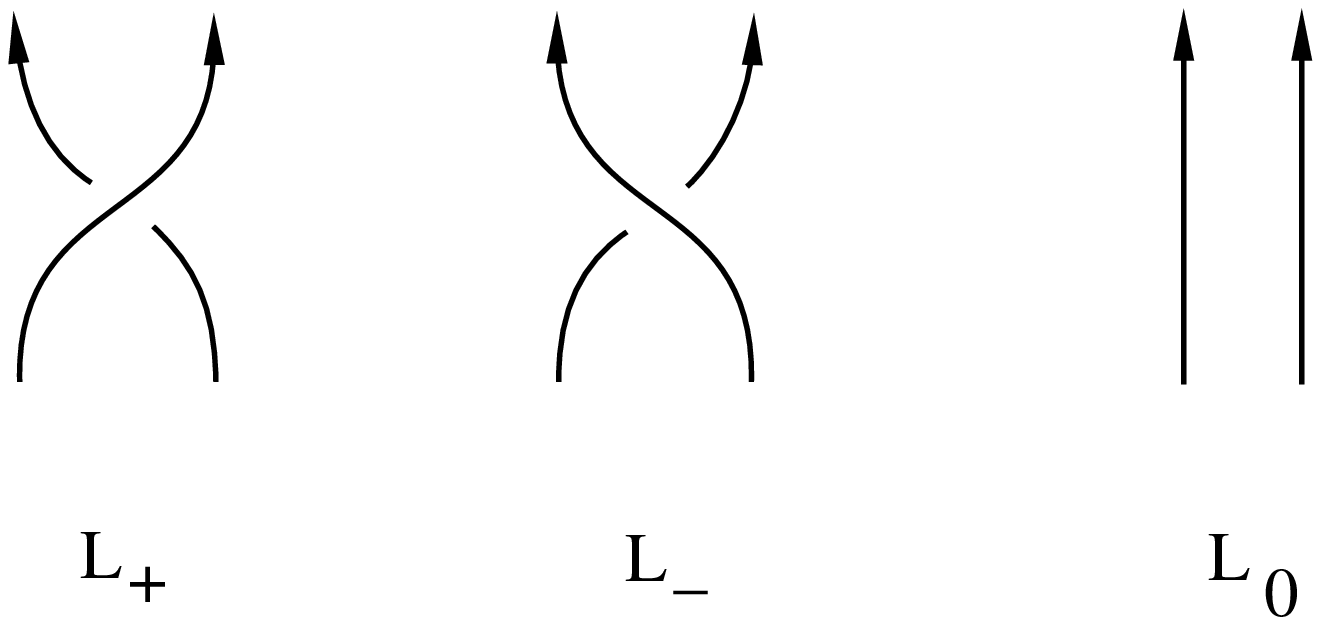,height=5cm,width=6cm}}
\vskip 0.5 truecm 
\centerline{{\bf Figure B4}}
\vskip 0.5 truecm 
\end{figure}

Another polynomial invariant naturally arises in the Chern-Simons field 
theory: the $S$ polynomial
$S \left (L; \alpha,\beta,z \right )$, which is essentially a generalization 
of the  Jones polynomial.
$S \left (L; \alpha,\beta,z \right )$ is defined in the following way 
\cite{gm3}
\begin{description}
\item{i)} regular isotopy invariance;
\item{ii)} $S \left (C_u \right ) \; = \; 1 $;
\item{iii)} $S \left (L^{(+)} \right ) \; = \; \alpha \, S \left (L^{(0)} \right) \;,  \qquad  S \left (L^{(-)} \right ) \; = \; \alpha^{-1} \: S \left (L^{(0)}
\right )$;
\item{iv)} $\beta \, S \left (L_+ \right ) \; - \; \beta^{-1} \, S \left (L_-
\right ) \; = \; z \: S \left (L_0\right ) $.
\end{description}
The skein relation corresponding to $L^{(+)}, \; L^{(-)}$ and  $L^{(0)}$ is 
shown in figure B5. Equations iii) and iv) satisfied by $S$ generalize
the Skein relation of the Jones polynomials.
It should be noted that, starting from the regular isotopy invariant $S$, 
an ambient isotopy invariant ${\cal S}$ can be introduced; indeed 
\beeq 
{\cal S} \left (L; \, \alpha, \, \beta, \, z \right ) \; = \; \alpha^{-WR \left (L \right
)} \, S \left (L;\alpha,\beta,z \right ).  
\end{equation}

\begin{figure}[h]
\vskip 0.9 truecm 
\centerline{\epsfig{file=\path 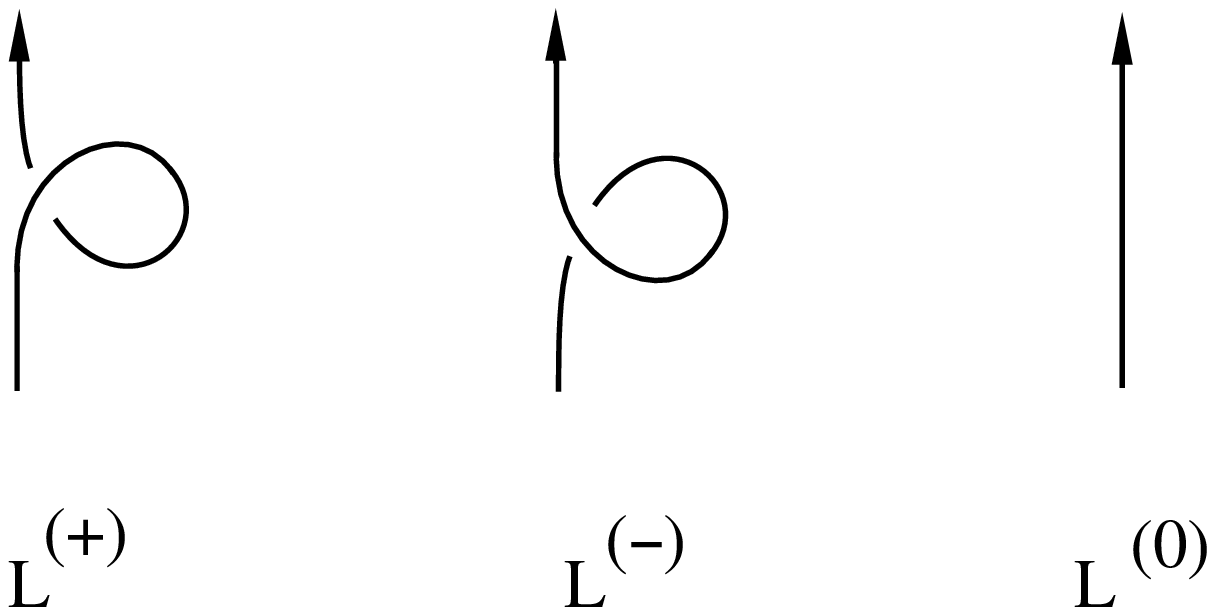,height=5.5cm,width=7cm}}
\vskip 0.5 truecm 
\centerline{{\bf Figure B5}}
\vskip 0.9 truecm 
\end{figure}

\section{\bf Artin braid group}
The construction and computation of link invariants can be put in an 
algebraic framework; one of the key ingredients toward this goal is the Artin
braid group \cite{art}.

The standard presentation of the Artin braid group $B_n$ is given in terms
of $(n-1)$ generators $\{g_1, \, \cdots , \, g_{n-1} \}$ satisfying the 
following relations
\bea
&&g_i \, g_j \; = \; g_j \, g_i \, , \qquad  |i-j| > 1 \quad , \label{br1} \\ 
&&g_i \, g_{i+1} \, g_i \; = \; g_{i+1} \, g_i \, g_{i+1} \, , \qquad i < n-1
\quad . \label{br2}
\ena
A very useful graphical representation of $B_n$ can be provided in terms of 
$n$ oriented strings. The effect of $g_i$ on the strings is represented by
an over-crossing between the $i^{\text{th}}$ and ${i+1}^{\text{th}}$ string 
originally parallel as 
shown in Fig.B7a. In the same way the effect of ${g_i}^{-1}$ is obtained by 
replacing the over-crossing by an under-crossing. For instance, in Fig.B7b 
is shown the elements $\sigma={g_2}^{-1}g_1$ of $B_n$. Clearly, the 
constraints (\ref{br1}) and (\ref{br2}) are satisfied.

\begin{figure}[h]
\vskip 0.5 truecm 
\centerline{\epsfig{file=\path 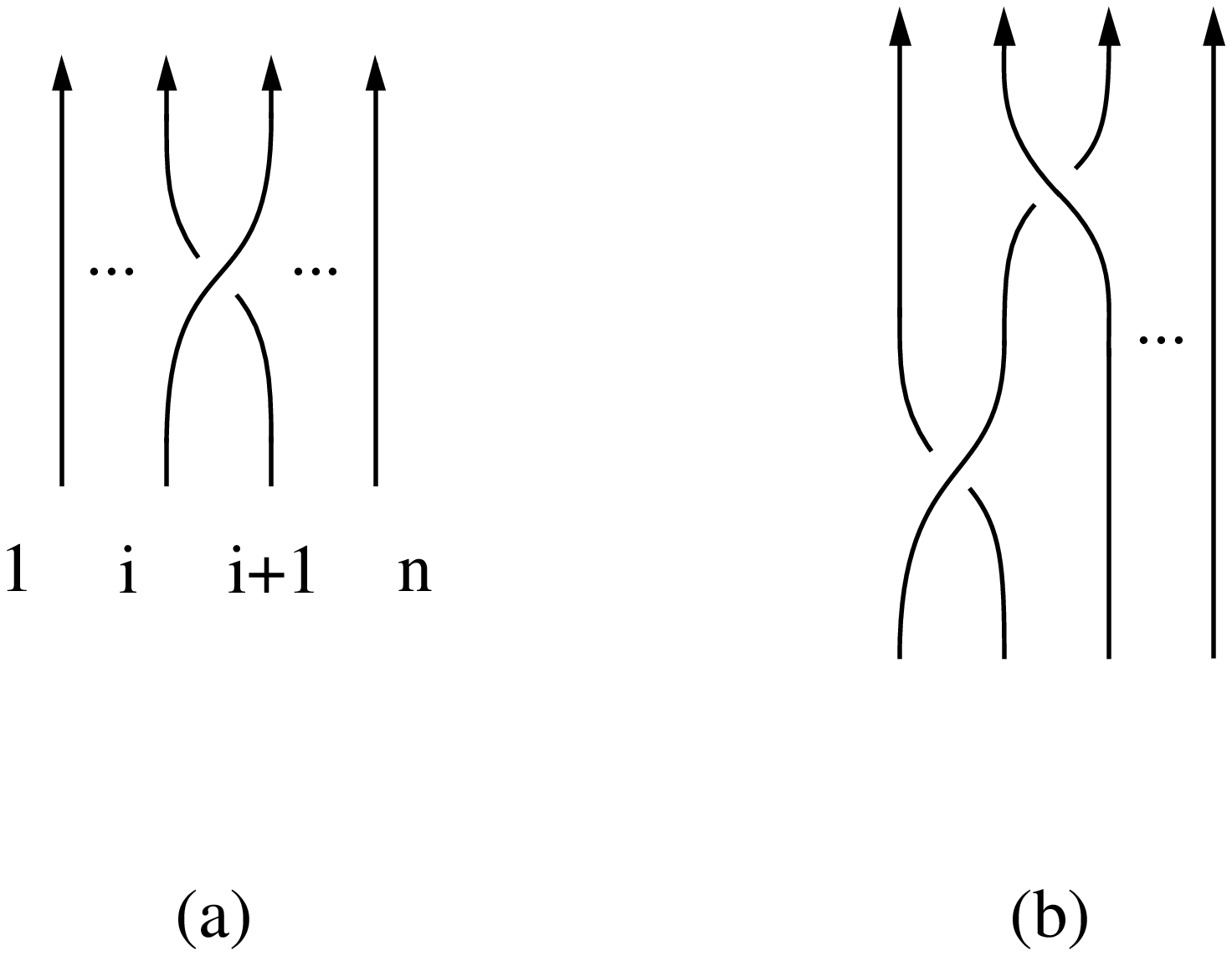,height=7cm,width=10cm}}
\vskip 0.5 truecm 
\centerline{{\bf Figure B7}}
\vskip 0.5 truecm 
\end{figure}

It should be evident that action of $g_i$, ${g_i}^{-1}$ and the identity on the
strings precisely corresponds to the $L_+, \; L_-,$ and $L_0$ admissible 
configurations encountered in the diagrams describing the projection of a 
link on a plane. The correspondence can be made more closely by noting that 
the invariance under Reidemeister moves of type II and III reproduces exactly
Eqs (\ref{br1}) and (\ref{br2}). In other words, the constraints among the
generators of $B_n$ are the algebraic statement of regular isotopy invariance.
Also Reidemeister moves of type I can be recovered by taking into account the
closure of the generic element $\sigma \in  B_n$. The closure $\hat{\sigma}$ of $\sigma
\in B_n$ is obtained by connecting the end points of the graph associated with
$\sigma$ in a orientation preserving way as shown in Fig.B8. 

\begin{figure}[h]
\vskip 0.5 truecm 
\centerline{\epsfig{file=\path 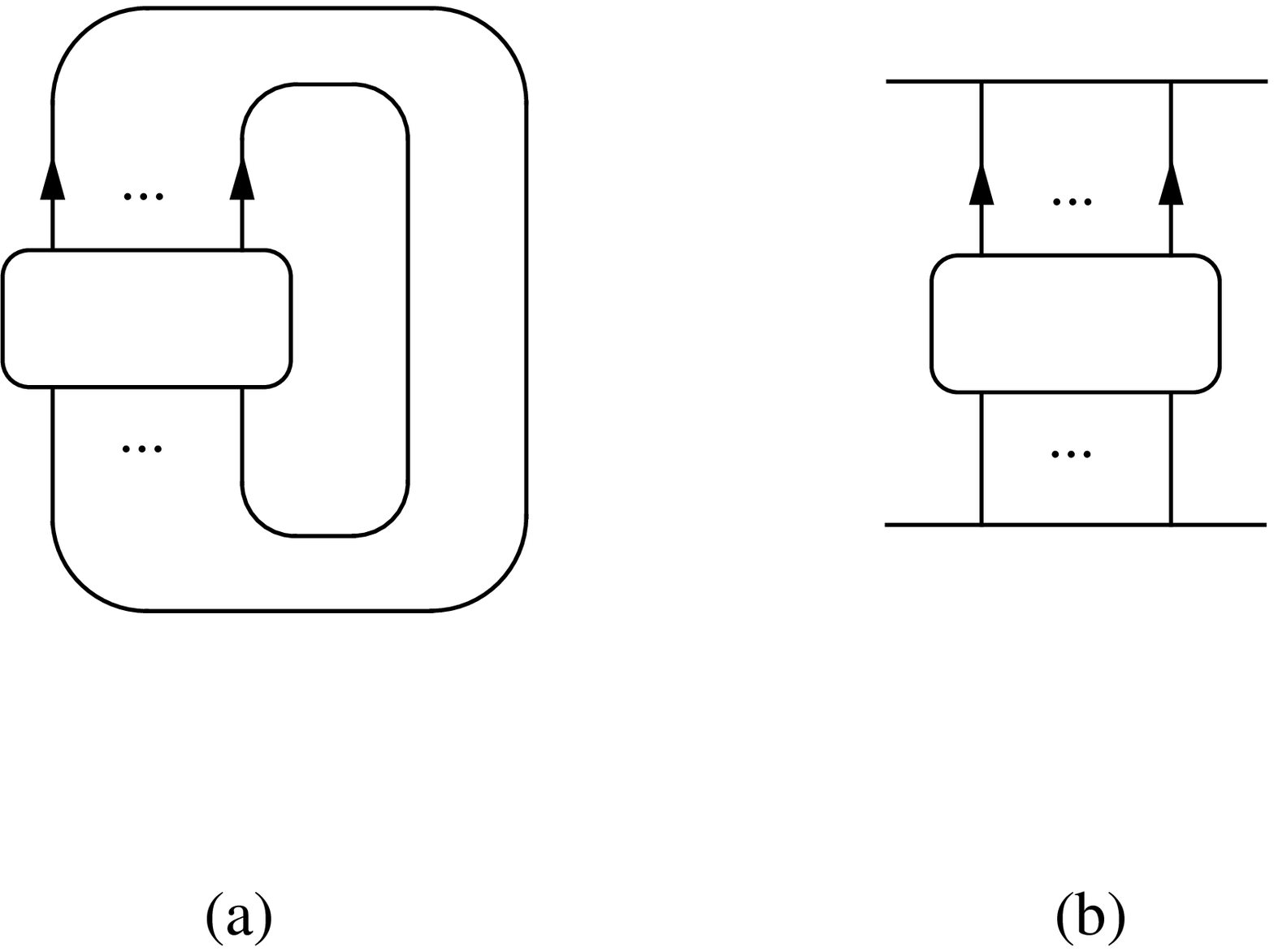,height=6cm,width=8cm}}
\vskip 0.5 truecm 
\centerline{{\bf Figure B8}}
\vskip 0.5 truecm 
\end{figure}
  
As results, the graph associated with the  closure of an element of $B_n$ corresponds
to a link diagram. However, care must be used in the
identification between the closure of braids and link diagrams. Indeed, given
two braids $\sigma_1, \, \sigma_2 \in B_n$ that belong to the same conjugacy
class, i.e. there exists $\sigma \in B_n$ such that $\sigma_1 = \sigma^{-1} 
\sigma_2 \sigma$, the corresponding closures $\hat{\sigma_1}$ and 
$\hat{\sigma_2}$ give link diagrams connected by  Reidemeister moves of 
type II. Therefore, in
order to make contact with link invariants, conjugate elements of $B_n$ must be
identified. The conjugacy class of an element $\sigma$ of $B_n$ will be 
denoted by $\langle \sigma \rangle_n$ and its closure with $\hat{\langle \sigma
 \rangle}_n$. Clearly, $B_n$ admits a natural inclusion in $B_{n+1}$. Any link
diagram can be obtained as the closure of direct sum $B_{\infty} = \bigoplus_{i=1}^\infty B_i$ of braid groups, the natural inclusion is understood.
By construction, any function defined on the conjugacy class of $B_\infty$
gives rise to a regular isotopy link invariant.

\no
{\bf Example}

\no
Consider the map $f: B_{\infty} \ra \mathbb{Z}$ such that
\beeq
f(g_i) \; = \; 1 \quad .
\end{equation} 
One can verify that the resulting regular isotopy link invariant coincides
with the writhe.

When ambient isotopy invariance is concerned, one must provide a way to 
take into account Reidemeister moves of type I. The key observation is that the
closure of $\langle \sigma \rangle_n$ and the closure of $\langle \sigma \,
{g_n}^{\pm} \rangle_{n+1}$ correspond actually to link diagrams related  by
a Reidemeister move of type I. Thus, the correspondence rules between the
closure of braids and diagrams of ambient isotopy links can be stated as follows.
Given two braids $\sigma_1$ and $\sigma_2$, their closures $\hat{\sigma_1}, \,
\hat{\sigma_2}$ are associated with link diagrams corresponding to ambient 
isotopic links if $\sigma_1$ and $\sigma_2$ are related by a sequence 
of Markov moves: M1, M2 \cite{mark}
\begin{description}
\item{M1)} $\sigma_1$ and $\sigma_2$ are conjugate elements of $B_n$ ,
\item{M2)} if $\sigma_1 \in B_n$ then $\sigma_2 = \sigma_1 \, {g_n}^{\pm 1} \in
B_{n+1}$ .
\end{description}
Clearly, any function on $B_\infty$ which is invariant under Markov moves,
a Markov trace, defines an ambient isotopy link invariant.
The Reshetikhin-Turaev \cite{retu,tur} invariant is defined starting  from 
an R-matrix representation of $B_n$ associated with a q-deformation of the 
universal enveloping algebra of $SU(2)$.

\chapter{\bf Miscellaneous results}
\section{\bf The SU(3) Hopf matrix is non singular}                                                           
For $G=SU(3)$ and for a fixed integer $k \geq 3$, let us consider the values $\{ \, H[(m,n) , (a,b)] \,
\}$ of the Hopf link for $(m,n) \in \Delta_k $ and  $(a,b) \in \Delta_k $.  
These complex numbers
can be understood as the matrix elements of a symmetric matrix called the Hopf matrix $H$. The
elements of the standard basis of ${\cal T}_{(k)}$ are $\{ \, \Psi [a,b] \, \}$ with 
$(a,b) \in \Delta_k$; to simplify the notation, we shall denote them simply by $\{ \, \psi_i
\, \}$ with the collective index $i$ running from 1 to the dimension of ${\cal T}_{(k)}$. 
The element $\psi_1$ corresponds to $\Psi [0,0]$  and, if $\psi_i$ represents 
$\Psi [a,b]$, then $\psi_{i^*}$ denotes $\Psi [b,a]$. The matrix elements of 
$H$ are  
\beeq
H_{ij} \; = \; H[ \psi_i \, , \, \psi_j \, ] \quad . 
\end{equation}
We shall prove that $H$ is invertible. 

First of all, let us evaluate $H^2$.  By definition, one has 
\beeq
\left ( \, H^2 \, \right )_{ij} \; = \; \sum_l\, H_{i l } \, H_{l j} \quad .  
\label{eqa2}
\end{equation}
By taking into account that the values of the Hopf link are  given in eq.(\ref{eq84}), one
finds that expression (\ref{eqa2}) consists of a combination of geometric finite sums. The direct 
computation of these geometric finite sums gives \cite{io,gp2} 
\beeq
\left ( \, H^2 \, \right )_{ij}\; = \; \left [ \frac{3k^2 }{ 256 \, \sin^6 (\pi /k) \, \cos^2 
(\pi /k) } \right ] \; \delta_{i \, j*}
\quad . 
\label{pare1}
\end{equation}
This equation shows that the Hopf matrix $H$ is invertible; indeed, 
\beeq
\left ( \, H^{-1} \, \right )_{ij}\; = \; \sum_{l } \; \left [ \frac{3k^2}{ 256 \, \sin^6 (\pi /k)
\, \cos^2  (\pi /k) } \right ]^{-1} \; H_{i l }\; \delta_{l^*  j} \quad . 
\end{equation}
A simple consequence of Eq.(\ref{pare1}) is the following
\beeq
\sum_m H_{{\mathbf 1}m} \, H_{mj} \; = \; \sum_m E_0[m] \, H_{mj} \; = \; 
b(k) \, \delta_{{\mathbf 1} j^\ast} \quad ,
\label{uhop}
\end{equation}
where $b(k)$ is defined as
\beeq
b(k) \; = \; \left [ \frac{3k^2 }{ 256 \, \sin^6 (\pi /k) \, \cos^2 
(\pi /k) } \right ] \qquad .
\end{equation}

\section{\bf On the $\mathbf{SU(3)}$ reduced tensor algebra structure 
constants} 

The main purpose of this appendix is to show that the structure constants of 
the $SU(3)$ reduced tensor algebra verify the relation $N_{ijm}= N_{i^* mj}$. 

Let us recall that the elements of the
standard basis of ${\cal T}_{(k)}$ are denoted by $\{ \, \psi_i \, \}$, and  the unit element 
of the algebra by $\psi_1$.  

\bigskip

\shabox{\noindent {\bf Property C1}} {\em ~The structure constants of 
$\, {\cal T}_{(k)}$ verify } 
\beeq
N_{ij \mathbf{1}} \; = \; \delta_{ij^*} \; = \; \delta_{i^* j} \quad . 
\label{A1}
\end{equation}

\bigskip

\noindent {\bf Proof.} ~When $k=1$ and $k=2$, the validity of eq.(\ref{A1}) 
can easily be verified by direct inspection of the expressions given in Sect.4.2.4 and in Sect.4.2.5. Let us concentrate then on the case $k\geq 3$. In Sect.C.1 we have proved that the Hopf matrix $H$ satisfies 
\beeq
\left ( \, H^2 \, \right )_{ij} \; = \; b(k) \; \delta_{ij^*}  \quad . 
\label{A2}  
\end{equation}
Since $H_{ij}$ represents the  value of the Hopf link,  by using the connected sum formula (\ref{eq:csr}), one finds 
\beeq
\left ( \, H^2 \, \right )_{ij} \; = \; \sum_m \, E_0[m] \; \langle \, W(C_1, C_2, C_3 ; 
\psi_i , \psi_j , \psi_m ) \, \rangle  |_{S^3} \quad , 
\end{equation}
where $\{ C_1 , C_2 , C_3 \, \}$ are the three (framed) components of the link shown in Fig.C1.  
By using the satellite relations (\ref{satg}) and eq.(\ref{A2}), one obtains 
\beeq
\left ( \, H^2 \, \right )_{ij} \; = \; \sum_{n , \, m} N_{ijm} E_0[m] \, 
H_{mn} \quad .
\label{cpa}
\end{equation}
Finally, Eqs.(\ref{cpa}), (\ref{pare1}) and (\ref{uhop}) lead to
\beeq
\left ( \, H^2 \, \right )_{ij} \; = \;
b(k) \; N_{ij \mathbf{1}} \; = \; b(k) \; \delta_{ij^*} \quad ,  
\end{equation}
which shows that eq.(\ref{A1}) is satisfied. {\hfill \ding{111}}

\vskip 0.5truecm

Let us introduce the complex valued linear function ${\cal F}$ which is defined on the elements
${\cal T}_{(k)}$. The action of $\cal F$ on the elements of the standard basis of ${\cal T}_{(k)}$ is
defined as 
\beeq
{\cal F} (\psi_i) \; = \; \delta_{\mathbf{1} j} \quad . 
\end{equation}

\begin{figure}[h]
\vskip 0.9 truecm 
\centerline{\epsfig{file=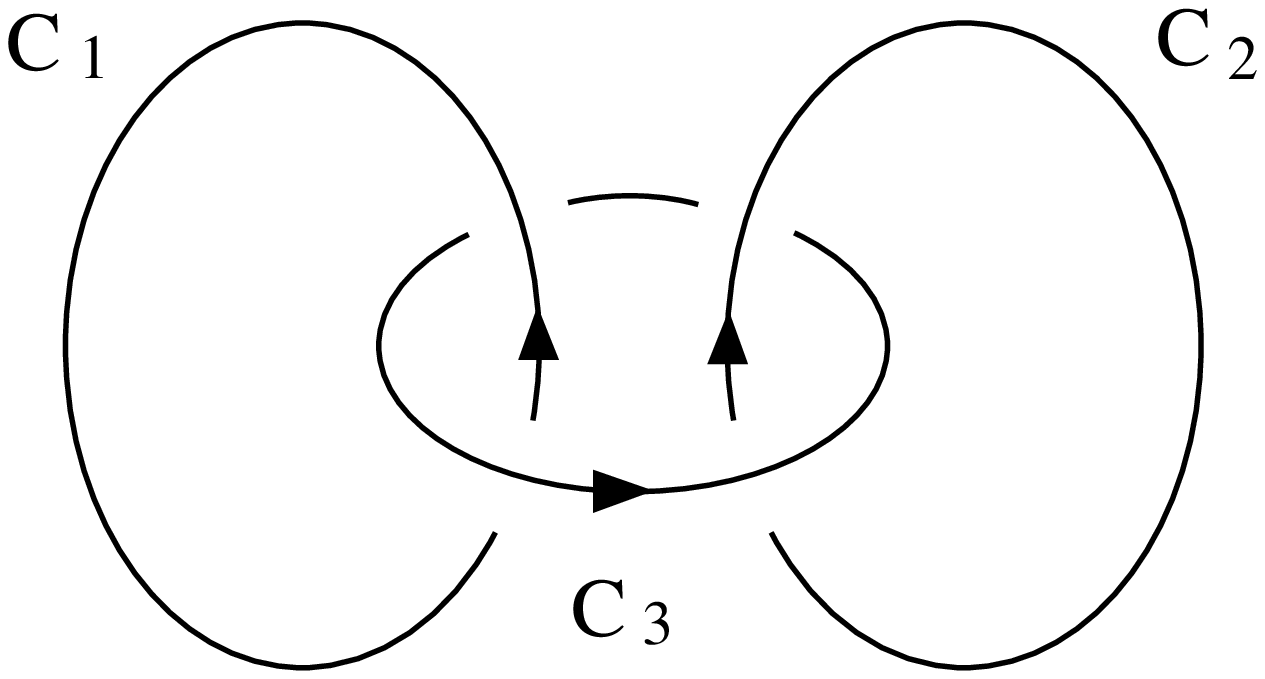,height=4cm,width=5cm}}
\vskip 0.9 truecm 
\centerline {{\bf Figure C.1}}
\vskip 0.9 truecm 
\end{figure}

\noindent From Property C1, it follows that 
\beeq
{\cal F} (\psi_i \, \psi_j ) \; = \; N_{ij1} \; = \; \delta_{ij^*} \quad . 
\end{equation}
Consequently, by means of $\cal F$, the structure constants $N_{ijm}$ can be 
written as  
\beeq
{\cal F}( \psi_i \psi_j \psi_{m^*} ) \; = \; \sum_n\, N_{ijn} \; {\cal F}( \psi_n \psi_{m^*} )\; =\; 
N_{ijm} \quad . 
\end{equation}
Since ${\cal T}_{(k)}$ is a commutative and associative algebra, one has  
\beeq
N_{ijm}\; = \; {\cal F}( \psi_i \psi_{m^*} \psi_j ) \; = \; N_{i m^*\, j^*} \quad . 
\end{equation}
At this point, one can use the relation $N_{ijm} = N_{i^* \, j^* \, m^*}$ 
(see eq.(\ref{real}) to get the final result 
\beeq
 N_{ijm} \; = \; N_{i^* \, mj} \quad . 
\end{equation}

\section{\bf Normalization of surgery operator for $\mathbf{SU(2), \; k \geq 
2}$}
The value of $a(k)$ and $\theta_k$ when $G=SU(2)$ is obtained by evaluating the
sum
\beeq
Z_{(+)} \; = \; \sum_{J=0}^{k/2-1} q^{Q(J)} \, {E_0[J]}^2 \quad .
\end{equation}
Alternatively, one can extend the sum over $0 \leq J \leq 2k-1$ simply
by introducing a factor $1/4$, i.e.
\beeq
Z_{(+)} \; = \; \frac{1}{4} \sum_{J=0}^{2k-2} q^{Q(J)} \, {E_0[J]}^2 \quad .
\label{sw2}
\end{equation}
Eq.(\ref{sw2}) is a trivial consequence of Property 4.4. From the expression
(\ref{unsu}) for $E_0[J]$ one gets
\beeq
Z_{(+)} \; = \; \frac{q^{-1/4}}{4 \left(q^{1/2} \, - \, q^{-1/2} \right)^2}
\sum_{s=1}^{4k-1} q^{-1} \left[q^{(s+2)^2/4} \; + \; q^{(s-2)^2/4} \; - \; 
2 q q^{s^2/4} \right] \quad ,
\end{equation}
with $s=2J+1$. A straightforward calculation gives
\beeq
Z_{(+)} \; = \; -\frac{q^{-3/4}}{2 \left(q^{1/2} \, - \, q^{-1/2} \right)}
\sum_{s=1}^{4k-1} q^{s^2/4} \quad .
\label{sw4}
\end{equation}
Let us now recall the definition of Gauss sum $S(n,m)$ \cite{cin}
\beeq
S(n, \, m) \; = \; \sum_{x=0}^{m-1} \exp \left(2 \pi i \frac{n}{m} x^2 \right)
\quad .
\label{gss}
\end{equation}
Thus, by using (\ref{gss}), Eq.(\ref{sw4}) can be written as
\beeq
Z_{(+)} \; = \; \frac{e^{i 3 \pi/(2k)}}{4i \sin(\pi /k)} \; S^\ast(1, \, 4k) 
\quad . \label{boh}
\end{equation} 
In particular, when $n=1$ one has \cite{cin}
\beeq
S(1, \, m) \; = \; \begin{cases}(1 \; + \; i) \sqrt{m} & \text{if } m \equiv 0  \mod \;4  \\
\sqrt{m} &\text{if } m \equiv 1  \mod \;4  \\
0 & \text{if } m \equiv 2  \mod \;4  \\
i \sqrt{m} & \text{if } m \equiv 3  \mod \; 4  
\end{cases} \quad .\label{vgs}
\end{equation}
Fron eq.(\ref{boh}) and eq.(\ref{vgs}) one has 
\beeq
Z_{(+)} \; = \; \frac{\sqrt{k}}{2} \left[\sin(\pi/k) \right]^{-1} \, e^{-i 
\frac{\pi}{4k} 3(k-2)} \quad .
\end{equation}
On the other hand we have
\beeq
Z_{(+)} \; = \; e^{i \theta_k} \, \frac{1}{a(k)} \quad ,
\end{equation}
thus
\bea
&&\theta_k \; = \; - \frac{\pi}{4k} 3(k-2) \quad ; \\
&&a(k) \; = \; \frac{2}{\sqrt{k}} \sin(\pi/k)  \quad .
\ena

\section{\bf Normalization of surgery operator for $\mathbf{SU(3), \, k \geq 3
}$} 
The value of $Z_0$, defined in eq.(\ref{A11}), has been computed in Sect.C.1.  
In this appendix we shall compute $Z_{+1}$ (see eq.(\ref{A12}))  when $k 
\geq 3$. 
Let us recall that the elements of the standard basis of ${\cal T}_{(k)}$ are 
$\{ \, \Psi [m,n] \, \}$ where the couples $(m,n)$ label the points of the fundamental domain $\Delta_k$ defined in Sect.4.2.3.
Since the value of the unknot vanishes on the points which belong to the 
boundary of $\Delta_k$,  eq.(\ref{A12}) can be written as  
\beeq
Z_{(+)} \; = \; \sum_{m=0}^{k-2} \, \sum_{n=0}^{k-2-m}\,  q^{Q(m,n)}\;  E_0^2[m,n]  \quad . 
\label{B3}
\end{equation}
By using the correspondence rules given in Sect.4.2.3, it is easy to verify 
that the sum appearing in eq.(\ref{B3}) can be extended to the region $0 \leq m \leq (4k -1)$ and  $0 \leq n \leq
(3k-1)$ provided that we divide by a factor $24 = 4 \times 3 \times 2$.  
Thus, 
\beeq
Z_{(+)} \; = \; \frac{1}{ 24} \, \sum_{m=0}^{4k-1} \sum_{n=0}^{3k-1}\,  q^{Q(m,n)}\;  E_0^2[m,n]  
\quad . 
\label{B4}
\end{equation}
By inserting in expression (\ref{B4}) the values $Q(m,n)$ of the quadratic 
Casimir operator and
the values $E_0 [m,n]$ of the unknot, one finds \cite{io,gp2}
\bea
&&Z_{(+)} \; = \; \frac{1}{ 24} \, \sum_{m=0}^{4k-1} \sum_{n=0}^{3k-1} \; 
\left(1-e^{-i \frac{2 \pi}{ k}}\right)^{-6}\; 
\left(1+e^{-i \frac{2 \pi}{  k}}\right)^{-2} \nb \\ 
&&\; \left \{ \; e^{-i \frac{2 \pi}{  3k} \left[n^2+n(m-3) \right]}e^{-i \frac
{2 \pi }{ 3k}(m^2-3m)}
-2e^{-i \frac{2 \pi}{  3k}\left[n^2+mn \right]}e^{-i
\frac{2\pi}{ 3k}(m^2-3m+3)} \right.\nb \\ 
&&\; +e^{-i \frac{2 \pi}{3k}\left[n^2+n(m+3)\right]}e^{-i \frac{2
\pi }{3k}(m^2-3m+6)}- 2e^{-i \frac{2 \pi}{  3k}\left[n^2+n(m-3)
\right]}e^{-i \frac{2 \pi}{ 3k}(m^2+3)} \nb \\ 
&&\; +2e^{-i \frac{2 \pi}{  3k}\left[n^2+mn\right]}e^{-i \frac{2
\pi }{ 3k}(m^2+6)}+2e^{-i \frac{2 \pi}{ 3k}\left[n^2+n(m+3) \right]}e^{-i \frac{2 \pi}{  3k}(m^2+9)} \nb \\ 
&&\; -2e^{-i \frac{2 \pi}{ 3k}\left[n^2+n(m+6) \right]}e^{-i \frac{2 
\pi}{  3k}(m^2+12)}+e^{-i \frac{2 \pi}{ 3k}\left[n^2+n(m-3) \right]}e^{-i \frac{2 \pi}{  3k}(m^2+3m+6)} \nb \\ 
&&\; +2e^{-i \frac{2 \pi }{ 3k}\left[n^2+mn \right]}e^{-i
\frac{2 \pi}{3k}(m^2+3m)+9}
-6e^{-i \frac{2 \pi}{  3k}\left[n^2+n(m+3) \right]}e^{-i \frac{2
\pi }{ 3k}(m^2+3m+12)} \nb \\  
&& \; +2e^{-i \frac{2 \pi}{ 3k}\left[n^2+n(m+6) \right]}e^{-i
\frac{2 \pi}{3k}(m^2+3m+15)}+e^{-i \frac{2 \pi}{  3k}\left[n^2+n(m+9)
\right]}e^{-i \frac{2 \pi}{ 3k}(m^2+3m+18)} \nb \\ 
&&\; -2e^{-i \frac{2 \pi}{ 3k}\left[n^2+mn \right]}e^{-i \frac{2
\pi}{3k}(m^2+6m+12)}
+2e^{-i \frac{2 \pi}{ 3k}\left[n^2+n(m+3)\right]}e^{-i \frac{2
\pi}{ 3k}(m^2+6m+15)} \nb \\
&&\; +e^{-i \frac{2 \pi}{  3k}\left[n^2+n(m+6) \right]}e^{-i 
\frac{2 \pi}{ 3k}(m^2+6m+18)}-2e^{-i \frac{2
\pi}{ 3k}\left[n^2+n(m+9) \right]}e^{-i \frac{2 \pi}{3k}(m^2+6m+21)} \nb \\ 
&&\; +e^{-i \frac{2\pi}{ 3k}\left[n^2+n(m+3) \right]} e^{-i \frac{2
\pi}{ 3k}(m^2+9m+18)}-2 e^{-i \frac{2 \pi}{  3k}\left[n^2+n(m+6)
\right]}e^{-i \frac{2 \pi}{3k}(m^2+9m+21)} \nb \\  
&&\left. \; +e^{-i \frac{2 \pi}{3k}\left[n^2+n(m+9) \right]}e^{-i \frac{2
\pi}{3k}(m^2+9m+24)}\; \right \}   \; \; . 
\label{B5}
\ena
The sums appearing in expression (\ref{B5}) have the form of generalized 
double Gauss sums. In order to
evaluate this expression, we need to recall a few properties of the Gauss 
sums. 

Let $F(x,y,\alpha) $ be the function (generalized Gauss sum) 
\beeq
F(y,\alpha) \; = \; \sum_{n=0}^{y-1} e^{-{i 2 \pi \over y} ( n^2+\alpha n)} 
\quad , 
\end{equation}
where  $y$ (with $y >1 $) and $\alpha$ are integers. 
At this point we can use the reciprocity formula \cite{deu} for the 
generalized Gauss sums 
\beeq
\sum_{n=0}^{|c|-1}e^{i \frac{\pi}{ c}(an^2+bn)}\; = \; \sqrt{\left| \frac{c}{a}
\right|}\; e^{i \frac{\pi}{ 4ac}(|ac|-b^2)} \; \sum_{n=0}^{|a|-1}e^{-i \frac{
y\pi }{a} (cn^2+bn)} \quad , 
\label{B7}
\end{equation}
where the integers $\, a, \, b, \, c\, $ satisfy the relations 
\beeq
 ac \, \neq \; 0 \quad \; \qquad , \qquad   ac \; + \; b \mbox{ ~is even} 
\qquad . 
\end{equation}
With the help of (\ref{B7}), one gets  
\beeq
F(y, \alpha ) \; = \; \sqrt { \frac{y}{ 2}} \; e^{\frac{ i \pi}{  4} \left ( 
\frac{2\alpha^2}{ y} - 1
\right ) } \; \left [ \, 1 + e^{ \frac{i \pi }{2} \left ( y + 2 \alpha \right 
)} \, \right ] \quad . 
\label{B8}
\end{equation}
Eq.(\ref{B8}) can be used to compute $Z_{(+1)}$. Indeed, 
each term entering expression (\ref{B5}) consists of a double generalized 
Gauss sum.  By using eq.(\ref{B8})
twice, each term can be evaluated explicitly. The final result is 
\beeq
Z_{(+)} \; = \; e^{i 6 \pi /k} \; \frac{k \sqrt{3}}{  16 \cos \left( {\pi/ k} \right) \sin^3
 \left( {\pi /k} \right)}\quad . 
\label{B9}
\end{equation}
A nontrivial check of eq.(\ref{B9}) is the following. The product 
$Z_{(+)}\, Z_{(-)} =  | \, Z_{(+)}\,  \bigr |^2  $ coincides with 
$Z_0$ given in eq.(\ref{A11}).
This means that the result (\ref{B9}) is in agreement with Property C2. 

\section{\bf The value of $\mathbf{I(\Sigma_g \times S^1)}$}
In this section, we give the explicit derivation of eq.(\ref{11.15}). The  
manifold $\Sigma_g \times S^1$ can be obtained from $S^2 \times S^1$ by 
``adding $g$ handles".  As we have shown in Sect.6.7.3, 
 each handle admits the decomposition (\ref{10.1}). When $k=5$, the 
non-vanishing 
values of the
$\eta$-coefficients are shown in eq.(\ref{10.5}). Therefore, in order to compute 
$I(\Sigma_g \times S^1) $, we need to consider the manifold $S^2 \times S^1$ 
with $g$ punctures on $S^2$ where each puncture has colour $\Psi_h = 6 
\Psi [0,0] + 3 \Psi [1,1]$. 
We shall decompose the resulting  colour state 
\beeq
\Psi_{gh} \; = \; \left ( \, \Psi_h \, \right )^g \; = \; 3^g \; \left ( \,  \Psi [0,0] + \Psi [1,1] \, \right )^g \quad ,   
\label{C1}
\end{equation}
and, because of Eq.(\ref{8.6}) or rather by using Theorem 6.5, we need to find the 
coefficient $\eta_{gh}  (0)$ of $\Psi [0,0]$ in this
decomposition.  From (\ref{10.4}) one has 
\begin{align}
\Psi [0,0]\, \Psi [0,0] \; = \; \Psi [0,0] \quad &, \quad  
\Psi [0,0]\, \Psi [1,1] \; = \; \Psi [1,1] \nb \\  
\Psi [1,1]\, \Psi [1,1] \; & = \; \Psi [0,0] + \Psi [1,1] \quad . 
\end{align}
Therefore, if $\Psi_b$ denotes the state 
\beeq
 \Psi_b \; = \; \Psi [0,0]\, + \, x \, \Psi [1,1] \quad , 
\end{equation}
where $x$ is a root of the equation $x^2 = x+1$, one gets (for $n \geq 2$) 
\beeq
\left ( \,  \Psi_b \, \right )^n \; = \; \left ( \, x+2 \, \right )^{n-1} \;  \Psi_b \quad . 
\end{equation}
The state $2 \Psi [0,0] + \Psi [1,1]$ can be written as 
\beeq
2 \Psi [0,0] + \Psi [1,1] \; = \; x^{-1} \, \left ( \, (2x+1) \Psi [0,0] + 
\Psi_b \, \right ) \quad . 
\end{equation}
Consequently, from Eq.(\ref{C1}) it follows that 
\beeq
\eta_{gh}  (0) \; = \; {3^g \, \left ( 2x-1 \right )^g \over x^g \, (x+2)} \; \left [ 
\, x+1 + \left ( 1 + \frac{x+2}{  2x-1} \right )^g \, \right ] \quad . 
\label{C6}
\end{equation}
\noindent The two possible values of $x$ are $ \left ( 1 \pm \sqrt 5 \right )/2$. Therefore, Eq.(\ref{C6})
becomes 
\beeq
\eta_{gh}  (0) \; = \; 3^g \, 5^{(g-1)/2} \, \left [ 
\,  \left (  \frac{\sqrt 5 -1}{  2 }\right )^{g-1} \, + \, 
 \left (  \frac{\sqrt 5 + 1}{  2 }\right )^{g-1} \, \right ] \quad . 
\end{equation}
Since the equation $x^2 = x+1$ defines the recursive relation 
\beeq
x^g \; =  \; F(g) \, x \; + \; F(g-1) \quad ,
\end{equation}
for the Fibonacci numbers $\{ \, F(g)\, \}$, the value of
the invariant 
\beeq
I(\Sigma_g \times S^1) \; = \; \eta_{gh}  (0) \; \sqrt{ 3  \left(  \sqrt 5+1 \right) / 2 }
\end{equation}
can be written in the final form
\beeq
I(\Sigma_g \times S^1) \; = \; \begin{cases}
3^g  \>  5^{g/2} \; F(g-1) \; \sqrt{ 3  \left(  \sqrt 5+1 \right) / 2 }
& \text{for }g \text{ even} \; ; \\
3^g  \>  5^{(g-1)/2} \> \left[ F(g)+F(g-2) \right] \> \sqrt{ 3  \left(  \sqrt 5+1 \right) / 2 }
& \text{for } g \text{ odd} \; , \end{cases}
\end{equation}
which is the equation reported in section 7.3. 

\section{\bf Proof of (\ref{numb})}

\shabox{{\bf Lemma~C2}}~{\em Let $\, a, \, b \, $ two integers, with $\, (a,b) = 1 \, $ and $\, b >2 \, 
$ even;  one  has}
\beeq
a^{\phi(b)} \, \equiv \, 1 \quad (\, mod \; 2b) \qquad . \label{999} 
\end{equation}

\bigskip

\no {\bf Proof}~The proof consists of two parts: firstly, it is shown by induction that Lemma~C2 holds when $\, b
= 2^m\, $ with $\, m >1 \, $ integer. Secondly, equation (\ref{999})  is proved when $\, b = 2^m \,
c \, $ with   $\, m \geq 1 \, $  and $\, c \, $ odd integer. 

Since $\, b \, $ is even, $\, a \, $ is clearly odd and can be written in the form $\, a = (2f +1)
\, $. When $ \, b \, $ is of the type $\, b = 2^m\, $, the condition $ \, b >2 \, $ implies that $
\, m \geq 2 \, $. Let us now consider the case   $\, m = 2 \, $; one has $ \, \phi (b) = \phi (\,
2^2\, ) = 2\, $, therefore 
\beeq
a^{\phi(b)} \; = \; (2f \; + \; 1)^2 \; = \; 1 \, + \, 4 f ( \, f+1 \, )  \; \equiv \; 1 \quad (\,
mod \; 2^3 \, ) \qquad .
\end{equation}
Thus, Lemma~C2 is satisfied when $\, b= 2^2 \, $. Suppose now that equation (\ref{999})
holds when $\, b = 2^n \,$ for a certain $ \, n \, $.  We need to prove that  (\ref{999}) is true
also for  $\, b= 2^{(n +1 )}  \,$. Indeed, $\, \phi(2^{n+1}\, ) = 2^n\, $ and one gets
\beeq
(2f \; + \; 1)^{\phi(2^{n+1})} \; = \; \left[\, (2f \;+ \; 1 )^{\phi ( 2^n \, ) } \, \right]^2
 \qquad .
\end{equation}
By using the induction hypothesis
\beeq 
(2f \;+ \; 1 )^{\phi ( 2^n \, ) } \; = \; 1 \, + \, N \, 2^{n+1} \qquad , 
\end{equation}
one finds 
\beeq
\left[\, (2f \;+ \; 1 )^{\phi ( 2^n \, ) } \, \right]^2 \; = \; 1 \, + \, 2^{n+2} \, N \,
\left(\, 1\, + \, 2^n \, N \, \right) \; \equiv \; 1 \quad (\, mod \; 2^{n+2} \,) \; .
\end{equation}
Therefore, equation  (\ref{999}) is also satisfied when $\, b= 2^{(n +1 )}  \,$. To sum up, for $
\, m > 1 \, $ and $ \, a \,  $ odd, one has 
\beeq
a^{\phi(2^m)}  \; \equiv \; 1 \quad (\, mod \; 2^{m+1} \, ) \qquad . \label{first}
\end{equation}
Let us now consider the general case in which $\, b = 2^m \, c \, $ with $\, c \, $ odd integer. 
From Euler's Theorem \cite{cin} it follows that 
\beeq
a^{\phi(b)} \; \equiv \; 1 \quad (\, mod \; b) \, \quad  \Rightarrow \,  \quad a^{\phi(b)}  \;  
\equiv \; 1 \quad (\, mod \; c) \quad .
\label{part1} 
\end{equation}
On the other hand, $ \, \phi( 2^m  c \, ) =  \phi( 2^m \, ) \phi (  c  ) \, $ and, for $ \, m > 1
\, $,  equation  (\ref{first}) implies
\beeq
a^{\phi(b)} \; = \; a^{ \phi(c)\, \phi(2^m\, )} \; \equiv \; 1 \quad (mod \; 2^{m+1}) \qquad . 
\label{part2} 
\end{equation}
Since $\, (2^{m+1}, c)  =1 \, $, from equations (\ref{part1}) and (\ref{part2}) one gets
\beeq
a^{\phi(b)} \; \equiv \; 1 \quad (mod \; 2^{m+1} \, c ) \; \equiv \; 1 \quad (mod \; 2b )\qquad .
\end{equation}
Finally, we need to consider the case $ \, b= 2 \, c \, $. Since $ \, \phi (c) \, $
is even, one gets 
\beeq
a^{\phi (2c)} \; = \; \left [ \, 1 \, + \, 4f(f+1) \, \right]^{\phi(c)/2} \; \equiv \; 1 
\quad (\, mod \; 2^{2} \, ) \qquad . \label{fine}
\end{equation}
Equations (\ref{first}) and (\ref{fine}) imply 
\beeq
a^{\phi (2c)} \; \equiv \; 1 \quad (\, mod \; 2^{2}c\, ) \qquad . 
\end{equation}
This concludes the proof. {\hfill \ding{111}}

\chapter{\bf Conformal field theory in two dimensions}
\section{\bf Introduction}
There is deep connection between Chern-Simons theory in three dimensions and 
two-dimensional conformal models. The '89 Field medal winning paper by Witten, which 
drew the attention of both mathematicians and physicists community
on CS theory, is based heavily on the properties of the conformal blocks 
of the Wess-Zumino-Witten model.  In this appendix we shall give a brief
introduction to some widely used techniques in the study of conformally 
invariant models in two dimensions; with these tools in hand, we shall review the 
original 
Witten's solution of the Chern-Simons theory. This should permit a critical 
comparison with the alternative method presented in this
thesis. Extensive reviews on CFT, with complete references, can be found for 
example in \cite{dms,itz,gin,card}, in particular in this appendix no attempt 
is made to touch important applications of conformal invariance to statistical 
mechanics.       
\section{\bf Conformal invariance}
Let us consider an n-dimensional manifold with a metric $g$. A diffeomorphism
 gives rise to a conformal transformation when
\beeq
g^\prime_{\mu \nu} (x^\prime) \; = \; e^{f(x)} \, g_{\mu \nu}(x) \quad ,
\label{conft}
\end{equation}
where $g^\prime_{\mu \nu} (x^\prime)$ is the transformed metric. The infinitesimal 
form of (\ref{conft}) leads to the conformal Killing's equation
\beeq
\nabla_\mu \epsilon_\nu \; + \; \nabla_\nu \epsilon_\mu \; = \; f \, g_{\mu \nu}
 \quad ,
\label{kill}
\end{equation}
the vector field $\epsilon^\mu$ is the generator of the transformation.
When $g$ is the flat n-dimensional Euclidean metric $\eta$, Eq.(\ref{kill}) can easily be
 solved. The conformal group for n-dimensional Euclidean spacetime is
isomorphic to $SO(n+1,1)$; the $(n+2)(n+1)/2$ generators are associated with the following 
transformations:
\begin{enumerate}
\item rigid translations ${x^\prime}^\mu \, = \, x^\mu \, + \, a^\mu$ ;
\item dilatations ${x^\prime}^\mu \, = \, e^\lambda \, x^\mu \qquad $ with 
$\lambda$ constant;
\item Lorentz transformations ${x^\prime}^\mu \, = \, {\Lambda^\mu}_\nu \, x^\nu$
;
\item special conformal transformations ${x^\prime}^\mu \, = \, \frac{x^\mu \, - \, b^\mu \, x^2}{1 \, - \, 2 b \cdot x \, + \, b^2 \, x^2}$, with $b^\mu$ a constant 
vector, and $x^2 \, = \, x^\mu x^\nu \eta_{\mu \nu}$.
\end{enumerate} 
Let us consider a Poincar\'e invariant classical field theory. By definition, 
the variation of the action under an infinitesimal local translation $
x^\mu \ra x^\mu + \epsilon(x)^\mu$ is proportional to the energy-momentum tensor
\beeq
\delta S \; = \; \frac{1}{2} \int d^nx T^{\mu \nu} \partial_\mu \epsilon_\nu \quad .
\end{equation}  
From the conformal Killing's equation (\ref{kill}), it follows that a sufficient
condition for a Poincar\'e invariant theory to be conformal invariant is  
a traceless energy-momentum tensor. A scale invariant theory can always be provided with 
a traceless energy-momentum tensor, at least when $n >2$. What happens when 
we try to quantize a classical scale invariant theory ? At the quantum level 
one is in trouble from the very beginning, in order to make a quantum field 
theory meaningful the renormalization procedure must be undertaken. However, no 
matter which scheme one uses, renormalization spoils the classical scale 
invariance. Indeed, the values of the renormalized parameters of the theory 
must be provided at some arbitrary energy scale. The breaking of the 
scale-invariance in the quantum theory is made explicit by  the Callan-Symanzik's 
equation for the renormalized correlation functions. In general, scale invariance at 
the quantum level is doomed to
fail, however in some particular case there is a way to escape. There exist models
for which one can fine tune the parameters in such a way that the beta function 
vanishes; in other words, the fixed point of the renormalization group in the 
space of the parameters is selected and scale invariance is recovered. 
Nevertheless, when $n \neq 2$, conformal invariance doesn't provide too stringent constraints: 
it enables one to fix the form of the two and three point function. 
On the contrary, in two dimensions something special happens: it turns out that 
the conformal group is infinite dimensional, or more precisely, the algebra of the 
infinitesimal conformal transformations is an infinite dimensional algebra, 
the celebrated Virasoro algebra.    

Let us consider $\mathbb{R}^2$ equipped with the flat Euclidean  metric. It is very useful
to use instead of the standard coordinates $t,x$ the following complex coordinates 
\beeq
z \; = \; t \; + \; i x \qquad \bar{z} \; = \; t \; - \; i x \quad .
\end{equation}
Any vector $v^\mu$ can be expressed as
\beeq
v^\alpha \; = \; (v^z, \, v^{\bar{z}}), \qquad v^z \; = v^0 \; + \; i v^1,  v^{\bar{z}} \; = \; v^0 \; - \; i v^1 \quad .
\end{equation}
In the new coordinates the metric becomes
\beeq
\eta_{\alpha \beta} \; = \; \left ( \begin{array}{cc} 0 & \frac{1}{2} \\
\frac{1}{2} & 0 \end{array} \right) \; , \qquad \eta^{\alpha \beta} \; = \; \left ( \begin{array}{cc} 0 & 2 \\ 2 & 0 \end{array} \right) \qquad \alpha, \, \beta \, = \, z, \, \bar{z} \quad .
\end{equation}
We shall often use the following differential operators
\beeq
\partial \; = \; \frac{1}{2} \left(\frac{\partial}{\partial t} \; - \; i \frac{\partial}{\partial x} \right) \; , \qquad \bar{\partial} \; = \; \frac{1}{2} \left(\frac{\partial}{\partial t} \; + \; i \frac{\partial}{\partial x} \right)\qquad .
\end{equation}
The conformal Killing's equation for $\eta_{\alpha \beta}$ assumes a simple form
\bea
&&\bad \epsilon \; = \; 0 , \qquad \di \bar{\epsilon} \; = \; 0 \quad , \nb \\
&&\di \epsilon \; + \; \bad \bar{\epsilon} \; = \; -\frac{1}{2} \, f \quad ,
\ena
with $\epsilon^z =\epsilon$ and $ \bar{\epsilon} = \epsilon^{\bar{z}}$. 
Thus, infinitesimal conformal transformations are generated by holomorphic and anti-holomorphic vector fields. Moreover, consider the transformation
\beeq
z \ra z^\prime \; = \; h(z), \qquad \bar{z} \ra {\bar{z}}^\prime \; = \; 
\bar{h}(\bar{z}) \quad ; 
\label{ctr}
\end{equation}
clearly it is a conformal transformation, indeed
\beeq
ds^2 \; = \; 2 \eta^\prime_{z^\prime {\bar{z}}^\prime} \, dz^\prime d{\bar{z}}
^\prime \; = \; 2 \eta_{z \bar{z}} \left(\di h \bad  \bar{h} \right) \, dz 
d \bar{z}
 \quad . 
\end{equation}
By adding the point at infinity, $\mathbb{R}^2 \cup \{ \infty \}$ can be identified with the Riemann sphere, equivalent, from a topological point of view, to $S^2$.
However, in general the transformations of the form (\ref{ctr}) cannot be regarded as the 
finite version of conformal transformations associated with holomorphic 
and anti-holomorphic vector fields; actually, the only global holomorphic 
maps of $S^2$ into $S^2$ are the Moebius transformations
\beeq
z \ra \frac{a z\; + \; b}{cz \; + \; d} \quad ,
\end{equation}
with $a,b,c,d$ are complex numbers satisfying $ad-cb=1$. Put differently, the 
group of global holomorphic maps of $S^2$ is isomorphic with $SL(2,
\mathbb{C})/\mathbb{Z}_2$. This result agrees with what we have found in the 
study of the general n-dimensional case, indeed $SO(3,1) \sim SL(2,\mathbb{C})/\mathbb{Z}_2$.
The existence of an infinite number of infinitesimal conformal transformations
constrains severely a conformal invariant local field theory. 
\section{\bf Conformal Ward identities}
Let us consider the correlation function of some local operator
$\Theta(x_1, \cdots, x_n)$
\beeq
\langle \, \Theta(x_1, \cdots, x_n) \, \rangle \; = \;  G_n(x_1, \cdots, x_n)
 \; = \; \int D[\varphi] \; e^{-S[\varphi]} \; \Theta(x_1, \cdots, x_n) \quad .
\end{equation}
The action functional $S[\varphi]$, which depends on a set of fields denoted
symbolically by $\varphi$, is supposed to be invariant under rigid translations. 
Consider a local infinitesimal translation, i.e. $x^\mu \ra {x^\prime}^\mu \, = \, x^\mu \, + \, \epsilon^\mu$, $\varphi^\prime(x) \, - \, \varphi(x) \, = \, \delta \varphi(x)$. The variation of the action is by definition the 
following
\beeq
\delta S \; = \; - \frac{1}{2 \pi} \int d^2x \; {T^\mu}_\nu \di_\mu \epsilon^\nu  \quad .
\end{equation}   
Clearly, because $\varphi$ is a dummy field in the path integral, the following identity holds
\beeq
 \int  D[\varphi^\prime] \; e^{-S[\varphi^\prime]} \; \Theta^\prime(x_1, \cdots, x_n) \; = 
\; \int D[\varphi] \; e^{-S[\varphi]} \; \Theta(x_1, \cdots, x_n) \quad .
\label{dummy}
\end{equation}
The statement of invariance at the quantum level of the theory  corresponds 
to the freedom in the path integral on the left hand side of eq.(\ref{dummy}) 
to integrate over the old field $\varphi$ instead of $\varphi^\prime$
\beeq
 \int D[\varphi] \; e^{-S[\varphi^\prime]} \; \Theta^\prime(x_1, \cdots, x_n) \; = 
\; \int D[\varphi] \; e^{-S[\varphi]} \; \Theta(x_1, \cdots, x_n) \quad .
\label{dum1}
\end{equation}
Eq.(\ref{dum1}), in the case of an infinitesimal local translation, leads to
\beeq
\int D[\varphi] \; e^{-S[\varphi]} \;  \frac{1}{2 \pi} \int d^2x \; {T^\mu}_\nu \di_\mu 
\epsilon^\nu \quad \Theta(x_1, \cdots, x_n) \; = \; - \langle \, \delta \Theta(x_1, \cdots, x_n)
\, \rangle \quad .
\end{equation}
Consider a vector field which is non-vanishing only in a region $D$, by using the 
conservation of the energy-momentum tensor and Stokes's theorem, one obtains 
the following conformal Ward identity
\beeq
\langle \, \frac{1}{2 \pi i} \int_C \left(T \, \epsilon \, dz \; - \; \bar{T} \, \bar{\epsilon} \, d \bar{z} \right) \; \Theta(z_1, \cdots, z_n; \, \bar{z}_1, \cdots, \bar{z}_n) \; = \; \langle \, \delta \Theta(z_1, \cdots, z_n; \, \bar{z}_1, \cdots, \bar{z}_n)
\, \rangle \quad ,
\label{cwi}
\end{equation}
where
\bea
&&T(z) \; = \; T_{z z}(z), \qquad \bar{T}(\bar{z}) \; = \; T_{\bar{z} \bar{z}}(\bar{z}) \quad , \nb \\
&&\epsilon(z) \; = \; \epsilon^z(z), \qquad \bar{\epsilon}(\bar{z}) \; = \; 
\epsilon^{\bar{z}}(z) \quad .
\ena
The curve $C$, anti-clockwise oriented, bounds the region in which the vector 
field $\epsilon^\mu$ is non-vanishing, and also contains the points $x_1, \cdots,
x_n$ as shown in Fig.D.1. Note that the conservation of the energy-momentum tensor and his property of being traceless are expressed as
\beeq
T^\mu_\mu \; = \; 0 \ra T_{z \bar{z}} \; = \; T_{\bar{z} z} \; = \; 0 \quad ,
\end{equation}
\beeq
\partial_\mu \, T^{\mu \nu} \; = \; 0 \ra \bad T \; = \; 0 \quad \mbox{and }
\quad \di \bar{T} \; = \; 0 \quad .
\end{equation} 

\begin{figure}[h]
\vskip 0.5 truecm 
\centerline{\epsfig{file=\path 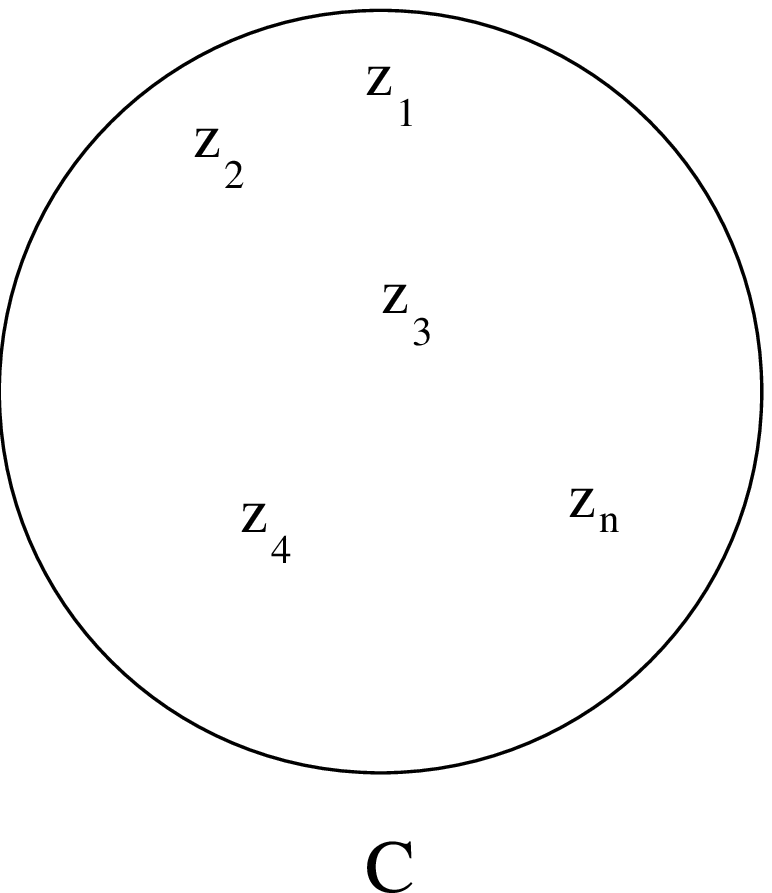,height=5cm,width=5cm}}
\vskip 0.5 truecm 
\centerline{{\bf Figure D.1}}
\vskip 0.5 truecm 
\end{figure}
\section{\bf Operator product expansion}
A distinct role in conformal field theories is played by primary fields. A primary field $\phi(z, \bar{z})$ transforms under $z \ra z^\prime(z), \;
\bar{z} \ra \bar{z}^\prime(\bar{z})$  as
\beeq
\phi^\prime(z^\prime, \,\bar{z}^\prime) \; = \; \left( \frac{d z^\prime}{d z}
\right)^{-h} \; \left(\frac{d \bar{z}^\prime}{d \bar{z}} \right)^{- \bar{h}}
\; \phi(z, \,\bar{z}) \quad ,
\label{prim}
\end{equation}
$h$ and $\bar{h}$ are by definition the conformal weights of the primary field $\phi$.
In particular, for an infinitesimal transformation  
\beeq
z^\prime(z) \; = \; z \; + \; \epsilon(z), \qquad \bar{z}^\prime(\bar{z}) \; = \; \bar{z} \; + \; \bar{\epsilon}(\bar{z}) \quad ,
\end{equation}
one gets
\beeq
\delta \phi(z, \,\bar{z}) \; = \; - \left(\epsilon \di \; + \; h \di \epsilon
\right)\phi(z, \,\bar{z}) \; - \; \left(\bar{\epsilon} \bad \; + \; \bar{h} 
\bad \bar{\epsilon} \right)\phi(z, \,\bar{z}) \quad .
\label{trp}
\end{equation}
Let us consider the consequences of eq.(\ref{cwi}) for a primary field. 
By using eq.(\ref{trp}), the conformal Ward identity 
can be written as
\beeq
\frac{1}{2 \pi i } \int_{C_w} \langle \, T \epsilon \, \phi(w, \bar{w}) dz
\rangle \; = \; \langle \,\left(\epsilon \di \; + \; h \di \epsilon
\right)\phi(w, \,\bar{w}) \, \rangle \quad .
\label{iope}
\end{equation}
For simplicity, we have written only the holomorphic part, the 
anti-holomorphic part being completely analogous. As a consequence of Cauchy's theorem,
eq.(\ref{iope}) determines the singular part, i.e the pole structure, of the product 
$T(z) \, \phi(w, \bar{w})$
\beeq
T(z) \, \phi(w, \bar{w}) \sim \frac{1}{(z \, - \, w)} \, \di \phi(w, \bar{w}) \; + \; 
\frac{h}{(z \, - \, w)^2} \, \phi(w, \bar{w}) \quad .
\label{opep}
\end{equation}
The symbol $\sim$ means that only the singular terms in the operator product 
expansion (OPE) are shown, on the other hand from eq.(\ref{iope}) only 
these singular terms can be determined. A similar expression is obtained 
for the OPE of the primary field $\phi$ with $\bar{T}$
\beeq
\bar{T}(\bar{z}) \, \phi(w, \bar{w}) \sim \frac{1}{(\bar{z} \, - \, \bar{w})} 
\, \bad \phi(w, \bar{w}) \; + \; 
\frac{\bar{h}}{(\bar{z} \, - \, \bar{w})^2} \, \phi(w, \bar{w}) \quad .
\end{equation}  
The consequences of the Ward identity are particular simple when a global
conformal transformation is considered: the action is invariant and, in the absence 
of anomalies, the correlation functions must be conformally invariant
\beeq
\langle \, \delta \Theta(z_1, \cdots, z_n; \, \bar{z}_1, \cdots, \bar{z}_n) 
\, \rangle \; = \; 0 \quad . 
\label{global}
\end{equation} 
Eq.(\ref{global}) fixes the form of both the two and three point function for 
primary fields
\bea
&&\langle \, \phi_1(z_1, \, \bar{z}_1) \; \phi_2(z_2, \, \bar{z}_2) \, \rangle \;
 = \; \frac{C_{12}}{(z_1 \, - \, z_2)^{2h} \; (\bar{z}_1 \, - \, 
\bar{z}_2)^{2\bar{h}}} \quad , \nb \\
&&C_{12} \; = \; \alpha_{12} \, \delta_{h_1, \, h_2} \; \delta_{\bar{h}_1, \, 
\bar{h}_2} \quad . 
\label{2point}
\ena
Unless the two primary fields have the same conformal weights, the two-point
function vanishes; one can also normalize the field in such way 
that $\alpha_{12}=1$. For the three-point function one has
\bea
\langle \, \phi_1(z_1, \, \bar{z}_1) \; \phi_2(z_2, \, \bar{z}_2) \phi_3(z_3, \,
 \bar{z}_3 ) \, \rangle \; && = \; C_{123} \; \frac{1}{z_{12}^{h_1 +h_2 -h_3} \;
z_{23}^{h_2 +h_3 -h_1} \; z_{13}^{h_1 +h_3 - h_2}} \nb \\
&&\frac{1}{\bar{z}_{12}^{\bar{h}_1 +\bar{h}_2 -\bar{h}_3} \;
\bar{z}_{23}^{\bar{h}_2 +\bar{h}_3 -\bar{h}_1} \; \bar{z}_{13}^{\bar{h}_1 +\bar{h}_3 - \bar{h}_2}} \quad .
\label{3point}
\ena 
\section{Central charge}
As we have seen, the singular part of the OPE of a generic field with the 
energy-momentum tensor is determined by the behaviour of that field under an 
infinitesimal conformal transformation. In particular, it is interesting to study
the OPE of $T$ with itself. 

By setting $\Theta =1$ in eq.(\ref{cwi}), it follows that 
\beeq
\langle \, \int_C \, T(z) \, \epsilon(z) \; dz \, \rangle \; = \; 0 \quad .
\label{cinf}
\end{equation}
In particular, when a global conformal transformation is considered, $\epsilon(z) 
\sim z^2$ and the curve $C$ can be taken as a circle surrounding the origin with an 
arbitrary radius. In order that eq.(\ref{cinf}) holds, 
one needs that  $T(z) \sim z^4$ as $|z| \ra 
\infty$. Moreover, by using translation and dilatation invariance, one can show 
that 
\bea
&&\langle \, T(z) \, T(0) \, \rangle \; = \; \frac{c}{2} \; \frac{1}{z^4} \quad , \nb \\
&&\langle \, \bar{T}(\bar{z}) \, \bar{T}(0) \, \rangle \; = \; \frac{c}{2} \; 
\frac{1}{\bar{z}^4} \quad ,
\label{2con}
\ena
where $c$ is a constant.
From eq.(\ref{2con}) and from eq.(\ref{cwi}), with $\Theta = T(0)$, one has
that
\beeq
\frac{1}{2 \pi i}  \int_C \frac{c}{2} \; \frac{\epsilon(z)}{z^4} \; dz \; = \; 
- \langle \, \delta T(0) \, \rangle \quad ,
\end{equation} 
thus 
\beeq
\langle \, \delta T(0) \, \rangle \; = \; -  \frac{c}{12} \; \frac{d^3 \epsilon}{d z^3} \bigr |_{z=0} \quad .
\label{anomaly}
\end{equation} 
On the one hand the energy-momentum tensor transforms as a primary field with weight $2$ under 
a global conformal transformation, on the other hand eq.(\ref{anomaly}) shows a non 
primary behaviour under a generic infinitesimal conformal transformation. 
The minimal assumption, consistent both with eq.(\ref{anomaly}) and with the 
primary behaviour of $T(z)$ under  $SL(2, \mathbb{C})$ is the following
\beeq
\delta T(z) \; = \;  -  \frac{c}{12} \; \frac{d^3 \epsilon}{d z^3} \; - \; 2 \, 
T(z) \, \di \epsilon(z) \; - \; - \epsilon(z) \, \di T(z) \quad .
\label{momb}
\end{equation}
The conformal Ward identity and  eq.(\ref{momb}) lead to the OPE
\beeq
T(z) \, T(0) \sim \frac{c/2}{z^4} \; + \; \frac{2}{z^2} \; T(0) \; + \; 
\frac{1}{z} \; \di T(0) \quad ,
\label{opet}
\end{equation}
\beeq
\bar{T}(\bar{z}) \, \bar{T}(0) \sim \frac{c/2}{\bar{z}^4} \; + \; \frac{2}{\bar{z}^2} \; \bar{T}(0) \; + \; 
\frac{1}{\bar{z}} \; \bad \bar{T}(0) \quad .
\label{opebt}
\end{equation}
Finally, the independent character of the transformations on $z$ and $\bar{z}$ leads to
\beeq
T(z) \, \bar{T}(0) \sim 0 \quad .
\end{equation} 
The constant $c$ in the OPE is the central charge.
Eq.(\ref{anomaly}) can be integrated to obtain the behaviour of the 
energy-momentum tensor under finite transformations, the result is 
\beeq
T^\prime(z^\prime) \; = \; \left(\frac{d z^\prime}{dz} \right)^{-2} \; \left[ 
T(z) \; - \; \frac{c}{12} \; \{z^\prime, \, z \} \right] \quad .
\label{momf}
\end{equation}
Where $\{f(z), \, z \}$ is the Schwarzian derivative, defined as
\beeq
\{f(z), \, z \} \; = \; \left( \frac{d^3 f}{dz^3} \right)^2 \, \left( \frac{d f}
{dz} \right)^{-1} \; - \; \frac{3}{2} \left( \frac{d^2 f}{dz^2} \right)^2 \, 
\left( \frac{d f}{dz} \right)^{-2} \quad .
\end{equation}
One can verify that $\{w(z), \, z \}= 0 $, when 
\beeq
w(z) \; = \; \frac{az \; + \; b}{cz \; + \; d} \qquad ad \; - \; bc \; = \; 1 
\quad .
\label{glo}
\end{equation} 
In general, a field transforming as a primary one under a global conformal 
transformation is called quasi-primary.

As a simple example, one can consider a free scalar field; it is straightforward 
to  verify that $c=1$; the energy momentum-tensor is defined by considering 
the normal ordered classical expression. Also the case of a free Majorana fermion can be  
easily worked out; one finds $c=1/2$. Because the energy-momentum tensor for non-interacting
fields is the sum of the energy-momentum tensors associated with the single fields,
the central charge is additive for these models. From a physical point of view,
the central charge probes the behaviour of the theory when a macroscopic scale is 
introduced. As an example, let us consider a theory on $\mathbb{R}^2$, and define 
the energy-momentum tensor such that 
\beeq
\langle \, T(z) \, \rangle \bigr |_{plane} \; = \; 0 \quad .
\end{equation} 
One can map the complex plane into a cylinder of radius $L$ by the 
following conformal transformation (note that is not a globally holomorphic map ) 
\beeq
z \ra  w \; = \; \frac{L}{2 \pi} \, \ln z \quad .
\end{equation}
By using eq.(\ref{momf}), the expectation value of the energy momentum-tensor in 
the cylinder is given by  
\beeq
\langle \, T(z) \, \rangle \bigr |_{cyl.} \; = \; - \frac{c \pi^2}{6 L^2} \quad .
\end{equation}
A Casimir energy, proportional to the central charge  has come out. 
\section{\bf Virasoro algebra}
Up to now we have used the path integral formalism; no operator acting on 
some Hilbert space has been introduced; in this section we shall fill the gap.
In conformal field theory there exists a peculiar way to introduce the canonical 
formalism: the radial quantization. The idea is to choose as monotonic ``time''
coordinate on the complex plane simply $\ln|z|$; as a result, the equal time lines
will be circles around the origin. Given two bosonic operators 
$\varphi_1(z_1)$ and $\varphi_2(z_2)$, the radial ordering is naturally defined as 
\beeq
{\cal R}  \left(\varphi_1(z_1) \; \varphi_2(z_2) \right)
\left \{ \begin{array}{cc} \varphi_1(z_1) \; \varphi_2(z_2) & \mbox{if }
\; |z_1| > |z_2| \\
\varphi_2(z_2) \; \varphi_1(z_1) & \mbox{if }
\; |z_2| > |z_1| \end{array} \right. \quad .
\end{equation}  
If no specified, operators are understood to be radial  ordered.
The radial quantization is particularly useful because one can represent the 
commutator between the  ``charge''operator  associated with a symmetry and a local 
operator as a contour integral. Indeed, define the operator $\cal A$ as
\beeq
{\cal A} \; = \; \oint a(z) \, dz \quad .
\end{equation} 
For contour integrals we shall use the following notation: when no specified, 
the contour surround the origin, otherwise $\oint_w$ denotes a contour 
integral around the point $w$. From the definition of radial order 
and from the holomorphic 
character of the fields, it follows that  
\beeq
\left [ {\cal A}, \; \varphi(w) \right] \; = \; \oint_w a(z) \, \varphi(w) \, dz 
\quad .
\label{comc}
\end{equation}
Eq.(\ref{comc}) is remarkable; given the charge operator $\cal A$, the action of 
the associated symmetry on a local field is determined only by 
the singular terms appearing in the OPE between the current from which the charge 
is derived and the local operator. A key role is played in conformal field theory 
by the current associated with local conformal symmetries: the 
energy-momentum tensor. Let us consider the charge
\beeq
Q_{\epsilon} \; = \; \oint \epsilon(z) \, T(z) \, dz
\end{equation}
From eq.(\ref{trp}) one has
\beeq
\delta \phi(z) \; = \;  - \left(\epsilon \di \; + \; h \di \epsilon
\right)\phi(z) \; = \; \left [ Q_\epsilon, \; \phi(z) \right]
\quad .
\end{equation}
As usual, a similar relation for the anti-holomorphic part is understood.
One can Laurent expand $T(z)$ around the origin in terms of modes
\bea
&&T(z) \; = \; \sum_{n \in \mathbb{Z}} \, z^{-n-2} \, L_n \quad , \\
&&L_n \; = \; \frac{1}{2 \pi i } \oint z^{n+1} \, T(z) \, dz \quad .
\label{mot}
\ena
Similarly for the anti-holomorphic part one has
\bea
&&\bar{T}(\bar{z}) \; = \; \sum_{n \in \mathbb{Z}} \, \bar{z}^{-n-2} \, \bar{L}_n
 \quad , \\
&&\bar{L}_n \; = \; \frac{1}{2 \pi i } \oint \bar{z}^{n+1} \, \bar{T}(\bar{z}) \, dz \quad .
\label{motb}
\ena
The charge $Q_\epsilon$ can be expressed in terms of modes $L_n$ as
\beeq
Q_\epsilon \; = \; \sum_{n \in \mathbb{Z}} \, \epsilon_n \, L_n \quad ,
\end{equation}
with 
\beeq
\epsilon(z) \; = \; \sum_{n \in \mathbb{Z}} \, \epsilon_n \, z^{n+1} \quad .
\end{equation}
By using Eq.(\ref{opet}), (\ref{opebt}) and the mode expansion of the energy momentum 
tensor, one can easily derive the commutation relations between $\{L_n \}$ and 
$\{\bar{L}_n \}$. The result is
\bea
&&\left[\, L_n , \, L_m \, \right] \; = \; (n-m) \, L_{n+m} \; + \; \frac{c}{12} \, n(n^2 \; - \; 1) \; \delta_{n+m, \, 0} \quad , \\
&&\left[\, \bar{L}_n , \, \bar{L}_m \, \right] \; = \; (n-m) \, \bar{L}_{n+m} \; + \; \frac{c}{12} \, n(n^2 \; - \; 1) \; \delta_{n+m, \, 0}  \quad ; \\
&&\left[\, L_n , \, \bar{L}_m \, \right] \; = \; 0 \quad .
\label{vir}
\ena
The operators $\{ L_n, \, \bar{L}_m \}$, together with the commutation relations
of eq.(\ref{vir}), represent a  pair of infinite dimensional algebras $Vir \times \bar{Vir}$. 
The algebra $Vir$ is the Virasoro algebra. The sub-algebra generated by 
$\{ L_{ \pm 1}, \, L_0 \}$ 
is associated with global conformal transformations. In particular $L_0$ is the 
generator of dilatations, as one can infer from
\beeq
\left[\, L_n, \, \phi(w, \, \bar{w}) \, \right] \; = \; h(n+1) \, w^n \, 
\phi(w, \, \bar{w}) \; + \; w^{n+1} \di \phi(w, \, \bar{w}) \qquad n \geq -1 
\qquad .
\label{compr}
\end{equation}
In the radial quantization, dilatations correspond to ``time'' 
translations, thus $L_0 \, + \bar{L}_0$, apart from an additive constant, plays
the role of the Hamiltonian. In particular, if one sets to zero its vacuum 
expectation in the case of plane geometry, it follows that
\beeq
H \; = \; L_0 \; + \; \bar{L}_0 \; - \; \frac{c}{12} \quad .
\label{hami}
\end{equation}
\section{\bf The Hilbert space}
In quantum field theory one postulates the existence of a cyclic vacuum 
state $|\, 0 \, \rangle$ from which the entire Hilbert space is built up  by 
acting on $|\, 0 \, \rangle$ with local operators. To any primary field 
$\phi(z, \bar{z})$ one associates an asymptotic ``in'' state by
\beeq
| \, \phi \, \rangle \; = \; \lim_{z, \bar{z} \ra 0 } \; \phi(z, \bar{z}) \,   
| \, 0 \, \rangle \quad .
\end{equation}
Given a quasi-primary field $\phi(z, \bar{z})$ its Hermitian conjugate is defined 
as 
\beeq
\left[\phi(z, \, \bar{z}) \right]^\dagger \; = \; \bar{z}^{-2h} \, z^{-2 \bar{h}} \phi(1/\bar{z}, \, 1/z) \quad .
\label{dagg}
\end{equation}
The definition (\ref{dagg}) is chosen in such way that the scalar product 
$\langle \phi_{out}| \phi_{in} \rangle \, = \, \langle 0| \phi_{in}^\dagger \phi_{in}|0 \rangle$ is well defined, as one can verify from 
the explicit form (\ref{2point}) for the two-point function.

As we have seen, the algebra $Vir \times \bar{Vir}$ generates the fundamental symmetries
of a conformal invariant theory, then  it is natural to seek for a Hilbert space 
carrying a representation of this algebra. First of all, note that the  requirement of 
invariance of the vacuum state under global conformal transformations and that 
$T(z)|0 \rangle$ makes sense (as well as its anti-holomorphic counterpart), lead to 
\beeq
L_n |0 \, \rangle  \; = \; 0, \qquad \bar{L}_n |0 \, \rangle \; = \; 0 \qquad 
n \geq -1 \quad .
\label{anic}
\end{equation}
A highest weight representation of $Vir \times \bar{Vir}$, or more precisely a module,
is obtained starting from the  highest weight state 
\beeq
|h, \, \bar{h} \, \rangle \; = \; \phi(0,0)|0 \, \rangle \quad .
\end{equation}
Because the holomorphic and anti-holomorphic parts of the algebra commute, they 
can be analyzed separately; once a relation involving the holomorphic part is 
found, the same relation for the anti-holomorphic part holds, for this reason we 
shall pay attention to the holomorphic part only.   
From (\ref{compr}) it follows that $|h, \, \bar{h} \, \rangle$ is an eigenstate with 
eigenvalue $h$
\beeq
L_0 |h, \, \bar{h} \, \rangle \; = \; h \, |h, \, \bar{h} \, \rangle \quad .
\end{equation}
On the other hand, eq.(\ref{anic}), together with the Virasoro algebra show that
$L_{-n}$ with $n >0$ act as raising operators. Indeed
\beeq
L_0 \, L_{-k_1} \, \cdots \, L_{-k_n} |h, \, \bar{h} \, \rangle \; = \; 
(h \; + \; k_1 \; + \; \cdots \; + \; k_n) |h, \, \bar{h} \, \rangle  \quad .
\end{equation}
The state $L_{-k_1} \, \cdots \, L_{-k_n} |h, \, \bar{h} \, \rangle$ is called a 
descendent of level $N=\sum_i k_n$. Note that, as a consequence of the Virasoro algebra, 
states at different levels are orthogonal. The subset spanned by $|h, \, \bar{h} \, 
\rangle$ and its descendents is closed under the action of $Vir \times \bar{Vir}$, this
 subset is called a Verma module. Each Verma module is determined by the 
central charge $c$ and the conformal weight $h, \, \bar{h}$ of 
the primary field corresponding to the highest weight state. 
The full module is the direct sum of  
the various Verma modules ${\cal V}$ associated with the primary fields of the theory
\beeq
{\cal H} \; = \; \sum_{i} {\cal V}(c, \, h_i) \otimes \bar{{\cal V}}(c, \, \bar{h}_i)
\quad . 
\end{equation}
In general, given ${\cal V}(c, \, h)$, negative norm states can be present. The 
Verma modules in which all the states have positive norm are called unitary. 
From a physical point of view, unitary Verma modules are particular important, in 
this case ${\cal H}$ is a Hilbert space. Necessary conditions for unitary Verma 
modules are clearly $c \geq 0$ and $h, \, \bar{h} \geq 0$. In particular, when
$0 < c < 1 $, a discrete series of unitary Verma modules can be found 
(minimal models); remarkably, among them one can find well known statistical 
models like Ising, tricritical Ising, 3 state Potts. One of the crucial property of 
 minimal models, is the existence of states of null norm;  
we shall see that this fact gives rise to non trivial constraints on the operator 
algebra.

\section{\bf Operator algebra and conformal blocks}
Descendent states can be obtained from the highest weight state by
applying $L_{-n}$ with $ n>0$; it is natural to define the corresponding
descendent fields
\beeq
L_{-n} |h, \, \bar{h} \, \rangle \; = \; L_{-n} \, \phi(0,0) |0 \,
\rangle \; = \; \phi^{(-n)}(0,0)|0 \, \rangle \qquad  \qquad n > 0 \quad .
\end{equation}
A similar definition can be given for $\phi^{(k_1, \cdots, k_n)}$. In
general the holomorphic field $(L_n \phi)(w)$ is defined in terms of
the mode expansion of $T(z)\phi(w)$ around $w$
\beeq
T(z) \, \phi(w) \; = \; \sum_{n \in \mathbb{Z}} (z-w)^{-n-2} \; (L_n
\phi)(w)
\qquad . 
\label{desc}
\end{equation}
Singular terms can be determined by using the OPE between $T(z)$
and $\phi(w)$, the regular part is given by Taylor expansion. A
primary field $\phi$, together with all its descendents is called
the conformal family of $\phi$.  An important consequence of OPE is that
correlation functions involving descendent fields and primary fields
can be expressed in terms of correlation functions of primary fields
only. Indeed
\beeq
\langle \, \phi^{(-k_1)}\, \cdots \phi^{(-k_n)})(w) \, \phi_1(w_1) \, \cdots
\phi_n(w_n) \, \rangle \; = \; {\cal D}_{-k_1} \, \cdots {\cal D}_{-k_n} 
\langle \, \phi(w) \, \phi_1(w_1) \, \cdots
\phi_n(w_n) \, \rangle \quad ,
\label{coneq}
\end{equation}
where the differential operator ${\cal D}_{-k_m}$ is defined as
\beeq
{\cal D}_{-k_i} \; = \; \sum_{i=1}^m \left[ \frac{(m \; - \;
1)h_i}{(w_i \; - \; w)^n} \; - \; \frac{1}{(w_i \; - \; w)^{n-1}}
\di_{w_i} \right] \qquad .
\end{equation} 
Eq.(\ref{coneq}) implies that all the correlation functions of the
theory can be worked out from the knowledge of the primary fields
correlation functions. Moreover, because the form of the two-point
function is completely constrained by global conformal invariance (see
eq.(\ref{2point})), when one is able to produce the complete OPE (regular
term included) between primary fields, all the correlation functions
can be calculated and the theory is solved. The OPE between two primary
fields can be written as
\beeq
\phi_i(z, \, \bar{z}) \, \phi_j(0, \, 0) \;  = \; \sum_p \sum_{\{k, \bar{k} \}}
C_{ij}^{p \, \{k, \bar{k} \}} \; z^{h_p-h_i-h_j + K} \;
\bar{z}^{\bar{h}_p-
\bar{h}_i - \bar{h}_j + \bar{K}} \; \phi_p^{\{k, \bar{k} \}}(0, \, 0) \quad ,
\label{calg}
\end{equation}
where $\phi_p^{\{k, \bar{k} \}}$ represents a generic element of the
conformal family of $\phi_p$, and $K = \sum_i k_i$, $\bar{K} = \sum_i
\bar{k}_i$. The dependence on $z, \, \bar{z}$ is fixed by global
conformal invariance. Given a conformal theory, the set of all OPE's
between primary fields gives rise to the so called conformal algebra.
As we have seen, the knowledge of the conformal algebra amounts to
solve the theory. One can show that
\beeq
C_{ij}^{p \, \{k, \bar{k} \}} \; = \; C_{ij}^p \; \beta_{ij}^{ \{k \}}
\; \bar{\beta}_{ij}^{ \{\bar{k} \}} \quad ,
\end{equation}
where $C_{pij}$ are the same coefficients appearing in the three point
function in eq.(\ref{3point}). Moreover, $\beta_{ij}^{ \{k \}}$ and
$\bar{\beta}_{ij}^{ \{\bar{k} \}}$ can be computed from the definition
in terms of the conformal weights and the central charge. Thus, the
conformal algebra is determined by the coefficients $C_{pij}$ involved
in the three-point function.

Clearly, it would be nice to have some constraints on $C_{pij}$; with this goal 
in mind, let us consider the following matrix element
\beeq
G^{21}_{34}(X, \bar{X}) \; = \; \langle \, h_1, \bar{h}_1| \,
\phi_2(1,1) \, \phi_3(X, \bar{X}) \, | h_4, \bar{h}_4 \, \rangle \quad
.
\label{ma4}
\end{equation}
The matrix element can be expressed in terms of the four-point function
which, by global conformal invariance, depends only on the anharmonic
ratios
\beeq
X \; = \; \frac{(z_1 \; - \; z_2)(z_3 \; - \; z_4)}{(z_1 \; - \;
 z_3)(z_2 \; - \; z_4)} \; , \qquad \bar{X} \; = \; \frac{(\bar{z}_1 \; - \;
 \bar{z}_2)(\bar{z}_3 \; - \; \bar{z}_4)}{(\bar{z}_1 \; - \;
 \bar{z}_3)(\bar{z}_2 \; - \; z_4)} \quad .
\end{equation}
By an $SL(2, \mathbb{Z})$ transformation one can set $z_4=0, \;
z_1= \infty, \; z_2=1$, thus $z_3=X$. By using the conformal algebra to
contract $\phi_3$ with $\phi_2$, one gets
\beeq
G^{21}_{34}(X, \bar{X}) \; = \; \sum_p C^p_{12} \; C^p_{34} \; {\cal
A}^{21}_{34}(p|X, \bar{X}) \quad ,
\label{ints}
\end{equation}
in eq.(\ref{ints}) we have defined
\beeq
{\cal A}^{21}_{34}(p|X, \bar{X}) \; = \; \left(C_{12}^p \right)^{-1}
X^{h_p -h_3-h_4}
\, \bar{X}^{\bar{h}_p -\bar{h}_3-\bar{h}_4} \; \sum_{ \{ k, \bar{k} \}}
\beta_{34}^{\{ k \}} \, \bar{\beta}_{34}^{\{ \bar{k} \}} X^{K} \,  
\bar{X}^{\bar{K}} \; \langle \, h_1, \bar{h}_1|\phi_2(1,1) \, \phi_p^{\{ k, 
\bar{k}}(0,0)|0 \, \rangle .
\end{equation} 
As usual, the holomorphic and anti-holomorphic parts factorize
\beeq
{\cal A}^{21}_{34}(p|X, \bar{X}) \; = \; {\cal F}^{21}_{34}(p|X) \;
\bar{{\cal F}}^{21}_{34}(p|\bar{X}) \quad .
\label{conb}
\end{equation}
The holomorphic and anti-holomorphic parts of ${\cal A}$ are called
holomorphic conformal blocks and anti-holomorphic conformal blocks
respectively, and they can recursively be calculated by the definition in
terms of conformal weights and the central charge. When expressed in
terms of conformal blocks, eq.(\ref{ints}) becomes
\beeq
G^{21}_{34}(X, \bar{X}) \; = \; \sum_p C^p_{12} \; C^p_{34} \;{\cal
F}^{21}_{34}(p|X) \; \bar{{\cal F}}^{21}_{34}(p|\bar{X}) \quad .
\label{inde}
\end{equation}
Conformal blocks can be understood as intermediate states
entering the decomposition of the four-point function. We note that
by using global conformal invariance one can equivalently set $z_2=0, \; z_4=1,
\; z1=\infty$, thus $z_3=1-X$, and contract $\phi_3$ with $\phi_4$, put
differently, the conformal algebra must be associative. One obtains
\beeq
G^{21}_{34}(X, \bar{X}) \; = \; G^{41}_{32}(1-X, 1-\bar{X}) \quad .
\label{cros}
\end{equation}
Eq.(\ref{cros}) constrains the conformal algebra, namely
$C^p_{ij}$; indeed
\beeq
\sum_p C^p_{nm} \; C^p_{lk} \; {\cal F}^{kl}_{nm}(p|X) \; \bar{{\cal F}}^{kl}_{nm}(p|\bar{X}) \; = \sum_p C^p_{nk} \; C^p_{lm} \; {\cal F}^{ml}_{nk}(p|1-X) \; \bar{{\cal F}}^{ml}_{nk}(p|1-
\bar{X}) \quad .
\label{crossing}
\end{equation}
Eq.(\ref{crossing}) shows a crossing symmetry in the
decomposition of the four-point function by using the conformal
algebra. The program to determine the $\{C^p_{ij}\}$ from Eq.(\ref{crossing}) is 
called conformal bootstrap. Whether the bootstrap program is viable in a
generic conformal field theory is an open problem, however, notably, it
works for minimal models.

\section{\bf Fusion rules} 
The fundamental objects entering the OPE between two primary fields
are the coefficients $\{C_{ij}^p \}$, they determine which conformal
families appear in the decomposition. One can focus the attention on this aspect
 by defining the fusion coefficients as
\beeq
{\cal N}_{ij}^p \; = \; \left \{ \begin{array}{cc} 0 & \mbox{ if }
\; C_{ij}^p \, = \, 0
\\ 1 & \mbox{otherwise} \end{array} \right. \quad .
\label{fure}
\end{equation}
By using the fusion coefficients, the OPE between two primary fields
can be represented in a compact way as
\beeq
\phi_i \times \phi_j \; = \; \sum_p {\cal N}_{ij}^p \; \phi_p \quad .
\label{fude}
\end{equation} 
The rule given in Eq.(\ref{fude}) is used to define an associative
algebra called fusion algebra. In particular, the conformal family
corresponding to the unity operator $\boldmath{1}$ gives rise to trivial
fusion rules
\beeq
{\cal N}_{i1}^p \; = \; \delta^p_i \quad .
\end{equation}
At this level, the crossing symmetry of the conformal algebra
translates into the associativity of the fusion algebra, explicitly
\beeq
\sum_p {\cal N}_{jm}^p \; {\cal N}_{ip}^n \; = \; \sum_s {\cal N}_{ij}^s
 \; {\cal N}_{sm}^n \quad .
\label{asso}
\end{equation}
We note that defining the matrix $(N_i)$ by
\beeq
(N_i)_{jm} \;  = \; {\cal N}_{ij}^m \qquad ,
\end{equation}
eq.(\ref{asso}) can be interpreted as 
\beeq
N_i \cdot N_j \; = \; N_j \cdot N_i \qquad ,
\end{equation} 
the dot denotes the standard matrix multiplication. The set of matrices 
$\{N_i\}$ is Abelian. As a linear space, $\{N_i\}$, give rise 
to the  regular representation of the fusion algebra by means of 
Eq.(\ref{asso}); the action of $N_i$ on $N_i$ is defined as  
\beeq
N_i(N_j) \; = \; \sum_s {\cal N}_{ij}^s \; N_s  \quad .
\end{equation}
One of the non trivial consequences of the existence of a null vector
in a Verma module is a constraint on the possible values of the fusion
coefficients.

\section{\bf Wess-Zumino-Witten model}
As an important example of CFT we shall discuss in this section  the 
Wess-Zumino-Witten (WZW) with a compact and simple Lie group $G$. For our 
purpose, it turns out that the WZW 
model is paradigmatic, it has been conjectured that all rational conformal
field theories can be obtained from it by the coset construction.

The starting point is a bosonic $\sigma-$model with the group manifold of $G$ 
as the target space; the action is 
\beeq
S_\sigma[g] \; = \; \frac{1}{4 \lambda^2} \int d^2x {\rm Tr} 
\left(\partial^\mu g^{-1} \partial_\mu g \right) \quad .
\label{smod}
\end{equation}
The field $g$ takes values in $G$; the trace is defined as in Chap.1. The 
action (\ref{smod}) is invariant under a local $G \times G$ transformation
\beeq
S_\sigma \left[h \, g(x) \, {h^\prime}^{-1} \right] \; = \; S_\sigma[g] \; , 
\qquad \qquad h, \, h^\prime \in G \quad .
\end{equation} 
Unfortunately, the classical conformal invariance does not survive at the quantum
level; the $\beta-$function is different from zero, and the model is 
asymptotically free. To solve this problem, Witten proposed to add a Wess-Zumino 
term to  $S_\sigma$, the new action of the model is
\beeq
S_\sigma[g] \; - \; i \kappa \Gamma_{M_3}[\tilde{g}] \quad .
\end{equation}
The boundary $\Sigma$ of the 3-manifold $M_3$ is the world sheet of 
the $\sigma-$model; we mainly focus our attention on $\Sigma=S^2$. The field 
$\tilde{g}$, living in  $M_3$, is an extension of $g$ on $M_3$. The extension 
exists because $\pi_2(G)=0$, thus the maps $g:\,  S^2 \ra G$ are all 
homotopically equivalent. Moreover, one can find $g_t:\, [0,1]\times S^2 \ra G$ 
with $g_{t=0} =g$, and $g_{t=1}$ the constant map on $S^2$. The extension is 
defined by 
taking $M_3=B^3$, where $B^3$ is the three ball with $\partial B^3=S^2$. The 
extension is not unique; however by taking $k$ integer, the quantum partition
function is well defined. This can be understood as follows. The difference between two 
extensions can be written as
\beeq
\Delta = \Gamma_{M_3} \; - \; \Gamma_{M_3^\prime} \; = \; \Gamma_M \qquad ,
\end{equation}
where $\di M_3 = \di M_3^\prime = S^2$, and $M= M_3 \cup M_3^\prime$ is a compact 3-manifold 
homeomorphic to $S^3$. From the analysis of Chap. 1 it follows that 
$\Gamma_M = m$, where $m$ is an integer. Thus, the quantum theory will be independent by 
the chosen extension when $\kappa$ is an integer. 

The Wess-Zumino-Witten model is defined by setting 
$\lambda^2 =4 \pi/ \kappa$; this particular value of $\lambda$ corresponds to 
the vanishing of the one loop $\beta-$function. The WZW action can be written as
\beeq
S_{WZW} \; = \; \frac{\kappa}{16 \pi} \int d^2x {\rm Tr}\left(\partial^\mu g^{-1} \partial_\mu 
g \right) \; - \; i \kappa \Gamma[\tilde{g}] \quad .
\end{equation} 
As it is shown in App.A, the variation of the Wess-Zumino term is a boundary 
term; as a result, the variational principle for $S_{WZW}$ is well defined and it 
leads to the following equations of motion 
\bea
&&\di \bar{J}(\bar{z}) \; = \; 0 \quad , \\ \label{mahol} 
&&\bar{J}(\bar{z}) \; = \; \kappa \, g^{-1} \bad g \quad .
\ena
Eq.(\ref{mahol}) also implies the existence of a conserved 
holomorphic current:
\bea
&&\bad J(z) \; = \; 0 \quad , \\ \label{mhol} 
&&J(z) \; = \; - \kappa \, \di g g^{-1} \quad .
\ena  
\subsection{\bf The Kac-Moody algebra}
The holomorphic and anti-holomorphic currents that appear in the equations of 
motion are associated with a new invariance of the WZW model. Namely, the 
global $G \times G$ global invariance of the $\sigma-$model is enhanced to
\beeq
g \ra \Omega(z) \, g \, \bar{\Omega}^{-1}(\bar{z}) \; , \qquad \qquad 
\Omega(z), \, \bar{\Omega}(\bar{z}) \in G \; .
\label{incu}
\end{equation}   
The invariance of $S_{WZW}$ under the transformation (\ref{incu}) can be 
established by using the Polyakov-Wiegman identity
\beeq
S_{WZW} \left[gh^{-1} \right] \; = \; S_{WZW} \left[g \right] \; + \;
S_{WZW} \left[h^{-1} \right] \; = \; \frac{k}{2 \pi} \int d^2x {\rm Tr} 
\left(g^{-1} \bad g h^{-1} \di h \right) \quad ,
\label{pwi}
\end{equation}    
which can be derived from eq.(\ref{wpol}) in App.A.
The invariance (\ref{incu}) gives rise to the following Ward identity
\beeq
\langle \, \frac{1}{2 \pi i} \left[\oint dz \, \omega^a(z) J_a(z) \; + \; 
\oint d \bar{z} \, \bar{\omega}^a(\bar{z}) \bar{J}_a(\bar{z})  \right] 
\Theta \, \rangle \; = \; - \delta_{\omega, \bar{\omega}} \langle \,
\Theta \, \rangle \quad ,
\label{wcurr}
\end{equation} 
with 
\bea 
&& J(z) \; = \; t_a \, J^a(z), \qquad \bar{J}(\bar{z}) \; = \; t_a \, 
\bar{J}^a(\bar{z})  \quad , \nb \\
&& \Omega(z) \; = \; \boldmath{1} \; + \; \omega^a(z) t_a \; + \cdots \quad , 
\nb \\
&& \bar{\Omega}(\bar{z}) \; = \; \boldmath{1} \; + \; \bar{\omega}^a(\bar{z}) 
t_a \; + \cdots \quad . \label{boh1}
\ena
In eq.(\ref{boh1}) we have introduced the generators $\{t_a \}$ of $G$; we do not use capital
letters in order to avoid confusion with the energy-momentum tensor.
Setting in (\ref{wcurr}) $\Theta=J(w)$ or $\Theta=\bar{J}(\bar{z})$, one 
easily  gets the following OPE
\bea
&&J_a(z) \, J^b(w) \sim \frac{\kappa \delta_{ab}}{(z-w)^2} \; + \; 
i f_{ab}^c \frac{J_c(w)}{(z-w)} \quad , \nb \\
&&\bar{J}_a(\bar{z}) \, \bar{J}^b(\bar{w}) \sim \frac{\kappa \delta_{ab}}
{(\bar{z}-\bar{w})^2} \; + \; i f_{ab}^c \frac{\bar{J}_c(\bar{w})}
{(\bar{z}-\bar{w})} \quad , \nb \\
&&J_a(z) \, \bar{J}^b(\bar{w}) \sim 0 \quad .
\label{kmo}
\ena
In the operator formalism, the OPE (\ref{kmo}) translates into the Kac-Moody
algebra for the current modes
\bea
&&\left[ J_n^a, \, J_m^b \right] \; = \; i f_{abc} \, J^c_{n+m} \; + \; 
\kappa \, n \delta_{ab} \, \delta_{0, n+m} \quad , \nb \\
&&\left[ \bar{J}_n^a, \, \bar{J}_m^b \right] \; = \; i f_{abc} \, \bar{J}^c_{n+m} \; + \; \kappa \, n \delta_{ab} \, \delta_{0, n+m} \quad , \nb \\
&&\left[ J_n^a, \,\bar{J}_m^b \right] \; = \;  0 \quad ,
\label{kma}
\ena
with 
\beeq
J^a(z) \; = \; \sum_{n \in \mathbb{Z}} z^{-1-n} \, J_n^a, \qquad \qquad
\bar{J}^a(\bar{z}) \; = \; \sum_{n \in \mathbb{Z}} \bar{z}^{-1-n} \, 
\bar{J}_n^a \quad .
\end{equation}
The currents are associated with a pair of the affine  
Lie algebras $\hat{g}_\kappa \times \bar{\hat{g}}_\kappa$ at level $\kappa$; $\hat{g}_\kappa$ is
also known as Kac-Moody algebra in the physics literature.

\subsection{The Sugawara construction}
Beside the current algebra, the WZW model is also conformal invariant; to 
study this aspect one has to define the energy-momentum tensor. From the classical 
definition one finds that $T(z)$ is proportional to $J^a(z)J^a(z)$. At the quantum level this 
expression is ill defined; the usual normal ordering procedure fails because the WZW model 
is not a free theory;  moreover the normalization of $T(z)$ is fixed by the OPE, a consistent 
definition of $T(z)$ must be in agreement eq.(\ref{opet}). A suitable definition of $T(z)$ (a parallel
analysis can be given for $\bar{T}$) is the following 
\beeq
T(z) \; = \; \gamma \left(J^a J^a\right)_{reg.}(z) \qquad .
\label{sudef}
\end{equation}
where $( \; )_{reg.}$ means that in the OPE only the regular terms must be 
retained, the constant $\gamma$ is determined by imposing eq.(\ref{opet}) for 
the energy momentum tensor. In terms of modes, the prescription (\ref{sudef}) 
is equivalent to push all the $J_n$ with $n < 0$ to the left of the operators 
$J_m$ with $m>0$, this clearly resembles the normal ordering for non 
interacting field theories. The value of $\gamma$ is given by 
\beeq
\gamma \; = \; \frac{1}{2(\kappa \; + \; c_v)} \quad ,
\end{equation}
where  $c_v$ is the value of the quadratic Casimir in the adjoint representation, 
defined as $f_{acd}f_{bcd}  = c_v \, \delta_{ab}$. 
The form of the energy-momentum tensor, quadratic in the Kac-Moody currents is 
known as the Sugawara form. The central charge is given by 
\beeq
c \; = \; \frac{\kappa \, D}{\kappa \; + \; c_v} \qquad ,
\label{wzwc}
\end{equation}
where $D$ is the dimension of the Lie algebra.
The explicit form of the Virasoro generators is the following 
\beeq
L_n \; = \; \frac{1}{2(\kappa + c_v)} \left[ \sum_{m \leq -1} J_m^a \, 
J_{n-m}^a \; + \; \sum_{m \geq 0} J^a_{n-m} \, J^a_m \right] \quad .
\label{mvk}
\end{equation}

\subsection{\bf Primary fields and integrable representations}   
Conformal primary fields are the fundamental objects in CFT; as we have seen,
they encode all the wanted information. In the WZW model the situation is slightly 
different; the fundamental symmetry is contained in the Kac-Moody algebra; since the 
energy-momentum tensor is expressed in terms of the Kac-Moody currents, it is
natural to guess that the notion of primary field is related to the 
transformation property under the $G(z) \times G(\bar{z})$ symmetry. 
A WZW primary field $\phi_{\lambda, \mu}(z, \bar{z})$ in the WZW is defined 
to transform in a covariant way
under $G(z) \times G(\bar{z})$, namely
\beeq
\phi_{\lambda, \mu}(z, \bar{z}) \ra \Omega_{\lambda} \, \phi_{\lambda, \mu}
(z, \bar{z}) \, \bar{\Omega}_\mu(\bar{z}) \qquad ,
\label{wzwprim}
\end{equation}
$\lambda$ is the highest weight of the representation  $\Omega_\lambda(z)$, and
$\mu$ is the highest weight of the representation  $\bar{\Omega}_\mu(z)$. 
Because of the Ward identity (\ref{wcurr}), Eq.(\ref{wzwprim}) is equivalent to 
the following OPE 
\bea
&&J^a(z) \, \phi_{\lambda, \mu}(w, \bar{w}) \sim \frac{-t^a_\lambda \;   
\phi_{\lambda, \mu}(w, \bar{w})}{(z \, - \, w)} \qquad , \\
&&\bar{J}^a(\bar{z}) \, \phi_{\lambda, \mu}(w, \bar{w}) \sim \frac{   
\phi_{\lambda, \mu}(w, \bar{w}) \; t^a_\mu}{(\bar{z}\, - \, \bar{w})} \qquad .
\label{wpri}
\ena
In the previous equation matrix multiplication is understood. In particular 
the field $g$ is itself WZW primary. We shall see that restricting oneself to 
finite
dimensional representation of $G$ in the definition of WZW primaries is not a
limitation.
As usual, to each primary field one associates an ``in'' state
\beeq
| \phi_\lambda \, \rangle \; = \; \phi_\lambda(0) \, | 0 \, \rangle \qquad .
\end{equation}
From the OPE (\ref{wpri}) it is easy to derive the following commutation relations
\beeq
\left[J_n^a, \, \phi_\lambda(w) \right] \; = \; - \, t^a_\lambda \,  
\phi_\lambda(w) \, w^n \qquad n \geq 0 \qquad .
\label{pcomm}
\end{equation}
Thus, the definition of a primary field in eq.(\ref{wpri}) can be restated  as
\bea
&&J_0^a \, | \phi_\lambda \, > \; = \; -  t^a_\lambda \qquad , \\
&&J_n^a \, | \phi_\lambda \, > \; = \; 0 \qquad n >0 \quad .
\ena
By using eq.(\ref{pcomm}) and eq.(\ref{mvk}) it is straightforward to show that
a WZW primary is also a Virasoro primary; indeed
\beeq
L_0 \, | \phi_\lambda \, > \; = \; h_\lambda \; | \phi_\lambda \, >, 
\qquad  L_n \, | \phi_\lambda \, > \; = \; 0 \quad n>0 \quad ,
\end{equation}
with 
\beeq
h_\lambda \; = \; \frac{c_\lambda}{2(\kappa \; + \; c_v)} \quad ,
\label{confwzw}
\end{equation} 
where $c_\lambda$ is the value of the quadratic Casimir in the representation 
with highest weight $\lambda$. It should be noted that a Virasoro primary 
is not in general also a WZW primary.

As in the pure Virasoro case, the primary field $\phi_{\lambda, \mu}$ gives rise 
to a highest weight representation for the affine algebra $\hat{g}_\kappa 
\times \bar{\hat{g}}_\kappa$, the descendent states (or field) are obtained from
$\phi_{\lambda, \mu}$  by applying $J_{-n}^a, \; \bar{J}_{-n}^a $ and $L_{-m},
\; \bar{L}_{-m}$. Given any local field, it can be decomposed in terms of descendents 
of some primary field, corresponding to the highest weight state of a representation of 
the affine algebra. By means of Ward identities, by using the same strategy as in 
eq.(\ref{coneq}), all correlation functions can be inferred from the 
knowledge of primary field correlation functions. 
Beside the invariance under global conformal transformations,
that leads to equations of the type  (\ref{2point}) and (\ref{3point}), the WZW 
model is also invariant under global $G \times G$ transformations, this constraint fixes the 
two and three point function in terms of conformal weights and the Clebsch-Gordan 
coefficients of $G$.

A powerful tool in deriving the operator algebra of the theory is the existence 
of null vectors in the highest weight representation. This is not new; the Kac
analysis of Verma modules has revealed the existence of such vectors for a family
of representations with $c<1$ corresponding to minimal models whose
fusion rules can be worked out. In our case, we are interested in the Kac-Moody 
null vectors. The first example is the following null field
\beeq
\chi(z) \; = \; \left( L_{-1} \; + \; \frac{1}{\kappa \, + \, c_v} \, J^a_{-1}
t^a_i \right) \phi_i \quad ,
\end{equation}
where $\phi_i$ is a primary field.
By using the following representation for the insertion of $J_{-1}^a$: 
\beeq
\langle \, \phi_1(z_1) \, \cdots \, J_{-1}^a \phi_i(z_i) \, \cdots \phi_n(z_n)
\, \rangle \; = \; \sum_{j \neq i} \frac{t^a_j}{(z_i \, - \, z_j)} 
\; \langle \, \phi_1(z_1) \, \cdots \, \phi_n(z_n) \, \rangle \qquad ,
\end{equation}
one derives
\bea 
\langle \, \phi_1(z_1) \, \cdots \, \chi(z_i) \, \cdots \phi_n(z_n)
\, \rangle \; &&= \;  0 \nb \\
&&= \; \left[ \di_{z_i} \; + \; \frac{1}{\kappa \, + \, c_v} 
\sum_{j \neq i} \frac{t^a_i \otimes t^a_j}{(z_i \, - \, z_j)} \right]
\langle \, \phi_1(z_1) \, \cdots \, \phi_n(z_n) \, \rangle \quad .
\ena
As a result the Knizhnik-Zamolodchikov equation follows
\beeq
\left[ \di_{z_i} \; + \; \frac{1}{\kappa \, + \, c_v} 
\sum_{j \neq i} \frac{t^a_i \otimes t^a_j}{(z_i \, - \, z_j)} \right]
\langle \, \phi_1(z_1) \, \cdots \, \phi_n(z_n) \, \rangle \; = \; 0 \quad .
\label{kze}
\end{equation}
Eq.(\ref{kze}) is a partial differential equation for primary field correlation 
functions; although a general solution is not known, for $n=4$, by using 
global conformal invariance, it can be reduced to an ordinary differential 
equation for the four-point function which can be solved in terms of 
hyper-geometric functions. As we have pointed out in Chap.8, 
the Knizhnik-Zamolodchikov equation encodes the monodromy 
properties of the correlators related to 
braid group representation.       

Let us now look more closely to affine algebra representations. The crucial 
property is the existence of a null vector in each integrable representation; 
this fact gives rise to rather stringent selection rules.
Let $\theta$ be the highest root of the Lie algebra of $G$, and
\beeq
\left[ t^\theta, \, t^{- \theta} \right] \; = \; \frac{2}{(\theta, \, \theta)} 
\sum_i \theta_i \, h_i \; \equiv \; h_\theta \quad ,
\end{equation}
where $\{h_i \}$ is a Cartan subalgebra of the Lie algebra of $G$.
The elements of the affine algebra: $P_-=J_{-1}^\theta$, 
$P_+=J_{+1}^{-\theta}$ and $P_3=m - J_{+0}^{h_\theta}$ with $m=2 \kappa 
(\theta, \theta)$, generate an $su(2)$ subalgebra; indeed
\beeq
\left[P_+, \, P_- \right] \; = \;  P_3, \quad  \left[P_3, \, P_+ \right] \; = \;  
2 P_+, \quad \left[P_3, \, P_- \right] \; = \;  -2 P_- \; . 
\end{equation}
Thus, any representation of the affine algebra can be decomposed into a sum of 
representations of $su(2)$. Let $|\lambda >$ be the state of the representation
of $\hat{g}_k$ corresponding to the highest weight representation of the 
primary field $\phi_\lambda$, and ${\cal R}_\lambda[su(2)]$ the representation
of $su(2)$ containing  $|\lambda >$. Because $\phi_\lambda$ is a primary
field, the quantity $M/2$ defined as 
\beeq
P_3 \, |\lambda > \; = \; \left(m \; - \; 2 \frac{(\lambda, \, \theta)}{(\theta, \, \theta)} \right) \, |\lambda > \; \equiv \; M \, |\lambda > \qquad ,
\end{equation}
gives the ``angular momentum'' content of ${\cal R}_\lambda[su(2)]$. The representation of
$\hat{g}_k$, associated with $\phi_\lambda$, is called integrable when 
${\cal R}_\lambda[su(2)]$ is finite dimensional. A necessary and sufficient 
condition is that $M$ be an integer. It can be shown that 
${\cal R}_\lambda[su(2)]$ is
irreducible; thus
\beeq
(P_-)^{M+1} \, |\lambda > \; = \; 0 \qquad .
\label{wnull}
\end{equation}
For fixed $\kappa$ there is a finite number of integrable representations. 
For each integrable representation there exists a null vector 
$(P_-)^{M+1} \, |\lambda >$. It is a remarkable fact that one needs to consider 
integrable representation only; indeed, consider $\phi_{\lambda_{ni}}$ 
associated with a non integrable representation of $\hat{g}_k$, it turns out that 
\beeq
\langle \, \phi_1(z_1) \cdots \phi_{\lambda_{ni}}(z_r) \cdots \phi_n(z_n) 
\rangle \; = \; 0 \qquad .
\end{equation} 
As a consequence, only integrable representations are relevant from a physical 
point of view. The existence of the null vector (\ref{wnull}) also strongly
constrains the form of the three point function; one can show that
\beeq
\left(t^\theta_1 \right)^{l_1} \left(t^\theta_2 \right)^{l_2}
\langle \, \phi_\lambda(z) \, \phi_1(z_1) \, \phi_2(z_2) \rangle 
\; = \; 0  \qquad \mbox{for any } l_1 \, + \, l_2 \geq M+1 \qquad .
\label{fuco}
\end{equation}
For example, when $G=SU(2)$, by using Eq.(\ref{fuco}), the fusion rules 
can be determined completely.

\section{\bf Modular invariance}  
Up to now we have considered CFT on $\mathbb{R}^2$, or more precisely 
on the Riemann sphere. Modular invariance is a tool to study CFT on two 
dimensional manifold with
non trivial topology. Non trivial topologies come in naturally in many 
applications of CFT; two important examples are: statistical models at criticality with  
periodic boundary conditions in both directions and  string theory. 

For concreteness let us study a classical dilatation invariant scalar field 
theory
on a compact, orientable  two dimensional manifold $\Sigma_g$ 
(a Riemmanian surface of genus $g$). 
The action is     
\beeq
S[h, \, \varphi] \; = \; \int_{\Sigma_g} d^2 \, \sqrt{h} \; h^{\mu \nu} (\di_\mu \varphi)(\di_\nu \varphi) \qquad ,
\label{gback}
\end{equation}
where $h_{\mu \nu}$ is the background metric. The topology of $\Sigma_g$ is specified by its genus $g$, i.e the number of handles. In two dimensions, locally, the metric
can be put in the form $h_{\mu \nu} \; = \; \eta_{\mu \nu} \, \exp{2f(x)}$, i.e.
any metric on $\Sigma_g$ is locally conformally equivalent to the standard Euclidean metric.
This fact implies that, locally, the action (\ref{gback}) coincides with 
$S[\eta, \, \varphi]$. Thus, when one is concerned with local aspects, the analysis goes along the same lines as in Sect.A.2; in particular the holomorphic and 
anti-holomorphic sectors of the theory decouple. It should be noted that  
at the quantum level gravitational anomalies play an important role. The action 
(\ref{gback}) depends on the gravitational background only through the tensor 
density $\hat{h}_{\mu \nu } = \sqrt{h} h_{\mu \nu}$; at the quantum level, if one respects 
diffeomorphism invariance, Weyl invariance is anomalous, and
the Weyl anomaly is proportional to $c \, R$, where $R$ is the trace of the Ricci tensor and 
$c$ is the familiar central charge. Alternatively, one may keep the variable 
$\hat{h}_{\mu \nu }$ and shift the anomaly into a diffeomorphism anomaly \cite{gua2d}.
When $\Sigma$ is $S^2$, the conformal gauge $h_{\mu \nu} \; = \; \eta_{\mu \nu} \, 
\exp{2f(x)}$ exists globally; on the contrary, when $g \geq 1$, the conformal gauge exists only 
locally. The action (\ref{gback}) depends only on the conformal
structure of $M$, moreover a complex structure on $M$ is in one to one 
correspondence with a  conformal structure on $M$; as a result we expect, up to an anomalous 
factor, a dependence of the partition function on $\Sigma_g$ on a set of parameters 
that describes the space of inequivalent complex structures on $\Sigma_g$, the moduli space. 
For $S^2$ the moduli space is trivial, there is only one complex structure. When
$g=1$ (the torus) the moduli space is described by one complex parameter $\tau$
subject to the following constraints
\bea
&&\mbox{Im}(\tau) > 0 \quad , \nb \\
&&\tau^\prime \; = \; \frac{a \tau \; + \; b}{c \tau \; + \; d} \; , \qquad a,b,c,d,
\in \mathbb{Z} \; \; ad \, - \, bc \, = \, 1 
 \; , \qquad \tau \; \mbox{and } \tau^\prime \; \mbox{identified}. 
\label{moduli} 
\ena
Thus, $\tau$ and $\tau^\prime$ are identified when they are related by an 
element of the modular group $PSL(2, \mathbb{Z})$ which is generated by
\bea
&& t: \; \tau \ra \tau \; + \; 1 \quad , \nb \\
&& s: \; \tau \ra - \frac{1}{\tau} \quad , \nb \\
&&\mbox{with } \; (st)^3 \; = \; s^2 \; = \; 1 \qquad .
\label{psl}
\ena 
The modular group coincides with the mapping class group of the torus encountered
in Chap.5, this is not accidental. The moduli space of the torus is obtained from
the Teichmuller space $Tei(\Sigma_1)$ as $Tei(\Sigma_1)/PSL(2, \mathbb{Z})$. The 
Teichmuller space $Tei(\Sigma_g )$ is defined as the space of metrics of constant 
curvature on $\Sigma_g$, modulo diffeomorphisms connected with the identity. 
On the torus, the modular group corresponds exactly to those diffeomorphisms not connected with the identity.

The partition function on the torus will be a function of $\tau$, the requirement
of conformal invariance on the torus is formulated in terms of modular invariance
of the partition function:
\beeq
Z(\tau) \; = \; Z(\tau \; + \; 1), \qquad Z(\tau) \; = \; Z(-1 / \tau) \quad .
\label{moinv}
\end{equation}
Differently from the $g=0$, the holomorphic and anti-holomorphic parts do interact;
imposing eq.(\ref{moinv}) gives non trivial constraints on the spectrum of the 
theory. We shall consider here the case $g=1$: the torus topology.

In CFT the partition function admits the following representation
\beeq
Z(\tau) \; = \; {\rm Tr} \left(q^{L_0 \, - \, c/24} \; \bar{q}^{\bar{L}_0 \, - \, 
c/24} \right) \quad ,
\label{cpart}
\end{equation}
with $q=\exp(2 \pi i \tau)$ and $\bar{q}=\exp(-2 \pi i \tau^\ast)$ ($q$ is not to
be confused with deformation parameter of CS theory). The trace is taken over 
the full Hilbert space of the theory. The partition function is related  to the
formal character associated with a Verma module
\beeq
\chi_{(c,h)}(\tau) \; = \; {\rm Tr} \left(q^{L_0 \, - \, c/24} \right) \; = \; 
\sum_{n=0}^\infty \mbox{dim}(h \, + \, n) \, q^{n+k -c/24} \quad ,
\label{virc}
\end{equation}
$h$ is the weight of the highest state of the Verma module, and 
$\mbox{dim}(h \, + \, n)$ is the number of linearly independent states 
at level $n$. In the case of a CFT
with a current algebra symmetry, eqs. (\ref{cpart}) and (\ref{virc}) still 
hold, 
Verma modules are replaced by the modules associated with integrable representations
of the affine algebra, and $\chi_{(c,h)}(\tau)$ becomes the formal character of 
the affine algebra. 
Thus, as anticipated,
modular invariance ties together the holomorphic and anti-holomorphic sectors.
The first non trivial example is a free-fermion on a torus. There are two types of 
boundary conditions: periodic (Ramond sector) or anti-periodic (Neveu-Schwarz sector); 
on a torus we have two directions in which boundary conditions have to be imposed, 
thus we have four sectors: (R,R), (R,NS), (NS,R), (NS,NS). 
It turns out that there are only two 
possibilities to satisfy modular invariance: only the (R,R) sector is present ( the
partition function in this case vanishes), alternatively the sectors (R,NS), (NS,R), (NS,NS) 
must be present at the same time.

We shall now address the relation between modular invariance and fusion algebra.
The characters transform covariantly under modular 
transformation 
\bea 
&&\chi_i(-1/\tau) \; = \; \sum_m {\cal S}_{im} \chi_m(\tau)  \quad , \nb \\
&&\chi_i(\tau \, + \, 1) \; = \; \sum_m {\cal T}_{im} \chi_m(\tau) \quad .
\label{cmod}
\ena
Modular invariance permits to prove the so called Verlinde formula
\beeq
{\cal N}_{ij}^s \; = \; \sum_m \frac{{\cal S}_{im} \, {\cal S}_{jm} 
{\cal S}^\ast_{mk}}{{\cal S}_{0m}} \quad .
\label{veformu}
\end{equation}
The matrices ${\cal S}$ and ${\cal T}$ satisfies
\beeq
{\cal S}^2 \; = \; \left({\cal S} {\cal T} \right)^3 \; = \; {\cal C}
\quad ,
\end{equation}
with ${\cal C}_{ij} = \delta_{i j^\ast}$. Thus, in general,  the matrices ${\cal S}
, \, {\cal T}$ acting on the space of characters give rise to a representation of 
the double covering of the modular group. 
Notice that Eq.(\ref{veformu}) can be interpreted as an eigenvalue equation for 
the action of $N_i$ in the regular representation of the fusion algebra; in this 
sense the matrix ${\cal S}$ diagonalizes the fusion rules.
Let us consider what happens to conformal blocks when the topology is 
non-trivial.
Recall that a conformal block essentially describes the intermediate states entering 
the decomposition of the four-point function (see eq.(\ref{inde})), this
concept can be generalized to an arbitrary n-point correlator 
as describing the intermediate states entering $< \phi_1 \cdots \phi_n>$.
The generalized conformal block ${\cal F }^{I}$ ($I$ is a collective index) 
will be a holomorphic function of  $n-3$ variables describing the
insertions of the primary fields of the n-point correlator. Let us introduce the 
vector space $V_n$ of linearly independent n-point conformal blocks.  
What happens to conformal blocks  ${\cal F }^{I}$ on a Riemannian surface ? 
Locally there is  no difference; in a small neighbourhood  ${\cal F}_{lm}^{ij}(X)$
is a holomorphic function; however globally it becomes a section of a bundle. 
The vector space $V_n$ will in general  depend also on the genus of $\Sigma_g$ in 
which the CFT is defined; for this purpose we introduce the new notation $V(\Sigma_{g,n})$.
The vector space $V(\Sigma_{g,n})$ can be considered as the fiber of a holomorphic
vector bundle over the moduli space $M(\Sigma_{g,n})$ of a Riemann surface with 
$n$ punctures corresponding to the location of the primary fields entering the 
correlator. In particular the fusion coefficient represents the dimension of
$V(\Sigma_{g,n})$ when $n=3$. Thus, the Verlinde formula can be also interpreted 
as the computation of $\dim(V(\Sigma_{g,3}))$ in terms of the action of the modular 
group on the characters. 

\section{\bf Witten's approach to Chern-Simons theory}
After this little tour in the realm of CFT, we come back to CS theory, 
namely to the Witten's original solution given in his 89' paper \cite{witt1}.
The essential tool used by Witten is the relation between two dimensional 
conformal field theory and three dimensional CS theory; about this topic see also
\cite{fro}.

For reasons that will be clear later, the constant $k$ in the CS action 
will be denoted by $\ell$ in this section. Let us consider the CS theory in 
$S^3$; as shown in Chap.1, $\ell$ must be an integer. Following Witten, we 
shall use the following notations 
\beeq
Z({\cal M}, \, C_1, \, \cdots \, C_n) \; = \; Z({\cal M}, \, L) \; = \; \langle \, 0| 
W(C_1, \, \cdots \, C_n; \, \rho_1, \, \cdots \, \rho_n)| 0 \, \rangle \bigr 
|_{\cal M} \; , 
\end{equation}
and 
\beeq
\langle \,  W(C_1, \, \cdots \, C_n; \, \rho_1, \, \cdots \, \rho_n) \, \rangle
\bigr |_{\cal M} \; = \; \frac{Z({\cal M}, \, C_1, \, \cdots \, C_n)}{Z({\cal M})} 
\end{equation}
for the expectation value of $n$ Wilson lines associated with the  
non-intersecting knots $\{C_1, \cdots, \,. C_n\}= L$ in $\cal M$ with colours $\rho_1, 
\cdots , \rho_n$.  All the knots
are framed. We would like to answer the following questions:
\begin{itemize}
\item How to compute $Z(S^3, \, L)$ ?
\item How to compute $Z({\cal M}, \, L)$ ?
\end{itemize}
The crucial observation is that the Hilbert space of the quantum theory is
finite dimensional and is isomorphic to $V(\Sigma_{g,n})$ introduced in the 
previous section. The way to establish this connection is the canonical quantization of CS theory. One can cut the three manifold $M$ along a two dimensional 
Riemann surface $\Sigma_g$, the knots $\{C_i \}$ pierce $\Sigma$ in $n$ marked
points carrying non-Abelian quantum numbers corresponding to the 
representations 
$\{\rho_i \}$ of $G$, thus actually the surface has to be regarded as 
$\Sigma_{g,n}$.
In this picture, $\cal M$ looks like $\Sigma_{g,n} \times 
\mathbb{R}$,
where $\mathbb{R}$ will be interpreted as the ``time'' in the canonical framework.
The 2+1 splitting of the action is
\beeq
S_{CS} \; = \;  \frac{\ell}{8 \pi} \int dt \, \int_\Sigma d^2x \, \epsilon^{ij}
 {\rm Tr} \left( A_i \, \di_t A_j \; + \; A_0 \, F_{ij} \right) \quad .
\end{equation}  
The action is quadratic, apart from the non-dynamical field $A_0$ 
imposing the constraint
\beeq
F_{ij}^a \; = \; 0 \quad .
\end{equation}
For the moment we have ignored the Wilson lines piercing $\Sigma_g$ which act 
as sources.
The physical phase space (reduced) $Mod(\Sigma)$ is given by the flat connection on $\Sigma$,
modulo gauge transformations; this space was studied by Atiyah and Bott and they 
found \cite{atb} that it is compact and finite dimensional.  
The quantization of the compact reduced phase space $Mod(\Sigma)$ is a difficult 
problem; a solution is available when  $Mod(\Sigma)$ has a K\"ahler structure. 
One can get a K\"ahler structure on $Mod(\Sigma)$ by picking an arbitrary complex 
structure on $\Sigma_g$. The quantization of the classical
phase space, with the Poisson brackets promoted to commutators, produces a Hilbert
space ${\cal H}_\Sigma$ which depends on the complex structure picked on $\Sigma_g$;
one can show that actually ${\cal H}_\Sigma$ can be interpreted as a holomorphic
flat vector bundle on the moduli space of $\Sigma_g$. Witten's analysis strongly
suggested that actually the Hilbert space ${\cal H}_\Sigma$ must be identified 
with $V(\Sigma_g)$. From a CFT point of view, $V(\Sigma_g)$ is the space of 
conformal
blocks associated with the correlators of the identity conformal family, and
$\mbox{Dim}(V(\Sigma_g)= 1$. The previous picture can be extended to the case when 
Wilson lines piercing $\Sigma_g$ are present, provided that the corresponding representations $\rho_i$ of $G$ give rise to highest weight integrable representations 
of the underlying affine algebra $\hat{g}_\ell$; as consequence, there are restrictions on the 
quantum numbers for the  sources in $\Sigma$. From a three dimensional point of 
view, this is equivalent to the statement of the vanishing $Z(S^3, \, L)$ when one the $\{ \rho_i\}$ corresponds to a non-integrable representation of $g_\ell$. 
We have recovered the null elements of the tensor algebra entering the reduced 
tensor algebra construction of Chap.4. There is a  point to be clarified: the 
relation between $\ell$ and and the renormalized CS coupling constant $k$,
the answer to this question will be clear at the end of the calculation of CS
observables using Witten's approach.

As a first example, let us consider the distant union $L$ of two links $L_1$ and $L_2$ in ${\cal M}$. One can cut $M$ along a two sphere $S^2$, and imagines 
it as the connected sum of ${\cal M}_1$ and  ${\cal M}_2$ with $\di  {\cal M}_1 = 
\di {\cal M}_2 = S^2$. The two ``pieces'' ${\cal M}_1$ and ${\cal M}_2$ of 
${\cal M}$ are chosen in a such way that no component of $L$ pierces $S^2$.
The Feynman path integral on ${\cal M}$ with the link $L$ can interpreted as 
follows
\beeq
Z({\cal M}, \, L) \; = \; (\eta_1, \; \eta_2) \quad .
\label{fsplit}
\end{equation}
One first integrates over the fields living in ${\cal M}_1$ with fixed boundary 
conditions, say $\phi$, on $\di {\cal M}_2 = S^2$, the result is the path 
integral representation for a state $\eta_2$ of the Hilbert space ${ \cal H}_{S^2}$ of the CS
theory; the state $\eta_2$ will depend $L_1$. In the same way, the path 
integration on ${\cal M}_2$  with the same fixed boundary 
value $\phi$ for fields on  $\di {\cal M}_1 = S^2$ represents a state on the dual
space of ${ \cal H}_{S^2}$. Finally, the path integration over the boundary value 
$\phi$ simply produces the inner product between $\eta_1$ and $\eta_2$ on the
right hand side of eq.(\ref{fsplit}). In particular, $S^3$ can be chopped along 
$S^2$, producing a pair of three balls $B_1$ and $B_2$; thus for the $S^3$ partition 
function one has
\beeq
Z(S^3) \; = \; (\eta, \; \eta^\prime) \quad .
\label{fsplit2}
\end{equation}
From CFT one infers that actually $\mbox{Dim}({ \cal H}_{S^2})=1$, and the inner 
product is trivial; thus, from eqs. (\ref{fsplit}) and (\ref{fsplit2}) one has
\beeq
Z(S^3) \; Z({\cal M}, \, L) \; = \; (\eta, \; \eta^\prime) \; (\eta_1, \; \eta_2) \;
= \; (\eta_1, \; \eta) \; (\eta^\prime, \; \eta_2) \quad .
\label{gluing}
\end{equation}
The right hand side of eq.(\ref{gluing}) can be understood as the product of the
partition function on ${\cal M}^\prime_1$ obtained from ${\cal M}_1$ by gluing  
$B_2$ along $S^2$ with the partition function on ${\cal M}^\prime_2$ obtained 
from ${\cal M}_2$ by gluing $B_1$ along $S^2$ (see Fig.D.2). In other words
\beeq
Z(S^3) \; Z({\cal M}, \, L) \; = \; Z({\cal M}^\prime_1, \, L_1) \; 
Z({\cal M}^\prime_2, \, L_2) \quad .
\label{dsglu}
\end{equation}

\begin{figure}[h]
\vskip 0.5 truecm 
\centerline{\epsfig{file=\path 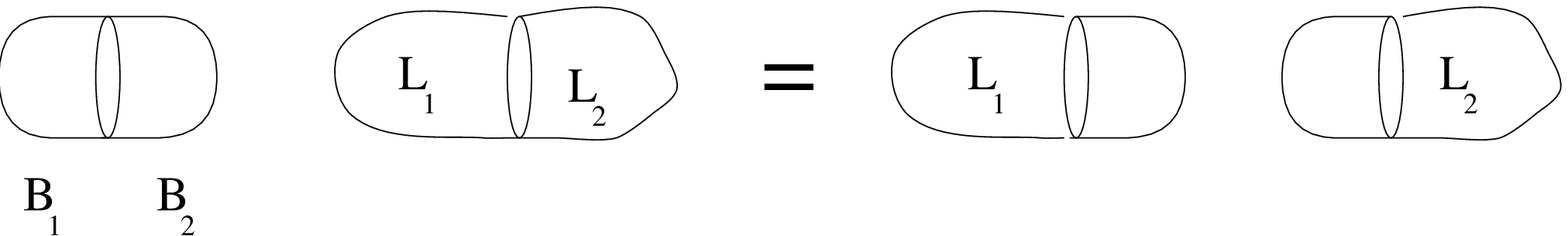,height=4cm,width=13cm}}
\vskip 0.5 truecm 
\centerline{{\bf Figure D.2}}
\vskip 0.5 truecm 
\end{figure} 
When ${\cal M}=S^3$, eq.(\ref{dsglu}) takes the form
\beeq
\langle \, W(L) \, \rangle \bigr|_{S^3} \; = \;  \langle \, W(L_1) \, \rangle 
\bigr|_{S^3} \; \langle \, W(L_2) \, \rangle \bigr|_{S^3} \quad ,
\label{difin}
\end{equation}
eq.(\ref{difin}) describes exactly the factorization property of Chap.2 (see 
eq.(\ref{dist})).

The decisive  step toward the calculability of CS observables in $S^3$ is to 
show that they satisfy a skein relation entering the definition of link 
invariants (see App.B). Let $L_+$ be a any link in $S^3$, and $G=SU(N)$ the gauge 
group, with all the components of $L$ associated with the fundamental $N$-dimensional 
representation $\rho_N$ of $SU(N)$. Consider a three ball
$B \subset S^3$ chosen in a such a way that it contains an  elementary 
over-crossing configuration between two strings of $L_+$; by cutting $S^3$ along $\di B$,
$Z(S^3,L_+)$ can be represented as
\beeq
Z(S^3,\, L_+) \; = \; (\chi, \, \eta_+) \quad.
\end{equation}    
The states $\chi$ and $\eta_+$ belong to the Hilbert space ${\cal H}_\Sigma$, with
$\Sigma$ a two sphere with four punctures associated with $\rho_N, \rho_N, 
\rho_N^\ast,\rho_N^\ast$. Following the same method, one can consider 
the states $\eta_0$ and 
$\eta_-$ corresponding to a no-crossing configuration and to an 
under-crossing configuration in $B$ respectively. In other words, we have 
chosen three links $L_+, L_0, L_-$ in $S^3$ that differ only in a region $B$ 
as shown in Fig.D.3.

\begin{figure}[h]
\vskip 0.5 truecm 
\centerline{\epsfig{file=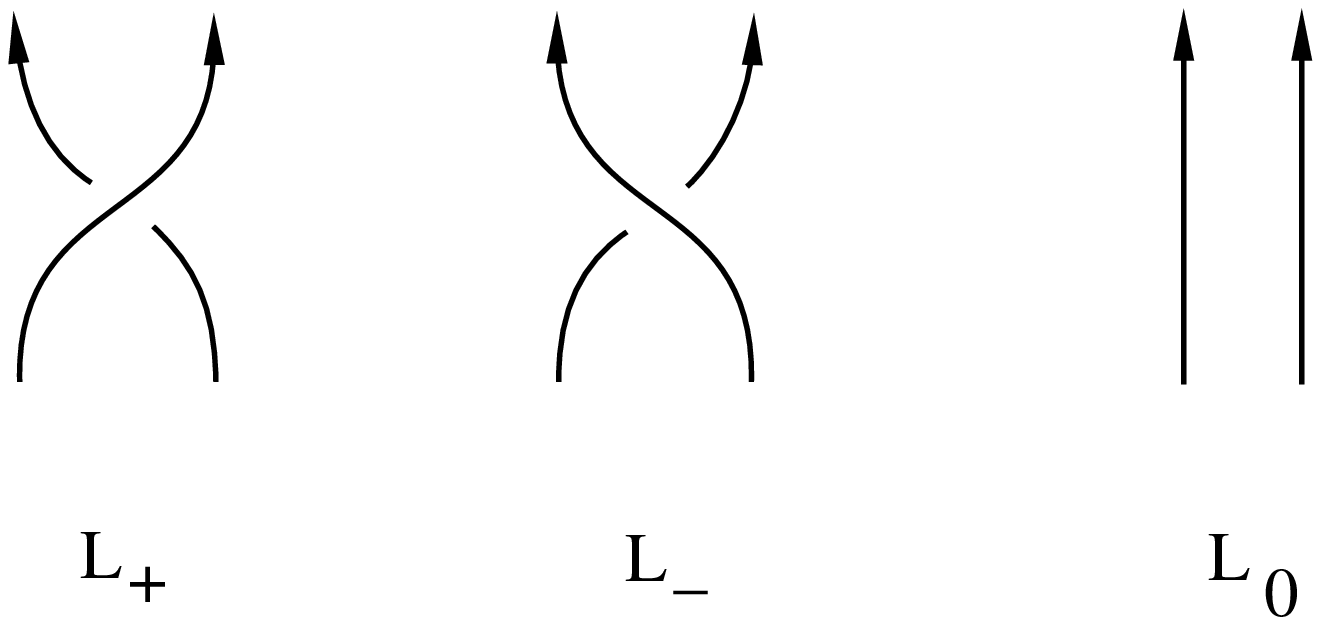,height=5cm,width=6cm}}
\vskip 0.5 truecm 
\centerline{{\bf Figure D.3}}
\vskip 0.5 truecm 
\end{figure} 

The skein relation simply follows from the fact that the states 
$\eta_+, \eta_0$ and $\eta_-$ are linear dependent. Indeed, in this case, 
$\mbox{Dim}({\cal H}_\Sigma)=2$; the four-point correlation function 
$< \phi_{\rho_N} \phi_{\rho_N} \phi_{\rho^\ast_N} \phi_{\rho^\ast_N} >$ of the WZW model must be invariant under 
the global group of transformations $SU(N)\times SU(N)$, and the trivial 
representation appears exactly twice in the tensor product $\rho_N \rho_N 
\rho^\ast_N \rho^\ast_N$. As a result, there exist complex numbers $\alpha, \beta$
and $\gamma$ such that
\beeq
\alpha \, \eta_+ \; +  \; \beta \, \eta_0 \; + \; \gamma \, \eta_- \; =  \; 0 \quad . \label{lidi}
\end{equation} 
From eq.(\ref{lidi}) it follows that 
\beeq
\langle \, \alpha \, W(L_+) \, \rangle \bigr |_{S^3} \; + \; \beta  \langle \, 
W(L_0) \, \rangle \bigr |_{S^3}\; + \;  \gamma \langle \, W(L_-) \, \rangle 
\bigr |_{S^3}\; = \; 0 \; , 
\label{skewit}
\end{equation}
which is the promised skein relation. The values of $\alpha, \beta$ and $\gamma$ are also
provided by the CFT; they can be determined by the monodromy properties of the 
WZW conformal blocks analyzed in \cite{mos} by Moore and Seiberg. The 
results is
\bea
&&\alpha \; = \; - w^{N/2} \quad , \nb \\
&&\beta \; = \; \left(w^{1/2} \; - \; w^{-1/2} \right) \quad , \nb \\
&& \gamma  \; = \; - w^{-N/2} \quad , 
\label{vaskein}
\ena
with $w=\exp(2 \pi i /(N + \ell))$.
It is understood that the links entering  the skein relation have preferred framing.  
By using the skein relation, it is easy to find the value of the Wilson line
associated with unknot $C$
\beeq
\langle \, W(C, \, \rho_N) \, \rangle \; = \; - \frac{\alpha \; + \; \beta}{\gamma}
\; = \; \frac{w^{N/2} \; - \; w^{-N/2}}{w^{1/2} \; - \; w^{-1/2}} \quad .
\label{unkwitt}
\end{equation}
The last ingredient is the behaviour of the observables under a change a 
framing.
From the CFT point of view, an elementary change of framing can be interpreted 
as a the action of an element of the modular group corresponding to a Dehn 
twist whose action on the space of conformal blocks is known; as result one 
has
\beeq
\langle \, W(C_{(1)}, \, \rho) \, \rangle \; = \; e^{2 \pi i h_{\rho}} \;
\langle \, W(C, \, \rho) \, \rangle \quad .
\label{framwitt}
\end{equation}  
The knot $C$ has preferred framing, and $h_\rho$ is the conformal weight of the
WZW primary field $\phi_\rho$, namely, from eq.(\ref{confwzw}) one has
\beeq
h_\rho \; = \; \frac{c_\rho}{2(\ell \; + \; c_v)} \quad .
\end{equation}
Finally, let us briefly address the problem of the calculation of 
$Z({\cal M})$. Consider a surgery presentation of a three manifolds $\cal M$ 
consisting of a knot $C$ in $S^3$. One cuts from $S^3$ a solid torus $N$ 
corresponding to the tubular neighbourhood of $C$, then the boundaries $N$ and
$(S^3 - \dot{N})$ are glued back according to a given homeomorphism. The 
boundaries: $\di N$ and $\di(S^3 - \dot{N})$ are tori, thus the gluing 
homeomorphism is determined by an element $K$ of the mapping class group of the
torus. From the CFT point of view, the action of the mapping class group
on the space of conformal blocks $V(\Sigma_{1,0})$ corresponds to a modular 
transformation. In particular, for a fixed level $\ell$, on can pick a basis
$\{v_i \}$ on  $V(\Sigma_{1,0})$. As studied by Verlinde in \cite{ver}, the 
action of $K$ is a linear transformation ${K_i}^j$ 
\beeq
K \, v_i \; = \; {K_i}^j \, v_j \quad .
\end{equation}     
Because in the starting manifold $S^3$ there is no Wilson line, the 
partition function on  $\cal M$ is represented by
\beeq
Z({\cal M}) \; = \; {K_0}^j  \; W(S^3, \, C \, \rho_i) \quad .
\end{equation}
In principle, more involved surgeries can be examined, although explicit 
calculations would be difficult. 
A comment must be added. From Chap.7 we have already known that $Z({\cal M})$ is not
a topological invariant; a suitable phase must be provided. Indeed, the element 
of the mapping class group is not uniquely determined; given an element $u$, $u$ 
and $u \cdot T^m$, with $m$ integer, give the same surgery, in the sense 
that the resulting manifolds are homeomorphic. This fact produces a phase
ambiguity $\exp(- 2 m \pi i c/ 24)$ in the partition function. Witten interpreted
the present ambiguity as the manifestation of the three manifold framing, 
materializing the central charge at the three dimensional level. 
For further discussion see \cite{wittc,atia}.  

Looking back at Chap.3 a problem in Witten's solution of CS
field theory appears: the expression of the unknot 
in eq.(\ref{unkwitt}) differs from eq.(\ref{eq74}) in Chap.3. The point is the
relation between the level $\kappa$ of the affine algebra and the Chern-Simons
coupling constant $\ell$; Witten sets $\ell = \kappa$. However, when $\ell$ is 
identified with the renormalized coupling constant, as Witten implicitly does,
the position $\ell = \kappa$ is inconsistent. From  
the explicit examples produced in this thesis, all in agreement with 
perturbative calculations, the correct identification is $\ell = \kappa +
c_v$; indeed for $SU(N)$ $c_v =N$, and  this is the only way to match 
eq.(\ref{eq74}) and eq.(\ref{unkwitt}). There is also a simple symmetry argument which shows that eq.(\ref{eq74}) cannot be trusted. Consider a change of the 
orientation of $S^3$, this corresponds in the  $CS$ action to send $\ell$ into
$- \ell$. The unknot and its mirror image are ambient isotopic; as a consequence,
the value of the unknot must be invariant under $\ell \ra - \ell$.
From direct inspection of  eq.(\ref{unkwitt}), it follows that the only 
possibility is $\ell = \kappa + c_v$.

\chapter{\bf Numerical results}
\centerline{
\begin{tabular}{|c|c|c|}
\hline
$\bullet$ & $L_{8/1}$ & $\bullet$ \\ \hline
$k$ & $I_k$ & $|I_k|$ \\ \hline
3 & 1.000000000 -  i 0.000000175 & 1.000000000 \\ \hline
4 & -1.000000000 +  i 0.000000012 & 1.000000000 \\ \hline
5 & -0.499999839 +  i 1.538841821 & 1.618033989 \\ \hline
6 & -2.000000000 +  i 0.000000175 & 2.000000000 \\ \hline
7 & -1.000000000 -  i 0.000000095 & 1.000000000 \\ \hline
8 & -6.828427084 +  i 6.828427165 & 9.656854249 \\ \hline
9 & -0.499999950 +  i 0.866025433 & 1.000000000 \\ \hline
10 & -4.236067816 +  i 3.077683759 & 5.236067977 \\ \hline
11 & -2.073846587 -  i 14.423920506 & 14.572244935 \\ \hline
12 & -7.464102180 -  i 12.928202904 & 14.928203230 \\ \hline
13 & -12.373524802 -  i 17.926145664 & 21.781891892 \\ \hline
14 & 18.195669358 +  i 0.000000868 & 18.195669358 \\ \hline
15 & 5.657005398 +  i 4.110054701 & 6.992443043 \\ \hline
16 & 63.431390926 +  i 26.274142180 & 68.657642707 \\ \hline
17 & 6.721172941 +  i 2.603796085 & 7.207906752 \\ \hline
18 & 15.581719525 +  i 26.988328071 & 31.163437478 \\ \hline
19 & -69.185356387 -  i 11.544994773 & 70.142001987 \\ \hline
20 & -66.118464248 +  i 0.000001734 & 66.118464248 \\ \hline
21 & -90.016155013 +  i 0.000000715 & 90.016155013 \\ \hline
22 & 43.367008373 -  i 50.048195295 & 66.223253224 \\ \hline
23 & 22.973052202 -  i 3.157573698 & 23.189036183 \\ \hline
24 & 219.054514271 -  i 58.695485329 & 226.781922164 \\ \hline
25 & 23.000602198 -  i 5.905551129 & 23.746646829 \\ \hline
26 & 84.434763344 +  i 44.314782640 & 95.357376335 \\ \hline
27 & -185.409752744 +  i 67.483626877 & 197.308936212 \\ \hline
28 & -161.491033989 +  i 77.769978391 & 179.241523084 \\ \hline
29 & -214.342425435 +  i 99.165358161 & 236.170369862 \\ \hline
30 & 50.544948881 -  i 155.561366312 & 163.566899299 \\ \hline
31 & 47.834849646 -  i 26.550506056 & 54.709251617 \\ \hline
32 & 443.615385766 -  i 296.414325177 & 533.531688523 \\ \hline
33 & 46.894822945 -  i 30.137471278 & 55.743982582 \\ \hline
34 & 211.662329441 +  i 39.566544008 & 215.328709440 \\ \hline
35 & -344.610722364 +  i 250.374335362 & 425.962272715 \\ \hline
36 & -290.376668930 +  i 243.654963011 & 379.059824907 \\ \hline
37 & -381.345093602 +  i 307.914657384 & 490.138262785 \\ \hline
38 & 27.064369107 -  i 326.618342223 & 327.737732877 \\ \hline
39 & 79.852315000 -  i 70.742976321 & 106.681586554 \\ \hline
40 & 734.287201006 -  i 734.287220264 & 1038.438931957 \\ \hline
41 & 78.053570702 -  i 75.119028091 & 108.329258655 \\ \hline
42 & 408.590935390 -  i 0.000001624 & 408.590935390 \\ \hline
43 & -545.541472393 +  i 565.843251992 & 785.998781122 \\ \hline
44 & -452.077098514 +  i 521.724781996 & 690.340822456 \\ \hline
45 & -590.050989168 +  i 655.318028318 & 881.817377951 \\ \hline
46 & -39.324647251 -  i 574.905929557 & 576.249299975 \\ \hline
47 & 118.947578576 -  i 140.691854636 & 184.235513433 \\ \hline
48 & 1090.316030520 -  i 1420.927547540 & 1791.039960963 \\ \hline
49 & 116.363242798 - i  145.914887153 & 186.632147733 \\ \hline
50 & 687.196328608 -  i 86.813088226 & 692.658145364 \\ \hline
\end{tabular}
}
\vskip 0.5truecm
\centerline{\bf Table 1}

\pagebreak

\null \vskip 1truecm
\centerline{
\begin{tabular}{|c|c|c|}
\hline
$\bullet$ & $L_{8/3}$ & $\bullet$ \\ \hline
$k$ & $I_k$ & $|I_k|$ \\ \hline
3 & 1.000000000 +  i 0.000000000 & 1.000000000 \\ \hline
4 & -1.000000000 +  i 0.000000000 & 1.000000000 \\ \hline
5 & 1.309016994 -  i 0.951056516 & 1.618033989 \\ \hline
6 & -2.000000000 +  i 0.000000000 & 2.000000000 \\ \hline
7 & -0.623489802 +  i 0.781831482 & 1.000000000 \\ \hline
8 & 0.000000000 +  i 0.000000000 & 0.000000000 \\ \hline
9 & -0.500000000 +  i 0.866025404 & 1.000000000 \\ \hline
10 & 1.618033989 +  i 4.979796570 & 5.236067977 \\ \hline
11 & 6.053529319 -  i 13.255380237 & 14.572244935 \\ \hline
12 & -7.464101615 +  i 12.928203230 & 14.928203230 \\ \hline
13 & 7.723965314 -  i 20.366422715 & 21.781891892 \\ \hline
14 & -16.393731622 +  i 7.894805057 & 18.195669358 \\ \hline
15 & -2.160783733 +  i 6.650208521 & 6.992443043 \\ \hline
16 & 0.000000000 +  i 0.000000000 & 0.000000000 \\ \hline
17 & -1.972537314 +  i 6.932749548 & 7.207906752 \\ \hline
18 & 15.581718739 +  i 26.988328525 & 31.163437478 \\ \hline
19 & 17.218843527 -  i 67.995675380 & 70.142001987 \\ \hline
20 & -20.431729095 +  i 62.882396270 & 66.118464248 \\ \hline
21 & 20.030478885 -  i 87.759262069 & 90.016155013 \\ \hline
22 & -55.710545730 +  i 35.802993757 & 66.223253224 \\ \hline
23 & -4.717948848 +  i 22.704016336 & 23.189036183 \\ \hline
24 & 0.000000000 +  i 0.000000000 & 0.000000000 \\ \hline
25 & -4.449677900 +  i 23.326028428 & 23.746646829 \\ \hline
26 & 54.169163837 +  i 78.477582217 & 95.357376335 \\ \hline
27 & 34.262337211 -  i 194.311370120 & 197.308936212 \\ \hline
28 & -39.884991120 +  i 174.747563877 & 179.241523084 \\ \hline
29 & 38.208113963 -  i 233.059184818 & 236.170369862 \\ \hline
30 & -132.328401250 +  i 96.142211171 & 163.566899299 \\ \hline
31 & -8.284500381 +  i 54.078362271 & 54.709251617 \\ \hline
32 & 0.000000000 +  i 0.000000000 & 0.000000000 \\ \hline
33 & -7.933195866 +  i 55.176589215 & 55.743982582 \\ \hline
34 & 129.764538515 +  i 171.836019661 & 215.328709440 \\ \hline
35 & 57.178306982 -  i 422.107212669 & 425.962272715 \\ \hline
36 & -65.823047822 +  i 373.301054424 & 379.059824907 \\ \hline
37 & 62.256293514 -  i 486.168356194 & 490.138262785 \\ \hline
38 & -258.631121471 +  i 201.300681961 & 327.737732877 \\ \hline
39 & -12.859044288 +  i 105.903757675 & 106.681586554 \\ \hline
40 & 0.000000000 +  i 0.000000000 & 0.000000000 \\ \hline
41 & -12.423570453 +  i 107.614511930 & 108.329258655 \\ \hline
42 & 254.752281347 +  i 319.449256739 & 408.590935390 \\ \hline
43 & 85.965636475 -  i 781.283554973 & 785.998781122 \\ \hline
44 & -98.245742501 +  i 683.314148273 & 690.340822456 \\ \hline
45 & 92.175015400 -  i 876.986690089 & 881.817377951 \\ \hline
46 & -447.003088251 +  i 363.663986140 & 576.249299975 \\ \hline
47 & -18.441181139 +  i 183.310248618 & 184.235513433 \\ \hline
48 & 0.000000000 +  i 0.000000000 & 0.000000000 \\ \hline
49 & -17.920983557 +  i 185.769741658 & 186.632147733 \\ \hline
50 & 441.516918550 +  i 533.702273720 & 692.658145364 \\ \hline
\end{tabular}
}
\vskip 0.5truecm
\centerline{\bf Table 2}

\pagebreak
\null \vskip 1truecm
\centerline{
\begin{tabular}{|c|c|c|}
\hline
$\bullet$ & $L_{15/1}$ & $\bullet$ \\ \hline
$k$ & $I_k$ & $|I_k|$ \\ \hline
3 & 1.000000000 -  i 0.000000175 & 1.000000000 \\ \hline
4 & 0.000000021 +  i 1.732050808 & 1.732050808 \\ \hline
5 & -2.665351925 +  i 1.936491953 & 3.294556414 \\ \hline
6 & 3.000000303 +  i 3.464101353 & 4.582575695 \\ \hline
7 & -0.000000165 +  i 1.732050808 & 1.732050808 \\ \hline
8 & -1.732050808 +  i 0.000000010 & 1.732050808 \\ \hline
9 & -0.907604426 -  i 11.866568847 & 11.901226911 \\ \hline
10 & -5.959909504 -  i 18.342712091 & 19.286669182 \\ \hline
11 & -17.245203220 +  i 14.943053456 & 22.818673947 \\ \hline
12 & -11.196152781 -  i 8.196151933 & 13.875544804 \\ \hline
13 & 0.000000084 -  i 15.347547346 & 15.347547346 \\ \hline
14 & 0.000000083 -  i 1.732050808 & 1.732050808 \\ \hline
15 & -101.423006915 -  i 32.954328677 & 106.642459228 \\ \hline
16 & -1.224744868 +  i 1.224744875 & 1.732050808 \\ \hline
17 & -20.645987906 +  i 15.591127116 & 25.871607244 \\ \hline
18 & -40.127945922 -  i 30.808776018 & 50.590836361 \\ \hline
19 & 65.239497521 -  i 83.819706393 & 106.216454548 \\ \hline
20 & -115.811330427 -  i 84.141852139 & 143.150674244 \\ \hline
21 & -53.061367118 -  i 135.569663547 & 145.583798394 \\ \hline
22 & 36.010672960 +  i 16.445523443 & 39.588177633 \\ \hline
23 & -40.070955963 -  i 2.740927872 & 40.164588848 \\ \hline
24 & 164.290883844 -  i 129.064834808 & 208.923972053 \\ \hline
25 & 274.923532195 -  i 34.730921473 & 277.108616721 \\ \hline
26 & 0.000000762 -  i 277.056941014 & 277.056941014 \\ \hline
27 & 152.288761117 +  i 30.720691710 & 155.356453557 \\ \hline
28 & -0.000003253 +  i 136.433611353 & 136.433611353 \\ \hline
29 & -4.270422302 +  i 12.674160517 & 13.374260781 \\ \hline
30 & 485.145409358 +  i 667.745345688 & 825.378649414 \\ \hline
31 & 13.416605095 +  i 0.680413518 & 13.433847358 \\ \hline
32 & 164.960256203 +  i 68.328775084 & 178.551694562 \\ \hline
33 & -82.729230504 +  i 287.059281883 & 298.742626511 \\ \hline
34 & -538.981501170 -  i 268.380890132 & 602.104111256 \\ \hline
35 & -415.659227492 +  i 629.696787632 & 754.513510650 \\ \hline
36 & -681.771583909 +  i 223.335013764 & 717.419696551 \\ \hline
37 & 101.630605413 -  i 150.367371847 & 181.491395038 \\ \hline
38 & -59.482939675 +  i 173.267881065 & 183.193828283 \\ \hline
39 & -279.481438371 -  i 847.784737546 & 892.663898458 \\ \hline
40 & 347.379313920 -  i 1069.123643311 & 1124.143119192 \\ \hline
41 & -941.842709163 -  i 512.157840150 & 1072.088308877 \\ \hline
42 & 399.545316356 -  i 429.453517605 & 586.572061733 \\ \hline
43 & 390.071778264 +  i 279.072428091 & 479.622155784 \\ \hline
44 & 24.267771316 +  i 37.761389387 & 44.887049949 \\ \hline
45 & 2753.539308039 +  i 289.408610795 & 2768.706568944 \\ \hline
46 & 32.920210846 -  i 30.745331135 & 45.044596443 \\ \hline
47 & 536.002877882 -  i 206.437637985 & 574.382784800 \\ \hline
48 & 394.055675294 +  i 817.737030229 & 907.729985094 \\ \hline
49 & -1754.712300120 +  i 400.501606784 & 1799.837990828 \\ \hline
50 & 137.556258644 +  i 2186.393963508 & 2190.716843400 \\ \hline
\end{tabular}
}
\vskip 0.5truecm
\centerline{\bf Table 3}

\pagebreak
\null \vskip 1truecm
\centerline{
\begin{tabular}{|c|c|c|}
\hline
$\bullet$ & $L_{15/2}$ & $\bullet$ \\ \hline
$k$ & $I_k$ & $|I_k|$ \\ \hline
3 & 1.000000000 +  i 0.000000000 & 1.000000000 \\ \hline
4 & 0.000000000 -  i 1.732050808 & 1.732050808 \\ \hline
5 & 0.000000000 +  i 0.000000000 & 0.000000000 \\ \hline
6 & 3.000000000 -  i 3.464101615 & 4.582575695 \\ \hline
7 & 0.000000000 -  i 1.732050808 & 1.732050808 \\ \hline
8 & 1.732050808 +  i 0.000000000 & 1.732050808 \\ \hline
9 & 10.730551990 -  i 5.147276559 & 11.901226911 \\ \hline
10 & 0.000000000 +  i 0.000000000 & 0.000000000 \\ \hline
11 & -12.336706536 +  i 19.196290072 & 22.818673947 \\ \hline
12 & -11.196152423 +  i 8.196152423 & 13.875544804 \\ \hline
13 & 14.350206054 -  i 5.442315293 & 15.347547346 \\ \hline
14 & 0.000000000 +  i 1.732050808 & 1.732050808 \\ \hline
15 & 0.000000000 +  i 0.000000000 & 0.000000000 \\ \hline
16 & 1.224744871 -  i 1.224744871 & 1.732050808 \\ \hline
17 & 9.345902508 +  i 24.124555285 & 25.871607244 \\ \hline
18 & -6.617211192 -  i 50.156208387 & 50.590836361 \\ \hline
19 & 50.553444640 -  i 93.414583721 & 106.216454548 \\ \hline
20 & 0.000000000 +  i 0.000000000 & 0.000000000 \\ \hline
21 & -53.061366041 +  i 135.569663969 & 145.583798394 \\ \hline
22 & -11.153278505 +  i 37.984578277 & 39.588177633 \\ \hline
23 & -29.353726021 -  i 27.414466365 & 40.164588848 \\ \hline
24 & -164.290886665 -  i 129.064831217 & 208.923972053 \\ \hline
25 & 0.000000000 +  i 0.000000000 & 0.000000000 \\ \hline
26 & 228.013392388 +  i 157.386281027 & 277.056941014 \\ \hline
27 & 56.698638283 +  i 144.640561664 & 155.356453557 \\ \hline
28 & 0.000000000 -  i 136.433611353 & 136.433611353 \\ \hline
29 & 11.816320486 +  i 6.264616638 & 13.374260781 \\ \hline
30 & 0.000000000 +  i 0.000000000 & 0.000000000 \\ \hline
31 & -1.359079795 -  i 13.364922631 & 13.433847358 \\ \hline
32 & 164.960256101 +  i 68.328775330 & 178.551694562 \\ \hline
33 & -224.792217910 -  i 196.762841162 & 298.742626511 \\ \hline
34 & -110.636340119 -  i 591.852144574 & 602.104111256 \\ \hline
35 & 0.000000000 +  i 0.000000000 & 0.000000000 \\ \hline
36 & 147.472006710 +  i 702.099015977 & 717.419696551 \\ \hline
37 & 45.731849880 +  i 175.635202563 & 181.491395038 \\ \hline
38 & -177.588145856 -  i 44.971426177 & 183.193828283 \\ \hline
39 & -875.291194999 -  i 175.254556482 & 892.663898458 \\ \hline
40 & 0.000000000 +  i 0.000000000 & 0.000000000 \\ \hline
41 & 1052.478012596 +  i 204.116082250 & 1072.088308877 \\ \hline
42 & 399.545318063 +  i 429.453516017 & 586.572061733 \\ \hline
43 & -171.335849479 -  i 447.974819607 & 479.622155784 \\ \hline
44 & 44.430164502 +  i 6.388093254 & 44.887049949 \\ \hline
45 & 0.000000000 +  i 0.000000000 & 0.000000000 \\ \hline
46 & -17.945816315 -  i 41.315412930 & 45.044596443 \\ \hline
47 & 571.498128245 +  i 57.493242102 & 574.382784800 \\ \hline
48 & -817.737034535 -  i 394.055666358 & 907.729985094 \\ \hline
49 & -780.920437266 -  i 1621.597997004 & 1799.837990828 \\ \hline
50 & 0.000000000 +  i 0.000000000 & 0.000000000 \\ \hline
\end{tabular}
}
\vskip 0.5truecm
\centerline{\bf Table 4}
\pagebreak

\null \vskip 1truecm
\centerline{
\begin{tabular}{|c|c|c|}
\hline
$\bullet$ & $L_{15/4}$ & $\bullet$ \\ \hline
$k$ & $I_k$ & $|I_k|$ \\ \hline
3 & 1.000000000 +  i 0.000000000 & 1.000000000 \\ \hline
4 & 0.000000000 +  i 1.732050808 & 1.732050808 \\ \hline
5 & -1.018073921 - i 3.133309346 & 3.294556414 \\ \hline
6 & 3.000000000 +  i 3.464101615 & 4.582575695 \\ \hline
7 & -0.751508681 -  i 1.560523855 & 1.732050808 \\ \hline
8 & -1.732050808 +  i 0.000000000 & 1.732050808 \\ \hline
9 & 10.730551990 +  i 5.147276559 & 11.901226911 \\ \hline
10 & -15.603243133 -  i 11.336419711 & 19.286669182 \\ \hline
11 & 17.245203123 +  i 14.943053568 & 22.818673947 \\ \hline
12 & -11.196152423 -  i 8.196152423 & 13.875544804 \\ \hline
13 & 3.672908488 +  i 14.901575513 & 15.347547346 \\ \hline
14 & -1.688624678 +  i 0.385417563 & 1.732050808 \\ \hline
15 & 0.000000000 +  i 0.000000000 & 0.000000000 \\ \hline
16 & -1.224744871 +  i 1.224744871 & 1.732050808 \\ \hline
17 & -17.429589094 -  i 19.119348456 & 25.871607244 \\ \hline
18 & -6.617211192 +  i 50.156208387 & 50.590836361 \\ \hline
19 & 0.000000000 -  i 106.216454548 & 106.216454548 \\ \hline
20 & -44.235991098 +  i 136.144381552 & 143.150674244 \\ \hline
21 & 72.909410760 -  i 126.011349400 & 145.583798394 \\ \hline
22 & -36.010673013 +  i 16.445523326 & 39.588177633 \\ \hline
23 & -10.836276386 +  i 38.675176941 & 40.164588848 \\ \hline
24 & 164.290886665 -  i 129.064831217 & 208.923972053 \\ \hline
25 & -257.649075864 +  i 102.010485575 & 277.108616721 \\ \hline
26 & 275.036883975 -  i 33.395523912 & 277.056941014 \\ \hline
27 & -153.611719961 +  i 23.217819719 & 155.356453557 \\ \hline
28 & 106.668092622 +  i 85.064965309 & 136.433611353 \\ \hline
29 & -9.197471902 +  i 9.709653035 & 13.374260781 \\ \hline
30 & 0.000000000 +  i 0.000000000 & 0.000000000 \\ \hline
31 & -4.021598499 +  i 12.817761129 & 13.433847358 \\ \hline
32 & -164.960256101 -  i 68.328775330 & 178.551694562 \\ \hline
33 & 85.599713692 +  i 286.216431937 & 298.742626511 \\ \hline
34 & -217.505092368 -  i 561.445362956 & 602.104111256 \\ \hline
35 & 33.851120653 +  i 753.753765751 & 754.513510650 \\ \hline
36 & 147.472006710 -  i 702.099015977 & 717.419696551 \\ \hline
37 & -136.240563896 +  i 119.906777214 & 181.491395038 \\ \hline
38 & 0.000000000 +  i 183.193828283 & 183.193828283 \\ \hline
39 & 538.949599873 -  i 711.605343156 & 892.663898458 \\ \hline
40 & -909.450887536 +  i 660.754746927 & 1124.143119192 \\ \hline
41 & 1008.985242455 -  i 362.383943545 & 1072.088308877 \\ \hline
42 & -546.310790198 +  i 213.568031595 & 586.572061733 \\ \hline
43 & 426.548198677 +  i 219.303548819 & 479.622155784 \\ \hline
44 & -24.267771377 +  i 37.761389348 & 44.887049949 \\ \hline
45 & 0.000000000 +  i 0.000000000 & 0.000000000 \\ \hline
46 & -6.133571758 +  i 44.625048641 & 45.044596443 \\ \hline
47 & -558.735830232 -  i 133.153503483 & 574.382784800 \\ \hline
48 & 394.055666358 +  i 817.737034536 & 907.729985094 \\ \hline
49 & -883.212092863 -  i 1568.232505801 & 1799.837990828 \\ \hline
50 & 410.499402001 +  i 2151.913225229 & 2190.716843400 \\ \hline
\end{tabular}
}
\vskip 0.5truecm
\centerline{\bf Table 5}
\end{appendix}


\begin{thebibliography}{99}
\bibitem{swa} A.S.~Schwarz, Lett. Math. Phys. 2 (1978) 247.
\bibitem{wittf} E.~Witten, J. Diff. Geom. 17 (1982) 661.
\bibitem{wittd}  E.~Witten, Comm. Math. Phys. 117 (1988) 353.
\bibitem{blau} D.~Birmingham, M.~Blau,M.~Rakowski and G.~Thompson, Phys. Rep. 4 \& 5 (1991) 129.
\bibitem{dij} R.~Dijkgraaf, {\em Les Houches lectures on fields, strings and duality}
, hep-th/9703136.
\bibitem{atw} M.F.~Atiyah, {\em New invariants of three- and four-dimensional 
manifolds}, in {\em The mathematical heritage of Hermann Weyl}, Proc. Symp. Pure
Math. 48, ed. R.~Wells; Amer. Math. Soc., 1988.
\bibitem{sch1} A.S.~Schwarz, {\em New topological invariants arising in the 
theory of quantized fields} in Abstracts of the International Topological
Conference, Baku 1997. 
\bibitem{dona} S.~Donaldson, J. Diff. Geom. 18 (1983) 269.
\bibitem{jon} V.F.R.~Jones, Bull. Amer. Math. Soc. 12 (1985) 103; 

\no
Ann. Math 126 (1987) 335.
\bibitem{witt1} E.~Witten, Comm. Math. Phys. 121 (1989) 351.
\bibitem{ata}  M.F.~Atiyah, Pubbl. Math. I.H.E.S. 68 (1988) 175.
\bibitem{tur} V.G. Turaev, {\em Quantum invariants of 3-manifolds}, Walter de Gruyter Studies in Mathematics 18 (Berlin 1994).
\bibitem{brs} C.~Becchi, A~Rouet and R.~Stora, Comm. Math. Phys. 42 (1975) 127; Ann. Phys. (NY) 98,287 (1976).
\bibitem{gm1} E.~Guadagnini, M.~Martellini and M.~Mintchev, Phys. Lett. B 227 
(1989) 111.
\bibitem{gau} L.~Alvarez-Gaum\`e, J.M.F.~Labastida and A.V.~Ramallo, Nucl. Phys.
B 334 (1990) 103.
\bibitem{del}  F.~Delduc, O.~Piguet, C.~Lucchesi and S.P.~Sorellla, Nucl. Phys.
 B 346 (1990) 313.
\bibitem {pol} A.M.~Polyakov, Mod. Phys. Lett.  A 3 (1988) 325. 
\bibitem{gm2} E.~Guadagnini, M.~Martellini and M.~Mintchev,  Nucl. Phys. B330 
(1990) 575.
\bibitem{guad1} E.~Guadagnini, Int. Journ. Mod. Phys. A7 (1992) 877.    
\bibitem{choq} Y.~Choqhet-Bruhat, C.~DeWitt-Morette and M.~Dillard-Bleick,
{\em Analysis, Manifolds and Physics, part1}; North-Holland, 1989.
\bibitem{bott} R.~Bott, Bull. Soc. Math. France, 84 (1956) 251. 
\bibitem{djt} G.V.~Dunne, R.~Jackiw and C.A.~Trugenberger, Ann. Phys. 194 
(1989) 197.  
\bibitem{rol} D.~Rolfsen, {\em Knots and Links}, Publish or Perish, 1976.
\bibitem{kau} L.H.~Kauffman, {\em On Knots}, Princeton Univ. Press, 1987.    
\bibitem{kel} W.H.~Thomson, Trans. R. Soc. Edin. 25 (1869) 217.
\bibitem{art} E.~Artin, Ann. of Math. 48 (1947) 101
\bibitem{mark} J~Birman, {\em Braids, Links and Mapping Class Group}, Ann. 
Math. Stud. 82 (1974);

Can. J. Math. 28 (1976) 264.

\bibitem{gm3} P.~Cotta-Ramusino, E.~Guadagnini, M.~Martellini and M.~Mintchev,
Nucl. Phys. B330 (1990) 557.
\bibitem{can}E.~Guadagnini, M.~Martellini and M.~Mintchev, Nucl. Phys. B336 (1990) 581; 

\no
Nucl. Phys. B18 (Proc. Suppl.) (1990) 121.
\bibitem{guad2}  E.~Guadagnini, Phys. Lett. B 260 (1991) 353. 
\bibitem{glib}  E.~Guadagnini, {\em The Link Invariant of the Chern-Simons
Field Theory}, {\em De Gruyter Exposition in Mathematics}, Vol. 10 (de Gruyter,
Berlin, 1992).
\bibitem{seil} E.~Seiler, {\em Gauge Theories as a Problem of Constructive Quantum
Field Theory and Statistical Mechanics}; Lecture Notes in Physics N. 159, Springer-Verlag 1982.
\bibitem{gp1} E.~Guadagnini and L.~Pilo, J. Geom. Phys. 14 (1994) 236. 
\bibitem{lic} W.B.R.~Lickorish and K.C. Millet, Topology 26 (1987) 107. 
\bibitem{io} L.~Pilo, {\em Teoria di Chern-Simons con gruppo di gauge SU(3) ed
invarianti di 3-variet\`a}, Thesis (1993). 
\bibitem{gp2} E.~Guadagnini and L.~Pilo , J. Geom. Phys. 14 (1994) 365;

\no Phys. Lett. B 319 (1993) 139.
\bibitem{guad3} E.~Guadagnini, Nucl. Phys. B375 (1992) 381;

\no
E.~Guadagnini and  S.~Panicucci, Nucl. Phys. B388 (1992) 159.
\bibitem{kir} R.~Kirby Invent. Math. 45 (1978) 35; 

\no
R.~Fenn and C.~Rourke, Topology 18 (1979) 1;

\no
D.~Rolfsen, Pacific J. Math. 110 (1984) 377.
\bibitem{lick}W.B.R.~Lickorish, Ann. Math. 76 (1962) 531.
\bibitem{deu} C.L.~Sigel, Nachr. Akad. Wiss. G\"ottingen Math. Phys. Kl. 1
(1960) 1.
\bibitem{km} R.~Kirby and P.~Melvin, Invent. Math. 105 (1991) 473.
\bibitem{lick1}W.B.R.~Lickorish, Pacific J. Math. 149 (1991) 337.
\bibitem{mor} H.R.~Morton and P.M.~Strickland, {\em Satellites and Surgery Invariants},
in  Knots 90, edit. A.~Kawauchi, (Walter de Gruyter, Berlin 1992). 
\bibitem{retu} N.Y.~Reshetikhin and V.G.~Turaev, Invent. Math. 103  (1991) 547.
\bibitem{tuvi}V.G.~Turaev and O.Y.~Viro, Topology 31 (1992) 865.
\bibitem{guad4} E.~Guadagnini, Nucl. Phys. B404 (1993) 385.
\bibitem{kau1} L.H.~Kauffman and L.L.~Sostenes {\sl Temperley-Lieb recoupling theory and invariants of
3-manifolds}, Princeton University Press (Princeton, 1994).
\bibitem{koh} T. Kohno, Topology 31 (1992) 203. 
\bibitem{koh1}T. Kohno, Ann. Inst. Fourier, Grenoble 37 (1987) 139. 
\bibitem{fregomp} D.S. Freed and R.E. Gompf, Commun. Math. Phys. 14X (1991) 
79. 
\bibitem{jef} L. C. Jeffrey, Commun. Math Phys. 147 (1992) 563
\bibitem{gp4} E.~Guadagnini and L.~Pilo, HEP-TH 9612090, to be published in 
Comm. Math. Phys.
\bibitem{cin} Hua~Loo-Keng, {\em Introduction to number theory\/}, Springer-Verlag 
(New York, 1982). 
\bibitem{arwil} F.~Archer and R.~Williams, Phys. Lett. B 273 (1991) 438. 
\bibitem{tom} E. Guadagnini and P. Tomassini, Phys. Lett. B 336 (1994) 330.
\bibitem{gp3} E.~Guadagnini and L.~Pilo, Nucl. Phys. B433, 597 (1995). 
\bibitem{wit} E. Witten, Commun. Math. Phys. 121 (1989) 351.
\bibitem{wzm} S.P. Novikov, Usp. Mat. Nauk 37 (1982) 3;

\no
A.M. Polyakov and P.B. Wiegmann, Phys. Lett. B131 (1983)  121;

\no
E. Witten, Commun. Math. Phys. 92 (1984) 455. 

\bibitem{kiz} V.G. Knizhnik and A.B. Zamolodchikov, Nucl. Phys. B247 
(1984) 83.
\bibitem{apz} A.A. Belavin, A.M. Polyakov, A.B. Zamolodchikov Nucl. Phys. B241 
(1984) 333. 
\bibitem{ver} E. Verlinde, Nucl. Phys. B300 (1988) 360. 
\bibitem{frd} D. Friedan and S.H. Shenker, Nucl. Phys. B281 (1987) 509.
\bibitem{mos} G. Moore and N. Seiberg, Commun. Math. Phys. 123 (1989) 177. 
\bibitem{drin} V.G.~Drinfeld, {\em Quasi-Hopf Algebras and
Knizhnik-Zamolodchikov Equations}, in  Problems o f Modern Quantum Field Theory,
ed. A.A.~Belavin,  A.V.~Klinyk and  A.B.~Zamolodchikov (Springer-Verlag, 1990).
\bibitem{dms} P.~Di~Francesco, P.~Mathieu and D.~Senechal, {\em Conformal Field 
Theory}, Springer-Verlag, 1997.
\bibitem{itz}C.~Itzykson and J.-M.~Drouffe, {\em Statistical Field Theory} Vol.2, Cambridge University Press.
\bibitem{gin}P.~Ginsparg, {\em Applied Conformal Field Theory}, in {\em Les Houches,
session XLIX, 1988,  Fields, strings and critical phenomena}, Eds. E.~Br\`ezin and J. Zinn-Justin, Elsevier 1989.
\bibitem{card} J.L.~Cardy, {\em Conformal Invariance and Statistical Mechanics},in {\em Les Houches,
session XLIX,1988,  Fields, strings and critical phenomena}, Eds. E.~Br\`ezin and J. Zinn-Justin, Elsevier 1989.ty Press 1989.
\bibitem{gua2d} E.~Guadagnini, Phys. Rev. D38 (1988) 2482.
\bibitem{fro} J.~Frohlich and C.~King Comm. Math. Phys. 126 (1989) 167 ;

\no
Int. Jour. Mod. Phys. A4 (1989) 5321.
\bibitem{atb} M.F.~Atiyah and R.~Bott, Phil. Trans. R. Soc. Lond. A308 (1982) 523.
\bibitem{wittc}E.~Witten, {\em The central charge in three dimensions}, IASSNS-HEP-89/38, June, 1989.
\bibitem{atia} M.F~Atiyah, Topology 29 (1990) 1.
\end{thebibliography}
\end{document}